\documentclass[prd,twocolumn,nofootinbib]{revtex4-1}

\usepackage{graphicx,verbatim,layout,textcomp}
\usepackage{subfig,mathrsfs}
\usepackage[fleqn]{amsmath}
\usepackage{placeins}
\usepackage{afterpage}
\usepackage{morefloats}
\usepackage{soul}
\usepackage{color}
\usepackage{hyperref}

\begin{document}

\newcommand{\twoF}{{2\cal{F}}}
\newcommand{\F}{{\cal{F}}}
\newcommand{\avgTwoF}{{\widehat{\twoF}}}
\newcommand{\avgF}{{\widehat{\F}}}
\def\mapcomment#1{{\bf{\color{cyan}#1}}}

\title{Optimal directed searches for continuous gravitational waves}

\author{Jing Ming, Badri Krishnan, Maria Alessandra Papa, Carsten Aulbert, Henning Fehrmann}

\affiliation{Max-Planck-Institute for Gravitational Physics, Albert
  Einstein Institute, Callinstrasse 38, 30167 Hannover}

\begin{abstract}
  Wide parameter space searches for long lived continuous
  gravitational wave signals are computationally limited.  It is
  therefore critically important that available computational
  resources are used rationally.  In this paper we consider directed
  searches, i.e. targets for which the sky position is known
  accurately but the frequency and spindown parameters are completely
  unknown.  Given a list of such potential astrophysical targets, we
  therefore need to prioritize.  On which target(s) should we spend
  scarce computing resources? What parameter space region in frequency
  and spindown should we search?  Finally, what is the optimal search
  set-up that we should use?  In this paper we present a general framework that allows to solve 
  all three of these problems. This framework is based on maximizing
  the probability of making a detection subject to a constraint on the
  maximum available computational cost. We illustrate the method for a simplified problem.
\end{abstract}

\maketitle

\section{Introduction}
\label{sec:intro}

A rapidly rotating non-axially symmetric neutron star is expected to
emit long lived periodic gravitational waves (GW) signals, also known
as ``continuous waves'' (CW). These signals could be
detectable by the second generation of ground based GW
observatories such as LIGO \cite{ligoref,ligoref3}, Virgo
\cite{virgo2}, GEO600 \cite{Geo}, KAGRA \cite{kagraref} and LIGO-India
\cite{ligoindia}.  These observatories are sensitive over a broad
frequency range, typically $\mathcal{O}(10)$-$\mathcal{O}(10^3)\,$Hz
which means that the neutron star needs to be rotating fairly rapidly.
The CW signal is parametrized by the sky-position of the neutron star,
a frequency $f$ and its time derivatives (the spindown
parameters) $\dot{f},\ddot{f}\ldots$, all measured at some fiducial
reference time.  If the neutron star is in a binary system, we also
need to consider the orbital parameters of the binary, though in this
paper we focus on isolated systems.

There are a number of interesting astrophysical targets for CW searches. The
known pulsars are particularly good examples for which we know all of
the afore-mentioned parameters.  Such searches, where the
sky-position, frequency and spindown are all known accurately, are
referred to as \emph{targeted} searches in the literature. In these cases it is fairly
straightforward, at least from a computational point of view, to
search for possible GW signals emitted by the neutron
star.  A number of such searches have been carried out (see
e.g. \cite{Aasi:2013sia}).  Most notably,
for the Crab and Vela pulsars, the upper-limit
on the GW amplitude is more constraining than the limit one derives by
assuming that all of the observed spindown is due to GW emission
\cite{Abbott:2008fx,Abadie:2011md,Aasi:2014jln}. 

At the other extreme we have the \emph{blind} searches where nothing
is known \textit{a priori} about the source parameters.  One has to
survey a data set which could span several months or years, a
frequency range of $\mathcal{O}(10^3)\,$Hz, the entire sky and a
reasonable choice of spindown parameters.  Such searches are
computationally limited and are, by far, the most computationally
challenging GW searches of all.  Results from a number
of such searches have been published (see
e.g. \cite{S2ScoX1,S2Hough,S4PSH,S5Powerflux2009,S5Powerflux2011,
  EatHS4R2,EatHS5R1,Abbott:2008uq}).  Some of these results
\cite{EatHS4R2,EatHS5R1,Abbott:2008uq}) have utilized the public
distributed computing project Einstein@Home \cite{Einstweb}.

Between these two extremes lie the \emph{directed} searches where one targets 
interesting astrophysical objects or regions. In this case the signals parameters 
are partially known.  In particular the
sky-position is known accurately but no information is available
on the spin frequency of the star and hence the GW frequency.  
Such searches are also computationally limited. 
A few such results have been published in the
literature so far : a search for CW signals from the
supernova remnant Cassiopeia A \cite{S5CasA}, from the
galactic center \cite{Aasi:2013jya} which could potentially harbor a
number of young and rapidly rotating neutron stars and from nine young supernova remnants \cite{owen2014}.  
A deep search for CW signals from Cassiopeia A (Cas A)using Einstein@Home was completed last year.  
All of these searches were computationally limited.  We
expect that similar directed searches will be of great interest in the
near future.

When the searches are computationally limited (the directed and blind
searches), the most commonly employed methods are
\emph{semi-coherent}: rather that matching the full $\sim$year long
data set with coherent signal templates, one splits up the data set
into $N$ shorter segments (stacks) typically $\sim$hours or $\sim$days long.
Each segment is matched with a set of signal templates coherently and
finally the results of these $N$ searches are combined incoherently.
Descriptions of such methods can be found in
\cite{ForMetricExprRes,HierarchP1,HierarchP2,Papa:2000wg,HoughP2,
  HierarchP3,Pletsch:2009uu,GlobCorr}.  Semi-coherent methods have
also been considered as parts of multi-stage hierarchical schemes for
surveying large parameter spaces, see for example
\cite{HierarchP3,HierarchP1,Shaltev:2014toa}.

It is very important to spend the computational resources wisely: what
search set-up to use, what astrophysical objects to target and for
each target what waveforms to search can make the difference between
making a detection or missing it.  

In this paper we shall focus on the directed searches though the
general scheme we propose is also applicable to the blind searches.
We assume that we have a list of $N_\mathrm{t}$ potential targets.  For each of
the targets we assume that we know how far and old it is.  We also
make assumptions on the likelihood for different values of the signal
amplitude, its frequency and the frequency-derivates, as discussed in Sec.~\ref{subsec:astroprior}.
Given these priors, the question we address
is: what sources should we target?  What is the optimal search set-up
and what is the search region in frequency and spindown that maximizes
our probability of making a detection?  Various parts of this problem
have been partially addressed previously and here we present a
complete solution. We explain how the key search pipeline parameters
can be determined, which source and which part of parameter space to
target taking into account the prior astrophysical knowledge, the
performance of our search software, the available computational
resources and the quality of the data from the GW
detectors.

Previous works have typically fixed the parameter space to be searched \emph{a priori} 
-- often based  on reasonable astrophysical arguments -- and then optimized for the search 
parameters (e.g. \cite{BC00,HierarchP3,Prix:2012yu}).  Conversely there are 
studies of what parameter space to search (see
e.g. \cite{Palomba:2005fa,Knispel:2008ue,Wade:2012qc,Owen:2009tj})
which have largely neglected the computational cost of the GW search.  One of
the aims of the present work is to integrate both aspects of the problem.

The plan for the rest of this paper is as follows. We start with
a review of the expected GW signal and search methods
in Sec.~\ref{sec:cw}.  The general scheme for optimizing the detection
probability is explained in Sec.~\ref{sec:ranking}.  Finally we 
illustrate the general scheme with specific examples in
Sec.~\ref{sec:examples} and present the application results of this scheme in Sec.~\ref{sec:Application}.

\section{The expected GW signal, astrophysical targets and search methods}
\label{sec:cw}

\subsection{The gravitational waveform}
\label{subsec:waveform}

We summarize the GW signal waveform from a rapidly rotating
neutron star; details can be found in \cite{JKSPaper}.  In the rest
frame of the neutron star, the GW signal is
elliptically polarized with constant amplitudes $A_{+,\times}$ for the
two polarizations $h_{+, \times}(t)$. Thus, we can find a frame in
the plane transverse to the direction of propagation such that
\begin{equation}
  \label{eq:h+x} h_+(t) = A_+ \cos\phi(t) \,,\qquad h_\times(t) =
A_\times\sin\phi(t)\,.
\end{equation} 
The two amplitudes are related to an overall amplitude $h_0$ and the
inclination angle $\iota$ between the line of sight from Earth to the
neutron star's rotation axis:
\begin{equation}
  \label{eq:A+x} 
  A_+ = \frac{1}{2}h_0(1+\cos^2\iota)\,, \qquad
  A_\times = h_0\cos\iota\,.
\end{equation} 
Numerous mechanisms can cause the GW frequency to
change.  These include energy loss due to the emission of
gravitational radiation, electromagnetic interactions, local
acceleration of the source and accretion for neutron stars which have
a companion star.  The spin frequency of the neutron star is assumed
to vary slowly and smoothly with time for the observation duration.
This assumption may not hold if the neutron star glitches, but we
shall not consider this complication here.  With this assumption, it
is useful to expand the frequency evolution in a Taylor series
expansion
\begin{equation}
  \label{eq:fhat} \hat{f}(\tau) = f + \dot{f}(\tau-\tau_0) + \frac{1}{2}\ddot{f}(\tau-\tau_0)^2 + \ldots\,,
\end{equation} 
where $\tau$ is the arrival time of a wavefront at the solar system
barycenter (SSB), $f$ is the frequency at a fiducial reference time
$\tau_0$, and $(\dot{f},\ddot{f})$ denote the first and second time
derivatives of the frequency at $\tau_0$. In this paper we shall
assume that the frequency change is sufficiently small that we will
not have to consider any terms beyond $\ddot{f}$.  This should suffice
for almost all plausible CW sources over the relevant observation
times.

As the detector on the Earth moves relative to the SSB, the arrival
time of a wavefront at the detector, $t$, differs from the SSB time
$\tau$\footnote{Proper motion of the source can safely be neglected
  for distances greater than $\sim 10$~pc.}:
\begin{equation}
  \label{eq:tau(t)} \tau(t) = t + \frac{\mathbf{r}(t) \cdot \mathbf{n}}{c} +
\Delta_{\mathrm{E}\odot} - \Delta_{\mathrm{S}\odot}\,.
\end{equation} 
Here $\mathbf{r}(t)$ is the position vector of the detector in the SSB
frame, $\mathbf{n}$ is the unit vector pointing to the neutron star,
and $c$ is the speed of light; $\Delta_{\mathrm{E}\odot}$ and $\Delta_{\mathrm{S}\odot}$
are respectively the relativistic Einstein and Shapiro time
delays. 

The phase of the signal as observed at the detector, $\phi(t)$, is the 
same as the phase observed at the source, $\varphi(\tau)$, at the corresponding time:
$\varphi(\tau)=\varphi(\tau(t))=\phi(t)$. 
Thus, except for an initial phase $\phi_0$, $\phi(t)$ depends only on
the sky position $\mathbf{n}$, and on $(f_0,\dot{f},\ddot{f})$. For this
reason $(\mathbf{n}, f,\dot{f},\ddot{f})$ are called the {\it {phase
    evolution parameters}}.

The received signal at the detector is
\begin{equation}
  \label{Eq:h(t)Signal} h(t) = F_+(t;\mathbf{n},\psi)h_+(t)
  + F_\times(t;\mathbf{n},\psi)h_\times(t),
\end{equation} 
where $F_{+,\times}$ are the detector beam pattern functions which
depend on the sky position $\mathbf{n}$ and on the polarization angle
$\psi$.  It is often useful to rewrite the signal as \cite{JKSPaper}:
\begin{equation}
  \label{eq:13}
  h_i(t) = \sum_{\mu=1}^4 A_\mu h_i^\mu(t)\,.
\end{equation}
with the index $i$ running over the different detectors whose data we
are considering. The four amplitudes $A_\mu$ depend only on
$(h_0,\iota,\psi,\phi_0)$ which, for this reason are often referred-to
as the {\it {amplitude parameters}}.  The four detector dependent
signals $h^\mu_i(t)$ depend on the phase evolution parameters.

\subsection{The expected GW amplitude}
\label{subsec:gwamplitude}

The value of $h_0$ that we expect depends on the emission mechanism
that we consider.  We refer to \cite{S2ScoX1} and references therein
for a review of emission mechanisms.  The most basic estimate of the
largest amplitude that we can expect ($h_0^\mathrm{sd}$ of
Eq.~(\ref{eq:spindownh0}) below) is given by energy conservation
arguments which we now briefly summarize.

Neutron stars typically spindown and thus lose their rotational
kinetic energy.  Most of this energy loss is due to electromagnetic
interactions.  The simplest models (see e.g. \cite{ShapiroTeukolsky})
take the neutron star to be a rotating magnetic dipole $\mathbf{m}$
which is mis-aligned with the rotation axis.  The system then loses
energy at a rate proportional to $|\ddot{\mathbf{m}}|^2$.  Consider
instead a hypothetical star, for which this energy is carried away
entirely in gravitational radiation.  Let $f$ be the instantaneous
frequency of the emitted GW signal, $G$ Newton's constant, and $D$ the
distance to the star.  Setting the loss in rotational kinetic energy
to the energy carried away by GWs leads to a limit on $h_0$ known as
the spindown limit $h_0^\mathrm{sd}$.  GWs carry energy away at a rate
$\dot{E}_\mathrm{gw}$ given by
\begin{equation}
  \label{eq:17}
  \langle\dot{E}_\mathrm{gw}\rangle = -\frac{c^3}{16\pi G} \oint_{S}\langle \dot{h}_+^2 + \dot{h}_\times^2 \rangle dS\,.
\end{equation}
Here $S$ is a large sphere of radius $D$ centered at the neutron star,
$dS$ is the area element on this sphere, and the brackets
$\langle\cdot\rangle$ denote a time average over a sufficiently large
number of GW cycles.  We can take the GW waveform given in
Eq.~(\ref{eq:h+x}) and calculate $\dot{E}_\mathrm{gw}$; it is clear that
$\langle\dot{E}_\mathrm{gw}\rangle \propto f^2h_0^2D^2$. Since $h_0$ is
proportional to $1/D$, $\dot{E}_\mathrm{gw}$ is independent of $D$ as it
should be.  On the other hand, the rotational kinetic energy of the
star is $E_\mathrm{rot} = I\pi^2f^2/2 $ (we assume that the GW signal
frequency is twice the rotational frequency) so that $\dot{E}_\mathrm{rot} =
\pi If\dot{f}$.  Setting $\langle\dot{E}_\mathrm{gw}\rangle = \dot{E}_\mathrm{rot}$
and averaging over the sphere $S$, the GW amplitude $h_0^\mathrm{sd}$ can be
shown to be
\begin{equation}
  \label{eq:spindownh0}
  h_0^\mathrm{sd} = \frac{1}{D}\sqrt{\frac{5G I}{2c^3}\frac{|\dot{f}|}{f}}\\ ~~~~{\text{and}}~~~~
  h_0\le h_0^\mathrm{sd}\,. 
\end{equation}
The value $h_0^\mathrm{sd}$ is based on energy conservation, and is independent of the
actual mechanism which causes the neutron star to emit gravitational
radiation. This is thus an upper limit on $h_0$.  The actual amplitude
of the emitted GWs from isolated neutron stars is expected to be much
smaller and depends on the emission mechanism.  If we assume that not
more than a fraction $x$ of the spindown energy is carried away in
gravitational waves, then the corresponding limit is smaller by
a factor $\sqrt{x}$:
\begin{equation}
  \label{eq:10}
  h_0 \le  \frac{1}{D}\sqrt{x\cdot \frac{5GI}{2c^3}\frac{|\dot{f}|}{f}}\,.
\end{equation}
Observational limits on GW emission from the Crab and Vela pulsars
constrain $x$ to less than $1\%$ and $10\%$ respectively for these two
objects \cite{Aasi:2013sia}.

We concentrate on a particular emission mechanism, that due to the presence 
of non-axisymmetric distortions in the neutron star.  
The CW amplitude $h_0$ then depends on the ellipticity
$\varepsilon$ of the star defined as
\begin{equation} 
  \label{eq:ellipticitydef}
  \varepsilon = \frac{\left| I_{xx}-I_{yy}\right|}{I_{zz}}\,.
\end{equation} Here $I_{zz}$ is the principal moment of inertia of the
star, and $I_{xx}$ and $I_{yy}$ are the moments of inertia about the
other axes.  A straightforward application of Einstein's quadrupole 
formula yields:
\begin{equation}
  \label{eq:GWampl} h_0 = \frac{4\pi^2G}{c^4}\frac{I_{zz}f^2\varepsilon}{D}\,.
\end{equation} 
The distribution of $\varepsilon$ for neutron stars is uncertain. In
fact predictions exist for the maximum strain that a neutron star
crust can sustain before breaking according to various neutron star
models. However these predictions are only upper limits to the allowed
ellipticities rather than predictions of the actual ellipticity values
(see e.g. \cite{NonAxNS2,Horowitz,JohnsonMcDaniel:2012wg}).

We can combine Eqs.~(\ref{eq:spindownh0}) and (\ref{eq:GWampl}) to get 
the value of $\varepsilon$ required for emitting at the spindown limit:
\begin{equation}
  \label{eq:spindownEllipticity}
   {\varepsilon^\mathrm{sd}}=\sqrt{ \frac{5c^5}{32\pi^4G}\frac{|\dot{f}|}{If^5}}\,
\end{equation}
and correspondingly for emitting in GW a fraction $x$ of the spindown
energy:
\begin{equation}
  \label{eq:spindownEllipticityX}
  {\varepsilon^\mathrm{sd}_x}=\sqrt{ \frac{5c^5}{32\pi^4G}\frac{x|\dot{f}|}{If^5}}\,.
\end{equation}
As already pointed out, neutron star crusts cannot sustain deformations with arbitrarily high
values of $\varepsilon$: see e.g. \cite{NonAxNS2,Horowitz} for
discussions on the possible upper limit on $\varepsilon$.  It is
important that we take this into account as we plan our searches.

\subsection{Coherent and semi-coherent search methods}
\label{subsec:searchmethods}

We now turn to techniques for detecting the CW signals described
above. We assume $N_\mathrm{det}$ GW detectors labeled by an
integer $i=1\ldots N_\mathrm{det}$. We denote the calibrated strain data from
the $i^{th}$ detector as $x_i(t)$, the detector noise
by $n_i(t)$ and a possible GW signal by $h_i(t)$.  In the absence of
a signal $x_i(t)=n_i(t)$ and in the presence of a signal, $x_i(t) =
n_i(t)+h_i(t)$.  We shall assume that the noise in the detector is
Gaussian and stationary with zero mean.  The noise is then well
described by a power-spectral-density function (PSD), 
$S^{(i)}_n(f)$ for the $i^{th}$ detector.  $T_\mathrm{obs}$ is 
the total observation duration.

If computational cost were not an issue, the optimal technique for
detecting the CW signal would be matched filtering; i.e. correlating
the data streams $x_i(t)$ coherently with the expected signal $h_i(t)$.
As shown in the previous section, $h_i(t)$ depend on both the phase 
and amplitude parameters. However one can eliminate the explicit dependance 
on the amplitude parameters analytically either by maximizing 
(\cite{JKSPaper,MultiIfoFstat}) or by marginalizing \cite{Bstat,Whelan:2013xka} 
the coherent detection statistic with respect to these.
We use the maximization procedure.

The detection statistic that we obtain is known as the $\F$-statistic.  In Gaussian data the 
distribution of $2\F$ is a
$\chi^2$ distribution with $4$ degrees of freedom: $\chi^2_4(2\F|\rho^2)$. $\rho^2$ is
the non-centrality parameter determined by the two amplitudes
$A_{+,\times}$ and the detector sensitivity and orientation:
\begin{equation}
  \label{eq:1}
  \rho^2 = \sum_{i=1}^{N_\mathrm{det}}\rho^2_i = 4 \sum_{i=1}^{N_\mathrm{det}}\int df\frac{|\widetilde{h}_i(f)|^2}{S^{(i)}_n(f)} \,,
\end{equation}
$\widetilde{h}_i$ is the Fourier transform of the GW signal
$h_i(t)$ given in Eq.~(\ref{eq:13}).  Since the signal is narrow-band
in frequency it is reasonable to assume that the PSD is constant over
the frequency band of interest. We can take $S^{(i)}_n$ outside the
integral, evaluate it at the signal frequency $f_0$, and use Parseval's
identity to replace the integral over frequency by an integral over
time:
\begin{equation}
  \label{eq:noncentrality}
  \rho_i^2 = \frac{2}{S^{(i)}_n(f_0)}\int_{-T_\mathrm{obs}/2}^{T_\mathrm{obs}/2}h^2_i(t)\,dt\,.
\end{equation}
We have chosen the observation duration to be placed symmetrically
about $t=0$.  We can substitute $h(t)$ from Eqs.~(\ref{Eq:h(t)Signal})
and (\ref{eq:h+x}) to get the explicit dependence of $\rho^2_i$ on the
amplitude parameters: 
\begin{equation}
  \rho_i^2 = \frac{2h_0^2T_\mathrm{obs}}{S_n^{(i)}(f_0)}\left[ \frac{(1+\cos^2\iota)^2}{4}\langle(F_+^{(i)})^2\rangle_t   + \cos^2\iota\langle(F_\times^{(i)})^2\rangle_t \right]\,. 
\end{equation}
Here the angle brackets $\langle\cdot\rangle_t$ refer to an average over time.

When dealing with a data set spanning a duration of several months or
a year, it is not possible to carry out a purely coherent (e.g. $\F$-statistic)
search over $\geq \, 10^{13}$ templates (waveforms) with the best sensitivity.  The number of templates required to
cover the parameter space grows rapidly with the observation time and
soon becomes unmanageable for a fully coherent search. 
Semi-coherent searches have thus been applied in these cases.  The
general technique is to break up the full data set into shorter
segments, search each segment coherently and combine the results of
these coherent analyses to produce the final detection statistic.  The
combination of the coherent analyses will not maintain phase coherence
between the segments and for this reason this method is often called
semi-coherent or incoherent.  There are several methods proposed for
this
\cite{Schutz:1999mb,HierarchP1,HierarchP2,GlobCorr,Pletsch:2009uu}.
We will not go into the details of any of these methods, but we will
use the notions of computational cost and sensitivity of such methods
in this context.

In semi-coherent methods the final resolution in the signal parameter
space is obtained in two steps: the coherent searches and the
incoherent combination of the results of the coherent searches. Both
these stages require template banks to be set-up. The template bank
used in the coherent analysis of the segments is called the
\emph{coarse} grid and comprises $N_\mathrm{c}$ points.  For directed searches
the sky position is fixed and the coarse grid consists of points in
$(f,\dot{f},\ddot{f})$.  The semi-coherent step requires a different
template grid, the \emph{fine} grid\footnote{It should be emphasized
  that the fine grid is still much coarser than the grid that would be
  required for a coherent search on the total data set.}, defined by
refinement factors for all parameters. From these an overall
refinement factor can be derived: $N_\mathrm{refine}$ which is the number of
fine grid points for each coarse grid point. The final grid will
consist of a total of $N_\mathrm{c}\times N_\mathrm{refine}$ points.

We consider a stack-slide-type of semi-coherent search where the
detection statistic is the average of the $\F$-statistic across the
$N$ segments:
\begin{equation}
  \label{eq:15}
  \avgF = \frac{1}{N}\sum_{\ell=1}^N\F_\ell\,.  
\end{equation}
Here, $\F_\ell$ is the $\F$-statistic for the $\ell^{th}$ segment.  The
average $\avgF$ is to be evaluated at a point on the fine grid, while
on the right hand side the $\F_\ell$ are evaluated at a coarse grid
point, ideally the closest to the chosen fine grid point.  Since
$2\F_\ell$ follows a non-central $\chi^2$ distribution with 4 degrees of
freedom, it follows that $N \avgTwoF$ follows a non-central $\chi^2$
distribution with $4N$ degrees of freedom.  The non-centrality
parameter is the sum of $\rho^2$ over the $N$ segments.  For our
purposes, $\rho^2$ is approximately constant over each segment, and
thus the non-centrality parameter is well approximated by $N\rho^2$.
  
The computational time for searching a data set that comprises
$N_\mathrm{sft}$ 1800-s Short time-baseline Fourier Transforms (SFTs) and is
divided in $N$ coherent segments, is
$N_\mathrm{c}\left(N_\mathrm{sft}\tau_{\F} +   N N_\mathrm{refine}\tau_\mathrm{S}\right)$, 
where $\tau_{\F}=7.4\times10^{-8}\,\mathrm{s}$ and $\tau_\mathrm{S}=4.7\times 10^{-9}\,\mathrm{s}$ are the timing constants  used here for
the $\F$-statistic and semi-coherent computations. $\tau_{\F}$ is the
time necessary to compute the $\F$-statistic for a single coarse-grid
template per SFT. $\tau_\mathrm{S}$\footnote{The subscript ``S" stands for
  ``segment" in this stack-slide type of search \cite{BC00}.} is the time
necessary to compute the final detection statistic per fine-grid
template point and per segment. Both timing constants are derived by
direct timing of the search software.

The false alarm probability, i.e. the probability of obtaining a value
of $\avgTwoF$ above a given threshold, say $\avgTwoF^\star$, in the
absence of a signal is
\begin{equation}
  \label{eq:falseAlarm}
  \alpha(\avgTwoF^\star) = \int_{N\avgTwoF^\star}^{\infty} \chi^2_{4N}(y|0)dy,.
\end{equation}
We set the detection criterion based on a false-alarm threshold
$\alpha^\star$ and, from Eq.~(\ref{eq:falseAlarm}), we find the
corresponding threshold $\avgTwoF^\star$. The probability $\eta$ of
detecting a signal with parameters $\lambda_j$ is the probability of
obtaining a value of the detection statistic $\avgTwoF$ higher than the
detection threshold $\avgTwoF^\star$ when $N\avgTwoF$ is drawn from a
distribution
$p(N\avgTwoF|\rho^2)=\chi^2_{4N}(N\avgTwoF|N\rho^2)$:
 \begin{equation}
  \label{eq:3}
  \eta(\avgTwoF^\star|\rho^2) = \int_{N\avgTwoF^\star}^\infty \chi^2_{4N}(y|N\rho^2)dy\,.
\end{equation}
We emphasize that even though we have eliminated the amplitude
parameters from the detection statistic $\avgTwoF$, its distribution
still depends on them through the non-centrality parameter.

\section{The general optimization scheme}
\label{sec:ranking}

We are now ready to tackle the problem set out in the introduction:
given a set of potential targets, what is the optimal choice of
parameter space in $(f,\dot{f},\ddot{f})$ we should search for each,
and what should be the search set-up (in this case the coherent
segment length)?  To this end, we begin by discretizing the whole
parameter space into many small cells such that:
\begin{itemize}
\item The cells are non-overlapping, and the union of all the cells
  covers the parameter space of interest
\item The computing costs and detection probabilities for each target
  vary smoothly from one cell to the next
\item The cost of searching any cell for any target is much smaller
  that the total computational cost budget available
\end{itemize}
As long as these conditions are satisfied our optimization method will
be largely insensitive to how fine the cell- discretization is chosen.

We associate to each cell, astrophysical target and search set-up a
probability to detect a signal with parameters in that cell and the
computing cost for searching over the cell waveform parameters with a
particular semi-coherent search set-up. The goal of this work is to
choose a collection of cells and targets such that:
\begin{itemize}
\item The sum of the computational cost for searching all the chosen
  cells is within our computational budget, and
\item The sum of the probability values for the chosen cells is
  maximized. By this we mean that other choices of cells, search
  set-ups or/and targest would yield a lower detection probability.
\end{itemize}

\subsection{Single set-up case}
\label{subsec:singlesetup}

To illustrate the procedure we begin by considering a single
astrophysical target, i.e. a source corresponding to a single
sky-position, unknown frequency and unknown spindowns.  We also
restrict ourselves to a single search set-up, i.e. we assume a
specific coherent segment length and number of segments.

Let us indicate the frequency-spindown parameter space as
$\mathcal{P}$.  Based on available astrophysical information we define
a prior probability density $P(f,\dot{f},\ddot{f})$ for different
frequency and spindown values.  We will later make a particular choice
for this prior but the general method we describe now is applicable
for any choice.

We break the space $\mathcal{P}$ into non-overlapping cells small
enough so that the conditions described above are satisfied.  It is
simplest to consider rectangular cells defined by frequency and
spindown widths $df, d\dot f, d\ddot f$.  Next we assign a detection
probability to each cell for a given data set from an arbitrary number
of detectors and spanning a total duration $T_\mathrm{obs}$.  We assume that
a semi-coherent method is applied with the data broken up onto $N$
segments and the detection statistic is $\avgTwoF$, the average value
of $\twoF$ over the segments.

We calculate the detection probability for each parameter space cell
$c$.  In order to do this we need to assume distributions (priors) for
the parameters of the population of signals in that cell. In
particular, we need priors for $(\alpha,\delta),f,\dot f, \ddot f,
\psi, \cos\iota, \phi_0$ and $h_0$.  We assume that a compact object
is present at the position of the astrophysical target and hence we
will take the priors on $(\alpha,\delta)$ to be 1. The standard
physical priors for $\psi, \cos\iota, \phi_0$ are uniform, leading to
an average detection probability for such population, having assumed a
specific value of $h_0$ and $f,\dot f, \ddot f$:
\begin{equation}
  \label{eq:etaAverageOverNuisanceParams}
\langle{\eta}\rangle_{\phi_0,\psi,\cos\iota}(h_0) := \frac{1}{8\pi^2}\int_{-1}^1d\cos\iota\int_0^{2\pi}d\psi\int_0^{2\pi}d\Phi_0 ~\eta\,. 
\end{equation}
For a population of signals with a prior distribution on the
amplitude, $p(h_0)$, the average detection probability having assumed
values of $f,\dot f, \ddot f$ is:
\begin{equation}
  \label{eq:11}
  \langle\eta\rangle_{h_0,\cos\iota,\psi,\phi_0} = \int_o^\infty p(h_0)\langle\eta\rangle_{\cos\iota,\psi,\phi_0}\,dh_0\,.
\end{equation}
Finally folding in the prior $P(f,\dot{f},\ddot{f}\ldots|I)$ with the detection
probability $\langle\eta\rangle$, we find the total probability of
detection for a cell:
\begin{equation}
  \label{eq:cellProbfddot}
  P_c=\langle\eta\rangle_{h_0,\cos\iota,\psi,\phi_0} P(f_c,\dot{f_c},\ddot{f_c})df\,d\dot{f}\,d\ddot{f}\,.
\end{equation}
We note the difference between $P_c$  and
$\langle\eta\rangle$. $\langle\eta\rangle$ is the detection probability in a cell
with an assumption that the signal is actually in that cell, and
$P_c$ is the real detection probability in a cell because it contains
the prior probability density for that cell.  We also note that
$\langle\eta\rangle_{h_0,\cos\iota,psi,\phi}$ is actually independent
of $\ddot{f}$.  In fact, it depends on $\dot{f}$ only through the
prior $p(h_0)$.  Thus, since the prior $(f,\dot{f},\ddot{f})$ is
normalised to unity, we can drop the $\ddot{f}$ dependence in the
above equation:
\begin{equation}
  \label{eq:cellProb}
  P_c=\langle\eta\rangle_{h_0,\cos\iota,\psi,\phi_0} P(f_c,\dot{f_c})df\,d\dot{f}\,.
\end{equation}
The detection probability over the whole parameter space is
\begin{equation}
  \label{eq:totalDetectionProb}
  P_\mathrm{D} = \int_{\mathcal{P}} \langle\eta\rangle_{h_0,\cos\iota,\psi,\phi_0}P(f,\dot{f})df\,d\dot{f}\,. 
\end{equation}

Computational cost is the other quantity of interest.  We define a
computational cost density $C(f,\dot{f},\ddot{f})$ such that the cost
of searching a cell is
\begin{equation}
  \label{eq:7}
  c_c=C(f_c,\dot{f_c},\ddot{f_c})df\,d\dot{f}\,d\ddot{f}\,.  
\end{equation}
In practice, the cost function is strictly speaking not a density
because of overhead and startup costs associated with a search which
make the cost not strictly proportional to the size of the parameter
space cell.  However we shall neglect this because in practice these overhead 
costs are controlled and can be kept to a minimum by an appropriate choice of cell size.

We want to define a ranking criterion on the cells such that when we
pick, according to that criterion, the top $n_{C_\mathrm{max}}$ cells that
exhaust the computing budget, the resulting total detection
probability ($P_\mathrm{sum}=\sum_i^{n_{C_\mathrm{max}}} P_i$) is maximum. In other
words any other choice of cells would yield a lower value of the total
detection probability $P_\mathrm{sum}$.  These top $n_{C_\mathrm{max}}$ cells are
then the ones that we should search.

We use the detection probability and the computational cost to define a
ranking for each cell.  We motivate this as follows.  As explained at the beginning of this Section, 
the cells are small enough so that the cost for
any cell is much smaller than the total available computational
budget $C_\mathrm{max}$.  If all the cells had the same cost, then clearly we would
use the detection probability to rank the cells and we would simply pick 
as many top cells as we can before exhausting the computing budget.  However, the
cells will generally have different costs associated with them hence the ranking by 
detection probability does not ensure that the total detection probability is maximized.  A way to fix this
would be to adjust the size of the cells so that they do have the same
cost.  A simpler method is to instead use the ratio between the
detection probability and the cost, which we call 
efficiency, to rank the cells. Thus for each cell we construct the
ratio
\begin{equation}
  \label{eq:8}
  e(f_c,\dot{f_c},\ddot{f_c}) = \frac{\langle\eta\rangle_{h_0,\cos\iota,\psi,\phi_0} P(f_c,\dot{f_c},\ddot{f_c}|I)}{C(f_c,\dot{f_c},\ddot{f_c})}\,.
\end{equation}
Note that the efficiency $e$ contains information about the search
set-up through $\langle{\eta}\rangle$ and $C$, the detector sensitivity
through $\langle{\eta}\rangle$, the astrophysical priors through
$P(f_c,\dot{f_c},\ddot{f_c})$, and the computational cost through $C$.

A more rigorous argument that the efficiency is the
correct ranking function can be modeled on the proof of the
Neyman-Pearson lemma found in most statistics textbooks.  We can
formulate the problem as finding a region $\mathcal{P}_0\subset
\mathcal{P}$ in the parameter space $(f,\dot{f},\ddot{f})$, such that
the cost of searching over $\mathcal{P}_0$ is a chosen value 
$C_\mathrm{max}$
\begin{equation}
  \label{eq:costconstraint}
  \int_{\mathcal{P}_0}C(f,\dot{f},\ddot{f})df\,d\dot{f}\,d\ddot{f} = C_\mathrm{max}\,. 
\end{equation}
and such that the detection probability over the region $\mathcal{P}_0$ is larger than over any other region that satisfies 
the computing budget requirement (Eq.~(\ref{eq:costconstraint})): 
\begin{equation}
  \label{eq:sumprob}
  \int_{\mathcal{P}_0} \langle\eta\rangle_{h_0,\cos\iota,\psi,\phi_0}P(f,\dot{f},\ddot{f})df\,d\dot{f}\,d\ddot{f}=P_\mathrm{max}\,.   
\end{equation}
With the problem formulated in this way, the Neyman-Pearson lemma is
directly applicable and it tells us that the optimal choice of the
region $\mathcal{P}_0$ is to consider level sets of the efficiency function.  
For a given threshold $e_\mathrm{min}$ on the efficiency, the
condition $e(f_c,\dot{f_c},\ddot{f_c})\geq e_\mathrm{min}$ defines a region
$\mathcal{P}_0$.  We choose $e_\mathrm{min}$ so that the region
$\mathcal{P}_0$ satisfies the computational cost constraint for a given
maximum budget $C_\mathrm{max}$ according to Eq.~(\ref{eq:costconstraint}).

The optimization procedure for a single source and a given set-up is
then straightforward.  For each cell we compute the efficiency $e_c$,
pick the cells starting from the one with the largest efficiency and
continue till we have used up all the computing power $C_\mathrm{max}$.  By
doing this, we maximise the probability with a limited computing power
budget. If we have more than a single astrophysical target we can
still use this same ranking criterion: we consider the parameter space
cells from the different targets all together and drop the distinction
between the different targets.  The same procedure described above
will yield the optimal detection probability.


\subsection{The general case}
\label{subsec:multiplesetups}

The efficiency ranking introduced in the previous subsection is
applicable when we constrain the realm of possible searches to a
single search set-up for all the cells and for all the sources.  It is
clear that this is not optimal and we would gain by allowing for
varying set-ups. If we do this we also need to impose the additional
constraint:
\begin{itemize}
\item Each parameter space cell for a given source, must be chosen
  only once,
\end{itemize}
because it would clearly be wasteful to search the same cell in
parameter space for the same source more than once with different
set-ups.  The Neyman-Pearson method used earlier cannot incorporate
this additional constraint and we must modify our optimization
algorithm.

We reformulate our optimization problem in such a way that the
widely used method of linear programming (LP) is applicable.
LP is a optimization method which extremizes a linear combination of the parameters
also fulfilling a set of inequalities \cite{GVK180926950}.

We start again with the discrete form of Eqs.~(\ref{eq:costconstraint})
and (\ref{eq:sumprob}) using the cells as constructed earlier.  Let an
integer $i$ label each cell: $1\le i \le N_f\times N_{\dot f}$. 
For simplicity we consider searches with a
fixed total observation time, 300 days, and use a varying number of segments
$N$ to indicate different coherent observation time-baselines. For each
cell, we can pick among different set-ups, i.e. different values of $N$.
Let another integer $s$ label the different set-ups: $1\le s \le n_s$. We now introduce an 
index $j$ that uniquely labels every different cell-set-up combination: $j\leftrightarrow (i,s)$ and 
$1\le j\le n_s\times N_f\times N_{\dot f}$. Finding the optimal solution for our 
problem means finding which \{cell,set-up\} should be picked and which should be 
discarded. We describe this choice with an occupation index $X_j$:
\[
X_j = \begin{cases}
  1, & \text{if $j$ is chosen}, \\ 
  0, &  \text{if $j$ is discarded}.
\label{eq:Xj}
\end{cases}
\]
The ordered set of $X_j$ values with $1\le j\le n_s\times N_f\times N_{\dot f}$ constitutes a binary number with 
$n_s\times N_f\times N_{\dot f}$ digits. 
The total probability over the set of chosen cells, which is the quantity
that we want to maximize, is
\begin{equation}
  \label{eq: discrete_sumprob}
 P_\mathrm{sum} = \sum\limits_{j} P_{j} X_{j}
\end{equation}
with $P_{j}$ being the probability of the cell/set-up $j\leftrightarrow (i,s)$.
The computational cost constrain can be expressed as:
\begin{equation}
\label{eq:SumXCleCmax}
  \sum\limits_{j} C_{j} X_{j}  \leq C_\mathrm{max}
\end{equation}
We use LP to find the values of $X_j$ that satisfy (\ref{eq:
  discrete_sumprob}) under the constraint (\ref{eq:SumXCleCmax}), and
taking the $X_j$ to be real numbers rather than integers. More details
on this method, and the reason why the optimization procedure yields
(mostly) integers rather than real numbers can be found in the Appendix
\ref{A:LP}.

If we consider more than a single target and want to optimize also over targets $t$,
the problem does not change in nature. We simply consider more (cell,set-ups)
combinations, each now also labelled by a target ``t" index:
\[
\begin{cases}
\label{eq:generalCaseWithMultipleTargets}
	X_{i,s} & \rightarrow  X_{(i,s)_t} \\
	C_{i,s} & \rightarrow C_{(i,s)_t}.
\end{cases}
\]
We want to find the combination of $X_{(i,s)_t}$ values that maximizes 
\begin{equation}
  \label{eq: discrete_sumprobGeneral}
 P_\mathrm{sum} = \sum\limits_t\sum\limits_{(i,s)_t} P_{(i,s)_t} X_{(i,s)_t}
\end{equation}
with the constraints
\[
\begin{cases}
\label{eq:generalCaseConstraints}
  \sum\limits_t\sum\limits_{i,s} C_{{i,s}_t} X_{{i,s}_t}  \leq C_\mathrm{max} & \\
  \sum\limits_t\sum\limits_{s}X_{{i,s} _t} \leq 1~\, \mathrm{and}\quad X_{{i,s}_t} \geq 0\qquad & \mathrm{for every} \quad (i,t).
\end{cases}
\]	
We emphasize that the solution to this optimization problem
$X_{(i,s)_t}$ solves the problem that we posed in the introduction: it
tells us what astrophysical targets ($t$) we should search; for each
target what frequency-spindown values ($i$) we should search and what
semi-coherent search set-up ($s$) to use in each parameter space
cell. Moreover, the scheme incorporates, through the priors, any
astrophysical information on the distribution of the relevant signal
parameters.

\section{Examples of the optimization scheme}
\label{sec:examples}

We now illustrate our optimization scheme with a very specific and
practical example, namely searching a list of potential targets on
the public distributed computing
project Einstein@Home \cite{Einstweb}. 

The sources that we consider are taken from \cite{owen2014} and
are listed in Table~\ref{tab:sources}. This list comprises 
supernova-remnants (SNR) whose position in the sky is very well 
known (better than sub-src second accuracy), and described by their equatorial 
sky coordinates $\alpha, \delta$. We associate with each source its estimated 
age $\tau_t\pm d\tau_t$  and an estimate of what we believe is
the maximum intrinsic GW amplitude that it could be
emitting\footnote{This is the age-related spindown limit defined 
for example in \cite{S5CasA}.}: ${h_{0~t}^\mathrm{max}}$.We label the 
different point sources with an index $t$, and $t = 1
\ldots N_t$. 
\begin{table*}
\caption{
  \label{tab:sources}
  Point source targets considered in this paper }
  \begin{center}
  \begin{tabular}{ccccccccc}
    \tableline
   
    SNR G name & Other name & Point source~J & $D_\mathrm{kpc}$ &
    $\tau_\mathrm{kyr}$ & $10^{25}h_0^\mathrm{age}$ &
    \\
    \tableline
    111.7\textminus2.1 & Cas~A & 232327.9+584842 & 3.3--3.7 & 0.31--0.35 & 12
    \\
    189.1+3.0 & IC~443 & 061705.3+222127 & 1.5 & 3--30 & 3 & &
    \\
    266.2\textminus1.2 & Vela~Jr & 085201.4\textminus461753 &  0.2--0.75 & 0.7--4.3 & 15--140
    \\
    347.3\textminus0.5 & & 171328.3\textminus394953 &  1.3 & 1.6 & 14
    \\
    350.1\textminus0.3 & & 172054.5\textminus372652 & 4.5 & 0.9 & 5.3
    \\
    \tableline
  \end{tabular}
\end{center}

\end{table*}

\subsection{Astrophysical priors}
\label{subsec:astroprior}

In order to compute the detection probability in every cell we have to
choose the prior on the signal amplitude $h_0$: $p(h_0)$ (see
Eqs.~(\ref{eq:cellProb}) and (\ref{eq:11})).  The most relevant
parameter that $h_0$ depends on, is the ellipticity $\varepsilon$ defined
in Eq.~(\ref{eq:ellipticitydef}). We thus recast the integral
(\ref{eq:11}) on $h_0$ as an integral on $\varepsilon$. Unfortunately the
ellipticity is also the least known parameter so reflecting our
ignorance we take a flat probability density on $\log\varepsilon$ within
a conservative range of values. Consider a cell $i$ centered at a
particular frequency $f_i$ and spindown $\dot{f_i}$ for a particular
source chosen from Table~\ref{tab:sources}.  For this cell we can
consider two upper limits on $\varepsilon$:
The first is the spindown ellipticity, $\varepsilon^\mathrm{sd}_x$ of 
Eq.~(\ref{eq:spindownEllipticity}), with $x=0.01$, which 
is consistent with the latest limits on the emission of gravitational waves from the Crab pulsar
\cite{Aasi:2013sia}.  The second limit is based on the results 
of \cite{Horowitz}, according to which it is unrealistic to expect
$\varepsilon$ to exceed $\sim 10^{-4}$.  We thus set a cell-dependent maximum acceptable 
value of $\varepsilon$ as (for ease of notation we drop the subscript ``x'' in $\varepsilon^\mathrm{sd}_x$):
\begin{equation}
  \label{eq:epsilonMax}
    \varepsilon^\mathrm{max}_i = \min ( {10^{-4},~\varepsilon^\mathrm{sd}_i} ) .
\end{equation}
We consider now the minimum value of $\varepsilon$.  If the neutron star were
perfectly axisymmetric then $\varepsilon=0$, $h_0=0$ and there would be no GW emission.  However 
deviations from this axisymmetric configuration are expected due to the 
internal magnetic field, at a level that should be at least $\varepsilon \sim 10^{-14}$
\cite{Andersson:2009yt}.  We hence take 
\begin{equation}
  \label{eq:epsilonMin}
    \varepsilon^\mathrm{min} = 10^{-14}.
\end{equation}


Based on the above discussion, our prior $p(\varepsilon)$ is: 
\begin{equation}
  \label{eq:epsilonPrior}
  p(\varepsilon) = \left\{ 
    \begin{array}{cc}
     {1\over\varepsilon} {1\over{\log(\varepsilon^\mathrm{max}/\varepsilon^\mathrm{min})}}\, &\quad  \varepsilon^\mathrm{min}<\varepsilon<\varepsilon^\mathrm{max}   \, \\
      0\,   & \mathrm{elsewhere} \,.
    \end{array}   
  \right. 
\end{equation}

\begin{figure}
\begin{centering}
\includegraphics[width=0.45\textwidth]{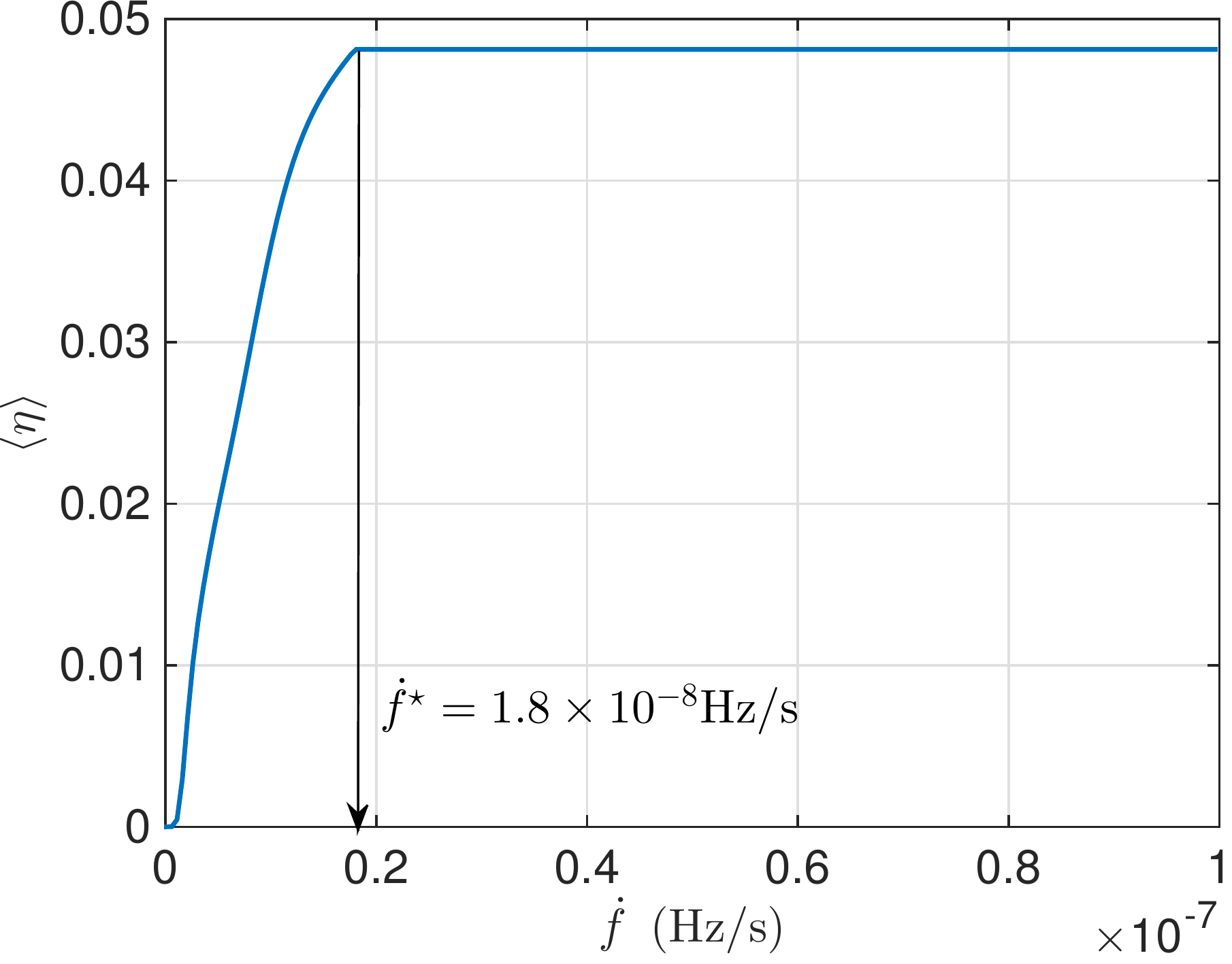}
\caption{Detection probability as a function of the source's spindown, for a source at the position of Cas A, emitting at $101$ Hz and being searched with 20-day coherent time-baseline segments over 300 days. Due to our prior on $h_0$ the detection probability stops increasing at $\dot{f}^\star=1.8\times 10^{-8}\mathrm{Hz/s}$.  }
 
\label{CasA_20days_limted_in_fdot}
\end{centering}
\end{figure}

As an illustration of this choice of prior, consider Cas A taken to be
a distance of 3.5 kpc from us.  Let us assume the star to be
emitting GWs at some frequency $f$, the fraction of
the rotational energy going into GWs to be $x=0.01$ and the standard
value of the moment of inertia $I$ to be $10^{38}\,\mathrm{kg \,m^2}$. At small
$|\dot{f}|$, $\varepsilon^\mathrm{max}$ is given by the spindown limit $\varepsilon^\mathrm{sd}(f,\dot{f})$.  As
$|\dot{f}|$ increases, the spindown limit $\varepsilon^\mathrm{sd}(f,\dot{f})$ also increases and with it also
$\varepsilon^\mathrm{max}$ until it reaches the value  
$10^{-4}$. This happens at a crossover spin down value of
\begin{equation}
  \dot{f}^\star = -1.71\times 10^{-8}\mathrm{Hz/s}\,\left(\frac{f}{100\mathrm{Hz}}\right)^5\,.
\end{equation}
For spindown values in absolute value larger than this crossover spindown value, 
$h_0^\mathrm{max}$ ceases to increase and remains constant at a value that corresponds to the maximum ellipticity value 
that we have set: $10^{-4}$. Correspondingly the
detection probability $\langle\eta\rangle$ at a fixed search frequency will cease to increase as a function of the spindown.  This is shown in
Fig.~\ref{CasA_20days_limted_in_fdot} where we assumed $f=101\,$Hz, a
20 day coherent integration time and 300 days observation time.

What about the prior $P(f,\dot{f})$? We consider uniform and log-uniform priors 
on all these variables with ranges sufficiently large to cover all possible values of these parameters\footnote{In other words, if the detectors were infinitely sensitive and $\langle\eta\rangle$ in Eq.~(\ref{eq:totalDetectionProb}) was equal to one, then, with our choice of priors also 
$P_\mathrm{D} = 1$. }:
\begin{equation}
\label{eq:Priors}
	\begin{cases}
	~~~~~~~~~~~~~~~0\,\mathrm{Hz} \le & f ~\le 1500\,\mathrm{Hz}\\
	-1\times10^{-7}\,\mathrm{Hz/s} \le & \dot{f} ~\le 0\\
	~~~~~~~~~~~0\,\mathrm{Hz/s}^2 \leq & \ddot{f} \leq ~5\dot{f}^2/f.
	\end{cases}
\end{equation}
For the second order spindown parameter we note that
if the frequency evolution follows $\dot{f}
\propto f^n$, where $n$ is the braking index,
then
\begin{equation}
  \label{eq:ddotf}
  \ddot{f} = n\dot{f}^2/f\,.
\end{equation}
For pure GW emission $n=5$, for all other possible mechanisms $n<5$ and 
in particular for pure electromagnetic emission $n=3$ (see
e.g. \cite{ShapiroTeukolsky}). Hence our range for $\ddot{f}$ in (\ref{eq:Priors}) encompasses all 
combinations of emission mechanisms.
 
Different ranges on $\dot{f}$, $\ddot{f}$ and $\varepsilon$ could have been set, based on 
the estimates of the age of the astrophysical targets. If we assume that the 
object has been spinning down 
 by $f$ at a spindown rate $\dot{f}$ 
during a time $\tau_\mathrm{c}$, its characteristic age, due to some mechanism with a braking index $n$, then 
\begin{equation}
\label{eq:tauc}
  \tau_\mathrm{c} = \frac{1}{n-1}\frac{f}{|\dot{f}|}\,.
\end{equation}
By maximising Eq.~(\ref{eq:tauc}) with respect to $n$ we derive a maximum range for $\dot{f}$. We then use that value in Eq.~(\ref{eq:ddotf}) to derive the largest range for $\ddot{f}$. The conservative search ranges are then 
\begin{equation}
  \label{eq:dotf_age}
  \begin{cases}
  |\dot{f}| < & f / \tau_\mathrm{c} \\
  |\ddot{f}| < & 5 f / \tau_\mathrm{c}^2\, ,
  \end{cases}
\end{equation}
having taken the estimated age of the object as a proxy for its characteristic age $\tau_\mathrm{c}$. Note that these maximum ranges for $\dot{f}$ and $\ddot{f}$ correspond to different $n$ values, namely 2 and 5. This is physically inconsistent for any single source but it ensures the broadest prior range over the search values now, allowing for deviations from the constant braking index model in the past evolution of the star.

Let us assume $n=5$, which means emission at the spindown limit, and recast the 
GW amplitude spindown upper limit (Eq.~(\ref{eq:spindownh0})) as well as the corresponding ellipticity (Eq.~(\ref{eq:spindownEllipticity})), in terms of $\tau_\mathrm{c}$ :
\begin{equation}
  \label{eq:agelimit}
  h_0^\mathrm{sd} =  \frac{1}{d}\sqrt{\frac{5GI}{8c^3\tau}}\,
\end{equation}
and
\begin{equation}
  \label{eq:eliimitage}
  {\varepsilon^\mathrm{age}}= \frac{c^2}{16\pi^2f^2}\sqrt{ \frac{10c}{GI\tau}}\,.
\end{equation}
When $n=5$ then $\tau_\mathrm{c} = f/4|\dot{f}|$ is the shortest lifetime compared to the characteristic ages for other emission mechanisms. Correspondingly the necessary spindown is the largest, and so are the GW amplitude upper limit and the ellipticity. Hence, choosing $n=5$ allows for the highest possible value of $\varepsilon$. Correspondingly, if we choose to fold in the prior information on the age of the object Eq.~(\ref{eq:epsilonMax}) becomes:
\begin{equation}
  \label{eq:epsilonMaxAge}
    \varepsilon^\mathrm{max}_i = \min ( {10^{-4},~\varepsilon^\mathrm{sd}_i,~\varepsilon^\mathrm{age}_i} ) .
\end{equation}
We remind the reader that the index $i$ labels a particular $f,\dot{f}$ cell in parameter space.


\subsection{Grid spacings} 
\label{subsec:grids}

Given the ranges for $f$, $\dot{f}$ and $\ddot{f}$, we now need to
specify the number of templates needed to cover the parameter space
covered by each cell. This is a pre-requisite for estimating the
computing cost for that cell. As discussed earlier, a semi-coherent
search requires a set set of templates for the coherent step and a set
for the semi-coherent steps. These are referred to, as the coarse and
fine grids respectively.  The grid spacings in each search parameter
can be parametrized in terms of nominal mismatches
$m_f,m_{\dot{f}},m_{\ddot{f}}$ in $f,\dot{f},\ddot{f}$
respectively\cite{Pletsch:2010xb}:
\begin{eqnarray}
\label{eq:gridSpacings}
  \delta f &=& \frac{\sqrt{12m_f}}{\pi T}\,,\\ 
  \delta\dot{f} &=& \frac{\sqrt{180m_{\dot{f}}}}{\pi T^2}\,,\\
  \delta\ddot{f} &=& \frac{\sqrt{25200 m}_{\ddot{f}}}{\pi T^2}\,.
\end{eqnarray}
Following that, the semi-coherent grid spacing in each dimension is taken a factor $\gamma^k$ finer that the coarse grid one:\begin{equation}
  (\delta f^{(k)})_\mathrm{semi-coh} = \frac{\delta f^{(k)}}{\gamma^{(k)}}\,,\quad k = 1,2\ldots\, 
\end{equation}
with the index $k$ labelling the frequency and spindown parameters:
$f^{(1)} = \dot{f}$ and $f^{(2)} = \ddot{f}$.  The refinement
factors $\gamma^{(k)}$ depend on the number of segments: 
\begin{eqnarray}
  \gamma^{(1)} &=& \sqrt{5N^2-4}\,,\\
  \gamma^{(2)} &=& \frac{\sqrt{35N^4-140N^2+108}}{\sqrt{3}}\,.
\end{eqnarray}
We note that in the simplified problem that we consider here we do not include the loss of signal-to-noise ratio associated with the given mismatches and we do not 
optimize with respect to searches with different grids.

\section{Application of the optimisation scheme under different assumptions}
\label{sec:Application}

We now apply our optimization scheme to a search for a CW signal from the sources
listed in Table~\ref{tab:sources}. We will consider different priors and show intermediate optimisation results: namely we 
firstly fix the search set-up and the target and eventually optimise also over these. In Sections ~\ref{subsec:ResSinglesetup}, ~\ref{subsec:diffsetups} and \ref{subsect:agepriorsresults} we will use uniform priors on $f,\dot{f}$ in order to illustrate the main features of this optimization scheme. In Section \ref{sec:LogUniformPriors} we will show the results for the more physically meaningful log-uniform priors.

\subsection{Uniform priors in $f$ and $\dot{f}$}
\label{sec:UniformPriors}

\subsubsection{Optimizing at fixed search set-up and separately for each target}
\label{subsec:ResSinglesetup}

For illustration purposes, we consider the
simplest case, namely when we have a pre-determined search set-up, i.e. a fixed value
for the number of coherent segments $N$.  The optimization scheme
will rank the parameter space cells of all the sources in decreasing order of
detection-promise and hence yield the parameter space regions that should be
searched for each source. We will consider the data to span
a total observation time of 300 days, to be from the LIGO Hanford and Livingston 
detectors at the best sensitivity level of the 
S6 science
run\footnote{https://github.com/gravitationalarry/LIGO-T1100338/blob/master/H1-SPECTRA-962268343-BEST.txt}. 
We assume as computing budget 12-Einstein@Home months (EMs). 
1EM corresponds to about 12,000 CPU cores round the clock.  Here and throughout the paper 
we use $m_f=m_{\dot{f}}=m_{\ddot{f}}$=0.18 in Eqs.~(\ref{eq:gridSpacings}) to (46) 
for the grid spacings. Note, these are arbitrary but reasonable choices of values illustrative of actual searches. We shall take the coherent segments to each be 10 days long, in this section.

The result will depend on what prior we choose.
We work with two choices: one that does not 
fold in the age information (Eq.~(\ref{eq:epsilonMax})) and one that does (Eq.~(\ref{eq:epsilonMaxAge})). 
We name these priors the ``distance-based prior'' and the ``age-based prior'', respectively.  In this
section we present results for the distance-based priors only
and the age-based prior results will be discussed later.

The source in Table~\ref{tab:sources} closest to us is vela Jr with
distance estimates ranging from 0.2 to 0.75 kpc.  Age and distance
estimates are highly uncertain due to the overlap of the SNR with the main Vela SNR and possible interaction between
them. Other targets such as IC 443 and G347.3 are relatively close to
Earth, with distances of 1.5 kpc and 1.3 kpc respectively.  The
estimated distance of Cas A is between 3.3 and 3.7 kpc, which is not
very close compared with the previous three source we mentioned
above. However, Cas A is the youngest source and we include it in our
list as a point of comparison.

We define a quantity $\mathcal{R}$ as the sum of detection probabilities
for the parameter space cells which are chosen for a given source:
\begin{equation}
  \label{eq:R}
  {\cal{R}} = \int_{\mathcal{P}_0} \langle\eta\rangle_{h_0,\cos\iota,\psi,\phi_0}P(f,\dot{f})df\,d\dot{f}\,. 
\end{equation}
Note that ${\cal{R}}$ is also the actual highest detection probability that one can 
obtain for that source with the given computing budget. Clearly the highest the 
${\cal{R}}$, the most promising is a search for the corresponding target.
For a given amount of computing power $C_\mathrm{max}$, sources with higher
$\mathcal{R}$ are more promising.

\begin{figure}%
 \centering
 \subfloat[${\mathcal{R}}$ versus distance]{{\includegraphics[width=.9\linewidth]{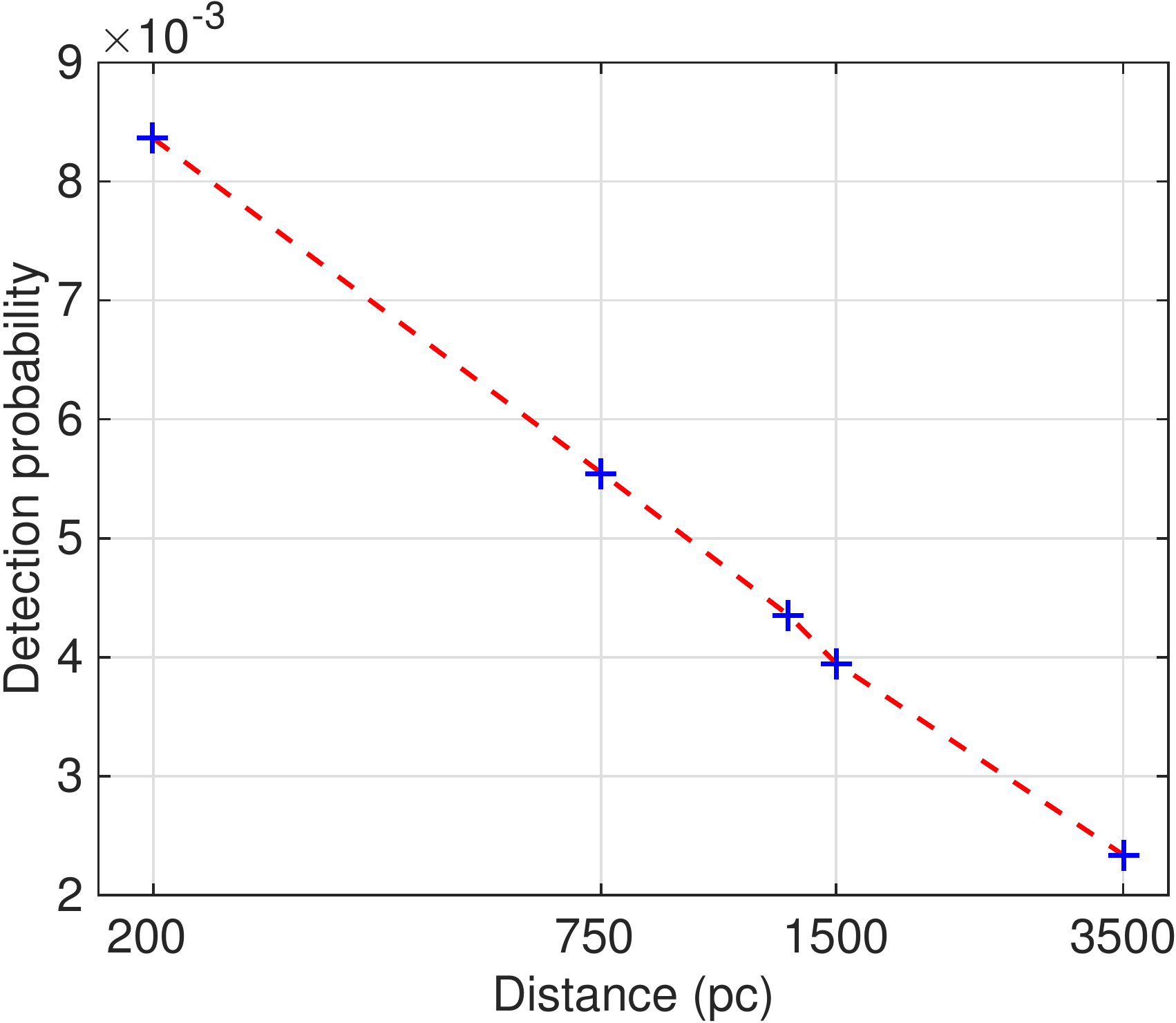}}}%
 \qquad
 \subfloat[${\mathcal{R}}$ versus $T_\mathrm{coh}$]{{\includegraphics[width=.9\linewidth]{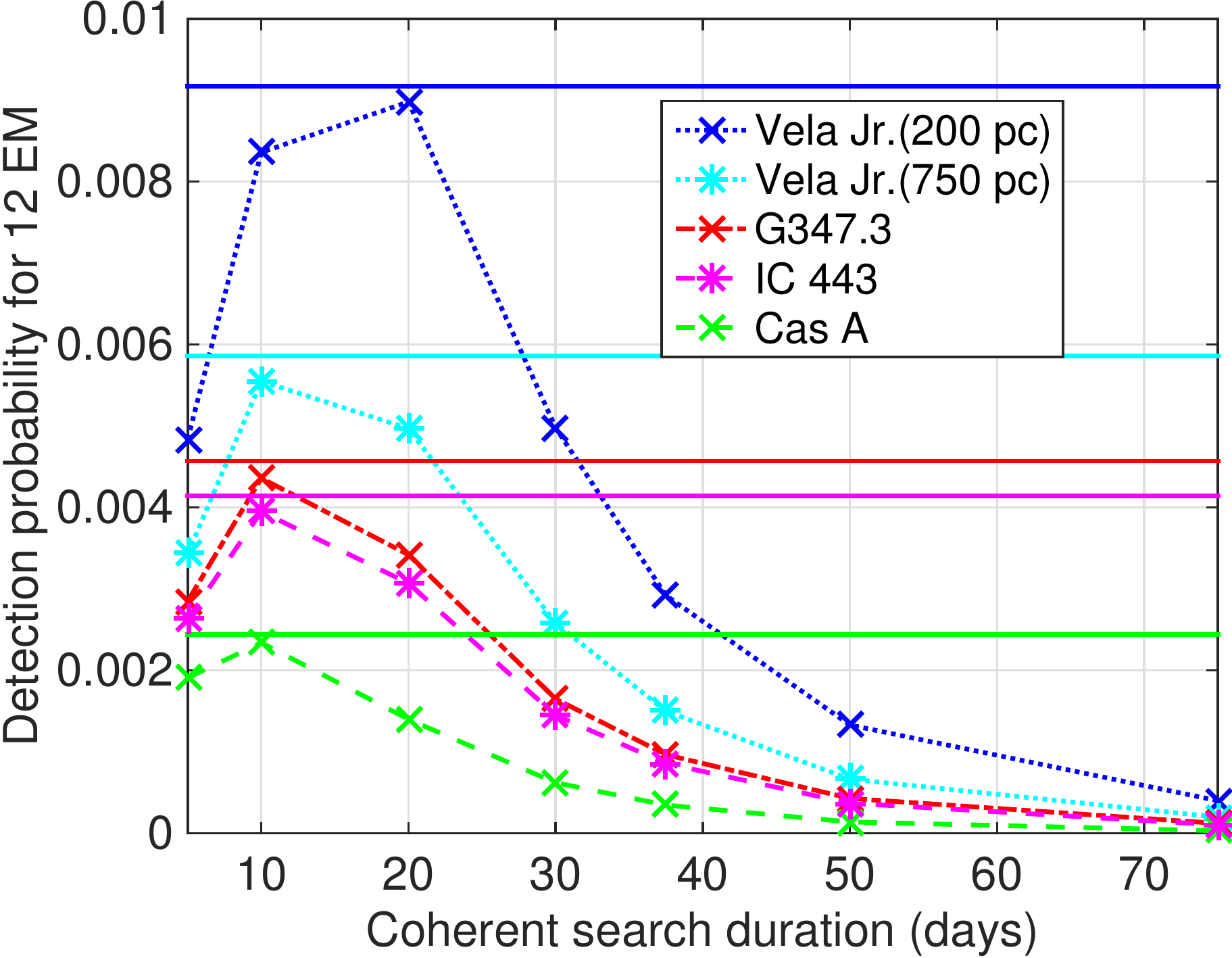}}}%
 \qquad
 \subfloat[${\mathcal{R}}$ versus $C_\mathrm{max}$]{{\includegraphics[width=.9\linewidth]{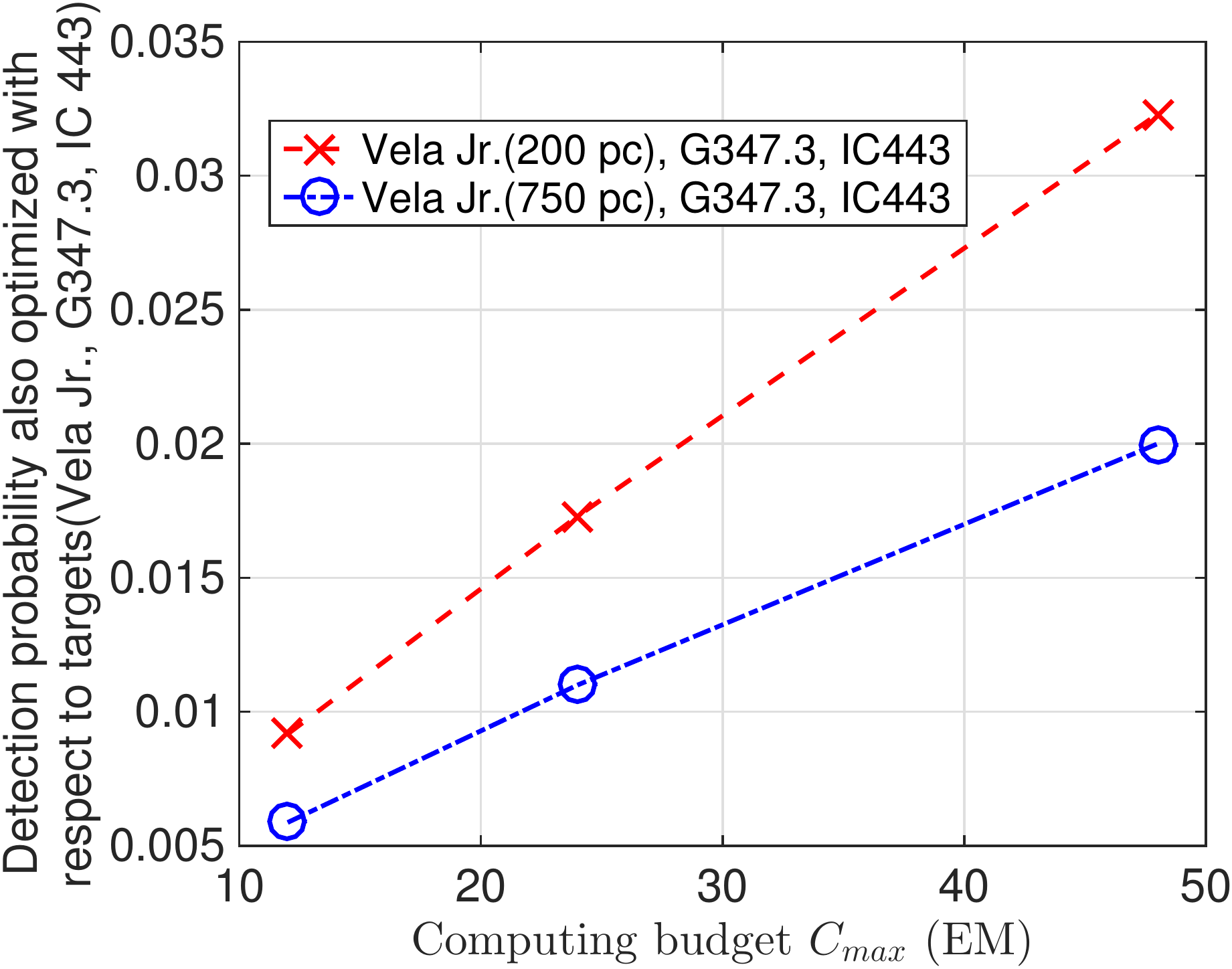}}}%
 \caption{Detection probability for various targets and search set-ups having assumed uniform and distance-based priors. The distances that were assumed for the targets are: Vela Jr (C) 200 pc, Vela Jr (F) 750 pc, G 347.3 1.3 kpc, IC443 1.5kpc, Cas A 3.5 kpc.}%
 \label{fig:RversusDistAndTcoh}
\end{figure}

\begin{table*}
\caption{
  \label{tab: R of nonage}
  $\mathcal{R}$ results with $f$ and $\dot{f}$ uniform priors and distance-based priors. The highest $\mathcal{R}$ with respect to set-up is in bold font.}
\begin{center}
  \begin{tabular}{ll|cccccccccc}
    \tableline
   
   && \multicolumn{10}{c}{$10^3\mathcal{R}$}
   \\
   \cline{3-12}
  
    Name & $~~D_\mathrm{kpc}$ & 5D & 10D&20D&30D&37.5D&50D&75D&\multicolumn{3}{c}{LP Optimized}\\
    
    && \multicolumn{7}{c}{Computing Budget: 12EM} &12EM&24EM&48EM\\
    \tableline

 Cas~A &~~3.5&1.91&\textbf{2.34}&1.40&0.622&0.347&0.140&0.032&2.44 &4.33&--\\
 IC~443 &~~1.5 &2.63&\textbf{3.95}&3.07&1.45&0.858&0.371&0.104&4.14&7.33&--\\
 G347.3&~~1.3 &2.85&\textbf{4.36}&3.42&1.65&0.966&0.429&0.122&4.57&8.19&--\\
 Vela~Jr &~~0.2&4.84&8.36&\textbf{8.98}&4.98&2.92&1.33&0.401&9.17&17.3&--\\
 Vela~Jr &~~0.75&3.43&\textbf{5.55}&4.97&2.58&1.50&0.666&0.199&5.86&10.9&--\\
 Top 3 (0.2 kpc)&~~--&--&--&--&--&--&--&--&9.17 &17.3 &32.3 \\
 Top 3 (0.75 kpc)&~~--&--&--&--&--&--&--&--&5.87 &11.0 &20.0\\

    \tableline
  \end{tabular}
\end{center}
\end{table*}

The highest value of $\mathcal{R}$, about $1\%$, is obtained with a search that targets the closest source, Vela Jr, at 
200 pc. For the other targets the detection probability is even lower and decreases with increasing 
distance as summarized in Fig.~\ref{fig:RversusDistAndTcoh}(a).

\begin{figure*}%
    \centering
    \subfloat[Efficiency]{{  \includegraphics[width=.38\linewidth]{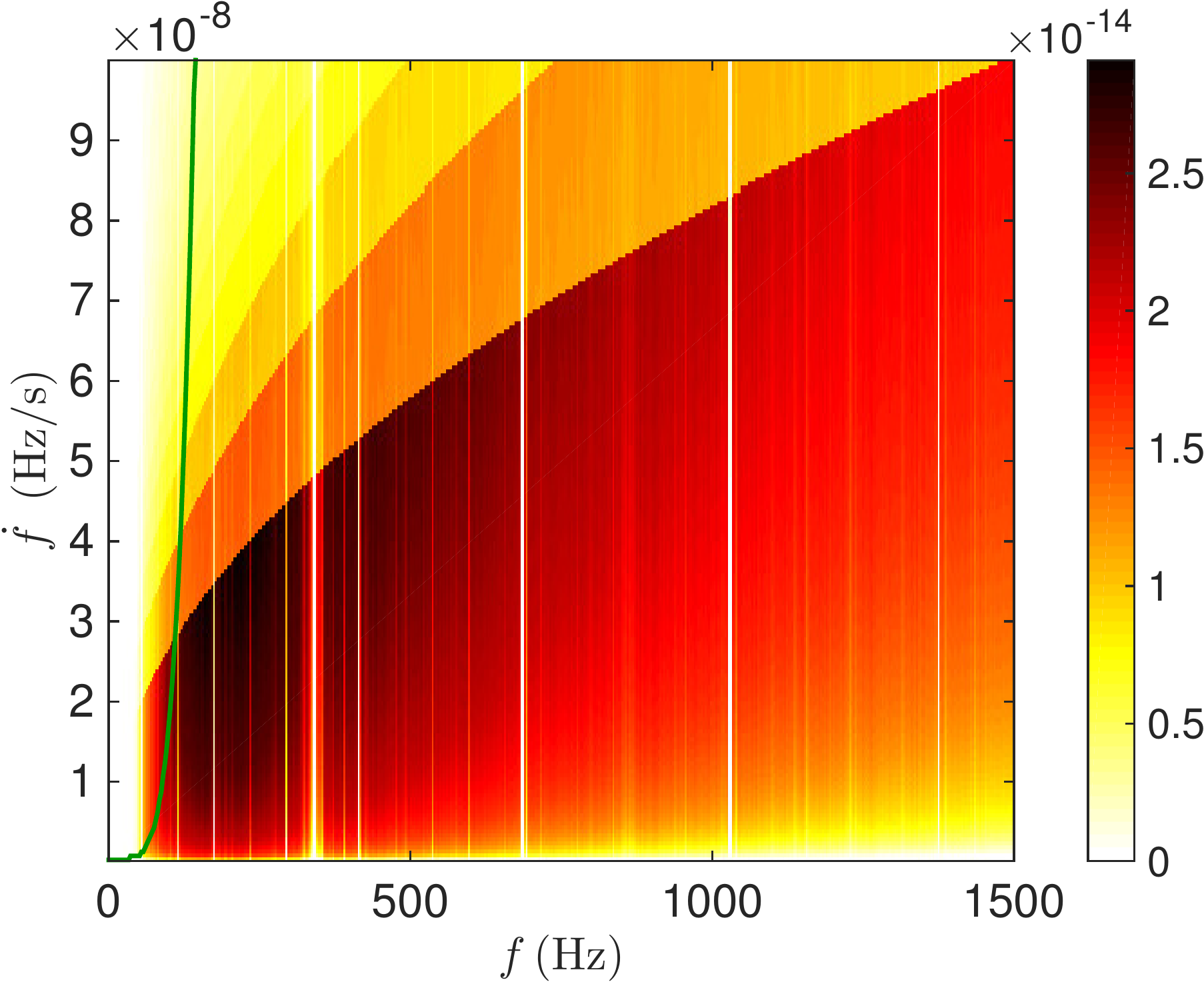}}}%
    \qquad
    \subfloat[Coverage]{{  \includegraphics[width=.38\linewidth]{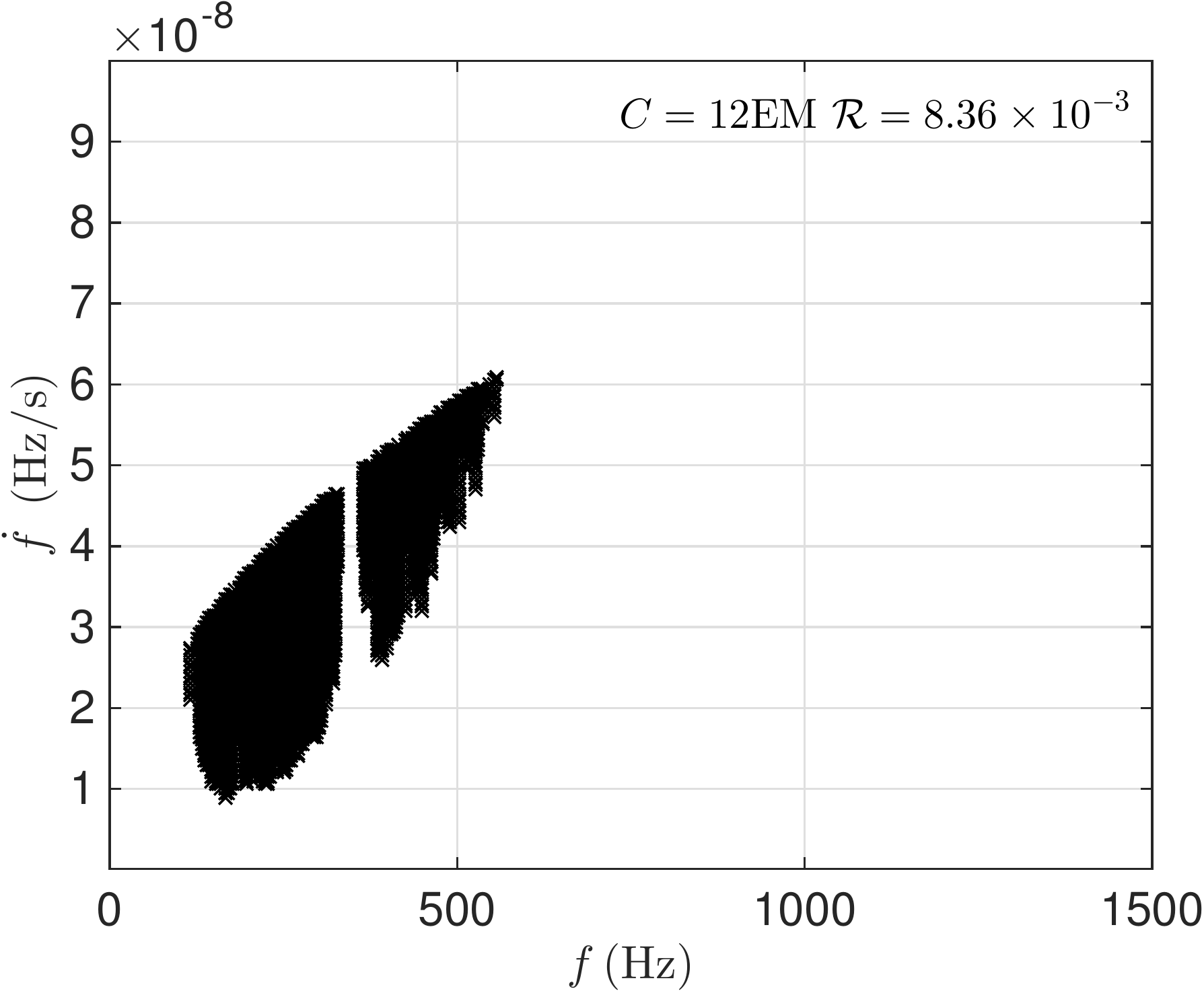}}}%
    \caption{Optimisation results for Vela Jr at 200 pc, assuming uniform and distance-based priors, and a 10-day coherent segment duration. The left plot shows the efficiency, color-coded, for each cell : $e(f,\dot{f})$. The green curve shows $\dot{f}^\star$  as a function of $f$. The right plot displays the cells selected by the optimization procedure with a computational budget of 12 EM. The detection probability $\mathcal{R}$ is $8.36\times10^{-3}$.}%
 \label{G2662_10days_noage}%
\end{figure*}

\begin{figure*}%
    \centering
    \subfloat[Efficiency]{{  \includegraphics[width=.4\linewidth]{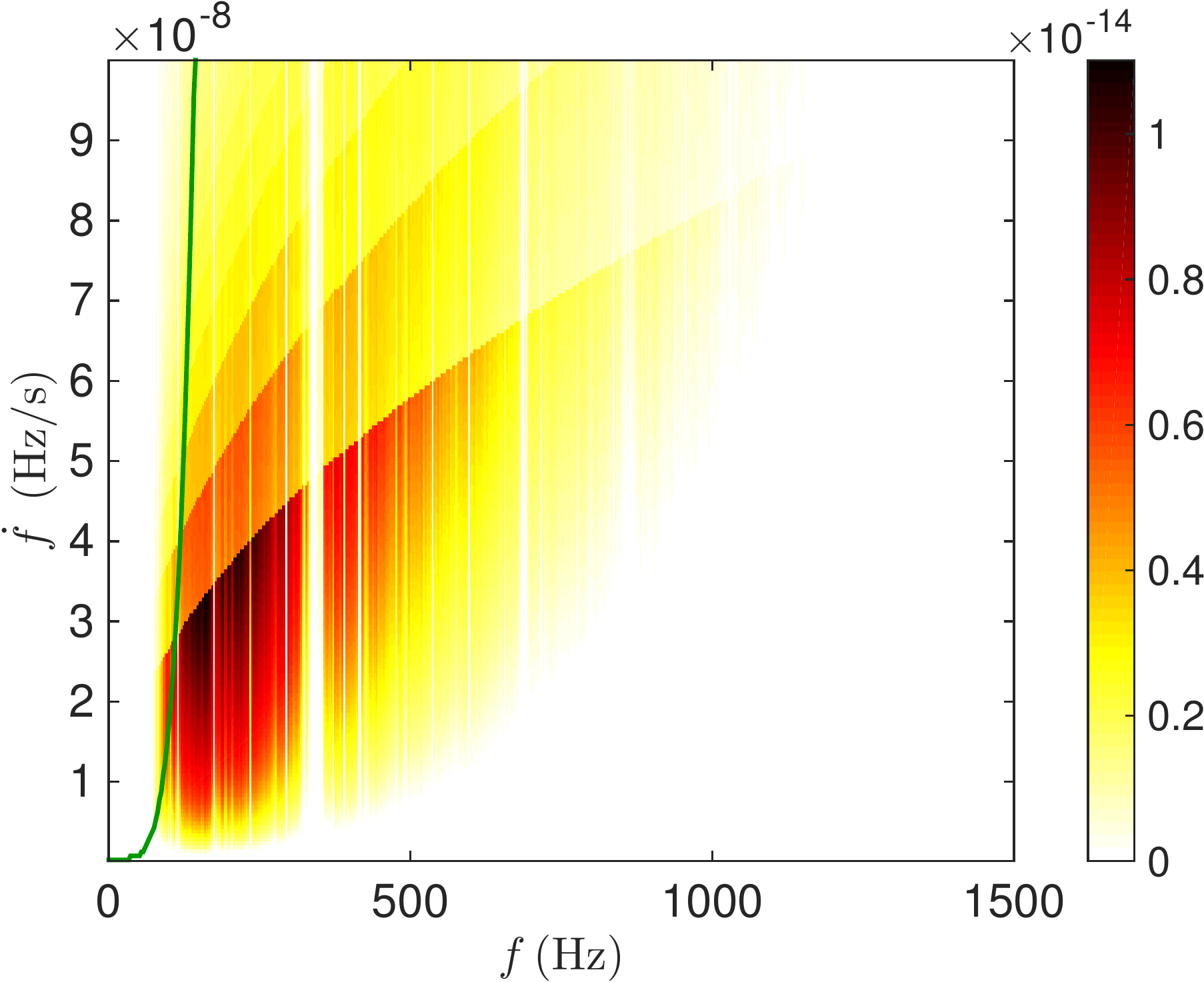}}}%
    \qquad
    \subfloat[Coverage]{{  \includegraphics[width=.4\linewidth]{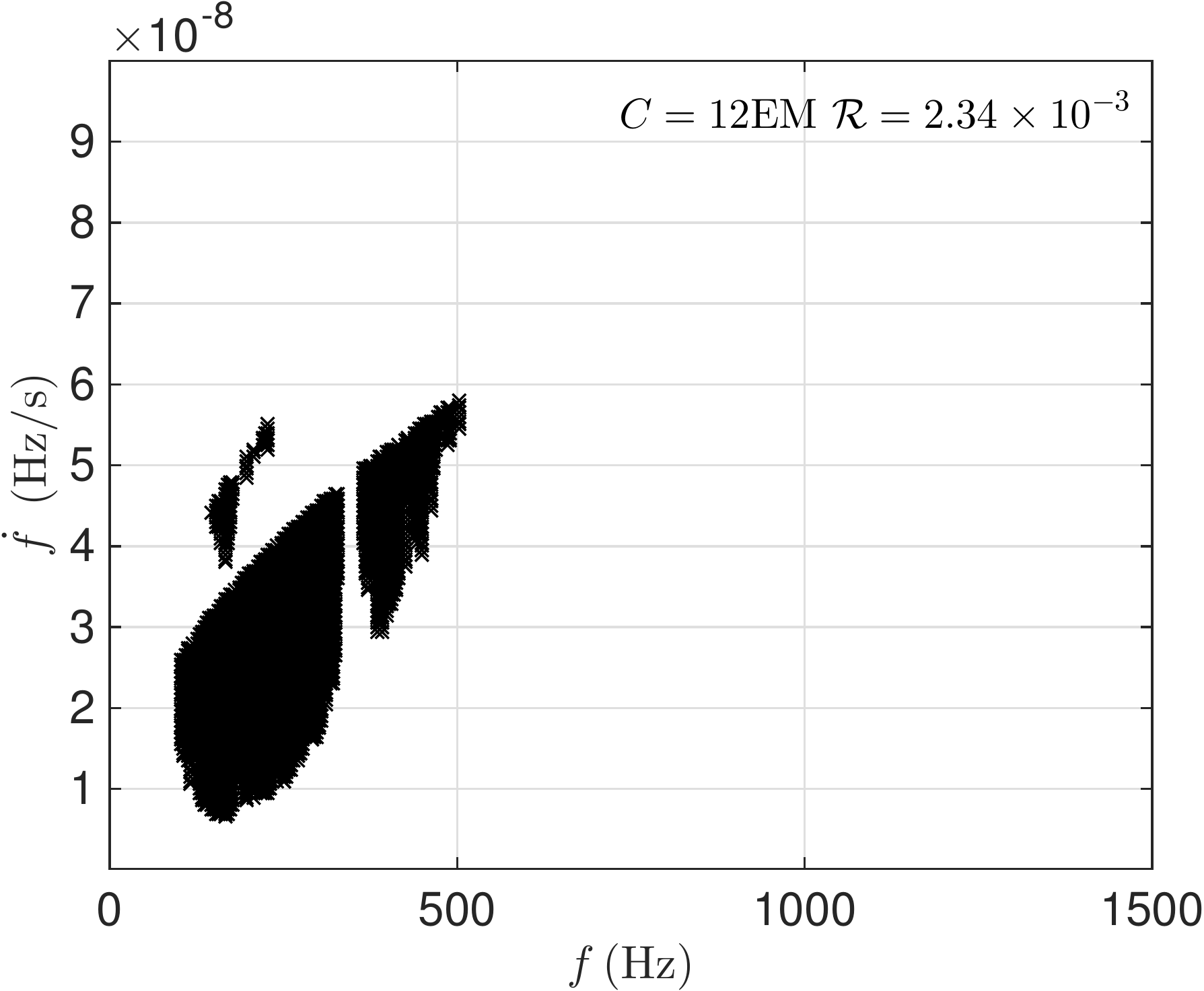}}}%
 \caption{Cas A at 3500 pc, assuming uniform and distance-based priors, and a 10-day coherent segment duration. The left plot shows the efficiency, color-coded, for each cell : $e(f,\dot{f})$. The green curve shows $\dot{f}^\star$  as a function of $f$. The right plot displays the cells selected by the optimization procedure to be searched with a computational budget 12 EM. The detection probability $\mathcal{R}$ is $2.34\times10^{-3}$.}%
    \label{CasA_10days_noage}
\end{figure*}

Figs.~\ref{G2662_10days_noage} and \ref{CasA_10days_noage}
display two plots for each target\footnote{Similar Figs.~\ref{G2662_10days_noage_longdis} to \ref{G1893_10days_noage} for other sources could be found in Appendix \ref{section:AppendixFigures}.}. The (a) plots shows the efficiency,
color-coded, for each cell : $e(f,\dot{f})$. The green curve in the
efficiency plots shows $\dot{f}^\star$ as a function of $f$. The (b)
plots display the cells selected by the optimization procedure to be
searched within the computational budget, i.e. the coverage that we
can afford.  It is interesting to note how the shape of the covered
parameter space changes as the source distance increases.  As the
distance decreases the detection probability per cell $P_c$ increases,
but it does so more slowly as the distance decreases because the
probability cannot exceed 1.  Below $\dot{f^\star}$, cells with higher
$\dot{f}$ have a higher detection probability through the maximum
allowed $h_0$ in Eq.~(\ref{eq:epsilonPrior}). However higher spindown also
means larger computational cost due to the broader range in
$\ddot{f}$. So, for the farther away sources like Cas A, the gain in
detection probability offsets the computational cost. However for
sources which are closer, and for which the gain is smaller, this is
not the case. This is the reason why in, say,
Fig.~\ref{CasA_10days_noage} more cells are picked from high
$\dot{f}$ regions than for Fig.~\ref{G2662_10days_noage}. In
general, given fixed-duration coherent segment, selected cells in
farther source (Cas A) are more likely from higher $\dot{f}$
region.  

\FloatBarrier

\subsubsection{optimizing with respect to search set-ups and targets}
\label{subsec:diffsetups}

We now vary the possible search set-ups and also optimize over these.
Again for illustration we consider seven representative choices of
coherent segment lengths: 5, 10, 20, 30, 37.5, 50 and 75 days.  As
before, the total observation time is 300 days. We present results for
the 3 sources Vela Jr (at 200 pc), G347.3 and Cas A.

A plot of the optimal detection probability as a function of the set-up is shown in Fig.~\ref{fig:RversusDistAndTcoh}(b), the non-solid lines. 
For Vela Jr, 20-day segment gives us the
best result where the detection probability $\mathcal{R}$ is $8.98\times10^{-3}$.

\begin{figure*}%
    \centering
    \subfloat[Efficiency, 5 days]{{  \includegraphics[width=.20\linewidth]{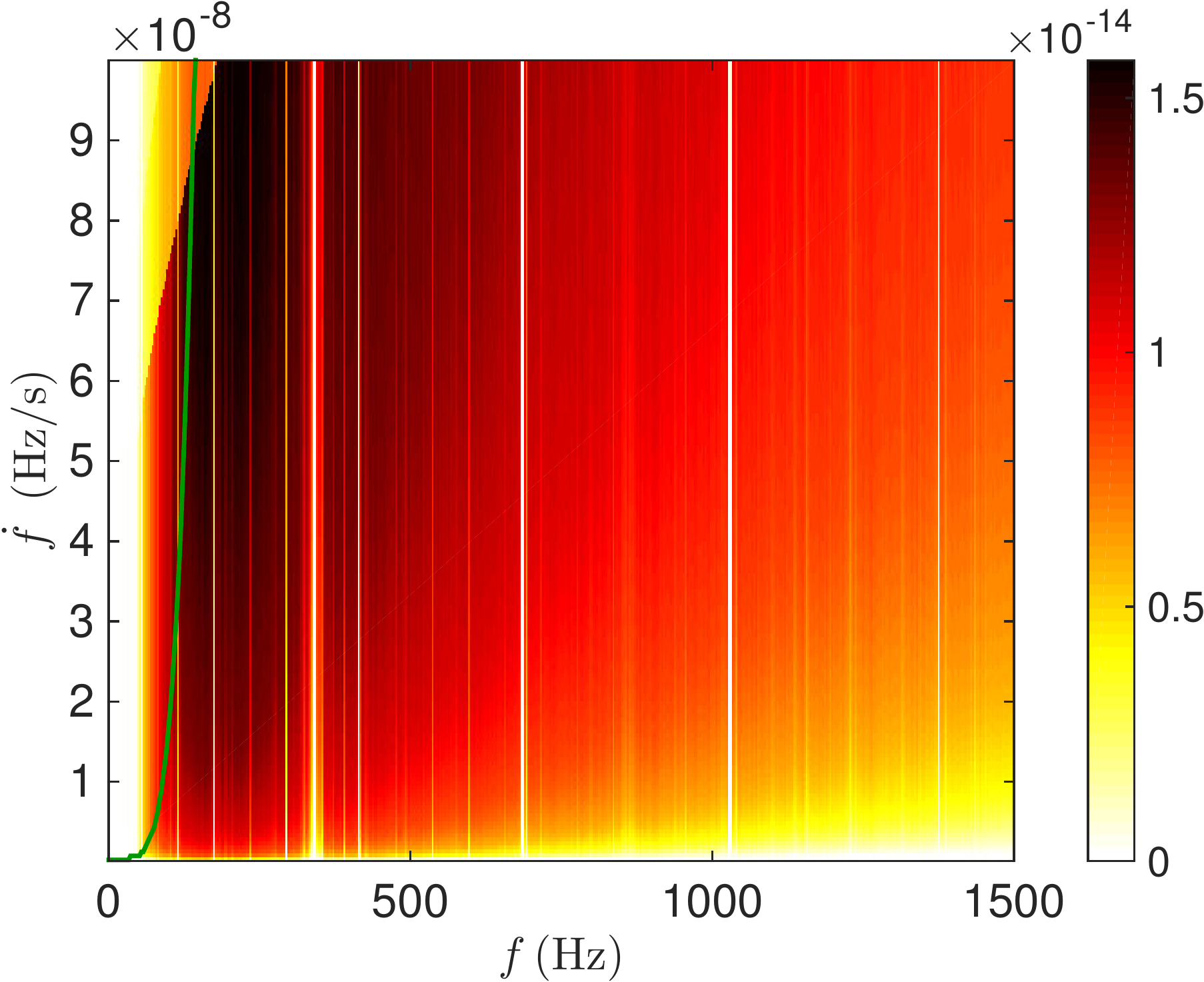}}}%
    \qquad
    \subfloat[Coverage, 5 days]{{  \includegraphics[width=.20\linewidth]{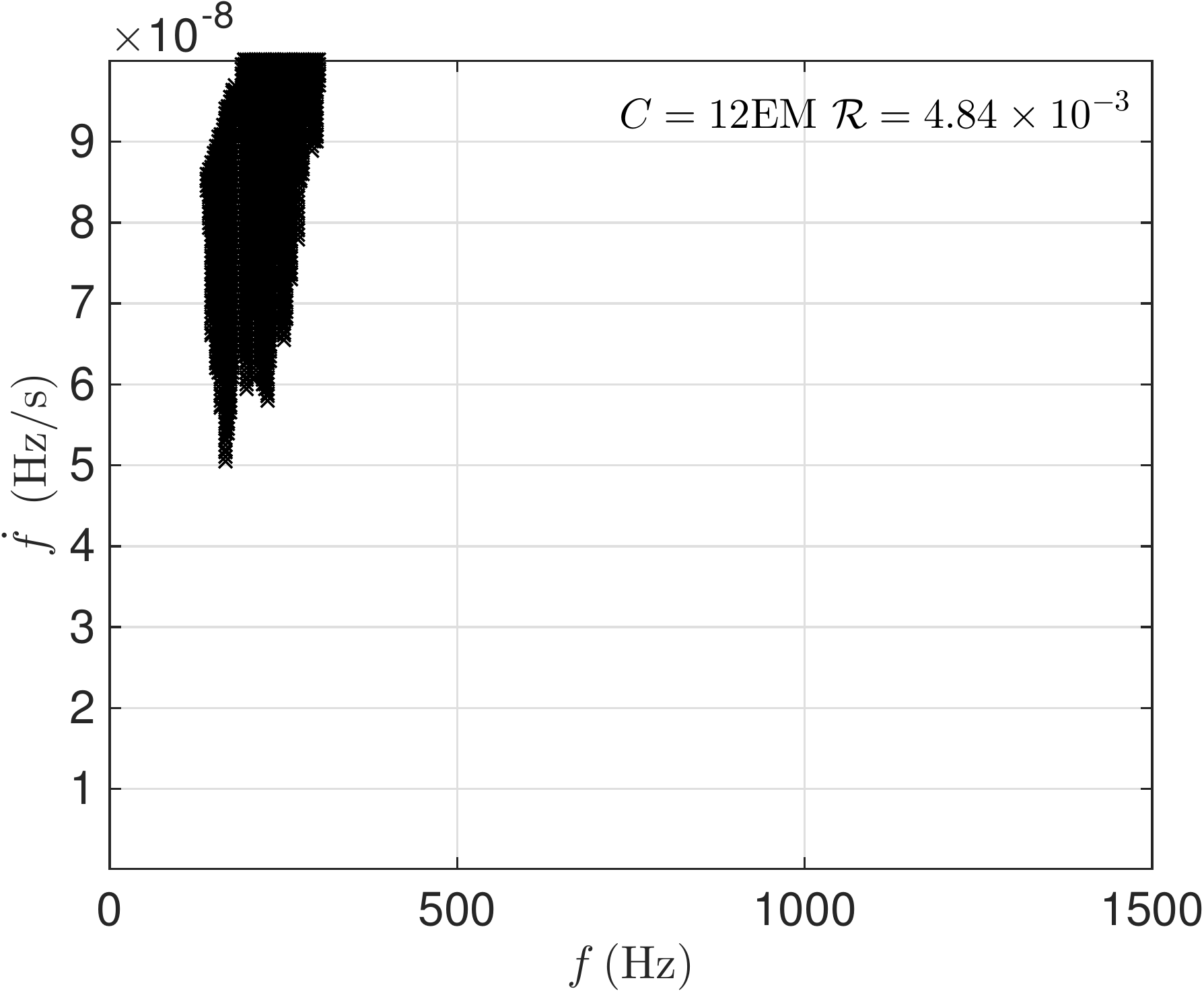}}}%
    \qquad
    \subfloat[Efficiency, 10 days]{{  \includegraphics[width=.20\linewidth]{figures/plots_nonage_eps/G2662_efficiency_10days_noage-eps-converted-to.pdf}}}%
    \qquad
    \subfloat[Coverage, 10 days]{{  \includegraphics[width=.20\linewidth]{figures/plots_nonage_eps/G2662_coverage_10days_noage-eps-converted-to.pdf}}}\\%
  
  \vspace{-0.3cm}
  
    \subfloat[Efficiency, 20 days]{{  \includegraphics[width=.20\linewidth]{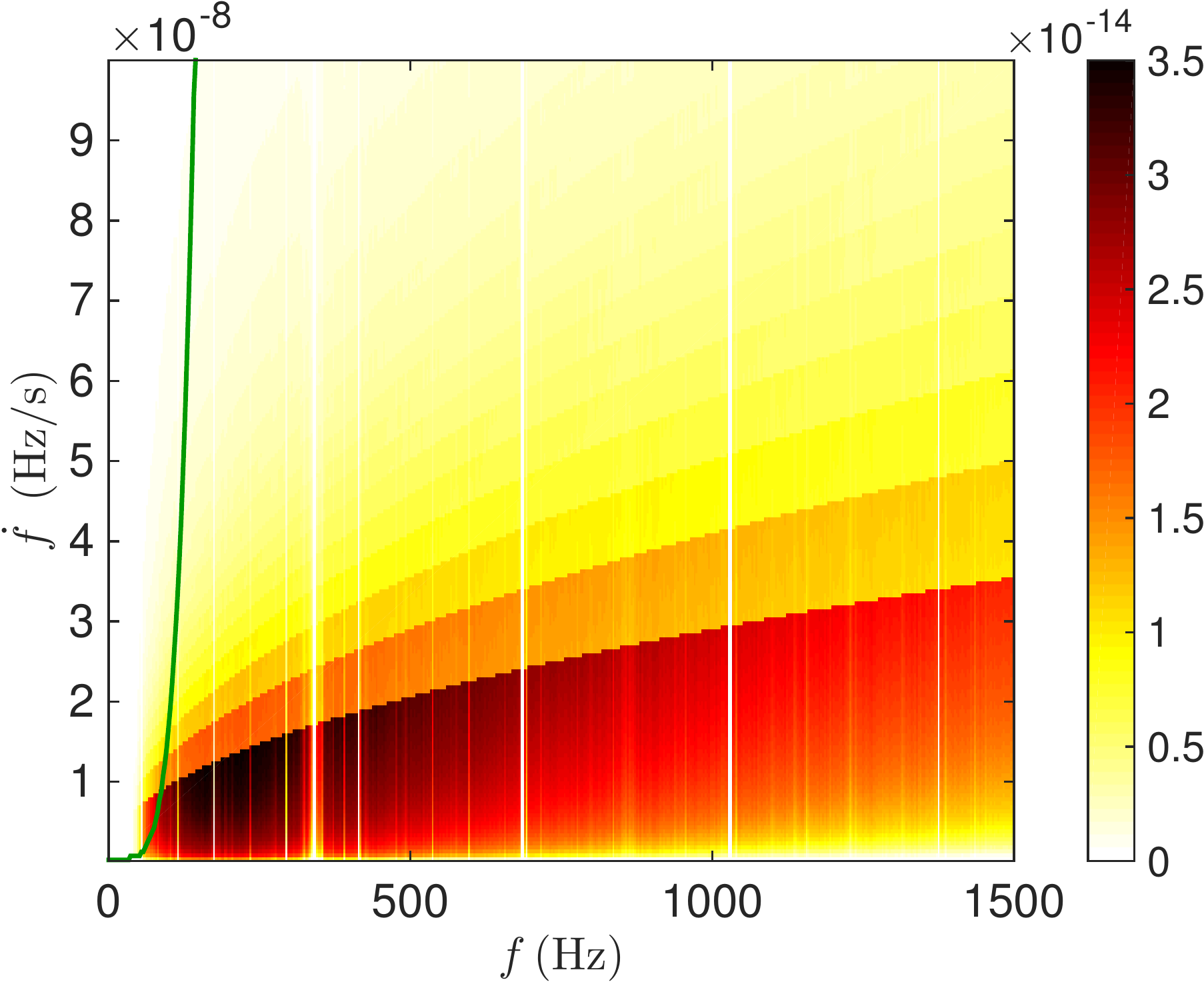}}}%
    \qquad
    \subfloat[Coverage, 20 days]{{  \includegraphics[width=.20\linewidth]{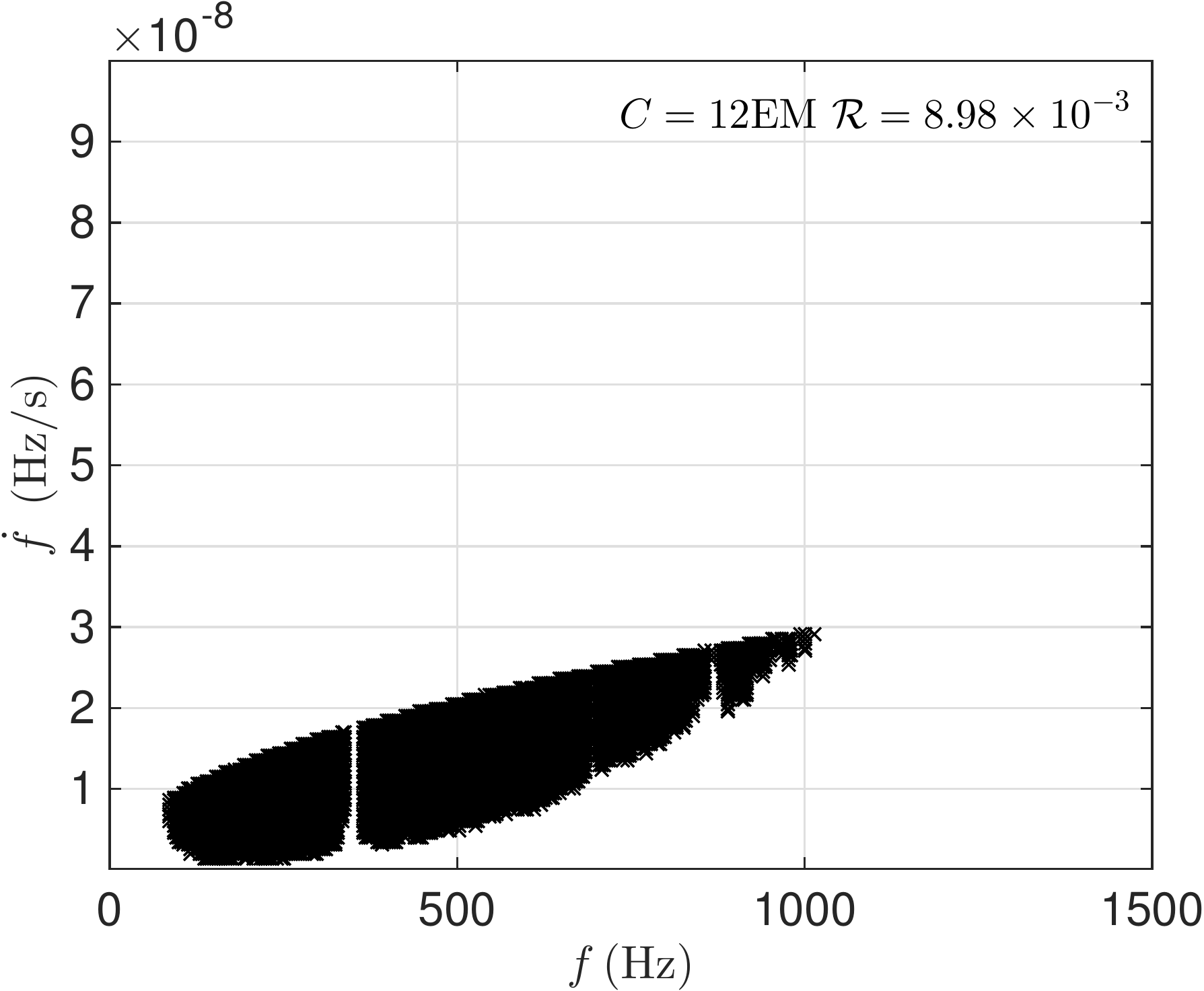}}}%
    \qquad
    \subfloat[Efficiency, 30 days]{{  \includegraphics[width=.20\linewidth]{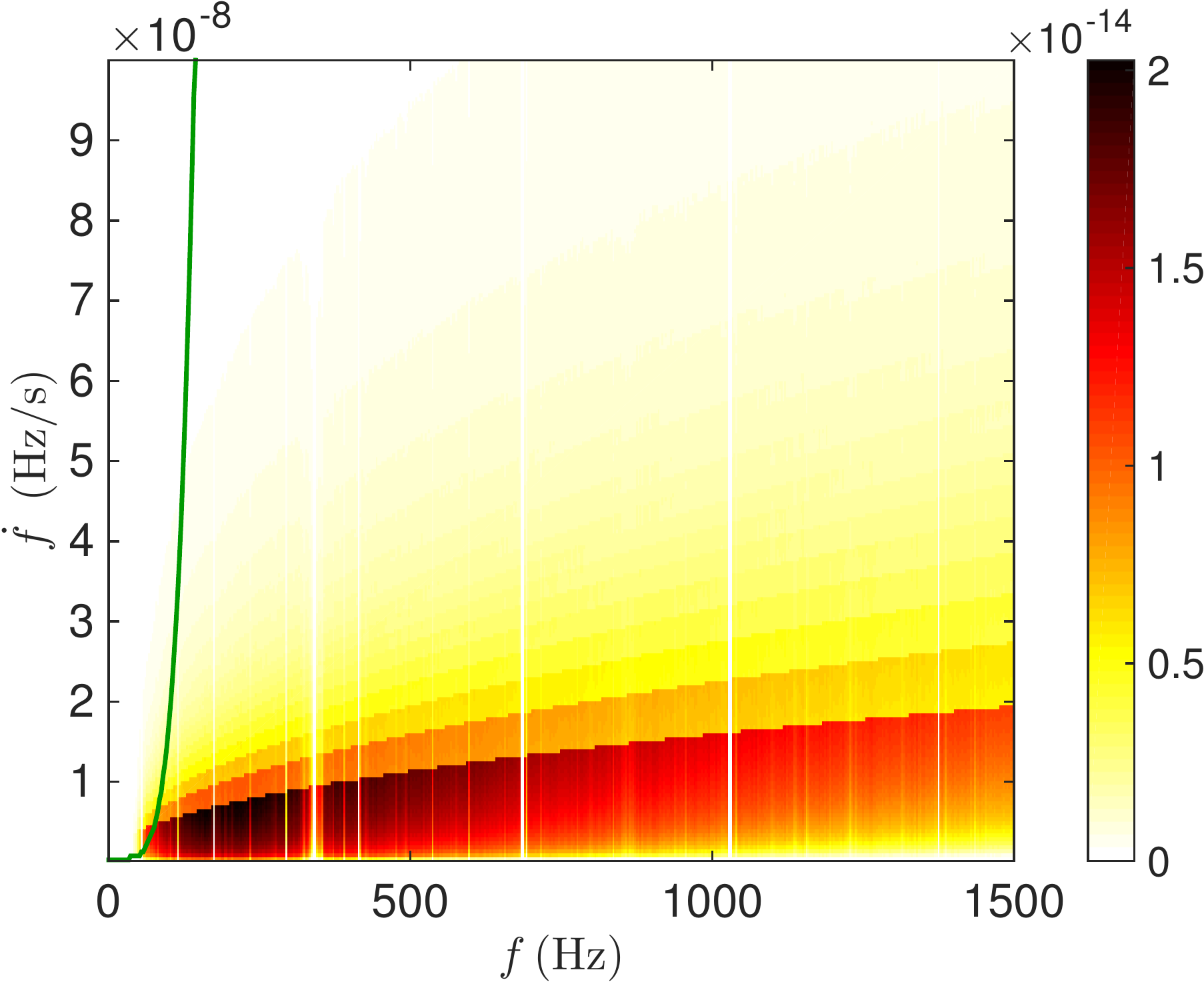}}}%
    \qquad
    \subfloat[Coverage, 30 days]{{  \includegraphics[width=.20\linewidth]{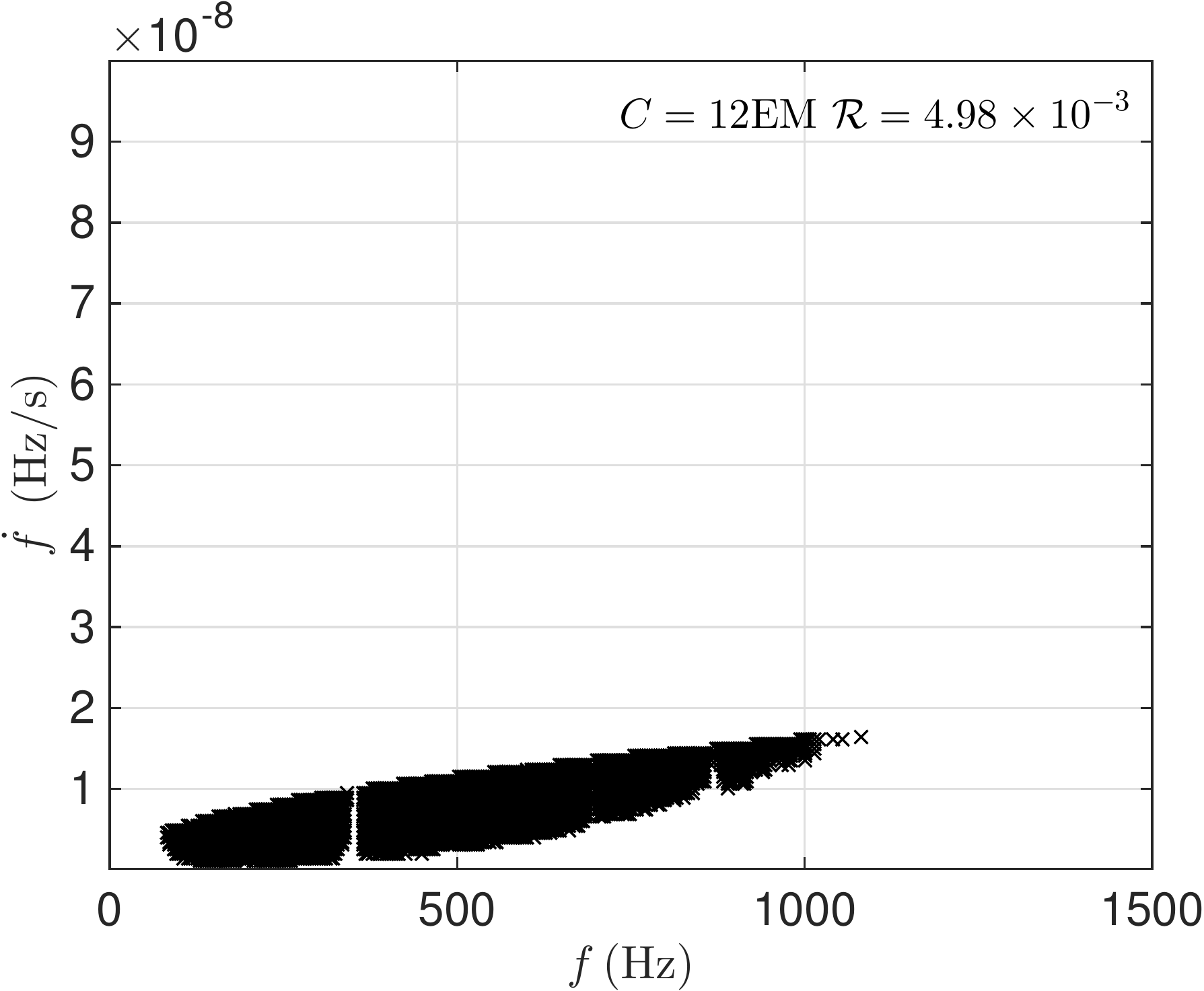}}}\\%
    
    \vspace{-0.3cm}
    
    \qquad
    \subfloat[Efficiency, 37.5 days]{{  \includegraphics[width=.20\linewidth]{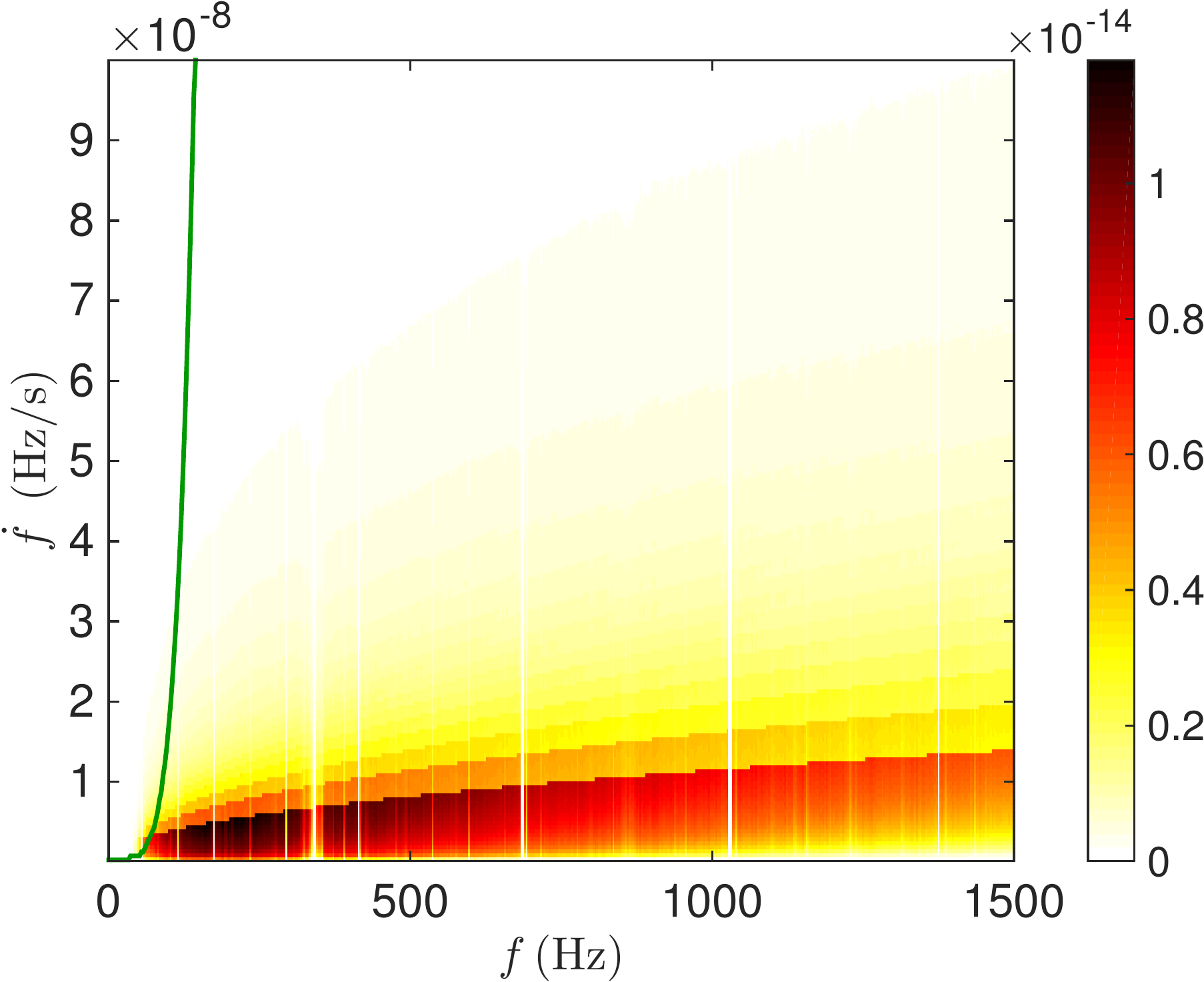}}}%
    \qquad
    \subfloat[Coverage, 37.5 days]{{  \includegraphics[width=.20\linewidth]{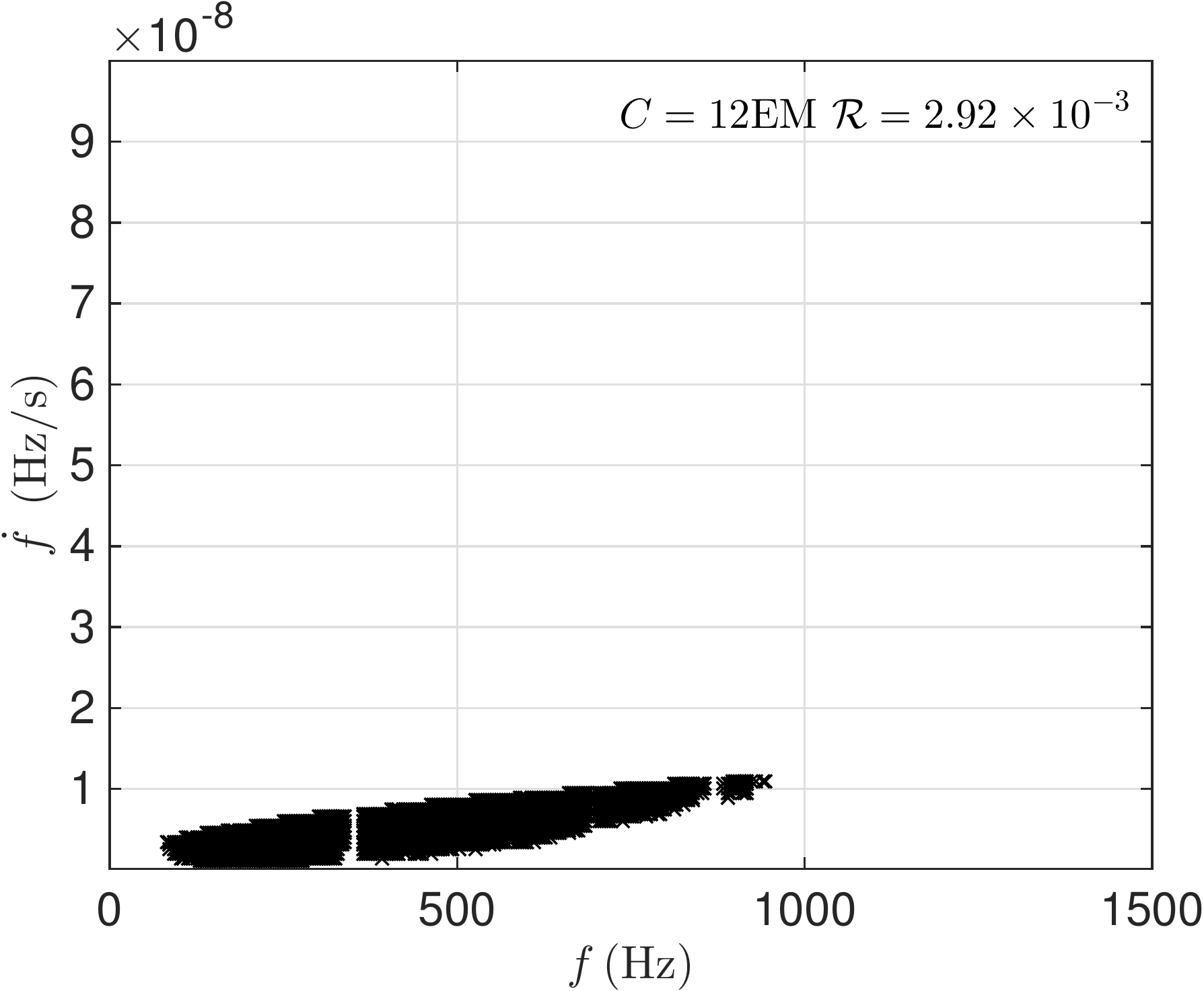}}}%
    \qquad
    \subfloat[Efficiency, 50 days]{{  \includegraphics[width=.20\linewidth]{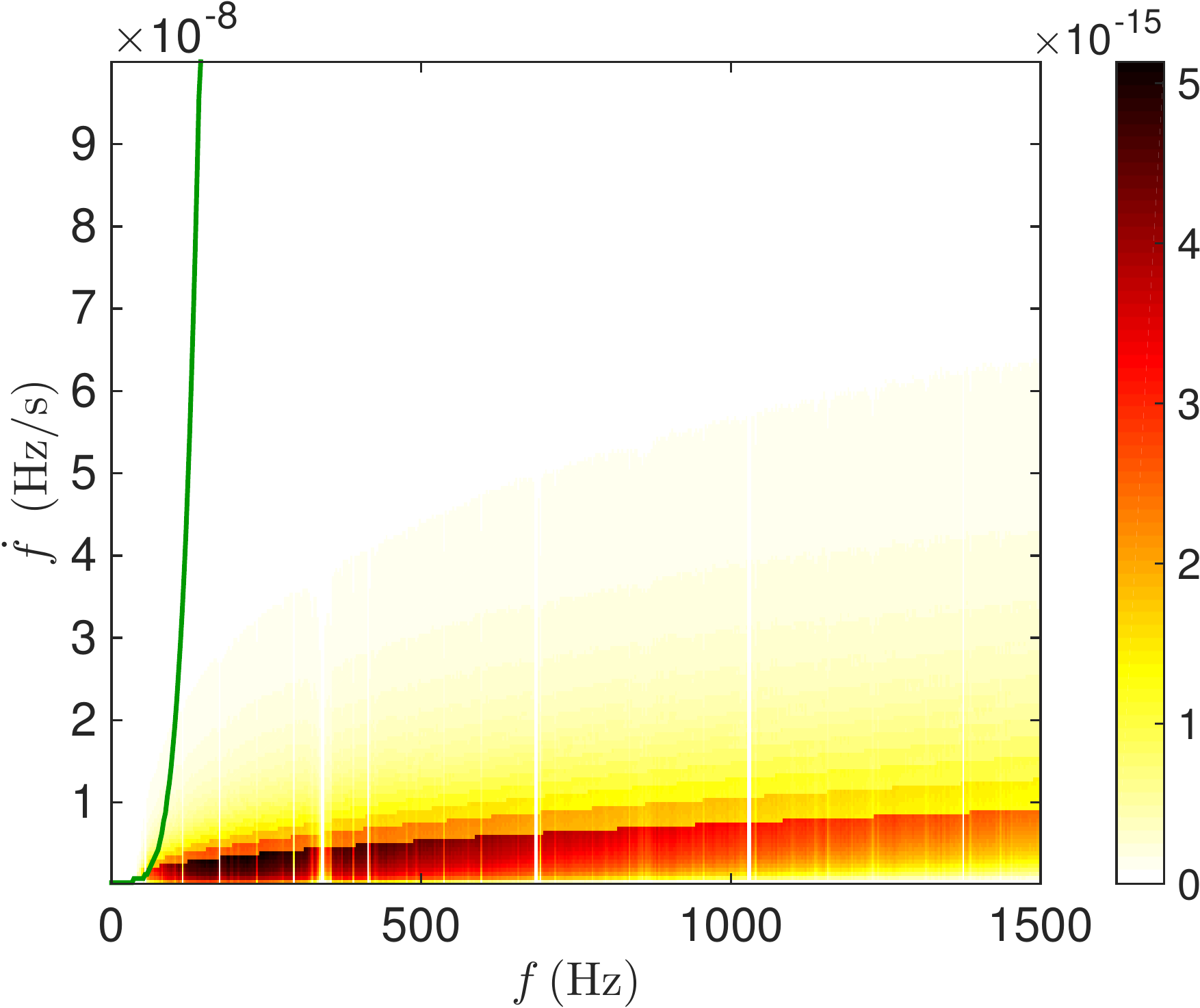}}}%
    \qquad
    \subfloat[Coverage, 50 days]{{  \includegraphics[width=.20\linewidth]{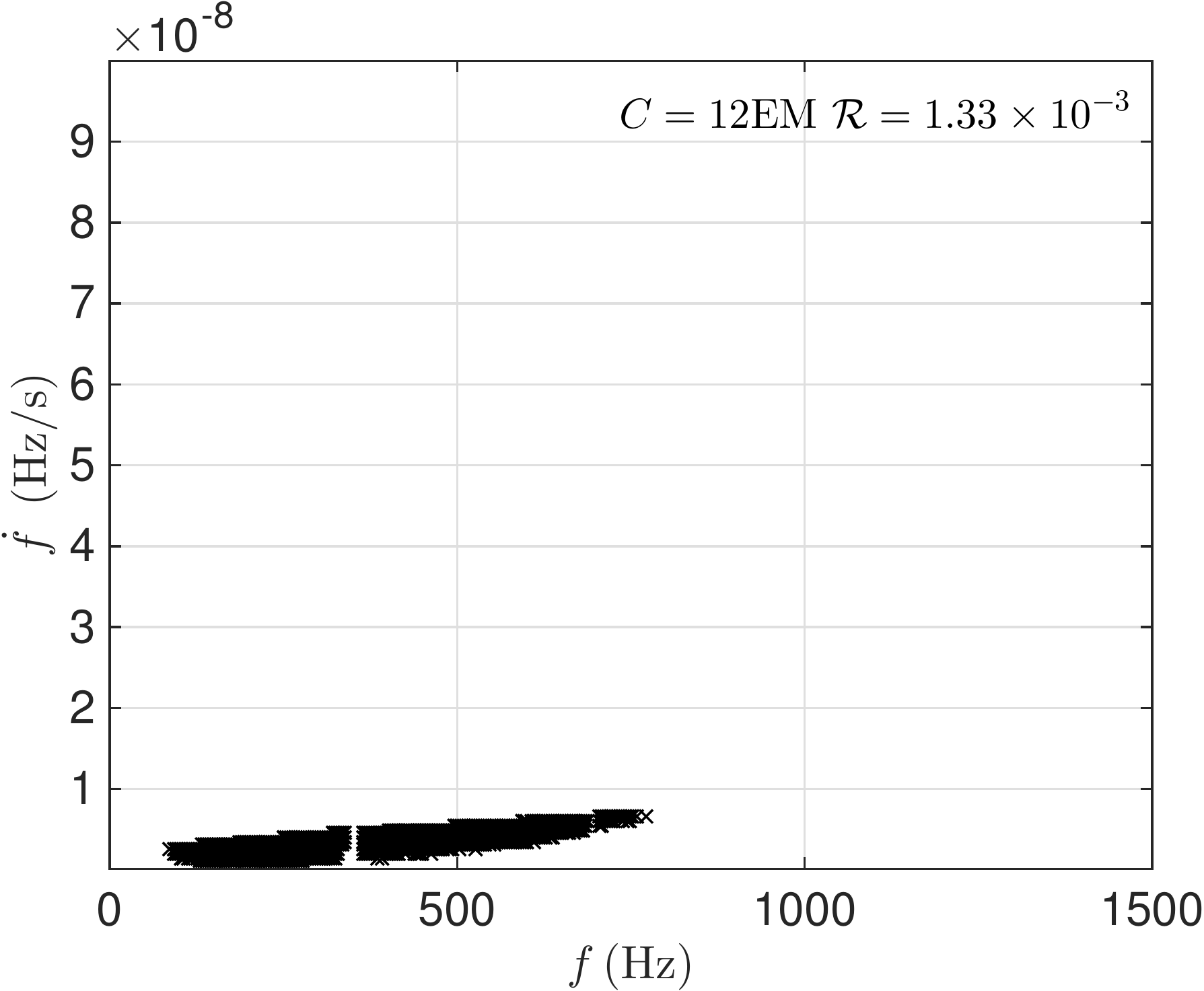}}}\\%
    
    \vspace{-0.3cm}
    
    \qquad
    \subfloat[Efficiency, 75 days]{{  \includegraphics[width=.20\linewidth]{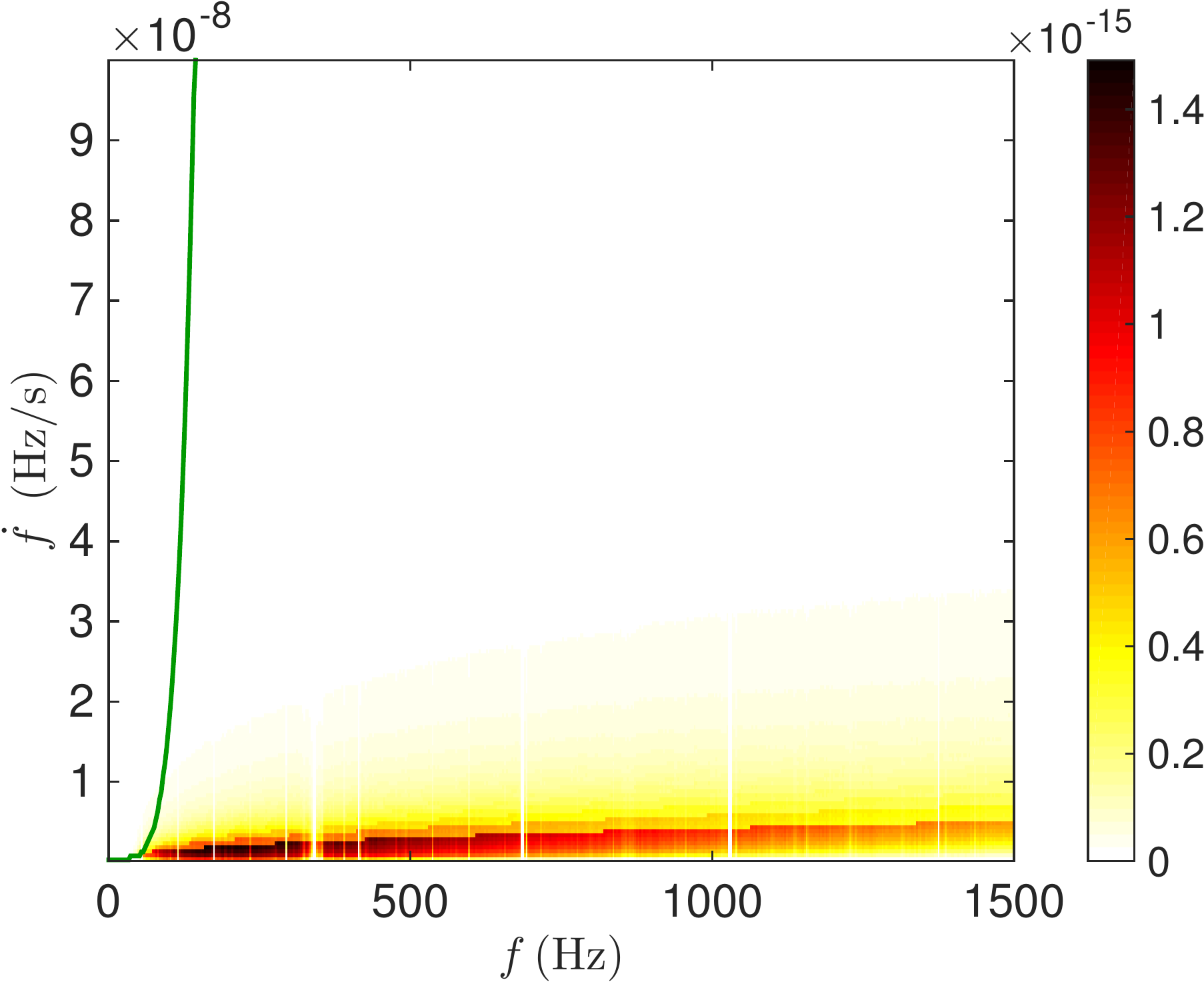}}}%
    \qquad
    \subfloat[Coverage, 75 days]{{  \includegraphics[width=.20\linewidth]{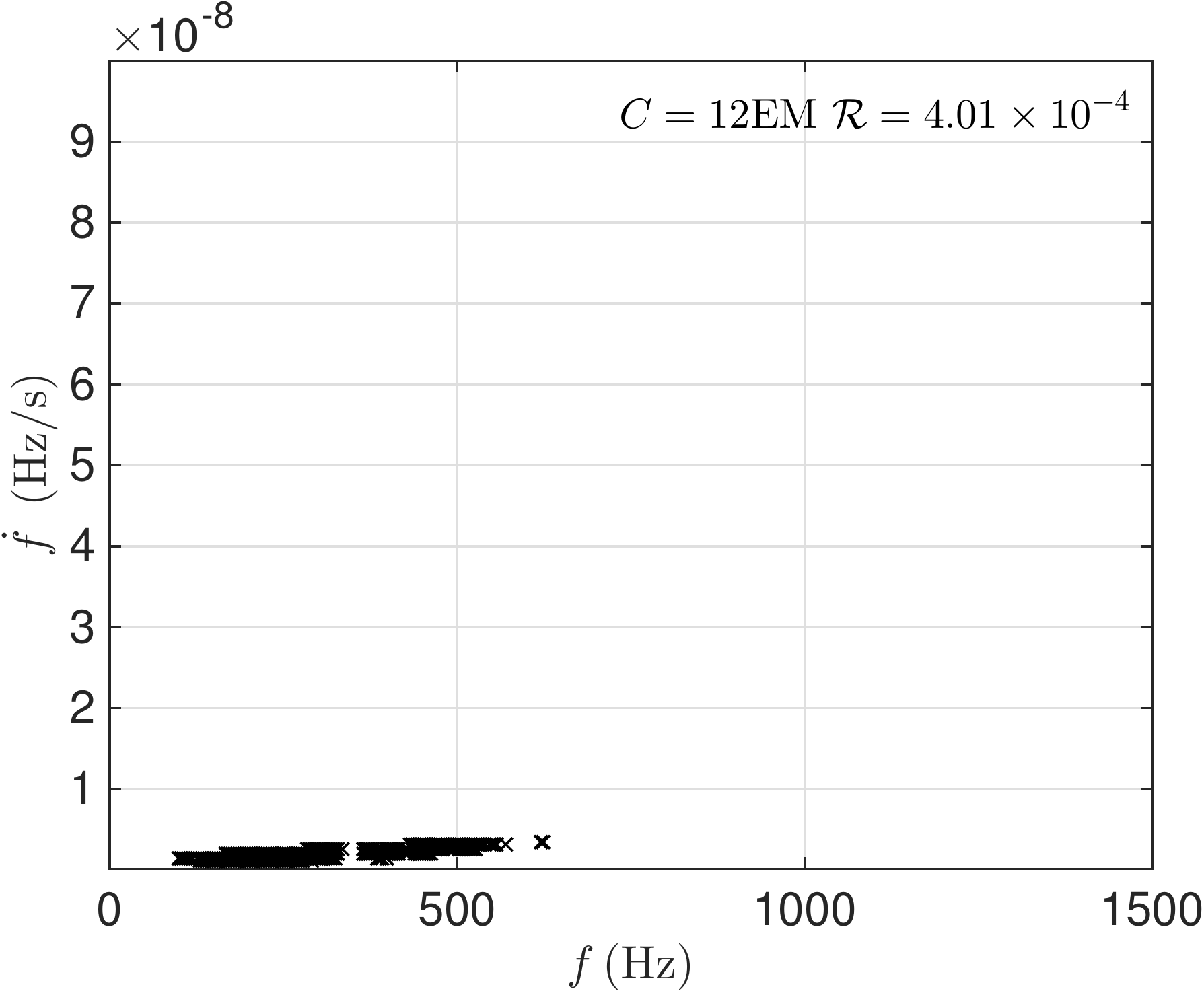}}}%
    \caption{Optimisation results for Vela Jr at 200 pc, assuming uniform and distance-based priors, for various coherent search durations: 5, 10, 20, 30, 37.5, 50 and 75 days. The total computing budget is assumed to be 12 EM.}%
    \label{G2662_51020days_noage}%
\end{figure*}

Fig.~\ref{G2662_51020days_noage} shows the efficiency and the parameter
space region that would be searched having optimized separately for every different set-up\footnote{Similar Figs.~\ref{G3473_51020days_noage} and
 \ref{CasA_51020days_noage} for sources G347.3 and Cas A respectively are listed in Appendix \ref{section:AppendixFigures}.}.
This plot shows that longer coherent segment lengths disfavour very high
values of $\dot{f}$ because the computing power grows more rapidly with increasing $\dot{f}$ than the gain in detection probability
due to the larger range in $\ddot{f}$.

\begin{figure*}%
    \centering
    \subfloat[Coverage, cost: 12 EM]{{  \includegraphics[width=.38\linewidth]{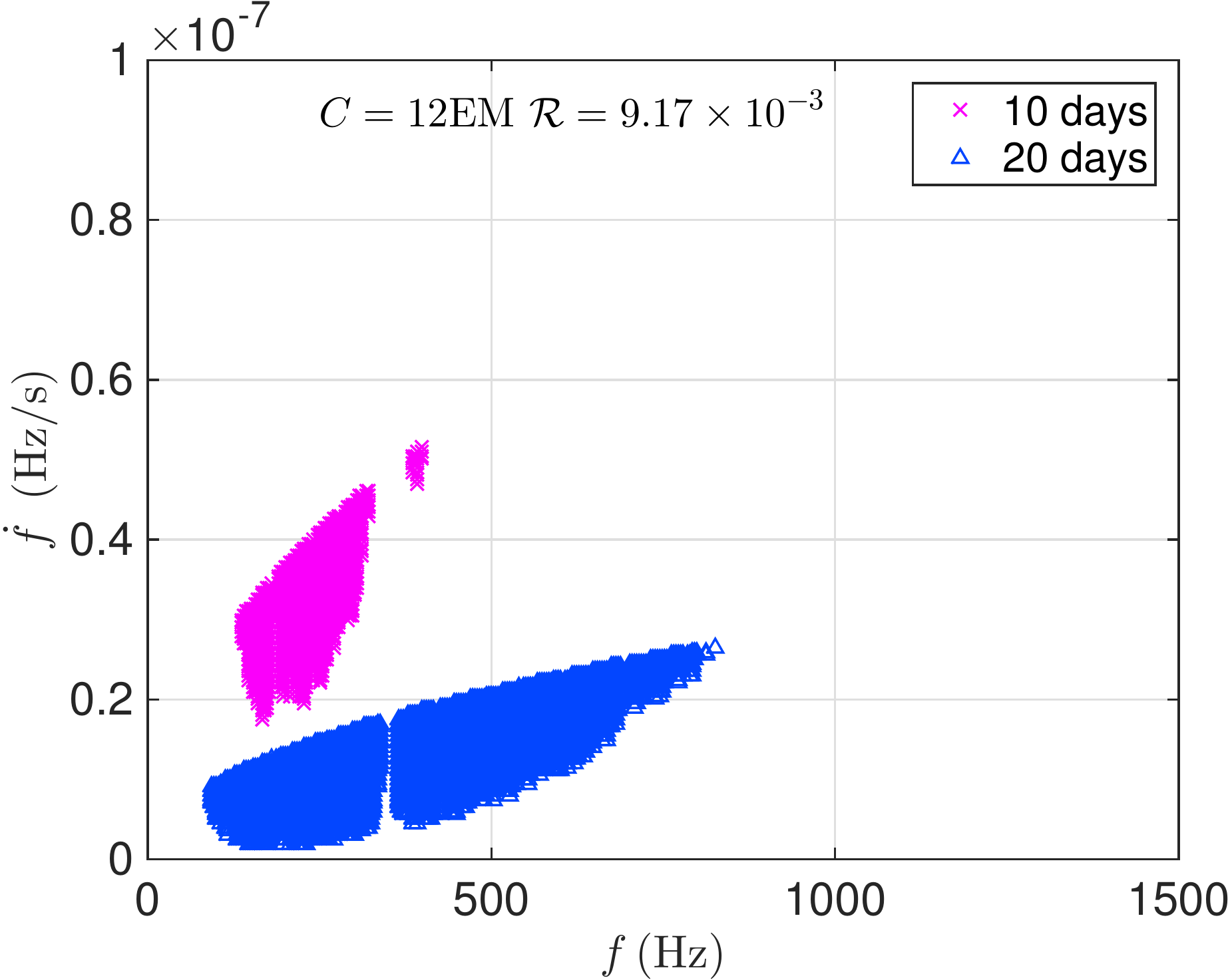}}}%
    \qquad
    \subfloat[Coverage,  cost: 24 EM]{{  \includegraphics[width=.38\linewidth]{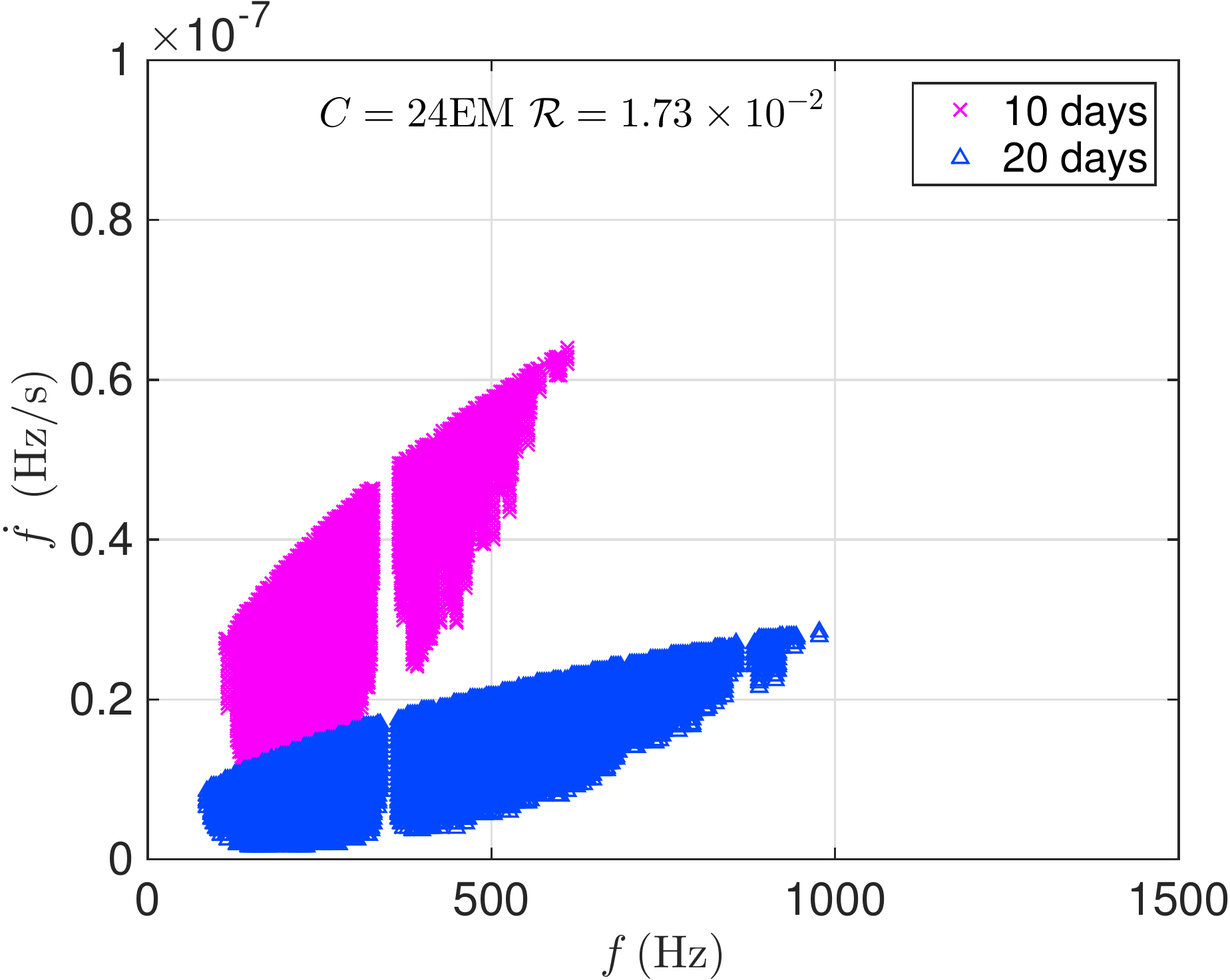}}}%
    \caption{Parameter space coverage for Vela Jr at 200 pc, assuming uniform and distance-based priors and optimizing over the 7  search set-ups also considered above at 12 EM (left plot) and 24 EM (right plot).}%
    \label{G2662_best_noage_shortdis}%
\end{figure*}

Using LP and optimizing also with respect to search set-ups we obtain the results 
shown in Fig.~\ref{G2662_best_noage_shortdis} for Vela Jr For illustration purposes we investigate different computing budgets: 12 EM and 24 EM\footnote{LP Figs.~\ref{G3473_best_noage} and \ref{CasA_best_noage} for G347.3 and Cas A respectively are in Appendix \ref{section:AppendixFigures}.}.

We note three points from these results:
\begin{itemize} 
\item For all the targets considered, doubling the computing cost increases the detection probability $\mathcal{R}$ by a factor of about 1.8 which means that the probability associated with the cells searched with the additional 12 EM is comparable with that associated to the cells searched by the first 12 EM.
\item Optimizing with respect to the set-up yields a higher  $\mathcal{R}$ as compared
to a fixed set-up. However this gain is relatively small when compared to the set-up that by itself gave the highest detection probability, as it is illustrated in  Fig.~\ref{fig:RversusDistAndTcoh}(b). There the  
solid lines show the detection probability attainable by combining different set-ups for every target and the non-continuous line show the detection probability optimized at fixed set-up. For example, the $\mathcal{R}$ for Vela Jr.
with the 20-day set-up is $8.98\times10^{-3}$ and it grows to $9.17\times10^{-3}$ by combining different set-ups;
G347.3 similarly increases from $4.36\times10^{-3}$ to $4.57\times10^{-3}$ and Cas A from $2.34\times10^{-3}$ to $2.44\times10^{-3}$. 
\item When optimizing for each target also with respect to set-up, the cells selected for Cas A's cells include the 5-day set-up (Fig.~\ref{CasA_best_noage} in Appendix \ref{section:AppendixFigures}), unlike for the other targets. The reason is that Cas A is the farthest of the considered targets and hence we gain more detection probability by searching higher spindowns, which in turn means higher maximum $h_0$ and hence higher detection probability, than by including more high-frequency cells. The cost of the high-spindown regions is higher than that of lower spindown ones and this is compensated by the optimization procedure by using a shorter time-baseline set-up. 
\end{itemize}

Since our goal is to optimize the probability of making a detection
from \emph{any} source, we do not want to restrict ourselves
\emph{a priori} to a particular source, and hence we optimize now also with respect 
to the targets.

Since the distance to Vela Jr is uncertain we consider two sets of three targets: Vela
Jr at 200 pc and at 750 pc, G347.3 and IC 443. We will show results for 12EM, 24EM and 48EM computing budget $C_\mathrm{max}$ (Fig.~\ref{all_noage} in Appendix \ref{section:AppendixFigures}). 
When Vela Jr is assumed at 200 pc, even when $C_\mathrm{max}=48$ EM, all the picked cells are from Vela Jr.
This is because Vela Jr is so much closer to us than the others that the detection probability is maximized by always targeting Vela Jr. 
Thus, if we really believe that Vela Jr is 200 pc away,
then we should concentrate all our computing budget on it.  
If in the optimization process we assume that Vela Jr is 750 pc away, then the result changes. 
With a 12 EM budget, cells both from Vela Jr and
G347.3 are picked. If we double the budget, some cells from IC 443  become worth searching and this effect becomes 
even more prominent if we quadruple the budget. However, even at $C_\mathrm{max}=48$ EM most of the searched parameter space targets 
Vela Jr.  Table~\ref{tab: R of nonage} lists the $\mathcal{R}$ numbers and Fig.~\ref{fig:RversusDistAndTcoh}(c) diplays them as a function of $C_\mathrm{max}$. Note that the highest $\mathcal{R}$ with respect to set-up is in bold font.

\subsubsection{Results with age-based priors}
\label{subsect:agepriorsresults}

\begin{figure}%

 \centering
 \subfloat[${\mathcal{R}}$ versus distance]{{\includegraphics[width=.9\linewidth]{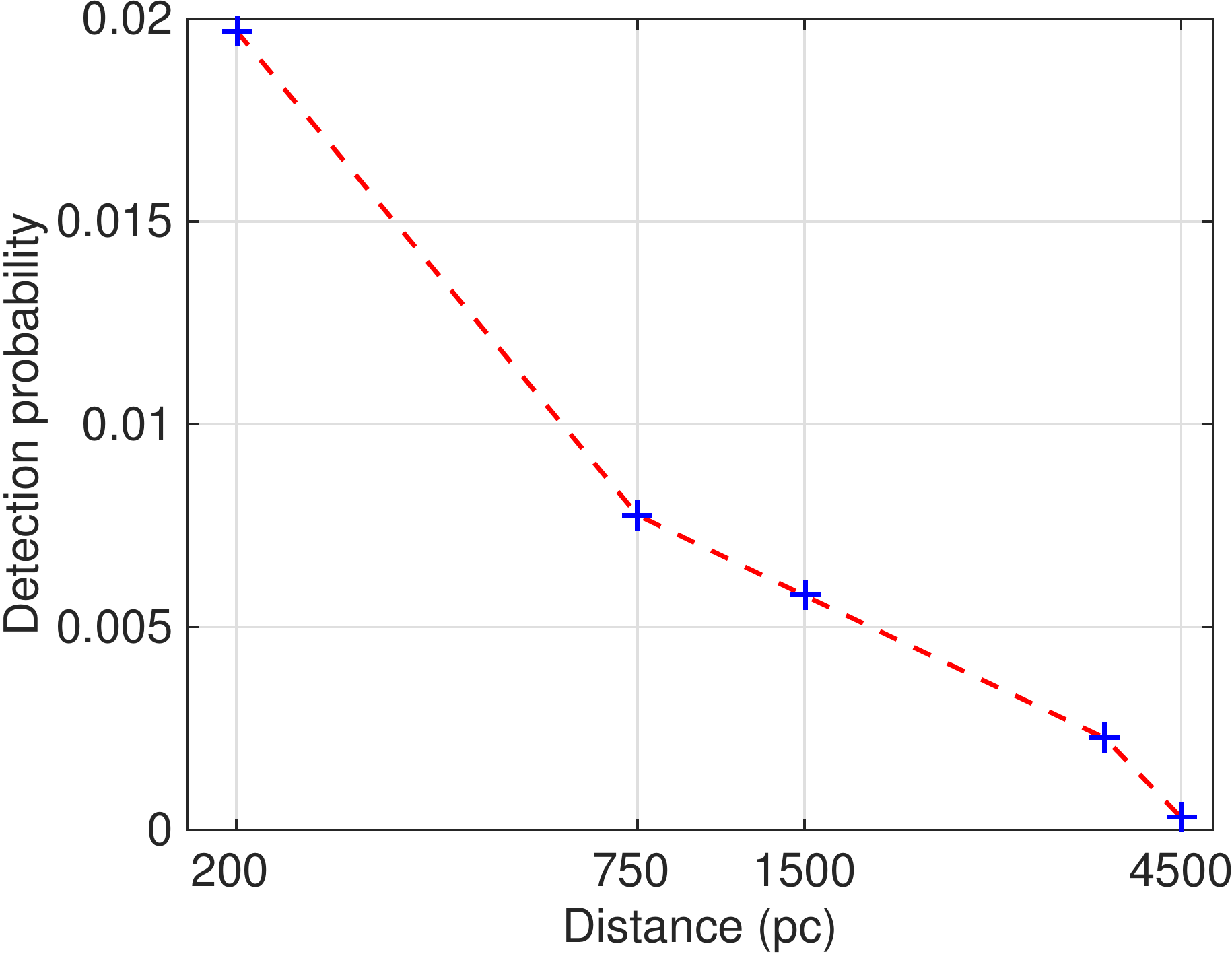}}}%
 \qquad
 \subfloat[${\mathcal{R}}$ versus $T_\mathrm{coh}$]{{\includegraphics[width=.9\linewidth]{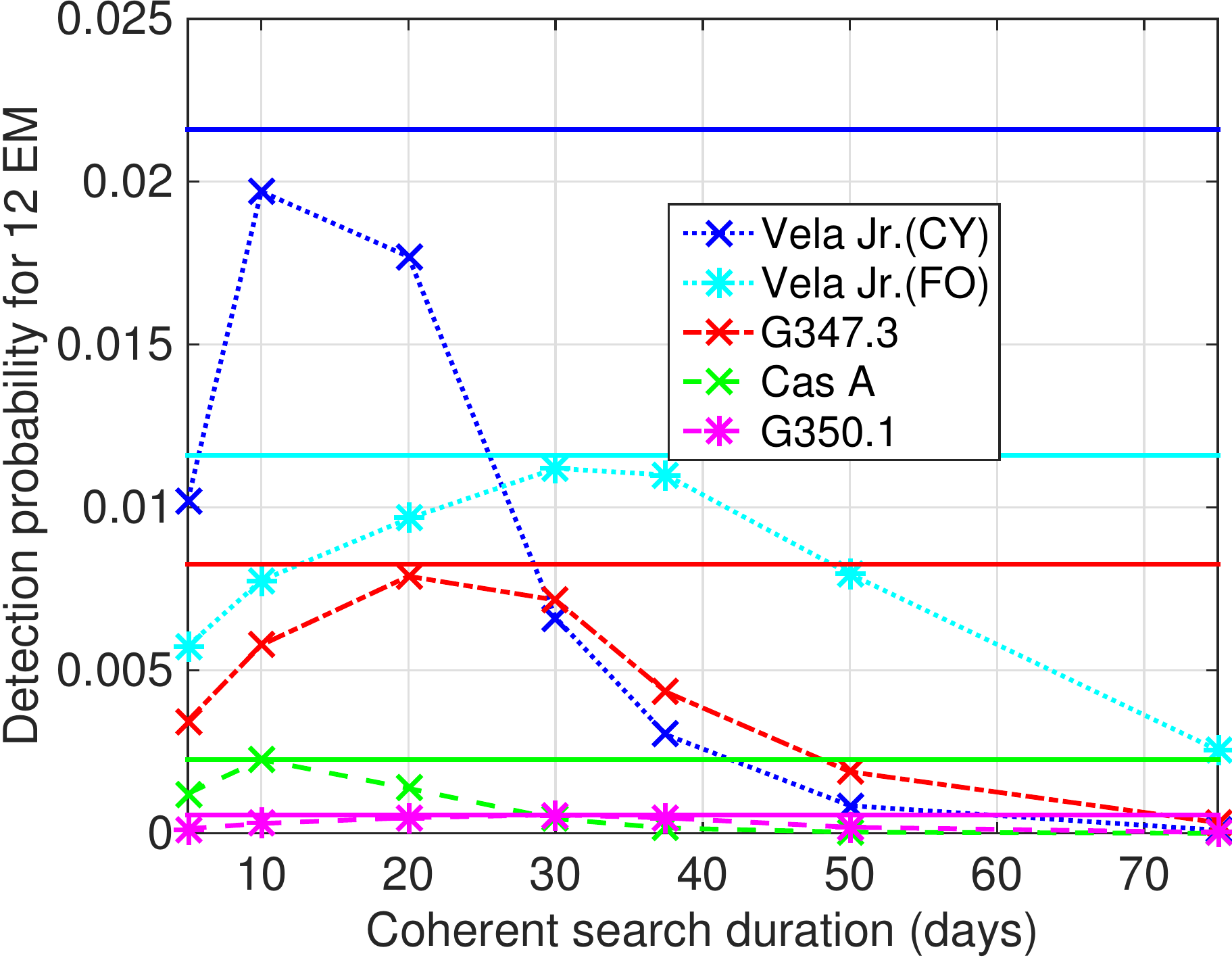}}}%
 \qquad
 \subfloat[${\mathcal{R}}$ versus $C_\mathrm{max}$]{{\includegraphics[width=.9\linewidth]{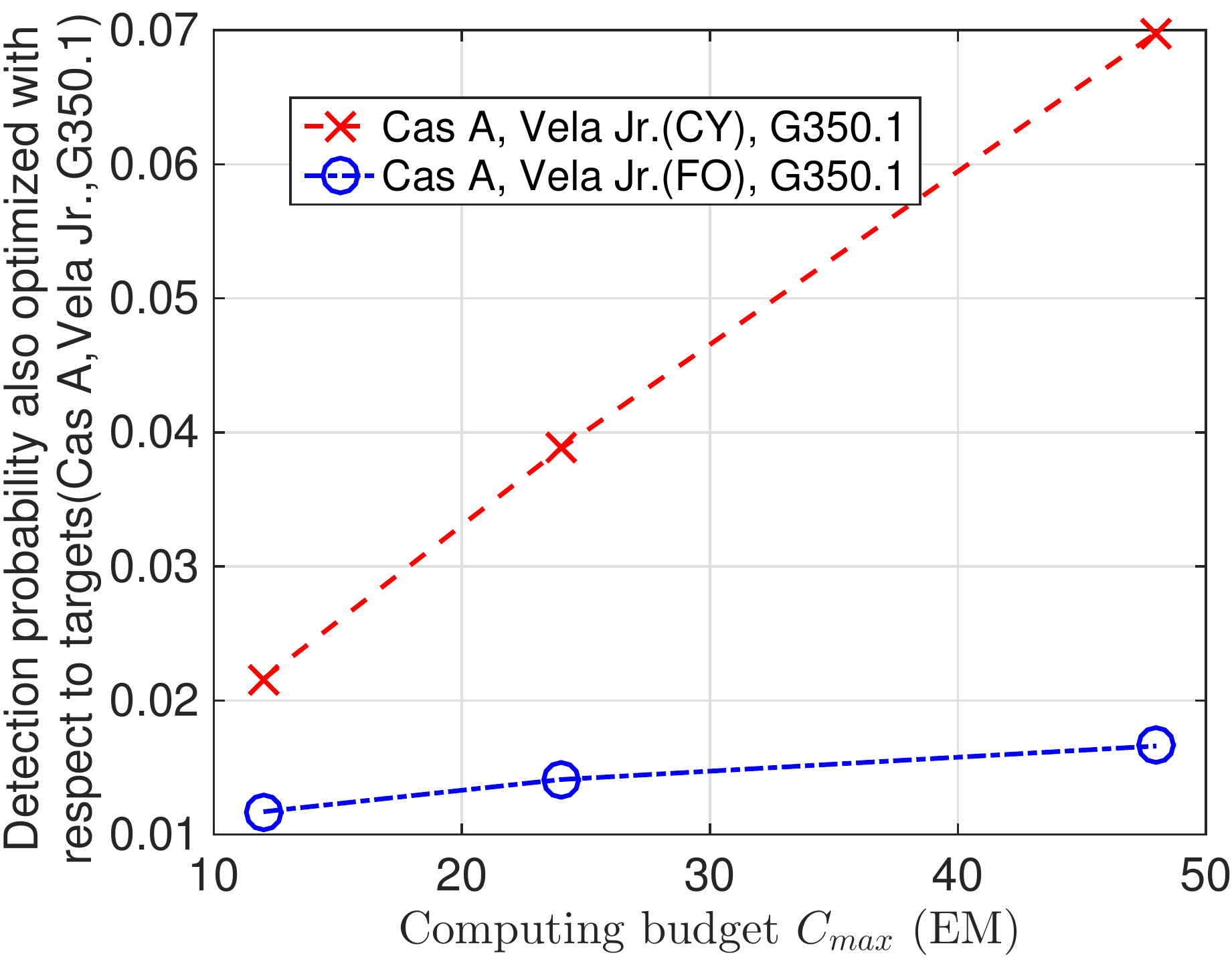}}}%

 \caption{Detection probability for various targets and search set-ups having assumed uniform and age-based priors. The distances that were assumed for the targets are: Vela Jr (C) 200 pc, Vela Jr (F) 750 pc, G 347.3 1.3 kpc, Cas A 3.5 kpc, G350.1 4.5 kpc.}%

 \label{fig:RversusDistAndTcohAge}
\end{figure}

\begin{table*}
\caption{
  \label{tab: R of age}
  $\mathcal{R}$ result with $f$ and $\dot{f}$ uniform priors and age-based priors. The highest $\mathcal{R}$ with respect to set-up is in bold font.}
\begin{center}

  \begin{tabular}{lll|cccccccccc}
    \tableline
   
   &&& \multicolumn{10}{c}{$10^3\mathcal{R}$}
   \\
   \cline{4-13}
  
    Name & $~~D_\mathrm{kpc}$ & $\tau_\mathrm{kyr}$ &5D & 10D&20D&30D&37.5D&50D&75D& \multicolumn{3}{c}{LP Optimized}
    \\
    &&& \multicolumn{7}{c}{Computing Budget: 12EM} &12EM&24EM&48EM\\
    \tableline

   Cas~A &~~3.5&0.35&1.22&\textbf{2.26}&1.38&0.446&0.164&0.036&0.003&2.26&3.32&--
    \\
    G350.1 &~~4.5 &0.9&0.142&0.303&0.480&\textbf{0.569}&0.474&0.187&0.027&0.559&0.640&--
    \\ 
     G347.3&~~1.3 &1.6&3.45&5.78&\textbf{7.89}&7.16&4.34&1.89&0.330&8.27&9.89&--
   \\
 Vela~Jr &~~0.2&0.7&10.2&\textbf{19.7}&17.7&6.62&3.03&0.850&0.097&21.6&38.8&--
 \\
  Vela~Jr &~~0.2&4.3&33.2&56.7&\textbf{67.1}(11.3EM)&56.2&38.5&20.3&5.35&--&--&--
\\
  Vela~Jr &~~0.75&4.3&5.75&7.76&9.66(11.3EM)&\textbf{11.2}&11.0&7.96&2.55&11.6&13.0&--
 \\
  Top 3 (CY) &~~--&--&--&--&--&--&--&--&--&21.6&38.8&69.8
 \\
  Top 3 (FO)&~~--&--&--&--&--&--&--&--&--&11.7&14.1&16.6
\\
    \tableline
  \end{tabular}
\end{center}
\end{table*}

\begin{figure*}%
    \centering
    \subfloat[Efficiency, 5 days]{{  \includegraphics[width=.20\linewidth]{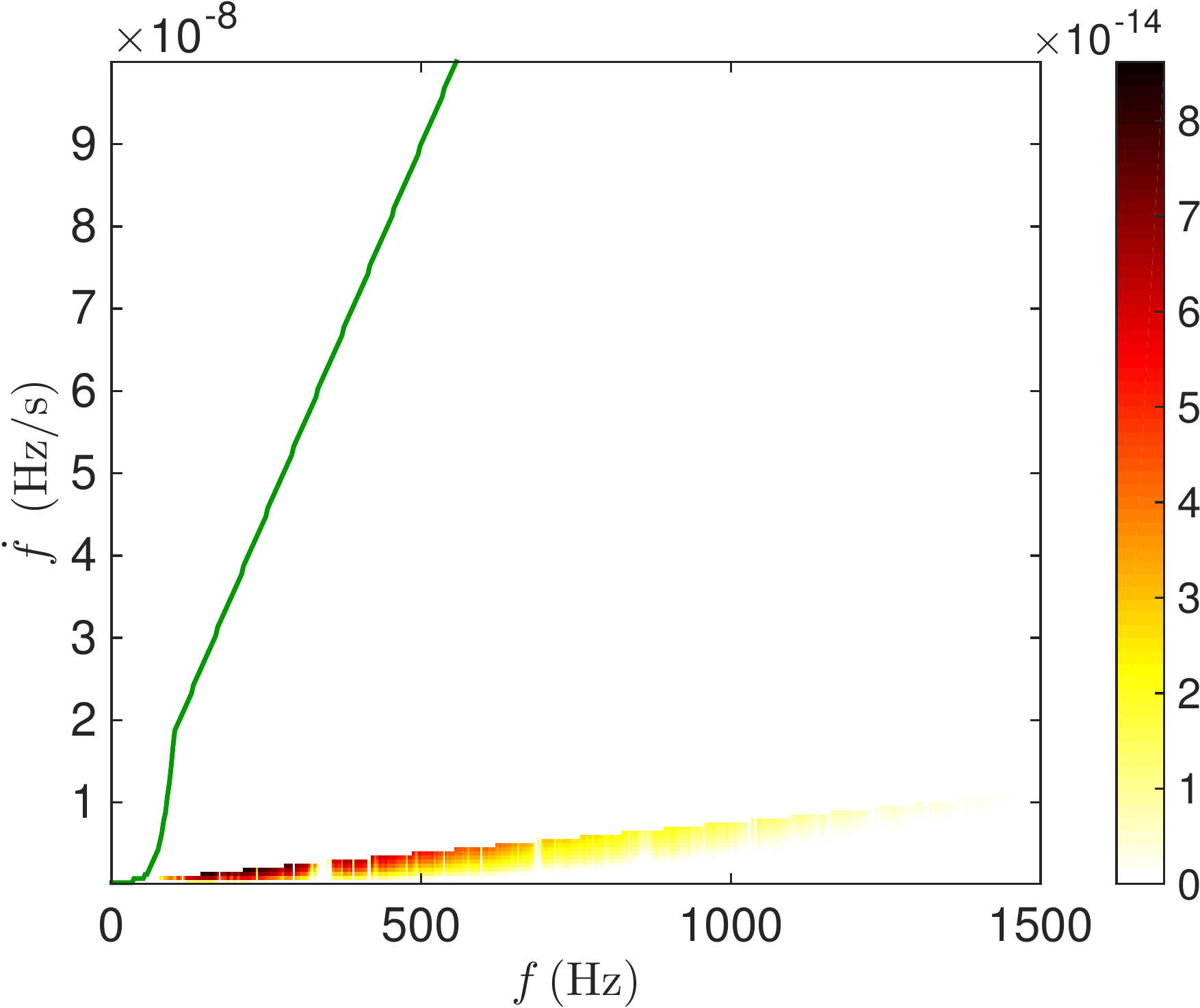}}}%
    \qquad
    \subfloat[Coverage, 5 days]{{  \includegraphics[width=.20\linewidth]{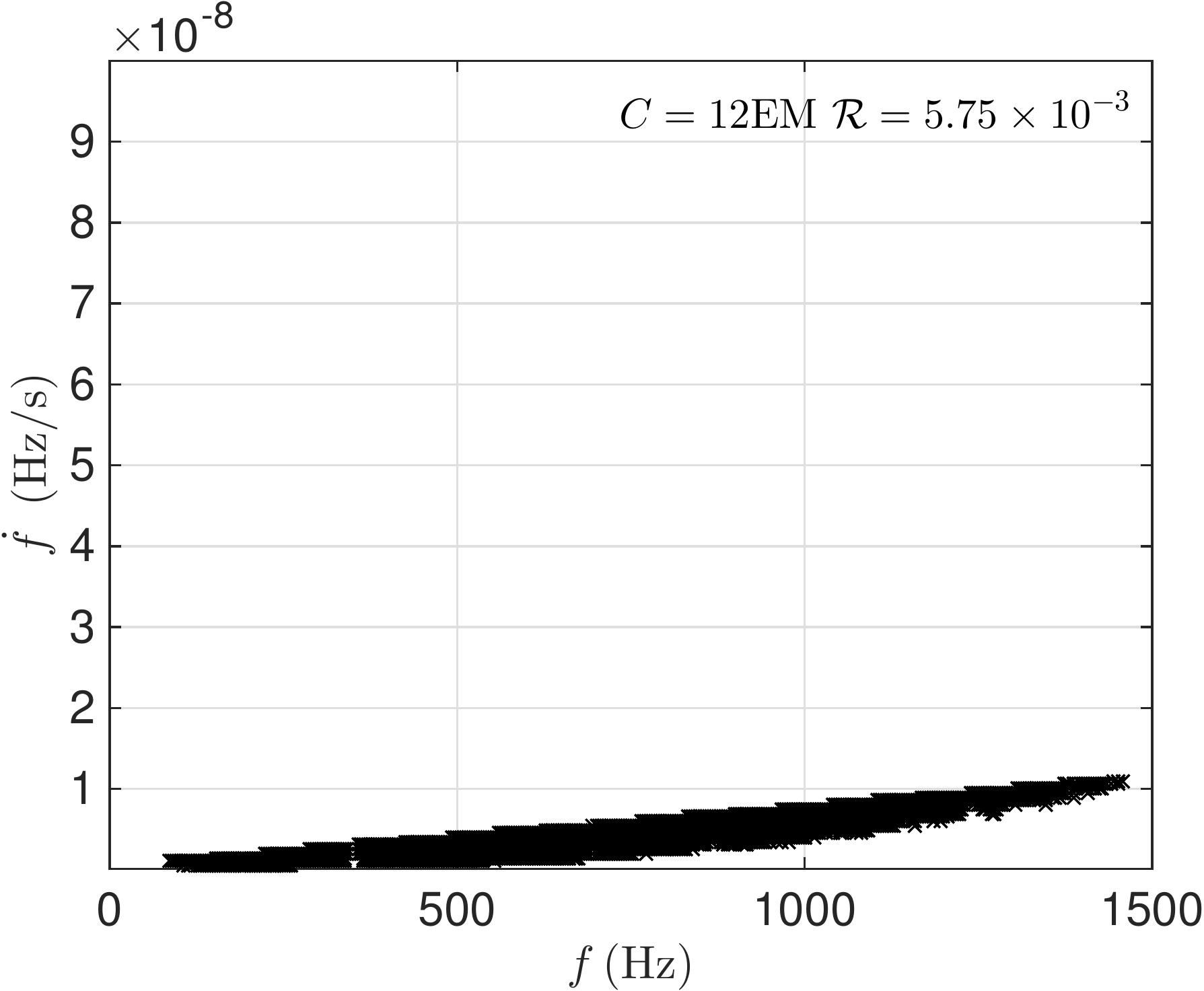}}}%
    \qquad
    \subfloat[Efficiency, 10 days]{{  \includegraphics[width=.20\linewidth]{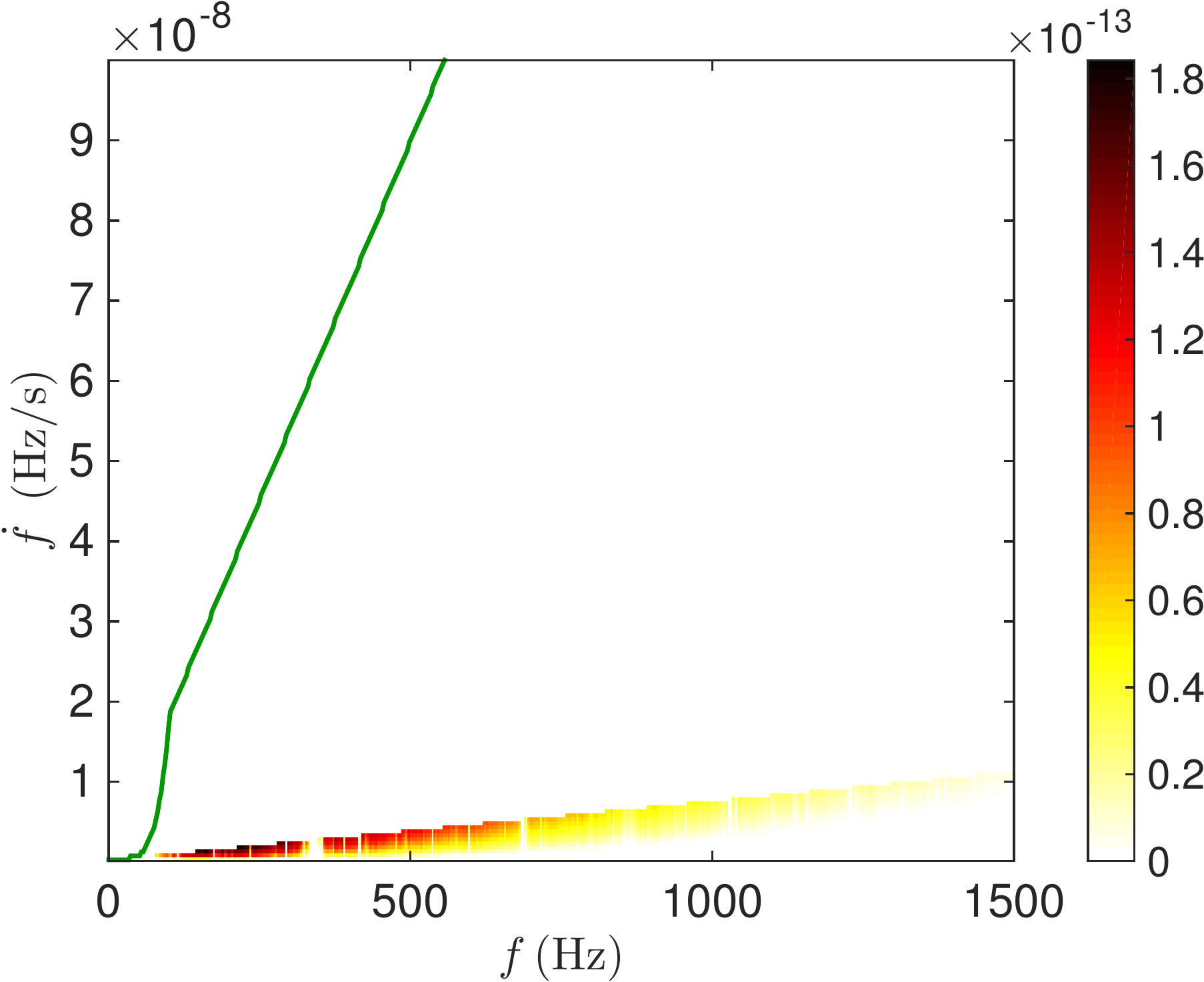}}}%
    \qquad
    \subfloat[Coverage, 10 days]{{  \includegraphics[width=.20\linewidth]{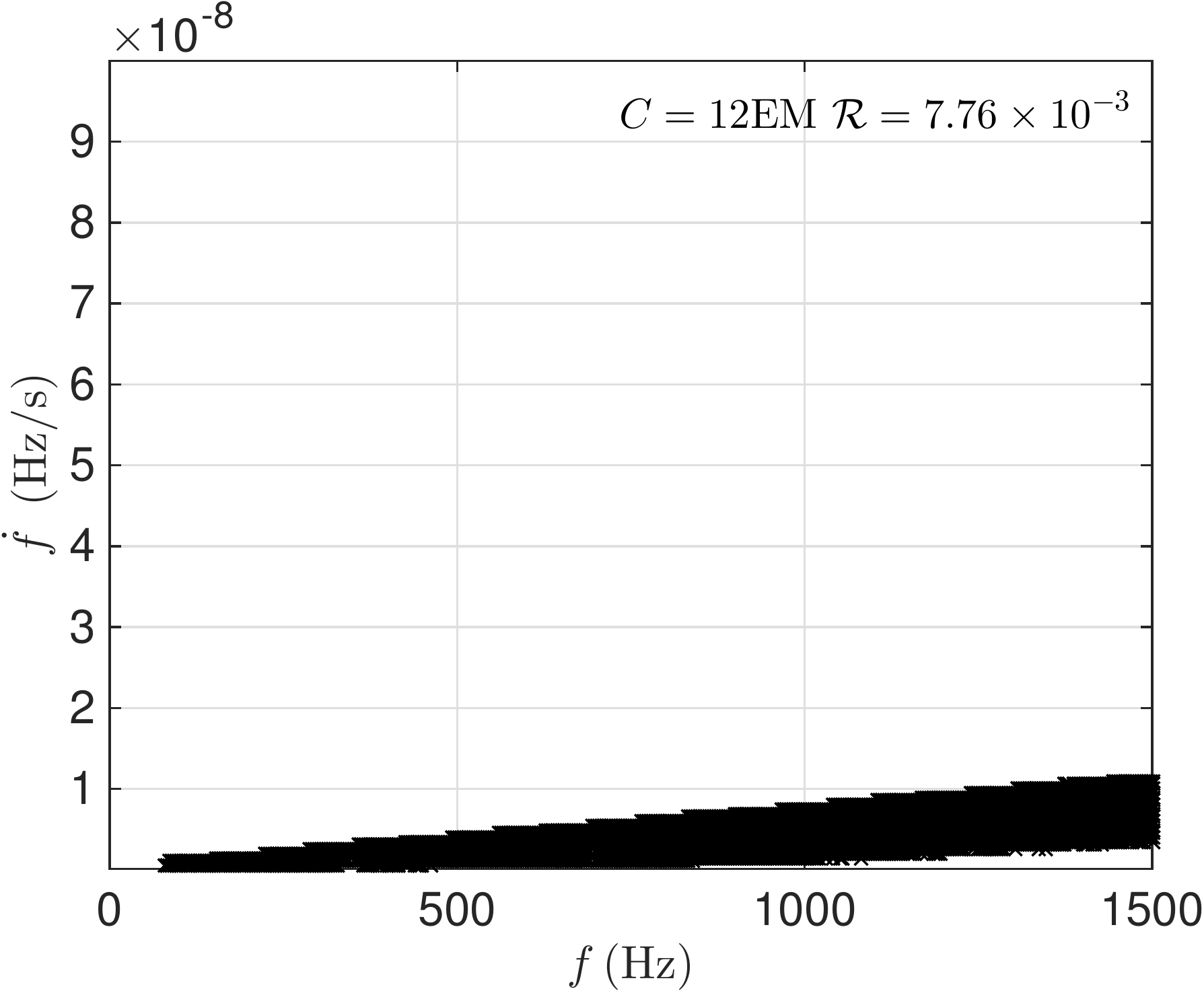}}}\\%
  
  \vspace{-0.3cm}
  
    \subfloat[Efficiency, 20 days]{{  \includegraphics[width=.20\linewidth]{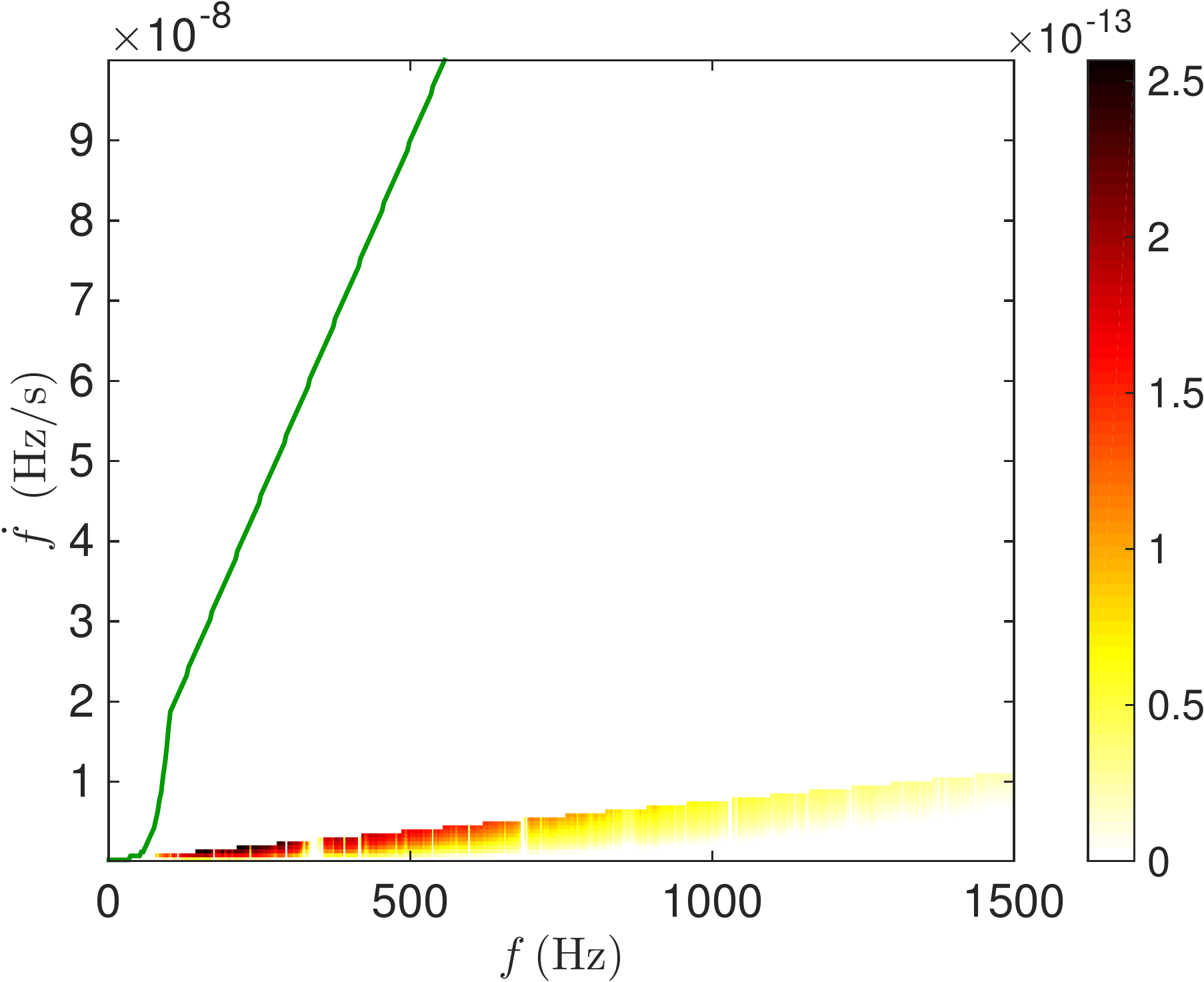}}}%
    \qquad
    \subfloat[Coverage, 20 days]{{  \includegraphics[width=.20\linewidth]{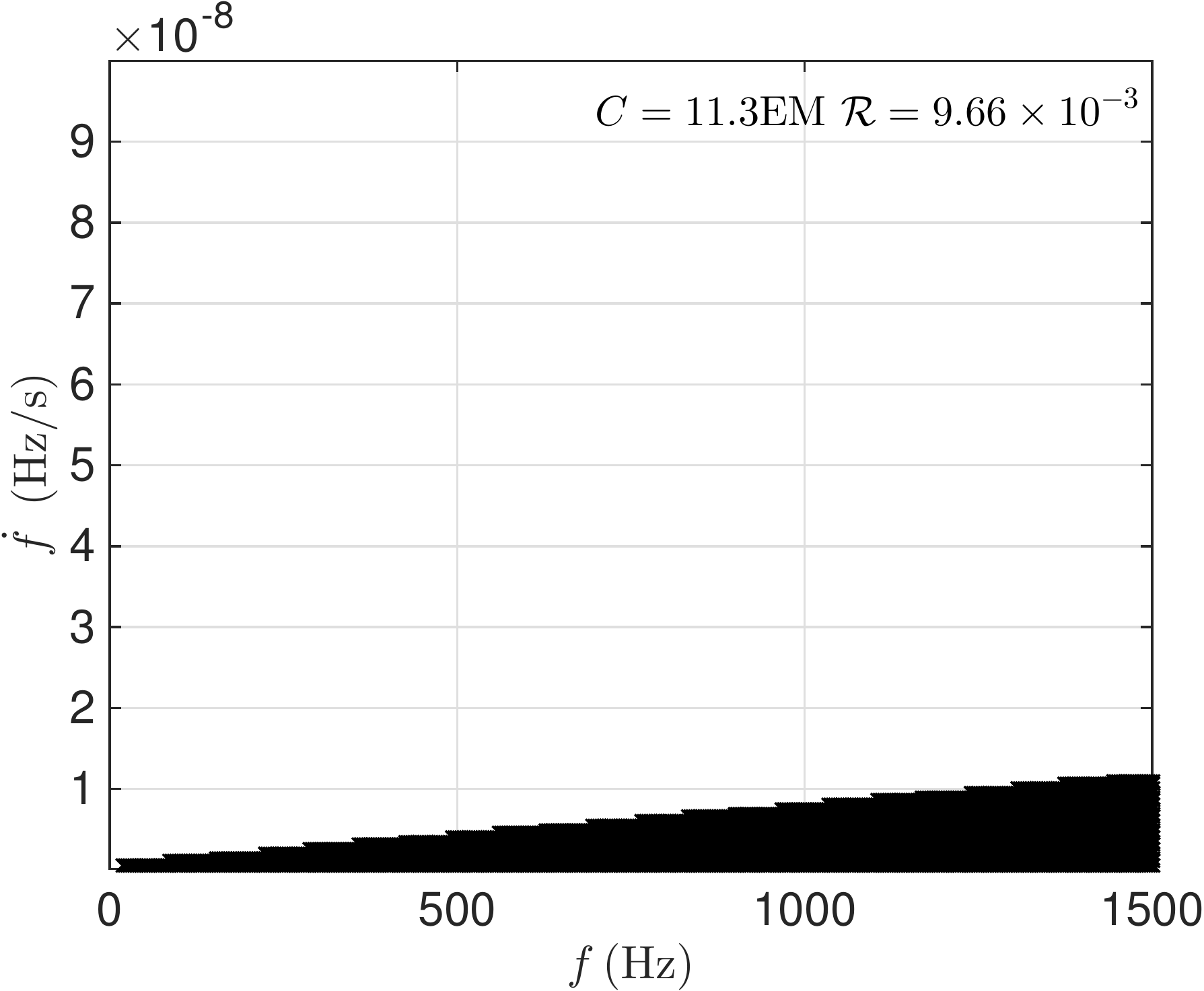}}}%
    \qquad
    \subfloat[Efficiency, 30 days]{{  \includegraphics[width=.20\linewidth]{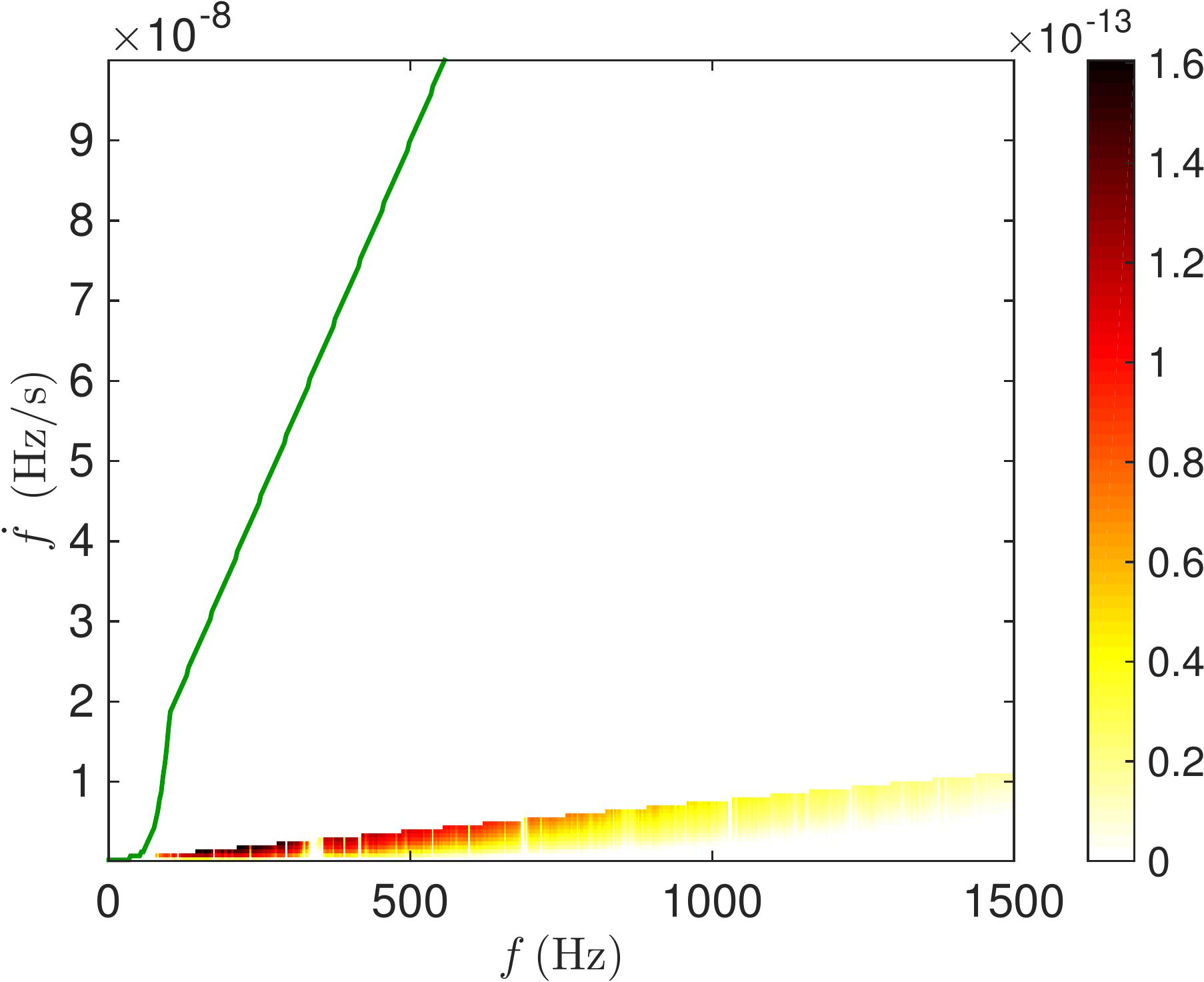}}}%
    \qquad
    \subfloat[Coverage, 30 days]{{  \includegraphics[width=.20\linewidth]{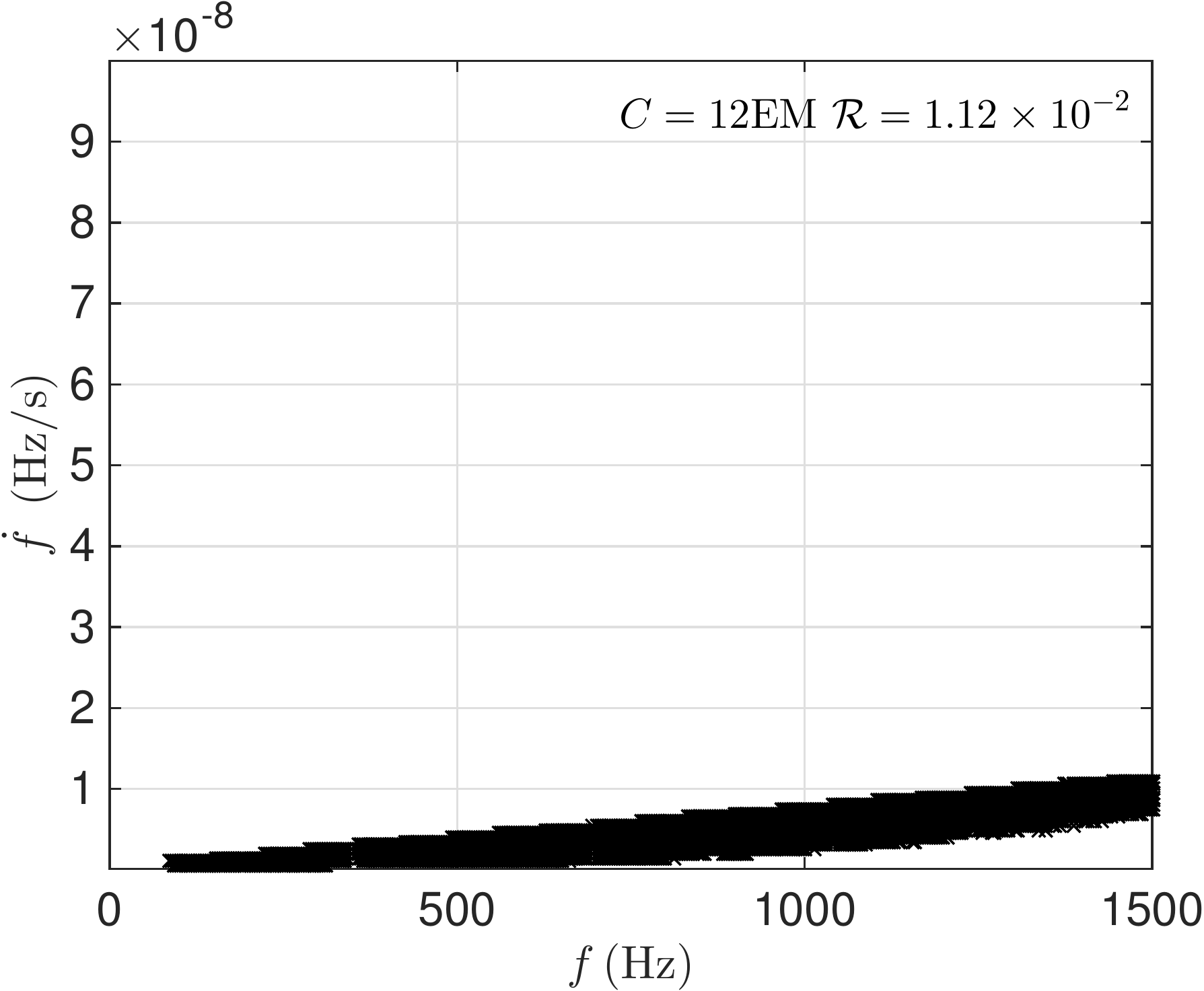}}}\\
    
    \vspace{-0.3cm}
    
    \subfloat[Efficiency, 37.5 days]{{  \includegraphics[width=.20\linewidth]{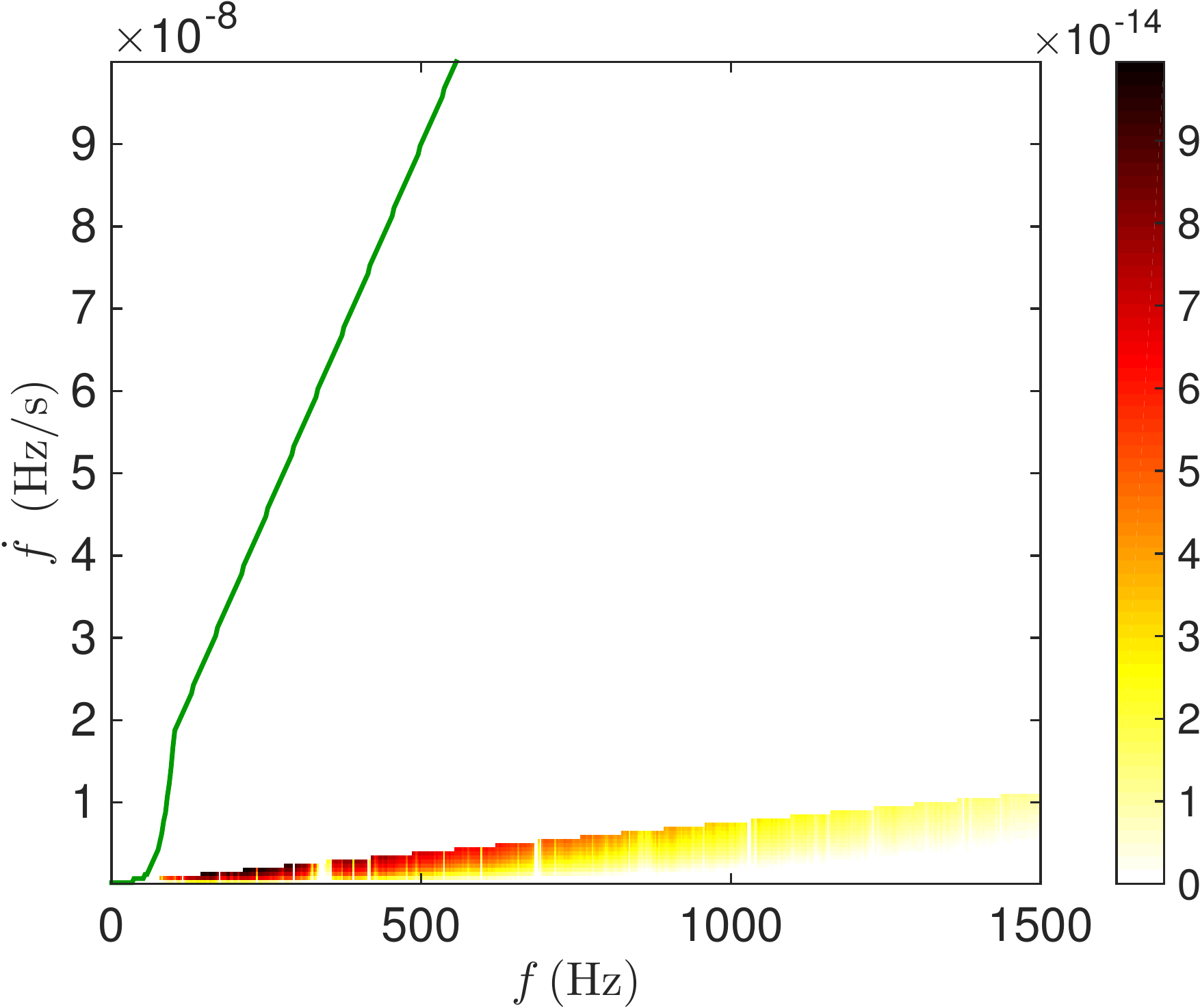}}}%
    \qquad
    \subfloat[Coverage, 37.5 days]{{  \includegraphics[width=.20\linewidth]{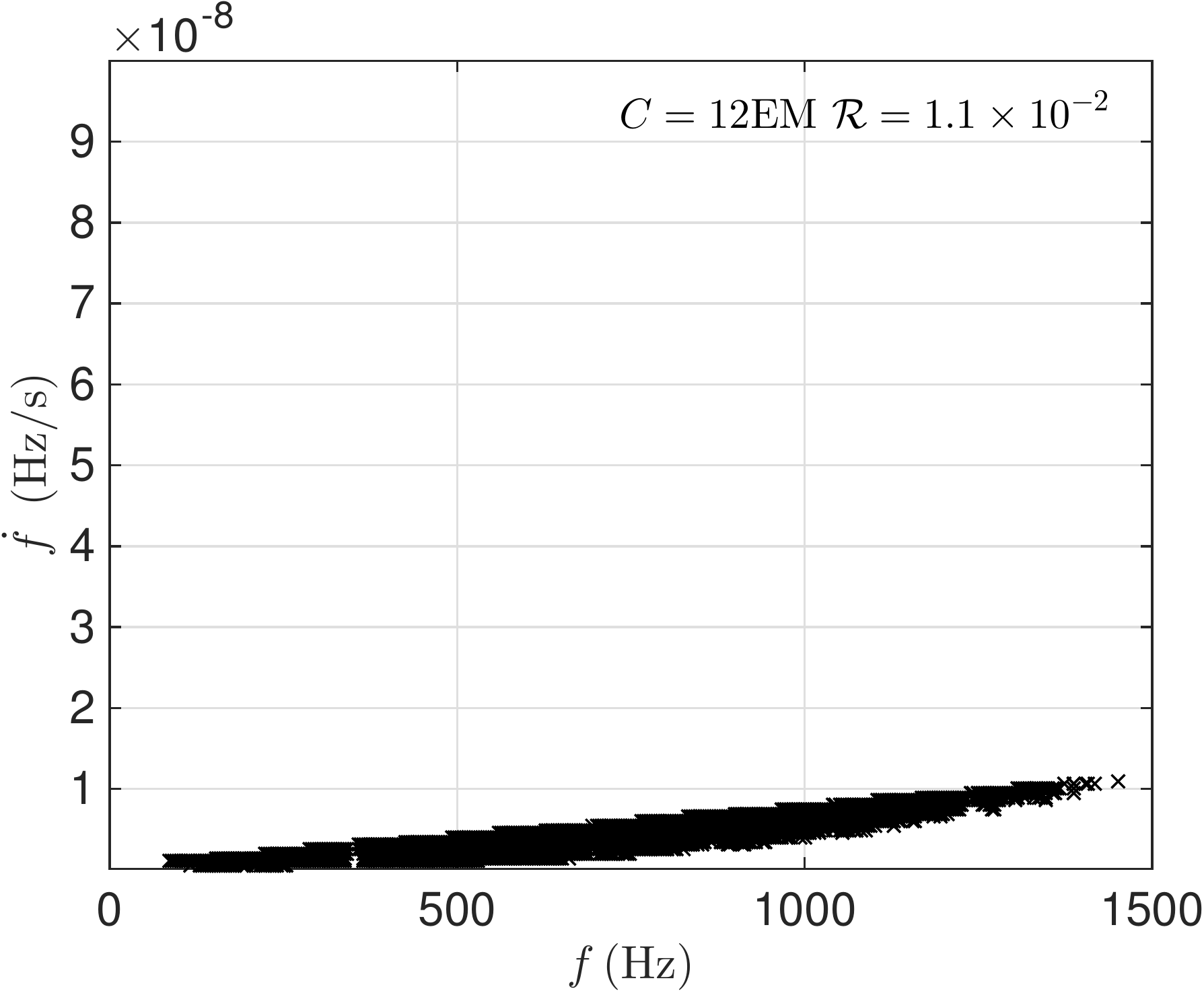}}}%
    \qquad
    \subfloat[Efficiency, 50 days]{{  \includegraphics[width=.20\linewidth]{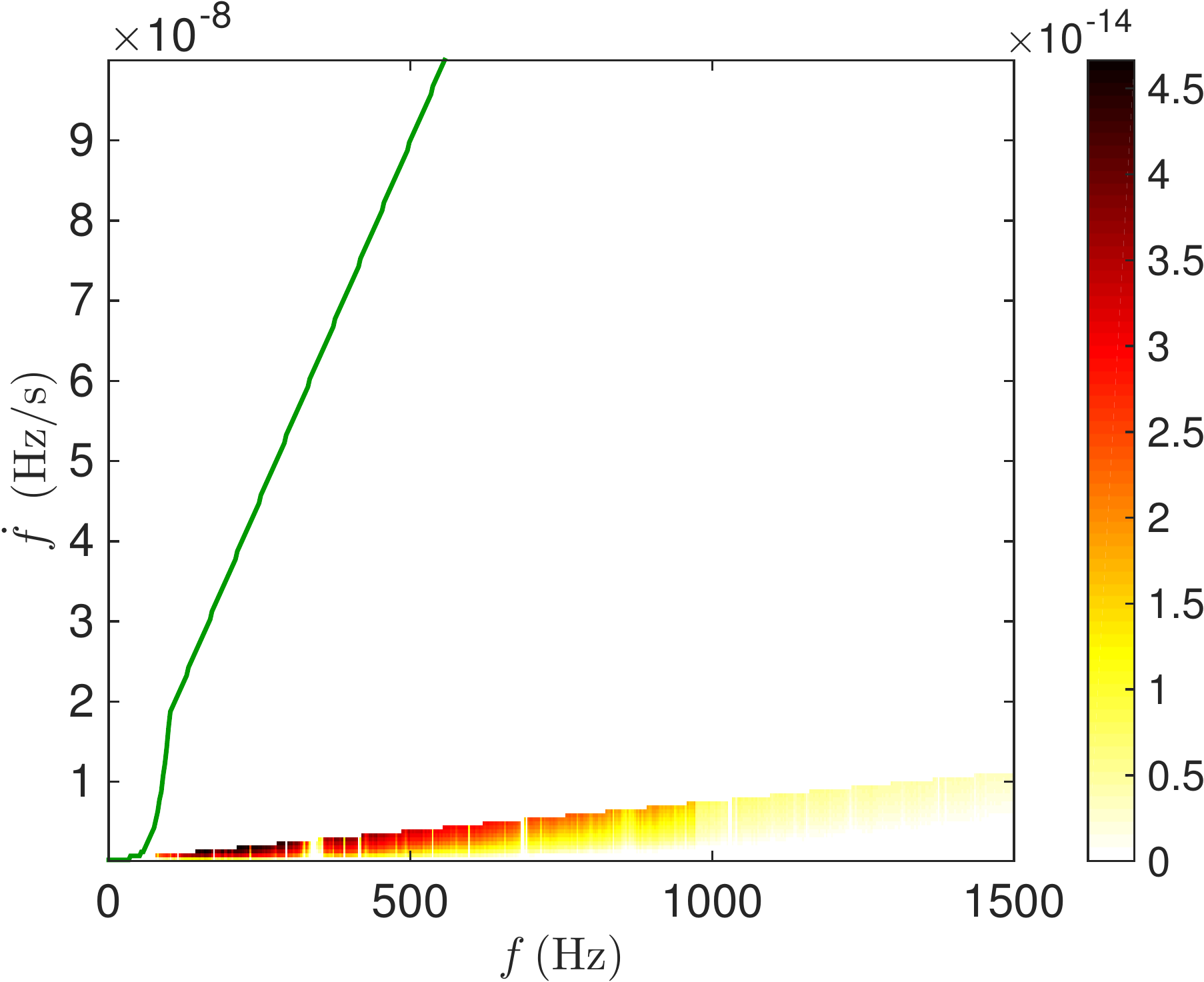}}}%
    \qquad
    \subfloat[Coverage, 50 days]{{  \includegraphics[width=.20\linewidth]{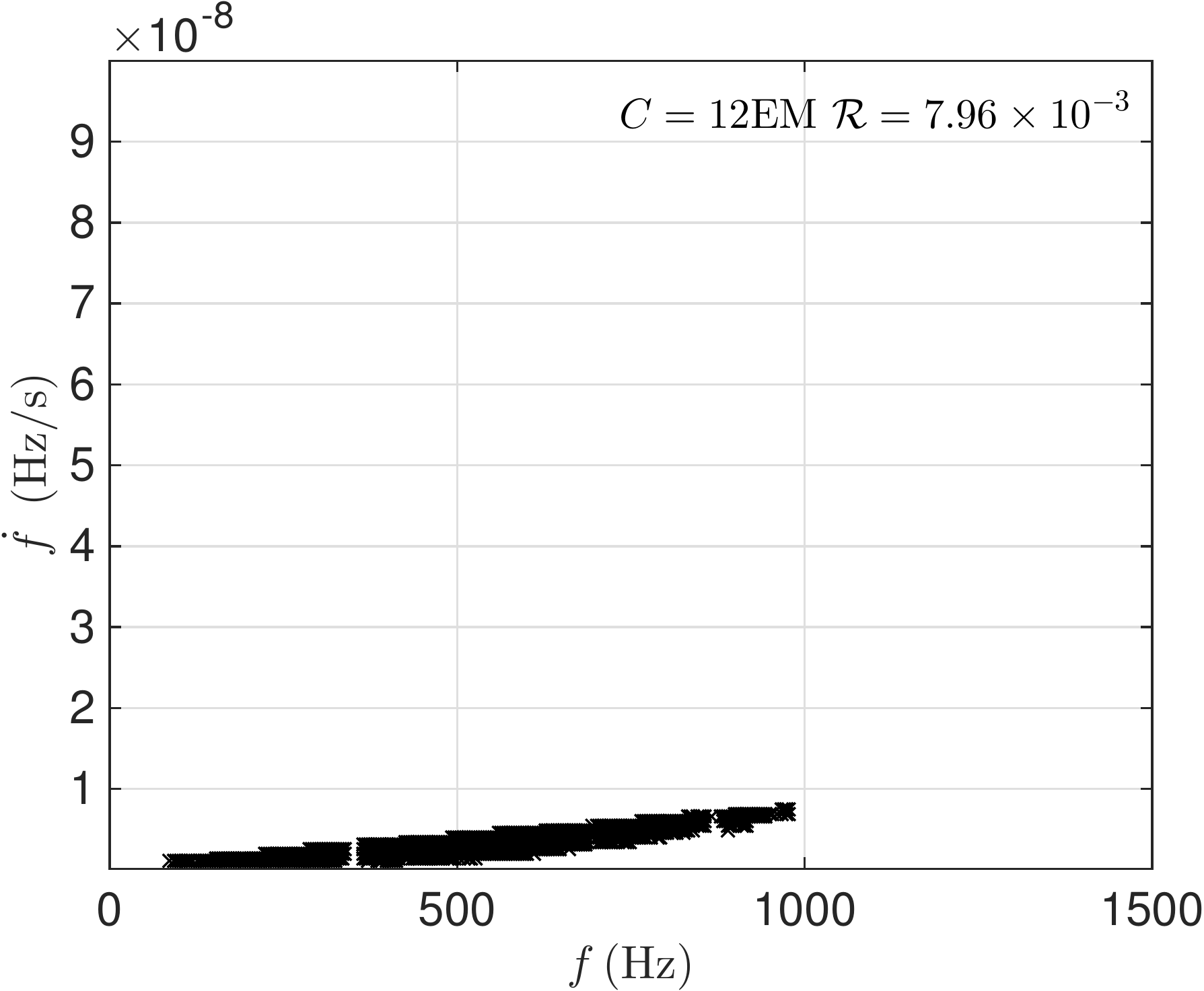}}}\\
    
   \vspace{-0.3cm} 
   
    \subfloat[Efficiency, 75 days]{{  \includegraphics[width=.20\linewidth]{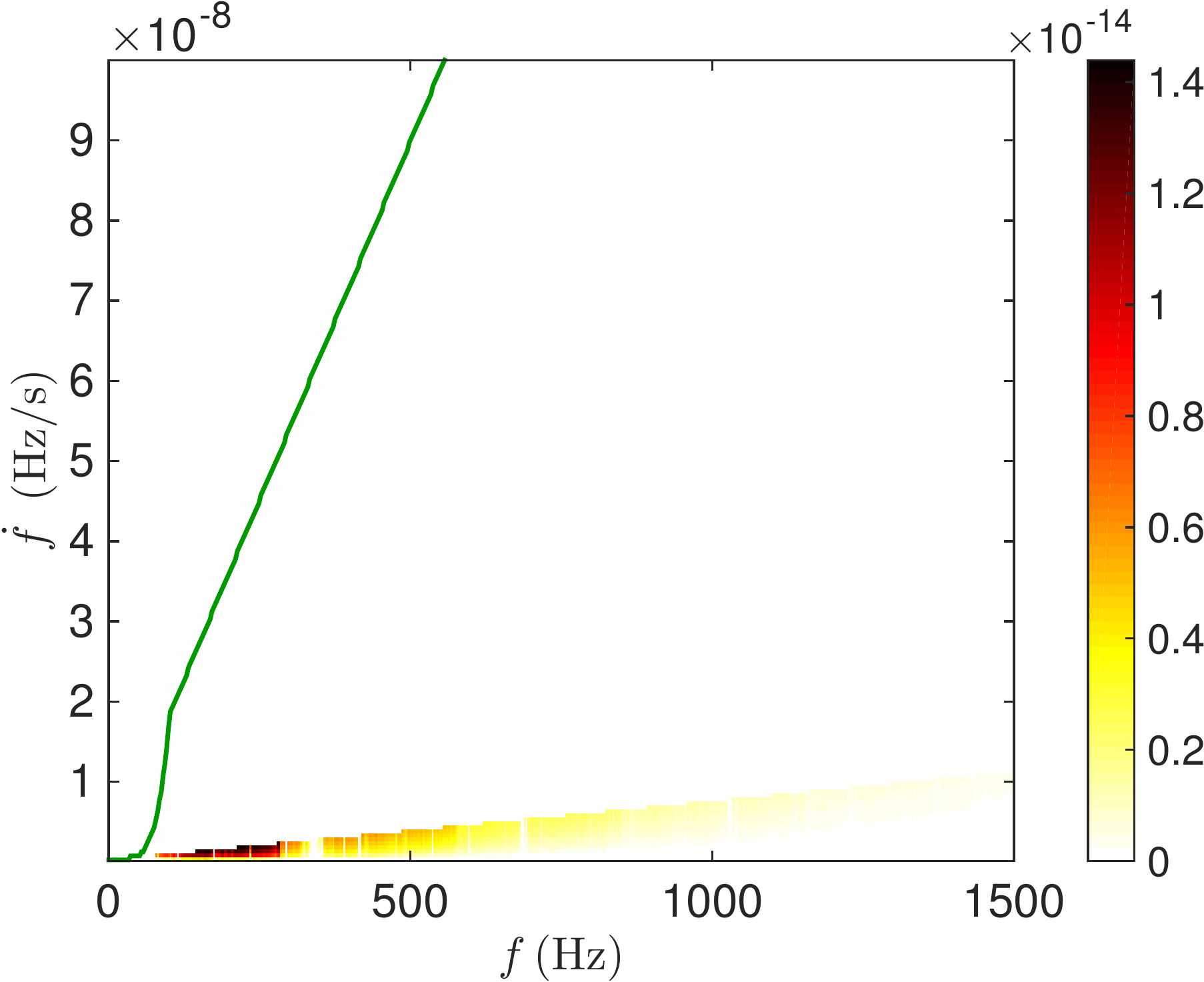}}}%
    \qquad
    \subfloat[Coverage, 75 days]{{  \includegraphics[width=.20\linewidth]{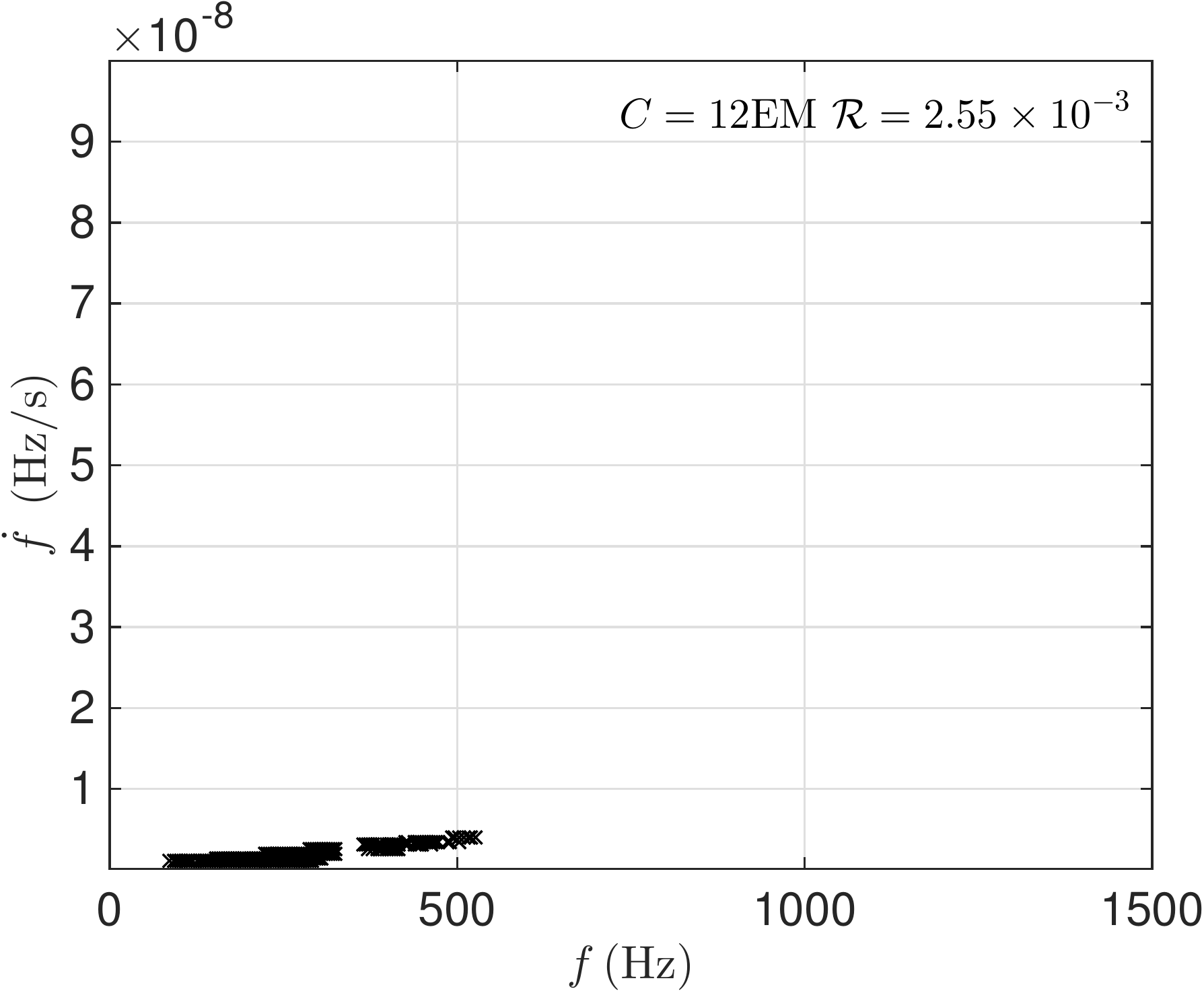}}}%
    \caption{Optimisation results for Vela Jr at 750 pc and 4300 years old, assuming uniform and age-based priors, for various coherent search durations: 5, 10, 20, 30, 37.5, 50 and 75 days. The total computing budget assumed is 12EM.}%
    \label{G2662_51020days_longage_longdist}%
\end{figure*}

We illustrate the results of the optimization when using the priors of Eq.~(\ref{eq:dotf_age}) that fold in the information on the age of the target. Fig.~\ref{G2662_51020days_longage_longdist} in this subsection and Figs.~\ref{CasA_51020days_age} to \ref{G3501_51020days} in Appendix \ref{section:AppendixFigures} show the color-coded $P_\mathrm{D}$-maps and the selected parameter space to target with a 12 EM computing budget allocated to each of the four targets Vela Jr (closest), Cas A (youngest),  G347.3 and G350.1 (close and young). Because of the uncertainty in the age and the distance of Vela Jr, we have investigated the two extreme scenarios: a close and young Vela Jr (CY) and an far and old Vela Jr (FO). As done in the previous section we also optimize the search with respect to set-ups and further with respect to targets. The complete set of results is summarized in Table ~\ref{tab: R of age} and Fig.~\ref{fig:RversusDistAndTcohAge}. We note the following:
\begin{itemize} 
\item The younger the target, the steeper is the slope that determines the prior $f-\dot{f}$ volume. This means that for younger targets higher values of $\dot{f}$ are allowed. At the same distance and frequency, more detection probability can be accumulated at higher ${\dot{f}}$ values because of the higher limit in $h_0$.
\item However, even when optimizing separately for every target the main factor that determines the detection probability at fixed computing cost is the distance. This is summarized in Fig.~\ref{fig:RversusDistAndTcohAge}(a). 
\item For the eldest target, Vela Jr with $\tau_\mathrm{c}=4300$ yrs which shows in Fig.~\ref{G2662_51020days_longage_longdist}, the prior $f-\dot{f}$ volume is small enough that with the 20-day set-up we do not exhaust the available computing budget. For shorter coherent time-baselines the computational cost is dominated by the incoherent step. As the coherent time-baseline increases the cost of the incoherent sum decreases because there are fewer segments to sum while the cost of coherent step increases rapidly, shifting the balance. 
\item Unlike in the case where we do not fold in the age information, doubling the computing budget does not bring a significant gain in detection probability. The reason is that the parameter space that is available for searching extends just to higher frequencies, not to higher spindowns, and there the sensitivity is lower and hence the increase in detection probability is marginal.  
\item  For the older sources the optimal search set-ups with age-based priors favour longer segment durations than those found with distance-based priors because their parameter space is limited to lower $\dot{f}$ regions.
\end{itemize}

 \begin{figure*}%
     \centering
     \subfloat[Coverage, cost: 12 EM]{{  \includegraphics[width=.38\linewidth]{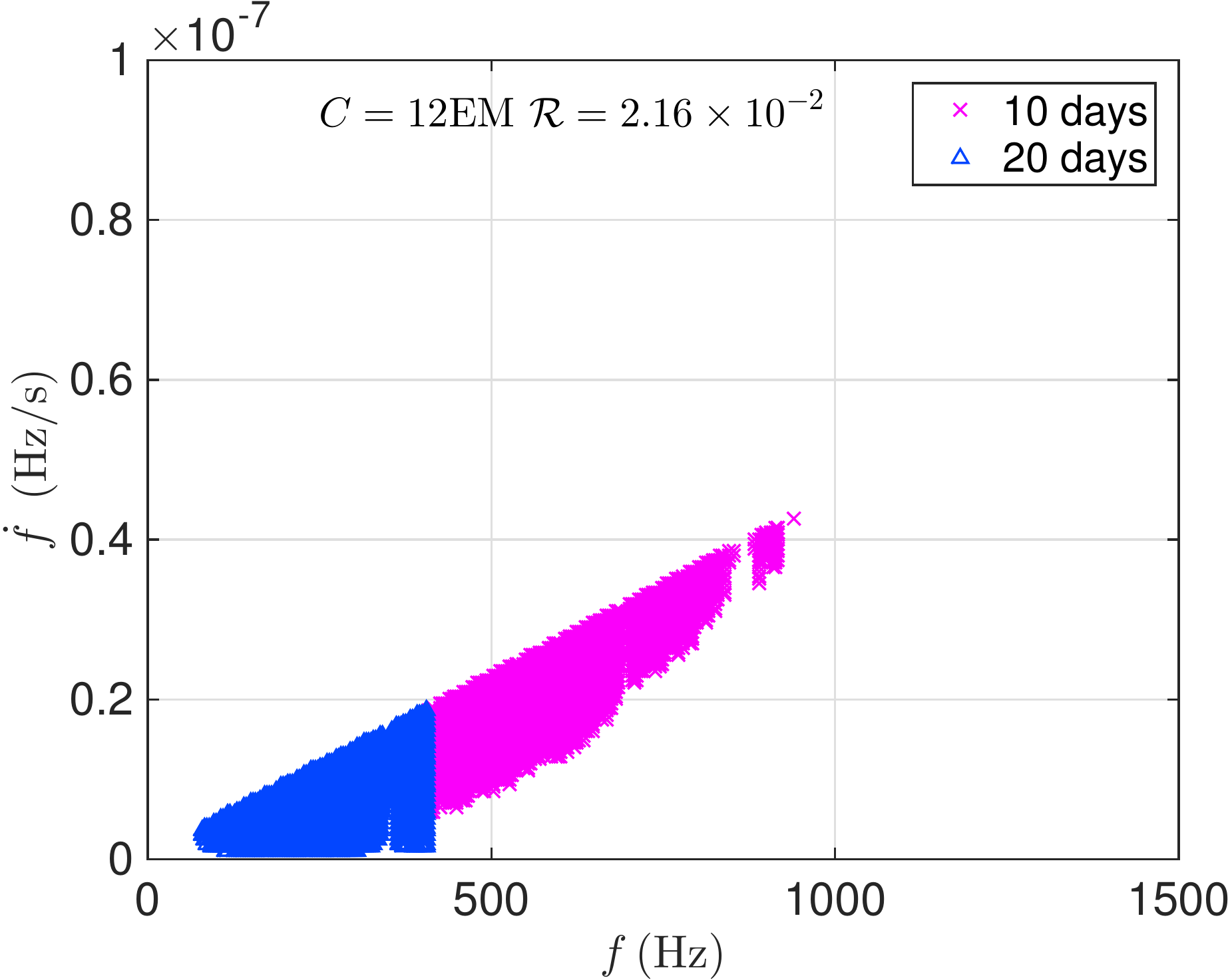}}}%
     \qquad
     \subfloat[Coverage,  cost: 24 EM]{{  \includegraphics[width=.38\linewidth]{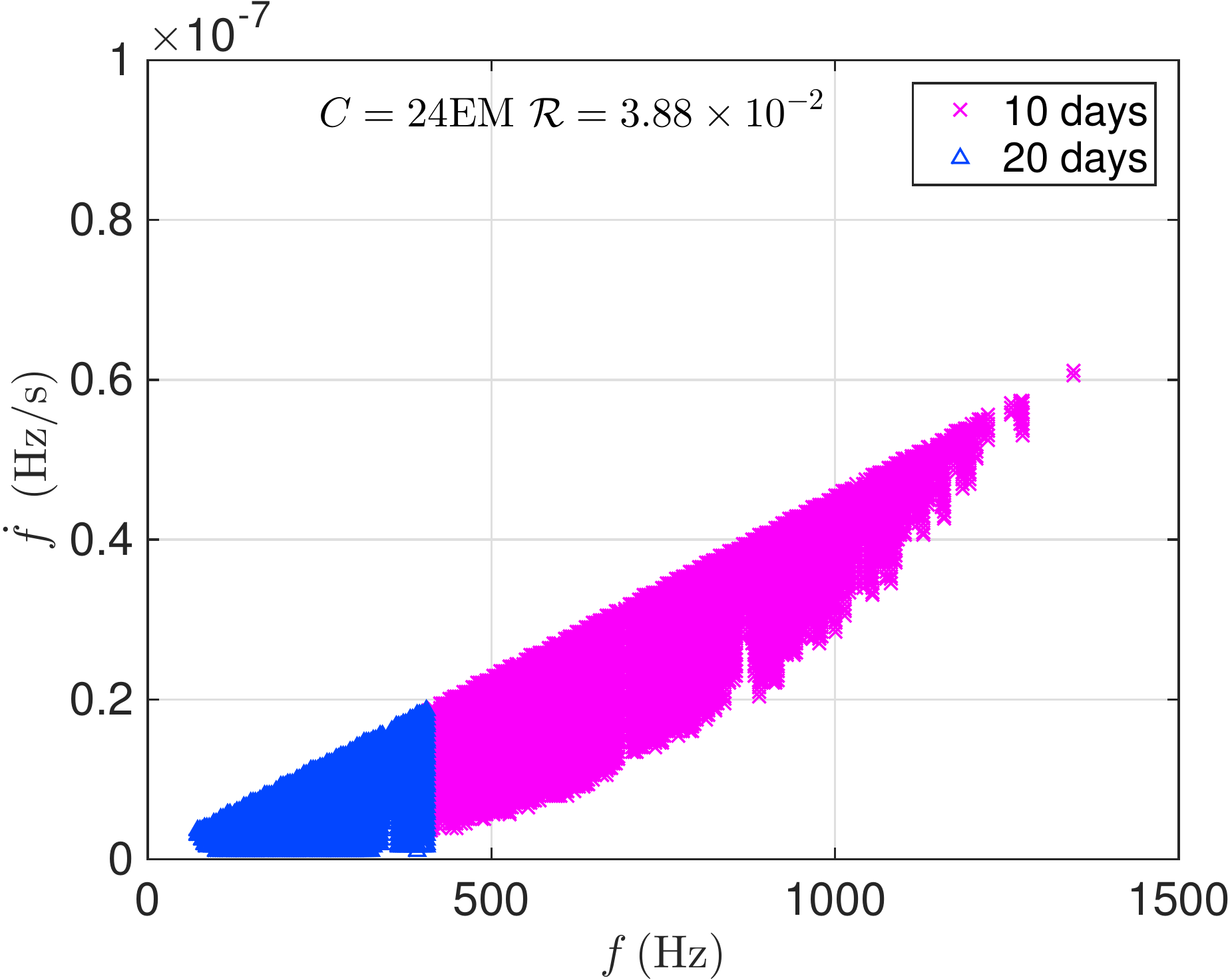}}}%
     \caption{Parameter space coverage for Vela Jr at 200 pc, 700 years old, assuming uniform and age-based priors and optimizing over the 7  search set-ups also considered above at 12 EM (left plot) and 24 EM (right plot).}%
     \label{2662_best_age}%
 \end{figure*}

\begin{figure*}%
     \centering
     \subfloat[Coverage of 3 sources, CY, Cost 12 EM ]{{  \includegraphics[width=.45\linewidth]{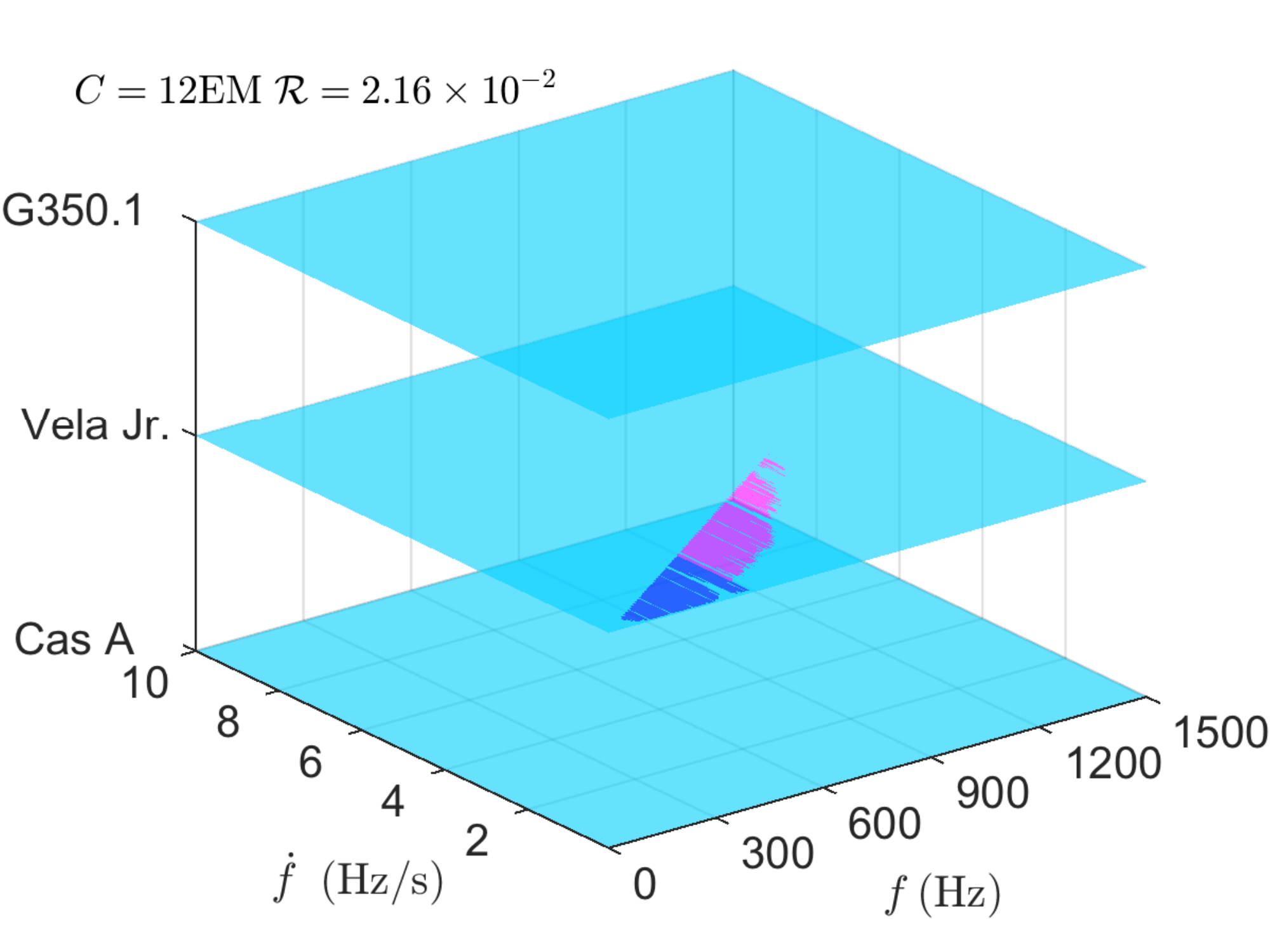}}}%
     \qquad
     \subfloat[Coverage of 3 sources, FO, Cost 12 EM ]{{  \includegraphics[width=.45\linewidth]{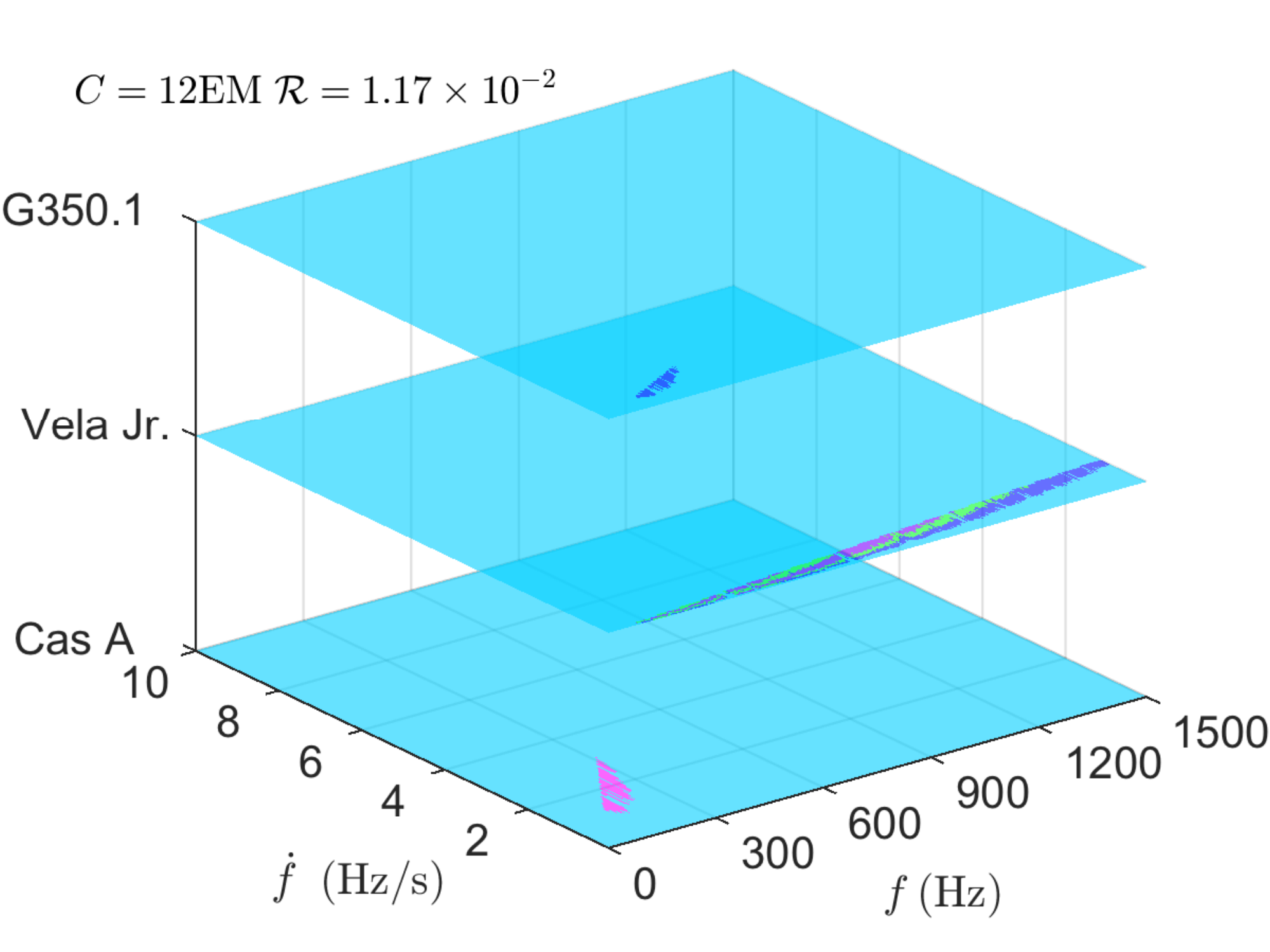}}}%
     \qquad
     \subfloat[Coverage of 3 sources, CY, Cost 24 EM ]{{  \includegraphics[width=.45\linewidth]{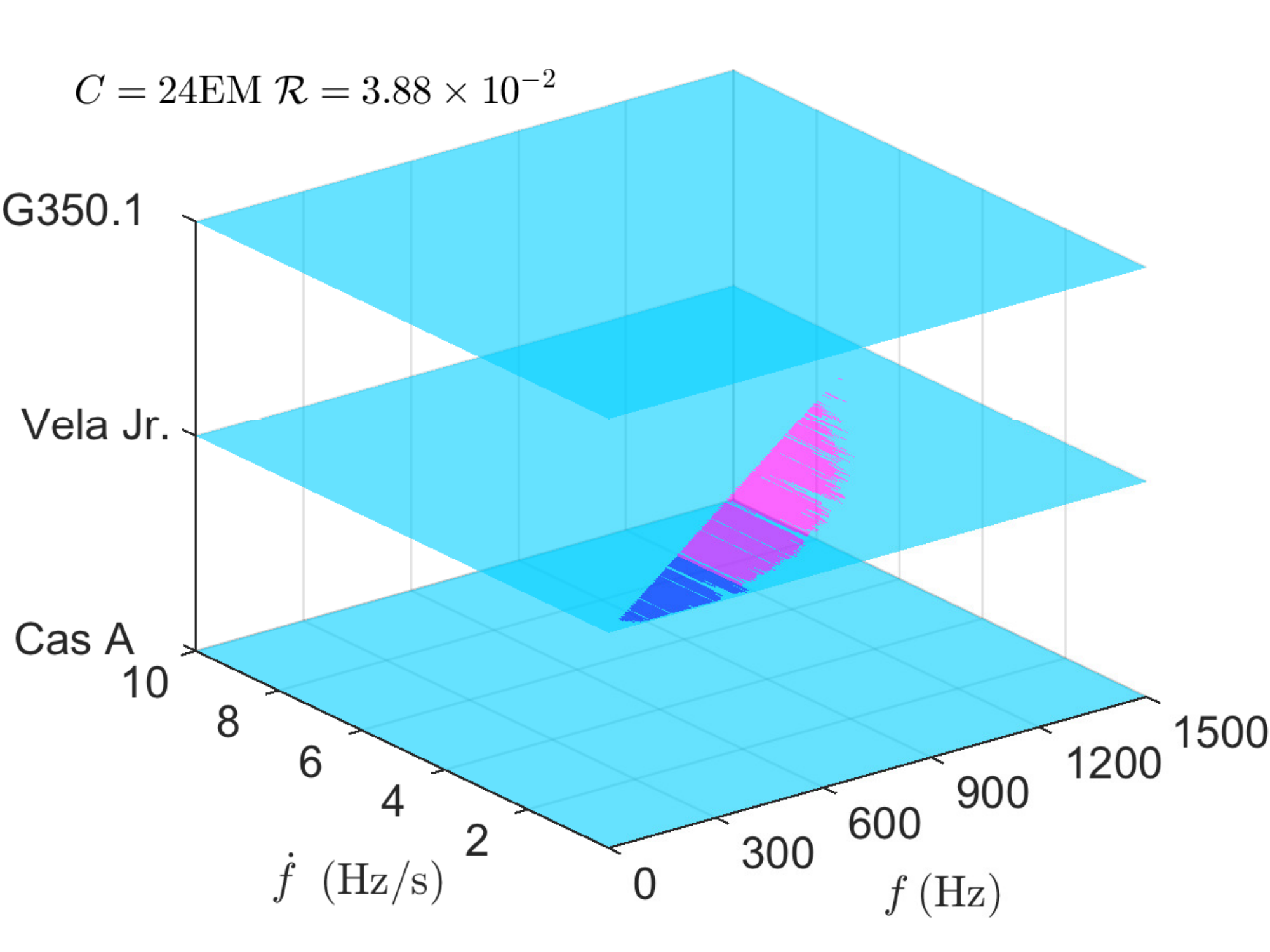}}}%
     \qquad
     \subfloat[Coverage of 3 sources, FO, Cost 24 EM ]{{  \includegraphics[width=.45\linewidth]{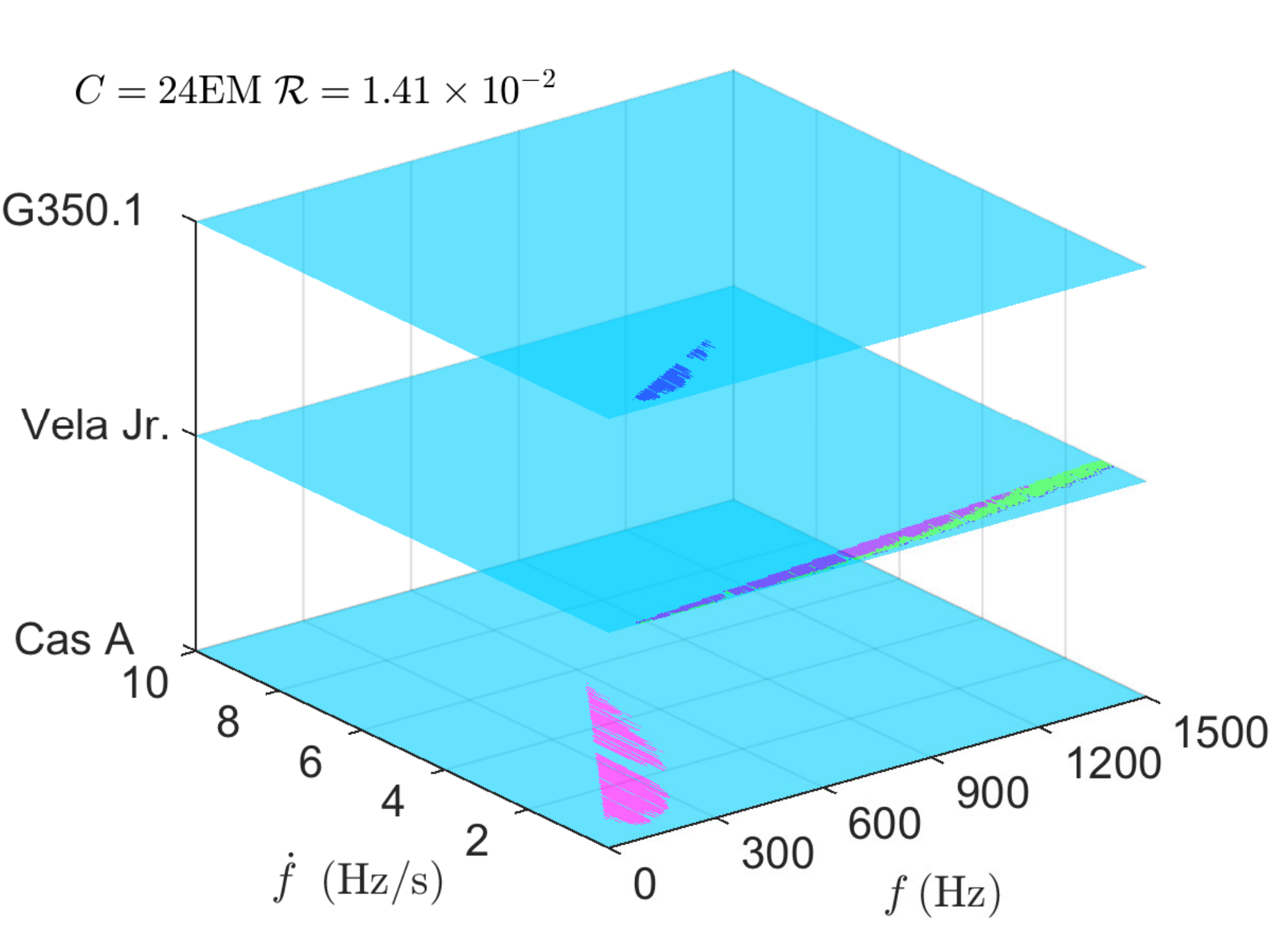}}}%
     \qquad
     \subfloat[Coverage of 3 sources, CY, Cost 48 EM ]{{  \includegraphics[width=.45\linewidth]{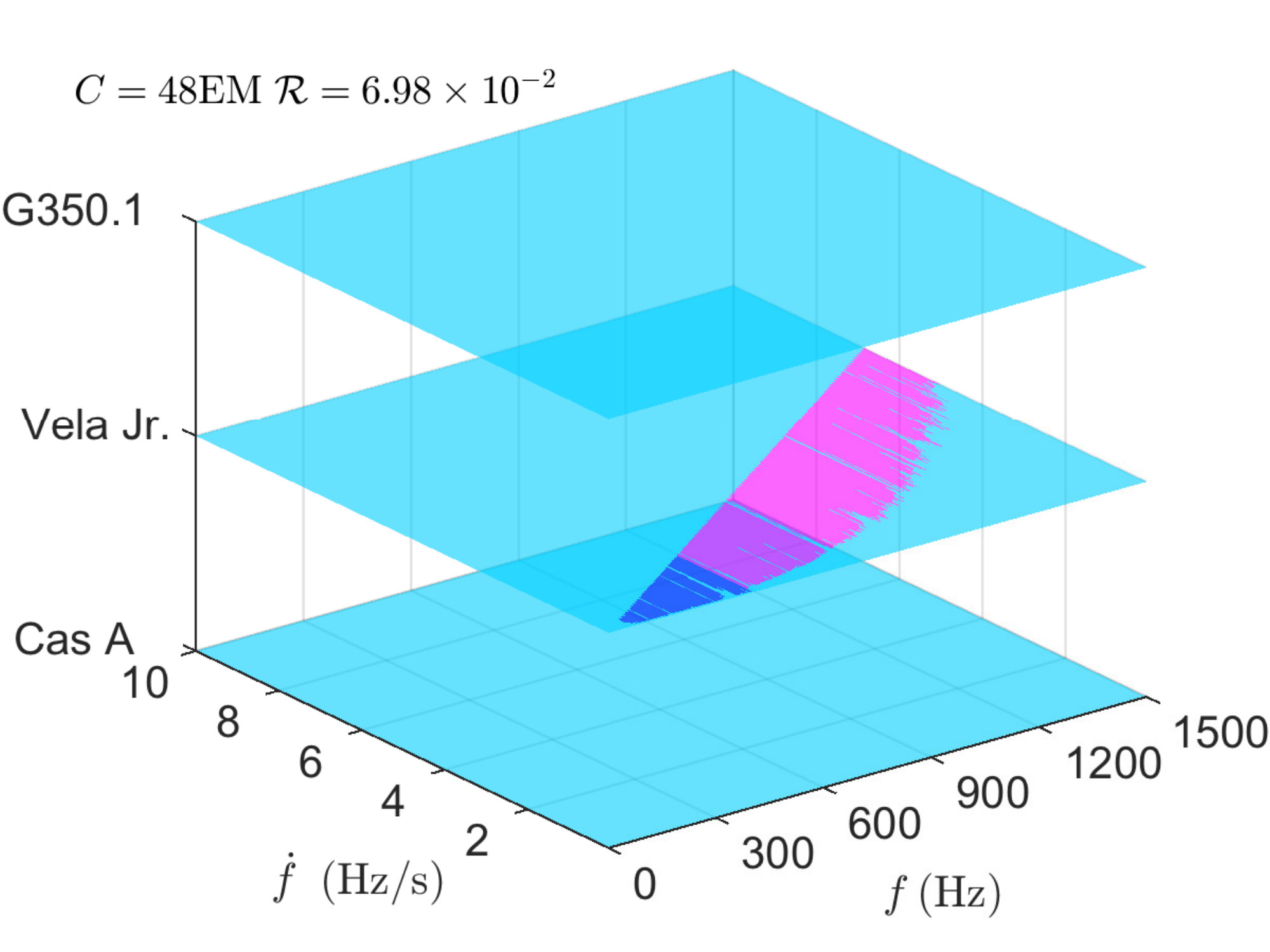}}}%
     \qquad
     \subfloat[Coverage of 3 sources, FO, Cost 48 EM ]{{  \includegraphics[width=.45\linewidth]{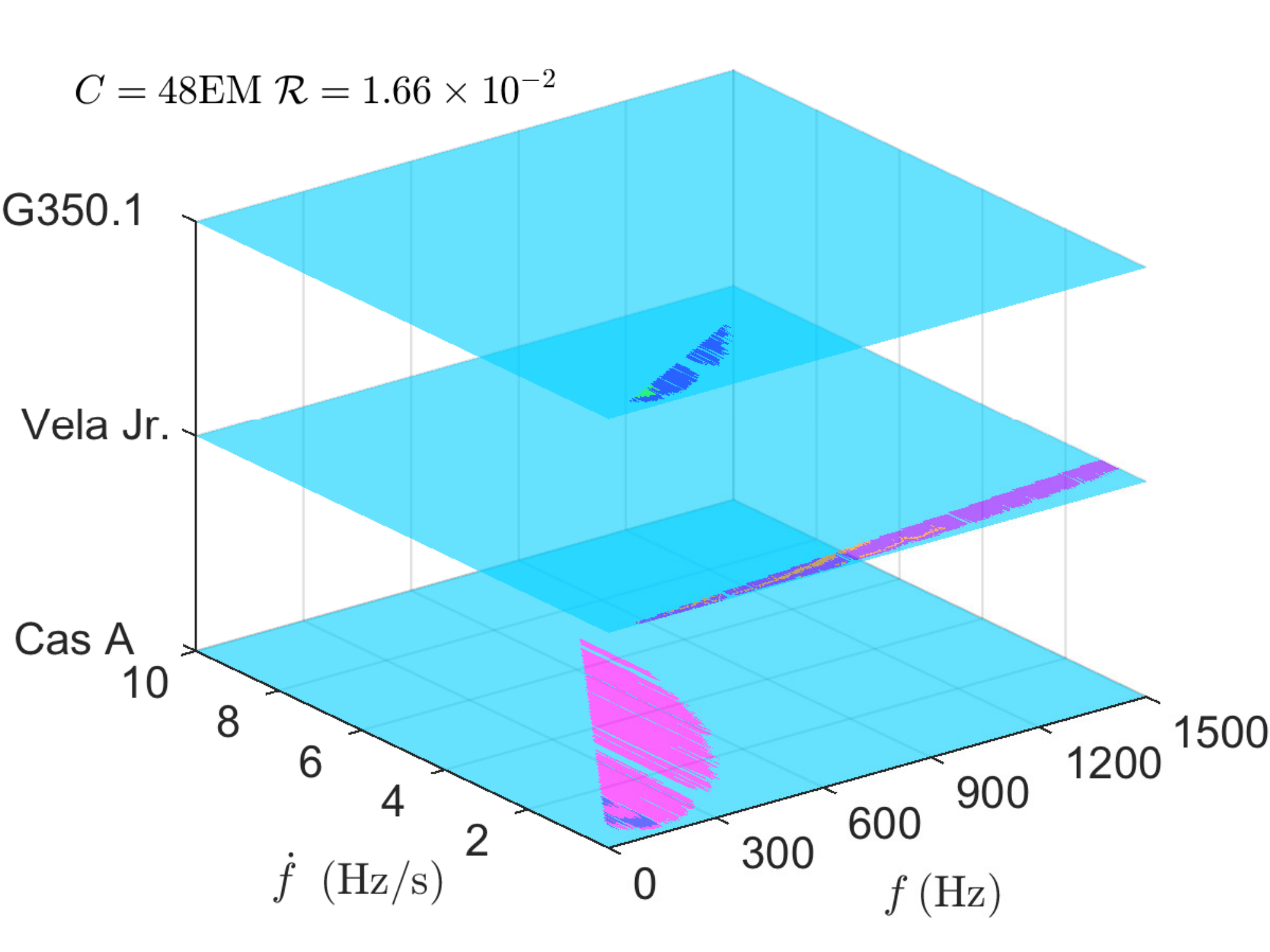}}}%

     \caption{Parameter space coverage assuming uniform and age-based priors and optimizing over the 7 search set-ups also considered above and over the three youngest targets (left plots: Cas A, Vela Jr at 200 pc and 700 years old (CY), G350.1 and right plots: Cas A, Vela Jr at 750 pc and 4300 years old (FO), G350.1) at 12, 24 and 48 EMs.}
     \label{all_age}%
 \end{figure*}

We now optimize also with respect to sources. 
Fig.~\ref{all_age} shows that the covered parameter space increases as computing cost
increases.  We note the following:
\begin{itemize}
\item At 12 EM the preferred target is Vela Jr solely, if we assume that it is CY. The cells picked by the optimization procedure are obviously the same as the cells picked in the Figure \ref{2662_best_age}(a), corresponding to 10 and 20-day set-ups that gave the highest $\mathcal{R}=2.16\times 10^{-2}$. 
\item If instead we assume that 
Vela Jr is FO then, at 12 EM, the detection probability is maximized by spending some fraction of the computing budget also on G350.1 and Cas A 
54\% of the prior parameter space of Vela Jr FO is searched leaving out high $f$ - low $\dot{f}$  cells which have a low detection probability. Again the optimal set-up is a combination of the 20, 30 and 37.5-day set-ups which yield the top three $\mathcal{R}$ in the fixed-set-up optimization, cfr. Figs.~\ref{G2662_51020days_longage_longdist}(f), (h) and (j). 10.2 EM (84.9\% of total) were spent to accumulate $1.12\times 10^{-2}$ (95.3\%) detection probability from Vela Jr FO and 1.8 EM (15.1\%) were spent to accumulate $5.48\times 10^{-4}$ (4.7\%) detection probability jointly from  G350.1 and Cas A. So one could say that the Vela Jr FO searched cells are, on average, a factor of 3.57 more efficient at accumulating detection probability per computing cost unit than the cells of the other two targets. The set-ups for the cells picked for Cas A are the same as those picked when optimizing with respect to set-up for Cas A only, cfr. Fig.~\ref{CasA_best_age}(a) in Appendix \ref{section:AppendixFigures}. This is not the case for G350.1: for the selected cells it turns out that it is more efficient to use the small computational budget on more cells with a shorter coherent time-baseline, than with a longer coherent baseline as when optimizing the 12 EM for G350.1 alone, cfr. Fig.~\ref{3501_best_age}(a) in Appendix \ref{section:AppendixFigures}. 
%
\item 24 EM buys more parameter space cells for Vela Jr CY, nearly doubling the detection probability with respect to the 12 EM case: from $2.16\times 10^{-2}$ to $3.88\times 10^{-2}$. 
\item Under the assumption that Vela Jr is FO, the additional 12 EM (total 24 EM) only increase the detection probability by less then $21\%$: from $1.17\times 10^{-2}$ to $1.41\times 10^{-2}$. This is reasonable: we know in fact that if we had 12 EM to spend just on Cas A the maximum probability that we could achieve is $2.26\times 10^{-3}$ and on G350.1 it is $5.59\times 10^{-4}$. So if we had an additional 24EM to spend, at most we could achieve an increase in detection probability of  $2.82\times 10^{-3}$ which amounts to 24\% of the $1.17\times 10^{-2}$. With half of that computing power we achieve just over half of this maximum gain. Regarding the set-ups picked for the different sources the same considerations hold as we made for the 12 EM case.

\item With 48 EM  more than half of the whole prior parameter space of Vela Jr CY is covered. With such a high amount of budget the highest sensitivity cells are still searched with the most efficient search set-up: 10 and 20-day. Detection probability is nearly doubled again and still no computing budget will be spent on Cas A and  G350.1.  This is because Vela Jr CY  has larger parameter space in $\dot{f}$ and more computing power could be spent on those cells in the higher $\dot{f}$ region.     

\item Under the assumption that Vela Jr is FO, the additional 24 EM (total 48 EM) only increase the detection probability by less than $18\%$. Not only more cells from Cas A and G350.1 are searched, but also cells in Vela Jr trend to use longer coherent segments. This is because rather than to spend more computing power on the  sources with less potential like  Cas A and G350.1, it could be better to use more expense and also more efficient set-ups for Vela Jr.

\end{itemize}

\begin{figure*}%
    \centering
    \subfloat[0.2 kpc, 12 EM budget]{{  \includegraphics[width=.45\linewidth]{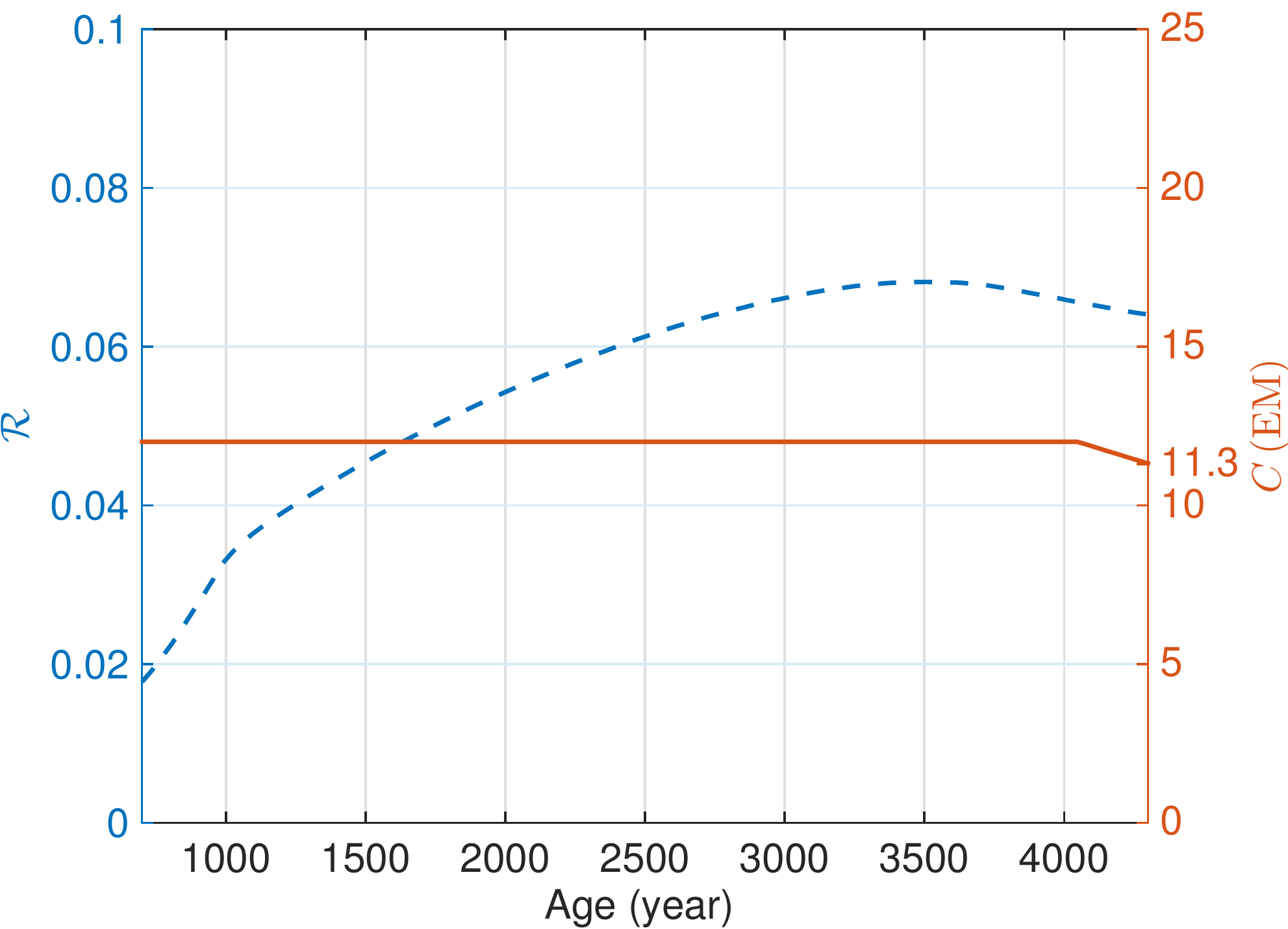}}}%
    \qquad
    \subfloat[0.75 kpc, 12 EM budget]{{  \includegraphics[width=.45\linewidth]{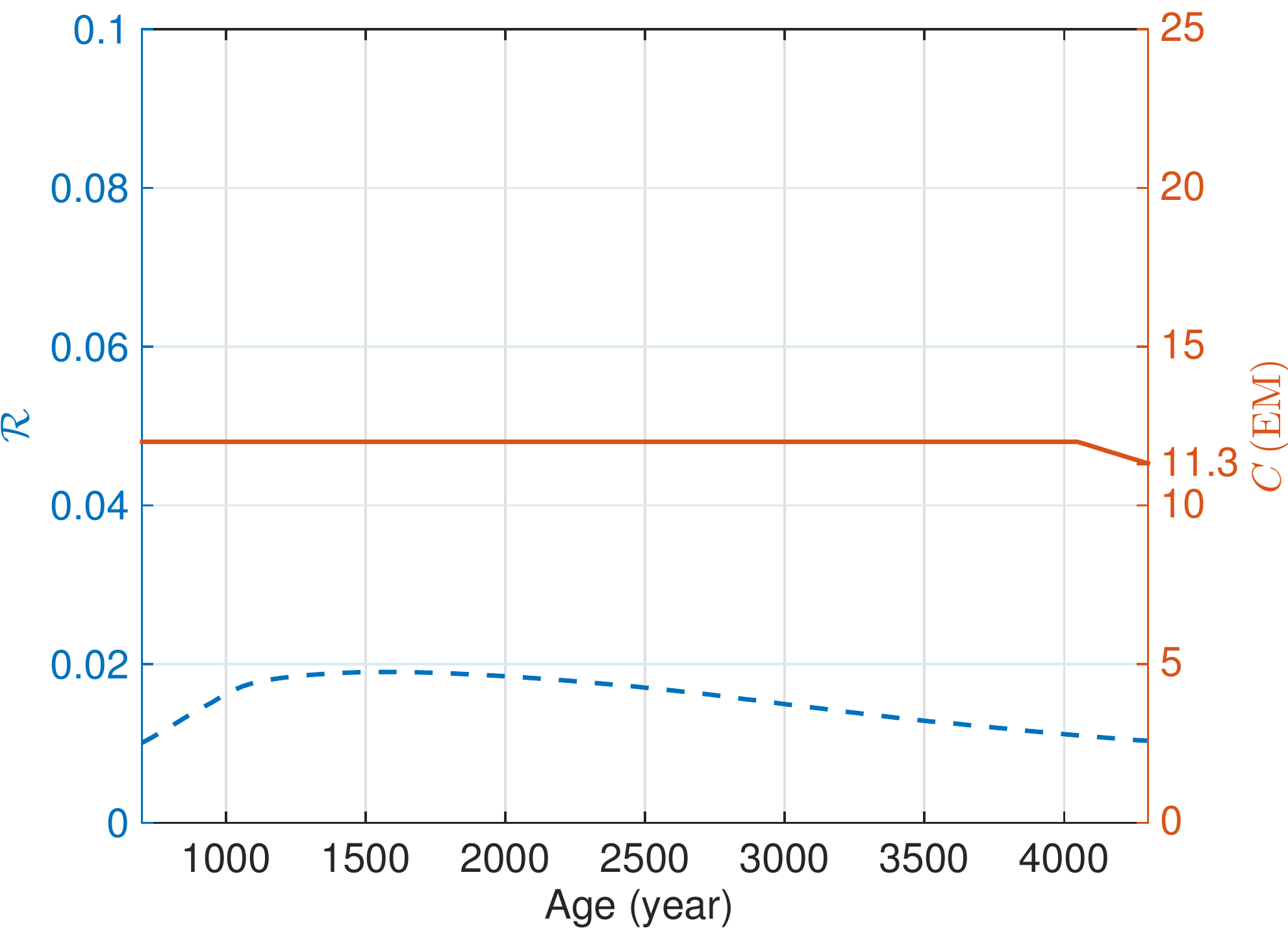}}}%
    \qquad
    \subfloat[0.2 kpc, 24 EM budget]{{  \includegraphics[width=.45\linewidth]{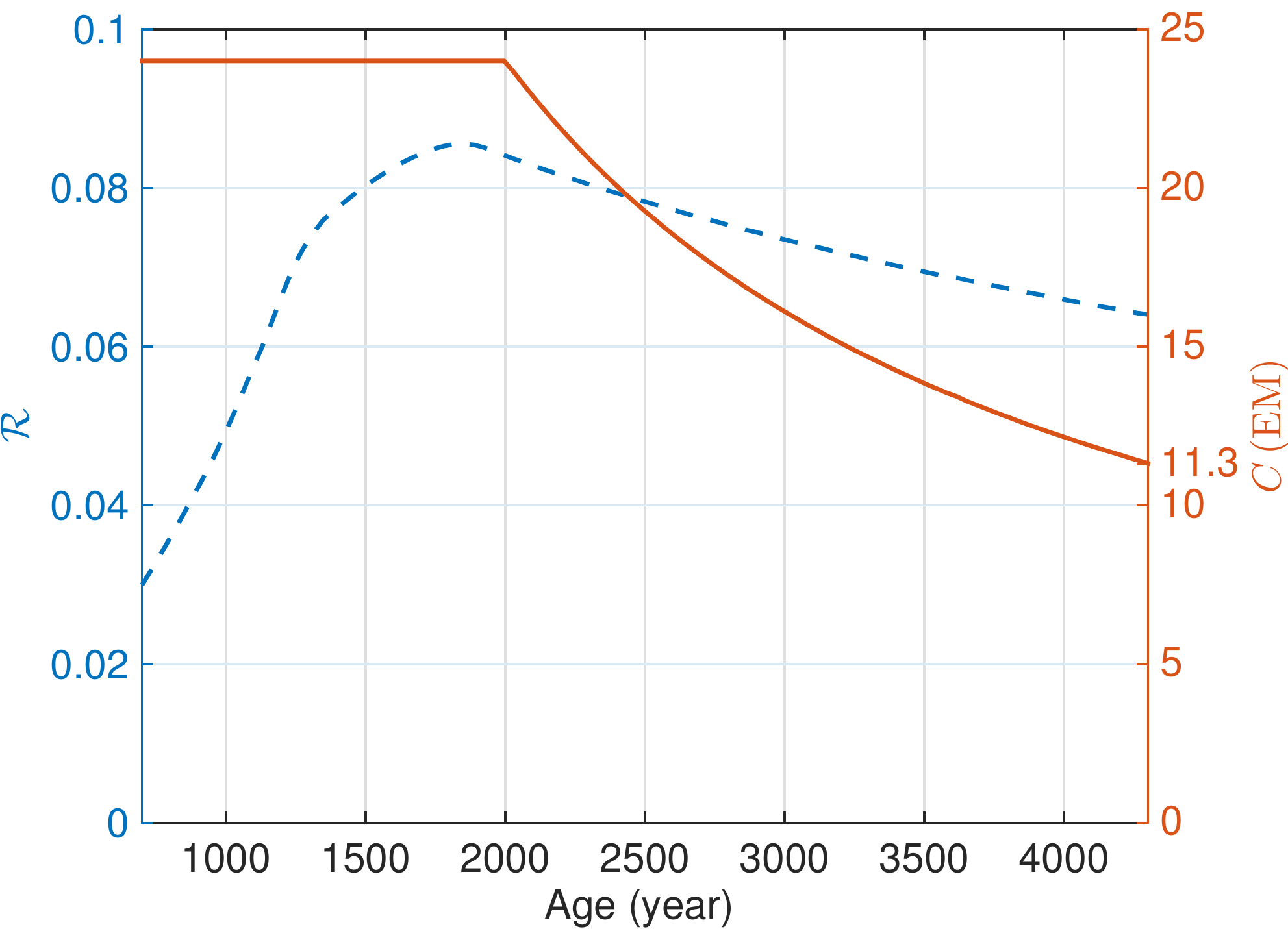}}}%
    \qquad
    \subfloat[0.75 kpc, 24 EM budget]{{  \includegraphics[width=.45\linewidth]{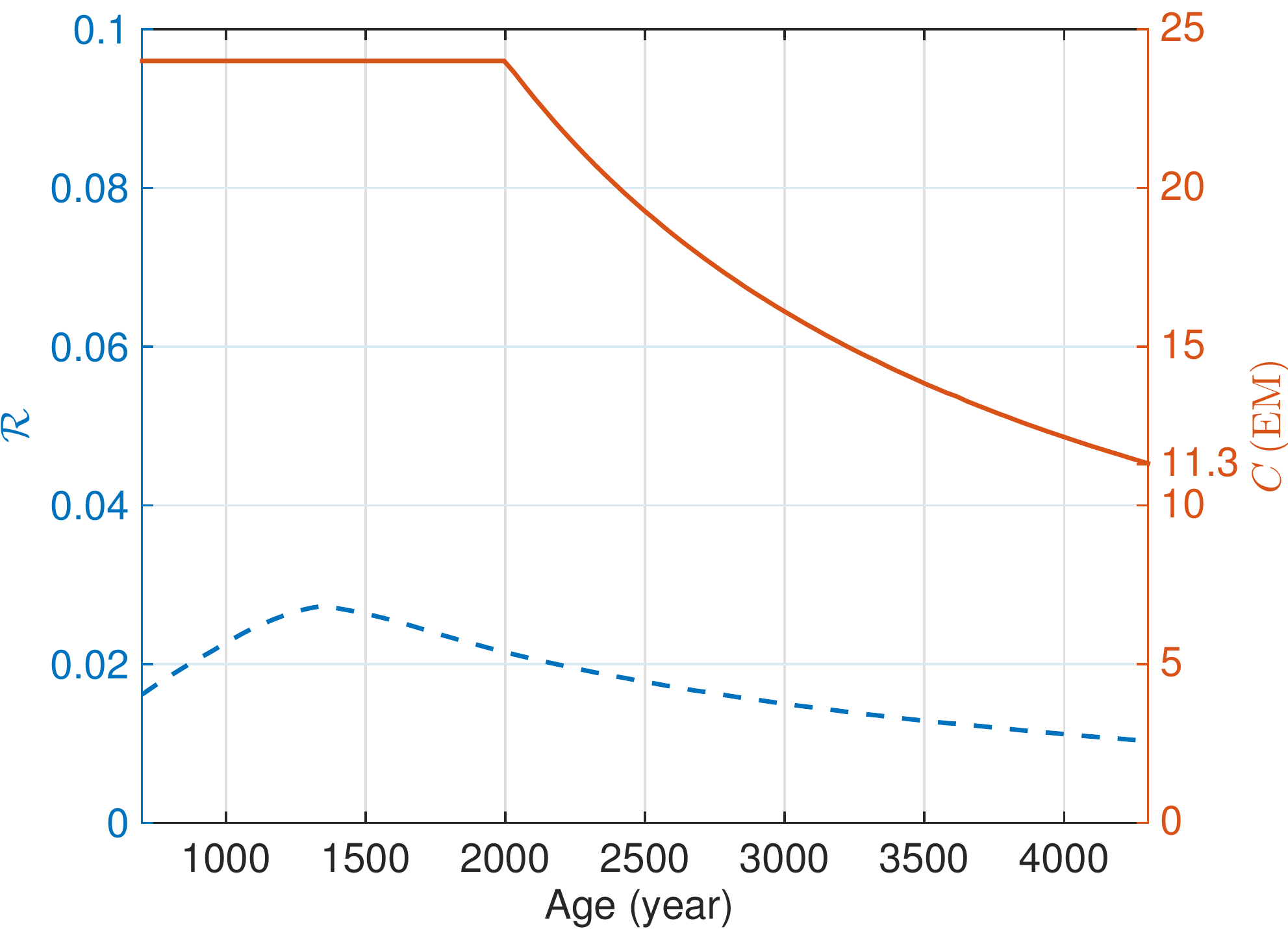}}}%
    \caption{$\mathcal{R}$ and $C$ as a function of age for an optimised search for Vela Jr, assuming uniform and age-based priors, a computing budget of 12 and 24 EMs and a 20-day coherent segment search set-up. }%
    \label{fig:investigation_age_plots}%
\end{figure*}

%
%
Fig.~\ref{fig:investigation_age_plots} shows how $\mathcal{R}$ and the used computing budget $C$ vary with age, having assumed a search for Vela Jr. at 200 and 750 pc, a coherent time-baseline of 20 days and a computing budget $C_{\mathrm{min}}$ of 12 and 24 EM. In the young age region, $C$ is always flat. That's because the older the object, the smaller is the prior $f-\dot{f}$ volume available for searching. Hence there is an age $\tau_{\mathrm{plat}}$ at which the allocated $C_{\mathrm{max}}$ is large enough to just cover such a volume. For higher values of the age the prior space shrinks and less computing power is needed to cover it. For lower values of the age the prior volume is larger and the optimization method will select what cells are the most promising to search while using up all the computational power, hence the plateau at low age values. 

Let us now look at $\mathcal{R}$. 
In all these four cases, $\mathcal{R}$ has a maximum value $\mathcal{R}_{\mathrm{peak}}$ at a certain age $\tau_{\mathrm{peak}}$. 
A larger computing budget gives a higher $\mathcal{R}_{\mathrm{peak}}$ and this happens at a lower age. However $\tau_{\mathrm{peak}}$ does not coincide with $\tau_{\mathrm{plat}}$ because even though as the age increases towards $\tau_{\mathrm{plat}}$ the fractional covered volume of parameter space is increasing, at the same time the total volume is shrinking and the cells that are not any more included are actually the ones contributing the most to the detection probability. This is because the dropped cells are the higher spindown ones which have the highest amplitude cut-off value ${h_0}_{\mathrm{max}}$.

%
%


\subsection{Log-uniform priors in $f$ and $\dot{f}$}
\label{sec:LogUniformPriors}

\begin{figure}%
 \centering
 \subfloat[${\mathcal{R}}$ versus distance]{{\includegraphics[width=.9\linewidth]{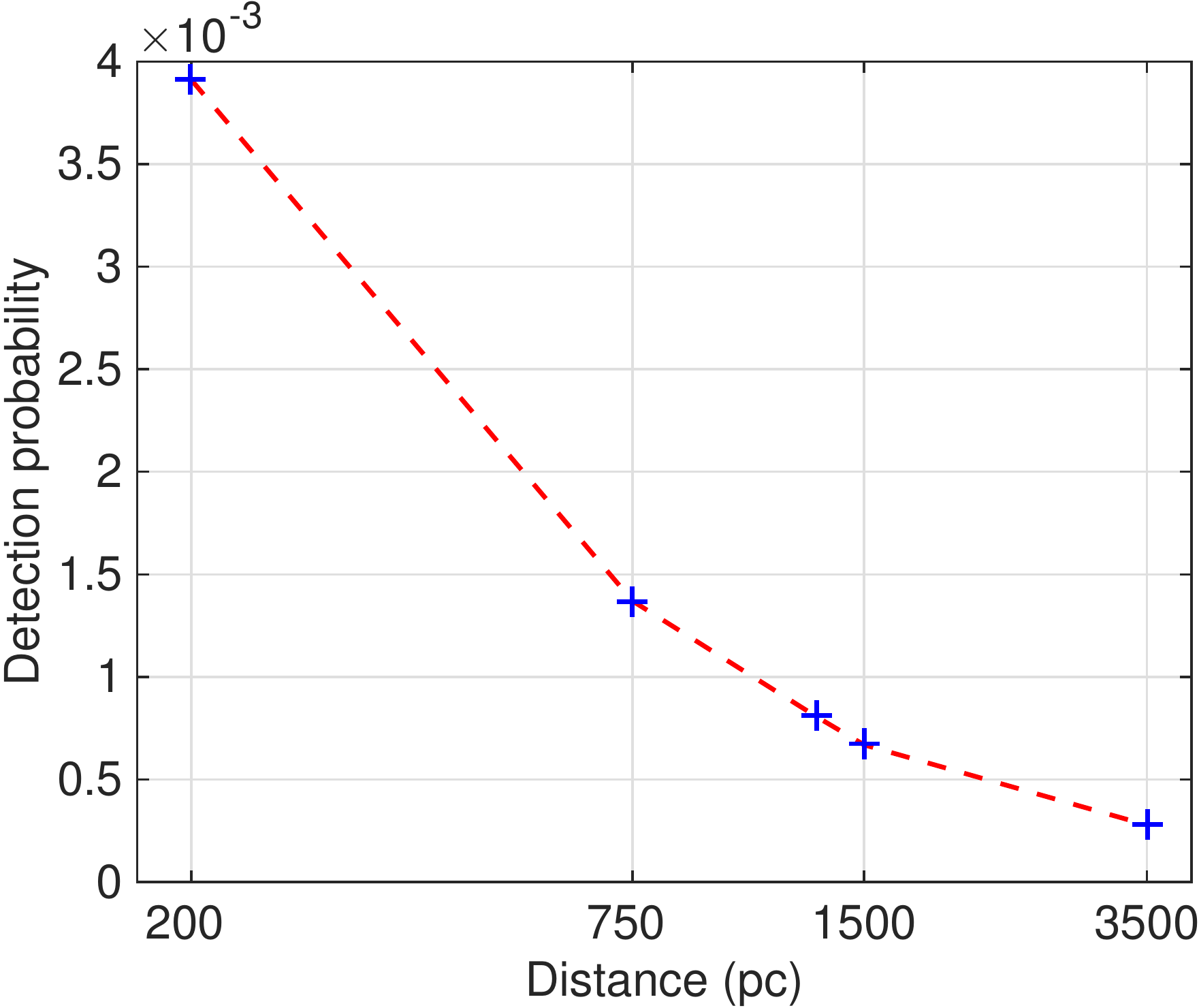}}}%
 \qquad
 \subfloat[${\mathcal{R}}$ versus $T_\mathrm{coh}$]{{\includegraphics[width=.9\linewidth]{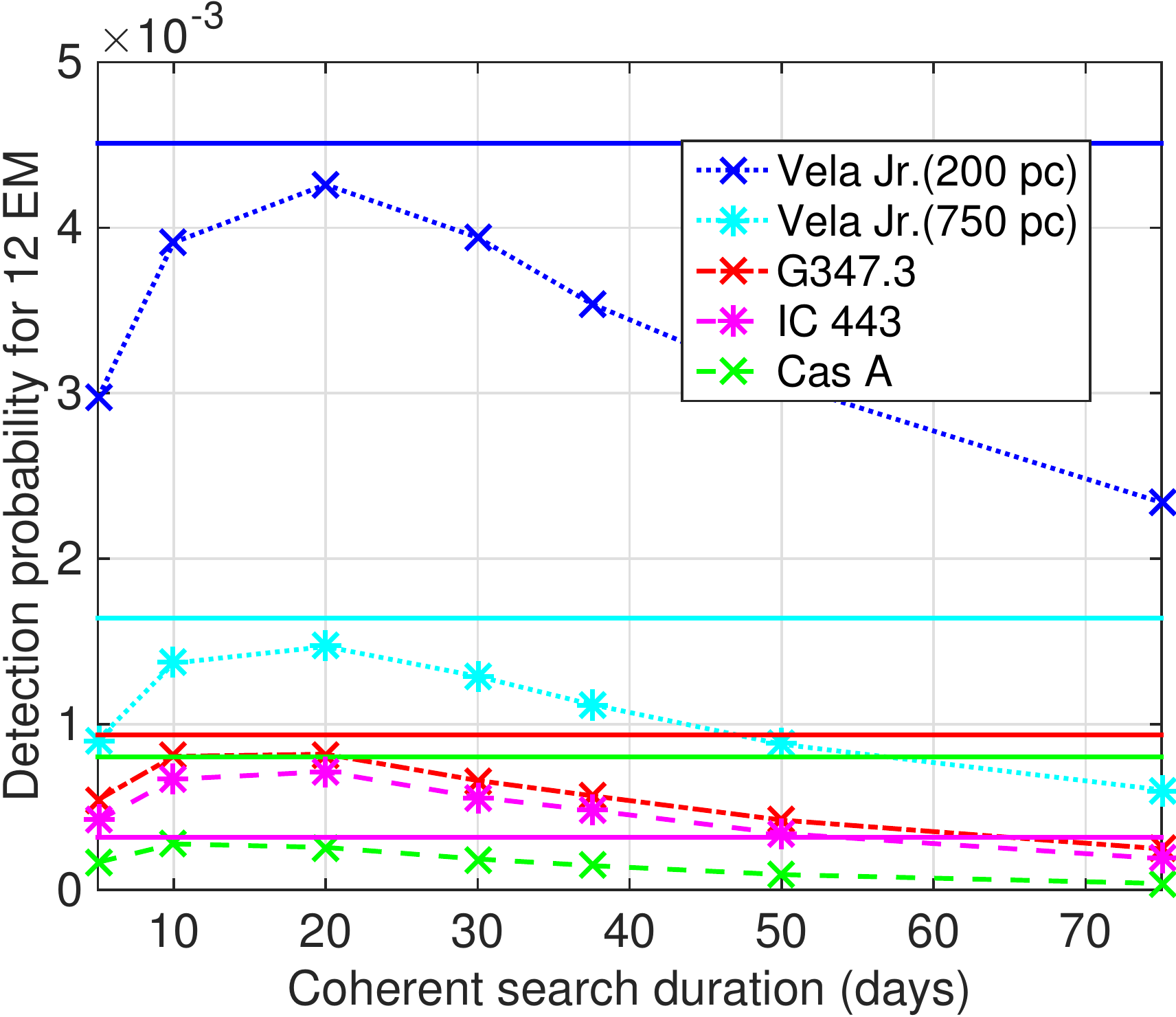}}}%
 \qquad
 \subfloat[${\mathcal{R}}$ versus $C_\mathrm{max}$]{{\includegraphics[width=.9\linewidth]{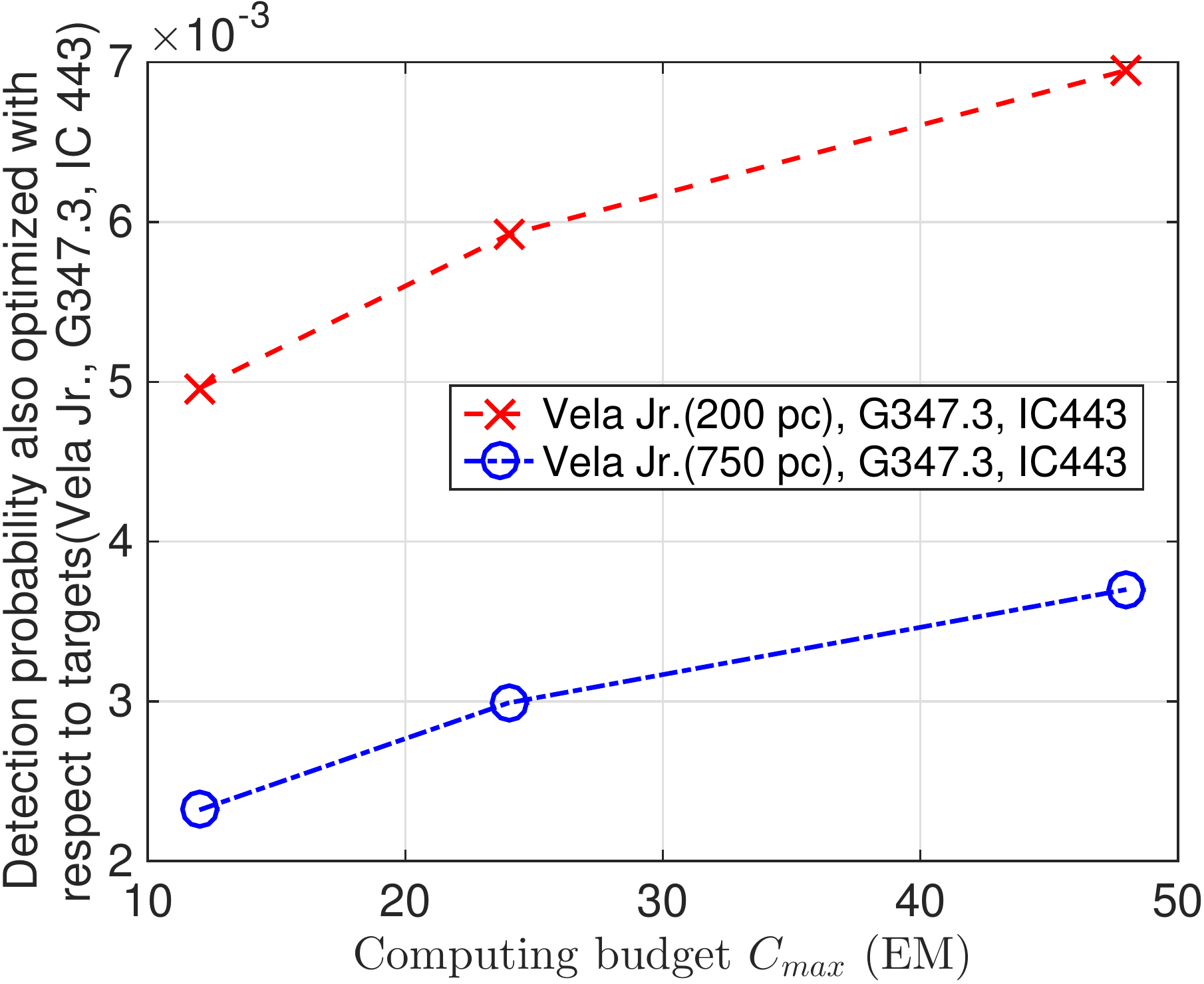}}}%
 \caption{Detection probability for various targets and search set-ups having assumed log-uniform and distance-based priors. The distances that were assumed for the targets are: Vela Jr (C) 200 pc, Vela Jr (F) 750 pc, G 347.3 1.3 kpc, IC443 1.5kpc, Cas A 3.5 kpc.}%
 \label{fig:RversusDistAndTcoh_log}
\end{figure}

\begin{figure}%
 \centering
 \subfloat[${\mathcal{R}}$ versus distance]{{\includegraphics[width=.9\linewidth]{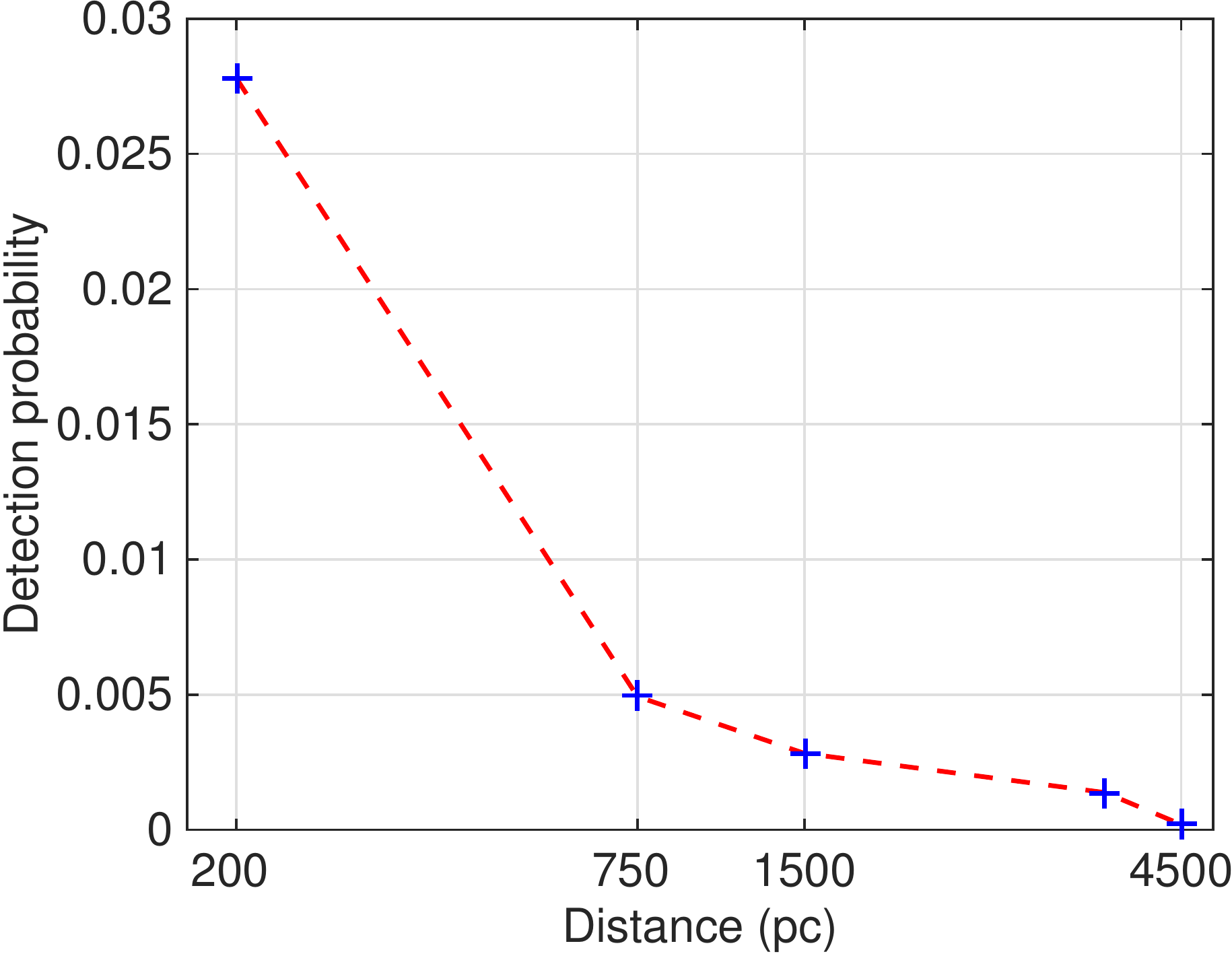}}}%
 \qquad
 \subfloat[${\mathcal{R}}$ versus $T_\mathrm{coh}$]{{\includegraphics[width=.9\linewidth]{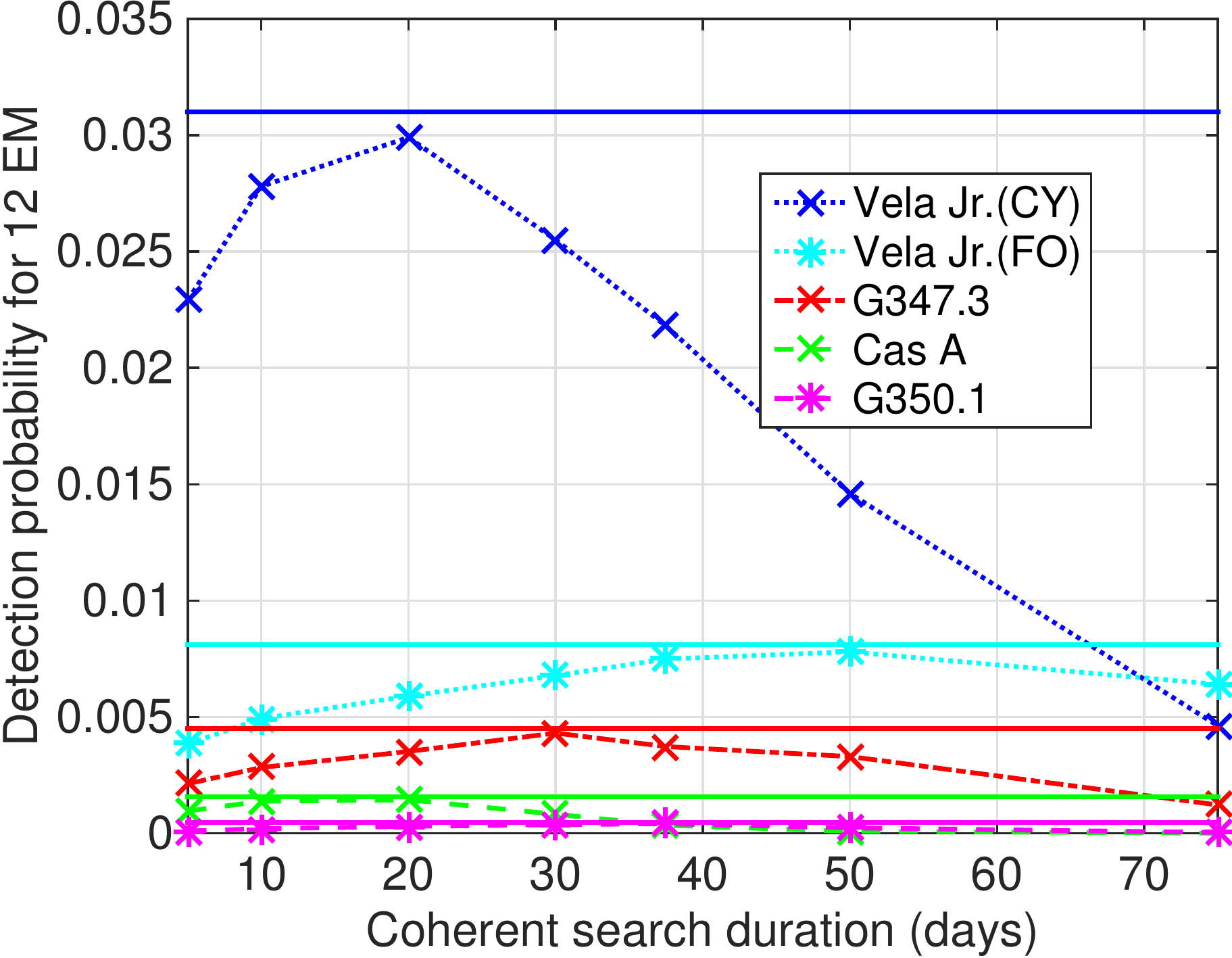}}}%
 \qquad
 \subfloat[${\mathcal{R}}$ versus $C_\mathrm{max}$]{{\includegraphics[width=.9\linewidth]{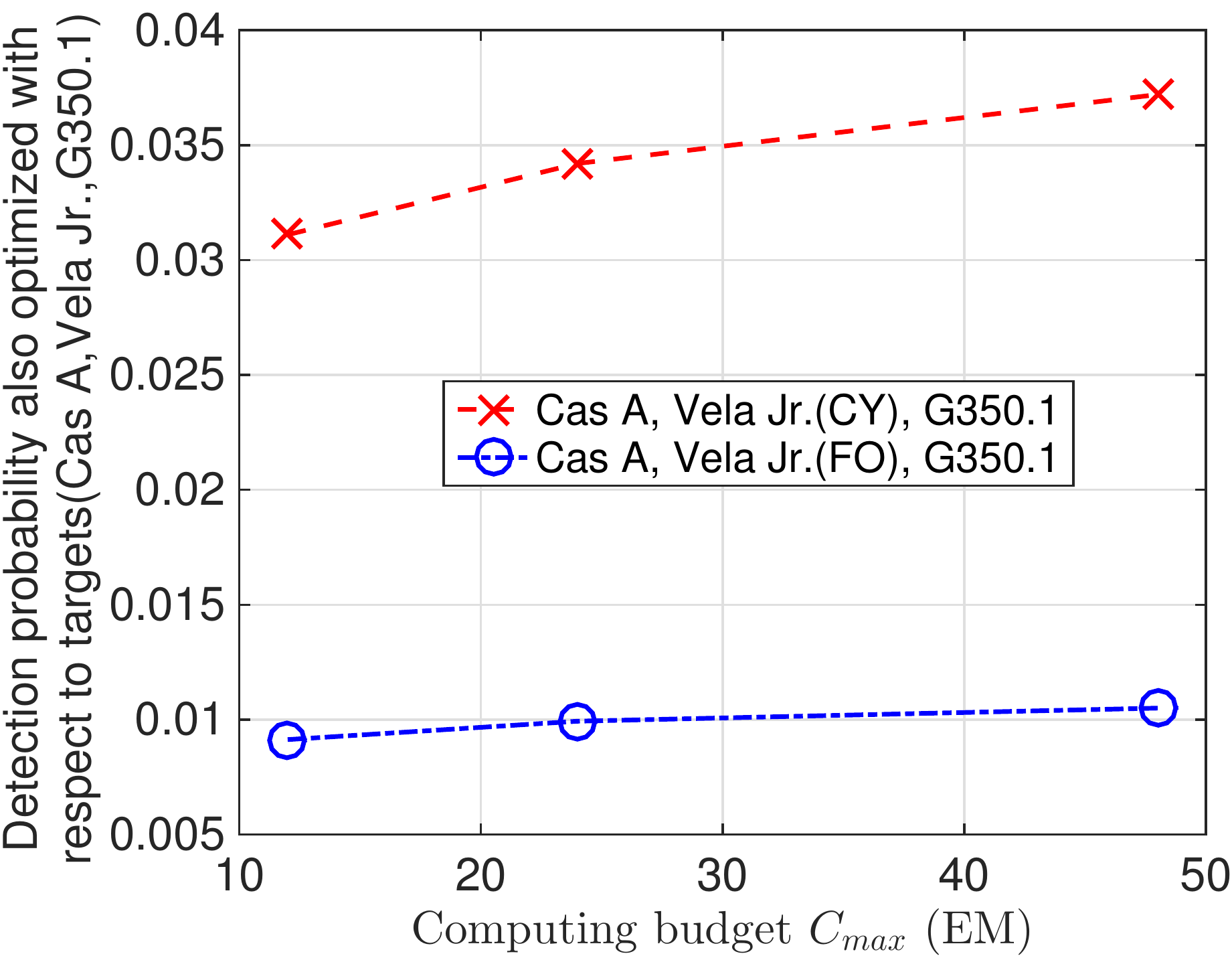}}}%
 \caption{Detection probability for various targets and search set-ups having assumed log-uniform and age-based priors. The distances that were assumed for the targets are: Vela Jr (C) 200 pc, Vela Jr (F) 750 pc, G 347.3 1.3 kpc, Cas A 3.5 kpc, G350.1 4.5 kpc.}%
 \label{fig:RversusDistAndTcohAge_log}
\end{figure}


\begin{table*}
\caption{
  \label{tab: R of nonage log}
  $\mathcal{R}$ result with $f$ and $\dot{f}$ log-uniform priors and distance-based priors. The highest $\mathcal{R}$ with respect to set-up is in bold font.}
\begin{center}
  \begin{tabular}{ll|cccccccccc}
    \tableline
   
   && \multicolumn{10}{c}{$10^3\mathcal{R}$}
   \\
   \cline{3-12}
  
    Name & $~~D_\mathrm{kpc}$ & 5D & 10D&20D&30D&37.5D&50D&75D& \multicolumn{3}{c}{LP Optimized}
    \\
    && \multicolumn{7}{c}{Computing Budget: 12EM} &12EM&24EM&48EM\\
    \tableline

Cas~A &~~3.5&0.167&\textbf{0.278}&0.257&0.187&0.147&0.093&0.039&0.319&0.414&--\\
   
IC~443 &~~1.5 &0.418&0.669&\textbf{0.714}&0.560&0.481&0.341&0.189&0.804&0.991&--\\
     
G347.3&~~1.3 &0.544&0.807&\textbf{0.820}&0.661&0.569&0.422&0.248&0.937&1.15&--\\
    
Vela~Jr &~~0.2&2.97&3.91&\textbf{4.26}&3.94&3.54&3.06&2.34&4.51&5.16&--\\
 
Vela~Jr &~~0.75&0.906&1.37&\textbf{1.47}&1.29&1.12&0.880&0.604&1.64&1.96&--\\
 
Top 3 (0.2 kpc)&~~--&--&--&--&--&--&--&--&4.96&5.92&6.95 \\
  
Top 3 (0.75 kpc)&~~--&--&--&--&--&--&--&--&2.32&2.99&3.70\\

\tableline
  \end{tabular}
\end{center}
\end{table*}

\begin{table*}
\caption{
  \label{tab: R of age log}
  $\mathcal{R}$ result with $f$ and $\dot{f}$ log-uniform priors and age-based  priors. The highest $\mathcal{R}$ with respect to set-up is in bold font.}
\begin{center}

  \begin{tabular}{lll|cccccccccc}
    \tableline
   
   &&& \multicolumn{10}{c}{$10^3\mathcal{R}$}
   \\
   \cline{4-13}
  
    Name & $~~D_\mathrm{kpc}$ & $\tau_\mathrm{kyr}$ &5D & 10D&20D&30D&37.5D&50D&75D& \multicolumn{3}{c}{LP Optimized}
    \\
    &&& \multicolumn{7}{c}{Computing Budget: 12EM} &12EM&24EM&48EM\\
    \tableline

   Cas~A &~~3.5&0.35&0.960&1.38&\textbf{1.44}&0.821&0.347&0.091&0.008&1.56&1.81&--\\
    
    G350.1 &~~4.5 &0.9&0.099&0.200&0.300&0.374&\textbf{0.414}&0.236&0.037&0.469&0.504&--\\
     
     G347.3&~~1.3 &1.6&2.13&2.83&3.52&\textbf{4.31}&3.73&3.30&1.22&4.50&4.75&--\\
   
 Vela~Jr &~~0.2&0.7&22.9&27.8&\textbf{29.9}&25.5&21.8&14.6&4.62&31.0&33.8&--\\
 
  Vela~Jr &~~0.2&4.3&26.9 &31.5&36.4(11.3EM)&\textbf{36.6}&36.4&31.0&33.8&--&--&--\\

  Vela~Jr &~~0.75&4.3&3.84&4.93&5.90(11.3EM)&6.78&7.49&\textbf{7.80}&6.40&8.12&8.36&--\\
 
  Top 3 (CY) &~~--&--&--&--&--&--&--&--&--&31.1&34.2&37.2\\
 
  Top 3 (FO)&~~--&--&--&--&--&--&--&--&--&9.13&9.93&10.5\\

    \tableline
  \end{tabular}
\end{center}
\end{table*}


\begin{figure*}%
     \centering
     \subfloat[Coverage of 3 sources, Cost 12 EM ]{{  \includegraphics[width=.45\linewidth]{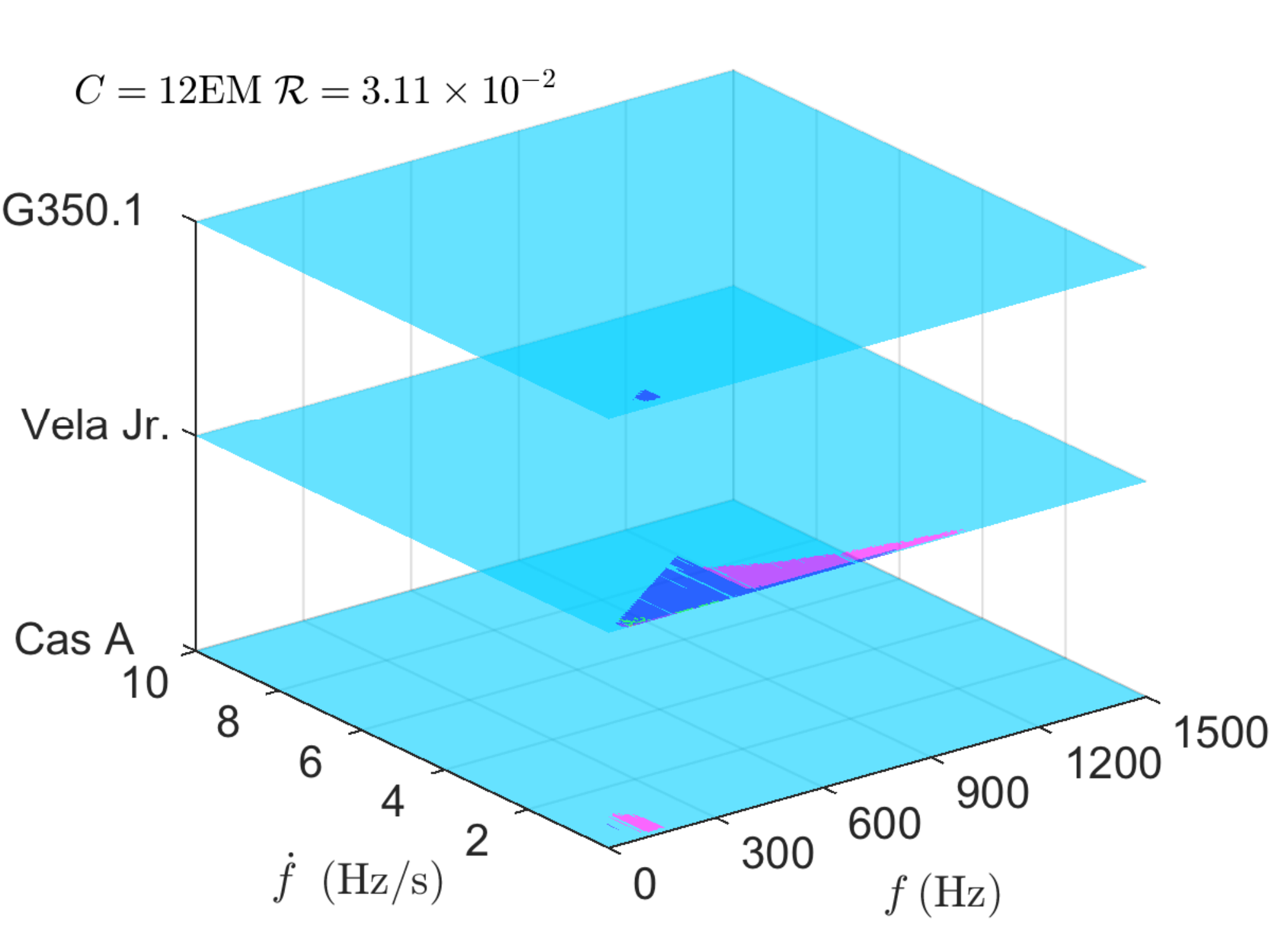}}}%
     \qquad
     \subfloat[Coverage of 3 sources, Cost 12 EM ]{{  \includegraphics[width=.45\linewidth]{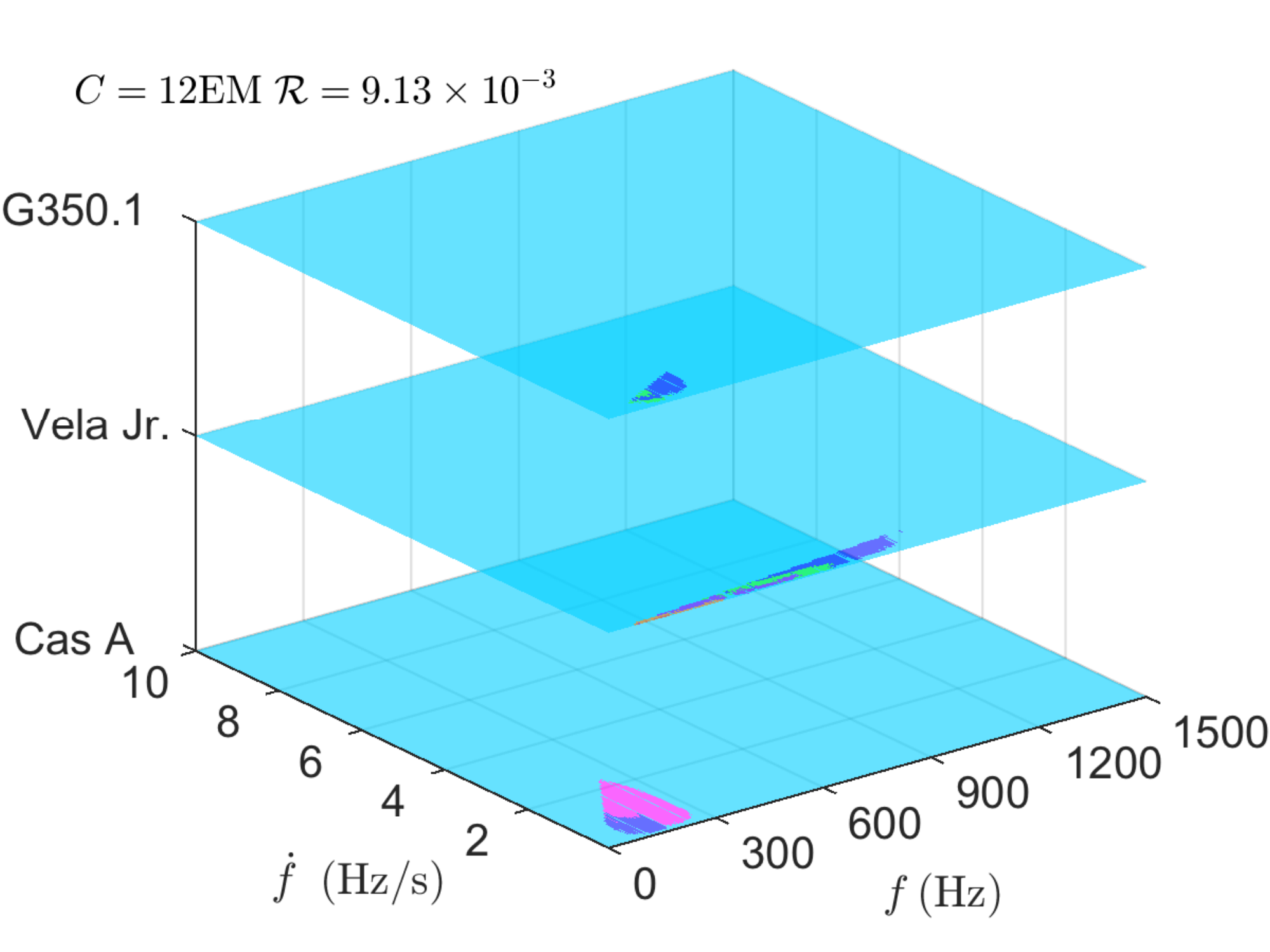}}}%
     \qquad
     \subfloat[Coverage of 3 sources, Cost 24 EM ]{{  \includegraphics[width=.45\linewidth]{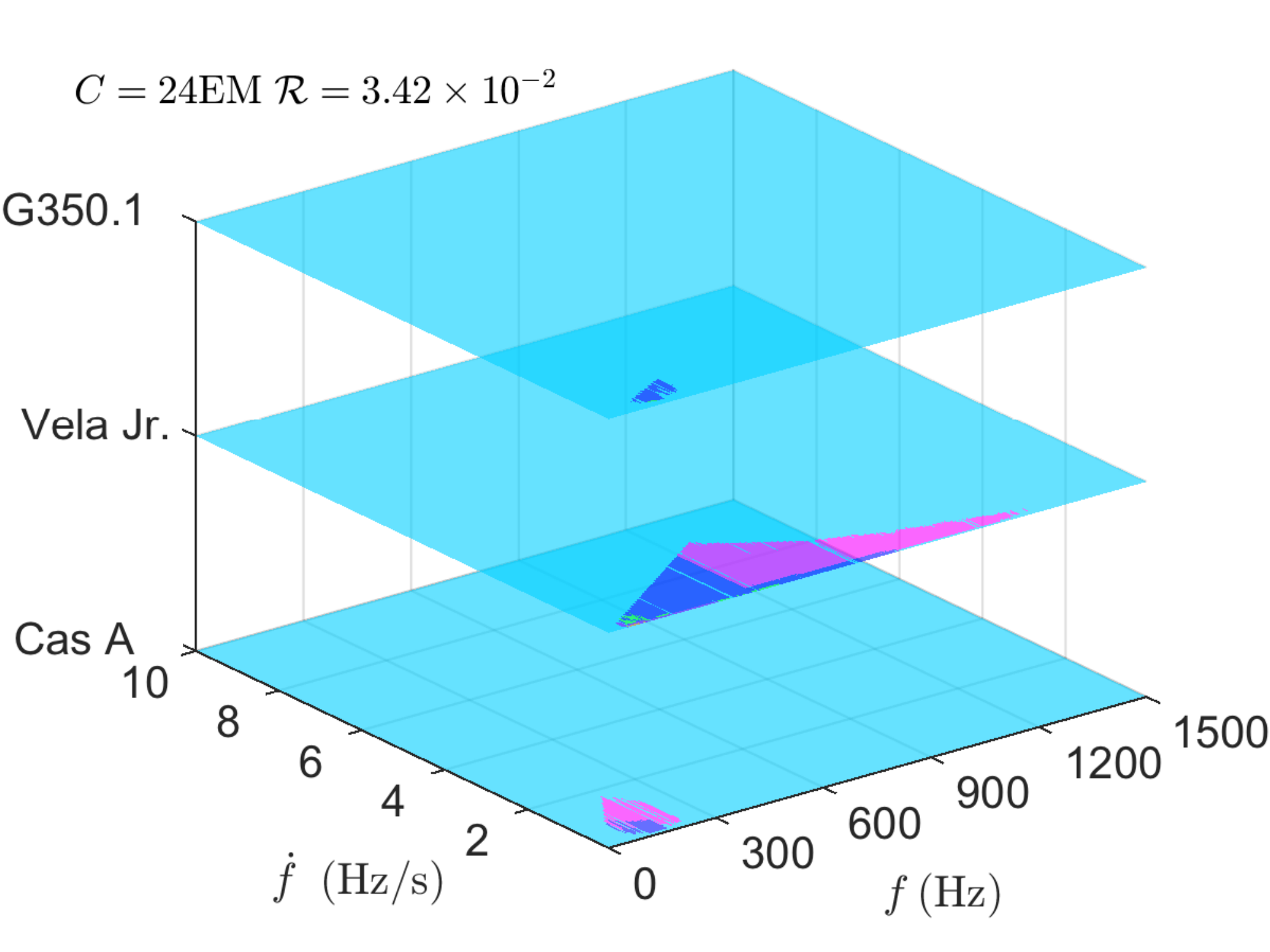}}}%
     \qquad
     \subfloat[Coverage of 3 sources, Cost 24 EM ]{{  \includegraphics[width=.45\linewidth]{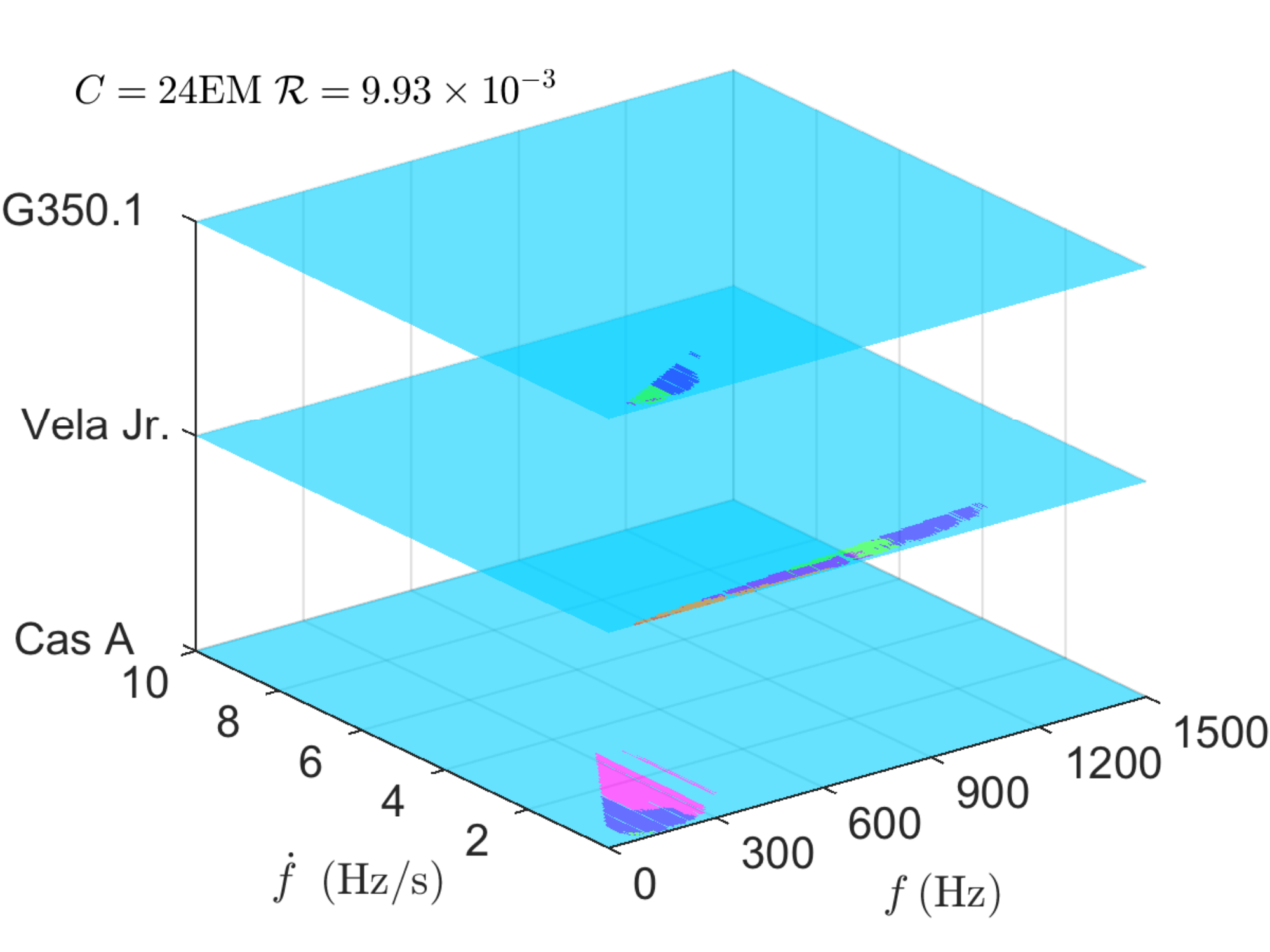}}}%
     \qquad
     \subfloat[Coverage of 3 sources, Cost 48 EM ]{{  \includegraphics[width=.45\linewidth]{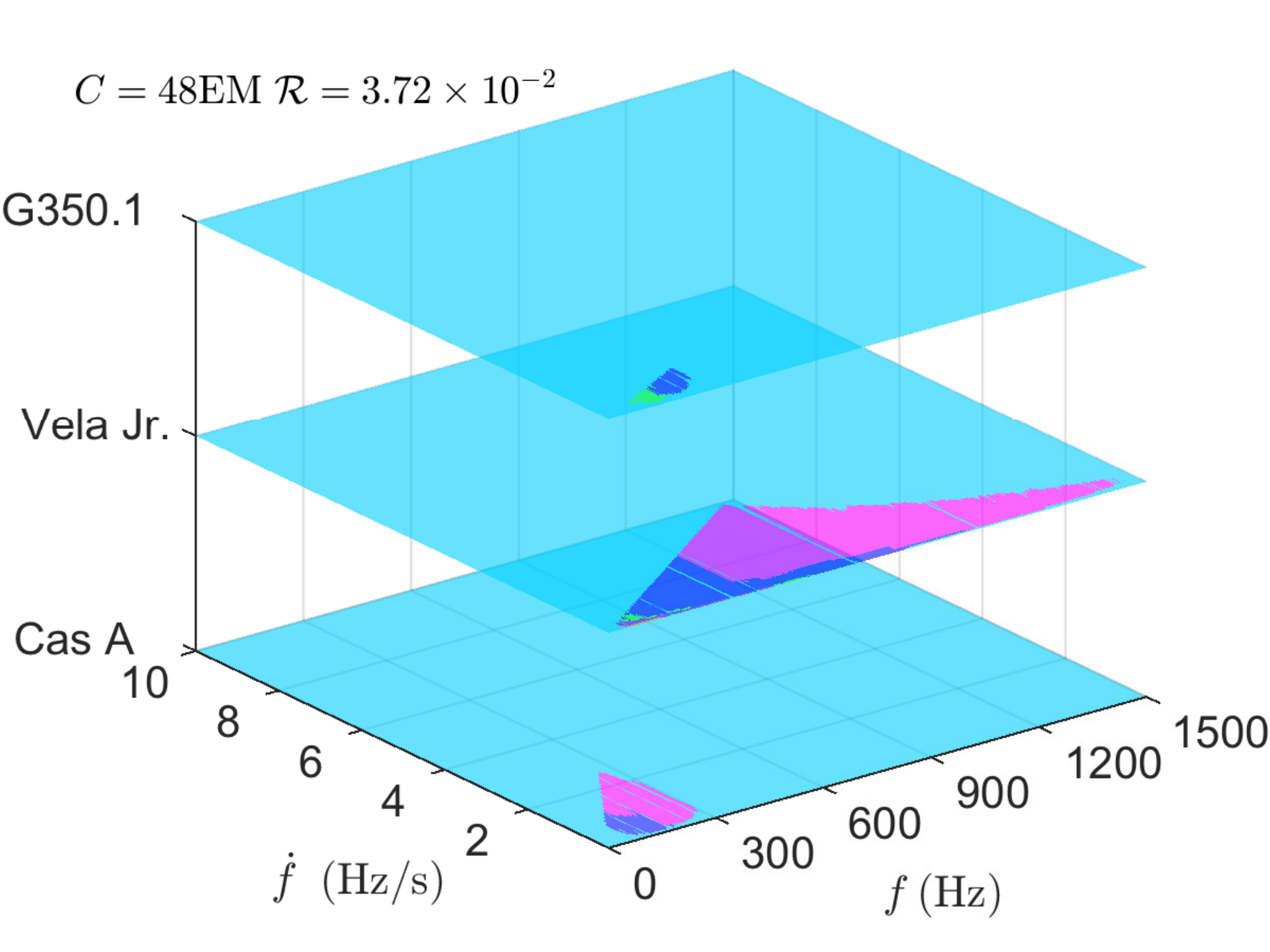}}}%
   \qquad
     \subfloat[Coverage of 3 sources, Cost 48 EM ]{{  \includegraphics[width=.45\linewidth]{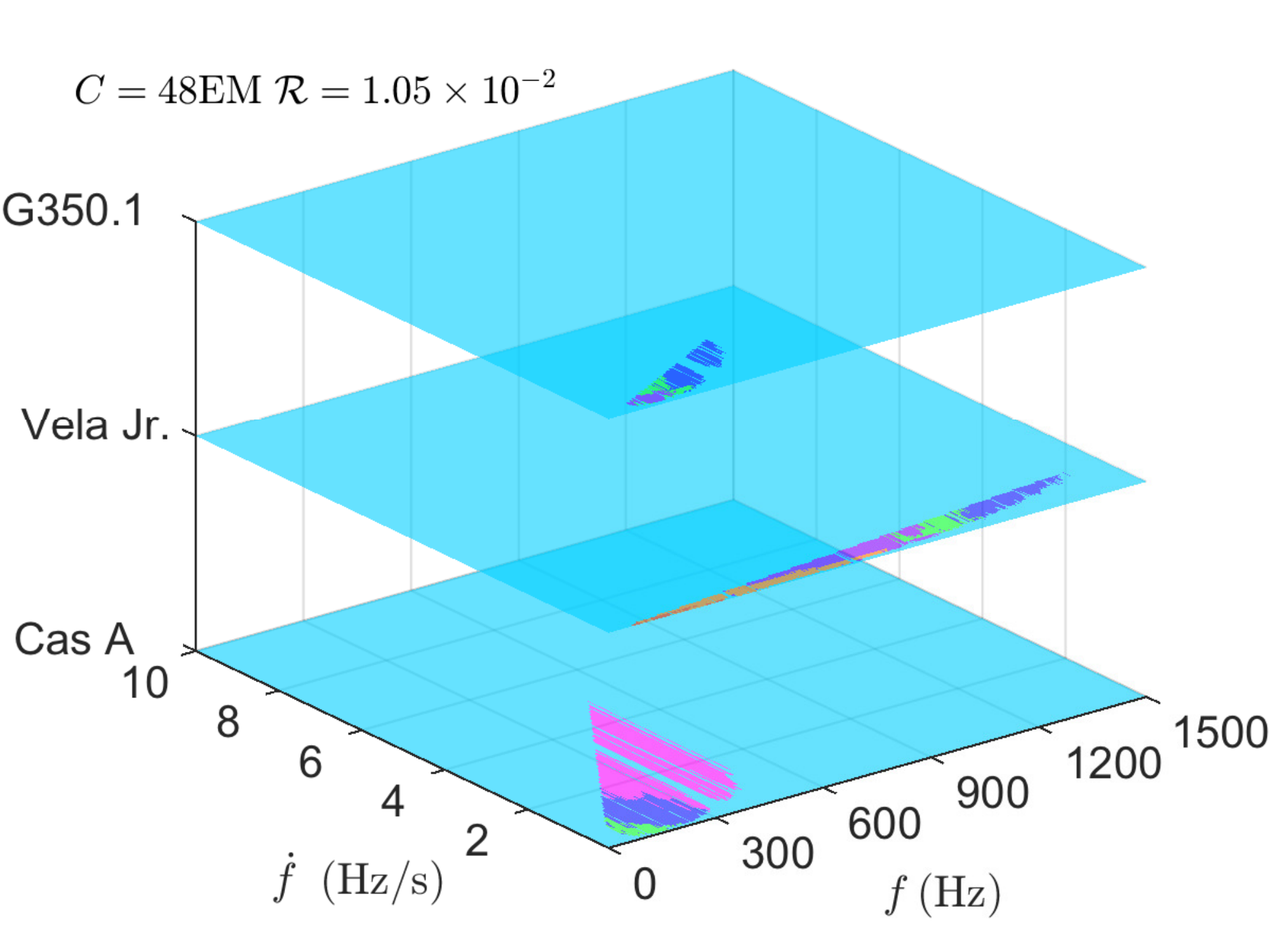}}}%

     \caption{Parameter space coverage assuming log-uniform and age-based priors and optimizing over the 7 search set-ups also considered above and over the three youngest targets (left plots: Cas A, Vela Jr at 200 pc and 700 years old (CY), G350.1 and right plots: Cas A, Vela Jr at 750 pc and 4300 years old (FO), G350.1) at 12, 24 and 48 EMs.}
     \label{all_age_log}%
 \end{figure*}

\begin{figure*}%
    \centering
    \subfloat[0.2 kpc, 12 EM budget]{{  \includegraphics[width=.45\linewidth]{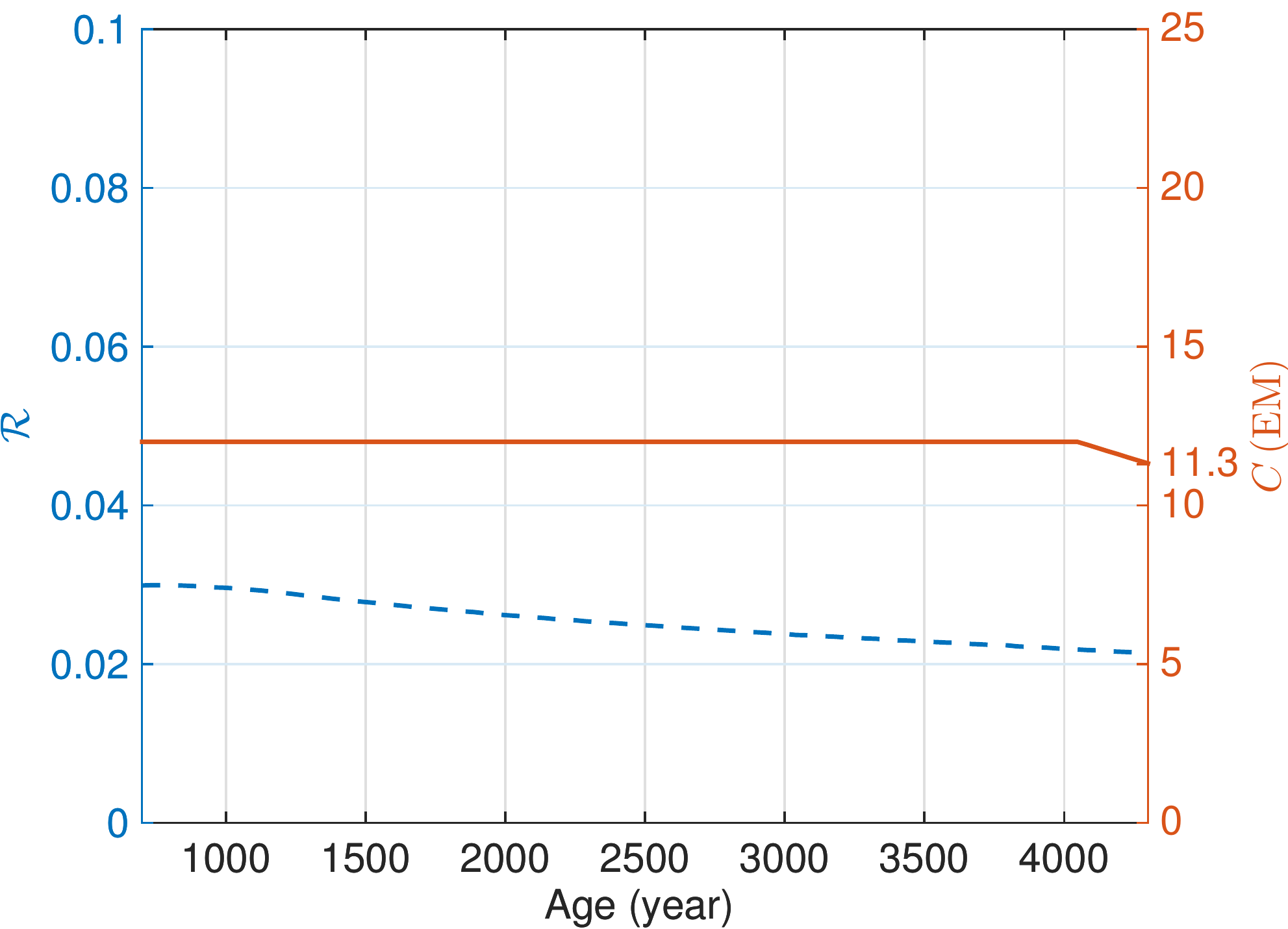}}}%
    \qquad
    \subfloat[0.75 kpc, 12 EM budget]{{  \includegraphics[width=.45\linewidth]{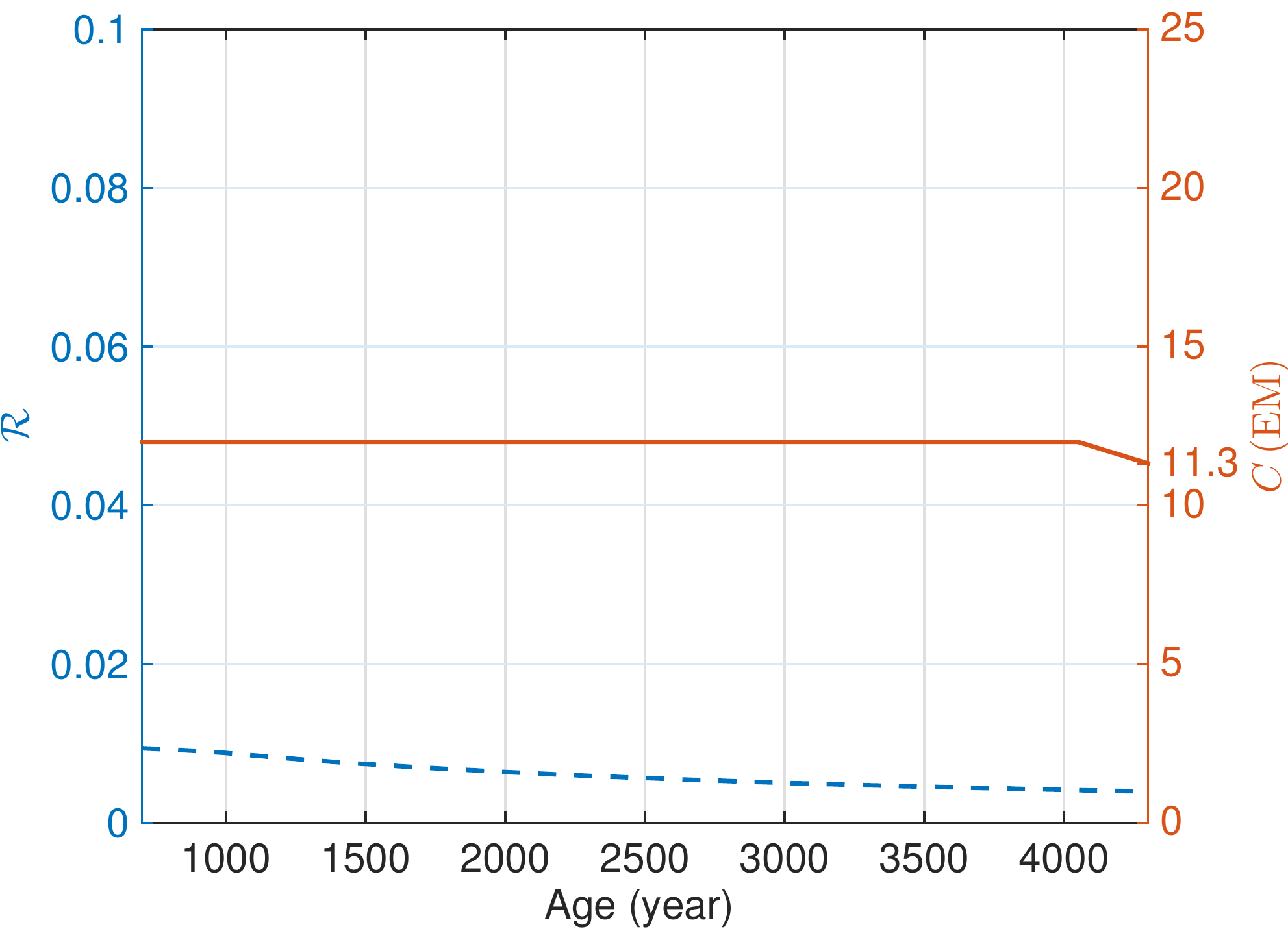}}}%
    \qquad
    \subfloat[0.2 kpc, 24 EM budget]{{  \includegraphics[width=.45\linewidth]{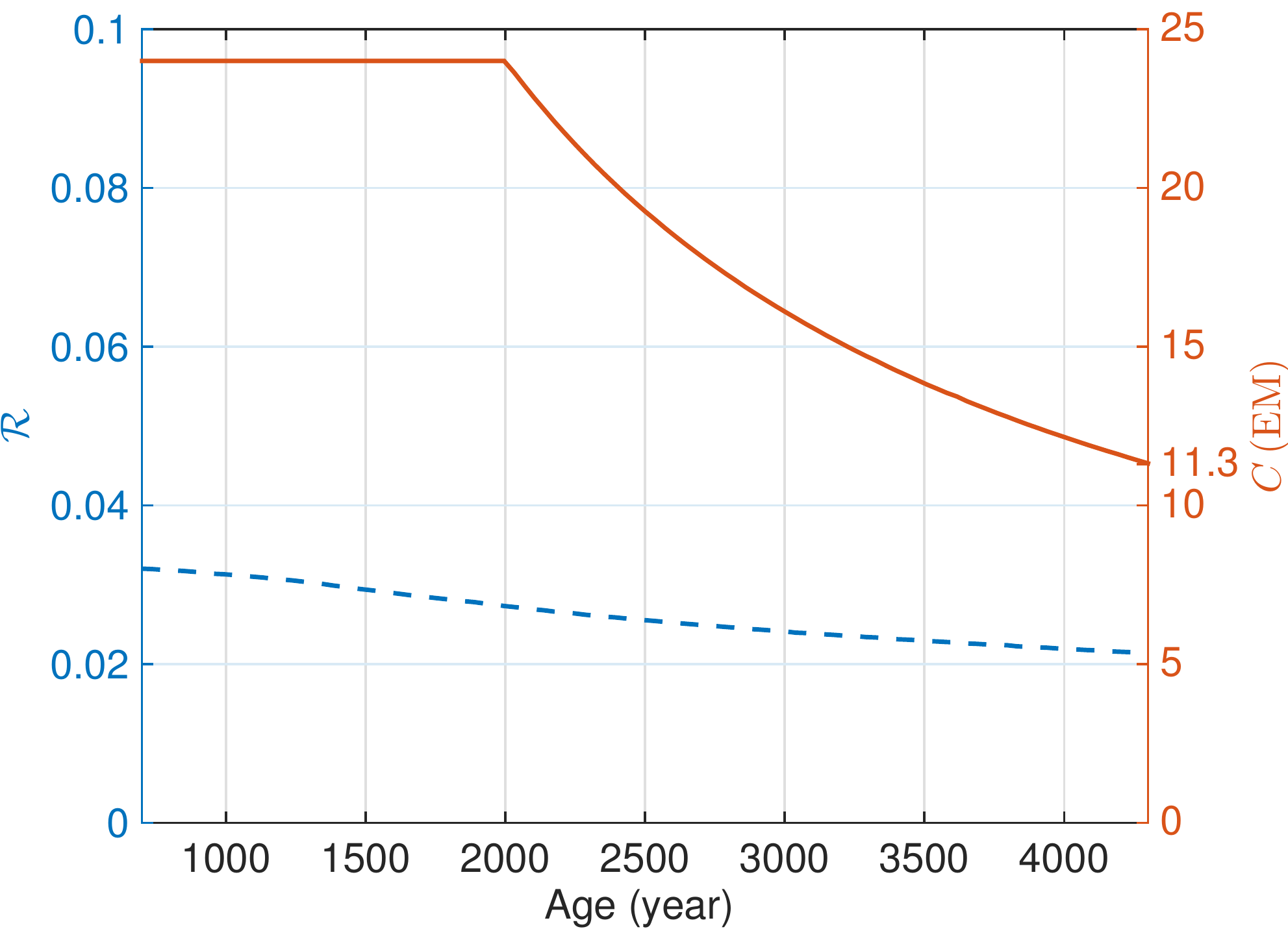}}}%
    \qquad
    \subfloat[0.75 kpc, 24 EM budget]{{  \includegraphics[width=.45\linewidth]{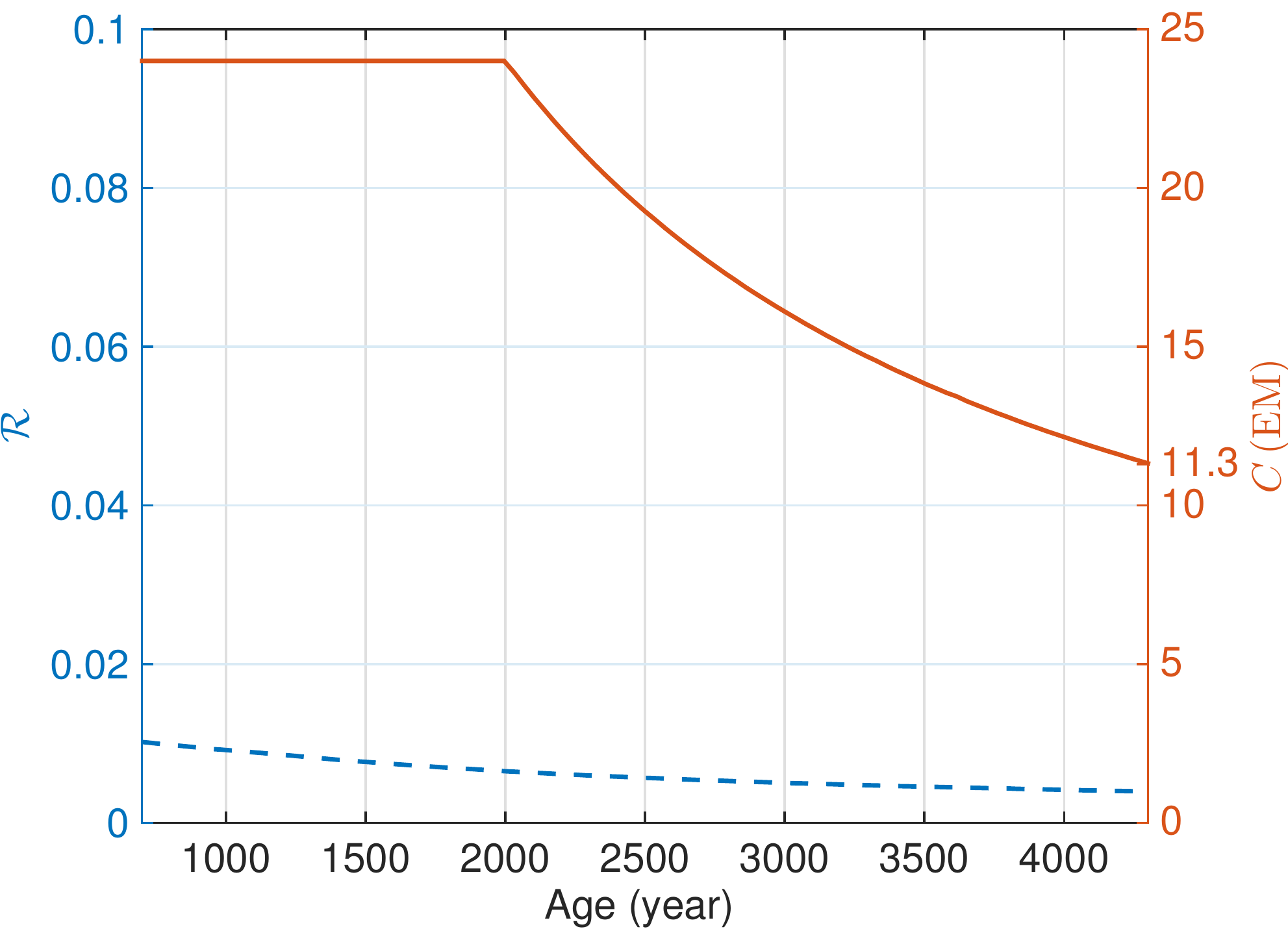}}}%
    \caption{$\mathcal{R}$ and $C$ as a function of age for an optimised search for Vela Jr, assuming log-uniform and age-based priors, a computing budget of 12 and 24 EMs and a 20-day coherent segment search set-up. }%
    \label{fig:investigation_age_plots_log}%
\end{figure*}

We do not comment here our findings with the same level of detail used in the previous section, as that was done in order to highlight the main factors contributing to the results\footnote{All results of assuming log-uniform prior are shown in Figs.~\ref{G2662_51020days_noage_log} to \ref{3501_best_age_log} in Appendix \ref{section:AppendixFigures}.}. Based on the material presented there, we are confident that the interested reader can do this himself/herself here. We highlight instead the following points:
\begin{itemize}
\item The log-uniform priors favour lower frequency and lower spindown values with respect to the uniform priors. 
\item Generally when assuming distance-based priors, this decreases the detection probability because the computing power is more eagerly invested in searching for signals with lower spindowns which typically have smaller maximum amplitudes (through Eqs. (\ref{eq:epsilonMax}) and (\ref{eq:epsilonMaxAge})). We note about a factor 2-7.6 decrease across the board.  
\item This is strictly not true when assuming age-based priors in fact  for Vela Jr (CY) the detection probability at 12 EM  increases from 2.16\%  (uniform and age-based) to 3.10\% (log-uniform and age-based). For all the other sources the detection probability decreases but not as much as in the distance-based case. 
\item At fixed source, the optimisation scheme prescribes segment lengths which are higher with respect to those of the uniform-priors searches. The reason for this is that longer duration segments can be more easily afforded at lower $\dot{f}$ regions, where the $\ddot{f}$ costs are lower.
\item The prescription for the parameter volumes to search is quite different: this is evident for all sources and set-ups. For example from the 3D plots shown in Figs.~\ref{all_age} and \ref{all_age_log} we see how markedly the log-uniform priors disfavour high $f-\dot{f}$ combinations with respect to the uniform priors.

\item Fig.~\ref{fig:investigation_age_plots_log} shows how $\mathcal{R}$ and $C$ vary with age. The cost curve is the same as the cost curve of Fig.~\ref{fig:investigation_age_plots}. This is quite obvious because the $f$ and $\dot{f}$ log-uniform prior does not  change the computing cost in each cell. However the log-uniform prior has a large influence on the detection probability: the $\tau_{\mathrm{peak}}$ is 700 years, the smallest. The reason is that the contribution to the detection probability from higher spindown cells that are excluded with respect to the uniform-prior case, is not compensated for by the higher fractional volume of searched parameter space. This means, for log-uniform priors, for the sources with the same distance, ``The younger, the better".

\end{itemize}

\section{Conclusions} 
\label{sec:conc}

Searches for continuous GWs, even the directed ones from sources with known
sky-positions, are computationally limited and decisions regarding the
parameter space, search set-up and the astrophysical target can make
the difference between making or missing a detection.  We have
described and implemented an optimization scheme with the goal of
maximizing the detection probability constrained by a limited
computing budget.  Specifically, we have addressed
the following questions:
\begin{itemize}
\item On which target(s) should we spend our computing resources?
\item What parameter space region in frequency and spindown should we search? 
\item What is the optimal search set-up that we should use?
\item What is the probability of making a detection, given prior assumptions on the signal parameters?
\end{itemize}

The crucial step in our procedure is that of choosing the priors on the 
frequency, spindown and ellipticity of the source. We choose the broadest range of 
plausible values under combinations of two different  assumptions: namely using or not the information 
on the age of  the object (age-based priors) and using uniform priors or log-uniform priors for the frequency and frequency derivative. 
The uniform priors are useful to illustrate the method. The log-uniform priors are more realistic. With these we find the following:
\begin{itemize}
\item Distance-based priors yield detection probabilities on average a factor of 4 smaller when used in conjunction with log-uniform priors than when used in conjunction with uniform priors. For age-based priors this difference is no more than a factor of two. 
\item The highest detection probability for a search at the LIGO S6 run sensitivity level, using about a year of data from two detectors with a duty factor of 50\% and assuming a computing budget of 12 EM, is  6.7\%. This is obtained under the assumption that Vela Jr is old, 200 pc away, a uniform distribution for $f$ and $\dot{f}$, an age-based prior, and assuming that these priors reflect reality. If the $f$ and $\dot{f}$ are instead log-uniformly distributed and we match our priors to this assumption, the detection probability drops to 3.6\%.
\item The optimisation over set-up for every cell in parameter space yields at most 15\% increase in detection probability with respect to single set-up search. Given the complexity of setting up and analysing the results of a search that uses different segment lengths for different areas of parameter space, this result is relevant because it indicates that using a single set-up or at most two (a practical solution), does not significantly impact the chances of making a detection.
\item Independently of all prior assumptions, all optimal searches cover the broadest fraction of the prior spin down range around the instruments' maximum sensitivity frequencies.
\end{itemize}

In forthcoming work we will investigate different priors, consider a range of search set-ups including different 
mismatch parameters, grids, number of segments and segment durations, and optimise over all these. We will fold in the mismatch distribution
arising from our choices of nominal mismatch values and of the grids, and not only work with the expected values as done here. 
Furthermore here we have not 
considered any uncertainty on the distance of the target and presented results separately having assumed 
different distances. A more general approach is to marginalize over the distance range using an appropriate prior, for 
example that given by \cite{Allen:2014yra}. The same applies to the age estimates.

What we want to stress with this paper is that the parameter space to be searched and the targets 
to be searched should be part of the search optimization procedure, as well as the search parameters themselves. In previous works these aspects have 
been considered separately: e.g. \cite{BC00,HierarchP3,Prix:2012yu} and \cite{Palomba:2005fa,Knispel:2008ue,Wade:2012qc,Owen:2009tj}.
The interplay between these quantities, for some assumed prior, is very difficult to intuitively predict and hence it is 
important to have a rational method to do so. The method that we propose here effectively achieves this goal and lends itself to further generalisations.

%

\section{Acknowledgements}
J.M. acknowledges support by the IMPRS on Gravitational Wave Astronomy at the Max Planck Institute for Gravitational
Physics in Hannover. M.A.P. gratefully acknowledges support from NSF PHY grant
1104902. The authors thank their colleagues Bruce Allen for useful
suggestions that were adopted in this work. We acknowledge
Reinhard Prix, Keith Riles, Curt Cutler, Ben Owen and Hyung Mok Lee for stimulating
discussions on this work.  We also thank Paola Leaci and David Keitel for their comments on the manuscript. 
This paper was assigned LIGO document number P1500188.

\appendix
\section{Linear programming}
\label{A:LP}


In this appendix we provide some further details of the method of
Linear Programming (LP) and its application to our problem.

Recall that the occupation numbers $X_{i,s}$ (or equivalently $X_j$)
were originally specified as binary numbers, i.e. $X_{i,s}$ could be
either 0 or 1.  It is however non-trivial to design an algorithm which
solves the optimization problem described in
Sec.~\ref{subsec:multiplesetups}. Rather than trying to do so, we have
formulated the problem by taking $X_{i,s}$ to be real and requiring
$0\leq X_{i,s}\leq 1$.  We have seen how the optimization problem can
be solved using linear programming (LP).

The first question that arises is: by allowing $X_{i,s}$ to be real,
do the solutions which maximize $P_\mathrm{sum}$ have the vast majority of
the $X_{i,s}$ as either 0 or 1?  This is observed empirically to be
true in all the cases that we have studied in this paper.  We shall
now demonstrate that this is in fact a more general feature.  We shall
restrict ourselves here to the case when there are two possible
set-ups for each cell.

To illustrate this we define the efficiency
$E_{i,s}=P_{i,s}/C_{i,s}$. LP yields a set of non trivially occupied
cells which can be ordered in decreasing values of $E_{i,s}$; thus,
$i=1$ corresponds to the cell with the largest efficiency, $i=2$ the
second largest and so on.  Consider first the non-degenerate case
where all $E_{i,s}$ are mutually different and we assume that in each
cell $i$ only one of the two $X_{i,s}$ is strictly bigger than $0$.
We will show in this case that the cell with the lowest efficiency
(let $j$ be the index for this cell) is the only one which can have a
fractional occupation: $0 < X_{j,s_j} \leq 1$. If any cell with index
$l$ with a higher efficiency would have a fractional occupation, the
total $P_\mathrm{sum}$ can be increased by decreasing $X_{j,s_j}$ and
increasing $X_{l,s_l}$ until either $X_{l,s_l}=1$ or $X_{j,s_j}=0$
such that the total cost $C_\mathrm{max}$ remains constant.  This argument
holds for all $l < j$, hence, all $X_{l,s_l}$ with $l < j$ must be
unity. If $X_{j,s_j}$ is set to be $0$ the cell with index $j-1$ is
now the one with the lowest efficiency among all non-trivially
occupied cells. Since $C_{i,s} \ll C_\mathrm{max}, \forall i$ the total cost
will be changed marginally if we set $X_{j,s_j}$ either to $0$ or to
$1$.

If a subset of non-trivially occupied cells has the same efficiency,
$P_\mathrm{sum}$ and $C_\mathrm{max}$ do not change if we decrease the occupation
$X_{i,s}$ by an amount $\Delta_{i,s}$ and increase the occupation of
another cell $X_{j,s_j}$ by $\Delta_{j,s_j}$ if both cells have the
same efficiency and if $\Delta_{j,s_j}/\Delta_{i,s} =
P_{i,s}/P_{j,s_j}$ is fulfilled. We can shift the occupation among
these cells such that one part has occupation $1$ and another part
$0$.  One cell of this subset will likely have a fractional occupation
which can be set as well to either $0$ or $1$ without changing the
total $C_\mathrm{max}$ significantly.  Following the previous argument, all
cells with higher efficiency must have the occupation number equal to
unity.

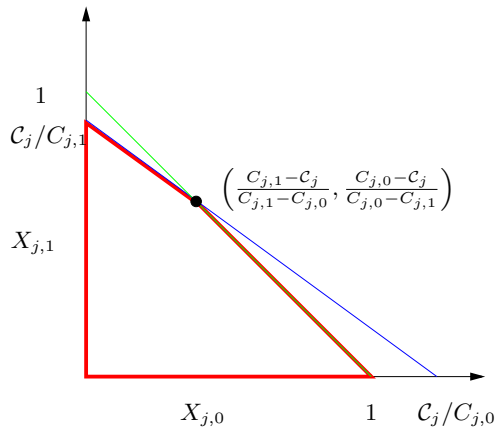
\begin{figure}
\begin{picture}(140,140)(0,0)
\put(0,15){
\includegraphics[width=0.30\textwidth]{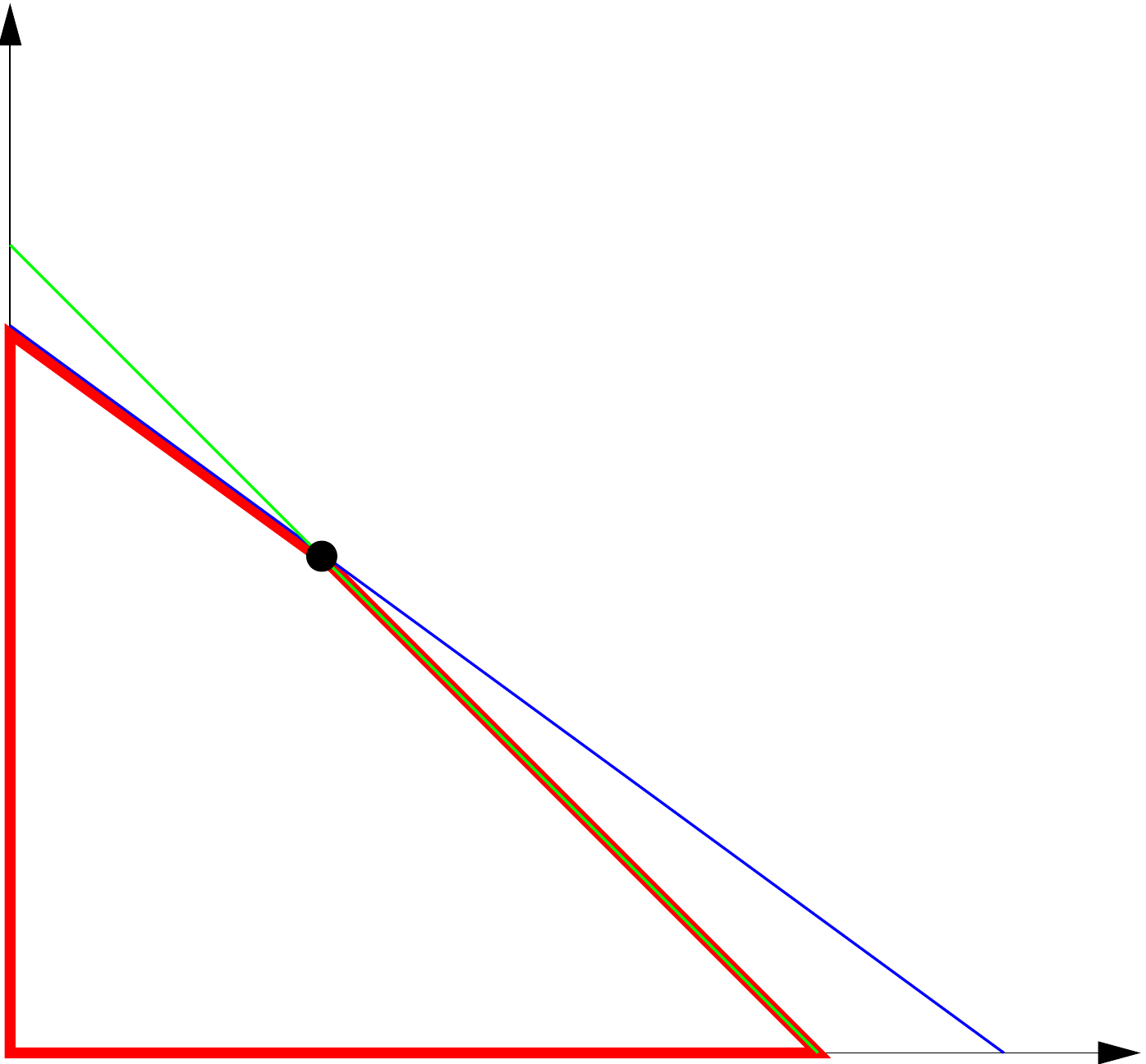}
}
\put(-24,105){$\mathcal{C}_j/ C_{j,1}$}
\put(-15,120){$1$}
\put(-24,65){$X_{j,1}$}
\put(40,0){$X_{j,0}$}
\put(110,0){$1$}
\put(130,0){$\mathcal{C}_j/ C_{j,0}$}
\put(55,85){$\left(\frac{C_{j,1}-\mathcal{C}_j}{C_{j,1}-C_{j,0}},\frac{C_{j,0}-\mathcal{C}_j}{C_{j,0}-C_{j,1}}\right)$}
\end{picture}
\caption{The red tetragon envelops a volume of variable space in which all
inequalities are fulfilled.}
\label{fig:LP_polygon}
\end{figure}

We would now like to show that for each cell $i$ with non-trivial
occupation numbers, only one of the two $X_{i,s} > 0$, unless the cell
has the lowest efficiency. We illustrate this by using the geometrical
interpretation of LP. The set of inequalities described earlier define
a polygon in the space of the $X_{i,s}$ in which valid solutions
exist. The set of inequalities can lead to either no solutions, an
unbounded problem, a unique solution or infinity many solutions. In
our situation only the latter two cases are possible. If only a single
solution is possible the optimal point lies in one corner of the
polygon. If more than one corner points were to lead to the same
optimal $P_\mathrm{sum}$ any point in the volume enveloped by these points
yield the same $P_\mathrm{sum}$.  The costs of our ordered set of
non-trivially occupied cells can be summed from cell 1 (the one with
the highest efficiency) up to the cell $j-1$. The remaining cost is
then $\mathcal{C}_j=C-\sum_{i=1}^{j-1} \sum_{s} C_{i,s} X_{i,s}$. We
consider now a subset of inequalities valid for cell $j$. There is
$X_{j,s_j} > 0$, $\sum_{s_j} X_{j,s_j} < 1$ and $\sum_{s_j} C_{j,s_j}
X_{j,s_j}< \mathcal{C}_j$.  The polygon is either a triangle, if
$\mathcal{C}_j/C_{j,s_j}$ is either bigger than 1 or smaller than 1
for both $s_j$. As depicted in Fig.~\ref{fig:LP_polygon}, the polygon
is a tetragon if one $\mathcal{C}_j/C_{j,s_j}$ is bigger than 1 for
one of the $s_j$ and smaller than 1 for the other. Both fractions
being bigger than 1 means that enough remaining cost is left to fully
occupy the cell with one of the two $X_{j,s_j}$.  Smaller than $1$
means, the cell is the non-trivially occupied cell with the smallest
efficiency. The remaining costs will be used in this cell.

If the enveloping polygon is a tetragon for the cell $j$ with the
lowest efficiency, $P_\mathrm{sum}$ is maximized if we chose the $X_{j,s_j}$
to be in one of the corners
$\left(\min(\mathcal{C}_j/C_{j,0},1),0\right)$,$\left(0,\min(\mathcal{C}_j/C_{j,0},1)\right)$
or $\left(
  (C_{j,1}-\mathcal{C}_j)/(C_{j,1}-C_{j,0}),(C_{j,0}-\mathcal{C}_j)/(C_{j,0}-C_{j,1})\right)$.
The latter case means that the LP optimization leads to a fractional
occupation of both $X_{j,s_j}$ simultaneously. Again, setting one
$X_{j,s_j}$ to $0$, the other one to $1$ or both to $0$ will not
change $C_\mathrm{max}$ significantly.  We can however not exclude that
pathological cases which are not covered by our assumptions. In
practice we only observed the cases decribed here and moreover, we can
always shift a few $X_{i,s}$ such that we only have integer
occupations for only a small change in the computational cost budget.

\bibliography{refs}

\begin{thebibliography}{50}%
\makeatletter
\providecommand \@ifxundefined [1]{%
 \@ifx{#1\undefined}
}%
\providecommand \@ifnum [1]{%
 \ifnum #1\expandafter \@firstoftwo
 \else \expandafter \@secondoftwo
 \fi
}%
\providecommand \@ifx [1]{%
 \ifx #1\expandafter \@firstoftwo
 \else \expandafter \@secondoftwo
 \fi
}%
\providecommand \natexlab [1]{#1}%
\providecommand \enquote  [1]{``#1''}%
\providecommand \bibnamefont  [1]{#1}%
\providecommand \bibfnamefont [1]{#1}%
\providecommand \citenamefont [1]{#1}%
\providecommand \href@noop [0]{\@secondoftwo}%
\providecommand \href [0]{\begingroup \@sanitize@url \@href}%
\providecommand \@href[1]{\@@startlink{#1}\@@href}%
\providecommand \@@href[1]{\endgroup#1\@@endlink}%
\providecommand \@sanitize@url [0]{\catcode `\\12\catcode `\$12\catcode
  `\&12\catcode `\#12\catcode `\^12\catcode `\_12\catcode `\%12\relax}%
\providecommand \@@startlink[1]{}%
\providecommand \@@endlink[0]{}%
\providecommand \url  [0]{\begingroup\@sanitize@url \@url }%
\providecommand \@url [1]{\endgroup\@href {#1}{\urlprefix }}%
\providecommand \urlprefix  [0]{URL }%
\providecommand \Eprint [0]{\href }%
\providecommand \doibase [0]{http://dx.doi.org/}%
\providecommand \selectlanguage [0]{\@gobble}%
\providecommand \bibinfo  [0]{\@secondoftwo}%
\providecommand \bibfield  [0]{\@secondoftwo}%
\providecommand \translation [1]{[#1]}%
\providecommand \BibitemOpen [0]{}%
\providecommand \bibitemStop [0]{}%
\providecommand \bibitemNoStop [0]{.\EOS\space}%
\providecommand \EOS [0]{\spacefactor3000\relax}%
\providecommand \BibitemShut  [1]{\csname bibitem#1\endcsname}%
\let\auto@bib@innerbib\@empty
\bibitem [{\citenamefont {Abramovici}\ \emph {et~al.}(1992)\citenamefont
  {Abramovici}, \citenamefont {Althouse}, \citenamefont {Drever}, \citenamefont
  {Gursel}, \citenamefont {Kawamura} \emph {et~al.}}]{ligoref}%
  \BibitemOpen
  \bibfield  {author} {\bibinfo {author} {\bibfnamefont {A.}~\bibnamefont
  {Abramovici}}, \bibinfo {author} {\bibfnamefont {W.~E.}\ \bibnamefont
  {Althouse}}, \bibinfo {author} {\bibfnamefont {R.~W.}\ \bibnamefont
  {Drever}}, \bibinfo {author} {\bibfnamefont {Y.}~\bibnamefont {Gursel}},
  \bibinfo {author} {\bibfnamefont {S.}~\bibnamefont {Kawamura}},  \emph
  {et~al.},\ }\href@noop {} {\bibfield  {journal} {\bibinfo  {journal}
  {Science}\ }\textbf {\bibinfo {volume} {256}},\ \bibinfo {pages} {325}
  (\bibinfo {year} {1992})}\BibitemShut {NoStop}%
\bibitem [{\citenamefont {Harry}(2010)}]{ligoref3}%
  \BibitemOpen
  \bibfield  {author} {\bibinfo {author} {\bibfnamefont {G.~M.}\ \bibnamefont
  {Harry}} (\bibinfo {collaboration} {LIGO Scientific Collaboration}),\ }\href
  {\doibase 10.1088/0264-9381/27/8/084006} {\bibfield  {journal} {\bibinfo
  {journal} {Class.Quant.Grav.}\ }\textbf {\bibinfo {volume} {27}},\ \bibinfo
  {pages} {084006} (\bibinfo {year} {2010})}\BibitemShut {NoStop}%
\bibitem [{\citenamefont {Accadia}\ \emph {et~al.}(2012)\citenamefont {Accadia}
  \emph {et~al.}}]{virgo2}%
  \BibitemOpen
  \bibfield  {author} {\bibinfo {author} {\bibfnamefont {T.}~\bibnamefont
  {Accadia}} \emph {et~al.},\ }\href@noop {} {\bibfield  {journal} {\bibinfo
  {journal} {Journal of Instrumentation}\ }\textbf {\bibinfo {volume} {7}},\
  \bibinfo {pages} {P03012} (\bibinfo {year} {2012})}\BibitemShut {NoStop}%
\bibitem [{\citenamefont {{Danzmann}}(1995)}]{Geo}%
  \BibitemOpen
  \bibfield  {author} {\bibinfo {author} {\bibfnamefont {K.}~\bibnamefont
  {{Danzmann}}},\ }in\ \href@noop {} {\emph {\bibinfo {booktitle} {First
  Edoardo Amaldi conference on gravitational wave experiments}}},\ \bibinfo
  {editor} {edited by\ \bibinfo {editor} {\bibnamefont {{E.~Coccia, G.~Pizzella
  \& F.~Ronga (World Scientific, Singapore)}}}}\ (\bibinfo {year} {1995})\ pp.\
  \bibinfo {pages} {100--111}\BibitemShut {NoStop}%
\bibitem [{\citenamefont {Somiya}(2012)}]{kagraref}%
  \BibitemOpen
  \bibfield  {author} {\bibinfo {author} {\bibfnamefont {K.}~\bibnamefont
  {Somiya}} (\bibinfo {collaboration} {KAGRA Collaboration}),\ }\href {\doibase
  10.1088/0264-9381/29/12/124007} {\bibfield  {journal} {\bibinfo  {journal}
  {Class.Quant.Grav.}\ }\textbf {\bibinfo {volume} {29}},\ \bibinfo {pages}
  {124007} (\bibinfo {year} {2012})},\ \Eprint {http://arxiv.org/abs/1111.7185}
  {arXiv:1111.7185 [gr-qc]} \BibitemShut {NoStop}%
\bibitem [{\citenamefont {Iyer}\ \emph {et~al.}(2011)\citenamefont {Iyer},
  \citenamefont {Souradeep}, \citenamefont {Unnikrishnan}, \citenamefont
  {Dhurandhar}, \citenamefont {Raja},\ and\ \citenamefont
  {Sengupta}}]{ligoindia}%
  \BibitemOpen
  \bibfield  {author} {\bibinfo {author} {\bibfnamefont {B.}~\bibnamefont
  {Iyer}}, \bibinfo {author} {\bibfnamefont {T.}~\bibnamefont {Souradeep}},
  \bibinfo {author} {\bibfnamefont {C.}~\bibnamefont {Unnikrishnan}}, \bibinfo
  {author} {\bibfnamefont {S.}~\bibnamefont {Dhurandhar}}, \bibinfo {author}
  {\bibfnamefont {S.}~\bibnamefont {Raja}}, \ and\ \bibinfo {author}
  {\bibfnamefont {A.}~\bibnamefont {Sengupta}},\ }\href@noop {} {\enquote
  {\bibinfo {title} {Ligo-india, proposal of the consortium for indian
  initiative in gravitational-wave observations (indigo)},}\ }\bibinfo
  {howpublished}
  {\url{https://dcc.ligo.org/cgi-bin/DocDB/ShowDocument?docid=75988}} (\bibinfo
  {year} {2011})\BibitemShut {NoStop}%
\bibitem [{\citenamefont {Aasi}\ \emph {et~al.}(2014)\citenamefont {Aasi} \emph
  {et~al.}}]{Aasi:2013sia}%
  \BibitemOpen
  \bibfield  {author} {\bibinfo {author} {\bibfnamefont {J.}~\bibnamefont
  {Aasi}} \emph {et~al.} (\bibinfo {collaboration} {LIGO Scientific}),\ }\href
  {\doibase 10.1088/0004-637X/785/2/119} {\bibfield  {journal} {\bibinfo
  {journal} {Astrophys. J.}\ }\textbf {\bibinfo {volume} {785}},\ \bibinfo
  {pages} {119} (\bibinfo {year} {2014})},\ \Eprint
  {http://arxiv.org/abs/1309.4027} {arXiv:1309.4027 [astro-ph.HE]} \BibitemShut
  {NoStop}%
\bibitem [{\citenamefont {Abbott}\ \emph
  {et~al.}(2008{\natexlab{a}})\citenamefont {Abbott} \emph
  {et~al.}}]{Abbott:2008fx}%
  \BibitemOpen
  \bibfield  {author} {\bibinfo {author} {\bibfnamefont {B.}~\bibnamefont
  {Abbott}} \emph {et~al.} (\bibinfo {collaboration} {LIGO Scientific
  Collaboration}),\ }\href {\doibase 10.1088/0004-637X/706/1/L203,
  10.1086/591526} {\bibfield  {journal} {\bibinfo  {journal} {Astrophys.J.}\
  }\textbf {\bibinfo {volume} {683}},\ \bibinfo {pages} {L45} (\bibinfo {year}
  {2008}{\natexlab{a}})},\ \Eprint {http://arxiv.org/abs/0805.4758}
  {arXiv:0805.4758 [astro-ph]} \BibitemShut {NoStop}%
\bibitem [{\citenamefont {Abadie}\ \emph {et~al.}(2011)\citenamefont {Abadie}
  \emph {et~al.}}]{Abadie:2011md}%
  \BibitemOpen
  \bibfield  {author} {\bibinfo {author} {\bibfnamefont {J.}~\bibnamefont
  {Abadie}} \emph {et~al.} (\bibinfo {collaboration} {LIGO Scientific
  Collaboration, Virgo Collaboration}),\ }\href {\doibase
  10.1088/0004-637X/737/2/93} {\bibfield  {journal} {\bibinfo  {journal}
  {Astrophys.J.}\ }\textbf {\bibinfo {volume} {737}},\ \bibinfo {pages} {93}
  (\bibinfo {year} {2011})},\ \Eprint {http://arxiv.org/abs/1104.2712}
  {arXiv:1104.2712 [astro-ph.HE]} \BibitemShut {NoStop}%
\bibitem [{\citenamefont {Aasi}\ \emph {et~al.}(2015)\citenamefont {Aasi} \emph
  {et~al.}}]{Aasi:2014jln}%
  \BibitemOpen
  \bibfield  {author} {\bibinfo {author} {\bibfnamefont {J.}~\bibnamefont
  {Aasi}} \emph {et~al.} (\bibinfo {collaboration} {VIRGO, LIGO Scientific}),\
  }\href {\doibase 10.1103/PhysRevD.91.022004} {\bibfield  {journal} {\bibinfo
  {journal} {Phys. Rev.}\ }\textbf {\bibinfo {volume} {D91}},\ \bibinfo {pages}
  {022004} (\bibinfo {year} {2015})},\ \Eprint {http://arxiv.org/abs/1410.8310}
  {arXiv:1410.8310 [astro-ph.IM]} \BibitemShut {NoStop}%
\bibitem [{\citenamefont {Abbott}\ \emph {et~al.}(2007)\citenamefont {Abbott}
  \emph {et~al.}}]{S2ScoX1}%
  \BibitemOpen
  \bibfield  {author} {\bibinfo {author} {\bibfnamefont {B.}~\bibnamefont
  {Abbott}} \emph {et~al.} (\bibinfo {collaboration} {LIGO Scientific
  Collaboration}),\ }\href {\doibase 10.1103/PhysRevD.76.082001} {\bibfield
  {journal} {\bibinfo  {journal} {Phys. Rev. D}\ }\textbf {\bibinfo {volume}
  {76}},\ \bibinfo {pages} {082001} (\bibinfo {year} {2007})},\ \Eprint
  {http://arxiv.org/abs/gr-qc/0605028} {arXiv:gr-qc/0605028} \BibitemShut
  {NoStop}%
\bibitem [{\citenamefont {Abbott}\ \emph {et~al.}(2005)\citenamefont {Abbott}
  \emph {et~al.}}]{S2Hough}%
  \BibitemOpen
  \bibfield  {author} {\bibinfo {author} {\bibfnamefont {B.}~\bibnamefont
  {Abbott}} \emph {et~al.} (\bibinfo {collaboration} {LIGO Scientific
  Collaboration}),\ }\href {\doibase 10.1103/PhysRevD.72.102004} {\bibfield
  {journal} {\bibinfo  {journal} {Phys. Rev. D}\ }\textbf {\bibinfo {volume}
  {72}},\ \bibinfo {pages} {102004} (\bibinfo {year} {2005})},\ \Eprint
  {http://arxiv.org/abs/gr-qc/0508065} {arXiv:gr-qc/0508065} \BibitemShut
  {NoStop}%
\bibitem [{\citenamefont {Abbott}\ \emph
  {et~al.}(2008{\natexlab{b}})\citenamefont {Abbott} \emph {et~al.}}]{S4PSH}%
  \BibitemOpen
  \bibfield  {author} {\bibinfo {author} {\bibfnamefont {B.}~\bibnamefont
  {Abbott}} \emph {et~al.} (\bibinfo {collaboration} {LIGO Scientific
  Collaboration}),\ }\href {\doibase 10.1103/PhysRevD.77.022001,
  10.1103/PhysRevD.80.129904, 10.1103/PhysRevD.77.022001,
  10.1103/PhysRevD.80.129904} {\bibfield  {journal} {\bibinfo  {journal} {Phys.
  Rev. D}\ }\textbf {\bibinfo {volume} {77}},\ \bibinfo {pages} {022001}
  (\bibinfo {year} {2008}{\natexlab{b}})},\ \Eprint
  {http://arxiv.org/abs/0708.3818} {arXiv:0708.3818 [gr-qc]} \BibitemShut
  {NoStop}%
\bibitem [{\citenamefont {Abbott}\ \emph
  {et~al.}(2009{\natexlab{a}})\citenamefont {Abbott} \emph
  {et~al.}}]{S5Powerflux2009}%
  \BibitemOpen
  \bibfield  {author} {\bibinfo {author} {\bibfnamefont {B.}~\bibnamefont
  {Abbott}} \emph {et~al.} (\bibinfo {collaboration} {LIGO Scientific
  Collaboration}),\ }\href {\doibase 10.1103/PhysRevLett.102.111102} {\bibfield
   {journal} {\bibinfo  {journal} {Phys. Rev. Lett.}\ }\textbf {\bibinfo
  {volume} {102}},\ \bibinfo {pages} {111102} (\bibinfo {year}
  {2009}{\natexlab{a}})},\ \Eprint {http://arxiv.org/abs/0810.0283}
  {arXiv:0810.0283 [gr-qc]} \BibitemShut {NoStop}%
\bibitem [{\citenamefont {Abadie}\ \emph {et~al.}(2012)\citenamefont {Abadie}
  \emph {et~al.}}]{S5Powerflux2011}%
  \BibitemOpen
  \bibfield  {author} {\bibinfo {author} {\bibfnamefont {J.}~\bibnamefont
  {Abadie}} \emph {et~al.} (\bibinfo {collaboration} {LIGO Scientific
  Collaboration and Virgo Collaboration}),\ }\href@noop {} {\bibfield
  {journal} {\bibinfo  {journal} {Phys. Rev. D}\ }\textbf {\bibinfo {volume}
  {85}},\ \bibinfo {pages} {022001} (\bibinfo {year} {2012})},\ \Eprint
  {http://arxiv.org/abs/1110.0208} {arXiv:1110.0208 [gr-qc]} \BibitemShut
  {NoStop}%
\bibitem [{\citenamefont {Abbott}\ \emph
  {et~al.}(2009{\natexlab{b}})\citenamefont {Abbott} \emph
  {et~al.}}]{EatHS4R2}%
  \BibitemOpen
  \bibfield  {author} {\bibinfo {author} {\bibfnamefont {B.}~\bibnamefont
  {Abbott}} \emph {et~al.} (\bibinfo {collaboration} {LIGO Scientific
  Collaboration}),\ }\href {\doibase 10.1103/PhysRevD.79.022001} {\bibfield
  {journal} {\bibinfo  {journal} {Phys. Rev. D}\ }\textbf {\bibinfo {volume}
  {79}},\ \bibinfo {pages} {022001} (\bibinfo {year} {2009}{\natexlab{b}})},\
  \Eprint {http://arxiv.org/abs/0804.1747} {arXiv:0804.1747 [gr-qc]}
  \BibitemShut {NoStop}%
\bibitem [{\citenamefont {Abbott}\ \emph
  {et~al.}(2009{\natexlab{c}})\citenamefont {Abbott} \emph
  {et~al.}}]{EatHS5R1}%
  \BibitemOpen
  \bibfield  {author} {\bibinfo {author} {\bibfnamefont {B.}~\bibnamefont
  {Abbott}} \emph {et~al.} (\bibinfo {collaboration} {LIGO Scientific
  Collaboration}),\ }\href {\doibase 10.1103/PhysRevD.80.042003} {\bibfield
  {journal} {\bibinfo  {journal} {Phys. Rev. D}\ }\textbf {\bibinfo {volume}
  {80}},\ \bibinfo {pages} {042003} (\bibinfo {year} {2009}{\natexlab{c}})},\
  \Eprint {http://arxiv.org/abs/0905.1705} {arXiv:0905.1705 [gr-qc]}
  \BibitemShut {NoStop}%
\bibitem [{\citenamefont {Abbott}\ \emph
  {et~al.}(2009{\natexlab{d}})\citenamefont {Abbott} \emph
  {et~al.}}]{Abbott:2008uq}%
  \BibitemOpen
  \bibfield  {author} {\bibinfo {author} {\bibfnamefont {B.}~\bibnamefont
  {Abbott}} \emph {et~al.} (\bibinfo {collaboration} {LIGO Scientific
  Collaboration}),\ }\href {\doibase 10.1103/PhysRevD.79.022001} {\bibfield
  {journal} {\bibinfo  {journal} {Phys.Rev.}\ }\textbf {\bibinfo {volume}
  {D79}},\ \bibinfo {pages} {022001} (\bibinfo {year} {2009}{\natexlab{d}})},\
  \Eprint {http://arxiv.org/abs/0804.1747} {arXiv:0804.1747 [gr-qc]}
  \BibitemShut {NoStop}%
\bibitem [{Ein()}]{Einstweb}%
  \BibitemOpen
  \href@noop {} {}\bibinfo {note} {Details of the Einstein@Home project can be
  found at \url{http://einstein.phys.uwm.edu}.}\BibitemShut {Stop}%
\bibitem [{\citenamefont {Abadie}\ \emph {et~al.}(2010)\citenamefont {Abadie}
  \emph {et~al.}}]{S5CasA}%
  \BibitemOpen
  \bibfield  {author} {\bibinfo {author} {\bibfnamefont {J.}~\bibnamefont
  {Abadie}} \emph {et~al.} (\bibinfo {collaboration} {LIGO Scientific
  Collaboration}),\ }\href {\doibase 10.1088/0004-637X/722/2/1504} {\bibfield
  {journal} {\bibinfo  {journal} {Astrophys. J.}\ }\textbf {\bibinfo {volume}
  {722}},\ \bibinfo {pages} {1504} (\bibinfo {year} {2010})},\ \Eprint
  {http://arxiv.org/abs/1006.2535} {arXiv:1006.2535 [gr-qc]} \BibitemShut
  {NoStop}%
\bibitem [{\citenamefont {Aasi}\ \emph {et~al.}(2013)\citenamefont {Aasi} \emph
  {et~al.}}]{Aasi:2013jya}%
  \BibitemOpen
  \bibfield  {author} {\bibinfo {author} {\bibfnamefont {J.}~\bibnamefont
  {Aasi}} \emph {et~al.} (\bibinfo {collaboration} {LIGO Scientific
  Collaboration, Virgo Collaboration}),\ }\href {\doibase
  10.1103/PhysRevD.88.102002} {\bibfield  {journal} {\bibinfo  {journal}
  {Phys.Rev.}\ }\textbf {\bibinfo {volume} {D88}},\ \bibinfo {pages} {102002}
  (\bibinfo {year} {2013})},\ \Eprint {http://arxiv.org/abs/1309.6221}
  {arXiv:1309.6221 [gr-qc]} \BibitemShut {NoStop}%
\bibitem [{\citenamefont {{Aasi}}\ \emph {et~al.}(2014)\citenamefont {{Aasi}},
  \citenamefont {{Abbott}}, \citenamefont {{Abbott}}, \citenamefont {{Abbott}},
  \citenamefont {{Abernathy}}, \citenamefont {{Acernese}}, \citenamefont
  {{Ackley}}, \citenamefont {{Adams}}, \citenamefont {{Adams}}, \citenamefont
  {{Adams}},\ and\ \citenamefont {et~al.}}]{owen2014}%
  \BibitemOpen
  \bibfield  {author} {\bibinfo {author} {\bibfnamefont {J.}~\bibnamefont
  {{Aasi}}}, \bibinfo {author} {\bibfnamefont {B.~P.}\ \bibnamefont
  {{Abbott}}}, \bibinfo {author} {\bibfnamefont {R.}~\bibnamefont {{Abbott}}},
  \bibinfo {author} {\bibfnamefont {T.}~\bibnamefont {{Abbott}}}, \bibinfo
  {author} {\bibfnamefont {M.~R.}\ \bibnamefont {{Abernathy}}}, \bibinfo
  {author} {\bibfnamefont {F.}~\bibnamefont {{Acernese}}}, \bibinfo {author}
  {\bibfnamefont {K.}~\bibnamefont {{Ackley}}}, \bibinfo {author}
  {\bibfnamefont {C.}~\bibnamefont {{Adams}}}, \bibinfo {author} {\bibfnamefont
  {T.}~\bibnamefont {{Adams}}}, \bibinfo {author} {\bibfnamefont
  {T.}~\bibnamefont {{Adams}}}, \ and\ \bibinfo {author} {\bibnamefont
  {et~al.}},\ }\href@noop {} {\bibfield  {journal} {\bibinfo  {journal} {ArXiv
  e-prints}\ } (\bibinfo {year} {2014})},\ \Eprint
  {http://arxiv.org/abs/1412.5942} {arXiv:1412.5942 [astro-ph.HE]} \BibitemShut
  {NoStop}%
\bibitem [{\citenamefont {Brady}\ \emph {et~al.}(1998)\citenamefont {Brady},
  \citenamefont {Creighton}, \citenamefont {Cutler},\ and\ \citenamefont
  {Schutz}}]{ForMetricExprRes}%
  \BibitemOpen
  \bibfield  {author} {\bibinfo {author} {\bibfnamefont {P.~R.}\ \bibnamefont
  {Brady}}, \bibinfo {author} {\bibfnamefont {T.}~\bibnamefont {Creighton}},
  \bibinfo {author} {\bibfnamefont {C.}~\bibnamefont {Cutler}}, \ and\ \bibinfo
  {author} {\bibfnamefont {B.~F.}\ \bibnamefont {Schutz}},\ }\href {\doibase
  10.1103/PhysRevD.57.2101} {\bibfield  {journal} {\bibinfo  {journal} {Phys.
  Rev. D}\ }\textbf {\bibinfo {volume} {57}},\ \bibinfo {pages} {2101}
  (\bibinfo {year} {1998})},\ \Eprint {http://arxiv.org/abs/gr-qc/9702050}
  {arXiv:gr-qc/9702050} \BibitemShut {NoStop}%
\bibitem [{\citenamefont {Brady}\ and\ \citenamefont
  {Creighton}(2000)}]{HierarchP1}%
  \BibitemOpen
  \bibfield  {author} {\bibinfo {author} {\bibfnamefont {P.~R.}\ \bibnamefont
  {Brady}}\ and\ \bibinfo {author} {\bibfnamefont {T.}~\bibnamefont
  {Creighton}},\ }\href {\doibase 10.1103/PhysRevD.61.082001} {\bibfield
  {journal} {\bibinfo  {journal} {Phys. Rev. D}\ }\textbf {\bibinfo {volume}
  {61}},\ \bibinfo {pages} {082001} (\bibinfo {year} {2000})},\ \Eprint
  {http://arxiv.org/abs/gr-qc/9812014} {arXiv:gr-qc/9812014} \BibitemShut
  {NoStop}%
\bibitem [{\citenamefont {Krishnan}\ \emph {et~al.}(2004)\citenamefont
  {Krishnan}, \citenamefont {Sintes}, \citenamefont {Papa}, \citenamefont
  {Schutz}, \citenamefont {Frasca} \emph {et~al.}}]{HierarchP2}%
  \BibitemOpen
  \bibfield  {author} {\bibinfo {author} {\bibfnamefont {B.}~\bibnamefont
  {Krishnan}}, \bibinfo {author} {\bibfnamefont {A.~M.}\ \bibnamefont
  {Sintes}}, \bibinfo {author} {\bibfnamefont {M.~A.}\ \bibnamefont {Papa}},
  \bibinfo {author} {\bibfnamefont {B.~F.}\ \bibnamefont {Schutz}}, \bibinfo
  {author} {\bibfnamefont {S.}~\bibnamefont {Frasca}},  \emph {et~al.},\ }\href
  {\doibase 10.1103/PhysRevD.70.082001} {\bibfield  {journal} {\bibinfo
  {journal} {Phys. Rev. D}\ }\textbf {\bibinfo {volume} {70}},\ \bibinfo
  {pages} {082001} (\bibinfo {year} {2004})},\ \Eprint
  {http://arxiv.org/abs/gr-qc/0407001} {arXiv:gr-qc/0407001} \BibitemShut
  {NoStop}%
\bibitem [{\citenamefont {Papa}\ \emph {et~al.}(2000)\citenamefont {Papa},
  \citenamefont {Schutz},\ and\ \citenamefont {Sintes}}]{Papa:2000wg}%
  \BibitemOpen
  \bibfield  {author} {\bibinfo {author} {\bibfnamefont {M.~A.}\ \bibnamefont
  {Papa}}, \bibinfo {author} {\bibfnamefont {B.~F.}\ \bibnamefont {Schutz}}, \
  and\ \bibinfo {author} {\bibfnamefont {A.~M.}\ \bibnamefont {Sintes}},\ }in\
  \href {http://www.ictp.trieste.it/~pub_off/lectures/lns003/Papa/Papa.pdf}
  {\emph {\bibinfo {booktitle} {{Gravitational waves: A challenge to
  theoretical astrophysics. Proceedings, Trieste, Italy, June 6-9, 2000}}}}\
  (\bibinfo {year} {2000})\ pp.\ \bibinfo {pages} {431--442},\ \Eprint
  {http://arxiv.org/abs/gr-qc/0011034} {arXiv:gr-qc/0011034 [gr-qc]}
  \BibitemShut {NoStop}%
\bibitem [{\citenamefont {Krishnan}(2005)}]{HoughP2}%
  \BibitemOpen
  \bibfield  {author} {\bibinfo {author} {\bibfnamefont {B.}~\bibnamefont
  {Krishnan}} (\bibinfo {collaboration} {LIGO Scientific Collaboration}),\
  }\href {\doibase 10.1088/0264-9381/22/18/S40} {\bibfield  {journal} {\bibinfo
   {journal} {Class. Quant. Grav.}\ }\textbf {\bibinfo {volume} {22}},\
  \bibinfo {pages} {S1265} (\bibinfo {year} {2005})},\ \Eprint
  {http://arxiv.org/abs/gr-qc/0506109} {arXiv:gr-qc/0506109} \BibitemShut
  {NoStop}%
\bibitem [{\citenamefont {Cutler}\ \emph {et~al.}(2005)\citenamefont {Cutler},
  \citenamefont {Gholami},\ and\ \citenamefont {Krishnan}}]{HierarchP3}%
  \BibitemOpen
  \bibfield  {author} {\bibinfo {author} {\bibfnamefont {C.}~\bibnamefont
  {Cutler}}, \bibinfo {author} {\bibfnamefont {I.}~\bibnamefont {Gholami}}, \
  and\ \bibinfo {author} {\bibfnamefont {B.}~\bibnamefont {Krishnan}},\ }\href
  {\doibase 10.1103/PhysRevD.72.042004} {\bibfield  {journal} {\bibinfo
  {journal} {Phys. Rev. D}\ }\textbf {\bibinfo {volume} {72}},\ \bibinfo
  {pages} {042004} (\bibinfo {year} {2005})},\ \Eprint
  {http://arxiv.org/abs/gr-qc/0505082} {arXiv:gr-qc/0505082} \BibitemShut
  {NoStop}%
\bibitem [{\citenamefont {Pletsch}\ and\ \citenamefont
  {Allen}(2009)}]{Pletsch:2009uu}%
  \BibitemOpen
  \bibfield  {author} {\bibinfo {author} {\bibfnamefont {H.~J.}\ \bibnamefont
  {Pletsch}}\ and\ \bibinfo {author} {\bibfnamefont {B.}~\bibnamefont
  {Allen}},\ }\href {\doibase 10.1103/PhysRevLett.103.181102} {\bibfield
  {journal} {\bibinfo  {journal} {Phys. Rev. Lett.}\ }\textbf {\bibinfo
  {volume} {103}},\ \bibinfo {pages} {181102} (\bibinfo {year} {2009})},\
  \Eprint {http://arxiv.org/abs/0906.0023} {arXiv:0906.0023 [gr-qc]}
  \BibitemShut {NoStop}%
\bibitem [{\citenamefont {Pletsch}(2008)}]{GlobCorr}%
  \BibitemOpen
  \bibfield  {author} {\bibinfo {author} {\bibfnamefont {H.~J.}\ \bibnamefont
  {Pletsch}},\ }\href {\doibase 10.1103/PhysRevD.78.102005} {\bibfield
  {journal} {\bibinfo  {journal} {Phys. Rev. D}\ }\textbf {\bibinfo {volume}
  {78}},\ \bibinfo {pages} {102005} (\bibinfo {year} {2008})},\ \Eprint
  {http://arxiv.org/abs/0807.1324} {arXiv:0807.1324 [gr-qc]} \BibitemShut
  {NoStop}%
\bibitem [{\citenamefont {Shaltev}\ \emph {et~al.}(2014)\citenamefont
  {Shaltev}, \citenamefont {Leaci}, \citenamefont {Papa},\ and\ \citenamefont
  {Prix}}]{Shaltev:2014toa}%
  \BibitemOpen
  \bibfield  {author} {\bibinfo {author} {\bibfnamefont {M.}~\bibnamefont
  {Shaltev}}, \bibinfo {author} {\bibfnamefont {P.}~\bibnamefont {Leaci}},
  \bibinfo {author} {\bibfnamefont {M.~A.}\ \bibnamefont {Papa}}, \ and\
  \bibinfo {author} {\bibfnamefont {R.}~\bibnamefont {Prix}},\ }\href {\doibase
  10.1103/PhysRevD.89.124030} {\bibfield  {journal} {\bibinfo  {journal} {Phys.
  Rev.}\ }\textbf {\bibinfo {volume} {D89}},\ \bibinfo {pages} {124030}
  (\bibinfo {year} {2014})},\ \Eprint {http://arxiv.org/abs/1405.1922}
  {arXiv:1405.1922 [gr-qc]} \BibitemShut {NoStop}%
\bibitem [{\citenamefont {{P. Brady \textit{et al.} }}(2000)}]{BC00}%
  \BibitemOpen
  \bibfield  {author} {\bibinfo {author} {\bibnamefont {{P. Brady \textit{et
  al.} }}},\ }\href@noop {} {\bibfield  {journal} {\bibinfo  {journal} {Phys.\
  Rev.\ D}\ }\textbf {\bibinfo {volume} {61}},\ \bibinfo {pages} {082001}
  (\bibinfo {year} {2000})}\BibitemShut {NoStop}%
\bibitem [{\citenamefont {Prix}\ and\ \citenamefont
  {Shaltev}(2012)}]{Prix:2012yu}%
  \BibitemOpen
  \bibfield  {author} {\bibinfo {author} {\bibfnamefont {R.}~\bibnamefont
  {Prix}}\ and\ \bibinfo {author} {\bibfnamefont {M.}~\bibnamefont {Shaltev}},\
  }\href {\doibase 10.1103/PhysRevD.85.084010} {\bibfield  {journal} {\bibinfo
  {journal} {Phys.Rev.}\ }\textbf {\bibinfo {volume} {D85}},\ \bibinfo {pages}
  {084010} (\bibinfo {year} {2012})},\ \Eprint {http://arxiv.org/abs/1201.4321}
  {arXiv:1201.4321 [gr-qc]} \BibitemShut {NoStop}%
\bibitem [{\citenamefont {Palomba}(2005)}]{Palomba:2005fa}%
  \BibitemOpen
  \bibfield  {author} {\bibinfo {author} {\bibfnamefont {C.}~\bibnamefont
  {Palomba}},\ }\href {\doibase 10.1088/0264-9381/22/18/S17} {\bibfield
  {journal} {\bibinfo  {journal} {Class.Quant.Grav.}\ }\textbf {\bibinfo
  {volume} {22}},\ \bibinfo {pages} {S1027} (\bibinfo {year}
  {2005})}\BibitemShut {NoStop}%
\bibitem [{\citenamefont {Knispel}\ and\ \citenamefont
  {Allen}(2008)}]{Knispel:2008ue}%
  \BibitemOpen
  \bibfield  {author} {\bibinfo {author} {\bibfnamefont {B.}~\bibnamefont
  {Knispel}}\ and\ \bibinfo {author} {\bibfnamefont {B.}~\bibnamefont
  {Allen}},\ }\href {\doibase 10.1103/PhysRevD.78.044031} {\bibfield  {journal}
  {\bibinfo  {journal} {Phys.Rev.}\ }\textbf {\bibinfo {volume} {D78}},\
  \bibinfo {pages} {044031} (\bibinfo {year} {2008})},\ \Eprint
  {http://arxiv.org/abs/0804.3075} {arXiv:0804.3075 [gr-qc]} \BibitemShut
  {NoStop}%
\bibitem [{\citenamefont {Wade}\ \emph {et~al.}(2012)\citenamefont {Wade},
  \citenamefont {Siemens}, \citenamefont {Kaplan}, \citenamefont {Knispel},\
  and\ \citenamefont {Allen}}]{Wade:2012qc}%
  \BibitemOpen
  \bibfield  {author} {\bibinfo {author} {\bibfnamefont {L.}~\bibnamefont
  {Wade}}, \bibinfo {author} {\bibfnamefont {X.}~\bibnamefont {Siemens}},
  \bibinfo {author} {\bibfnamefont {D.~L.}\ \bibnamefont {Kaplan}}, \bibinfo
  {author} {\bibfnamefont {B.}~\bibnamefont {Knispel}}, \ and\ \bibinfo
  {author} {\bibfnamefont {B.}~\bibnamefont {Allen}},\ }\href {\doibase
  10.1103/PhysRevD.86.124011} {\bibfield  {journal} {\bibinfo  {journal}
  {Phys.Rev.}\ }\textbf {\bibinfo {volume} {D86}},\ \bibinfo {pages} {124011}
  (\bibinfo {year} {2012})},\ \Eprint {http://arxiv.org/abs/1209.2971}
  {arXiv:1209.2971 [gr-qc]} \BibitemShut {NoStop}%
\bibitem [{\citenamefont {Owen}(2009)}]{Owen:2009tj}%
  \BibitemOpen
  \bibfield  {author} {\bibinfo {author} {\bibfnamefont {B.~J.}\ \bibnamefont
  {Owen}},\ }\href {\doibase 10.1088/0264-9381/26/20/204014} {\bibfield
  {journal} {\bibinfo  {journal} {Class.Quant.Grav.}\ }\textbf {\bibinfo
  {volume} {26}},\ \bibinfo {pages} {204014} (\bibinfo {year} {2009})},\
  \Eprint {http://arxiv.org/abs/0904.4848} {arXiv:0904.4848 [gr-qc]}
  \BibitemShut {NoStop}%
\bibitem [{\citenamefont {Jaranowski}\ \emph {et~al.}(1998)\citenamefont
  {Jaranowski}, \citenamefont {Kr{\'o}lak},\ and\ \citenamefont
  {Schutz}}]{JKSPaper}%
  \BibitemOpen
  \bibfield  {author} {\bibinfo {author} {\bibfnamefont {P.}~\bibnamefont
  {Jaranowski}}, \bibinfo {author} {\bibfnamefont {A.}~\bibnamefont
  {Kr{\'o}lak}}, \ and\ \bibinfo {author} {\bibfnamefont {B.~F.}\ \bibnamefont
  {Schutz}},\ }\href {\doibase 10.1103/PhysRevD.58.063001} {\bibfield
  {journal} {\bibinfo  {journal} {Phys. Rev. D}\ }\textbf {\bibinfo {volume}
  {58}},\ \bibinfo {pages} {063001} (\bibinfo {year} {1998})},\ \Eprint
  {http://arxiv.org/abs/gr-qc/9804014} {arXiv:gr-qc/9804014} \BibitemShut
  {NoStop}%
\bibitem [{\citenamefont {Shapiro}\ and\ \citenamefont
  {Teukolsky}(2004)}]{ShapiroTeukolsky}%
  \BibitemOpen
  \bibfield  {author} {\bibinfo {author} {\bibfnamefont {S.}~\bibnamefont
  {Shapiro}}\ and\ \bibinfo {author} {\bibfnamefont {S.}~\bibnamefont
  {Teukolsky}},\ }\href@noop {} {\emph {\bibinfo {title} {Black holes, white
  dwarfs, and neutron stars}}}\ (\bibinfo  {publisher} {Wiley-VCH Verlag GmbH
  \& Co. KgaA, Weinheim},\ \bibinfo {year} {2004})\BibitemShut {NoStop}%
\bibitem [{\citenamefont {Ushomirsky}\ \emph {et~al.}(2000)\citenamefont
  {Ushomirsky}, \citenamefont {Cutler},\ and\ \citenamefont
  {Bildsten}}]{NonAxNS2}%
  \BibitemOpen
  \bibfield  {author} {\bibinfo {author} {\bibfnamefont {G.}~\bibnamefont
  {Ushomirsky}}, \bibinfo {author} {\bibfnamefont {C.}~\bibnamefont {Cutler}},
  \ and\ \bibinfo {author} {\bibfnamefont {L.}~\bibnamefont {Bildsten}},\
  }\href {\doibase 10.1046/j.1365-8711.2000.03938.x} {\bibfield  {journal}
  {\bibinfo  {journal} {Mon. Not. Roy. Astron. Soc.}\ }\textbf {\bibinfo
  {volume} {319}},\ \bibinfo {pages} {902} (\bibinfo {year} {2000})},\ \Eprint
  {http://arxiv.org/abs/astro-ph/0001136} {arXiv:astro-ph/0001136} \BibitemShut
  {NoStop}%
\bibitem [{\citenamefont {Horowitz}\ and\ \citenamefont
  {Kadau}(2009)}]{Horowitz}%
  \BibitemOpen
  \bibfield  {author} {\bibinfo {author} {\bibfnamefont {C.}~\bibnamefont
  {Horowitz}}\ and\ \bibinfo {author} {\bibfnamefont {K.}~\bibnamefont
  {Kadau}},\ }\href {\doibase 10.1103/PhysRevLett.102.191102} {\bibfield
  {journal} {\bibinfo  {journal} {Phys. Rev. Lett.}\ }\textbf {\bibinfo
  {volume} {102}},\ \bibinfo {pages} {191102} (\bibinfo {year} {2009})},\
  \Eprint {http://arxiv.org/abs/0904.1986} {arXiv:0904.1986 [astro-ph.SR]}
  \BibitemShut {NoStop}%
\bibitem [{\citenamefont {Johnson-McDaniel}\ and\ \citenamefont
  {Owen}(2013)}]{JohnsonMcDaniel:2012wg}%
  \BibitemOpen
  \bibfield  {author} {\bibinfo {author} {\bibfnamefont {N.~K.}\ \bibnamefont
  {Johnson-McDaniel}}\ and\ \bibinfo {author} {\bibfnamefont {B.~J.}\
  \bibnamefont {Owen}},\ }\href {\doibase 10.1103/PhysRevD.88.044004}
  {\bibfield  {journal} {\bibinfo  {journal} {Phys. Rev.}\ }\textbf {\bibinfo
  {volume} {D88}},\ \bibinfo {pages} {044004} (\bibinfo {year} {2013})},\
  \Eprint {http://arxiv.org/abs/1208.5227} {arXiv:1208.5227 [astro-ph.SR]}
  \BibitemShut {NoStop}%
\bibitem [{\citenamefont {Cutler}\ and\ \citenamefont
  {Schutz}(2005)}]{MultiIfoFstat}%
  \BibitemOpen
  \bibfield  {author} {\bibinfo {author} {\bibfnamefont {C.}~\bibnamefont
  {Cutler}}\ and\ \bibinfo {author} {\bibfnamefont {B.~F.}\ \bibnamefont
  {Schutz}},\ }\href {\doibase 10.1103/PhysRevD.72.063006} {\bibfield
  {journal} {\bibinfo  {journal} {Phys. Rev. D}\ }\textbf {\bibinfo {volume}
  {72}},\ \bibinfo {pages} {063006} (\bibinfo {year} {2005})},\ \Eprint
  {http://arxiv.org/abs/gr-qc/0504011} {arXiv:gr-qc/0504011} \BibitemShut
  {NoStop}%
\bibitem [{\citenamefont {Prix}\ and\ \citenamefont {Krishnan}(2009)}]{Bstat}%
  \BibitemOpen
  \bibfield  {author} {\bibinfo {author} {\bibfnamefont {R.}~\bibnamefont
  {Prix}}\ and\ \bibinfo {author} {\bibfnamefont {B.}~\bibnamefont
  {Krishnan}},\ }\href {\doibase 10.1088/0264-9381/26/20/204013} {\bibfield
  {journal} {\bibinfo  {journal} {Class. Quant. Grav.}\ }\textbf {\bibinfo
  {volume} {26}},\ \bibinfo {pages} {204013} (\bibinfo {year} {2009})},\
  \Eprint {http://arxiv.org/abs/0907.2569} {arXiv:0907.2569 [gr-qc]}
  \BibitemShut {NoStop}%
\bibitem [{\citenamefont {Whelan}\ \emph {et~al.}(2014)\citenamefont {Whelan},
  \citenamefont {Prix}, \citenamefont {Cutler},\ and\ \citenamefont
  {Willis}}]{Whelan:2013xka}%
  \BibitemOpen
  \bibfield  {author} {\bibinfo {author} {\bibfnamefont {J.~T.}\ \bibnamefont
  {Whelan}}, \bibinfo {author} {\bibfnamefont {R.}~\bibnamefont {Prix}},
  \bibinfo {author} {\bibfnamefont {C.~J.}\ \bibnamefont {Cutler}}, \ and\
  \bibinfo {author} {\bibfnamefont {J.~L.}\ \bibnamefont {Willis}},\ }\href
  {\doibase 10.1088/0264-9381/31/6/065002} {\bibfield  {journal} {\bibinfo
  {journal} {Class.Quant.Grav.}\ }\textbf {\bibinfo {volume} {31}},\ \bibinfo
  {pages} {065002} (\bibinfo {year} {2014})},\ \Eprint
  {http://arxiv.org/abs/1311.0065} {arXiv:1311.0065 [gr-qc]} \BibitemShut
  {NoStop}%
\bibitem [{\citenamefont {Schutz}\ and\ \citenamefont
  {Papa}(1999)}]{Schutz:1999mb}%
  \BibitemOpen
  \bibfield  {author} {\bibinfo {author} {\bibfnamefont {B.~F.}\ \bibnamefont
  {Schutz}}\ and\ \bibinfo {author} {\bibfnamefont {M.~A.}\ \bibnamefont
  {Papa}},\ }in\ \href {http://alice.cern.ch/format/showfull?sysnb=0312851}
  {\emph {\bibinfo {booktitle} {{34th Rencontres de Moriond: Gravitational
  Waves and Experimental Gravity Les Arcs, France, January 23-30, 1999}}}}\
  (\bibinfo {year} {1999})\ \Eprint {http://arxiv.org/abs/gr-qc/9905018}
  {arXiv:gr-qc/9905018 [gr-qc]} \BibitemShut {NoStop}%
\bibitem [{\citenamefont {Dantzig}(1963)}]{GVK180926950}%
  \BibitemOpen
  \bibfield  {author} {\bibinfo {author} {\bibfnamefont {G.}~\bibnamefont
  {Dantzig}},\ }\href
  {http://gso.gbv.de/DB=2.1/CMD?ACT=SRCHA&SRT=YOP&IKT=1016&TRM=ppn+180926950&sourceid=fbw_bibsonomy}
  {\emph {\bibinfo {title} {Linear programming and extensions}}},\ Rand
  Corporation Research Study\ (\bibinfo  {publisher} {Princeton Univ. Press},\
  \bibinfo {address} {Princeton, NJ},\ \bibinfo {year} {1963})\BibitemShut
  {NoStop}%
\bibitem [{\citenamefont {Andersson}\ \emph {et~al.}(2011)\citenamefont
  {Andersson}, \citenamefont {Ferrari}, \citenamefont {Jones}, \citenamefont
  {Kokkotas}, \citenamefont {Krishnan} \emph {et~al.}}]{Andersson:2009yt}%
  \BibitemOpen
  \bibfield  {author} {\bibinfo {author} {\bibfnamefont {N.}~\bibnamefont
  {Andersson}}, \bibinfo {author} {\bibfnamefont {V.}~\bibnamefont {Ferrari}},
  \bibinfo {author} {\bibfnamefont {D.}~\bibnamefont {Jones}}, \bibinfo
  {author} {\bibfnamefont {K.}~\bibnamefont {Kokkotas}}, \bibinfo {author}
  {\bibfnamefont {B.}~\bibnamefont {Krishnan}},  \emph {et~al.},\ }\href
  {\doibase 10.1007/s10714-010-1059-4} {\bibfield  {journal} {\bibinfo
  {journal} {Gen.Rel.Grav.}\ }\textbf {\bibinfo {volume} {43}},\ \bibinfo
  {pages} {409} (\bibinfo {year} {2011})},\ \Eprint
  {http://arxiv.org/abs/0912.0384} {arXiv:0912.0384 [astro-ph.SR]} \BibitemShut
  {NoStop}%
\bibitem [{\citenamefont {Pletsch}(2010)}]{Pletsch:2010xb}%
  \BibitemOpen
  \bibfield  {author} {\bibinfo {author} {\bibfnamefont {H.~J.}\ \bibnamefont
  {Pletsch}},\ }\href {\doibase 10.1103/PhysRevD.82.042002} {\bibfield
  {journal} {\bibinfo  {journal} {Phys. Rev. D}\ }\textbf {\bibinfo {volume}
  {82}},\ \bibinfo {pages} {042002} (\bibinfo {year} {2010})},\ \Eprint
  {http://arxiv.org/abs/1005.0395} {arXiv:1005.0395 [gr-qc]} \BibitemShut
  {NoStop}%
\bibitem [{\citenamefont {Allen}\ \emph {et~al.}(2015)\citenamefont {Allen},
  \citenamefont {Chow}, \citenamefont {DeLaney}, \citenamefont {Filipovic},
  \citenamefont {Houck}, \citenamefont {Pannuti},\ and\ \citenamefont
  {Stage}}]{Allen:2014yra}%
  \BibitemOpen
  \bibfield  {author} {\bibinfo {author} {\bibfnamefont {G.~E.}\ \bibnamefont
  {Allen}}, \bibinfo {author} {\bibfnamefont {K.}~\bibnamefont {Chow}},
  \bibinfo {author} {\bibfnamefont {T.}~\bibnamefont {DeLaney}}, \bibinfo
  {author} {\bibfnamefont {M.~D.}\ \bibnamefont {Filipovic}}, \bibinfo {author}
  {\bibfnamefont {J.~C.}\ \bibnamefont {Houck}}, \bibinfo {author}
  {\bibfnamefont {T.~G.}\ \bibnamefont {Pannuti}}, \ and\ \bibinfo {author}
  {\bibfnamefont {M.~D.}\ \bibnamefont {Stage}},\ }\href {\doibase
  10.1088/0004-637X/798/2/82} {\bibfield  {journal} {\bibinfo  {journal}
  {Astrophys. J.}\ }\textbf {\bibinfo {volume} {798}},\ \bibinfo {pages} {82}
  (\bibinfo {year} {2015})},\ \Eprint {http://arxiv.org/abs/1410.7435}
  {arXiv:1410.7435 [astro-ph.HE]} \BibitemShut {NoStop}%
\end{thebibliography}%
\FloatBarrier

\section{Complete set of figures}
\label{section:AppendixFigures}

\begin{figure*}%
    \centering
    \subfloat[Efficiency]{{  \includegraphics[width=.4\linewidth]{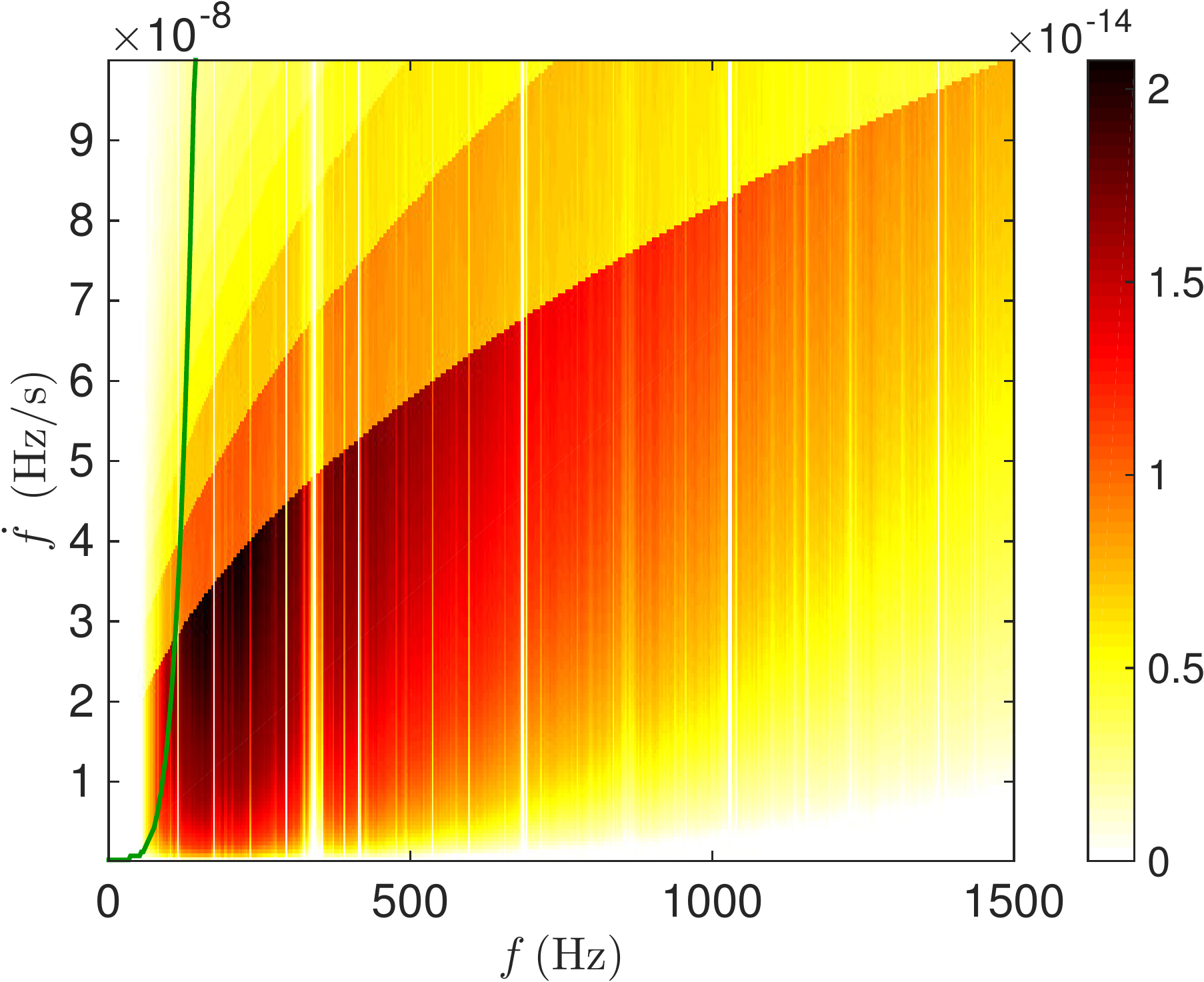}}}%
    \qquad
    \subfloat[Coverage]{{  \includegraphics[width=.4\linewidth]{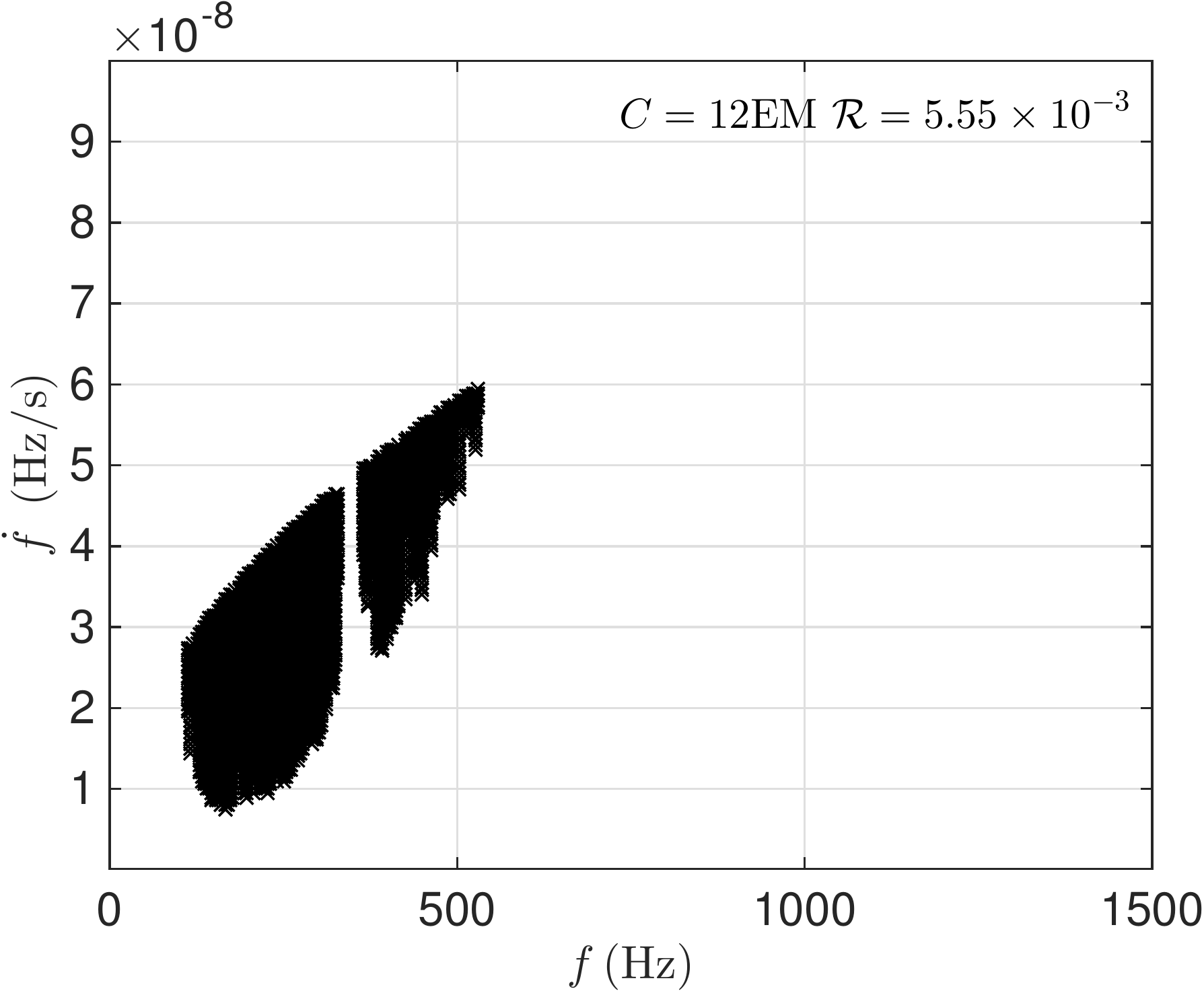}}}%
    \caption{Vela Jr at 750 pc, assuming uniform and distance-based priors, and a 10-day coherent segment duration. The left plot shows the efficiency, color-coded, for each cell : $e(f,\dot{f})$. The green curve shows $\dot{f}^\star$  as a function of $f$. The right plot displays the cells selected by the optimization procedure to be searched with a computational budget of 12 EM. The detection probability $\mathcal{R}$ is $5.55\times10^{-3}$.}%
    \label{G2662_10days_noage_longdis}%
\end{figure*}

\begin{figure*}%
    \centering
    \subfloat[Efficiency]{{  \includegraphics[width=.4\linewidth]{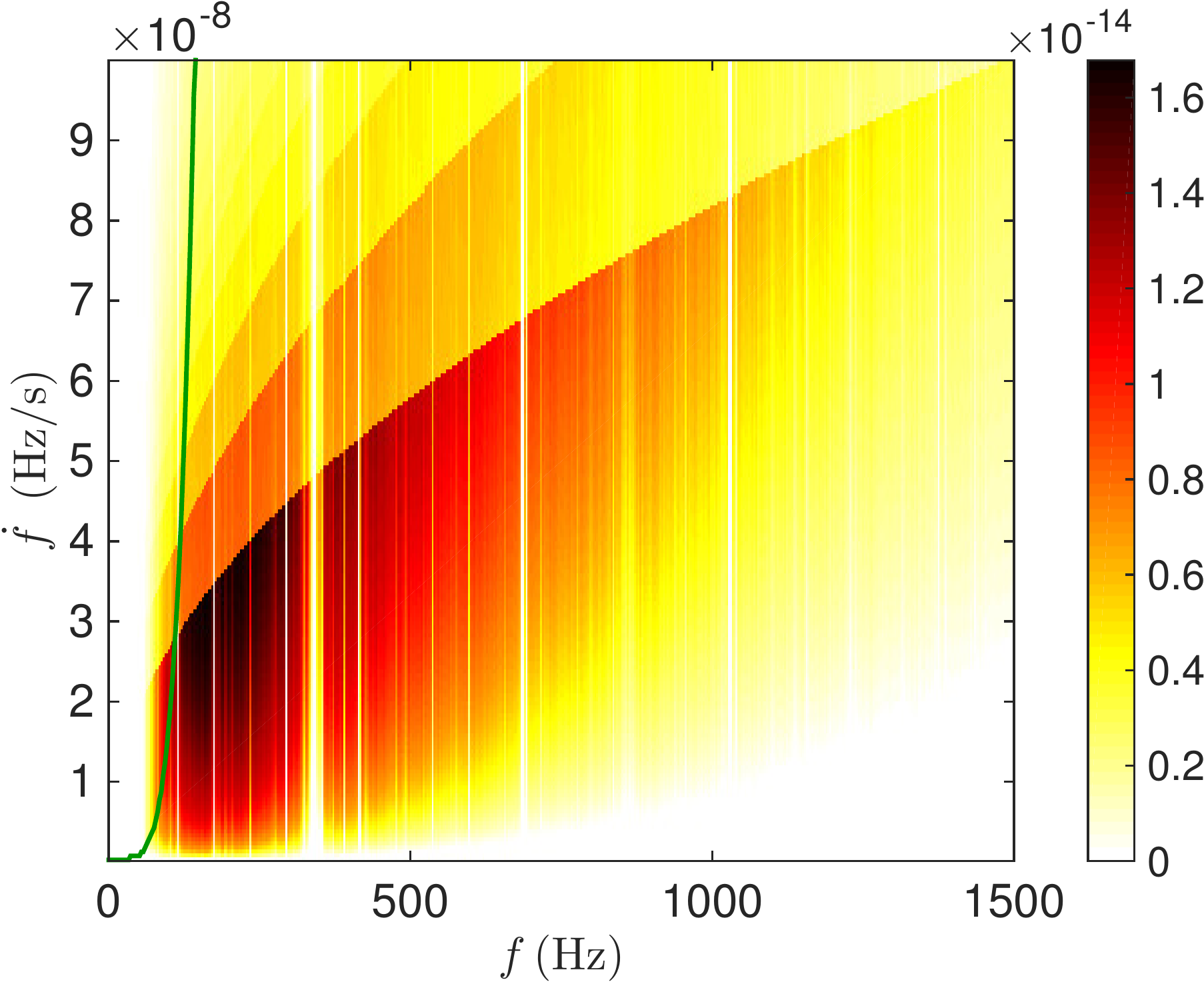}}}%
    \qquad
    \subfloat[Coverage]{{  \includegraphics[width=.4\linewidth]{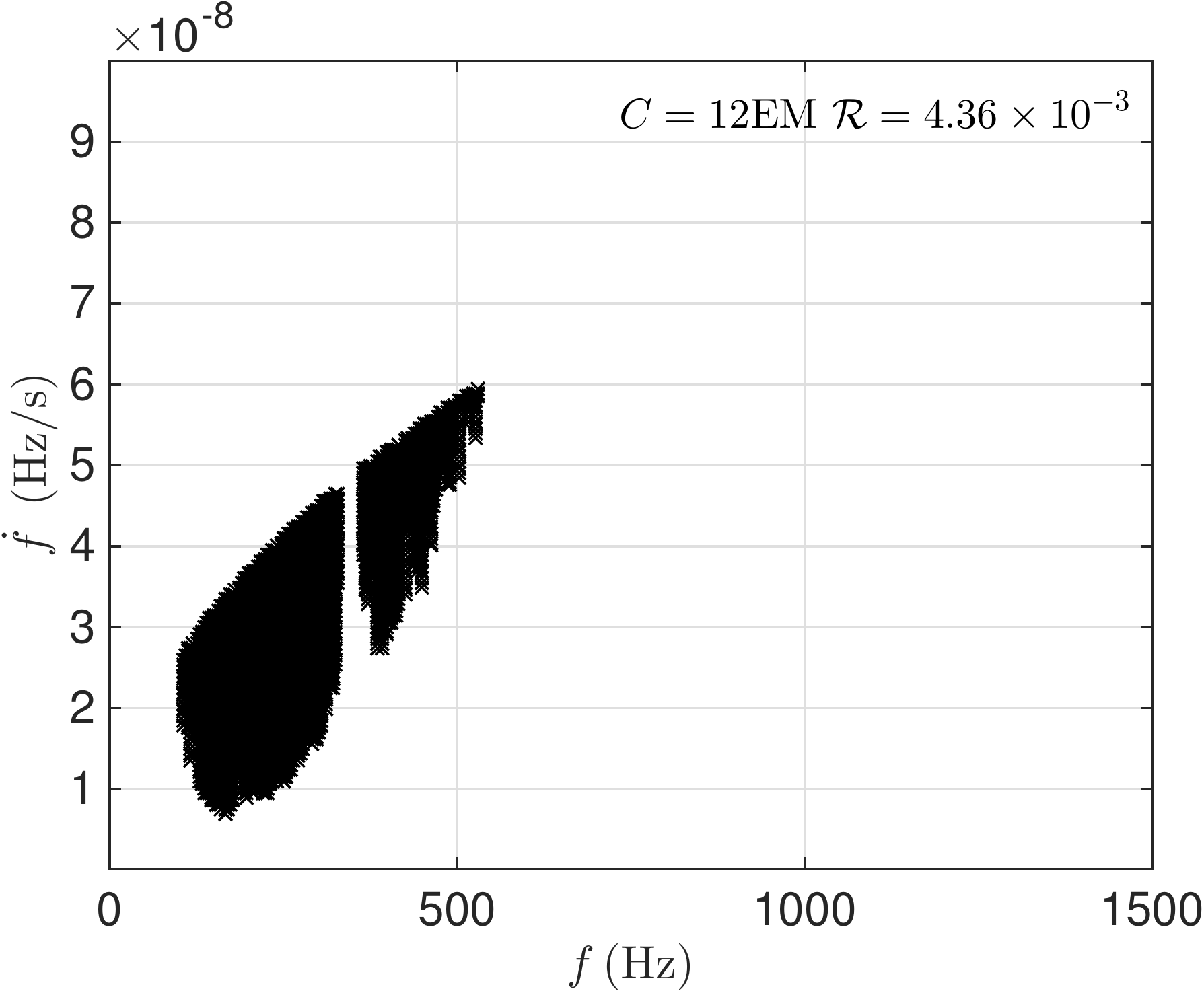}}}%
    \caption{G347.3 at 1300 pc, assuming uniform and distance-based priors, and a 10-day coherent segment duration. The left plot shows the efficiency, color-coded, for each cell : $e(f,\dot{f})$. The green curve shows $\dot{f}^\star$  as a function of $f$. The right plot displays the cells selected by the optimization procedure with a computational budget of 12 EM. The detection probability $\mathcal{R}$ is $4.36\times10^{-3}$.}%
    \label{G3473_10days_noage_longdis}%
\end{figure*}

\begin{figure*}%
    \centering
    \subfloat[Efficiency]{{  \includegraphics[width=.4\linewidth]{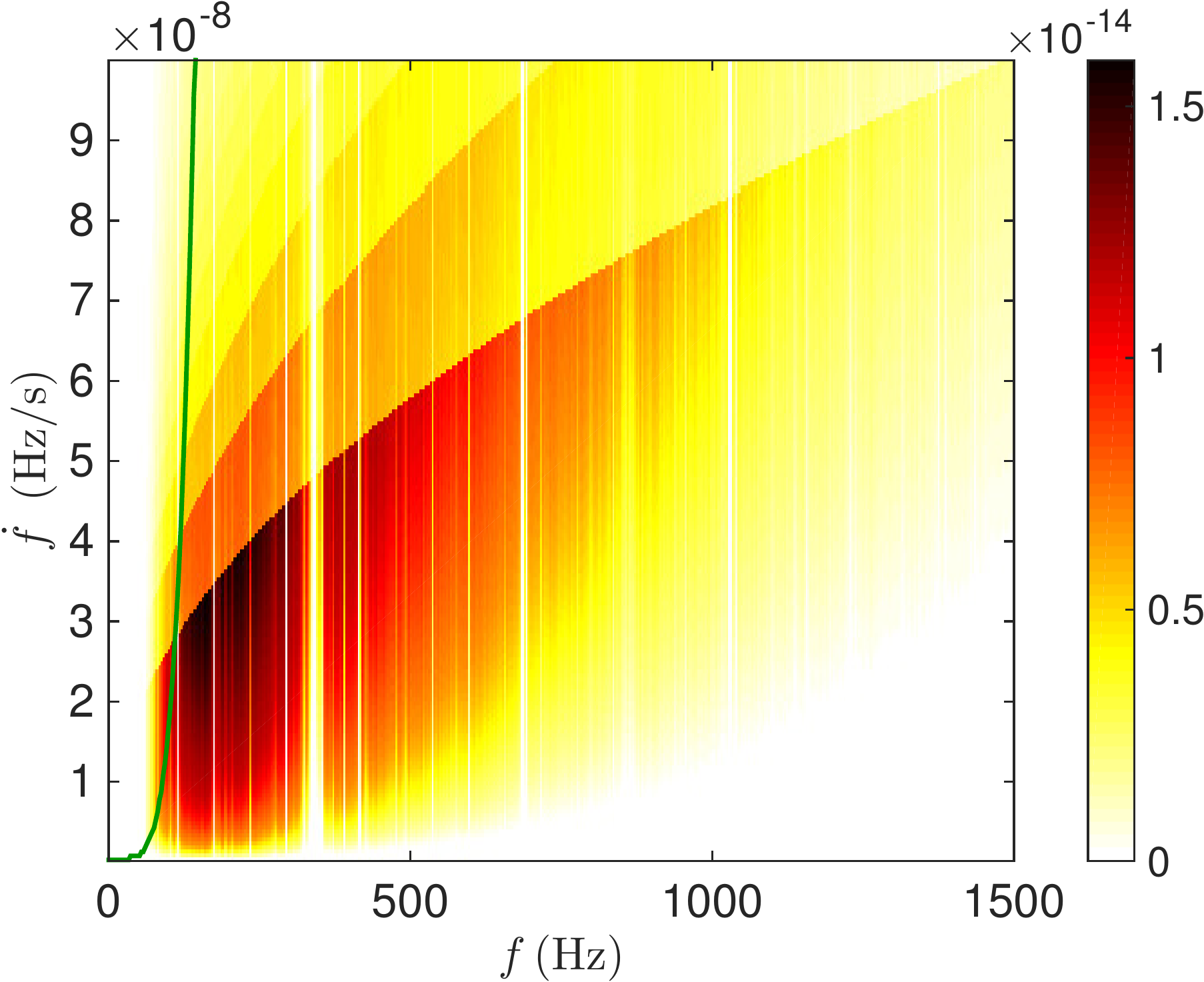}}}%
    \qquad
    \subfloat[Coverage]{{  \includegraphics[width=.4\linewidth]{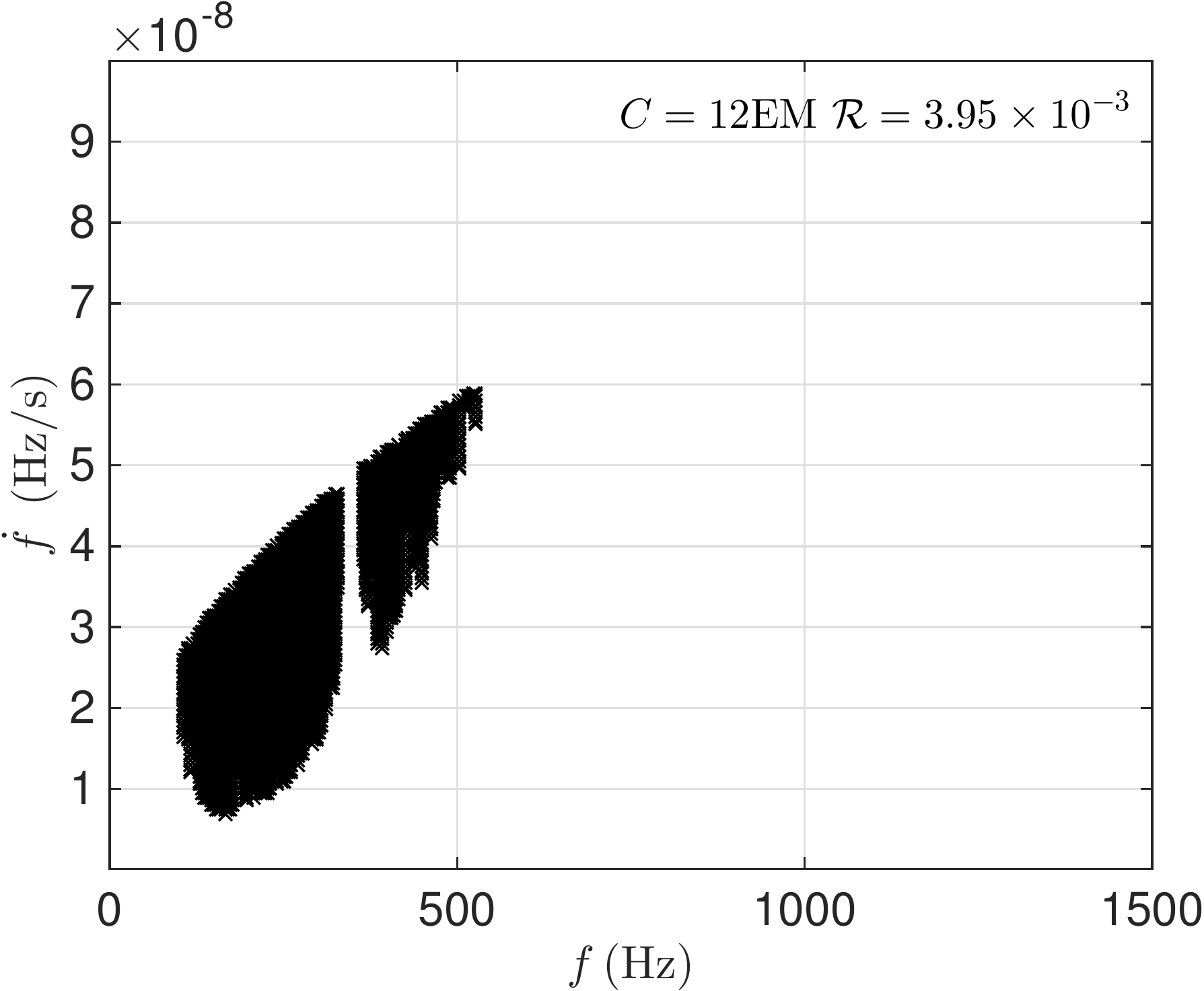}}}%
   \caption{IC 443 at 1500 pc, assuming uniform and distance-based priors, and a10-day coherent segment duration. The leftt plot shows the efficiency, color-coded, for each cell : $e(f,\dot{f})$. The green curve shows $\dot{f}^\star$  as a function of $f$. The right plot displays the cells selected by the optimization procedure with a computational budget of 12 EM. The detection probability $\mathcal{R}$ is $3.95\times10^{-3}$.}%
    \label{G1893_10days_noage}%
\end{figure*}

\begin{figure*}%
    \centering
    \subfloat[Efficiency, 5 days]{{  \includegraphics[width=.20\linewidth]{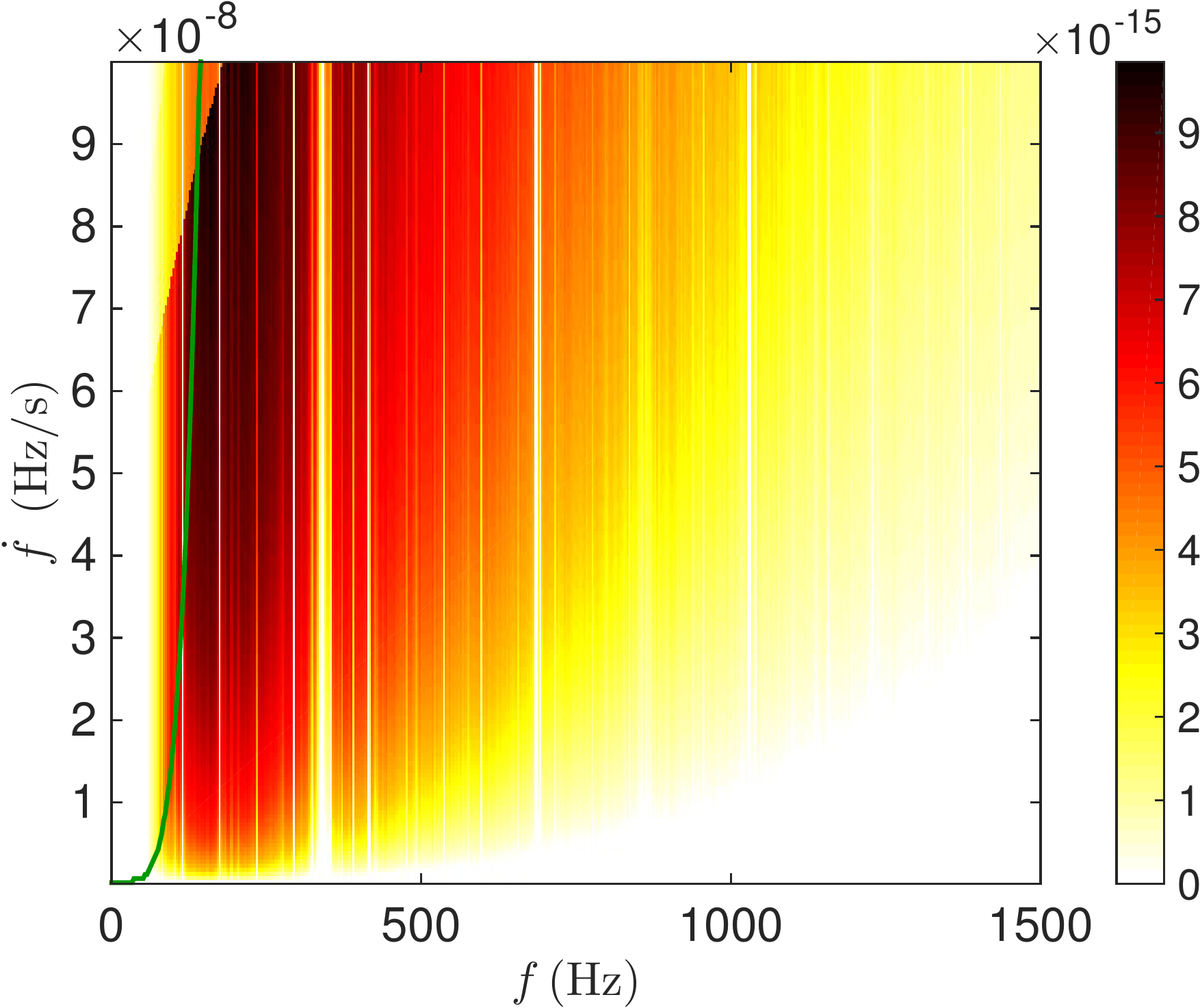}}}%
    \qquad
    \subfloat[Coverage, 5 days]{{  \includegraphics[width=.20\linewidth]{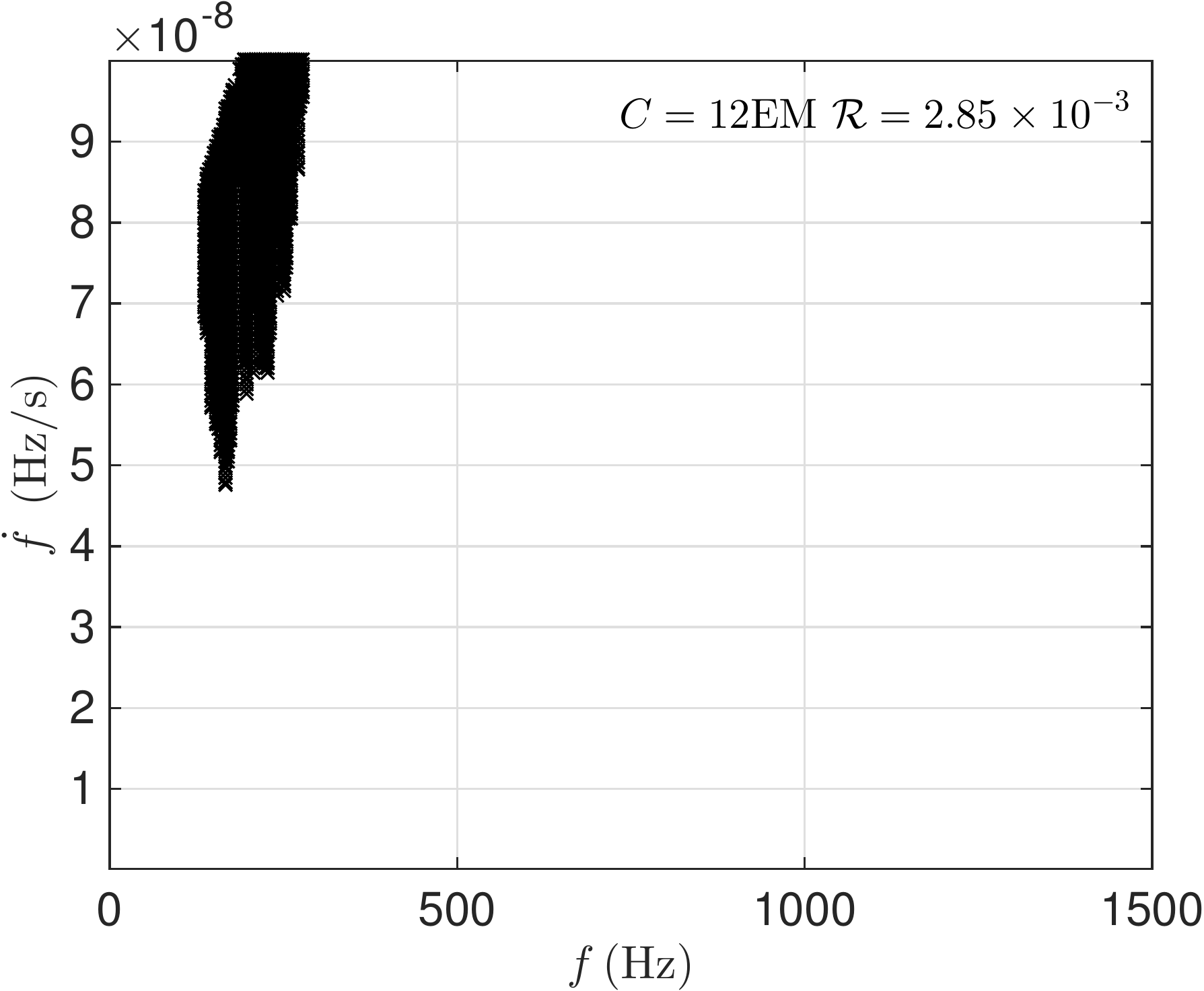}}}%
    \qquad
    \subfloat[Efficiency, 10 days]{{  \includegraphics[width=.20\linewidth]{figures/plots_nonage_eps/G3473_efficiency_10days_noage-eps-converted-to.pdf}}}%
    \qquad
    \subfloat[Coverage, 10 days]{{  \includegraphics[width=.20\linewidth]{figures/plots_nonage_eps/G3473_coverage_10days_noage-eps-converted-to.pdf}}}%
    \qquad
    \subfloat[Efficiency, 20 days]{{  \includegraphics[width=.20\linewidth]{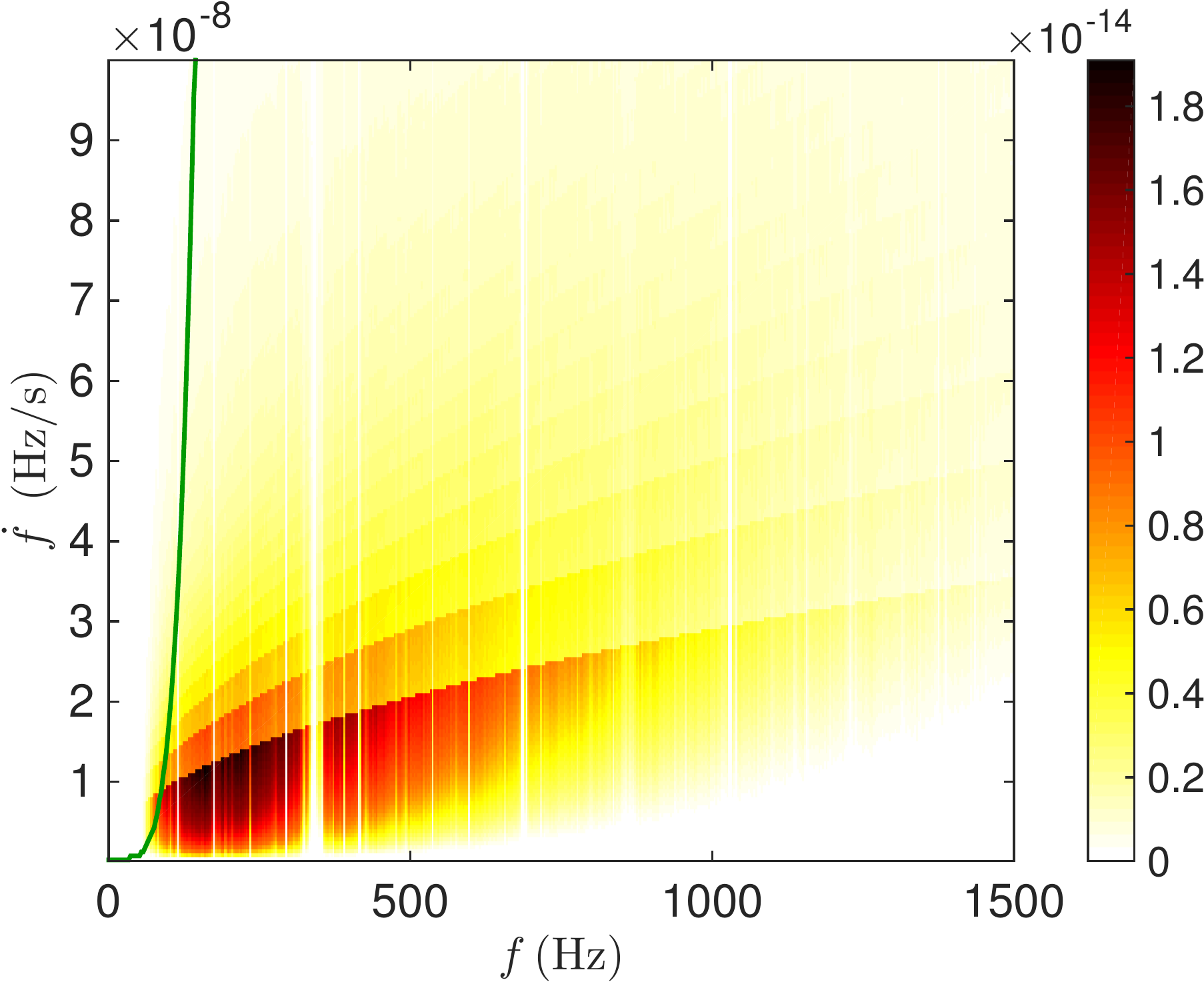}}}%
    \qquad
    \subfloat[Coverage, 20 days]{{  \includegraphics[width=.20\linewidth]{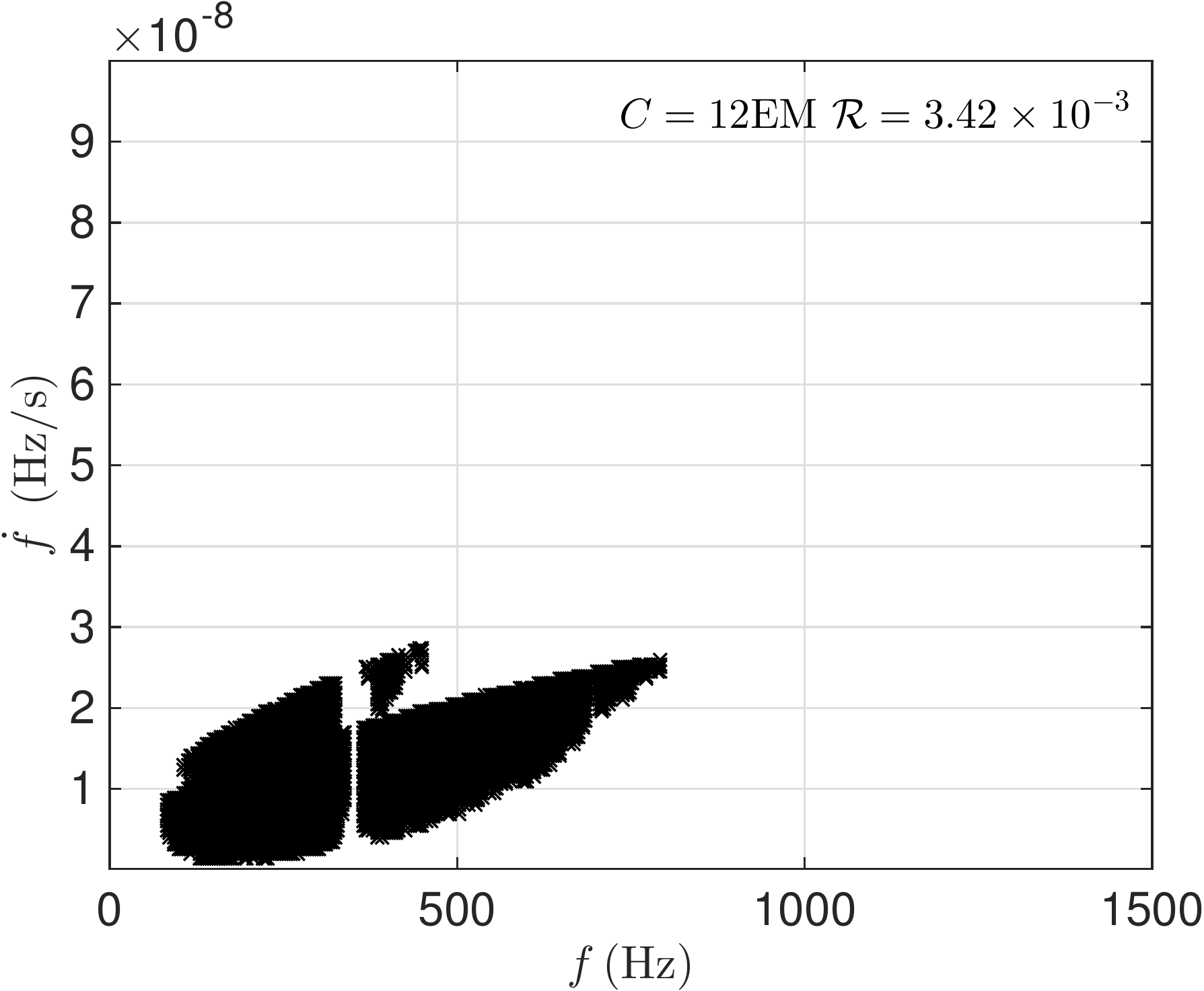}}}%
    \qquad
    \subfloat[Efficiency, 30 days]{{  \includegraphics[width=.20\linewidth]{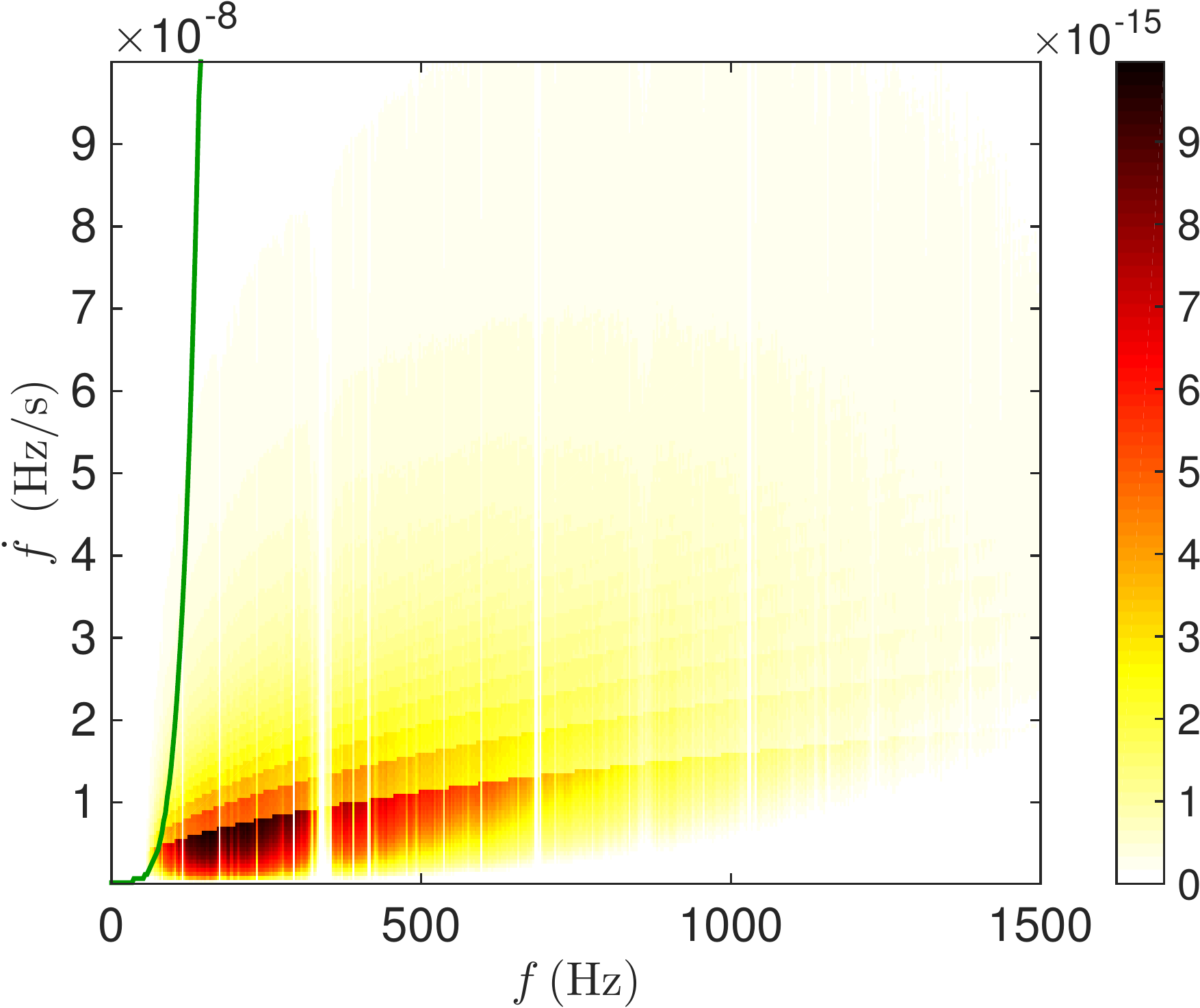}}}%
    \qquad
    \subfloat[Coverage, 30 days]{{  \includegraphics[width=.20\linewidth]{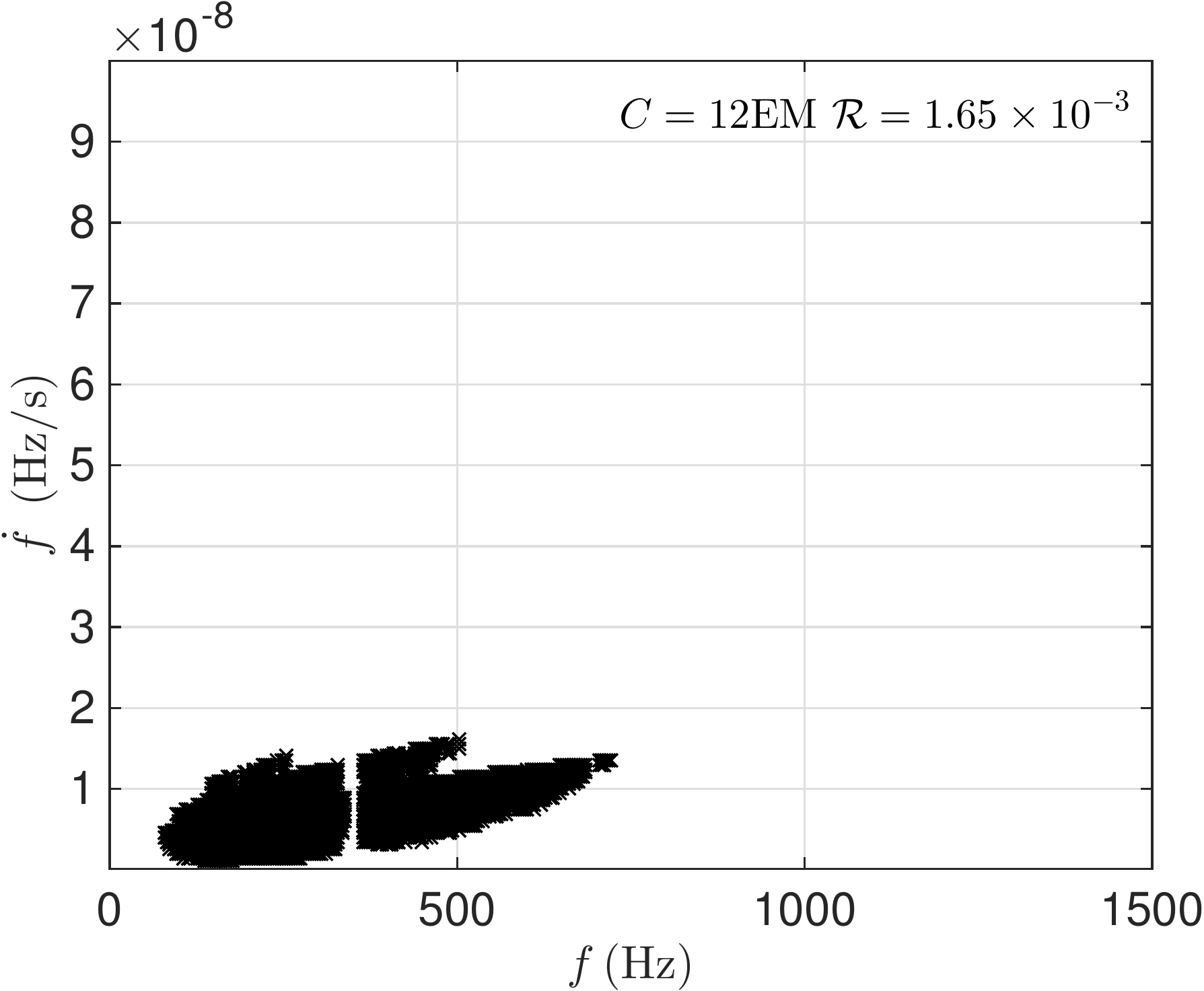}}}%
    \qquad
    \subfloat[Efficiency, 37.5 days]{{  \includegraphics[width=.20\linewidth]{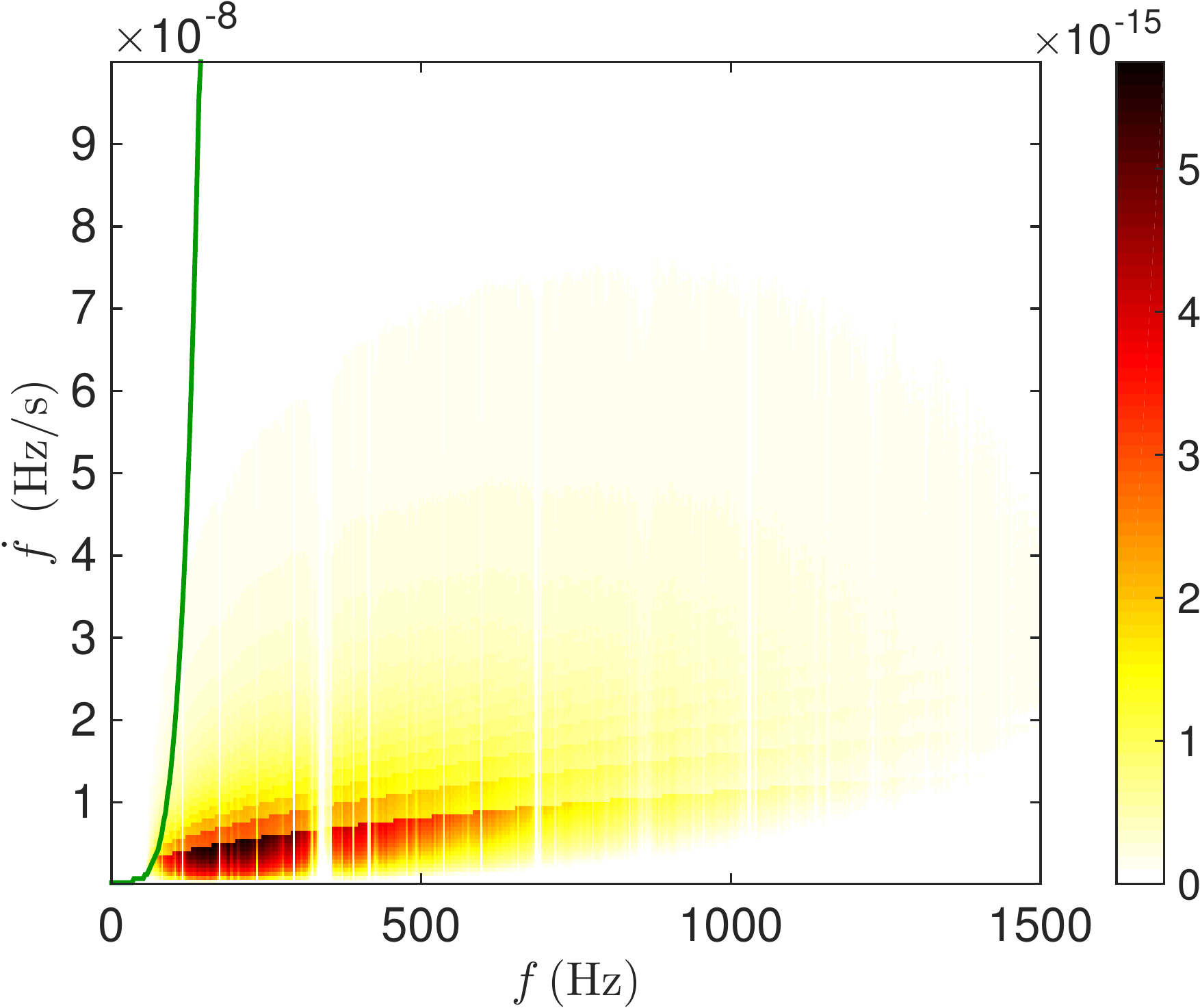}}}%
    \qquad
    \subfloat[Coverage, 37.5 days]{{  \includegraphics[width=.20\linewidth]{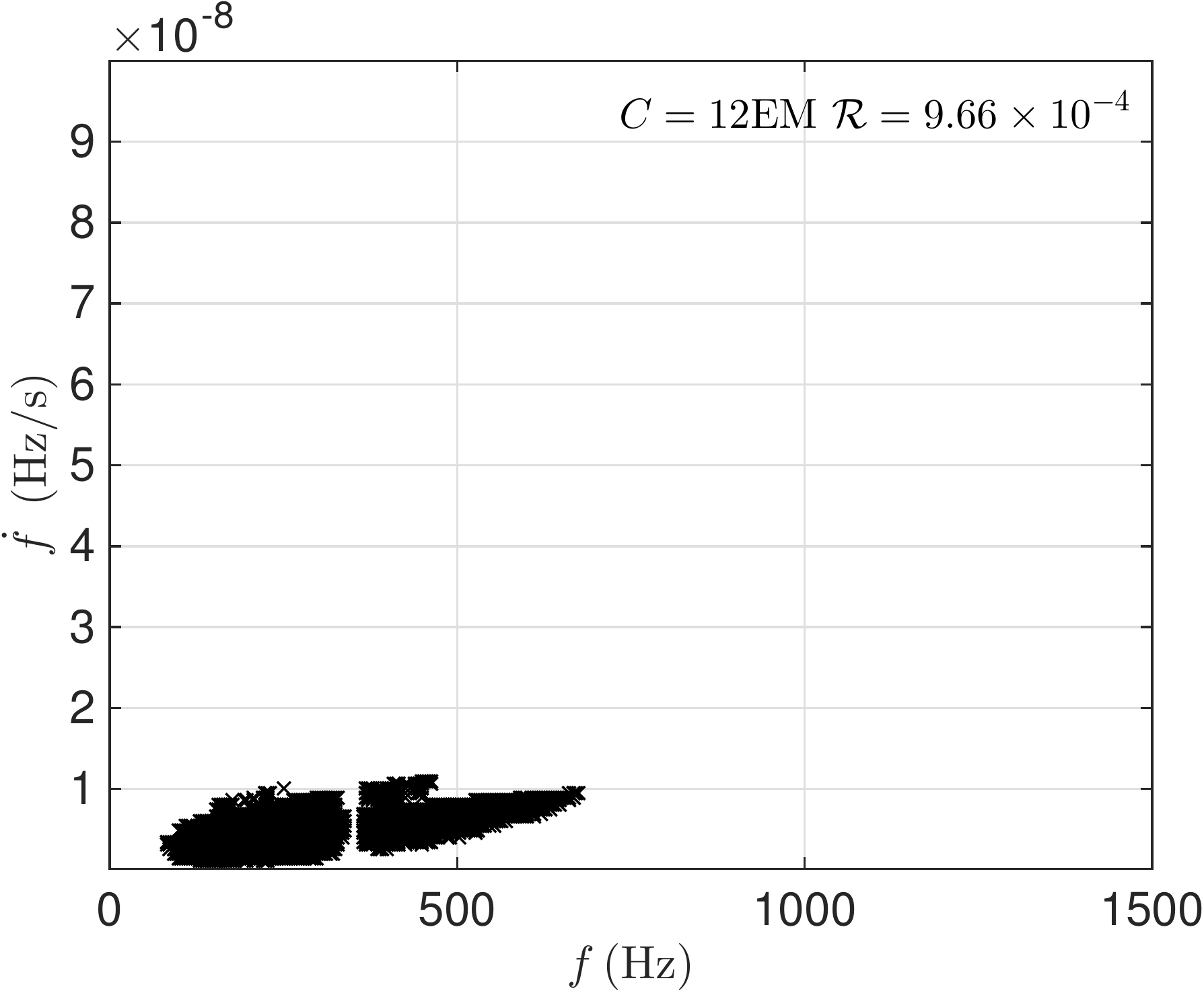}}}%
    \qquad
    \subfloat[Efficiency, 50 days]{{  \includegraphics[width=.20\linewidth]{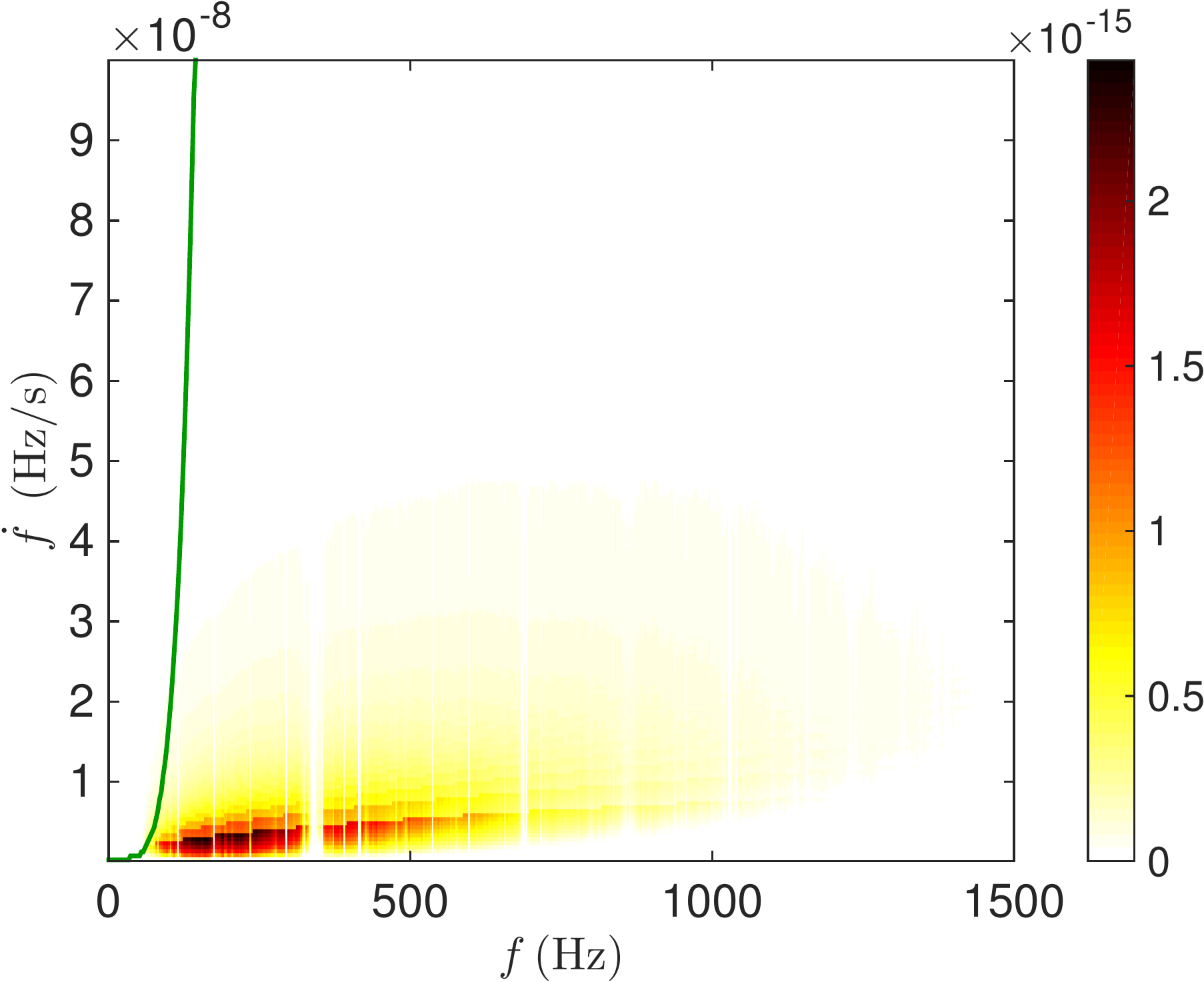}}}%
    \qquad
    \subfloat[Coverage, 50 days]{{  \includegraphics[width=.20\linewidth]{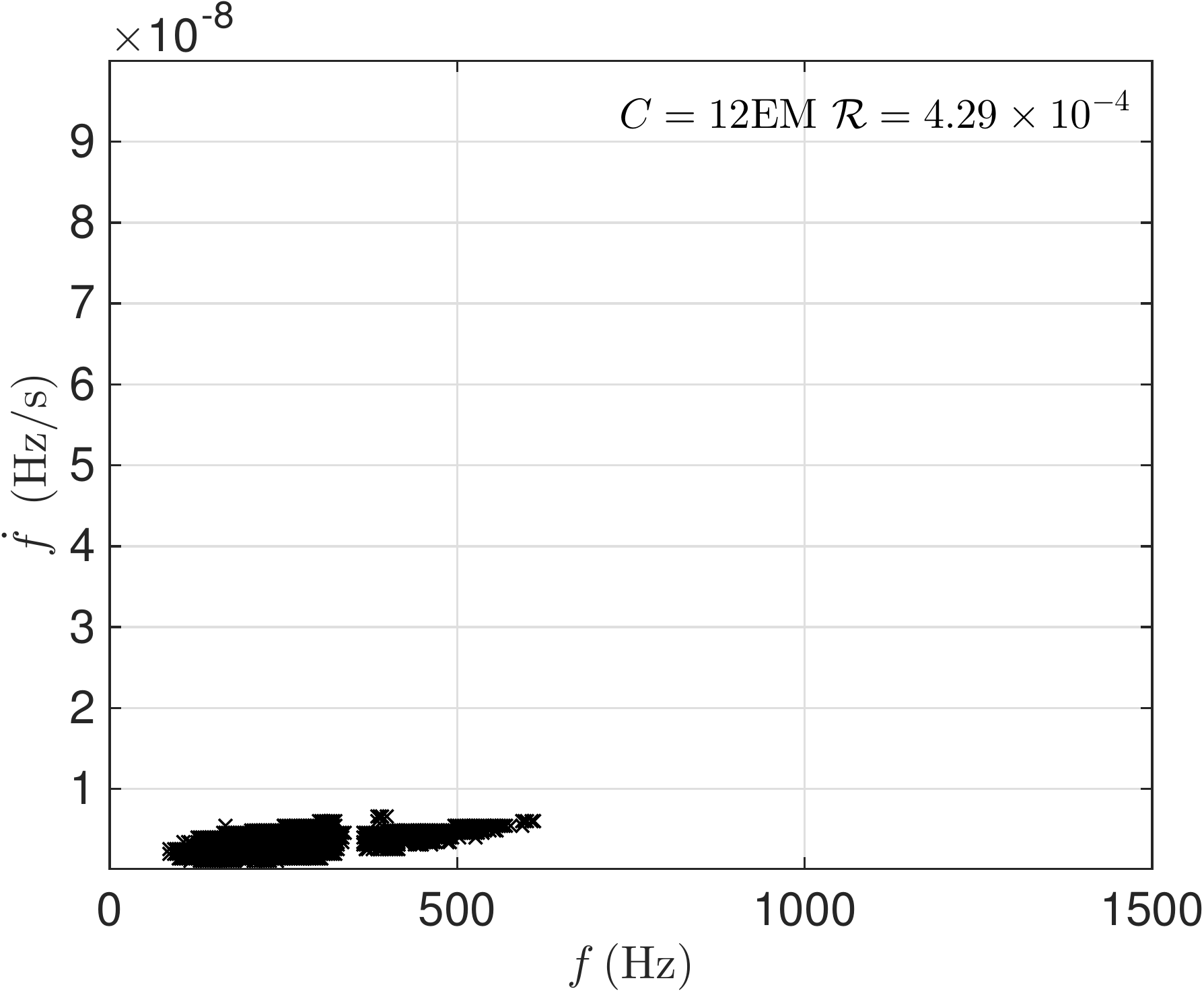}}}%
    \qquad
    \subfloat[Efficiency, 75 days]{{  \includegraphics[width=.20\linewidth]{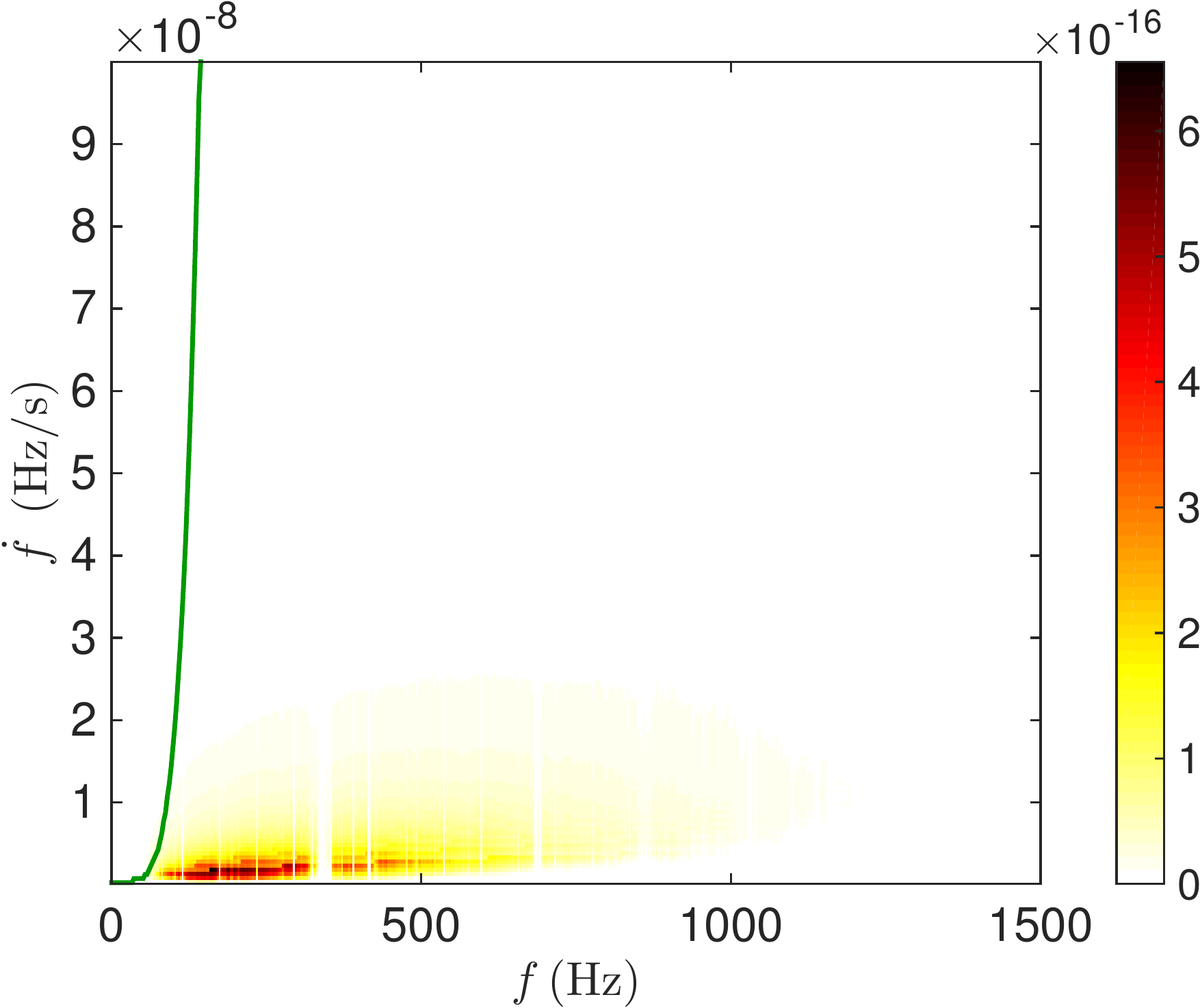}}}%
    \qquad
    \subfloat[Coverage, 75 days]{{  \includegraphics[width=.20\linewidth]{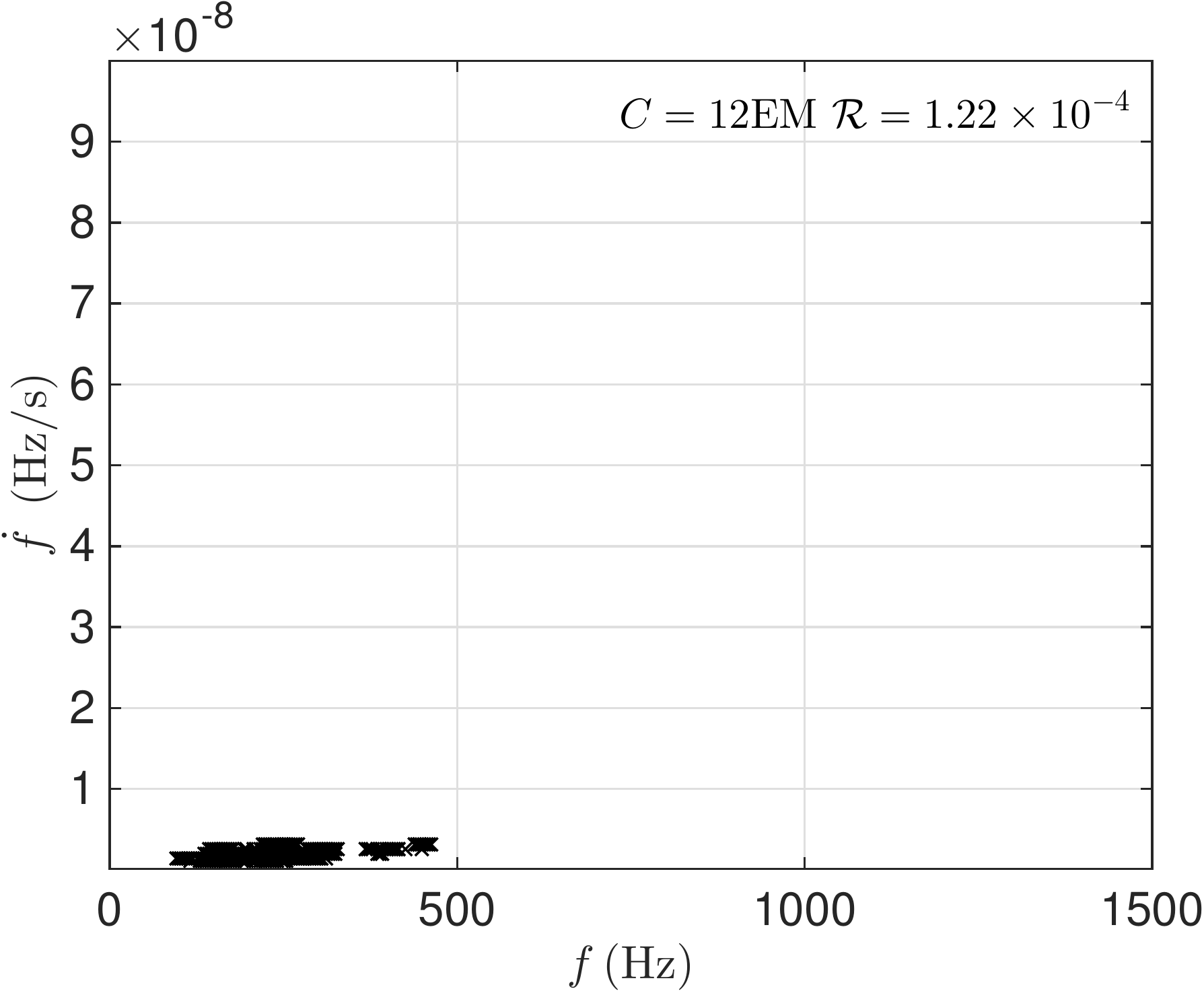}}}%
    \caption{Optimisation results for G347.3 at 1300 pc assuming uniform and distance-based priors, for various coherent search durations: 5, 10, 20, 30, 37.5, 50 and 75 days. The total computing budget is assumed to be 12 EM.}%
    \label{G3473_51020days_noage}%
\end{figure*}

\begin{figure*}%
    \centering
    \subfloat[Efficiency, 5 days]{{  \includegraphics[width=.20\linewidth]{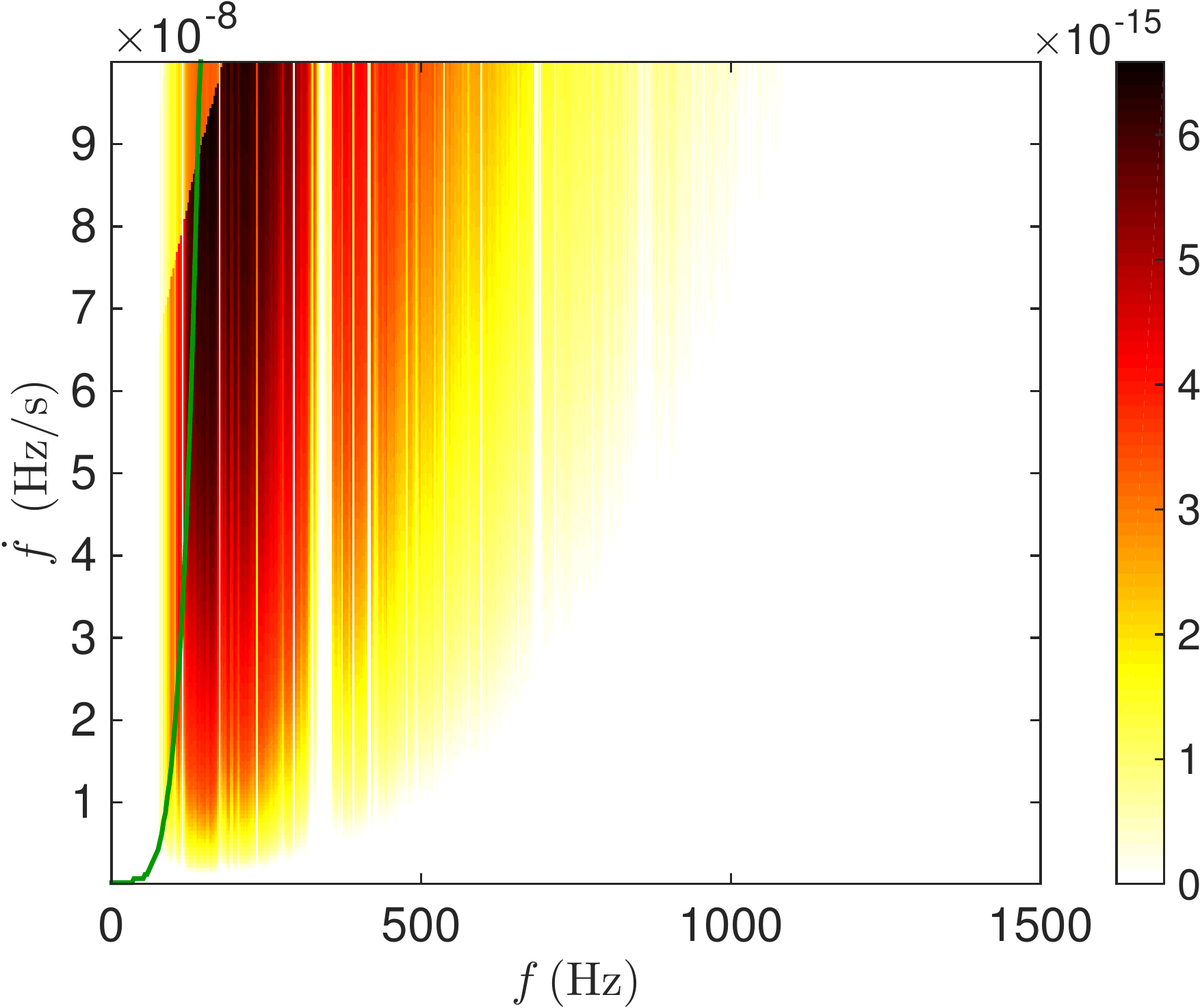}}}%
    \qquad
    \subfloat[Coverage, 5 days]{{  \includegraphics[width=.20\linewidth]{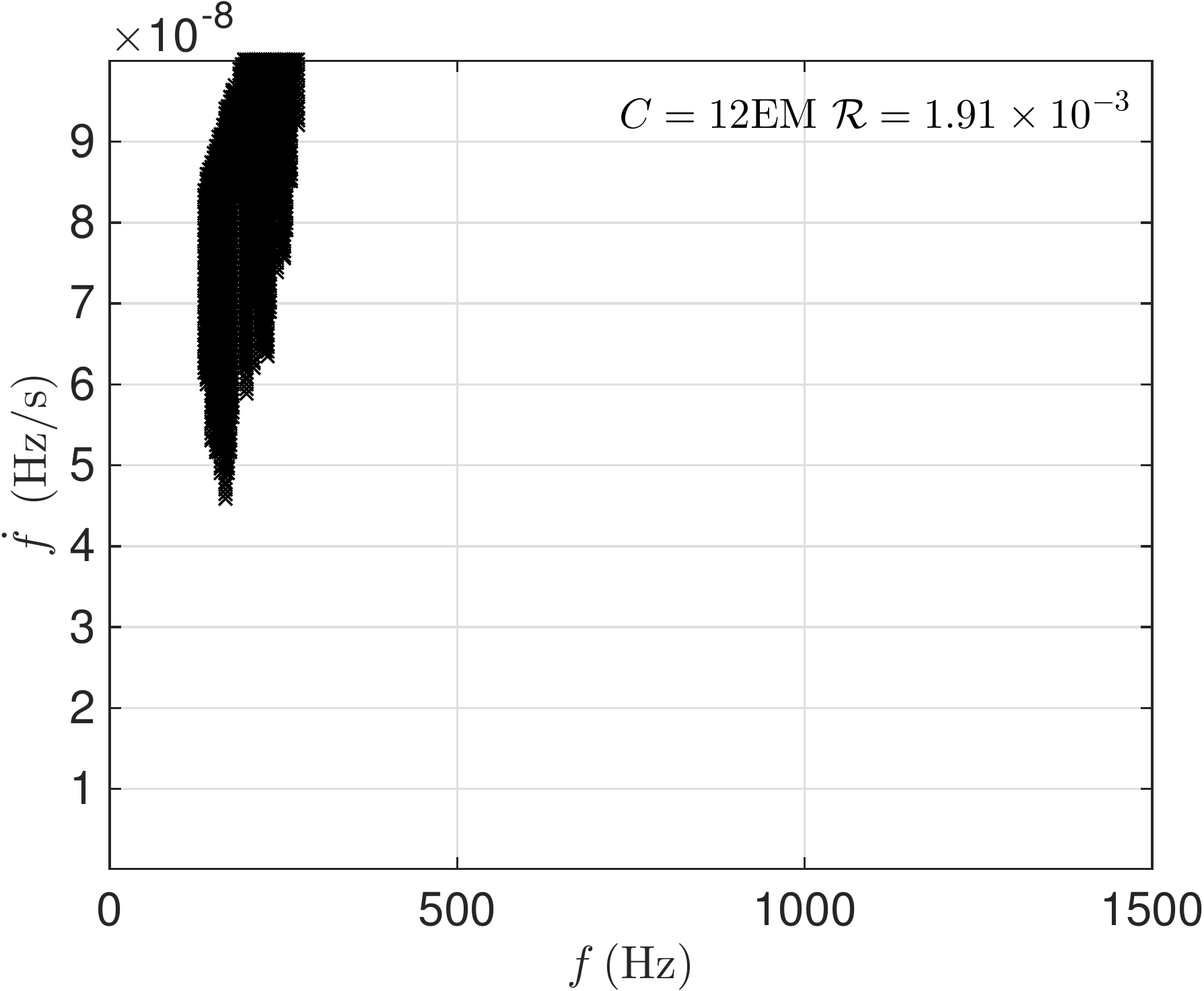}}}%
    \qquad
    \subfloat[Efficiency, 10 days]{{  \includegraphics[width=.20\linewidth]{figures/plots_nonage_eps/CasA_efficiency_10days_noage-eps-converted-to.pdf}}}%
    \qquad
    \subfloat[Coverage, 10 days]{{  \includegraphics[width=.20\linewidth]{figures/plots_nonage_eps/CasA_coverage_10days_noage-eps-converted-to.pdf}}}%
    \qquad
    \subfloat[Efficiency, 20 days]{{  \includegraphics[width=.20\linewidth]{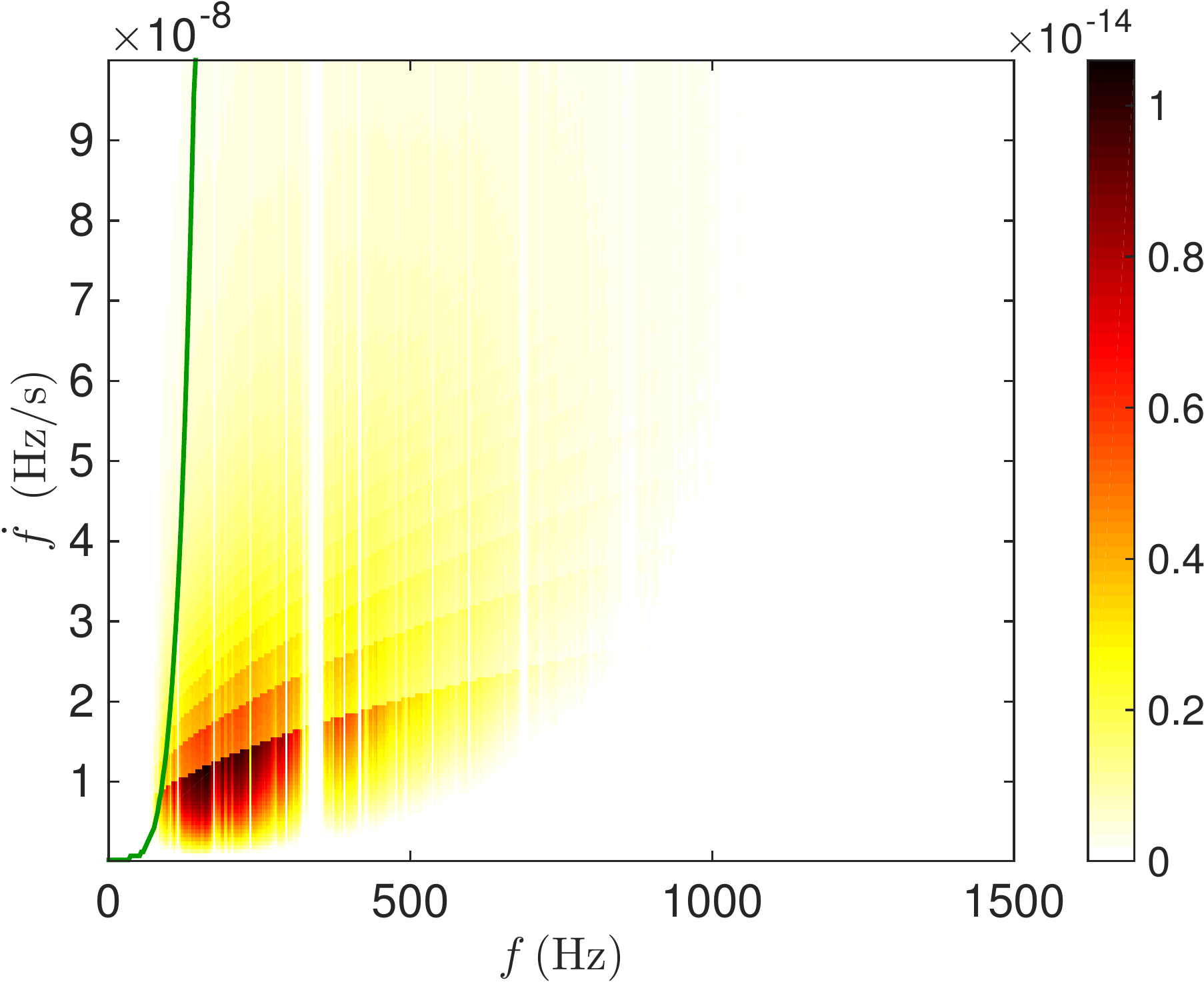}}}%
    \qquad
    \subfloat[Coverage, 20 days]{{  \includegraphics[width=.20\linewidth]{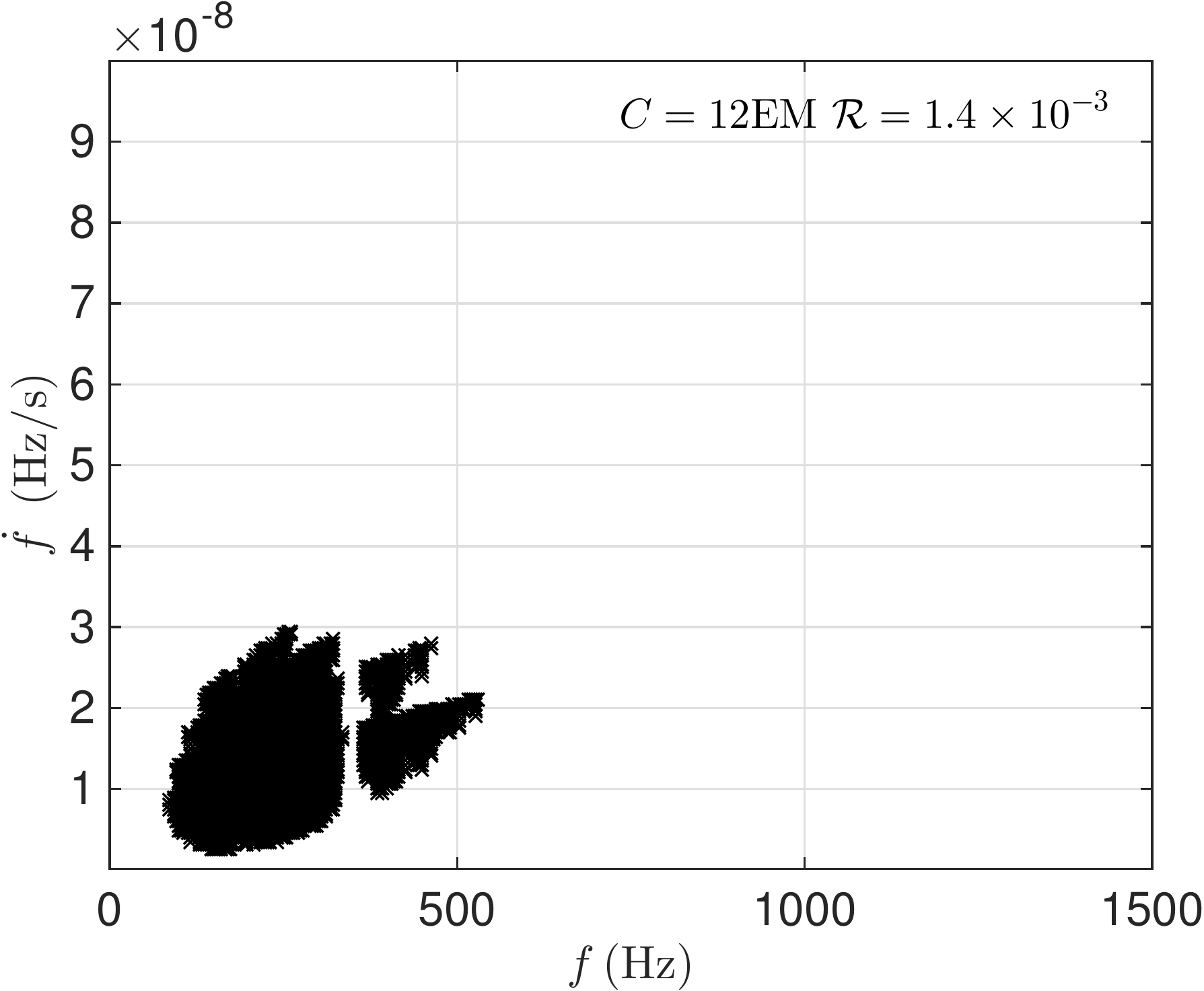}}}%
    \qquad
    \subfloat[Efficiency, 30 days]{{  \includegraphics[width=.20\linewidth]{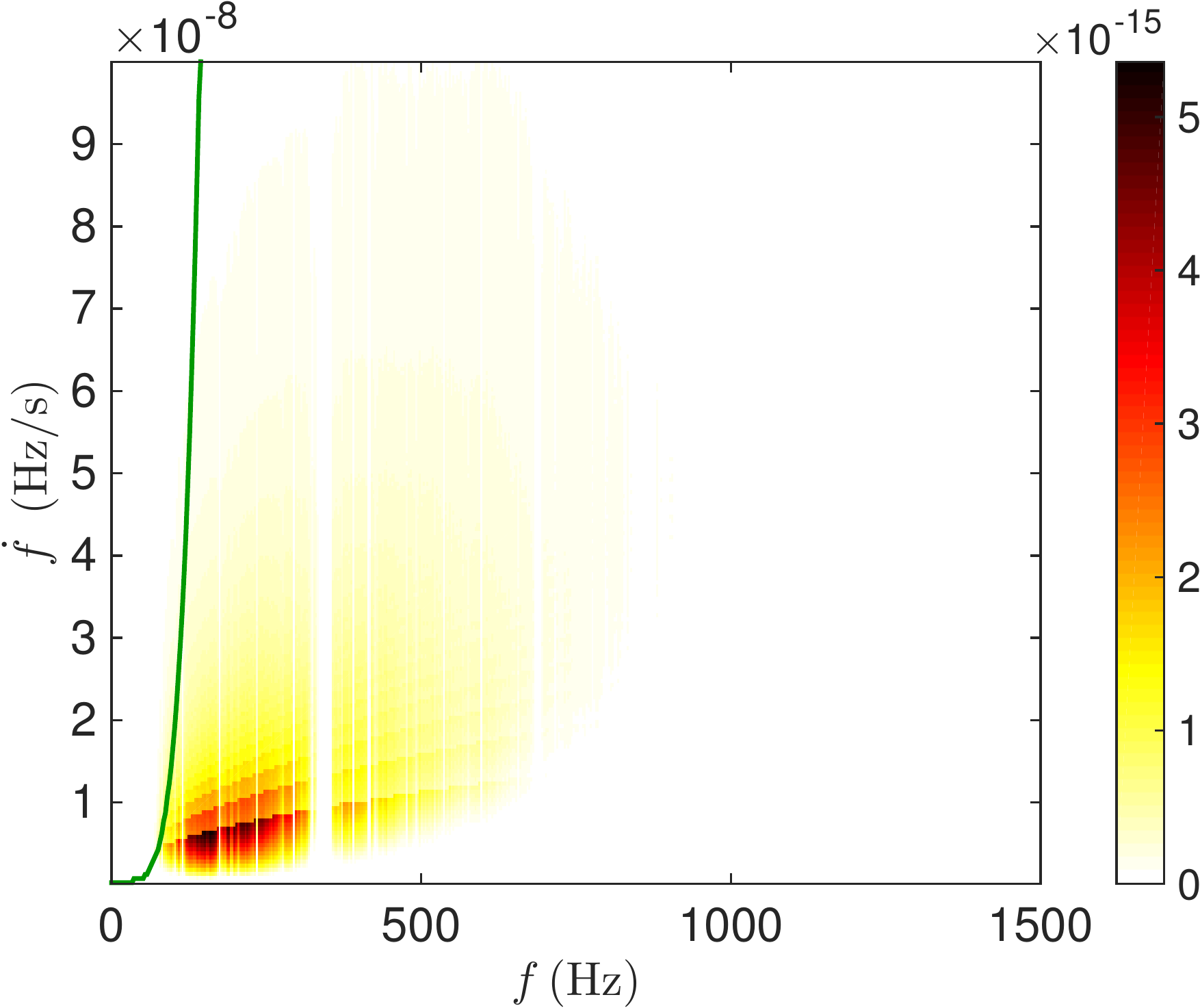}}}%
    \qquad
    \subfloat[Coverage, 30 days]{{  \includegraphics[width=.20\linewidth]{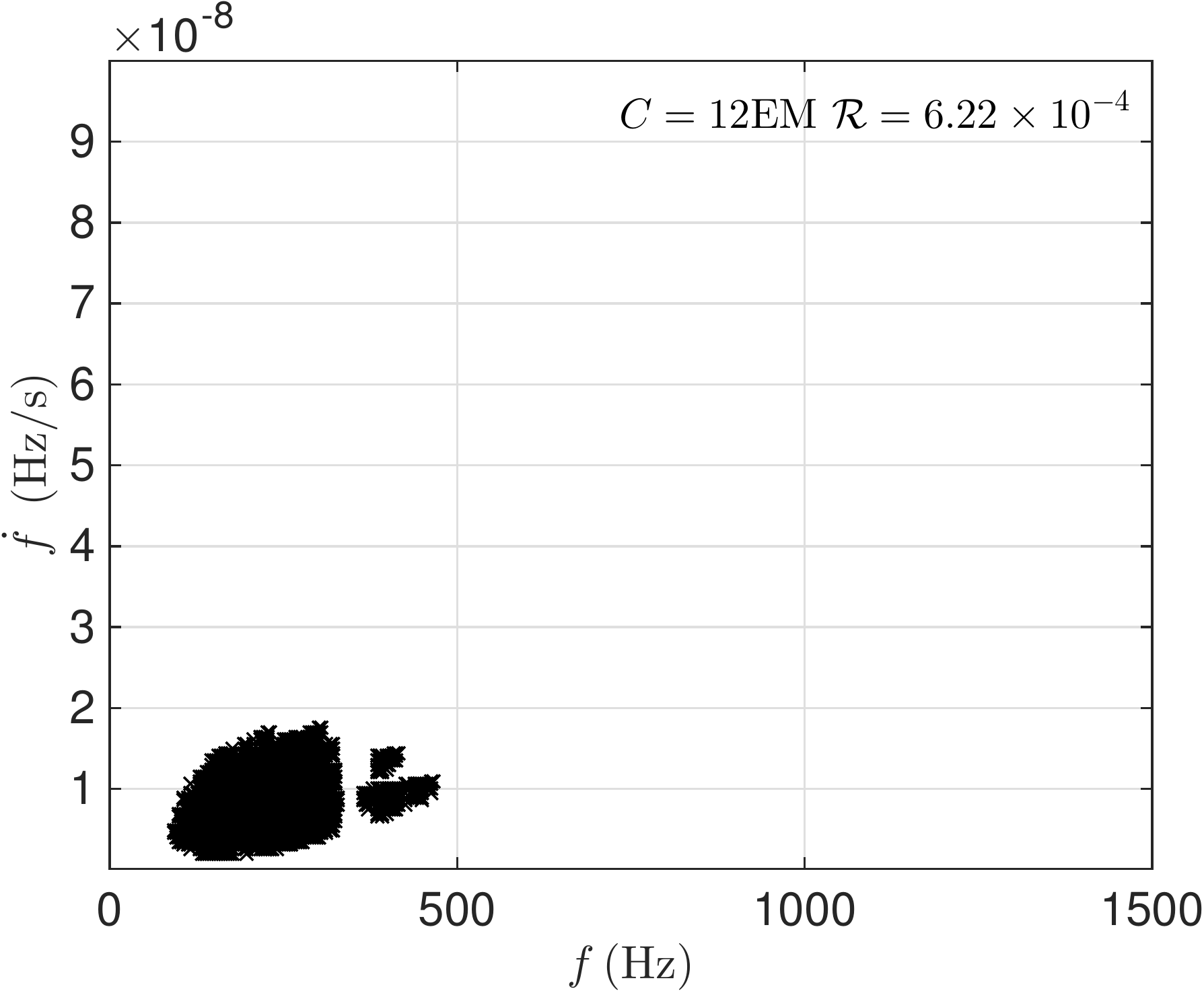}}}%
    \qquad
    \subfloat[Efficiency, 37.5 days]{{  \includegraphics[width=.20\linewidth]{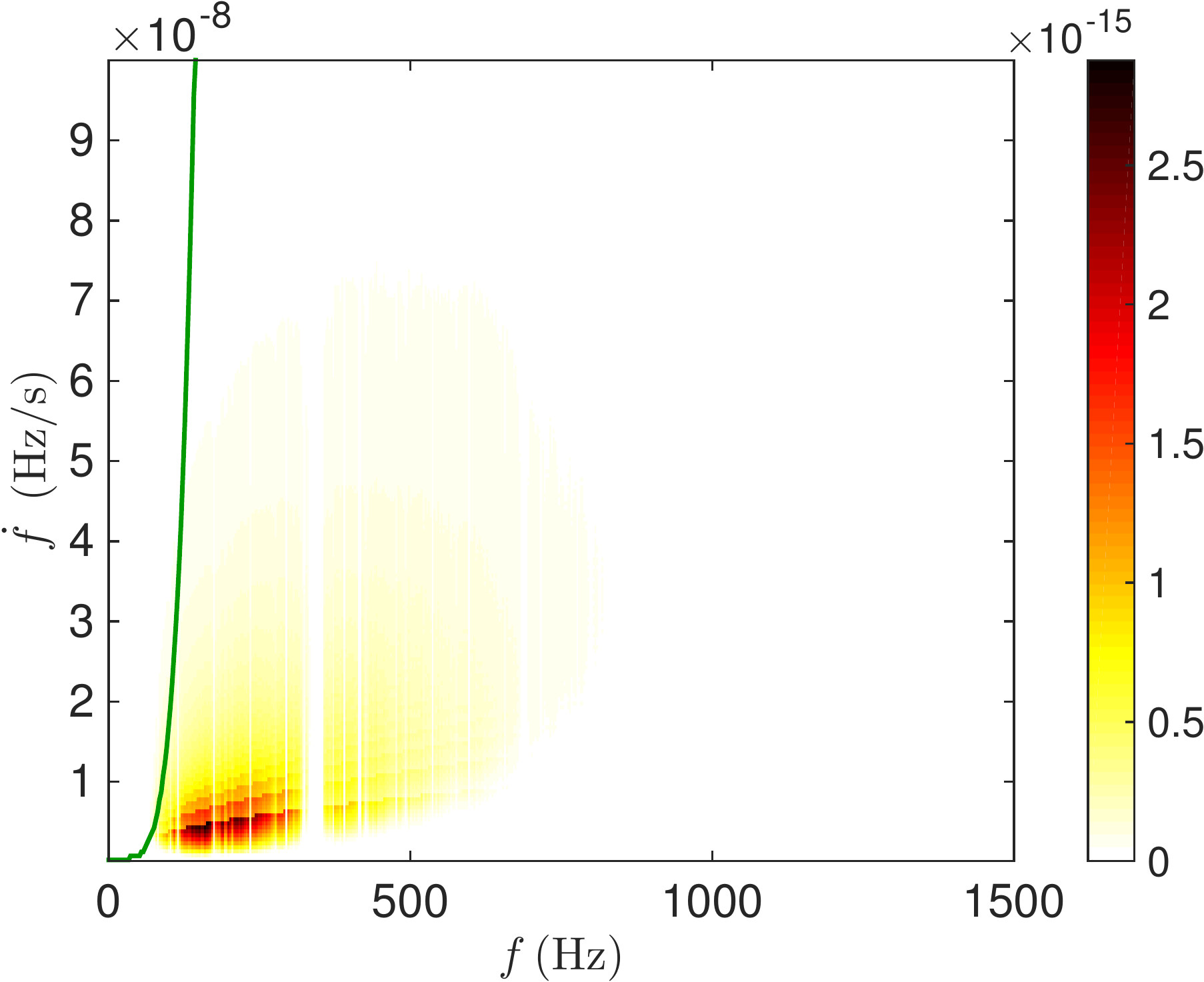}}}%
    \qquad
    \subfloat[Coverage, 37.5 days]{{  \includegraphics[width=.20\linewidth]{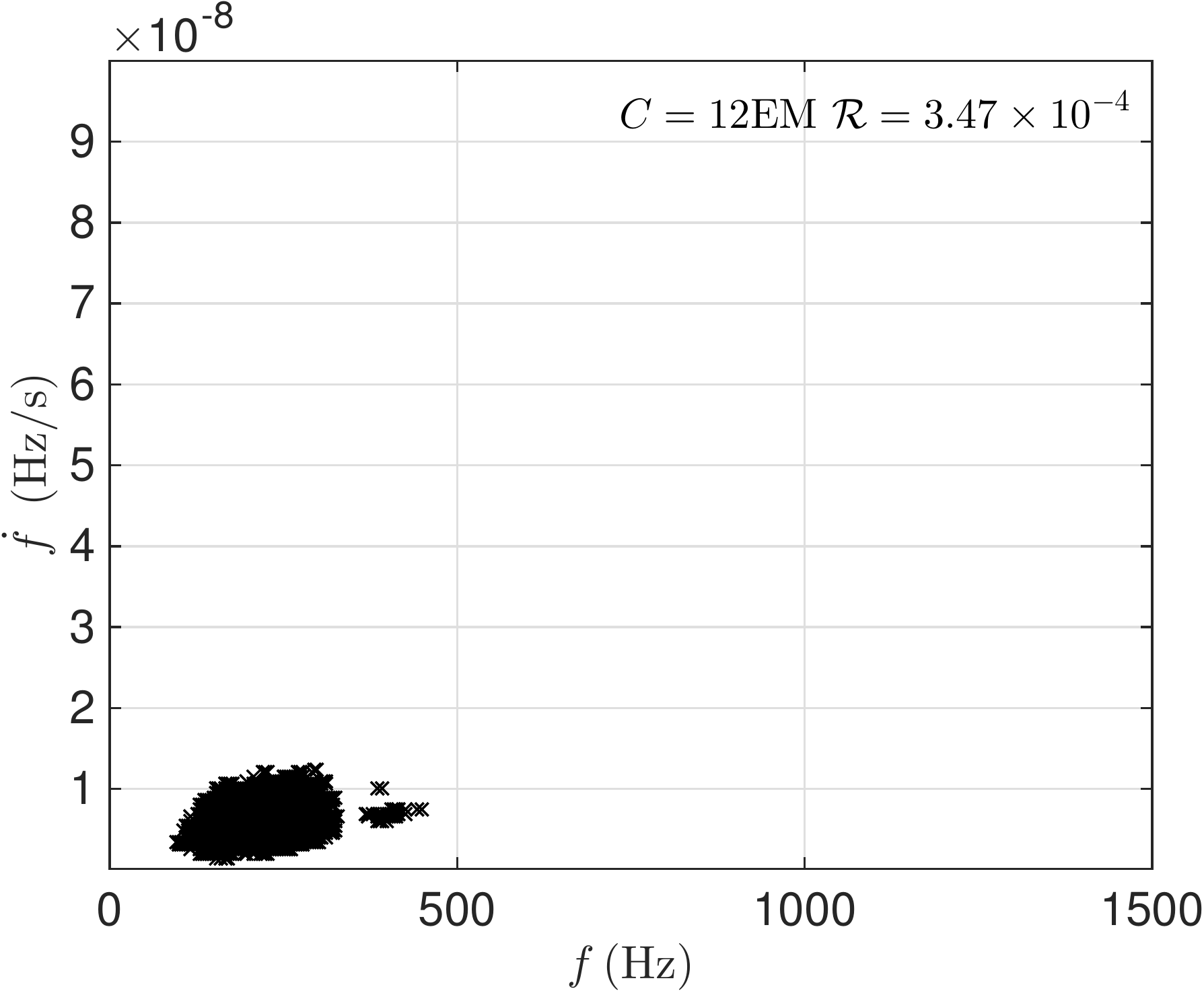}}}%
    \qquad
    \subfloat[Efficiency, 50 days]{{  \includegraphics[width=.20\linewidth]{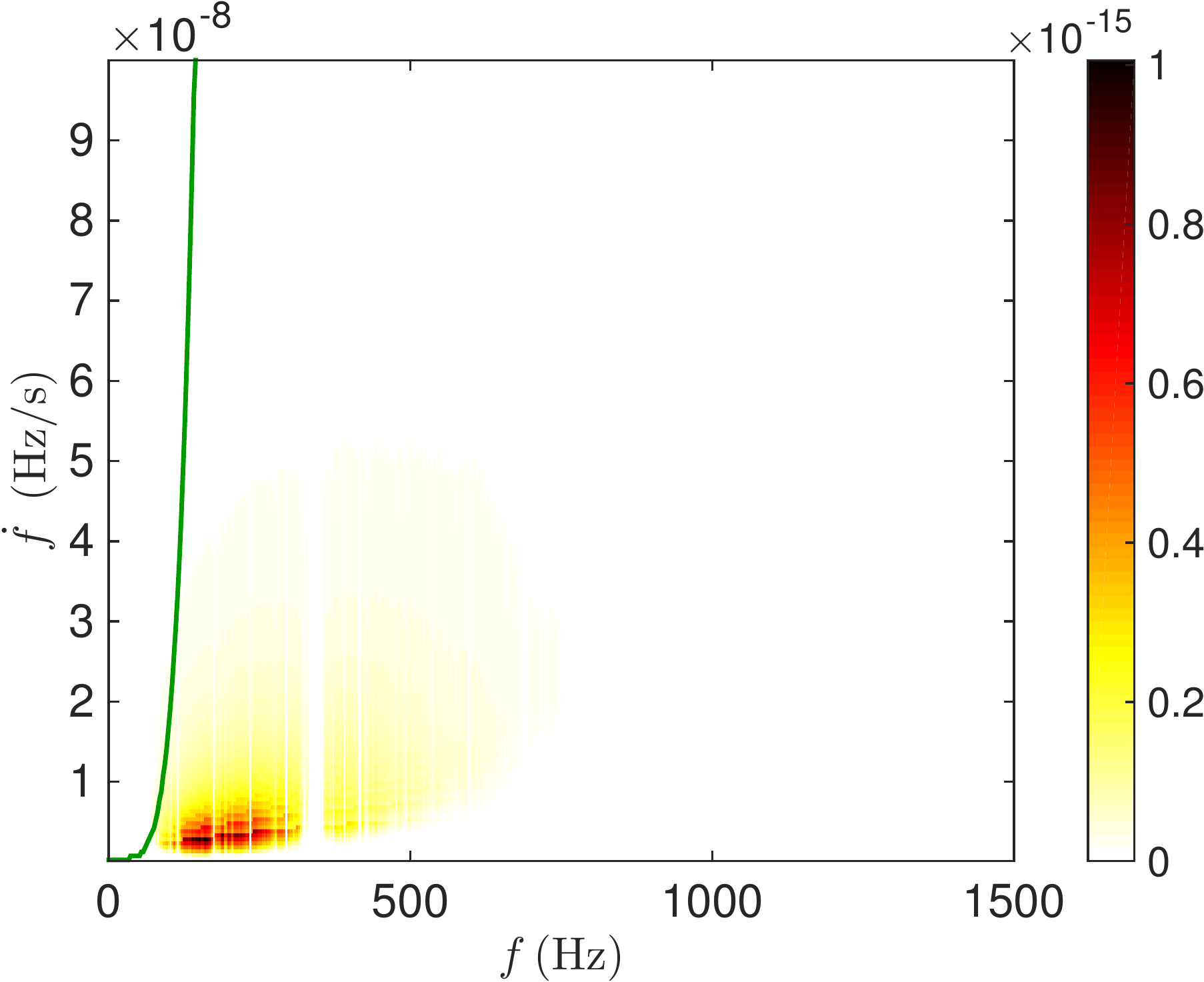}}}%
    \qquad
    \subfloat[Coverage, 50 days]{{  \includegraphics[width=.20\linewidth]{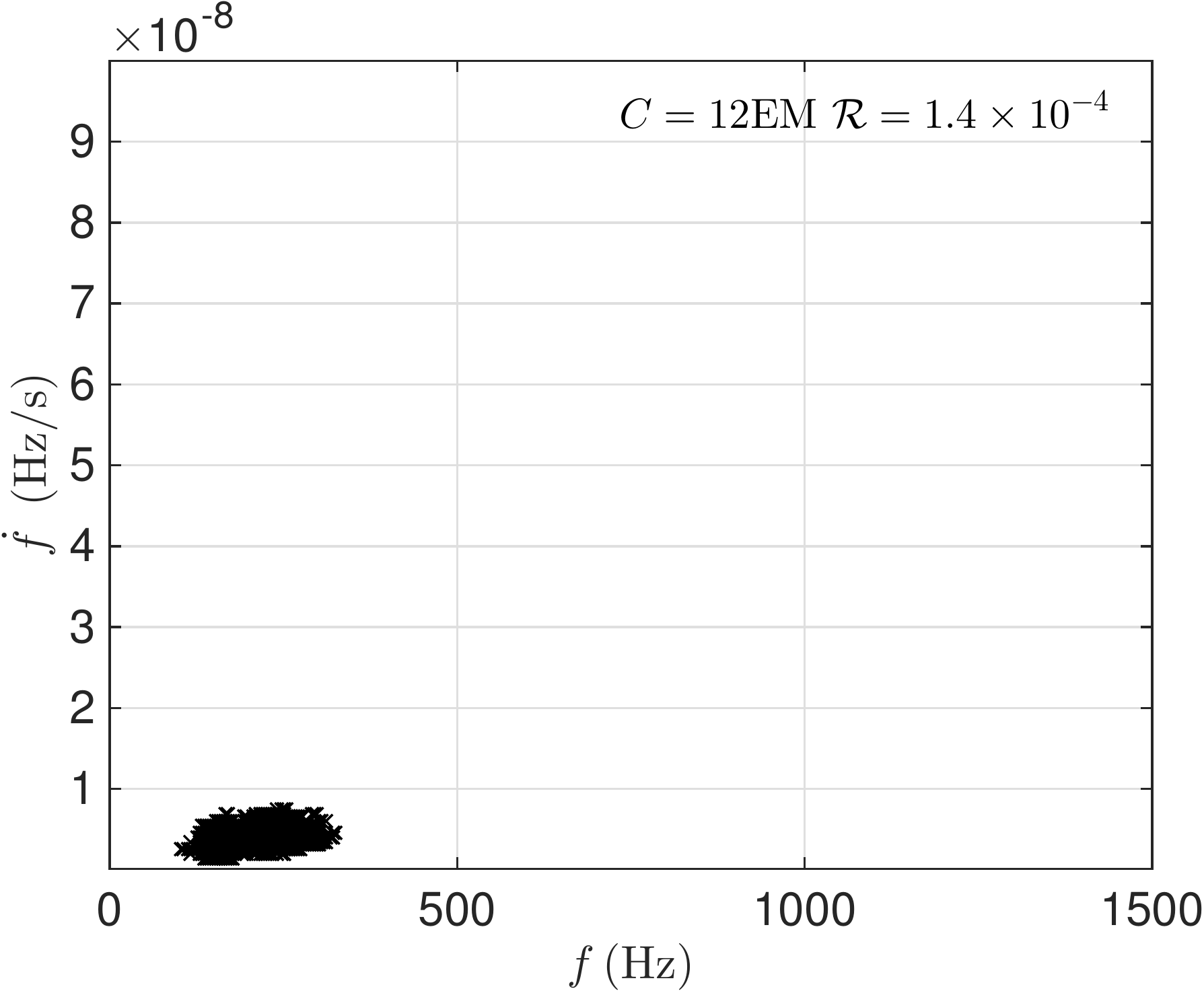}}}%
    \qquad
    \subfloat[Efficiency, 75 days]{{  \includegraphics[width=.20\linewidth]{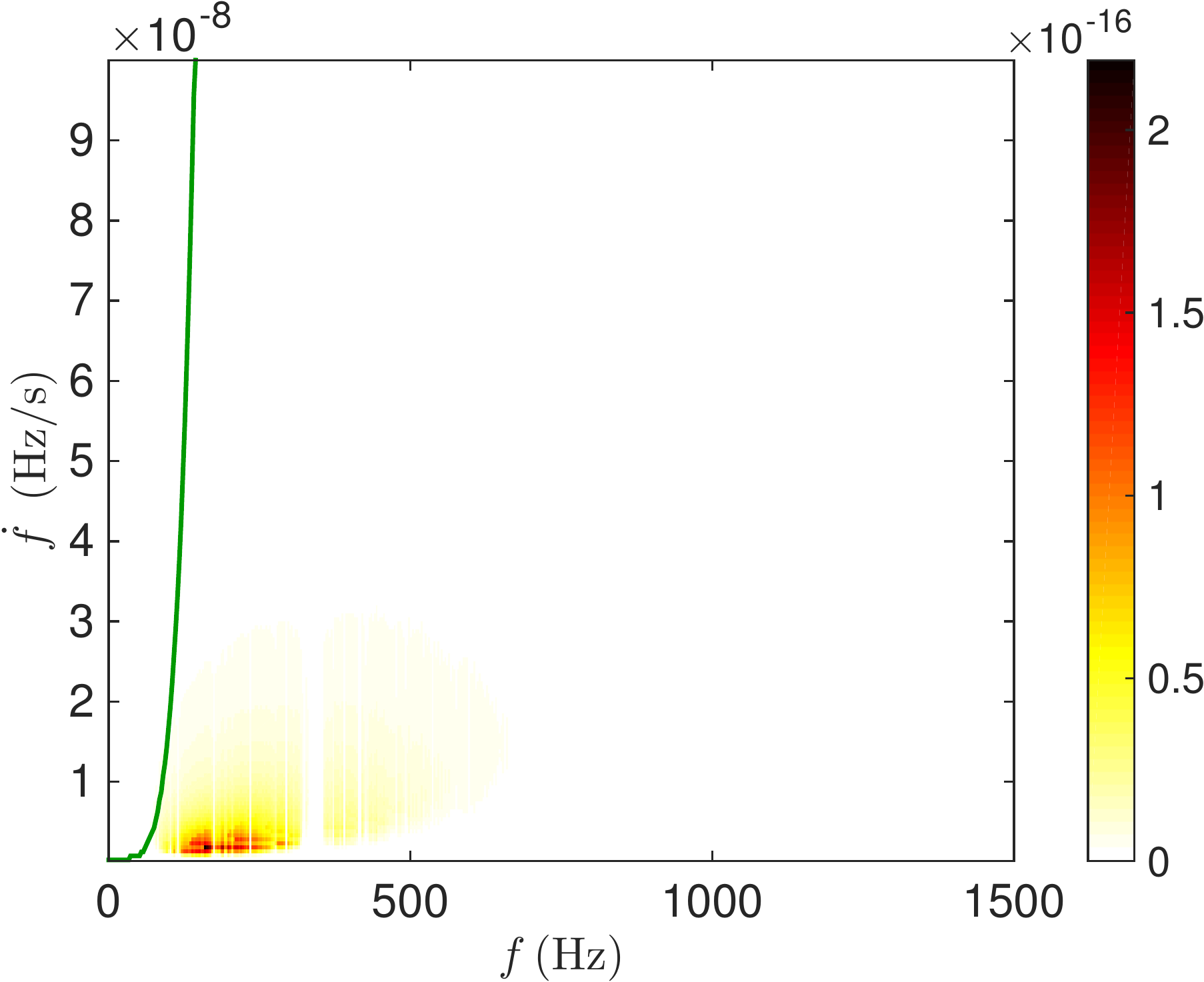}}}%
    \qquad
    \subfloat[Coverage, 75 days]{{  \includegraphics[width=.20\linewidth]{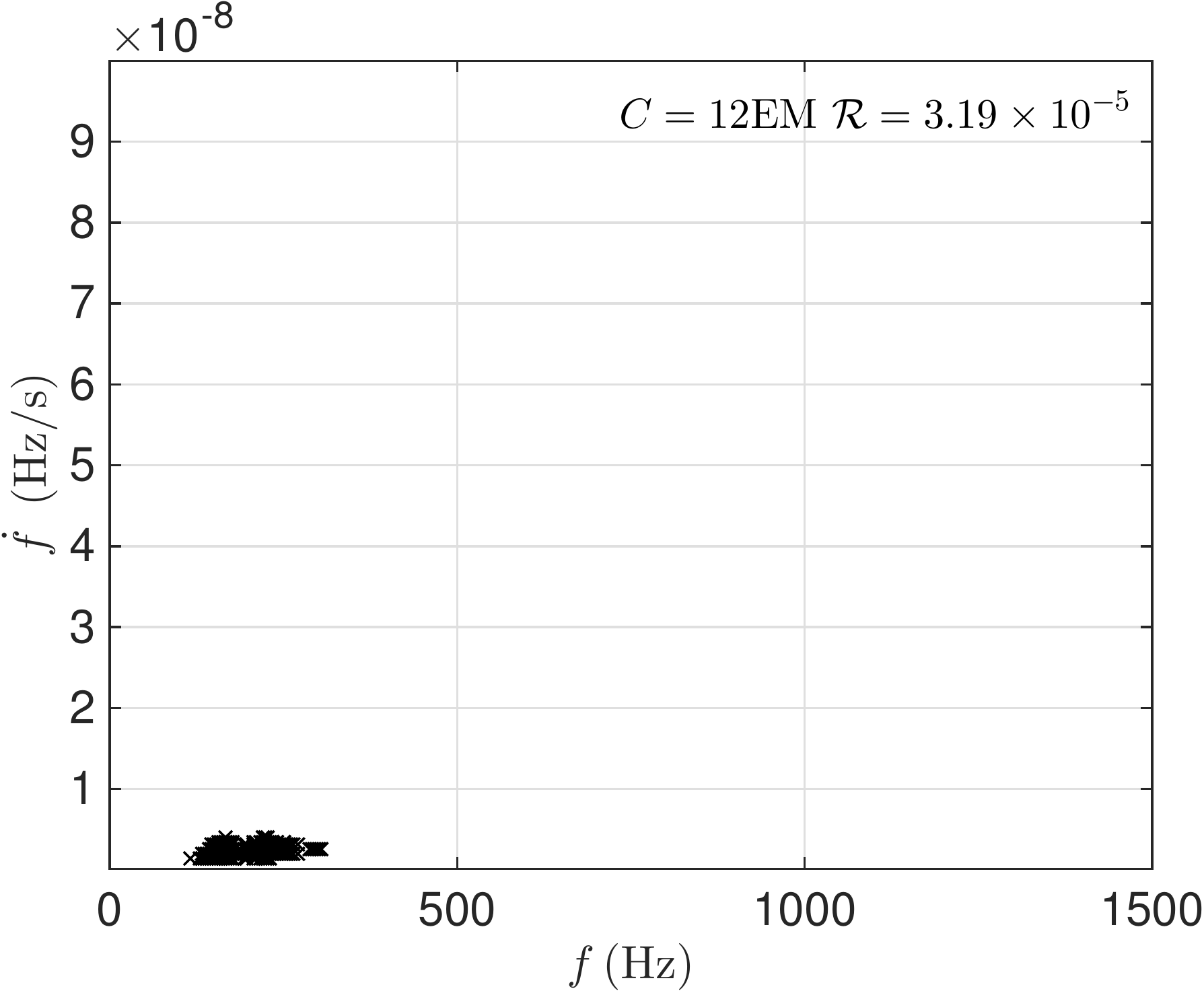}}}%
    \caption{Optimisation results for Cas A at 3500 pc assuming uniform and distance-based priors, for various coherent search durations: 5, 10, 20, 30, 37.5, 50 and 75 days. The total computing budget is assumed to be 12 EM.}%
    \label{CasA_51020days_noage}%
\end{figure*}

\begin{figure*}%
    \centering
    \subfloat[Coverage, cost: 12 EM]{{  \includegraphics[width=.45\linewidth]{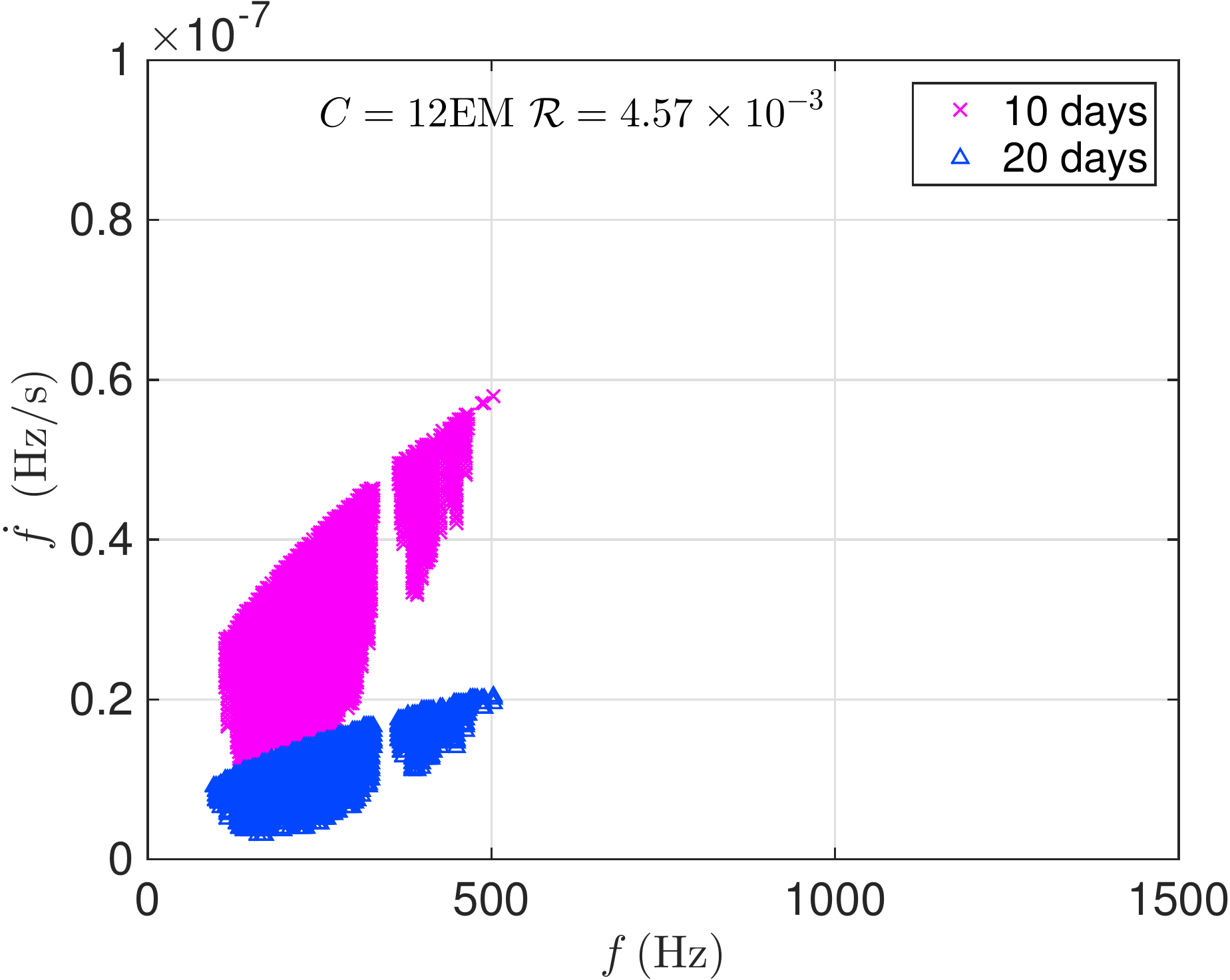}}}%
    \qquad
    \subfloat[Coverage,  cost: 24 EM]{{  \includegraphics[width=.45\linewidth]{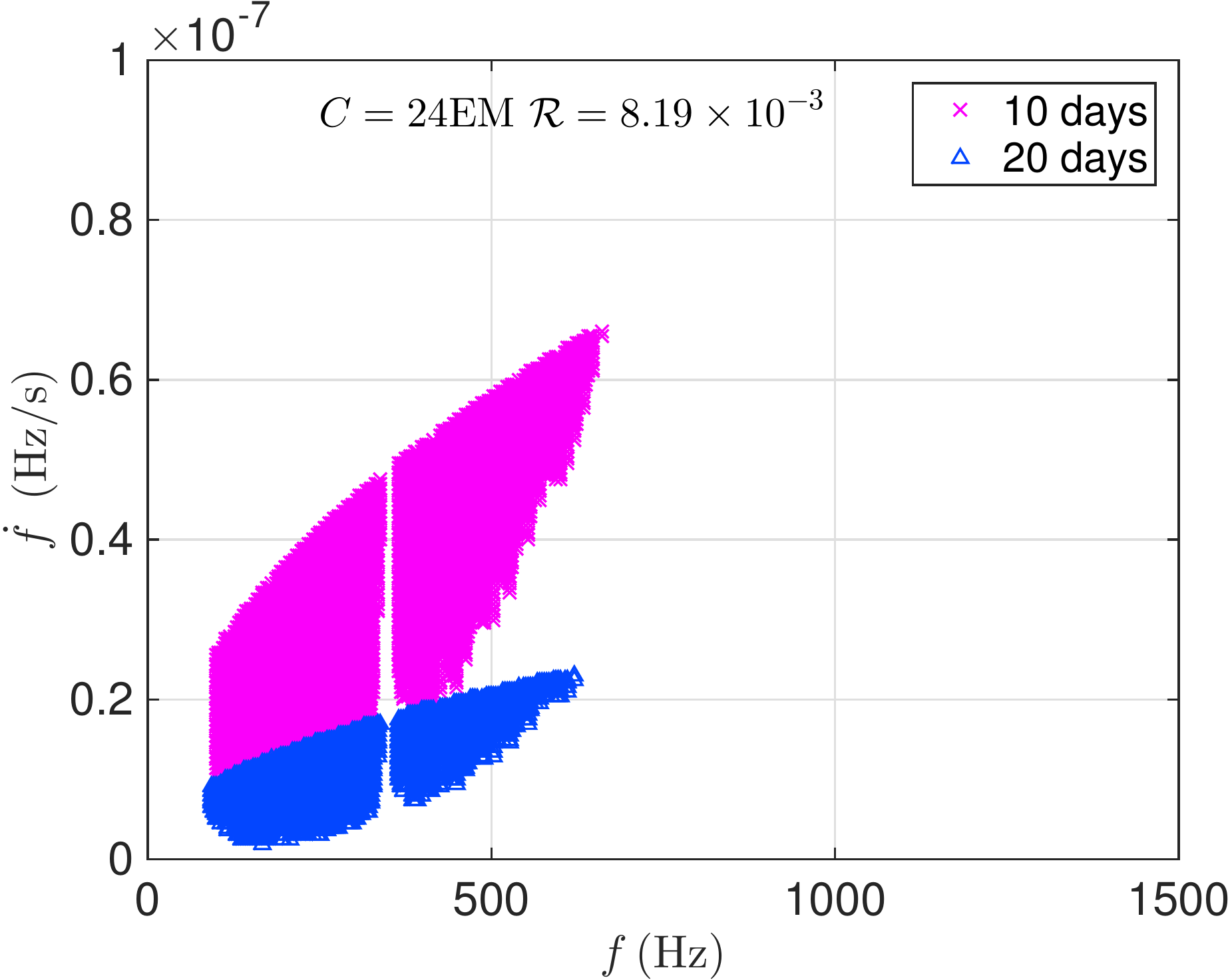}}}%
    \caption{Parameter space coverage for G347.3 at 1300 pc, assuming uniform and distance-based priors and optimizing over the 7  search set-ups also considered above at 12 EM (left plot) and 24 EM (right plot).}%
    \label{G3473_best_noage}%
\end{figure*}

\begin{figure*}%
    \centering
    \subfloat[Coverage, cost: 12 EM]{{  \includegraphics[width=.45\linewidth]{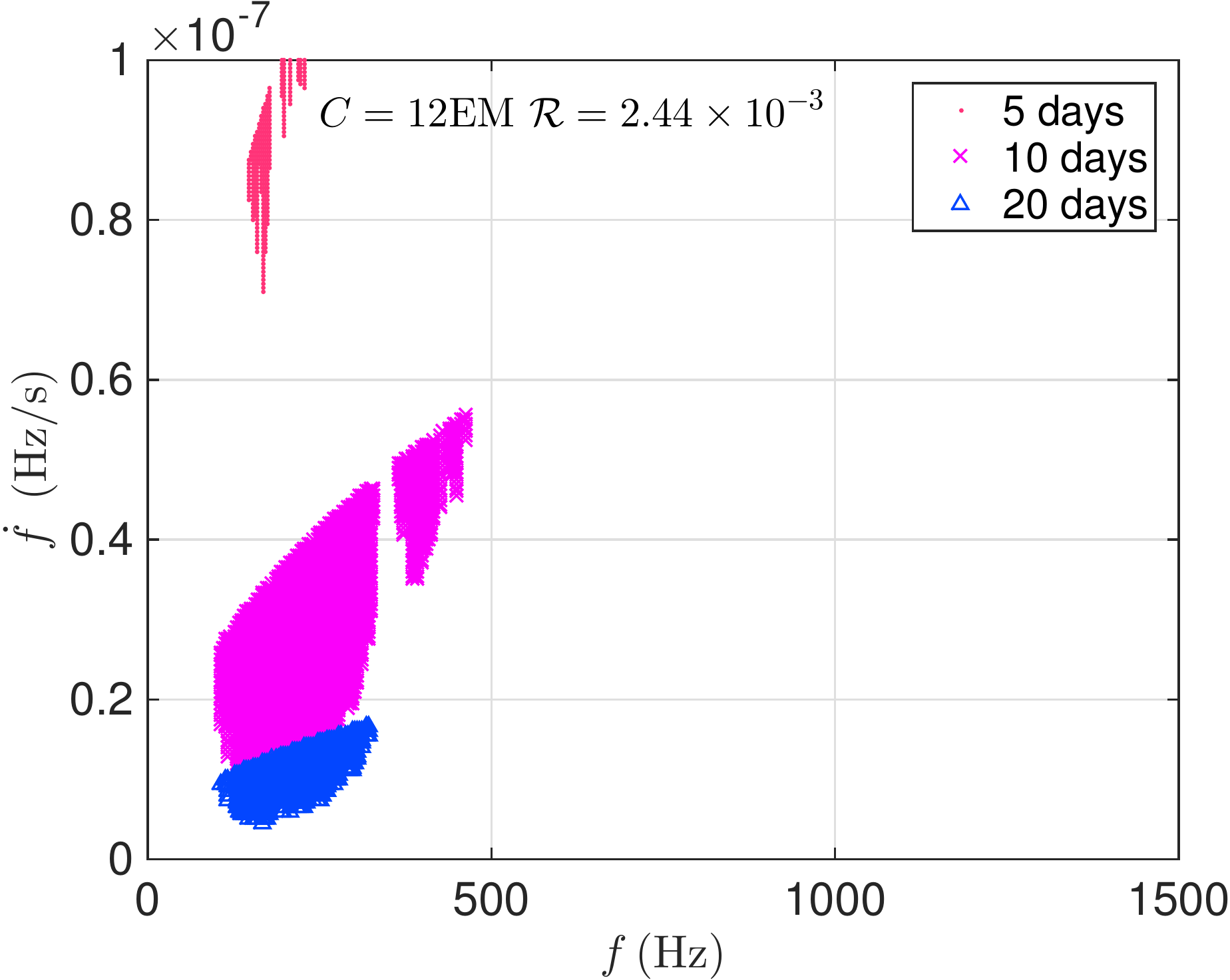}}}%
    \qquad
    \subfloat[Coverage,  cost: 24 EM]{{  \includegraphics[width=.45\linewidth]{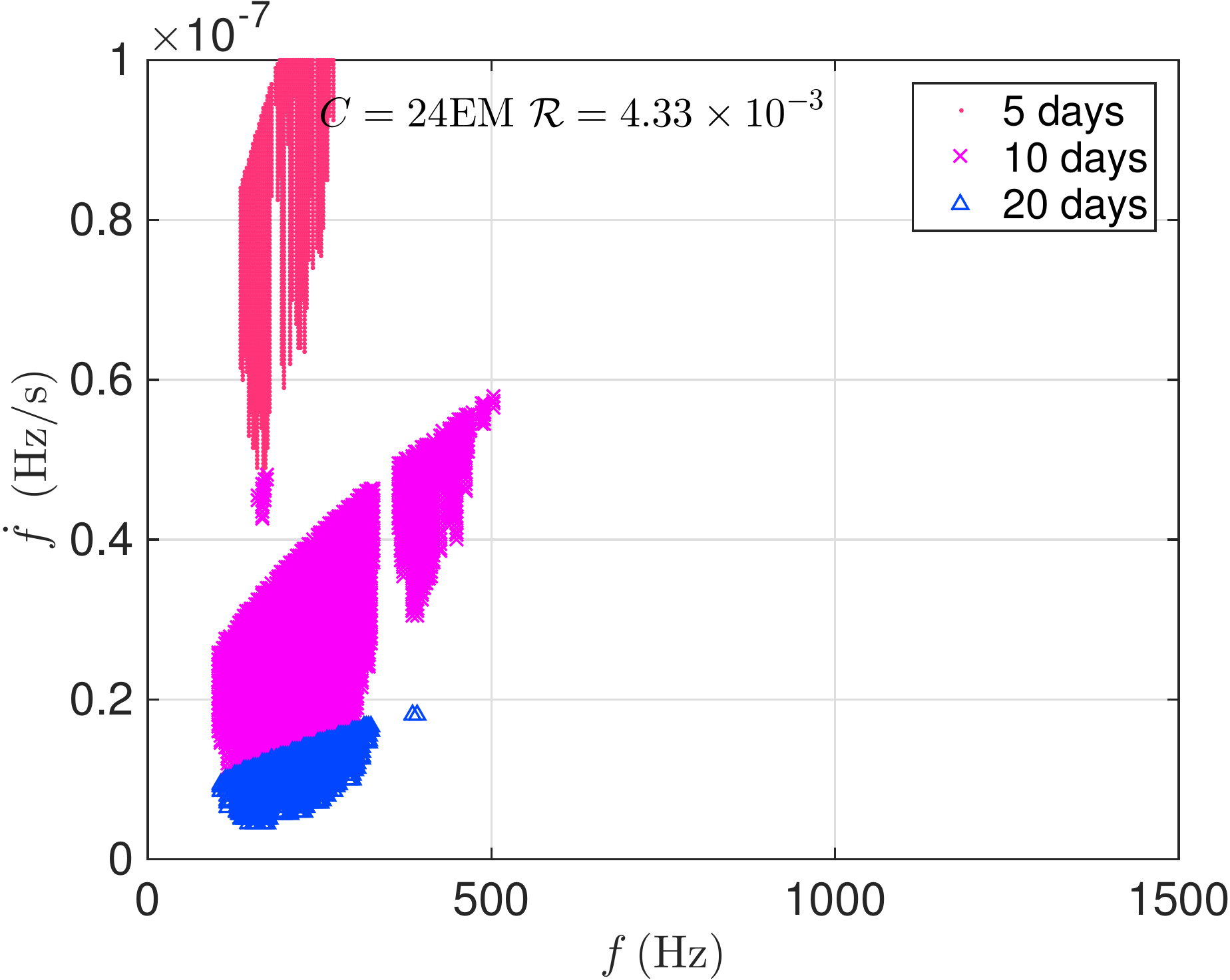}}}%
    \caption{Parameter space coverage for Cas A at 3500 pc, assuming uniform and distance-based priors and optimizing over the 7  search set-ups also considered above at 12 EM (left plot) and 24 EM (right plot).}%
    \label{CasA_best_noage}%
\end{figure*}

\begin{figure*}%
    \centering
    \subfloat[Coverage of 3 sources, Vela Jr at 200 pc, computing budget 12EM ]{{  \includegraphics[width=.45\linewidth]{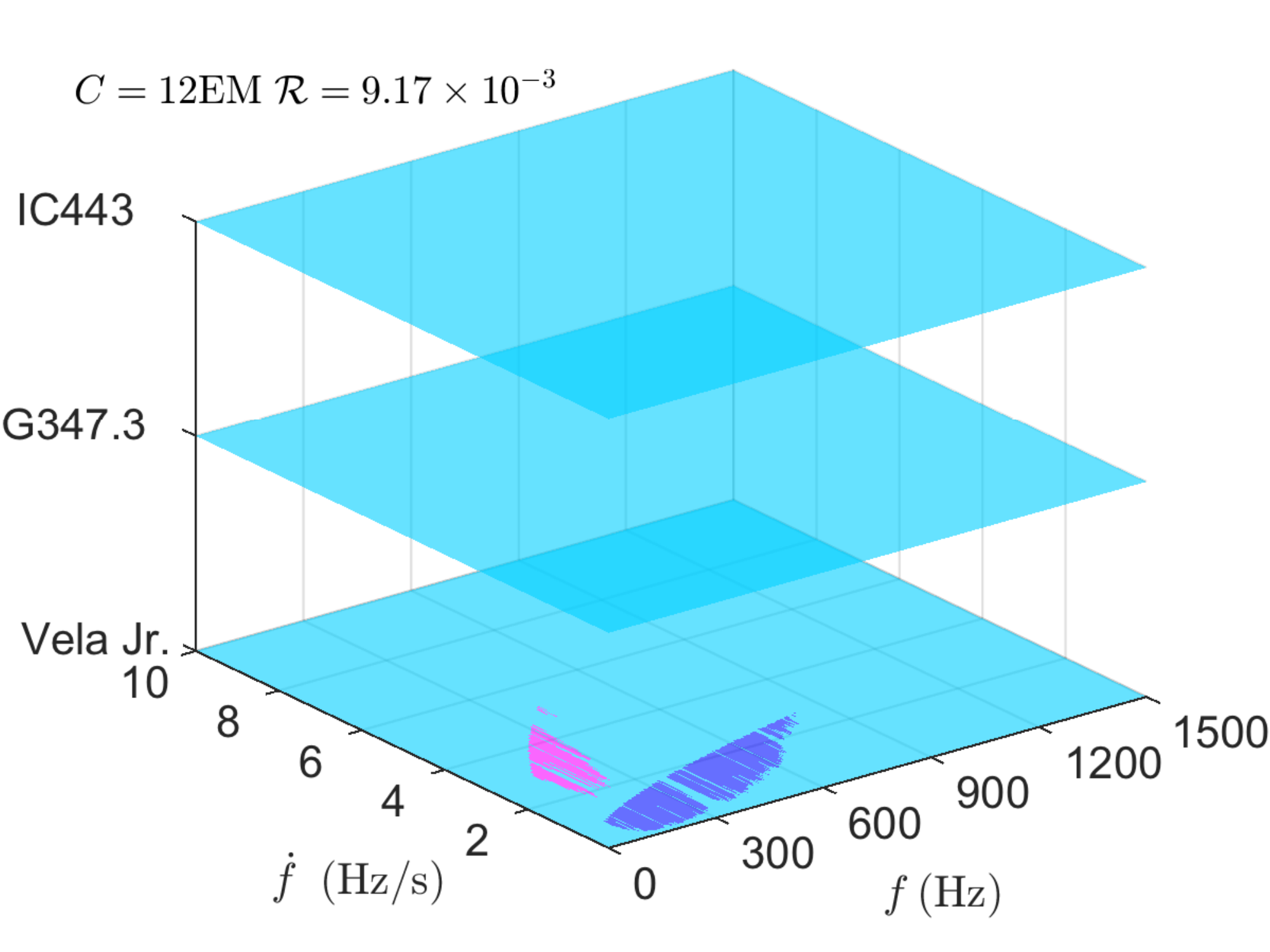}}}%
    \qquad
    \subfloat[Coverage of 3 sources, Vela Jr at 750 pc, computing budget 12EM ]{{  \includegraphics[width=.45\linewidth]{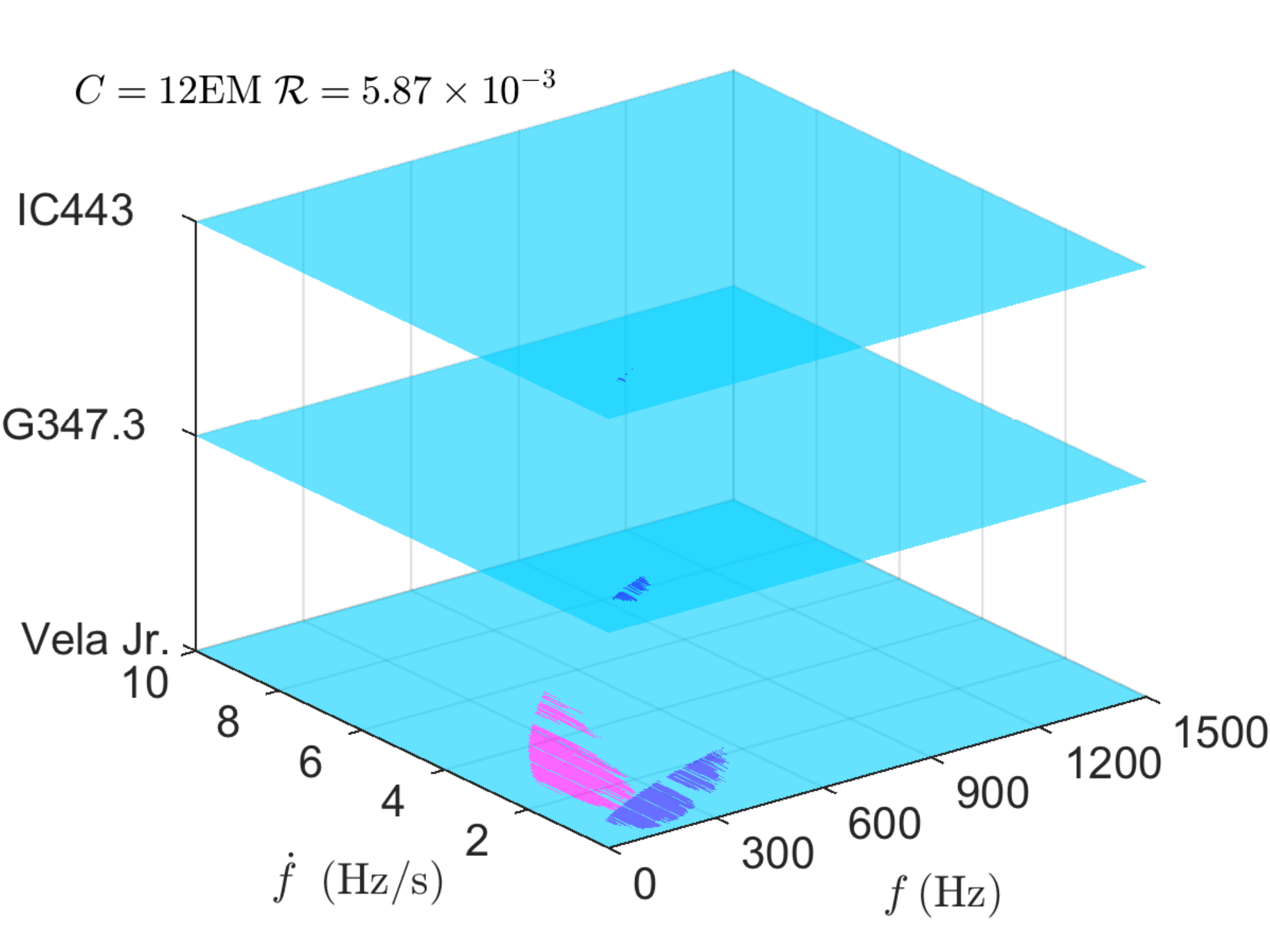}}}%
    \qquad
    \subfloat[Coverage of 3 sources, Vela Jr at 200 pc, computing budget 24 EM ]{{  \includegraphics[width=.45\linewidth]{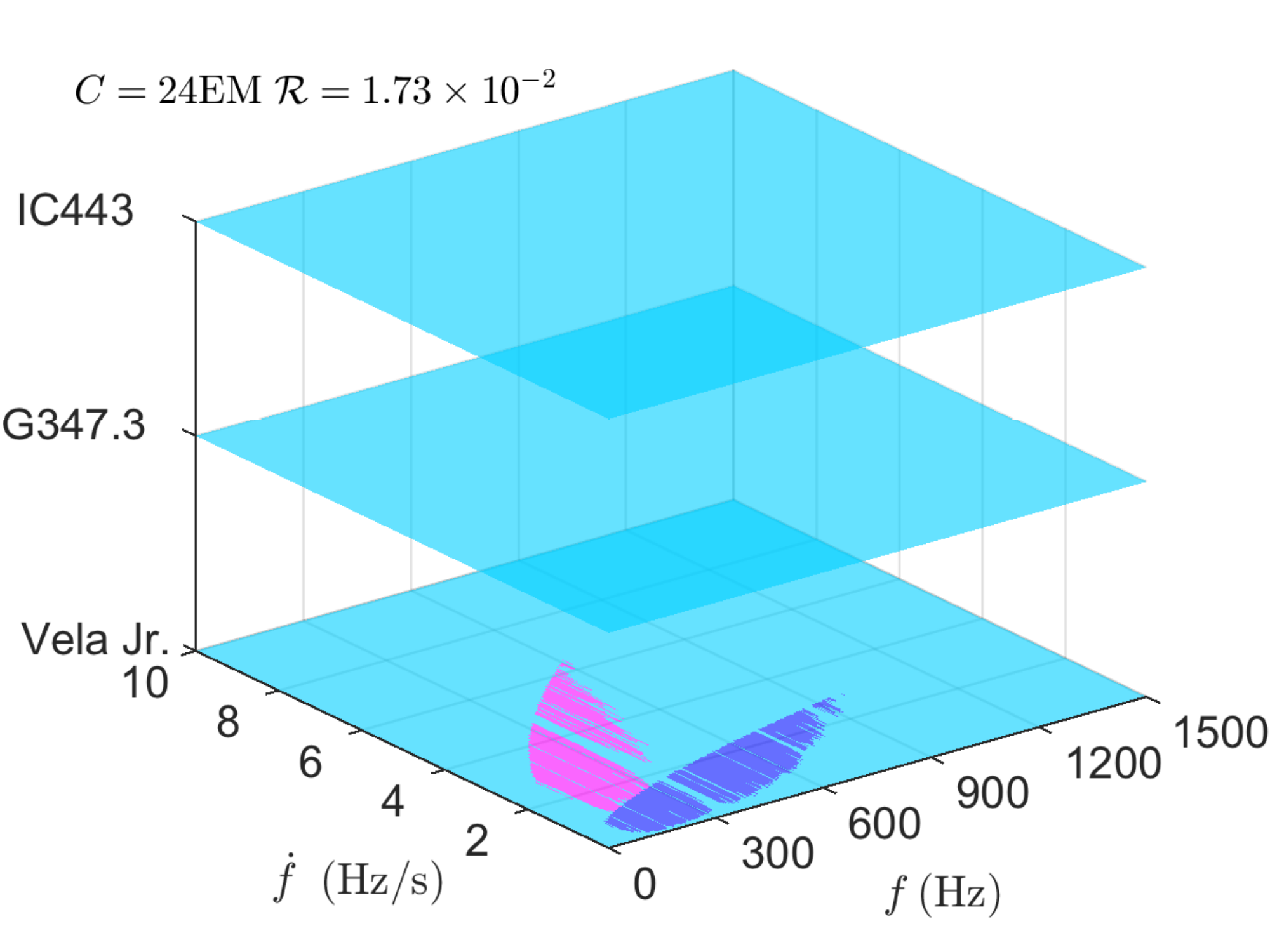}}}%
    \qquad
    \subfloat[Coverage of 3 sources, Vela Jr at 750 pc, computing budget 24 EM ]{{  \includegraphics[width=.45\linewidth]{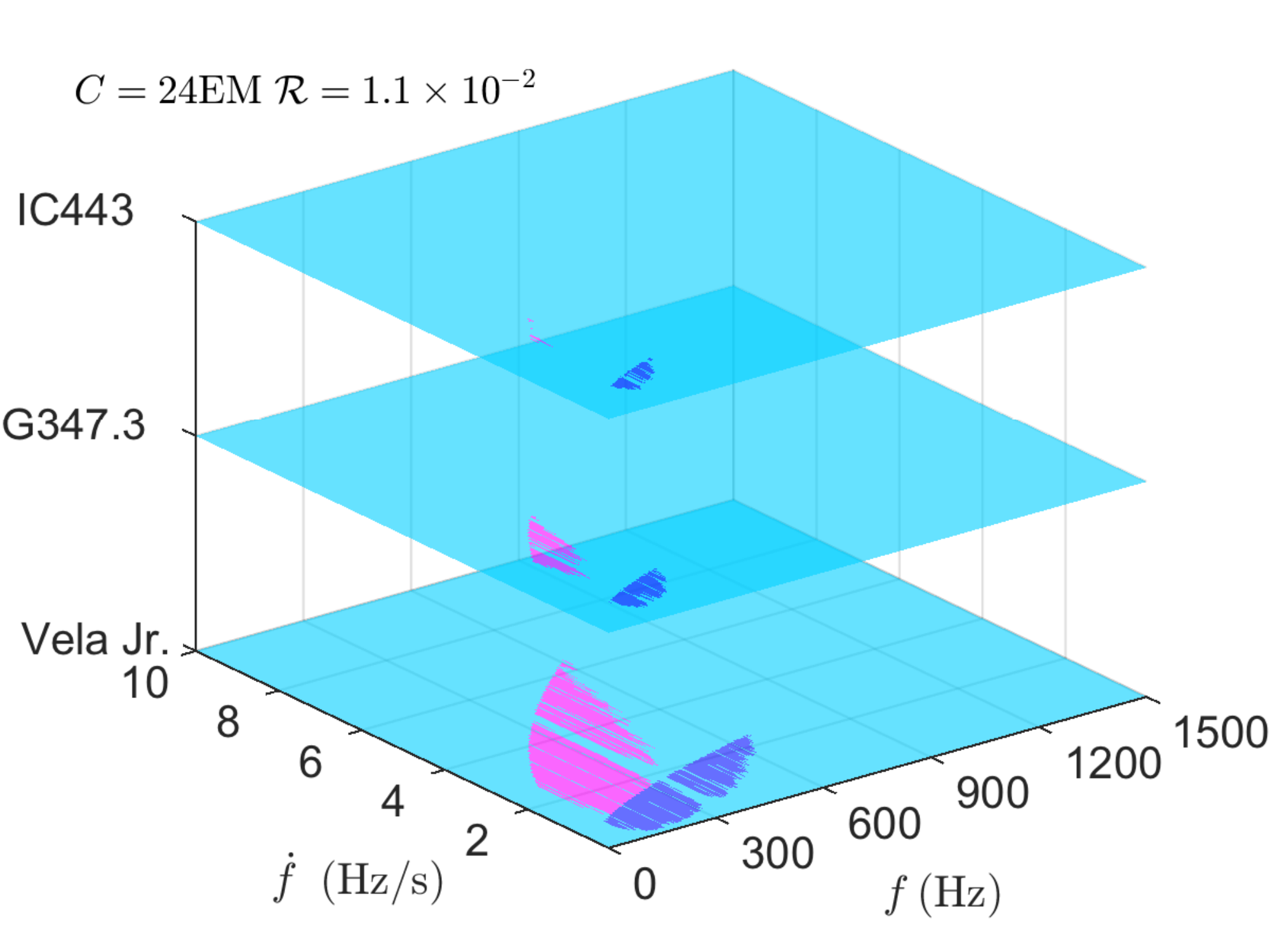}}}%
    \qquad
    \subfloat[Coverage of 3 sources, Vela Jr at 200 pc, computing budget 48 EM ]{{  \includegraphics[width=.45\linewidth]{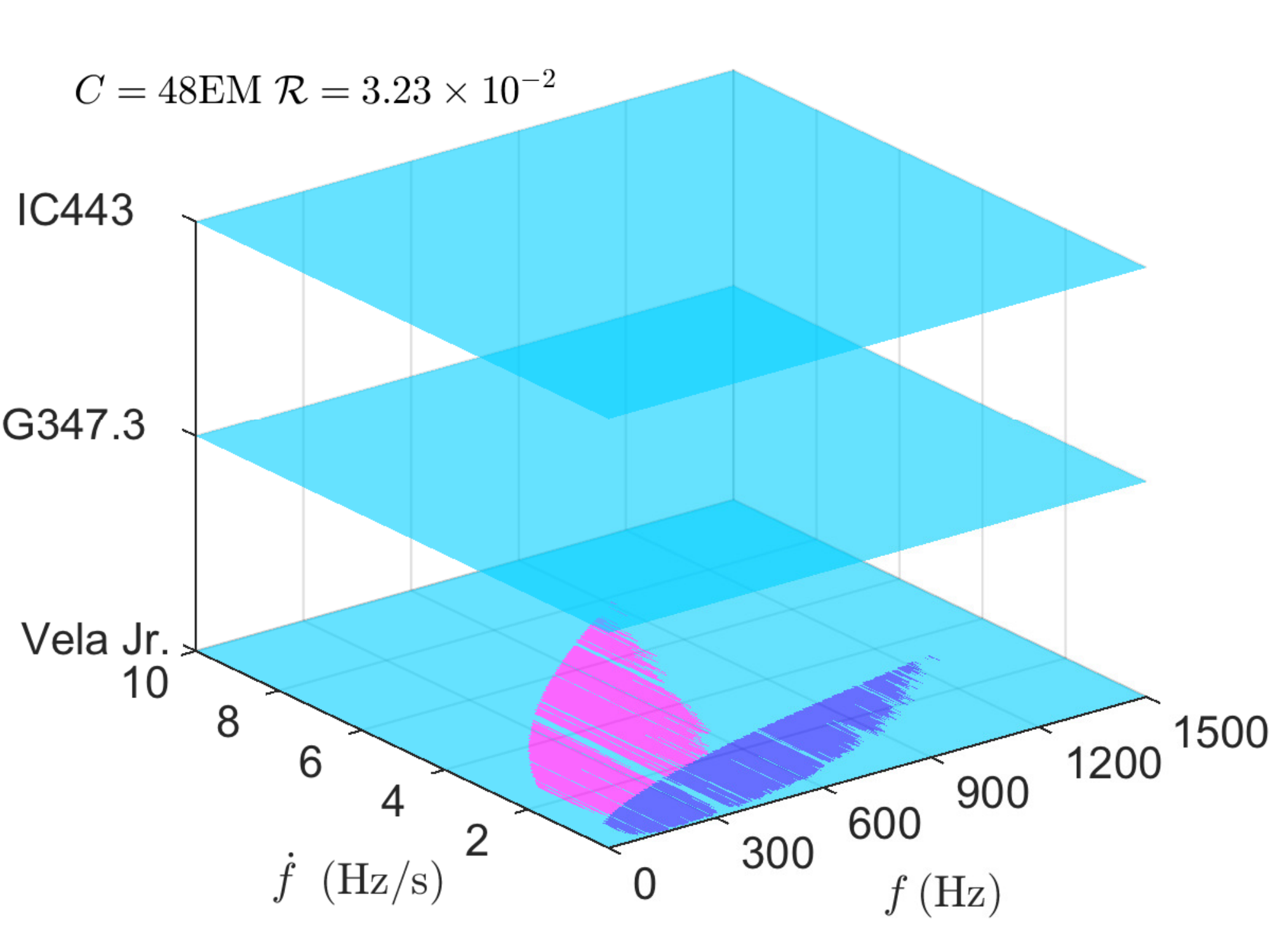}}}%
    \qquad
    \subfloat[Coverage of 3 sources, Vela Jr at 750 pc, computing budget 48 EM ]{{  \includegraphics[width=.45\linewidth]{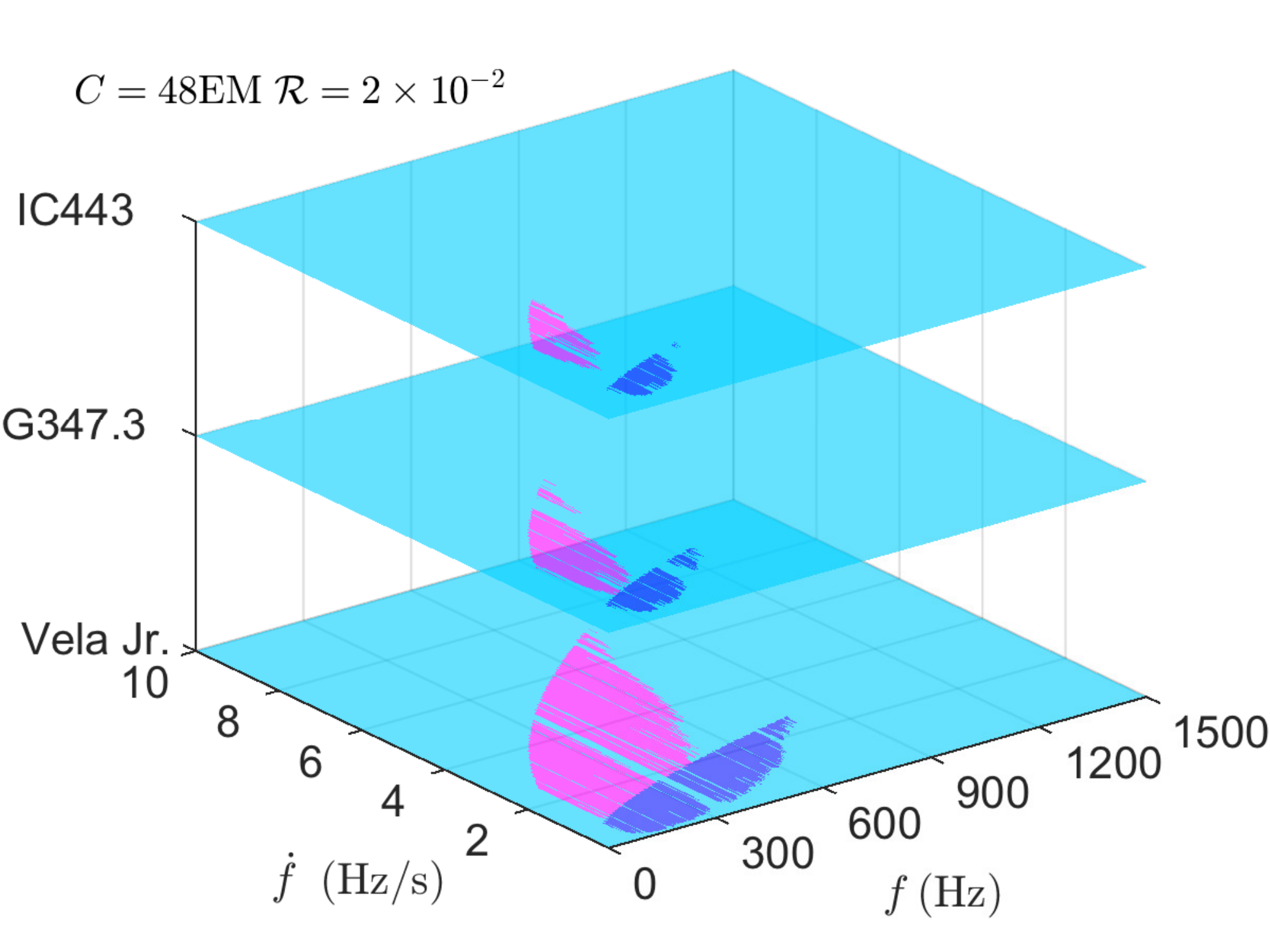}}}%
    
    \caption{Parameter space coverage assuming uniform and distance-based priors and optimizing over the 7 search set-ups also considered above and over the three closest targets (left plots: Vela Jr at 200 pc, G347.3, IC443 and right plots: Vela Jr at 750 pc, G347.3, IC443) at 12, 24 and 48 EMs.}%
    \label{all_noage}%
\end{figure*}


\begin{figure*}%
    \centering
    \subfloat[Efficiency, 5 days]{{  \includegraphics[width=.20\linewidth]{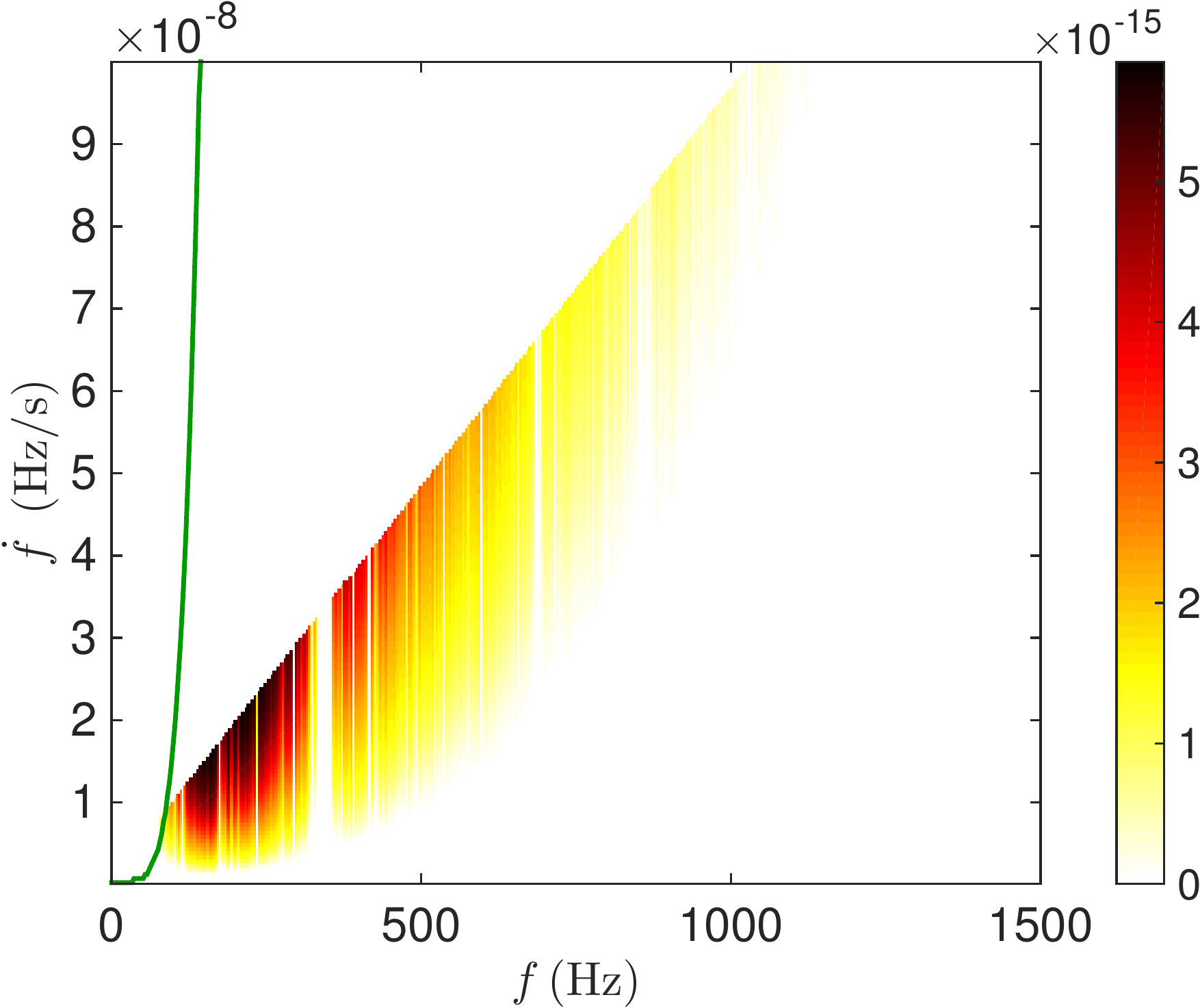}}}%
    \qquad
    \subfloat[Coverage, 5 days]{{  \includegraphics[width=.20\linewidth]{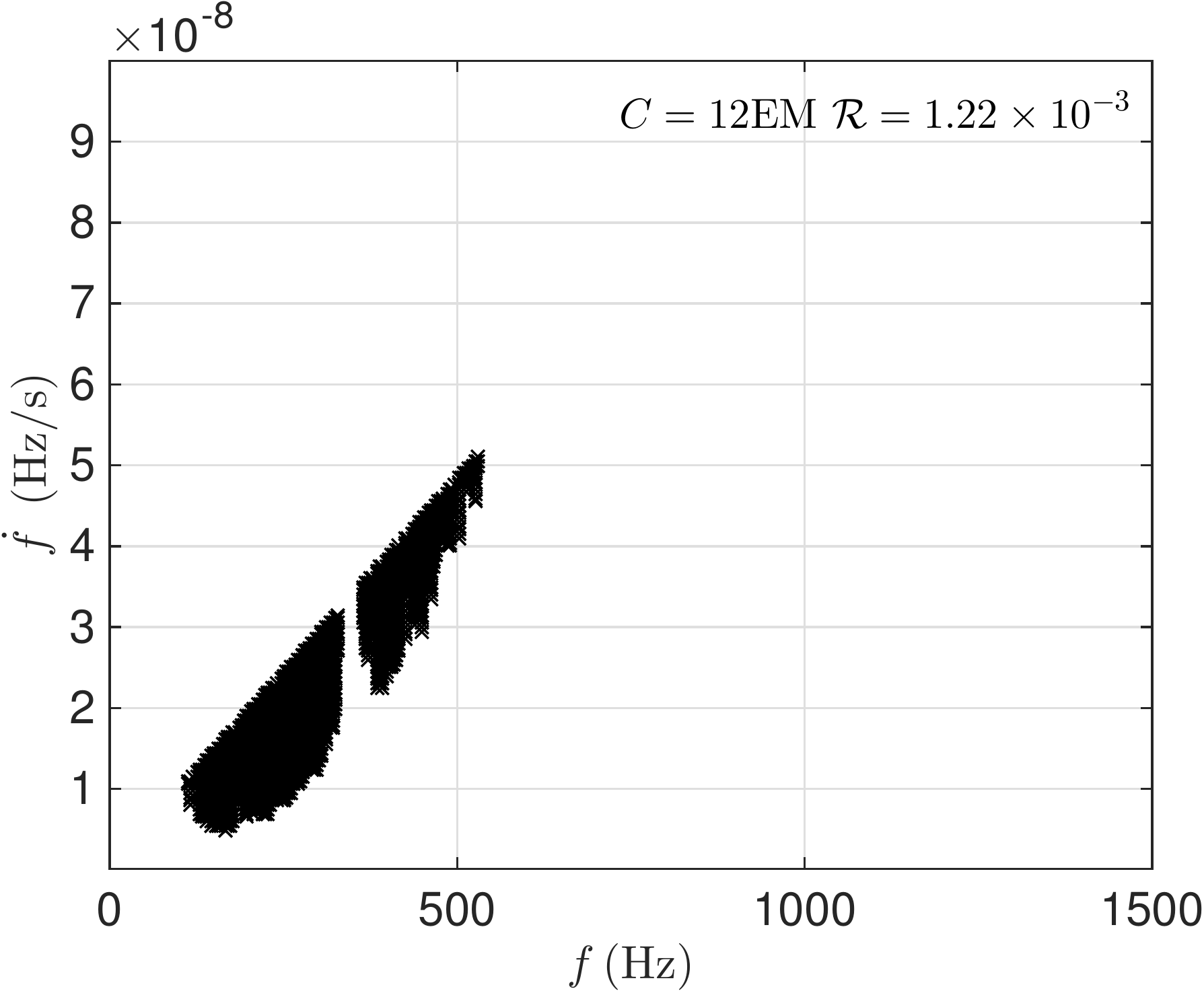}}}%
    \qquad
    \subfloat[Efficiency, 10 days]{{  \includegraphics[width=.20\linewidth]{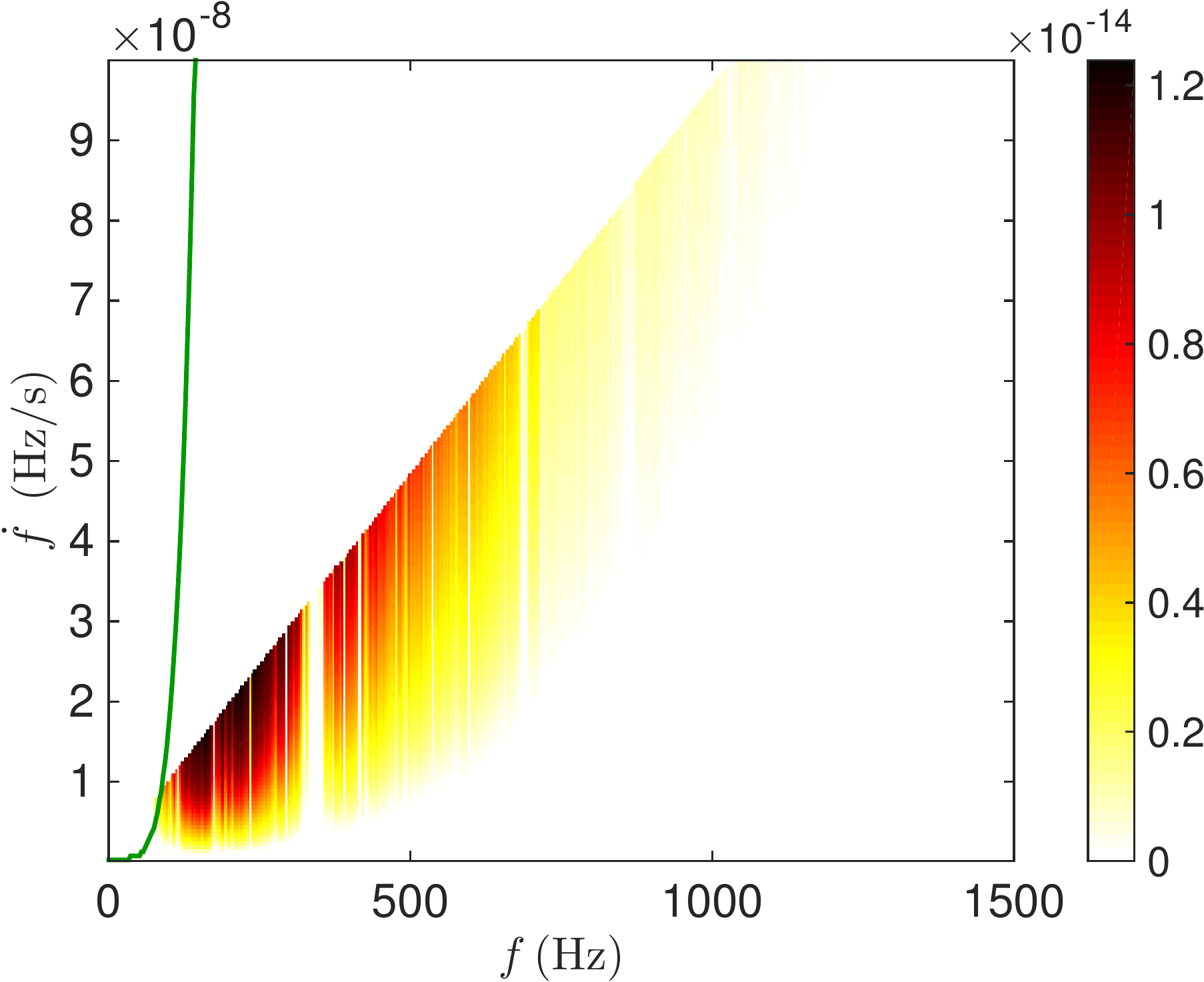}}}%
    \qquad
    \subfloat[Coverage, 10 days]{{  \includegraphics[width=.20\linewidth]{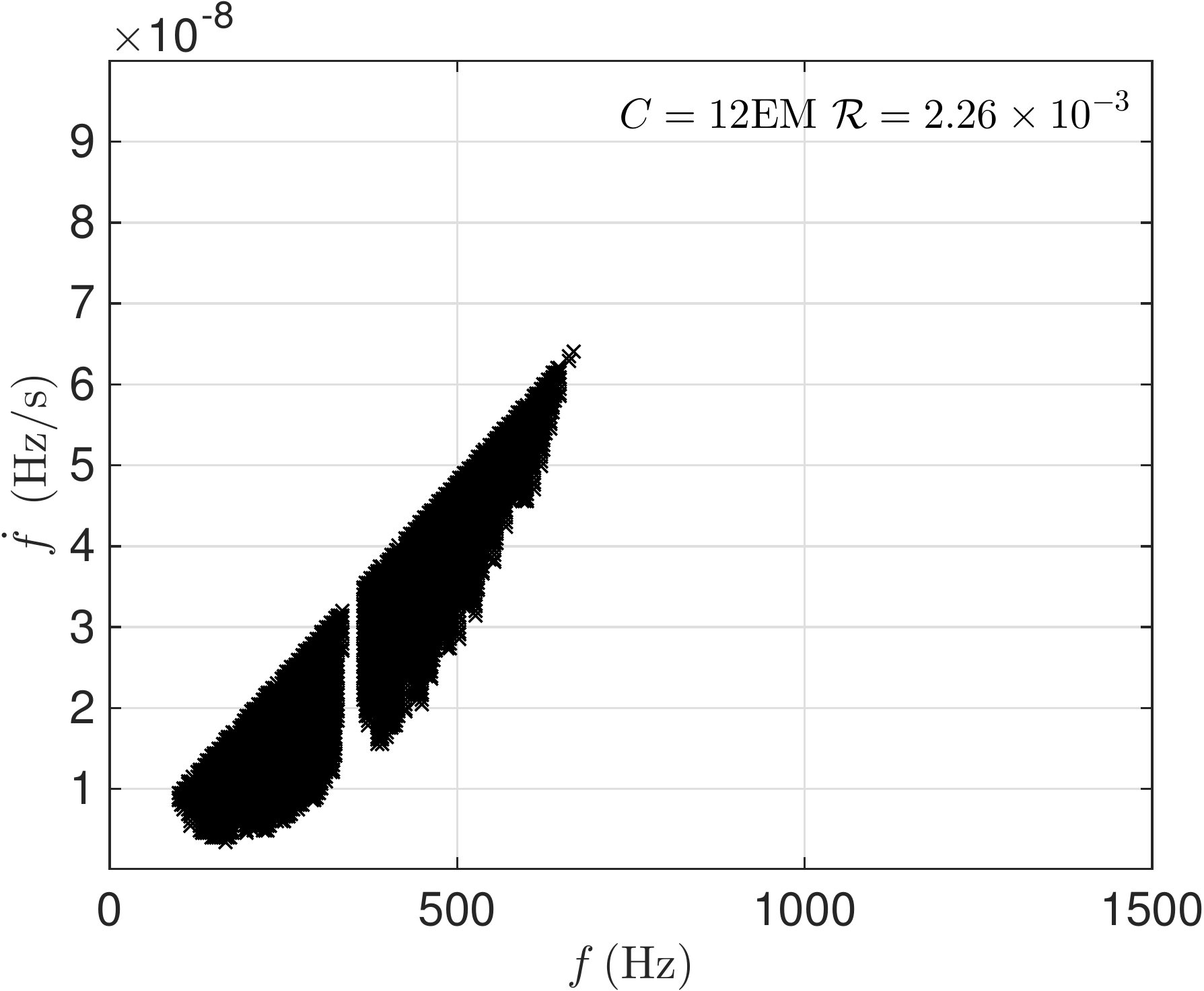}}}%
    \qquad
    \subfloat[Efficiency, 20 days]{{  \includegraphics[width=.20\linewidth]{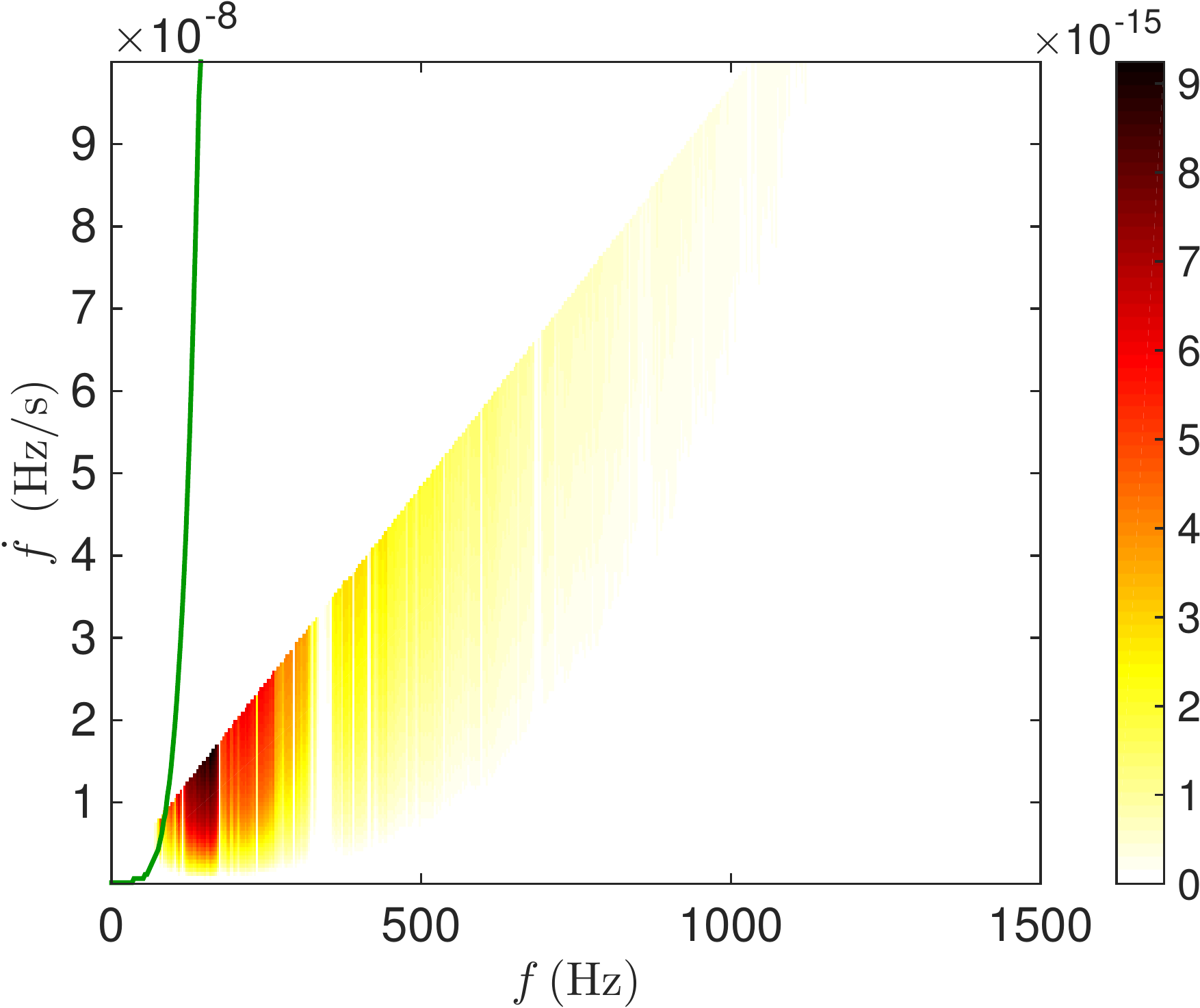}}}%
    \qquad
    \subfloat[Coverage, 20 days]{{  \includegraphics[width=.20\linewidth]{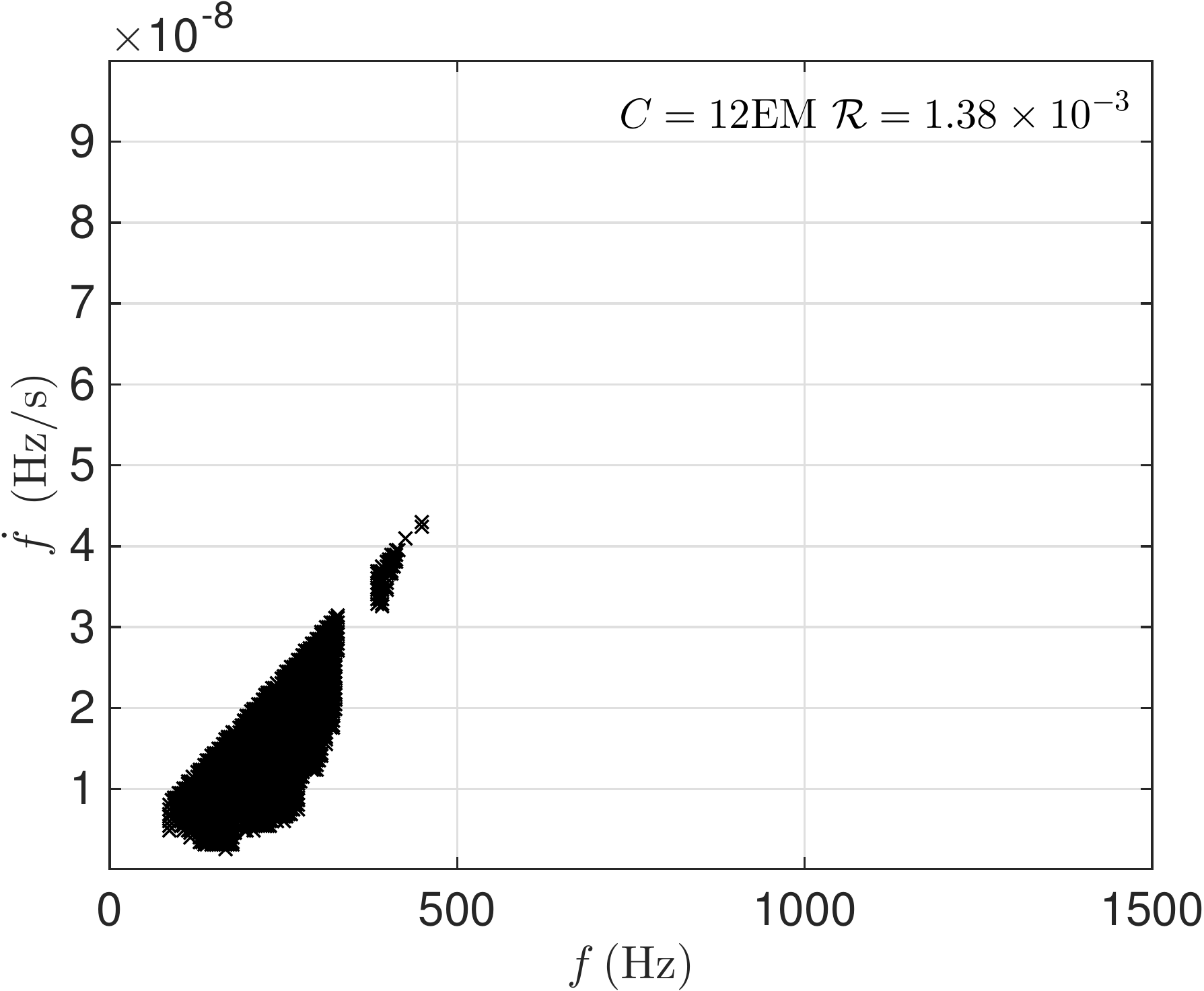}}}%
    \qquad
    \subfloat[Efficiency, 30 days]{{  \includegraphics[width=.20\linewidth]{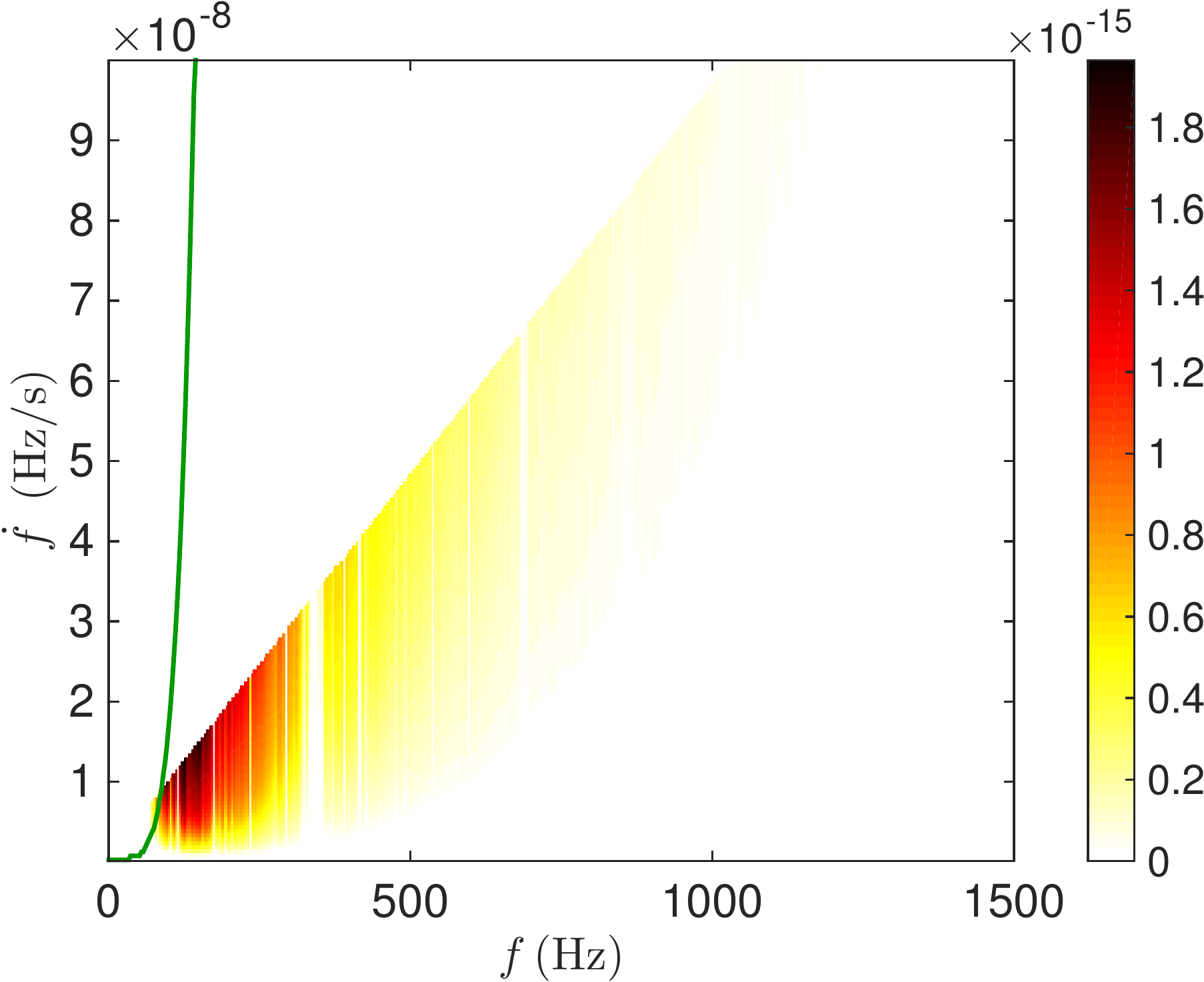}}}%
    \qquad
    \subfloat[Coverage, 30 days]{{  \includegraphics[width=.20\linewidth]{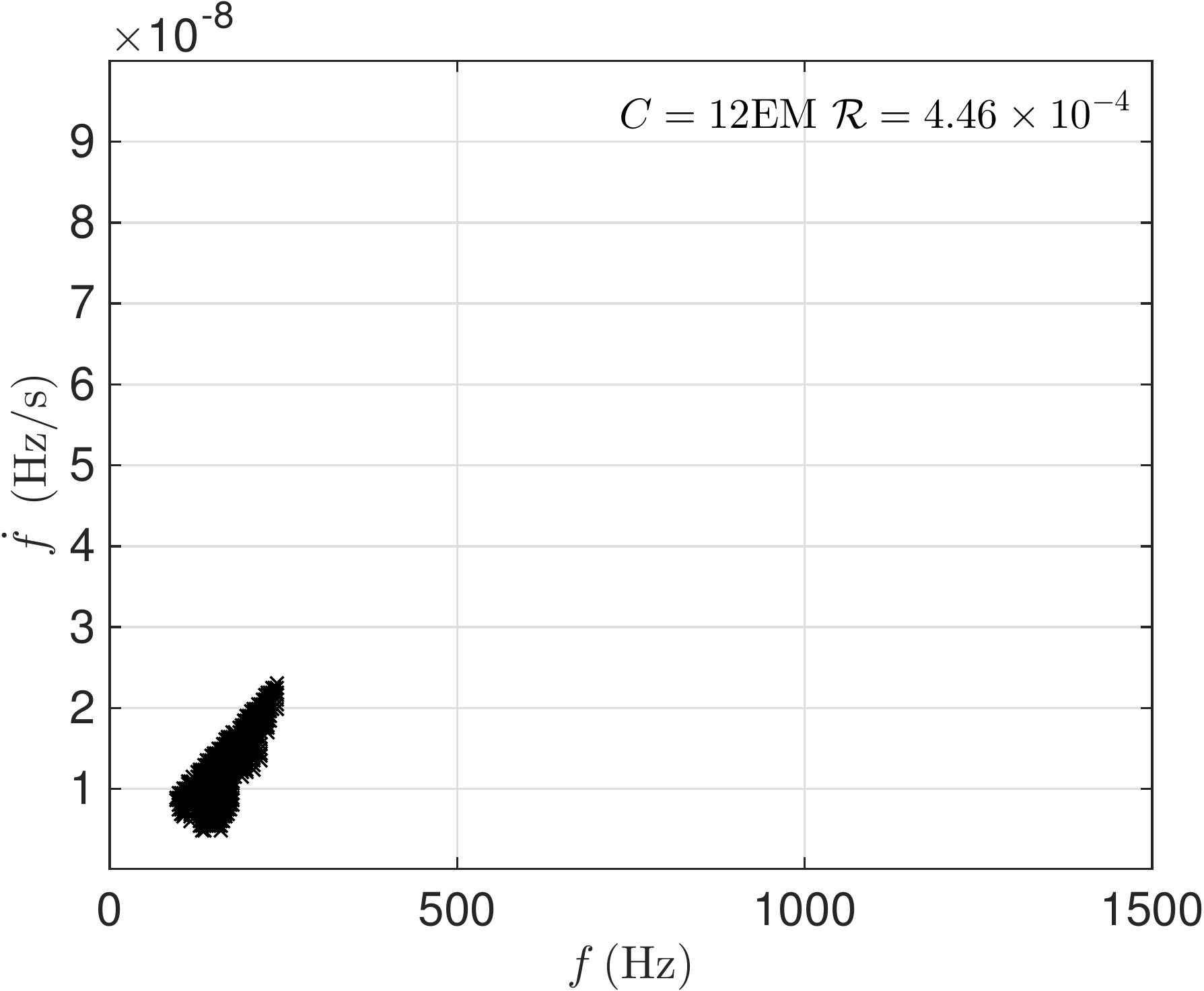}}}%
    \qquad
    \subfloat[Efficiency, 37.5 days]{{  \includegraphics[width=.20\linewidth]{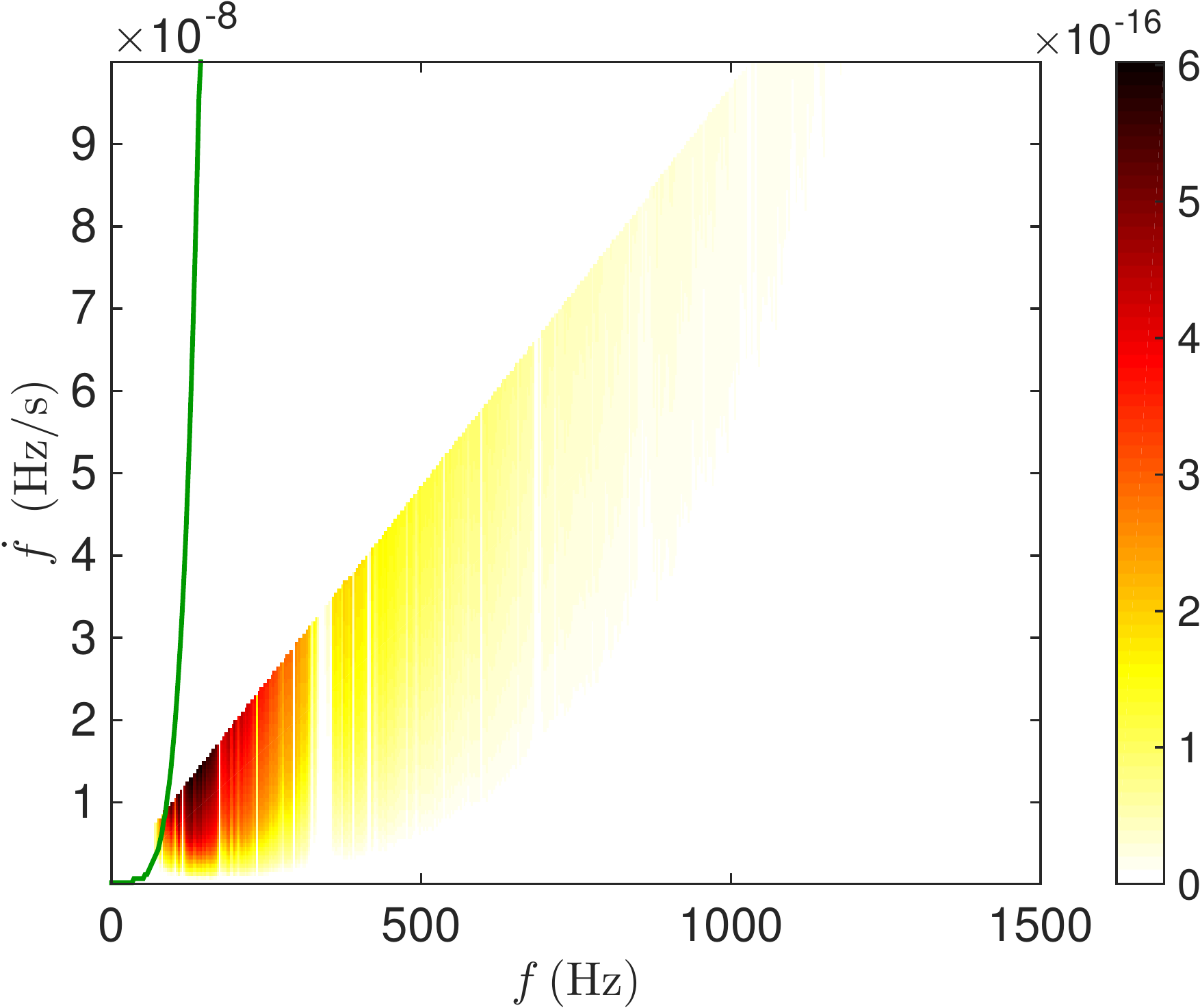}}}%
    \qquad
    \subfloat[Coverage, 37.5 days]{{  \includegraphics[width=.20\linewidth]{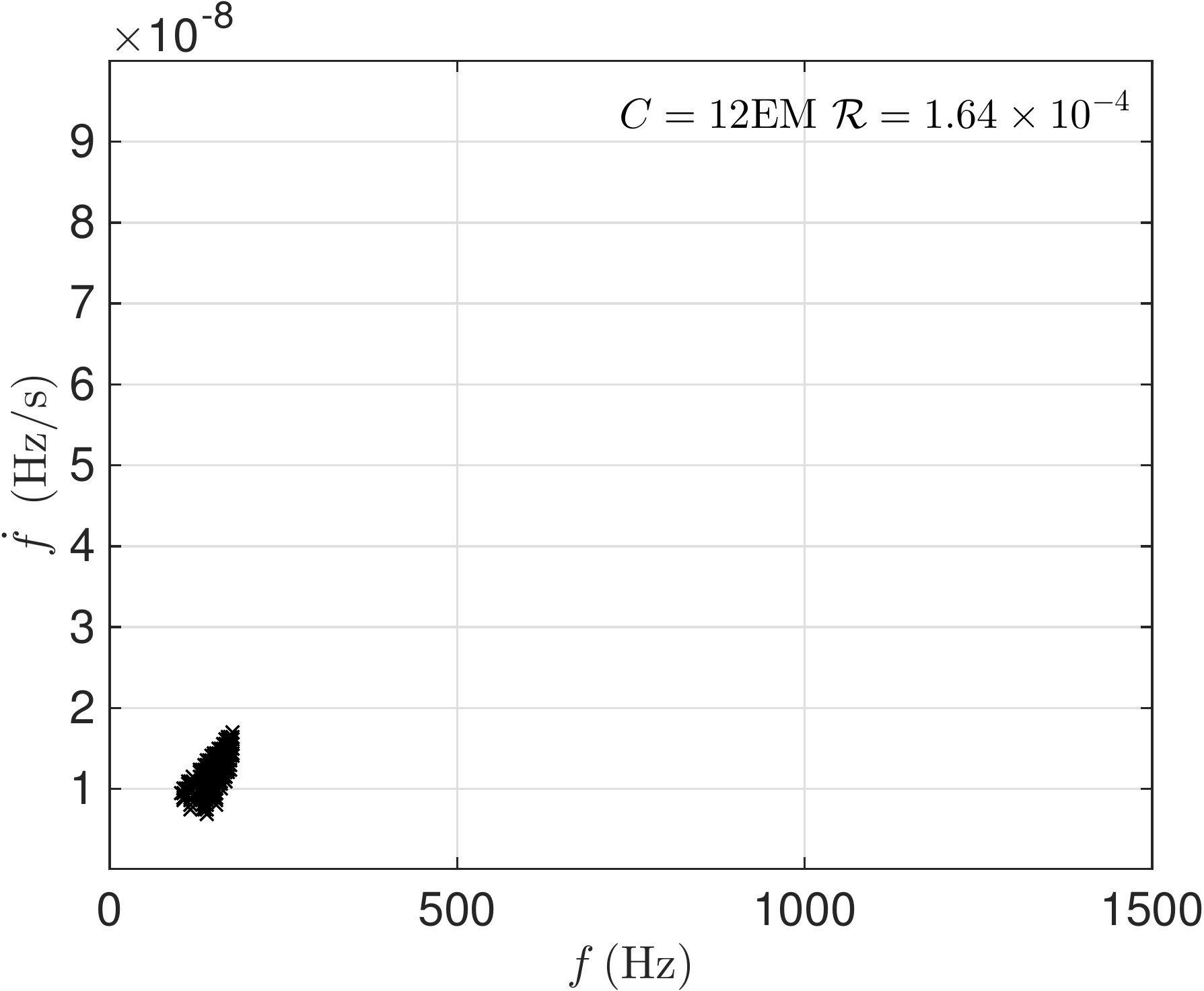}}}%
    \qquad
    \subfloat[Efficiency, 50 days]{{  \includegraphics[width=.20\linewidth]{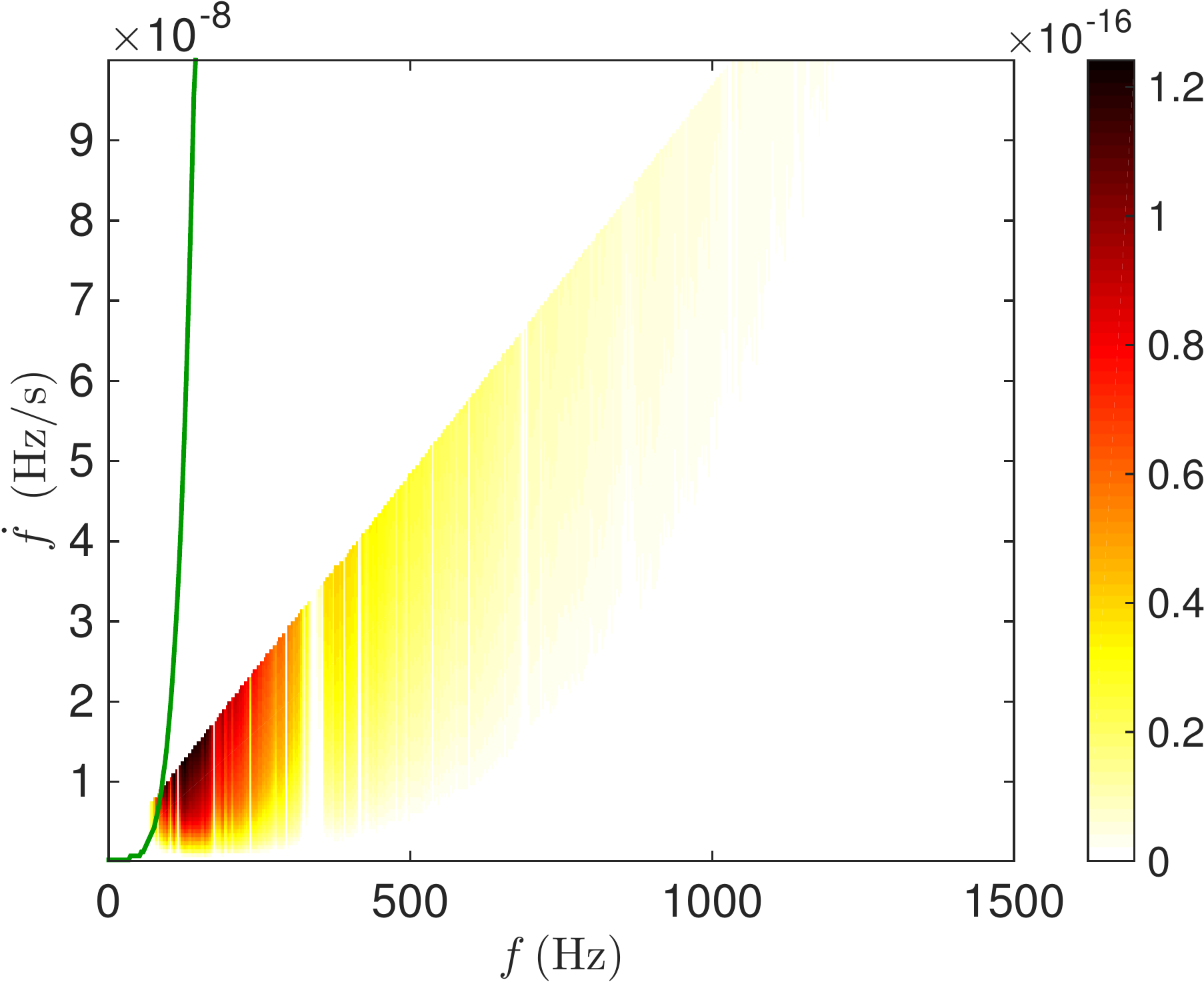}}}%
    \qquad
    \subfloat[Coverage, 50 days]{{  \includegraphics[width=.20\linewidth]{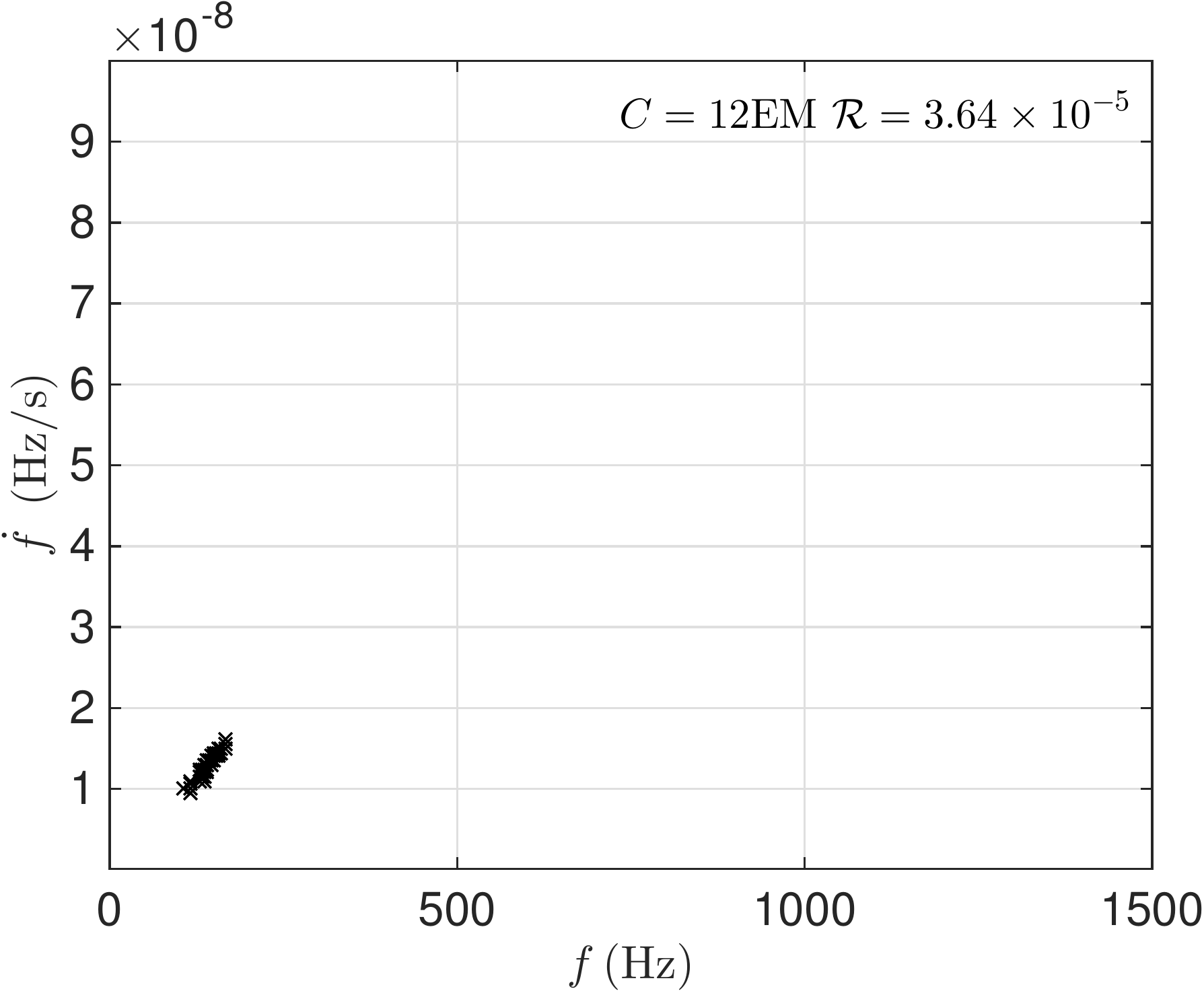}}}%
    \qquad
    \subfloat[Efficiency, 75 days]{{  \includegraphics[width=.20\linewidth]{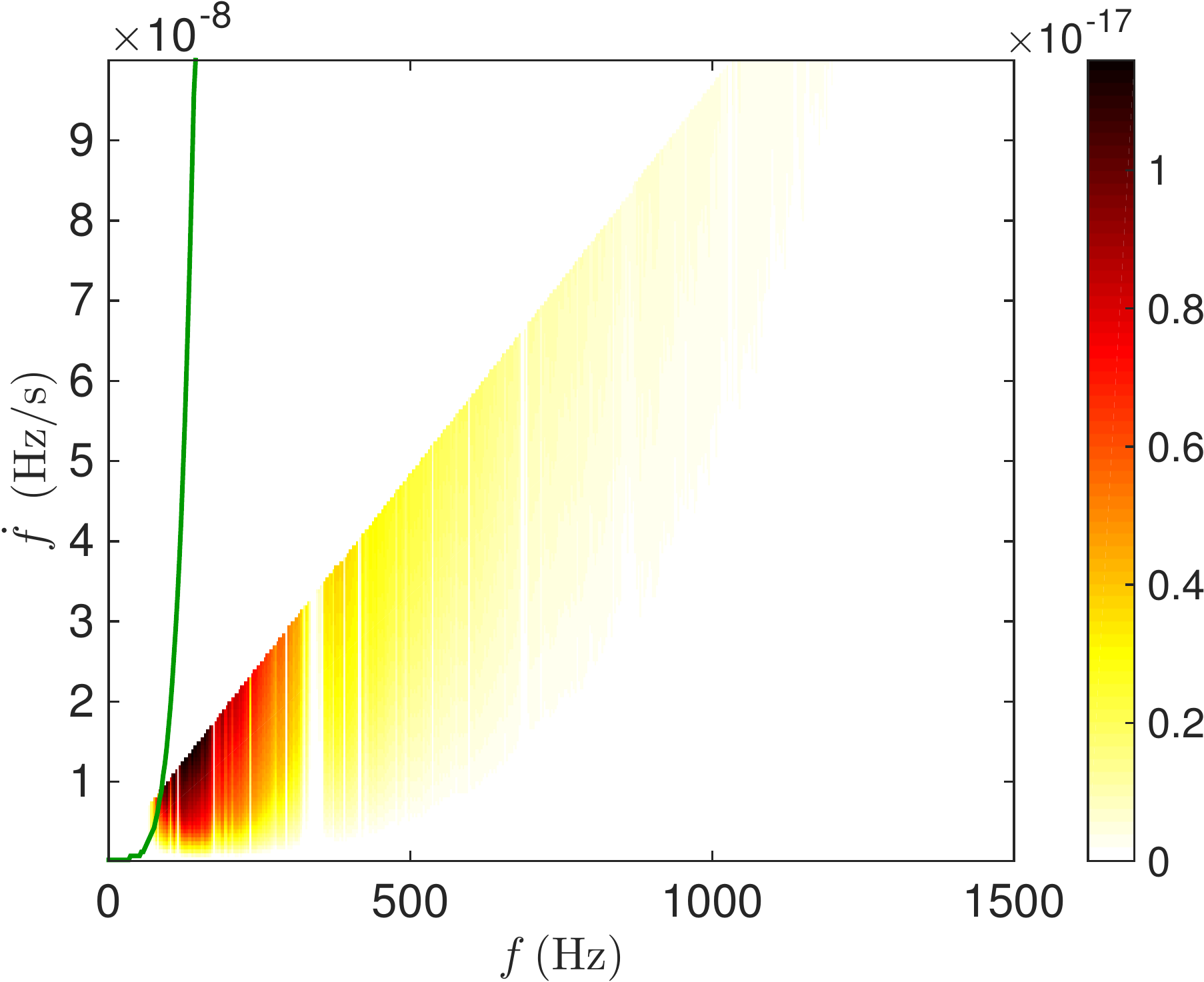}}}%
    \qquad
    \subfloat[Coverage, 75 days]{{  \includegraphics[width=.20\linewidth]{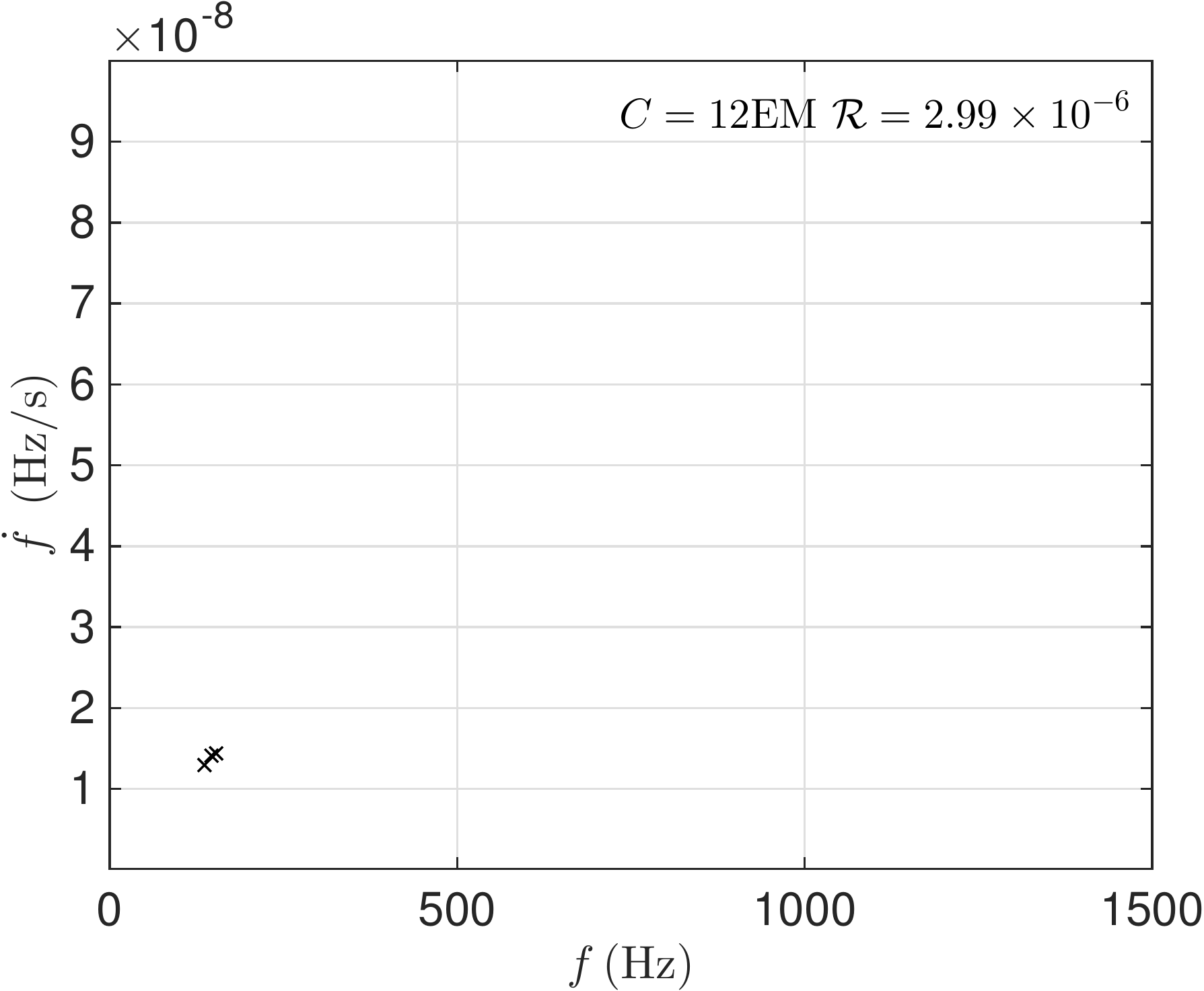}}}%
    
      \caption{Optimisation results for Cas A at 3500 pc, 330 years old, assuming uniform and age-based priors, for various coherent search durations: 5, 10, 20, 30, 37.5, 50 and 75 days. The total computing budget is assumed to be 12 EM. }%
    \label{CasA_51020days_age}%
\end{figure*}
%

\begin{figure*}%
    \centering
    \subfloat[Efficiency, 5 days]{{  \includegraphics[width=.20\linewidth]{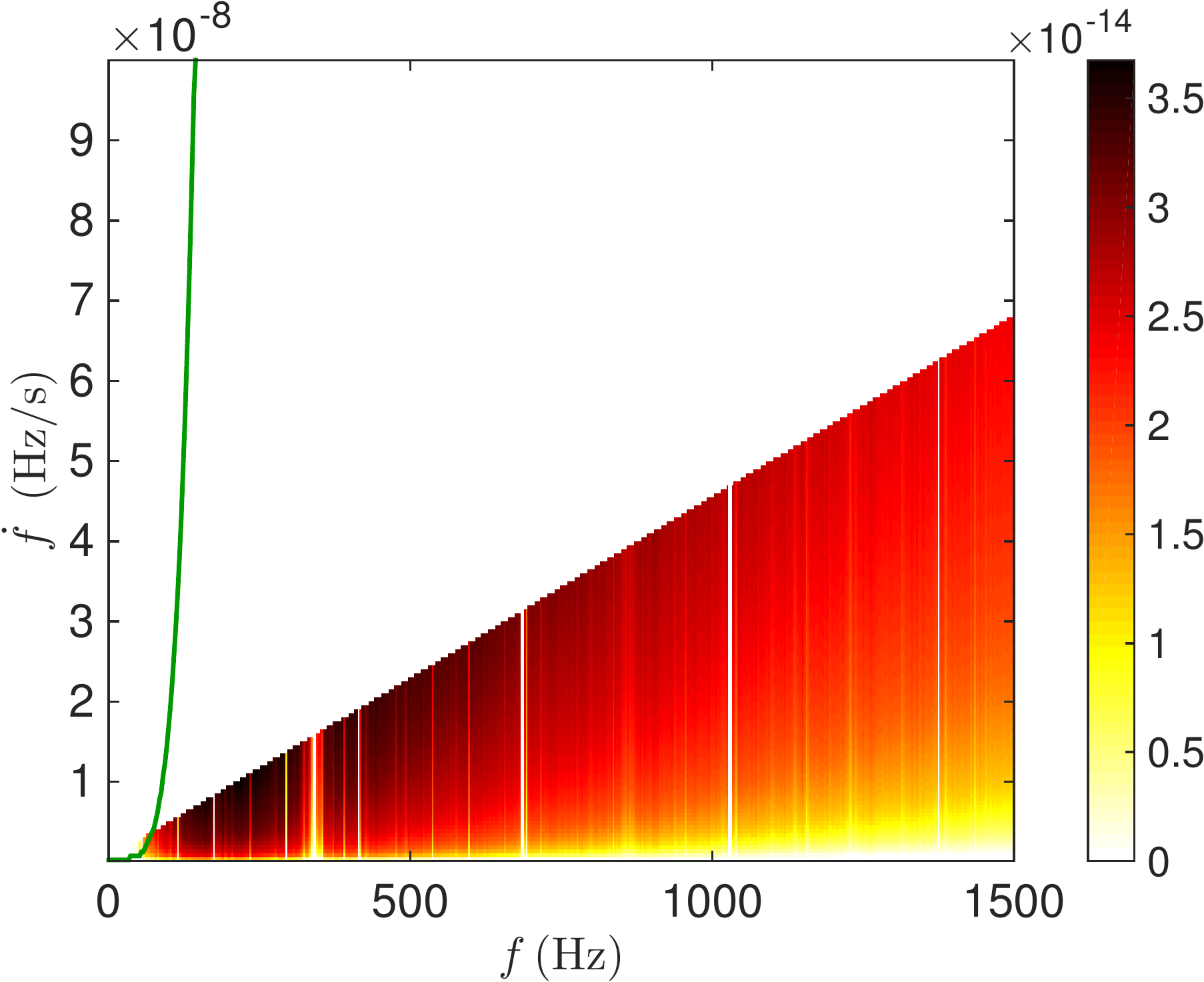}}}%
    \qquad
    \subfloat[Coverage, 5 days]{{  \includegraphics[width=.20\linewidth]{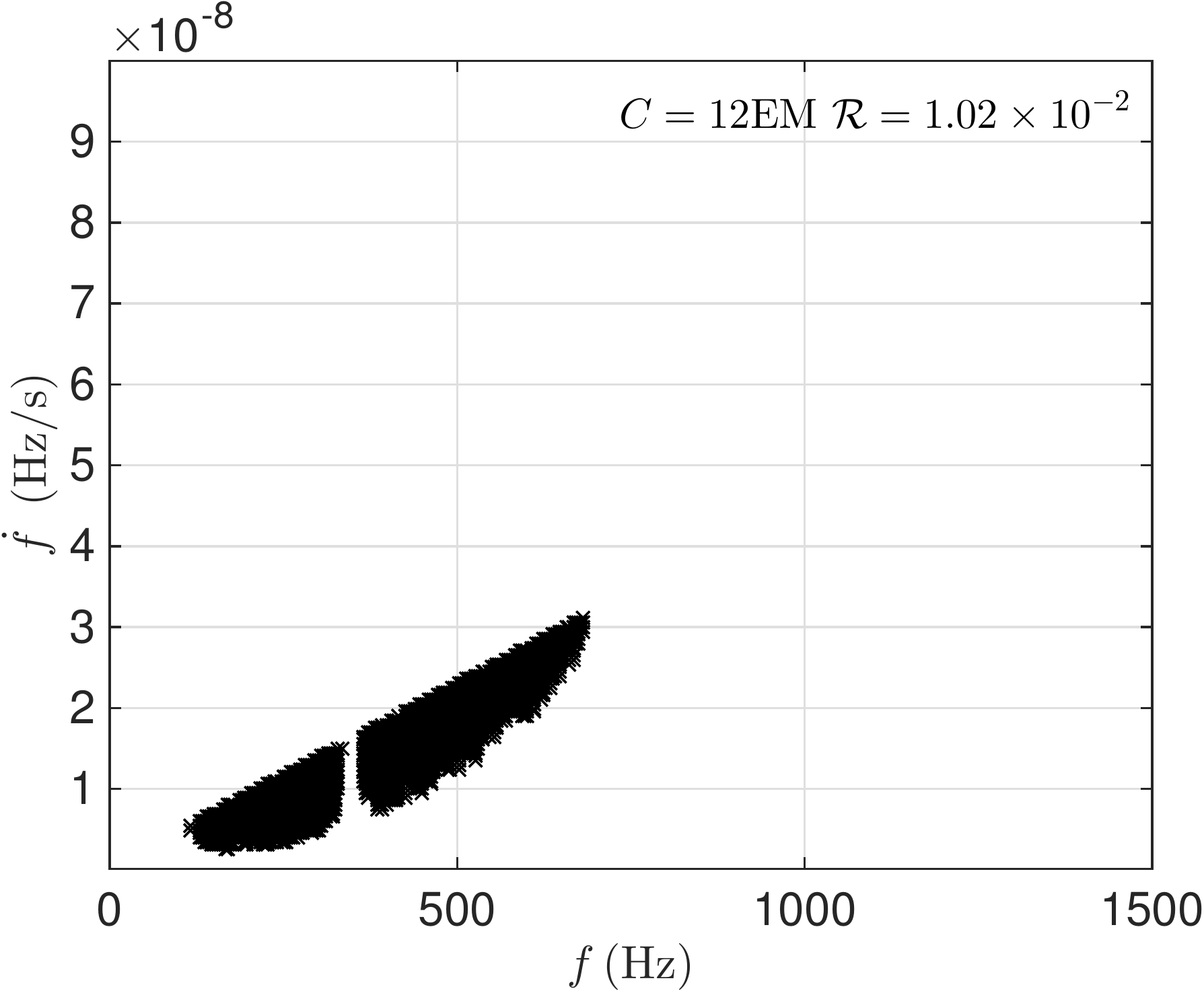}}}%
    \qquad
    \subfloat[Efficiency, 10 days]{{  \includegraphics[width=.20\linewidth]{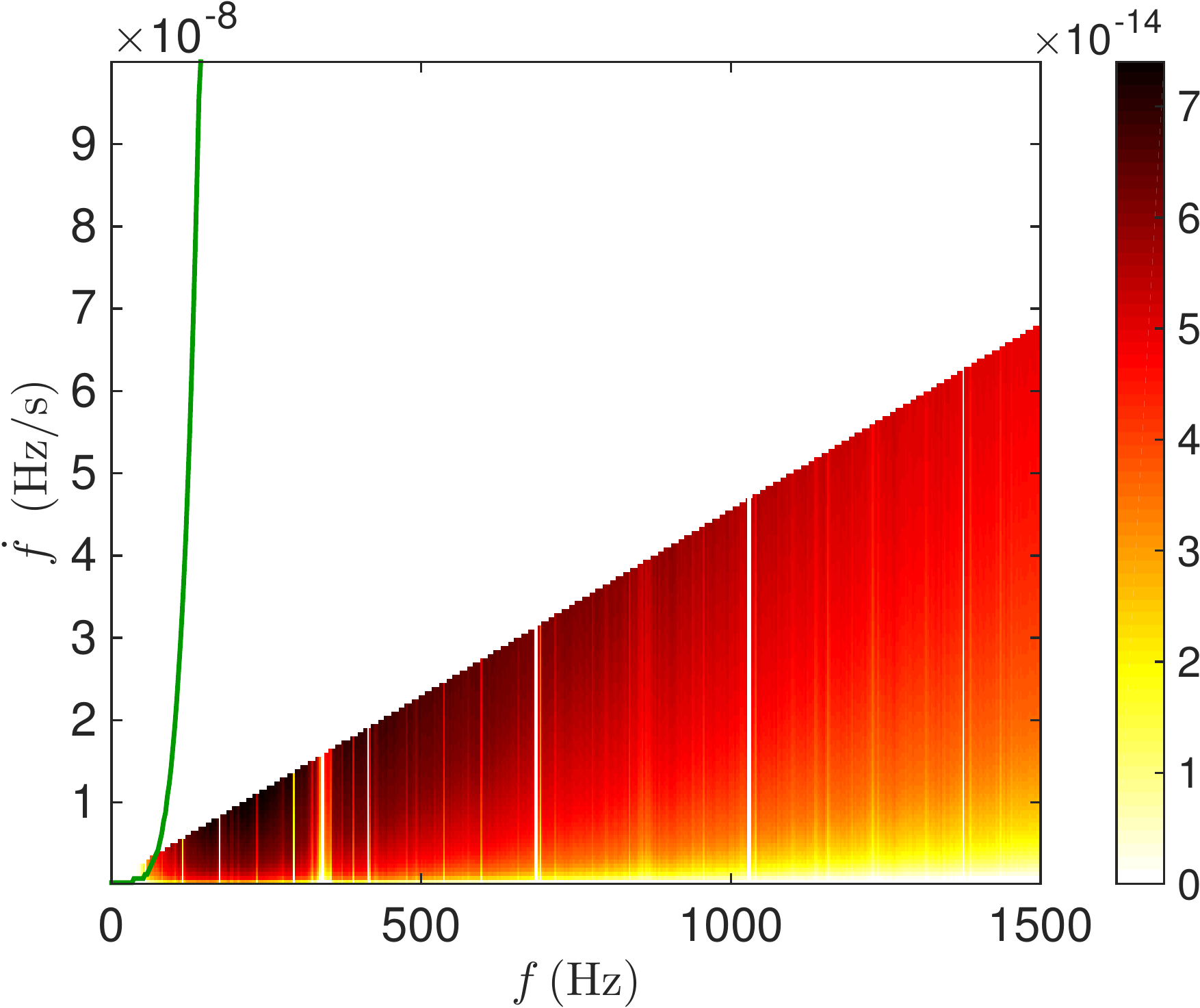}}}%
    \qquad
    \subfloat[Coverage,10 days]{{  \includegraphics[width=.20\linewidth]{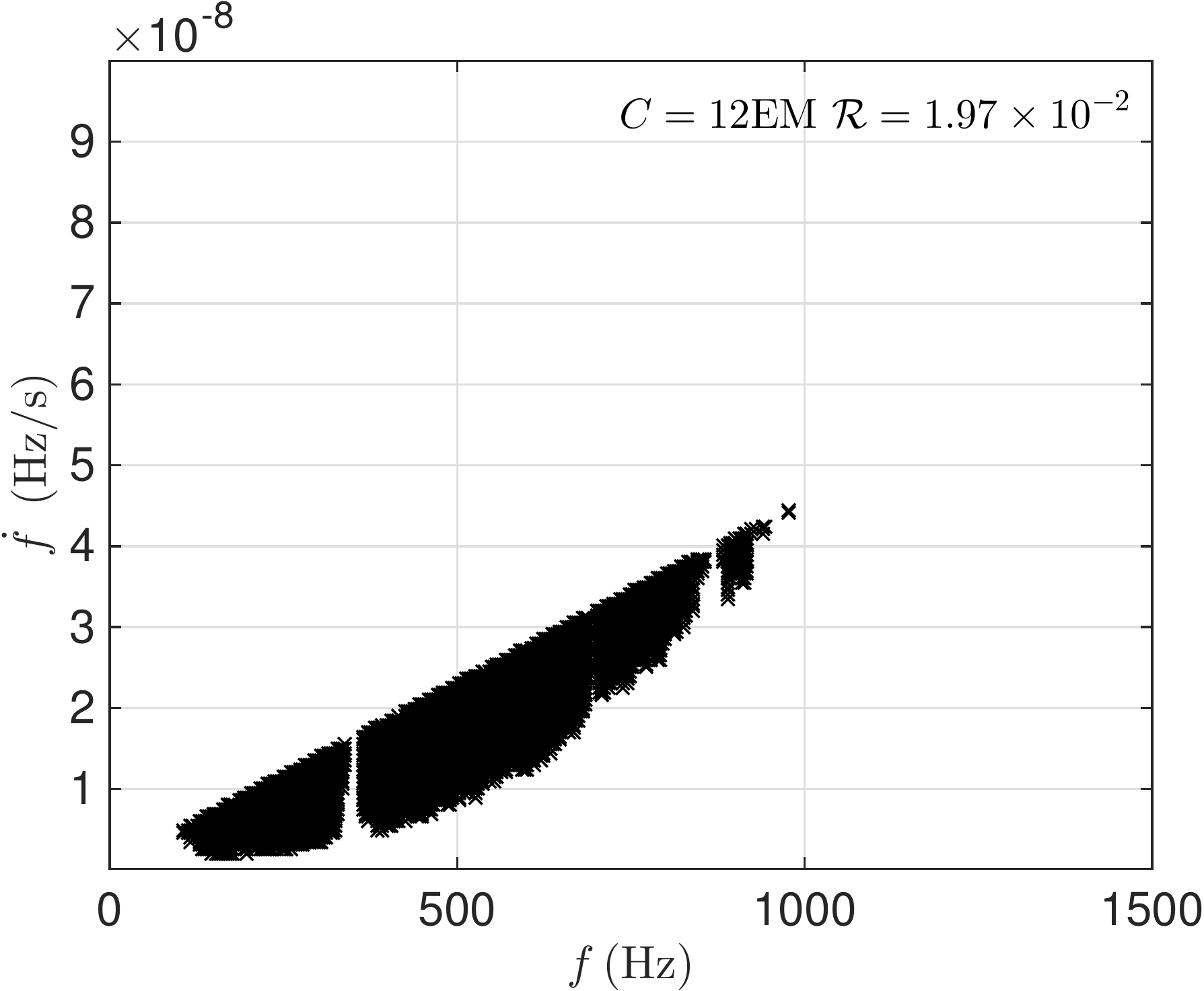}}}%
    \qquad
    \subfloat[Efficiency, 20 days]{{  \includegraphics[width=.20\linewidth]{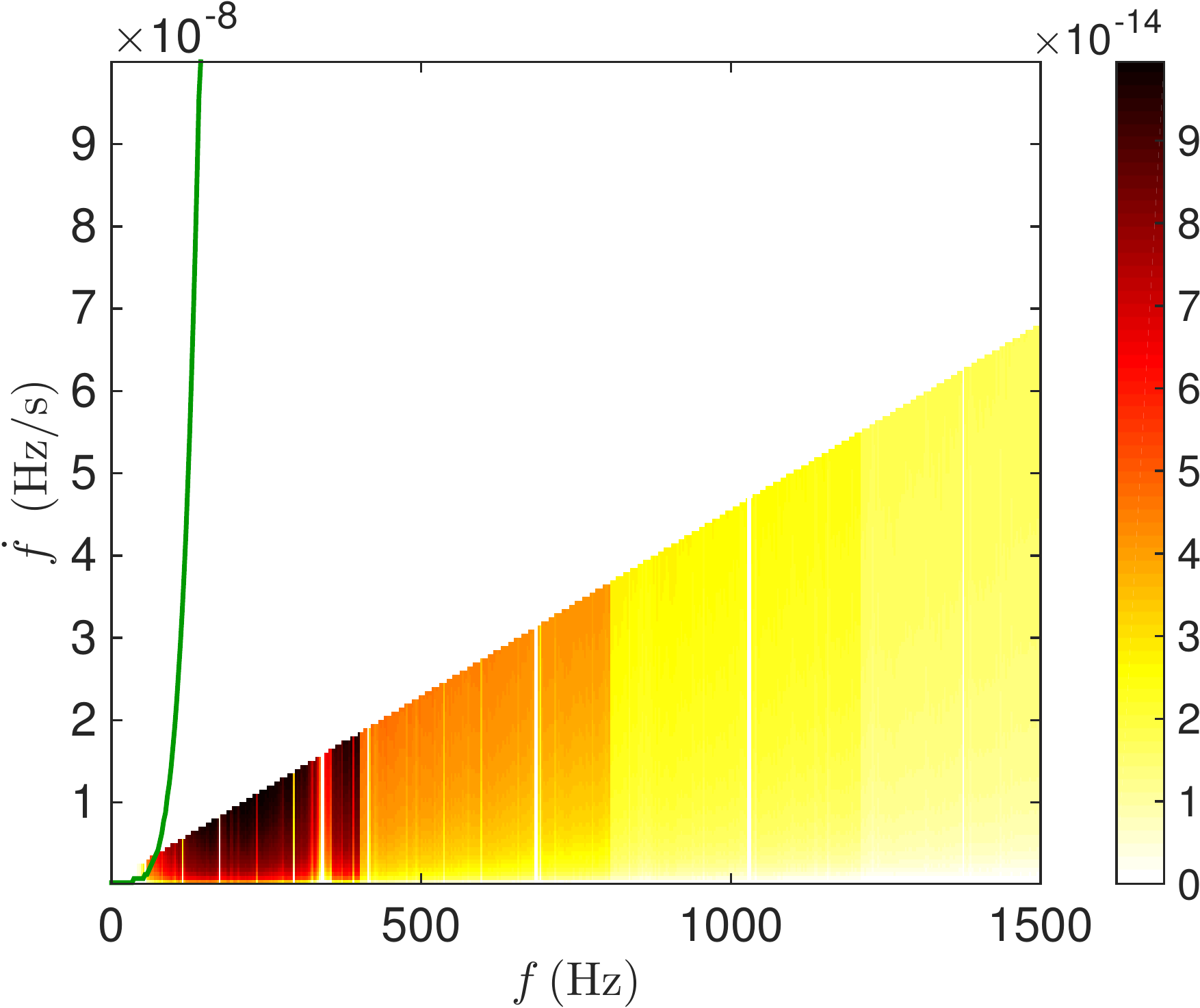}}}%
    \qquad
    \subfloat[Coverage, 20 days]{{  \includegraphics[width=.20\linewidth]{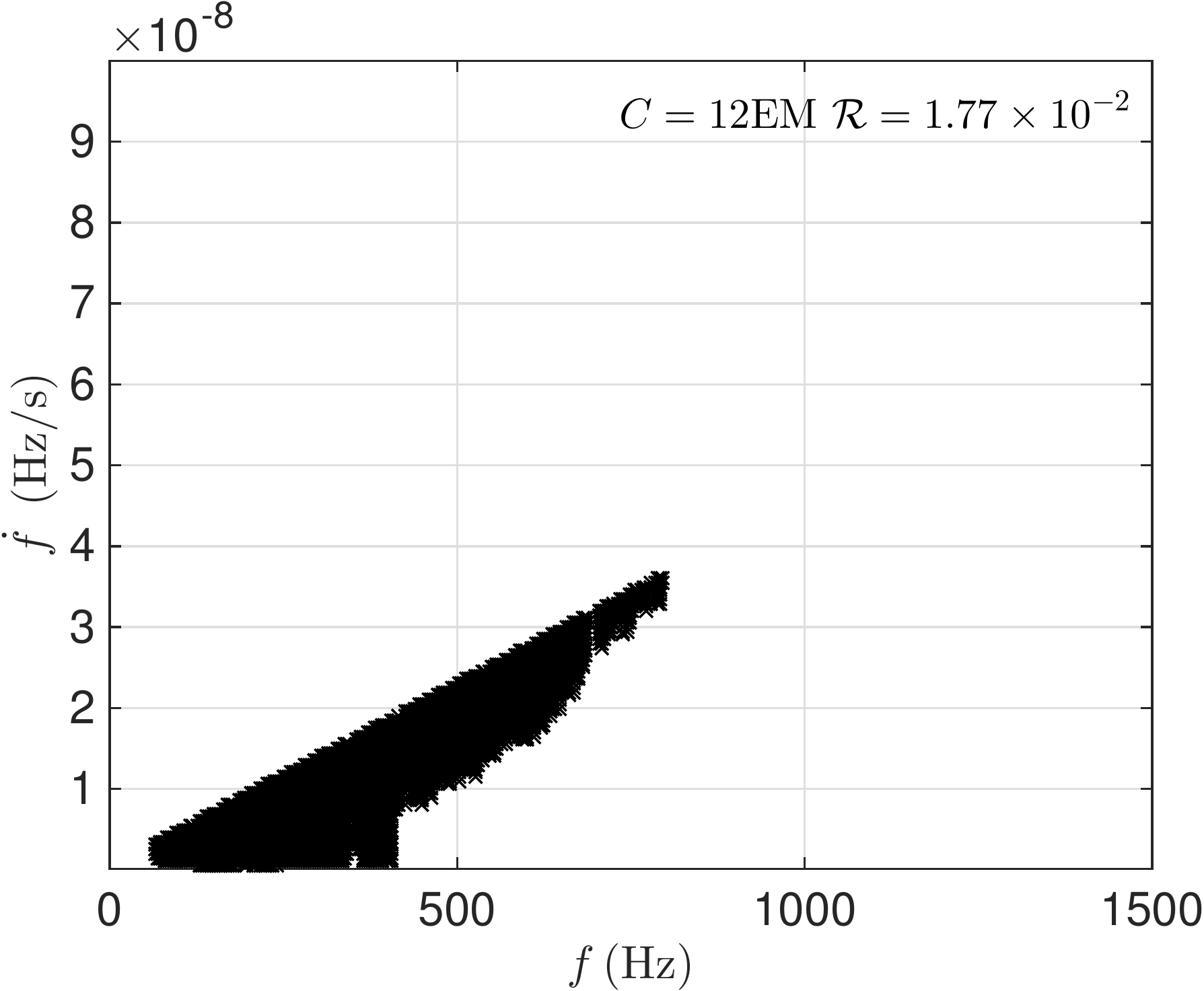}}}%
    \qquad
    \subfloat[Efficiency, 30 days]{{  \includegraphics[width=.20\linewidth]{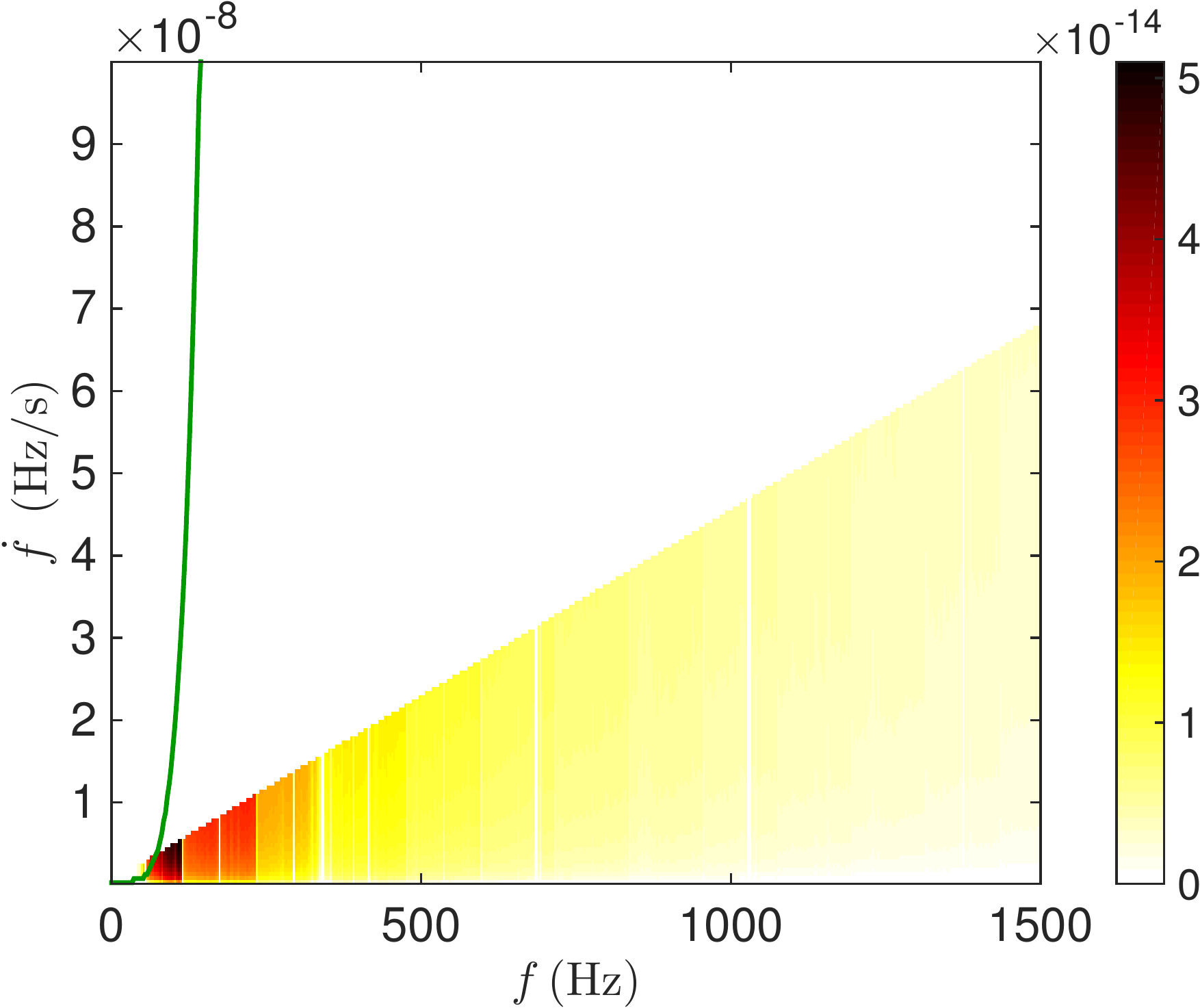}}}%
    \qquad
    \subfloat[Coverage, 30 days]{{  \includegraphics[width=.20\linewidth]{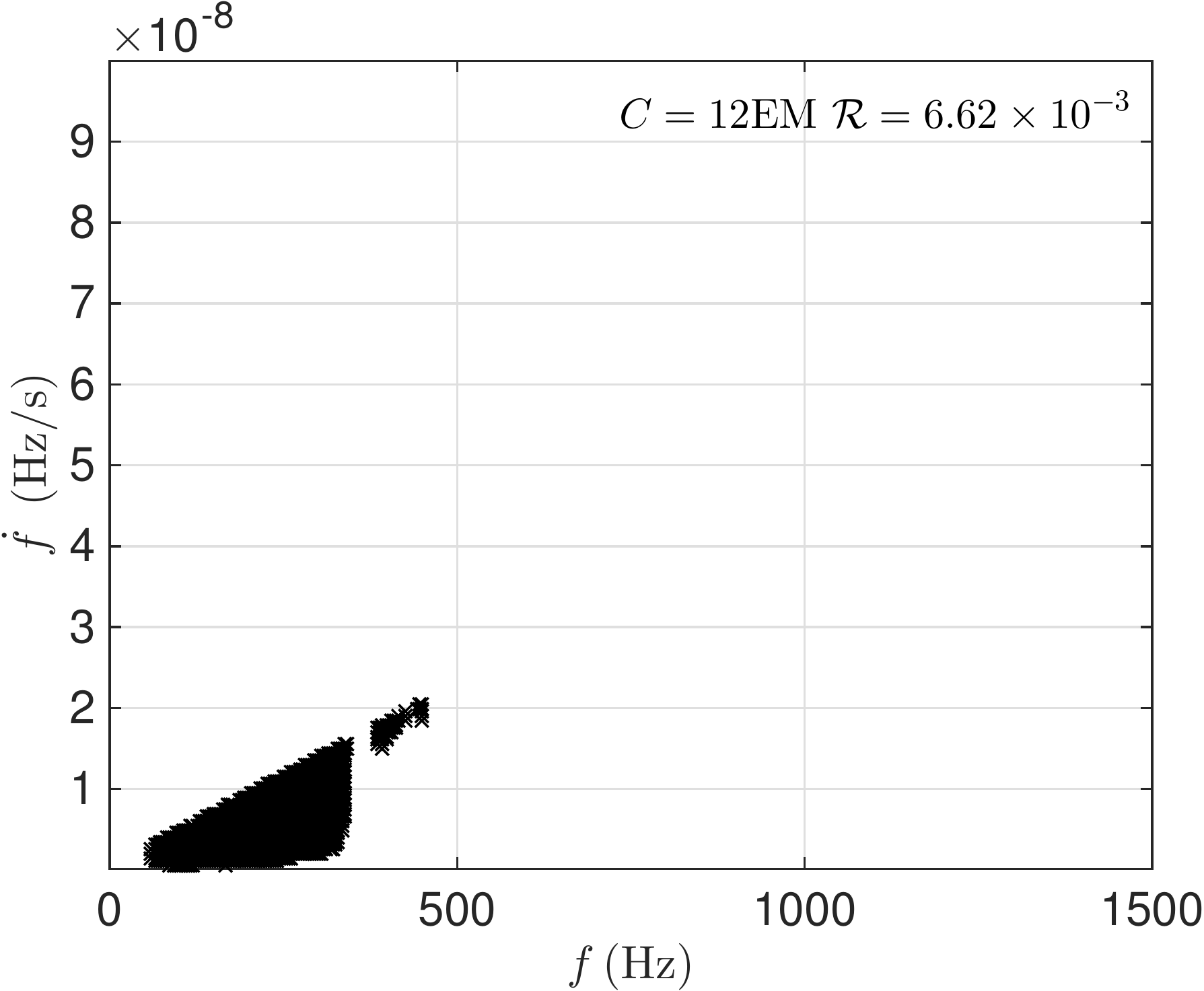}}}%
    \qquad
    \subfloat[Coverage, 37.5 days]{{  \includegraphics[width=.20\linewidth]{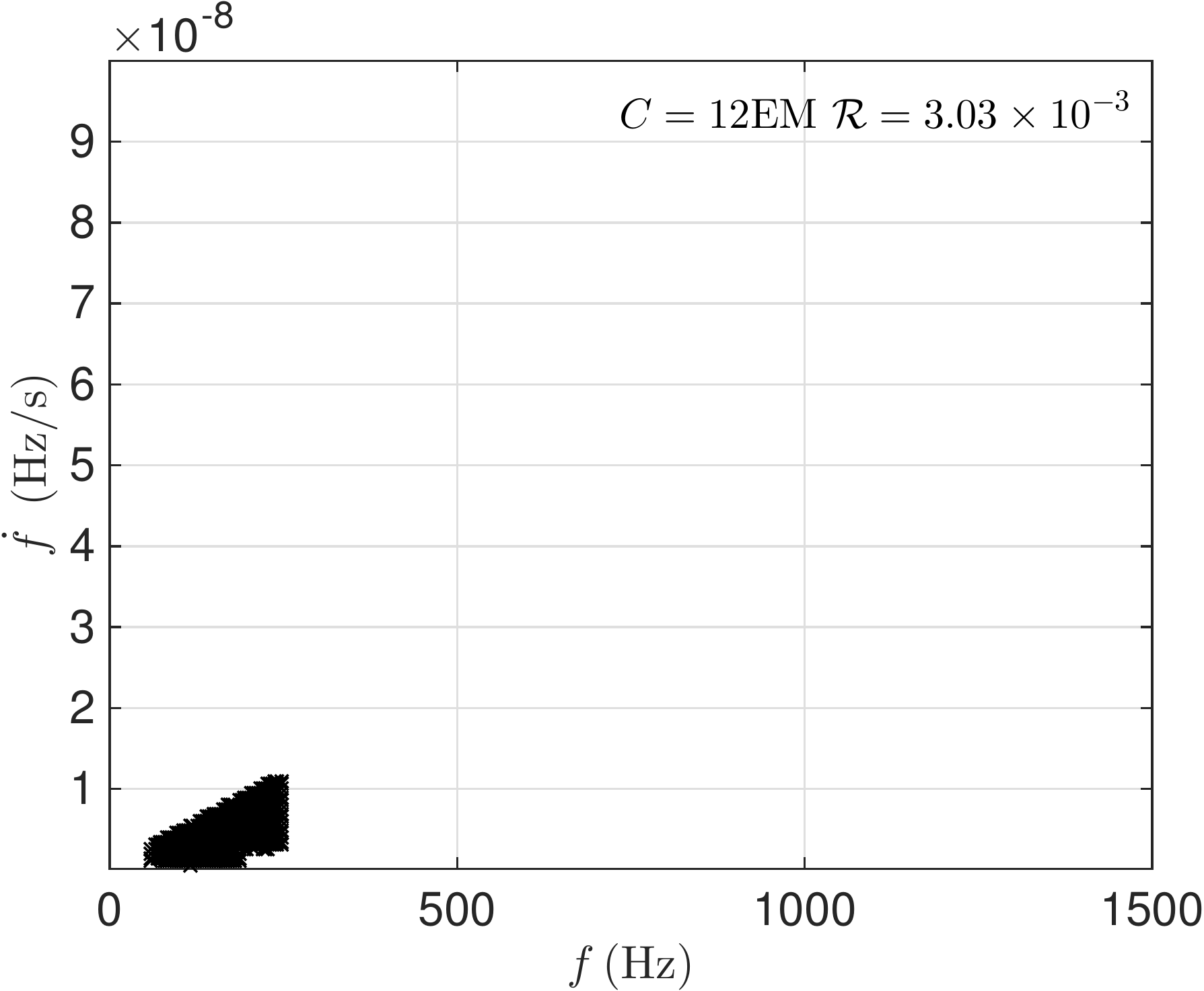}}}%
    \qquad
    \subfloat[Efficiency, 37.5 days]{{  \includegraphics[width=.20\linewidth]{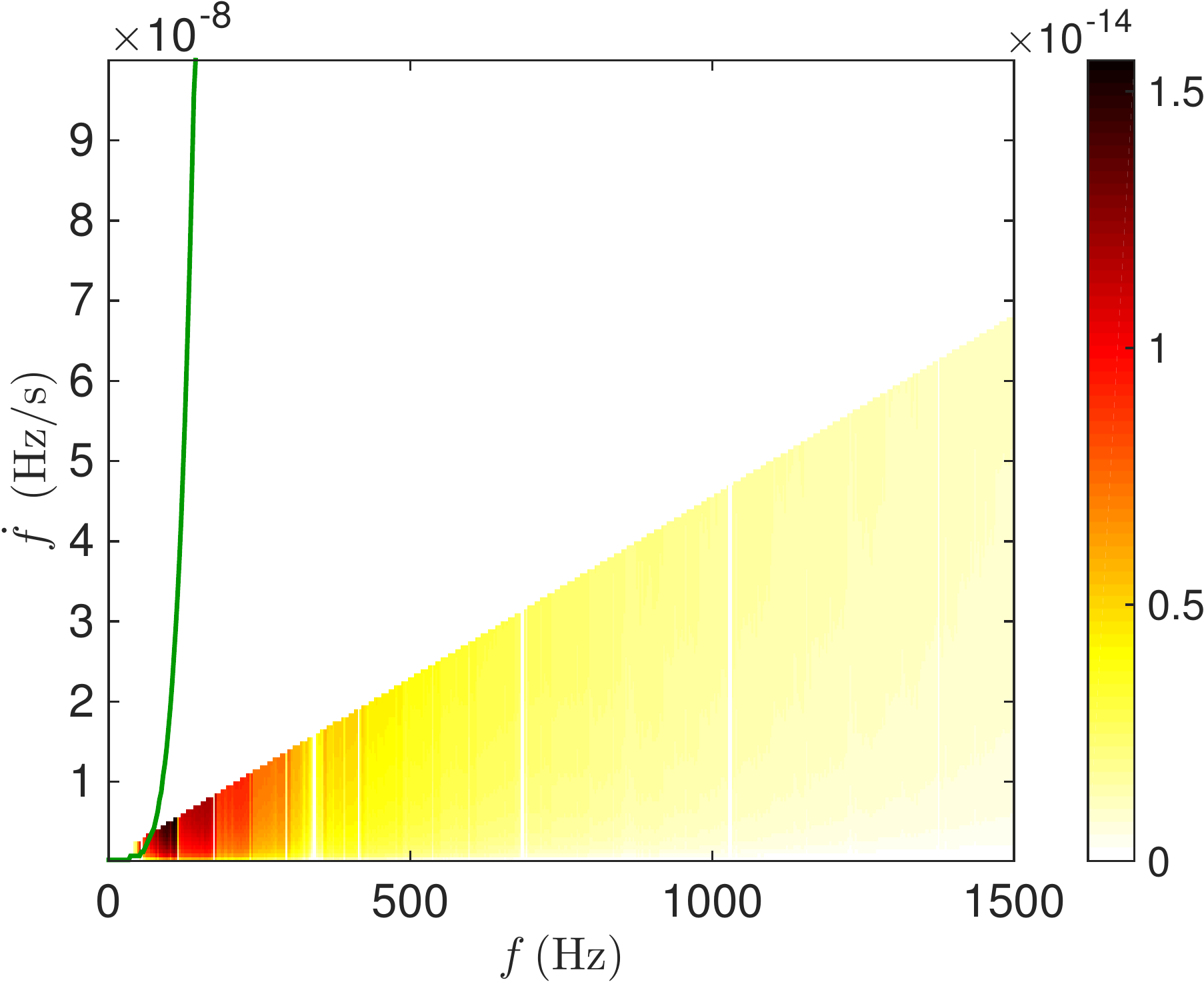}}}%
    \qquad
    \subfloat[Coverage, 50 days]{{  \includegraphics[width=.20\linewidth]{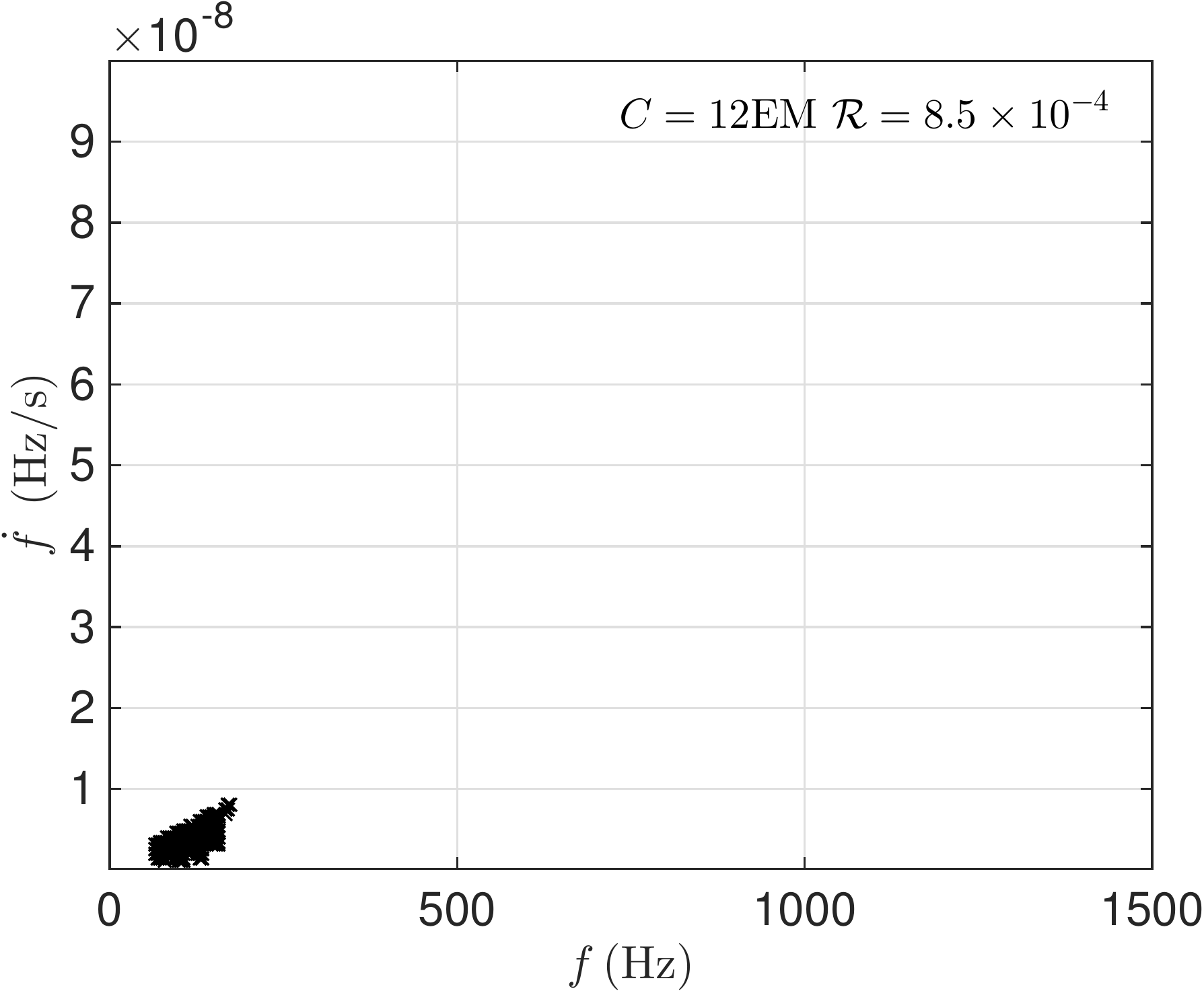}}}%
    \qquad
    \subfloat[Efficiency, 50 days]{{  \includegraphics[width=.20\linewidth]{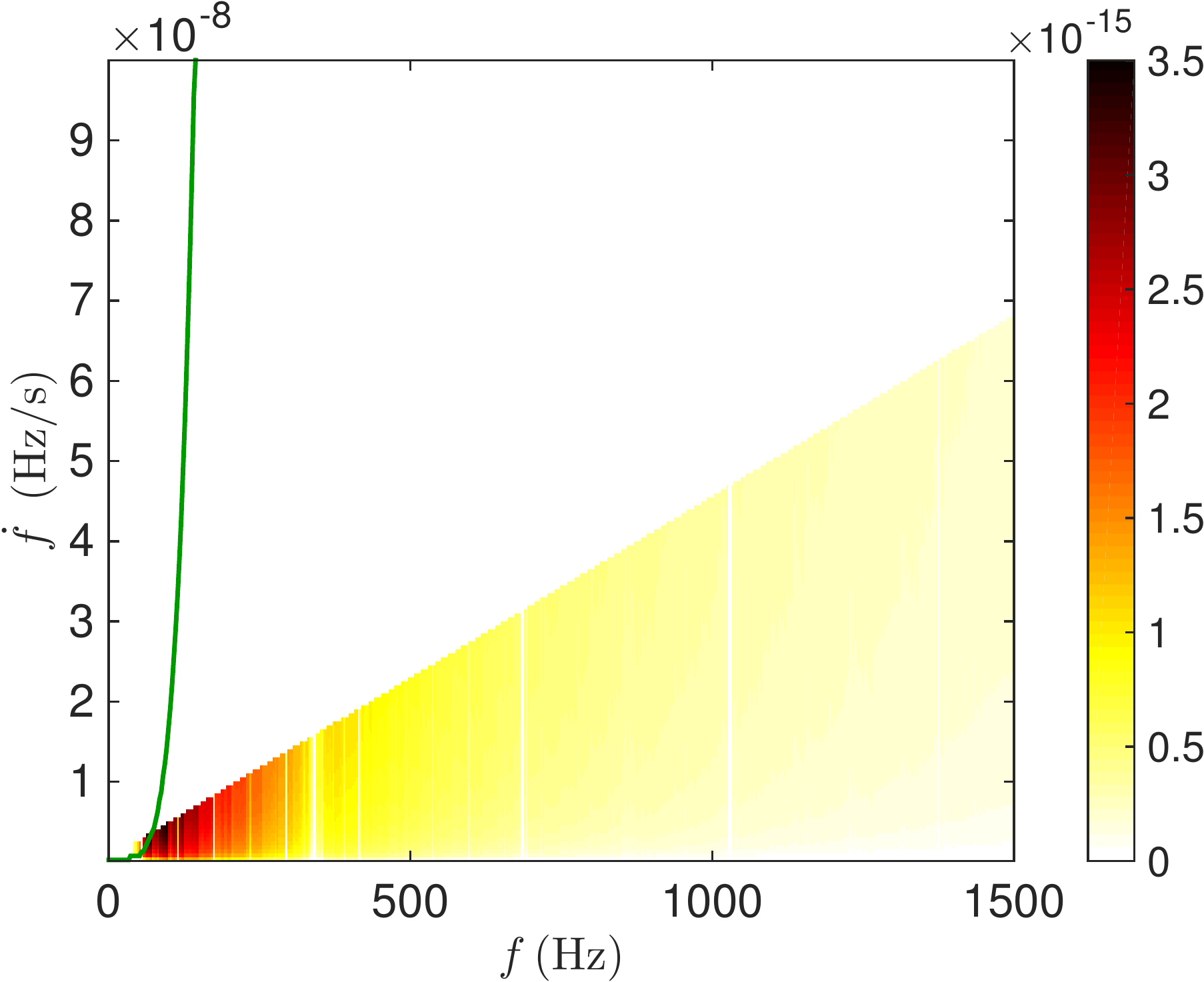}}}%
    \qquad
    \subfloat[Coverage, 75 days]{{  \includegraphics[width=.20\linewidth]{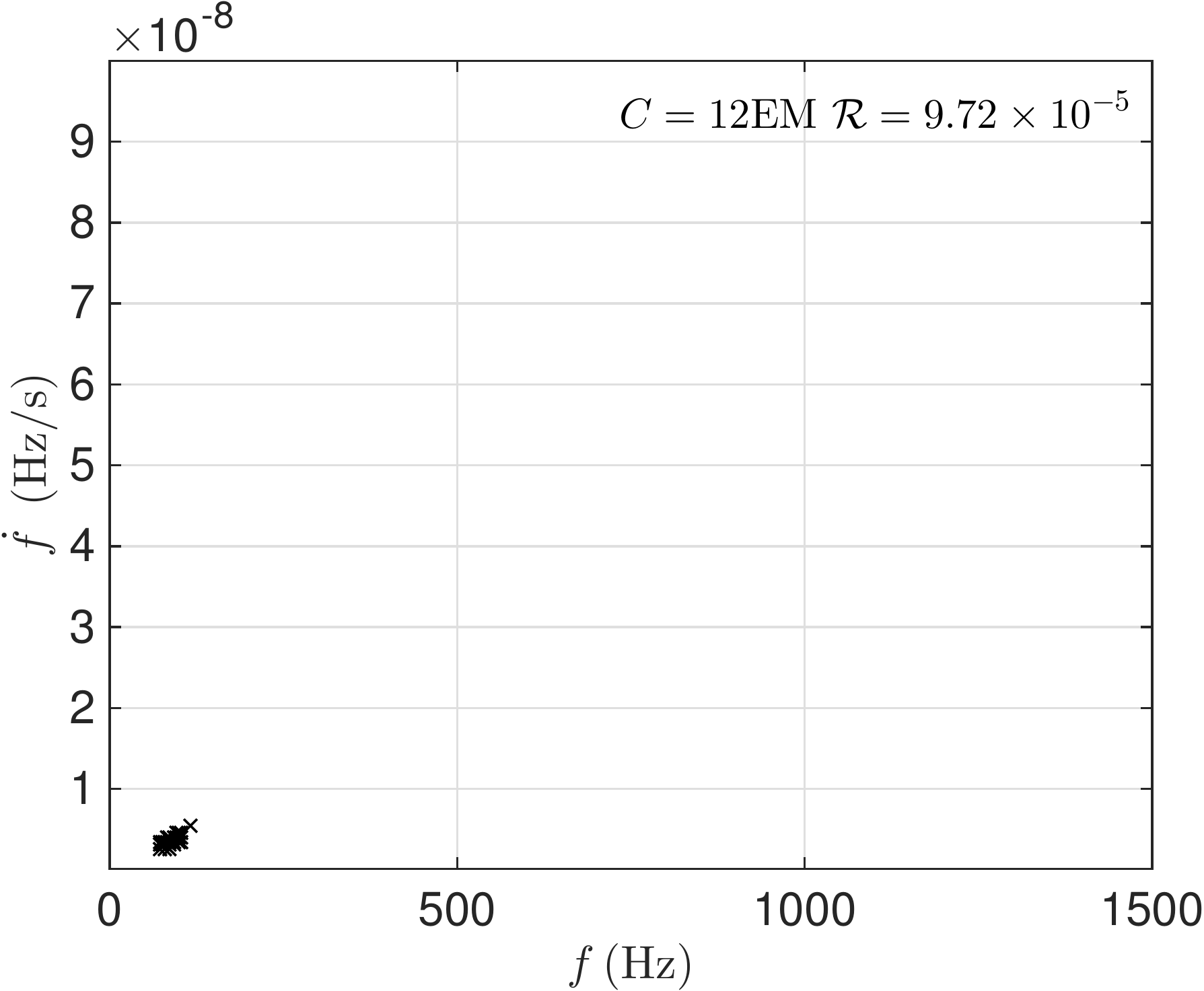}}}%
    \qquad
    \subfloat[Efficiency, 75 days]{{  \includegraphics[width=.20\linewidth]{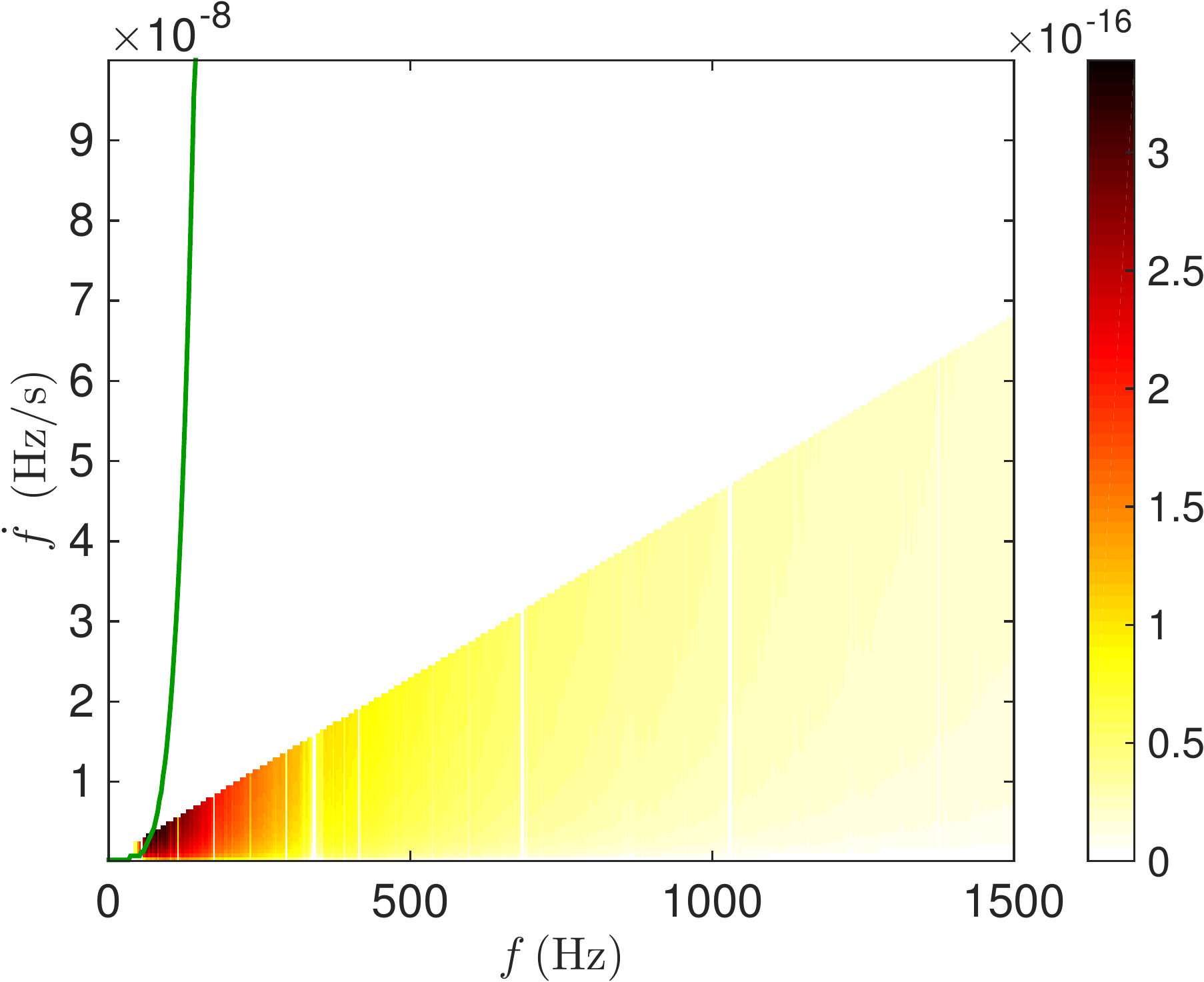}}}%
 
    \caption{Optimisation results for Vela Jr, at 200 pc and 700 years old (close and young, CY), assuming uniform and age-based priors, for various coherent search durations: 5, 10, 20, 30, 37.5, 50 and 75 days. The total computing budget is assumed to be 12 EM. }%
    \label{G2662_51020days_shortage}%
\end{figure*}


\begin{figure*}%
    \centering
    \subfloat[Efficiency, 5 days]{{  \includegraphics[width=.20\linewidth]{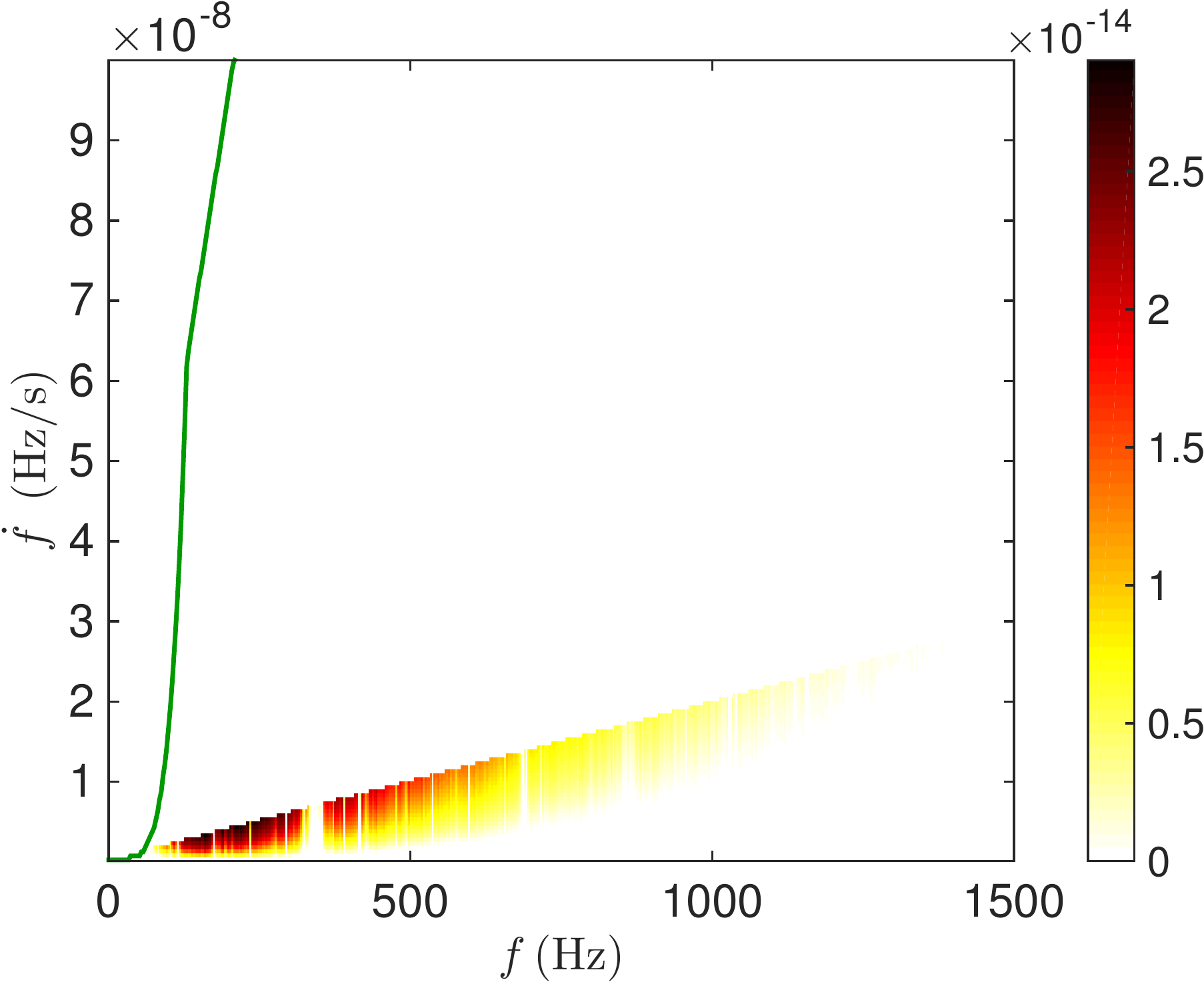}}}%
    \qquad
    \subfloat[Coverage, 5 days]{{  \includegraphics[width=.20\linewidth]{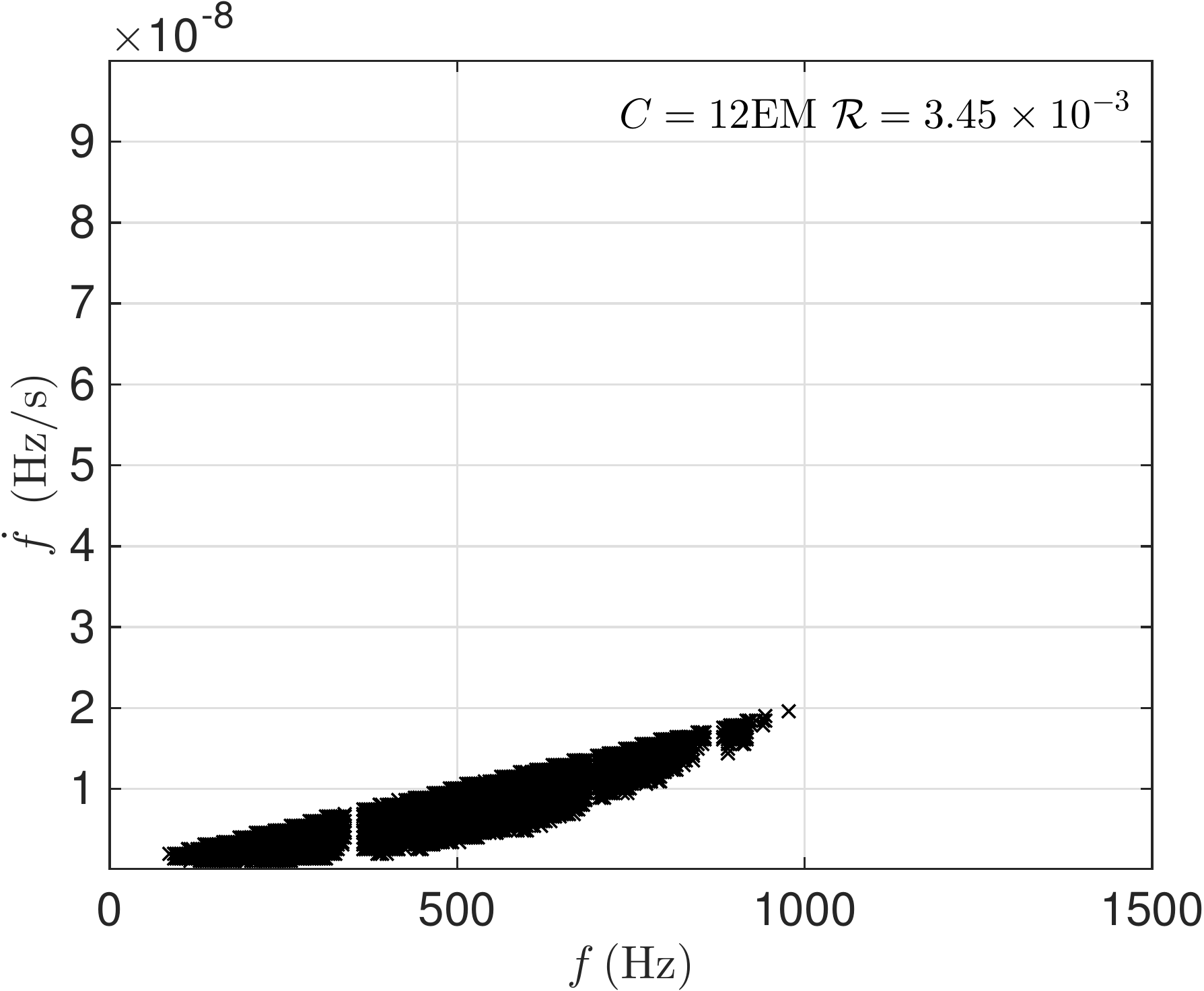}}}%
    \qquad
    \subfloat[Efficiency, 10 days]{{  \includegraphics[width=.20\linewidth]{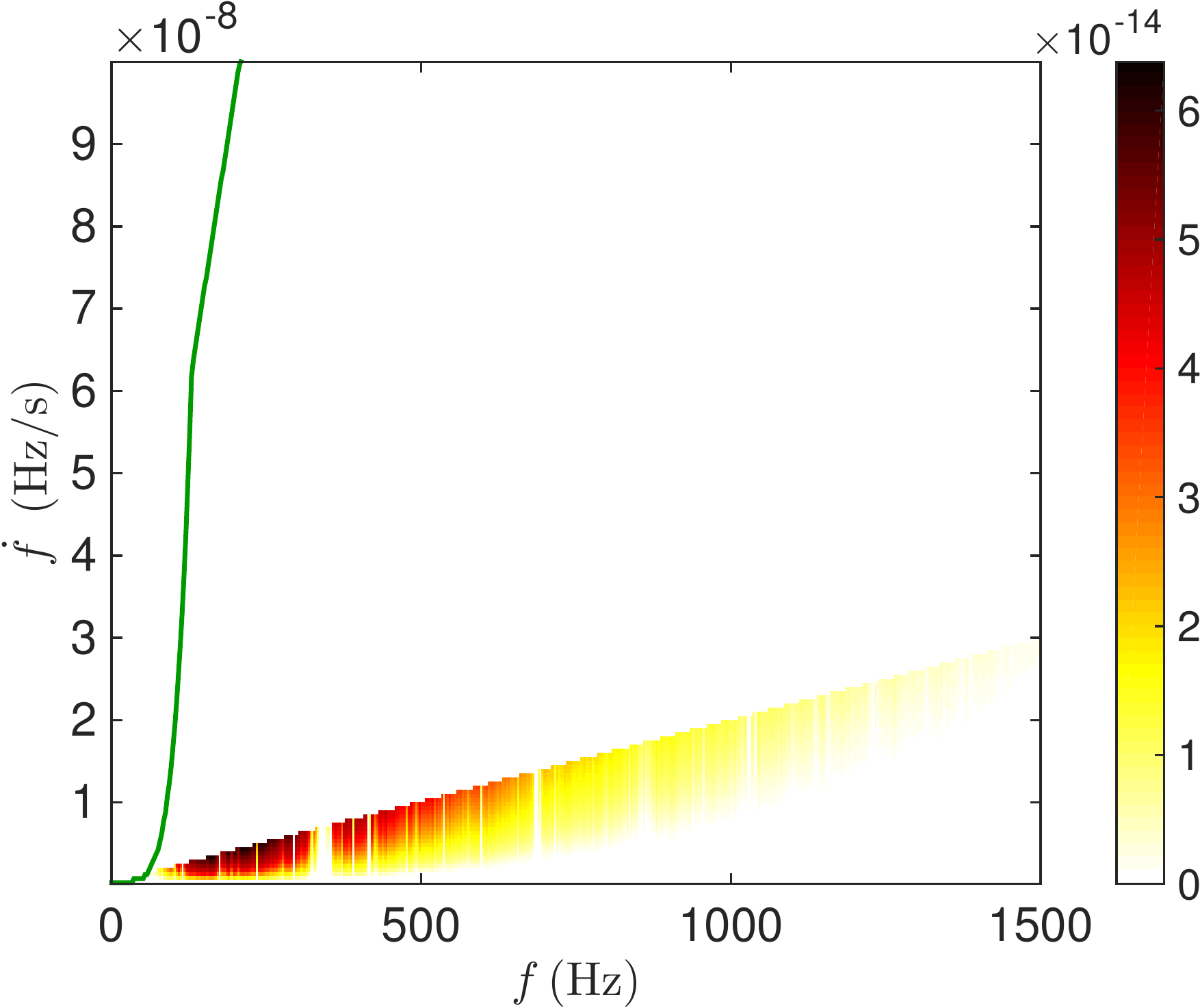}}}%
    \qquad
    \subfloat[Coverage, 10 days]{{  \includegraphics[width=.20\linewidth]{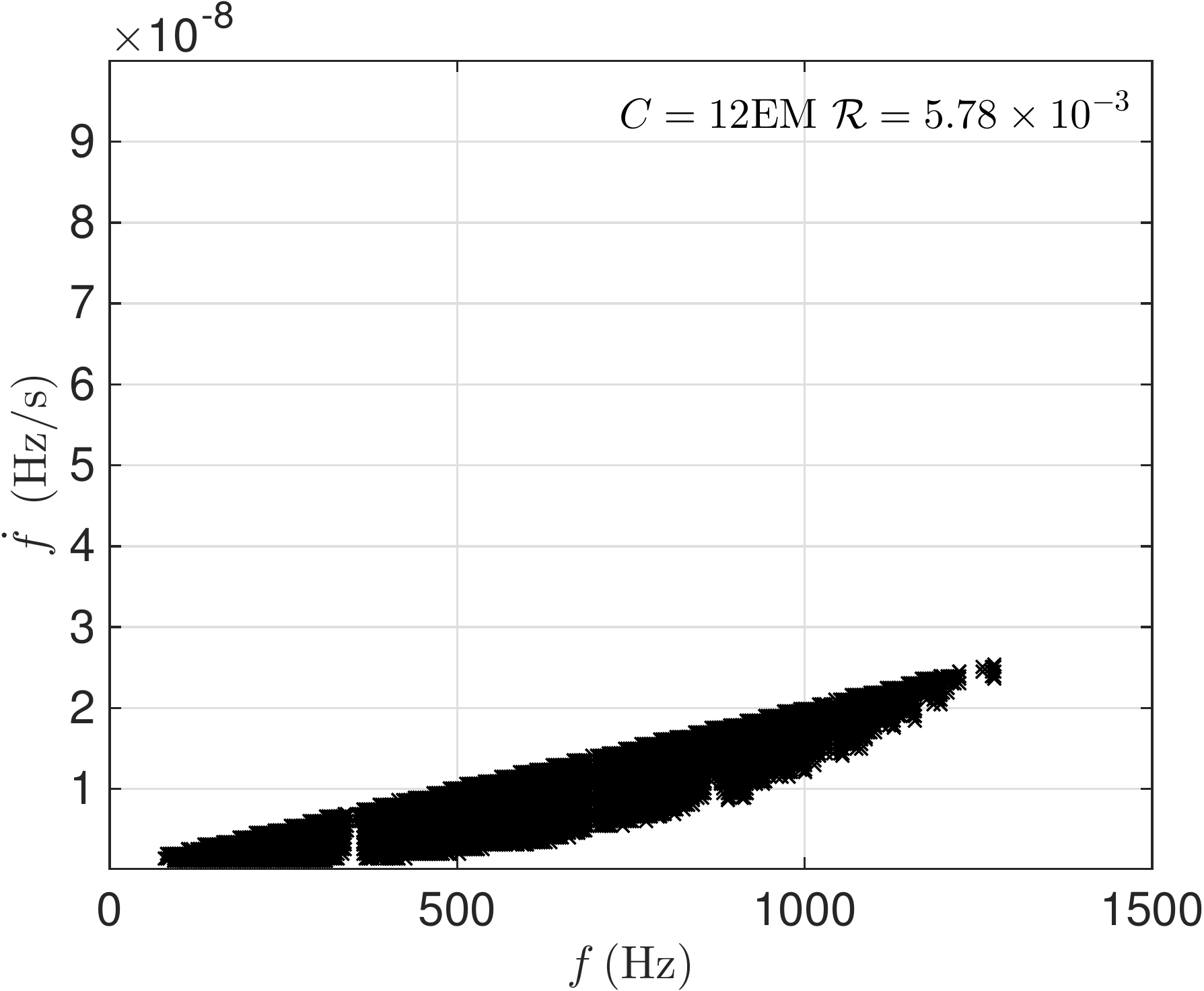}}}%
    \qquad
    \subfloat[Efficiency, 20 days]{{  \includegraphics[width=.20\linewidth]{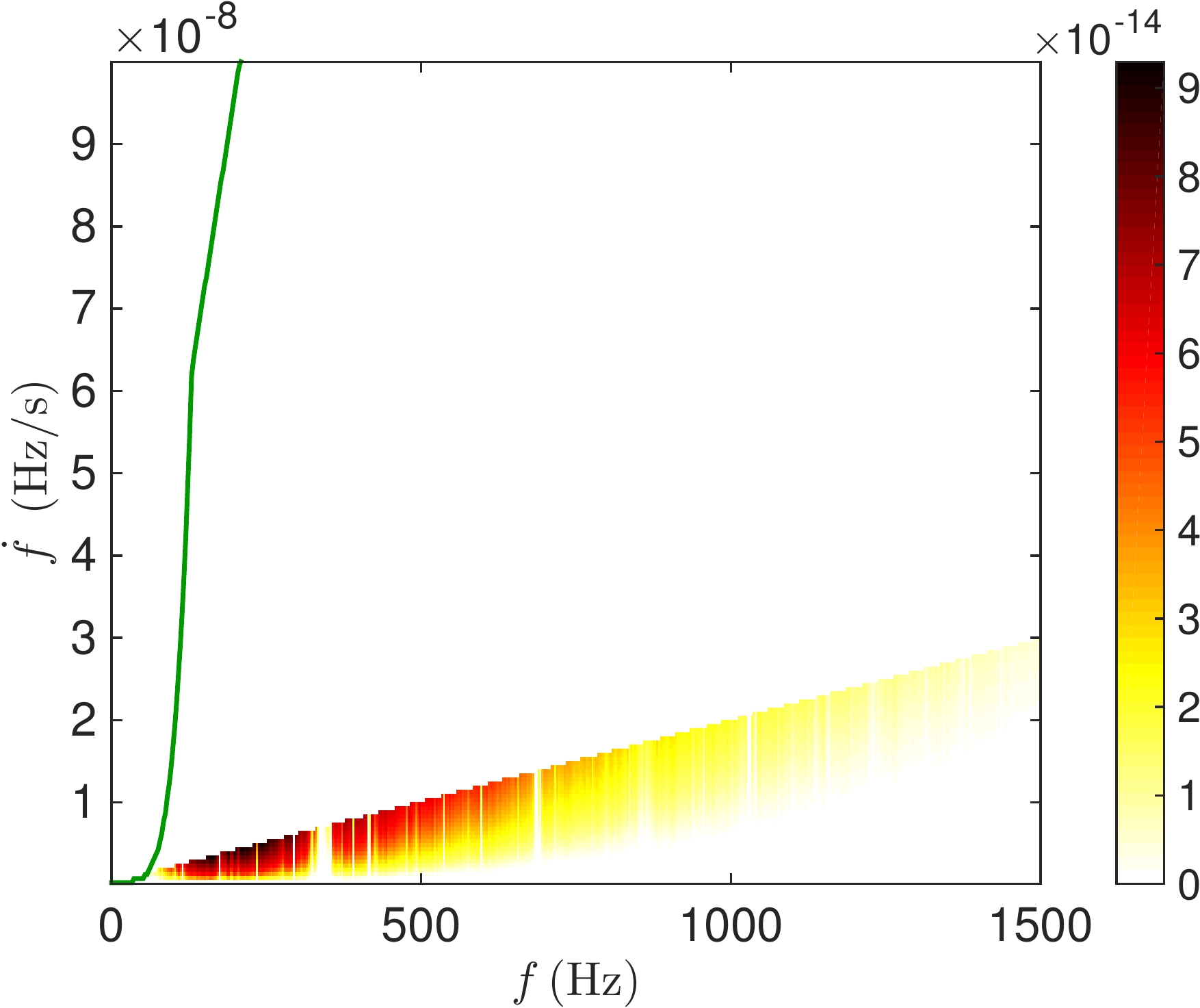}}}%
    \qquad
    \subfloat[Coverage, 20 days]{{  \includegraphics[width=.20\linewidth]{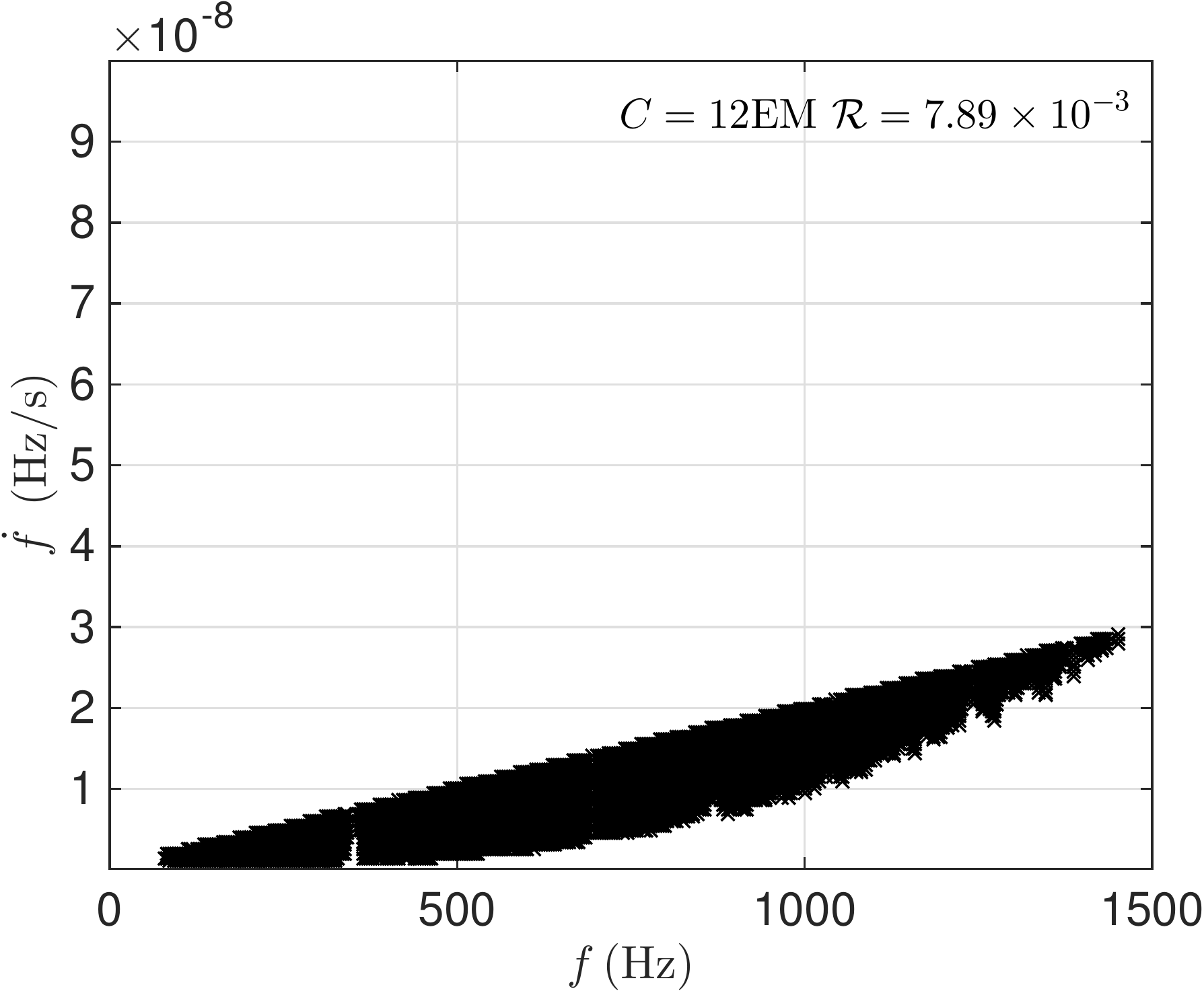}}}%
    \qquad
    \subfloat[Efficiency, 30 days]{{  \includegraphics[width=.20\linewidth]{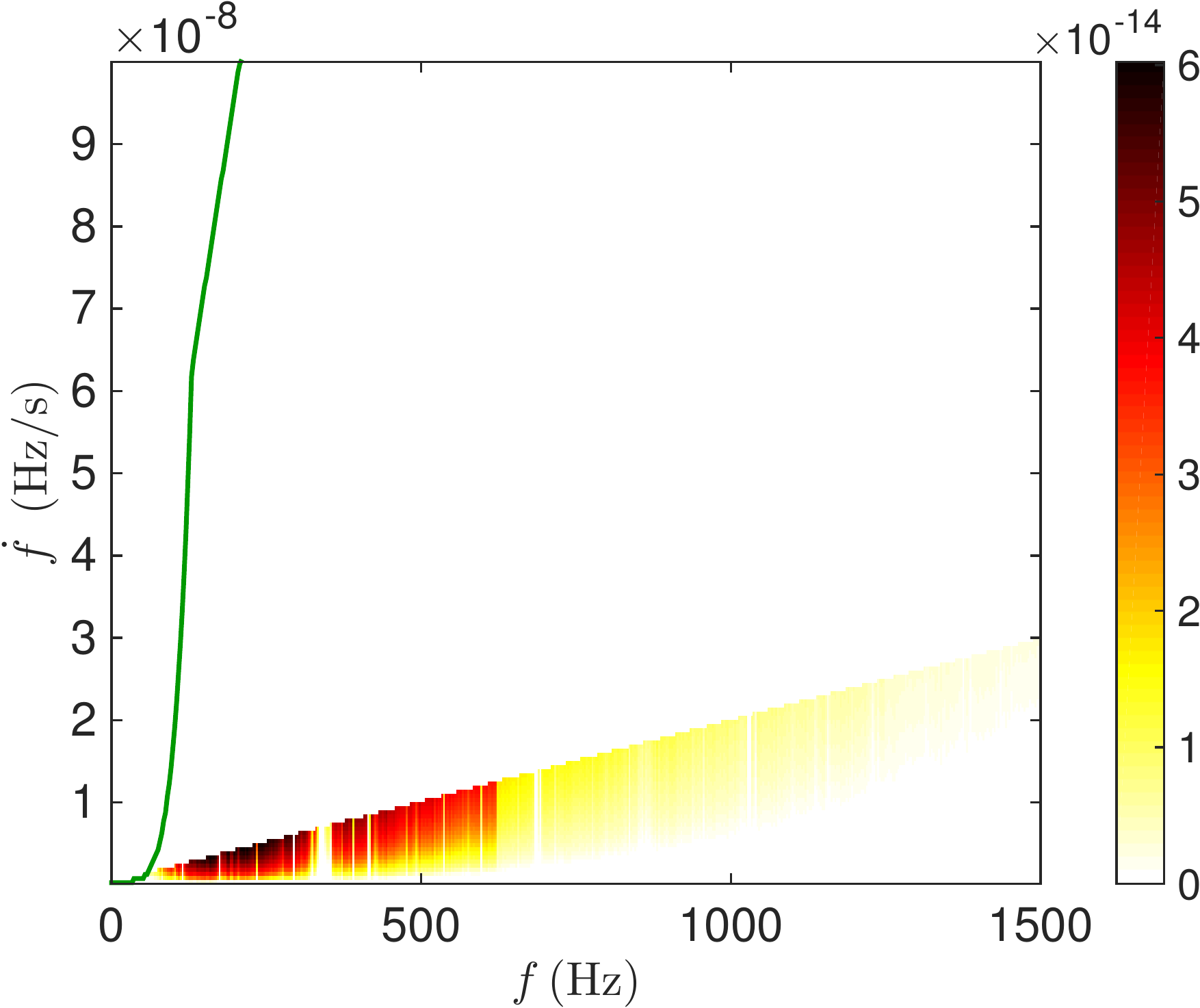}}}%
    \qquad
    \subfloat[Coverage, 30 days]{{  \includegraphics[width=.20\linewidth]{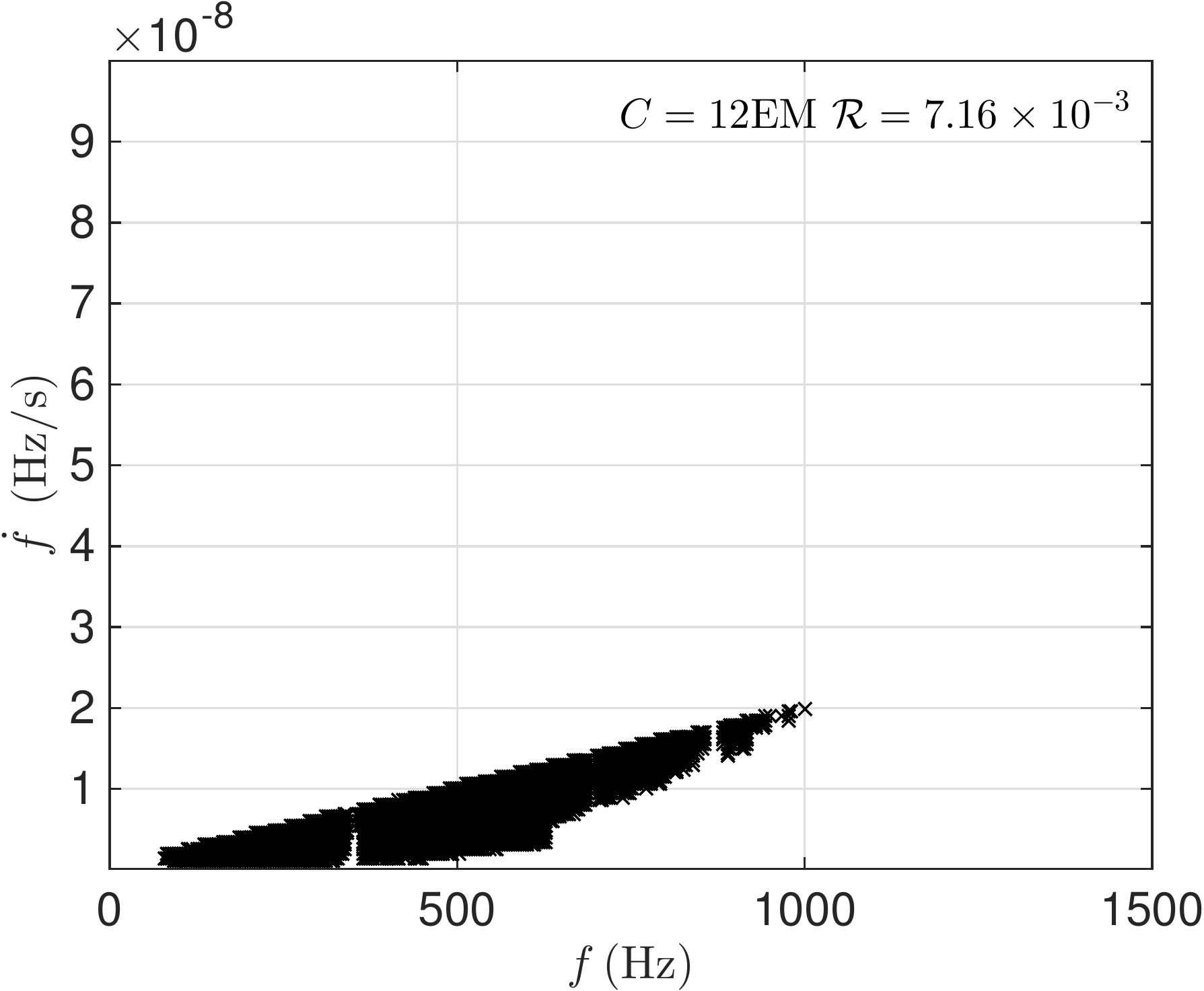}}}%
    \qquad
    \subfloat[Efficiency, 37.5 days]{{  \includegraphics[width=.20\linewidth]{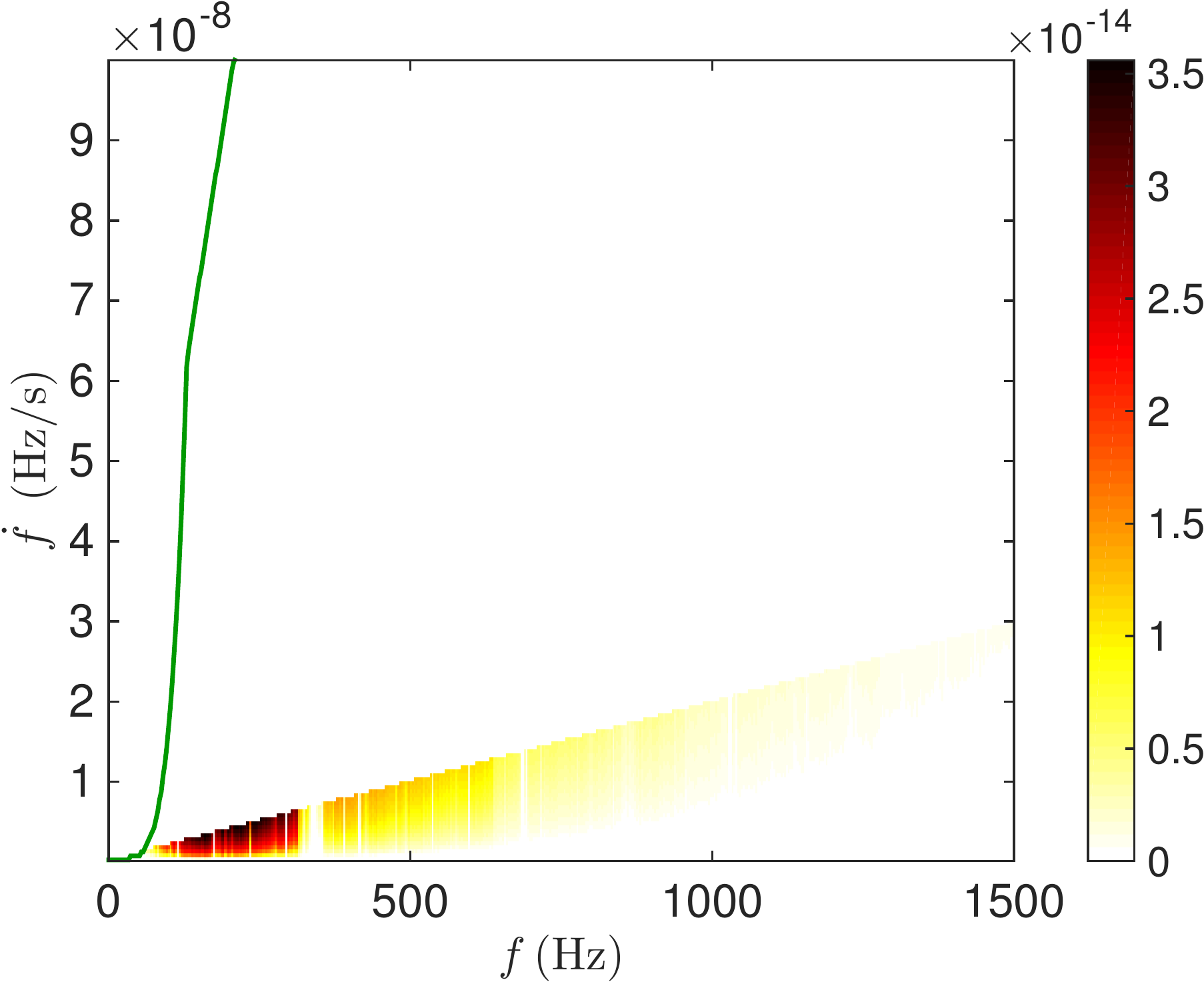}}}%
    \qquad
    \subfloat[Coverage, 37.5 days]{{  \includegraphics[width=.20\linewidth]{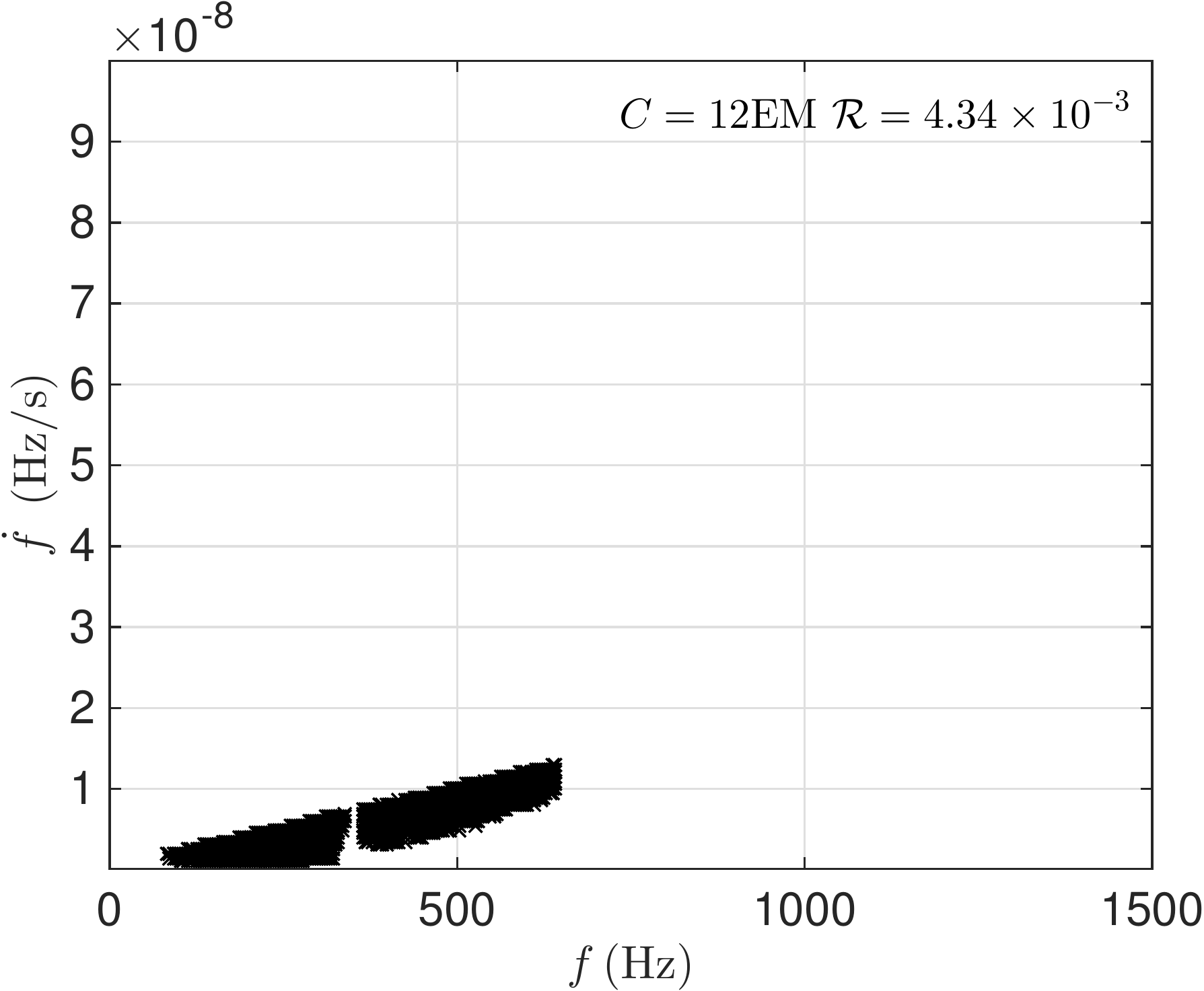}}}%
    \qquad
    \subfloat[Efficiency, 50 days]{{  \includegraphics[width=.20\linewidth]{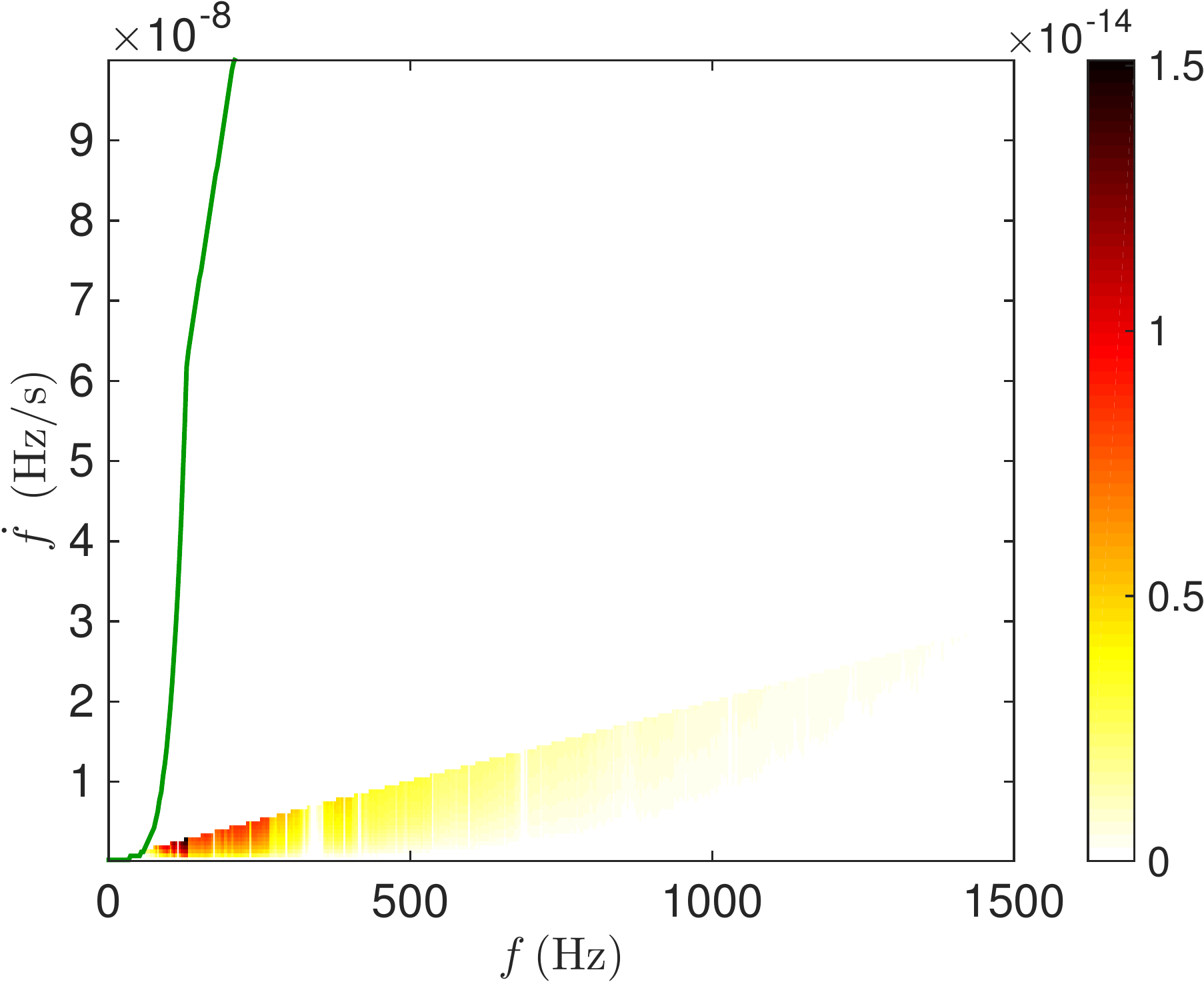}}}%
    \qquad
    \subfloat[Coverage, 50 days]{{  \includegraphics[width=.20\linewidth]{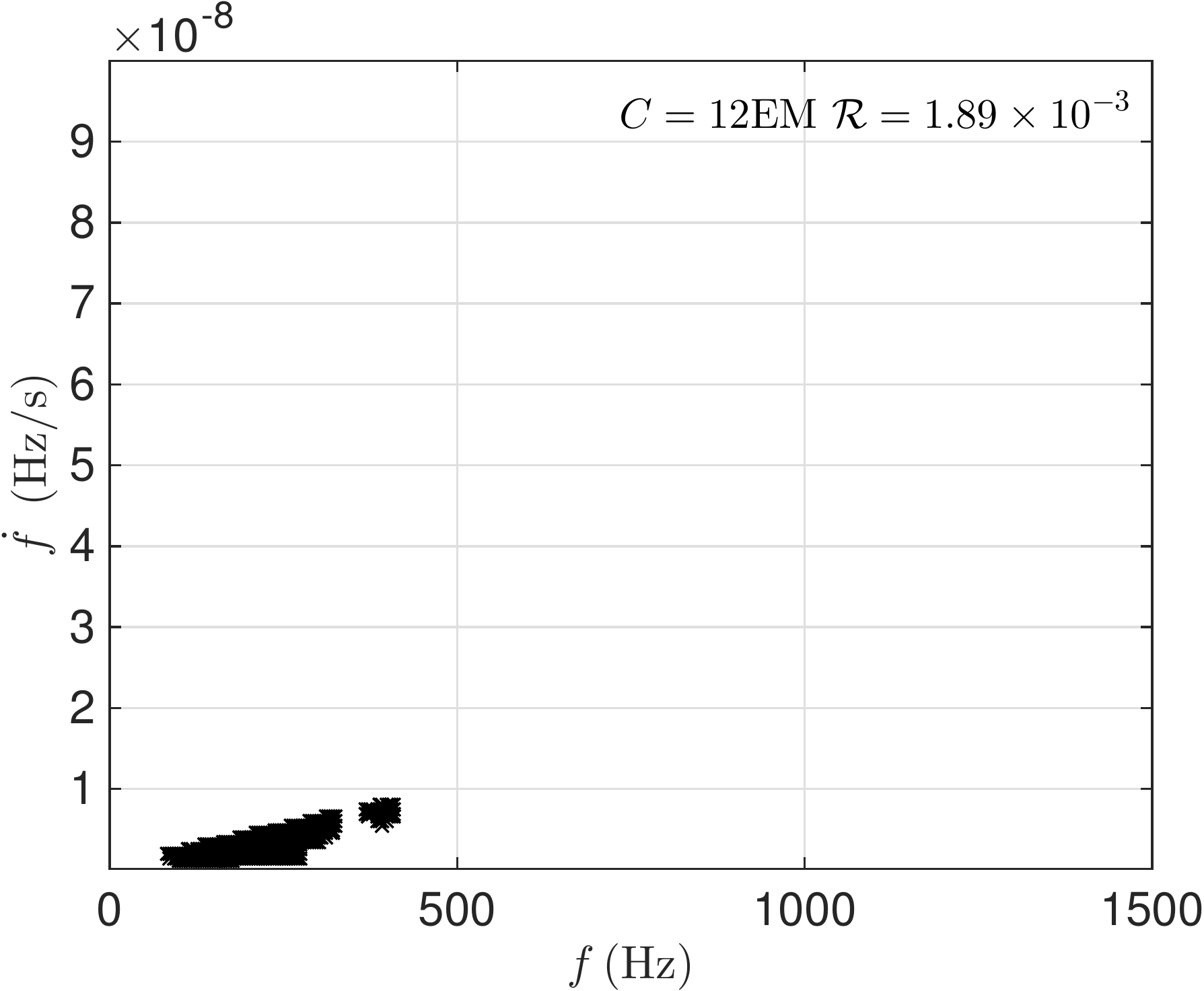}}}%
    \qquad
    \subfloat[Efficiency, 75 days]{{  \includegraphics[width=.20\linewidth]{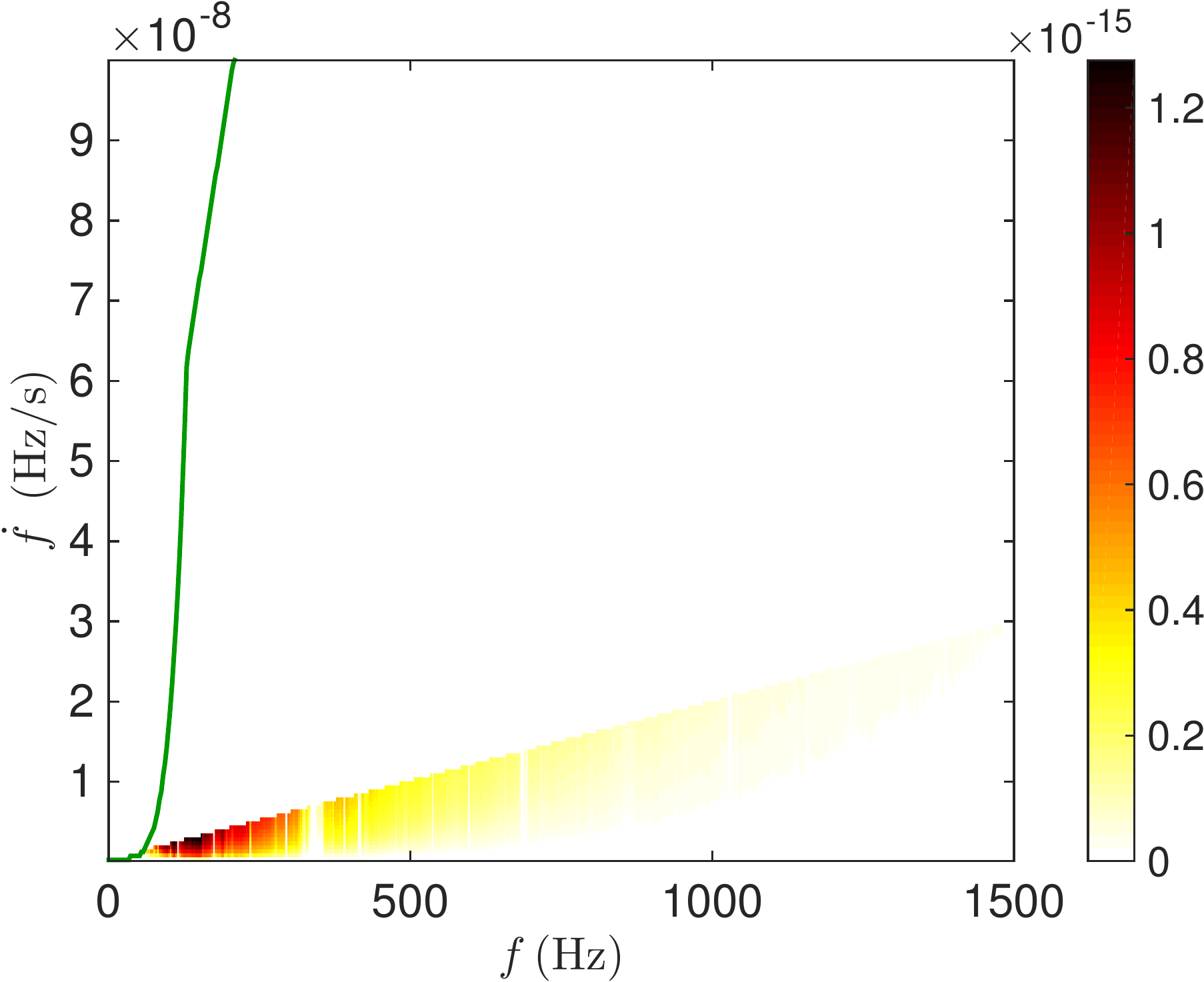}}}%
    \qquad
    \subfloat[Coverage, 75 days]{{  \includegraphics[width=.20\linewidth]{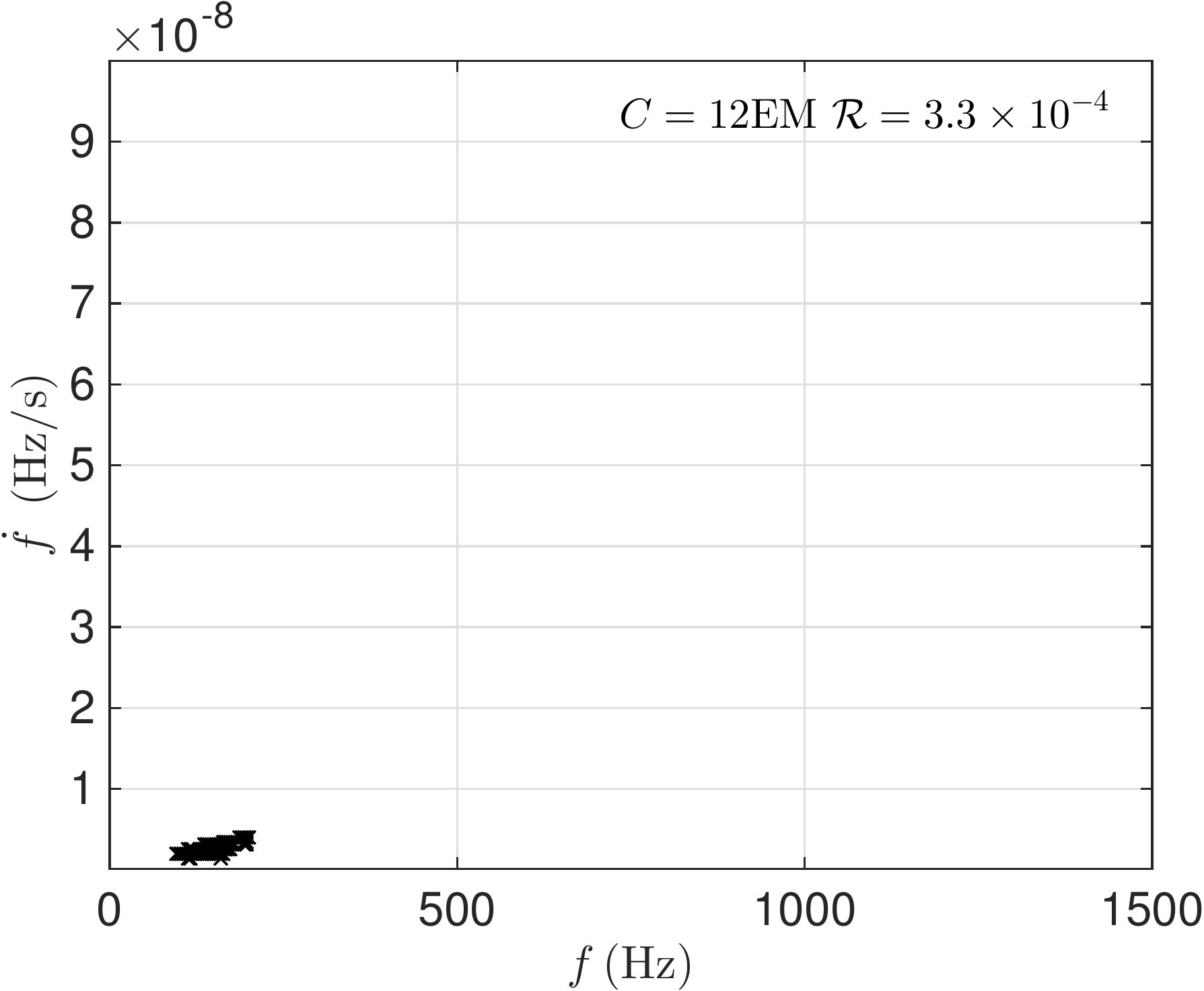}}}%

    \caption{Optimisation results for G347.3 Cas A at 1300 pc, 1600 years old, assuming uniform and age-based priors, for various coherent search durations: 5, 10, 20, 30, 37.5, 50 and 75 days. The total computing budget is assumed to be 12 EM. }%
    \label{G3473_51020days_shortage}%
\end{figure*}

\begin{figure*}%
    \centering
    \subfloat[Efficiency, 5 days]{{  \includegraphics[width=.20\linewidth]{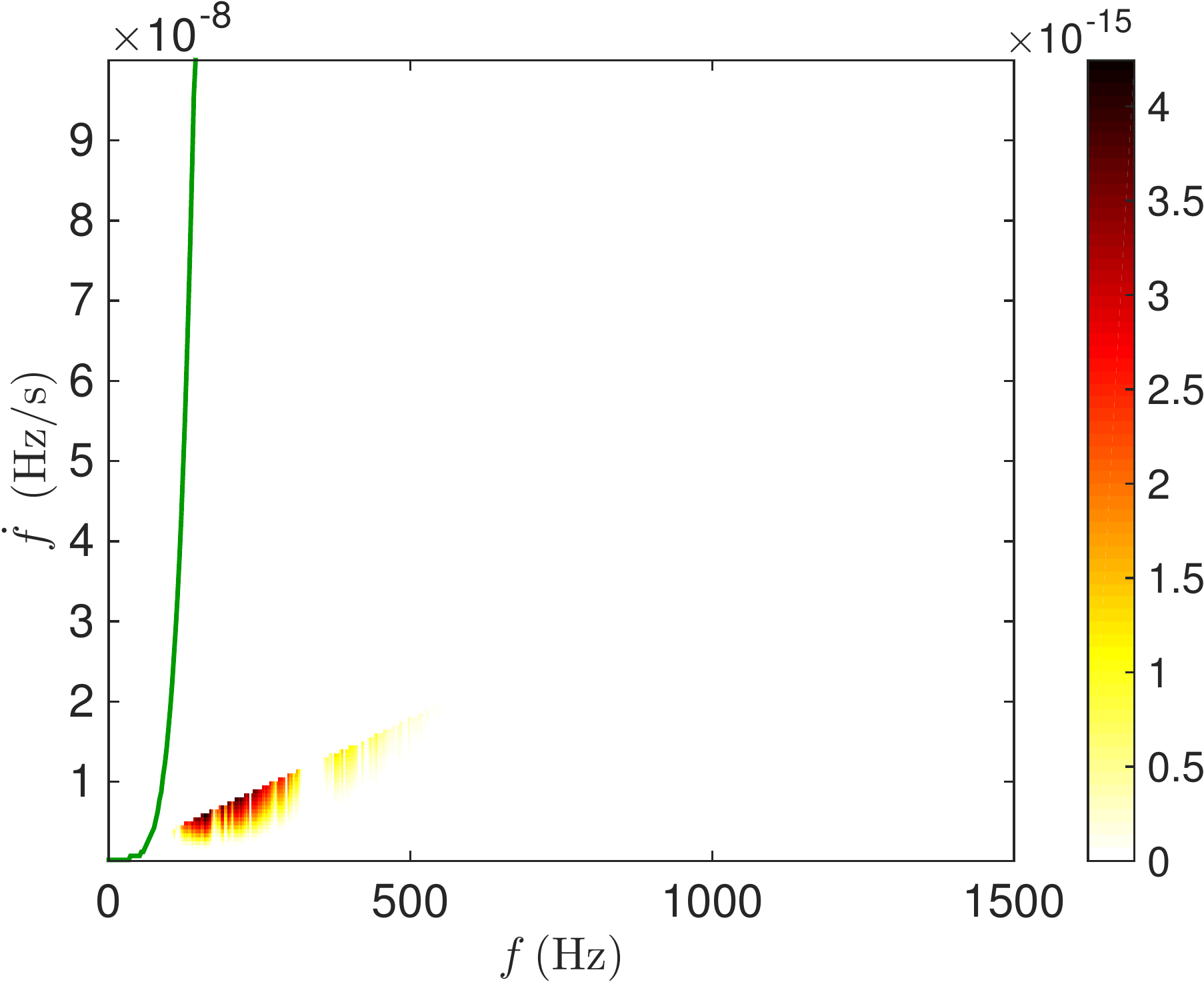}}}%
    \qquad
    \subfloat[Coverage, 5 days]{{  \includegraphics[width=.20\linewidth]{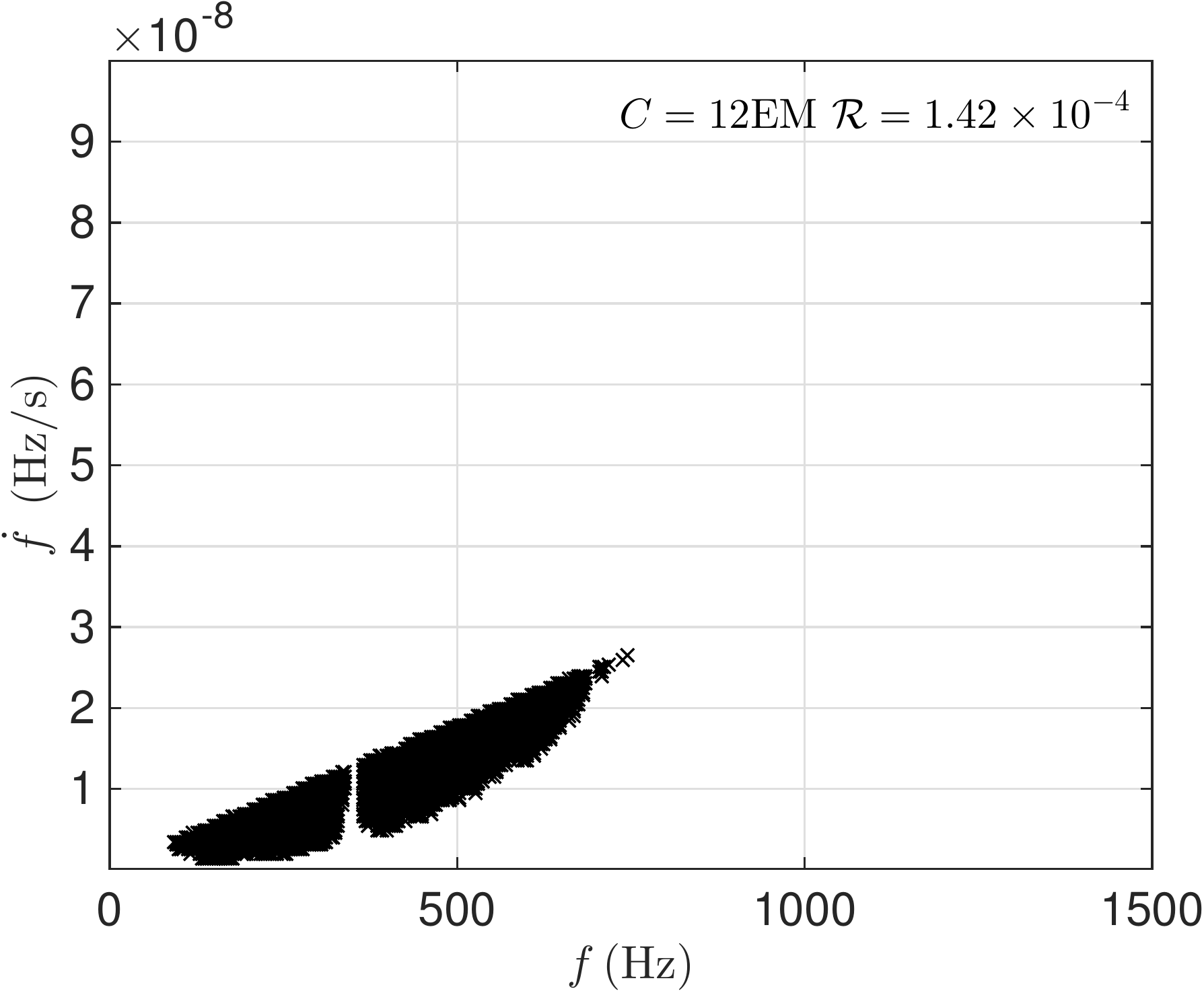}}}%
    \qquad
    \subfloat[Efficiency, 10 days]{{  \includegraphics[width=.20\linewidth]{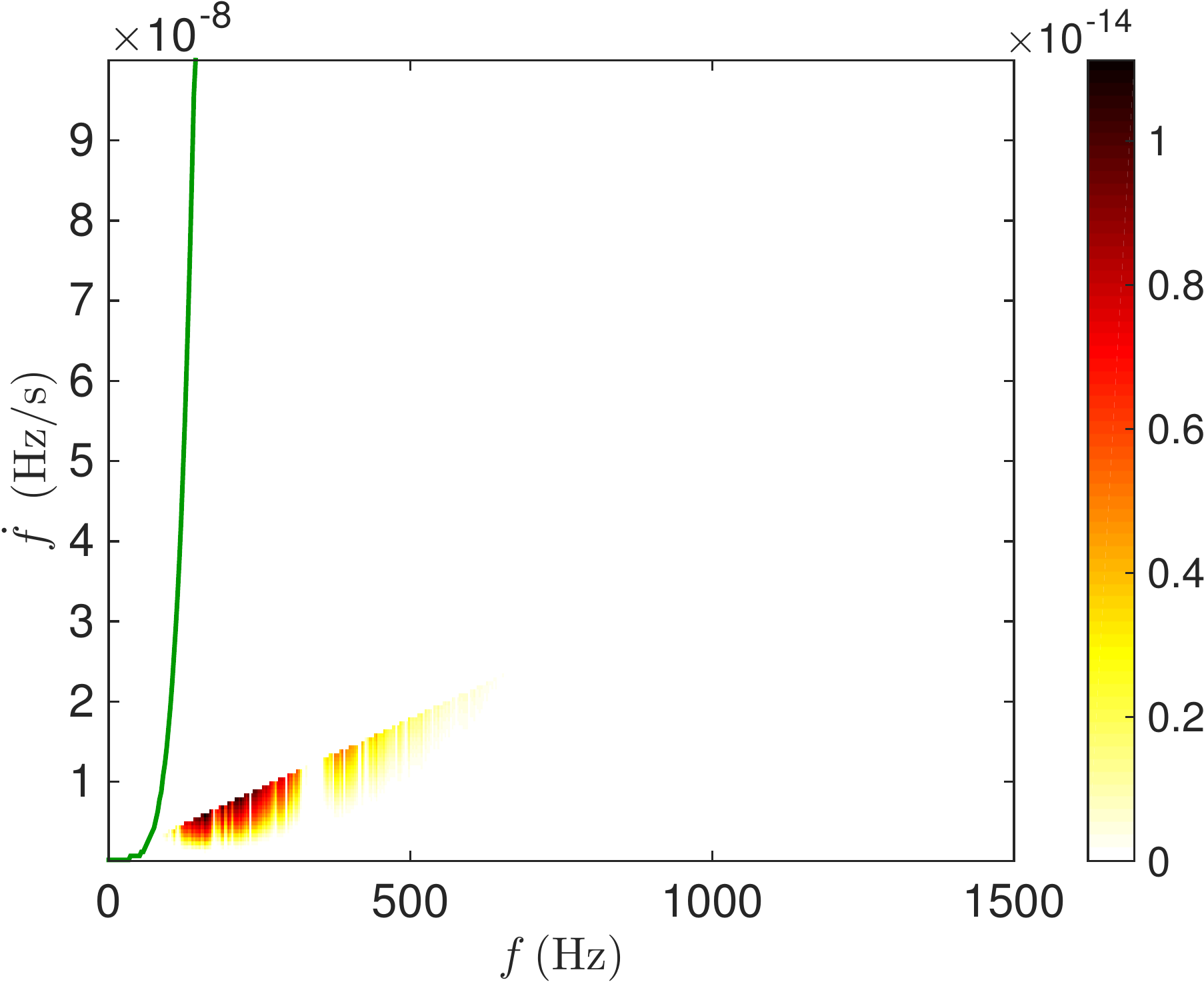}}}%
    \qquad
    \subfloat[Coverage, 10 days]{{  \includegraphics[width=.20\linewidth]{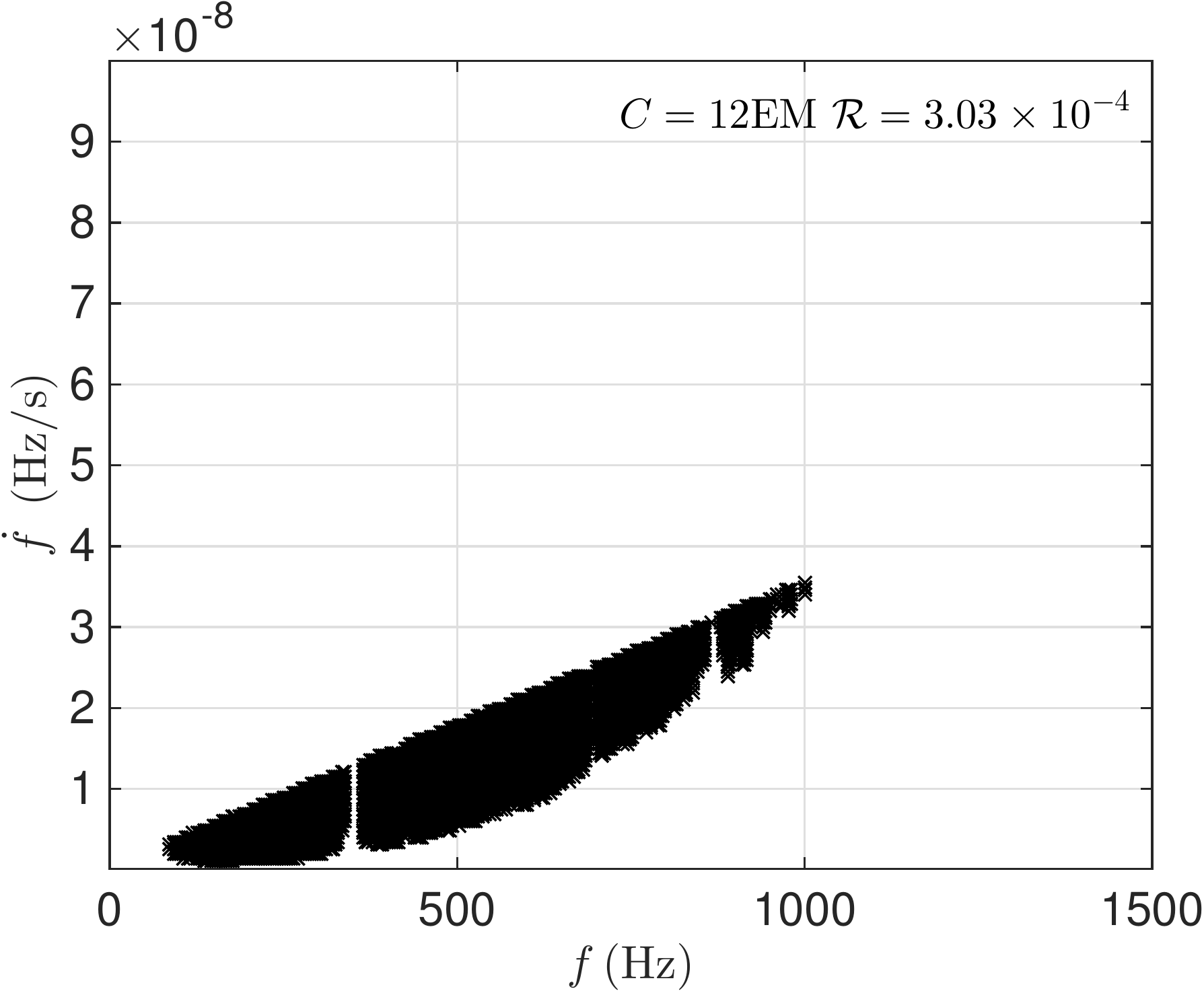}}}%
    \qquad
    \subfloat[Efficiency, 20 days]{{  \includegraphics[width=.20\linewidth]{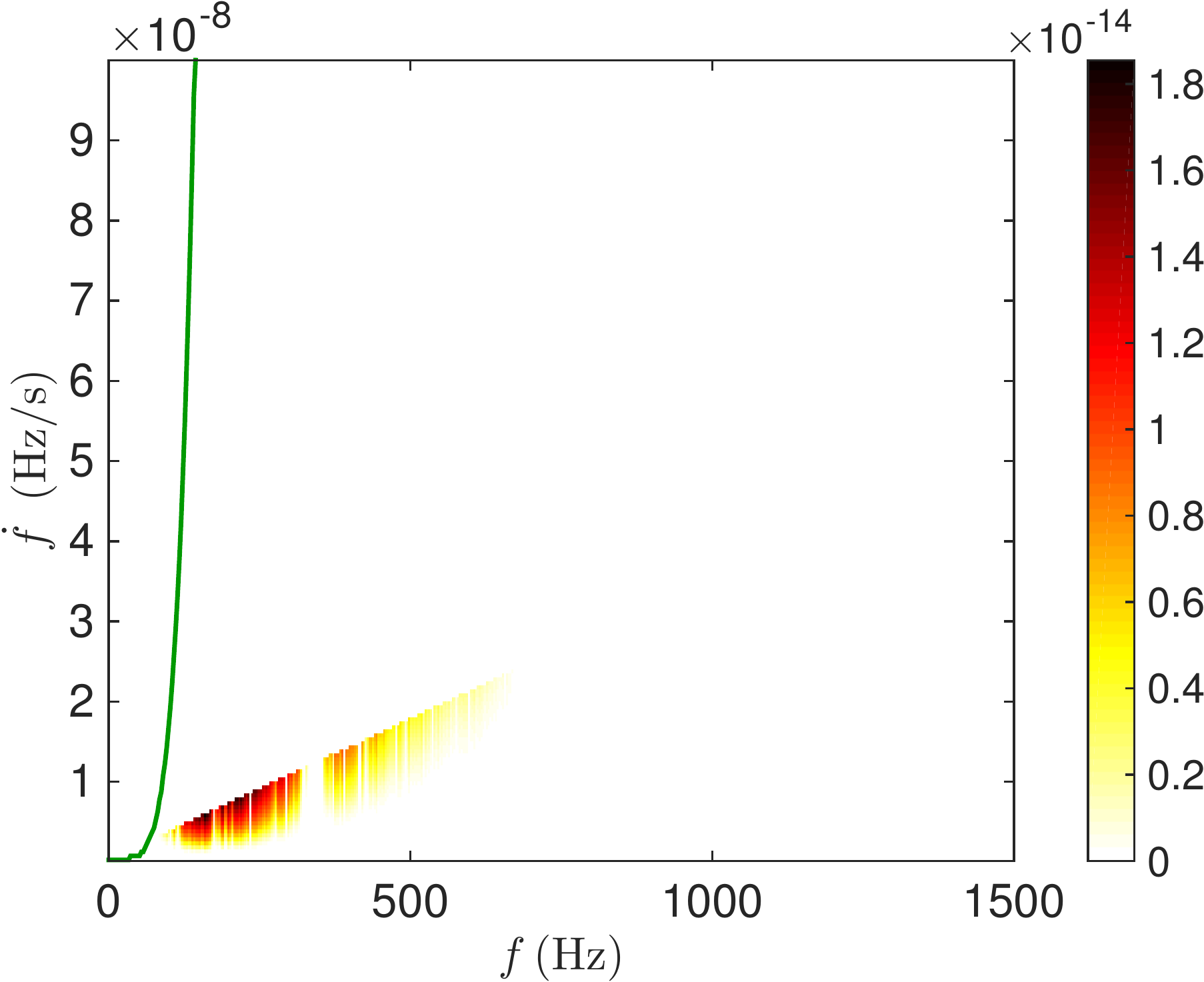}}}%
    \qquad
    \subfloat[Coverage, 20 days]{{  \includegraphics[width=.20\linewidth]{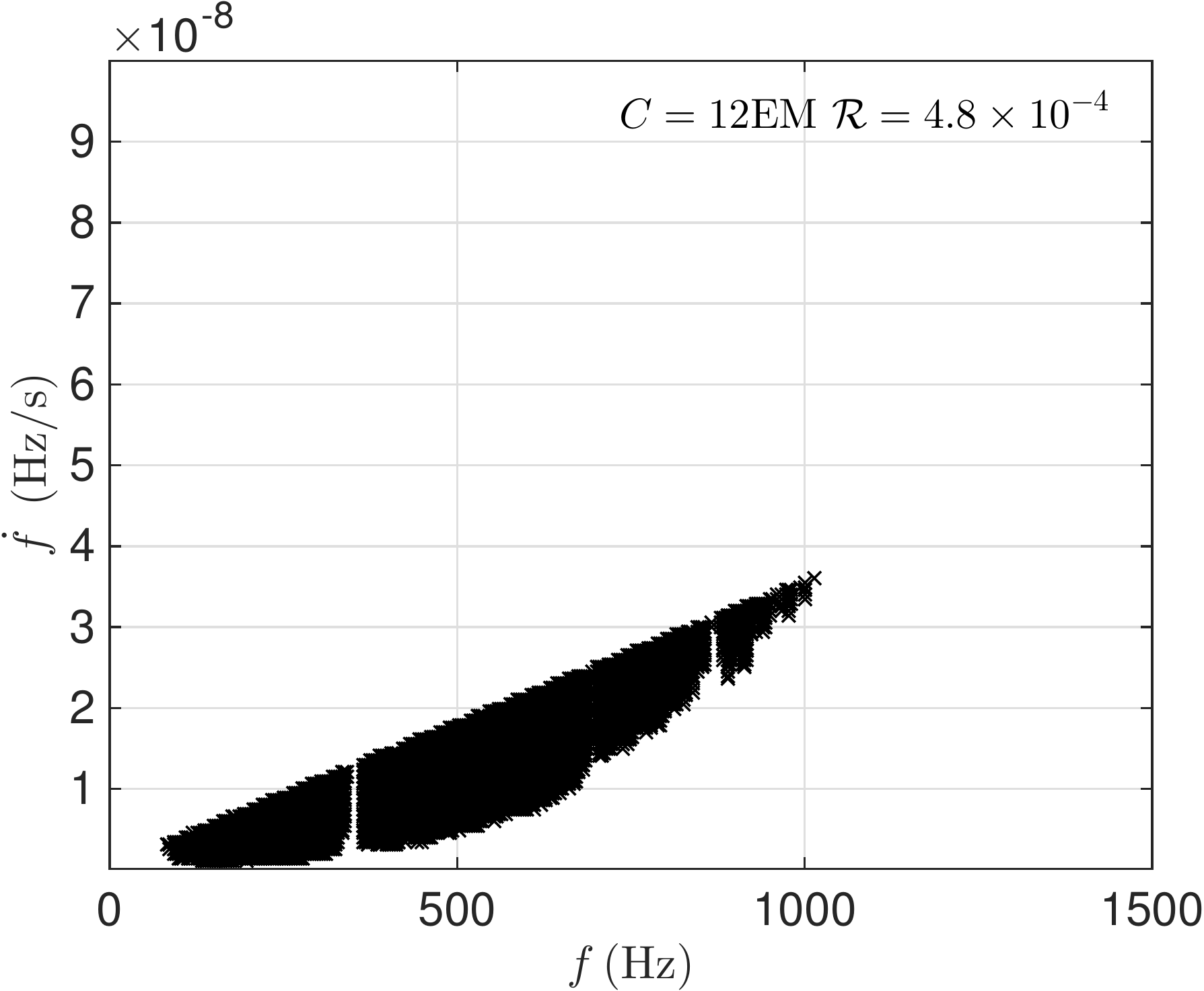}}}%
    \qquad
    \subfloat[Efficiency, 30 days]{{  \includegraphics[width=.20\linewidth]{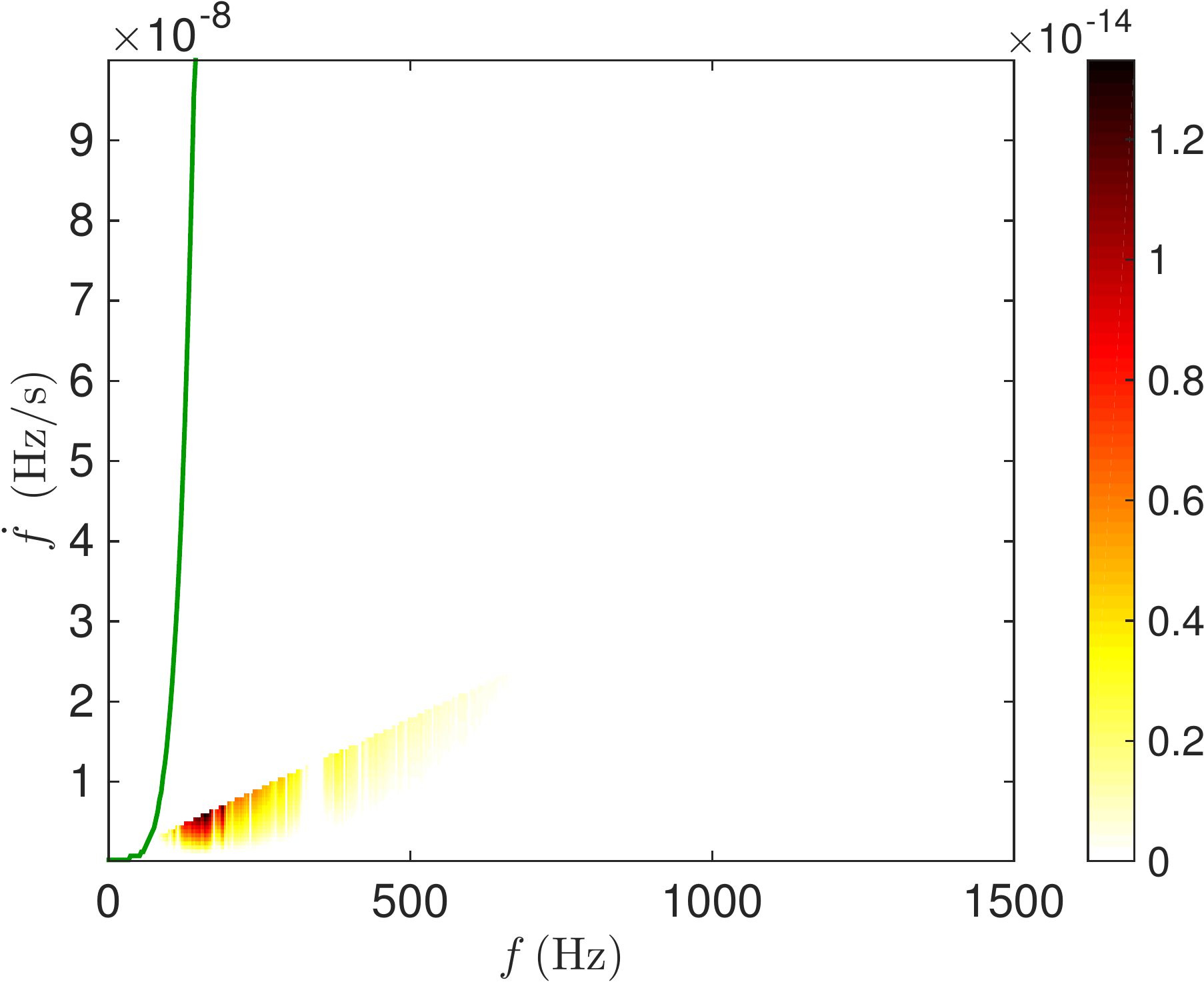}}}%
    \qquad
    \subfloat[Coverage, 30 days]{{  \includegraphics[width=.20\linewidth]{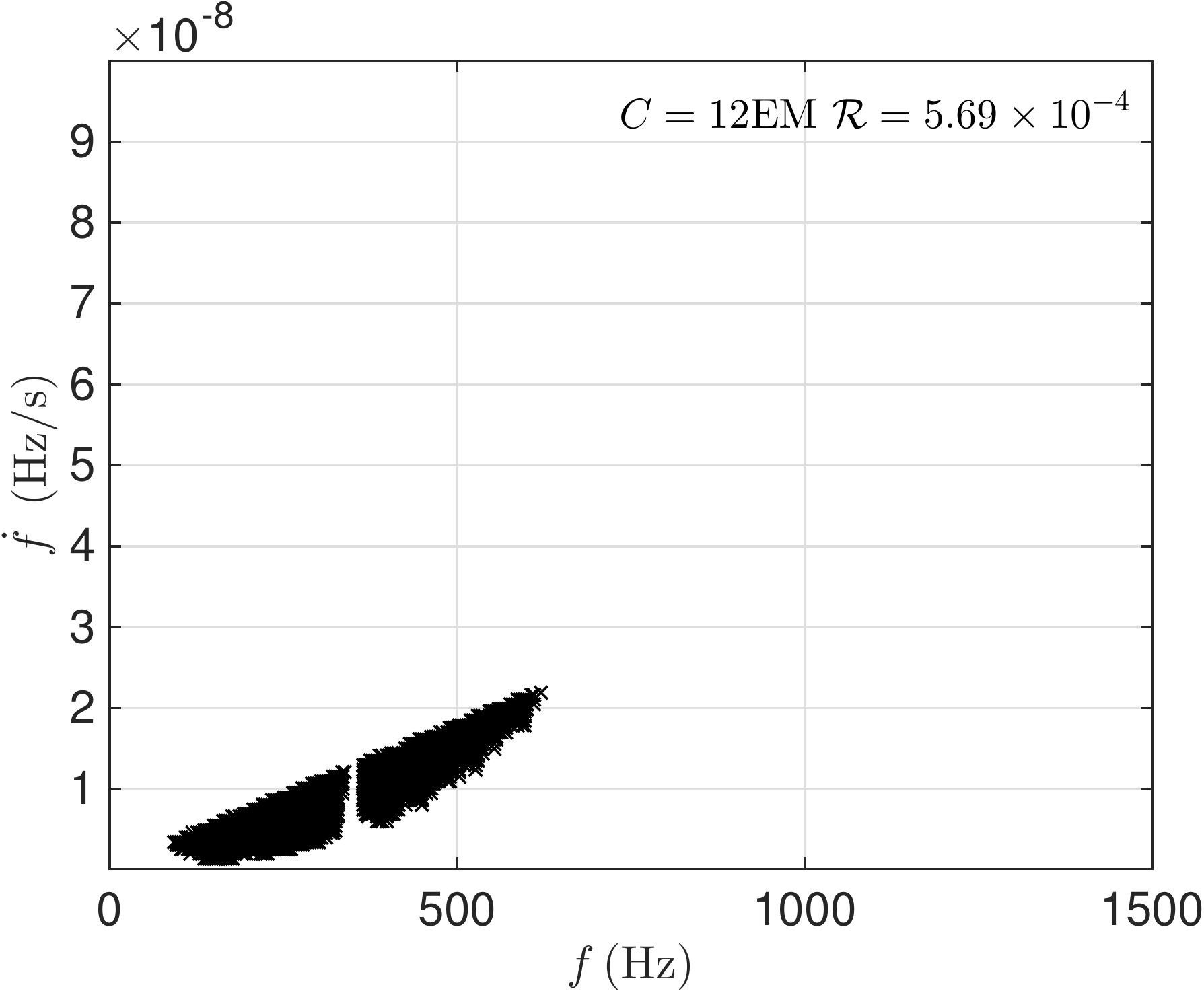}}}%
    \qquad
    \subfloat[Efficiency, 37.5 days]{{  \includegraphics[width=.20\linewidth]{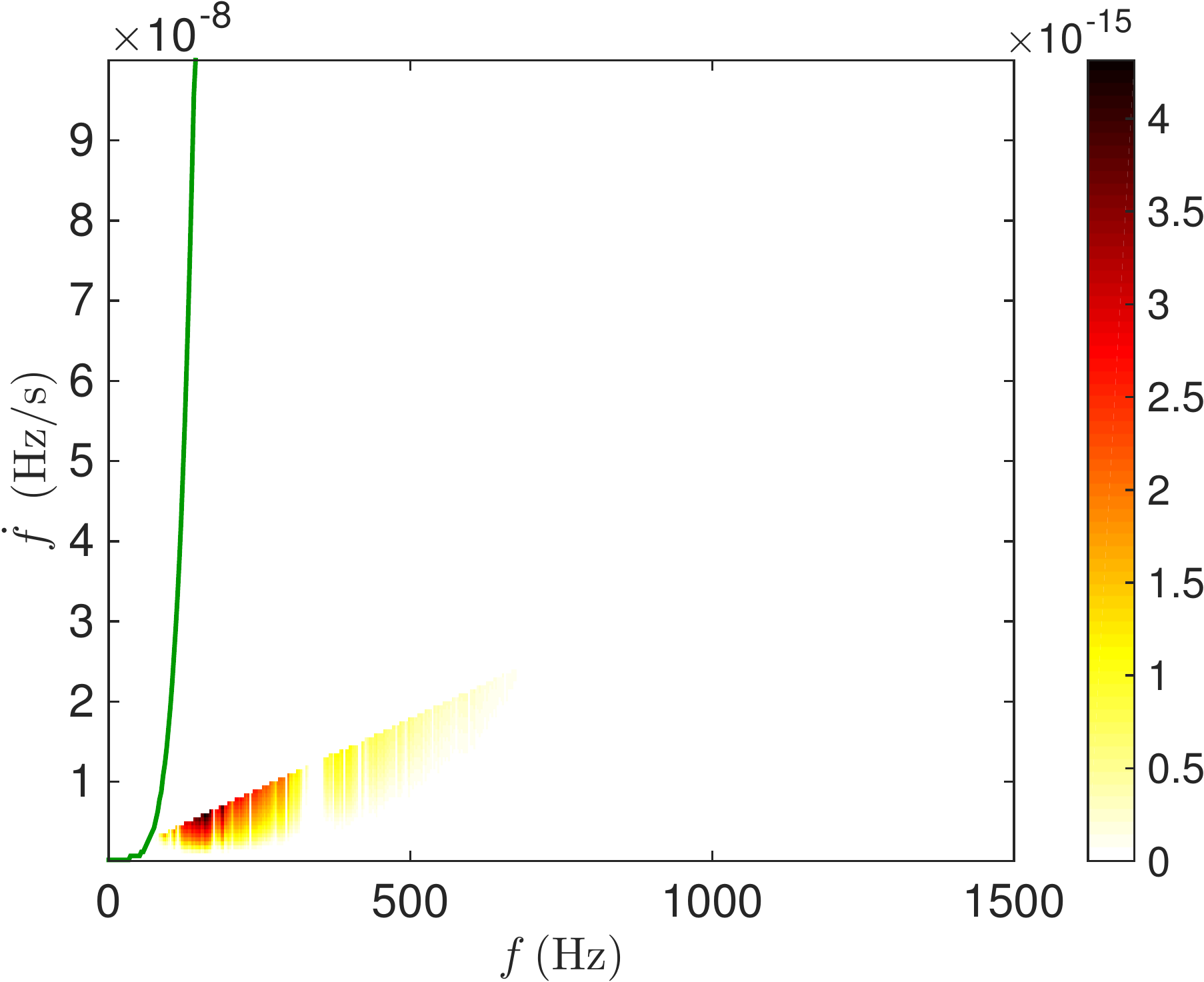}}}%
    \qquad
    \subfloat[Coverage, 37.5 days]{{  \includegraphics[width=.20\linewidth]{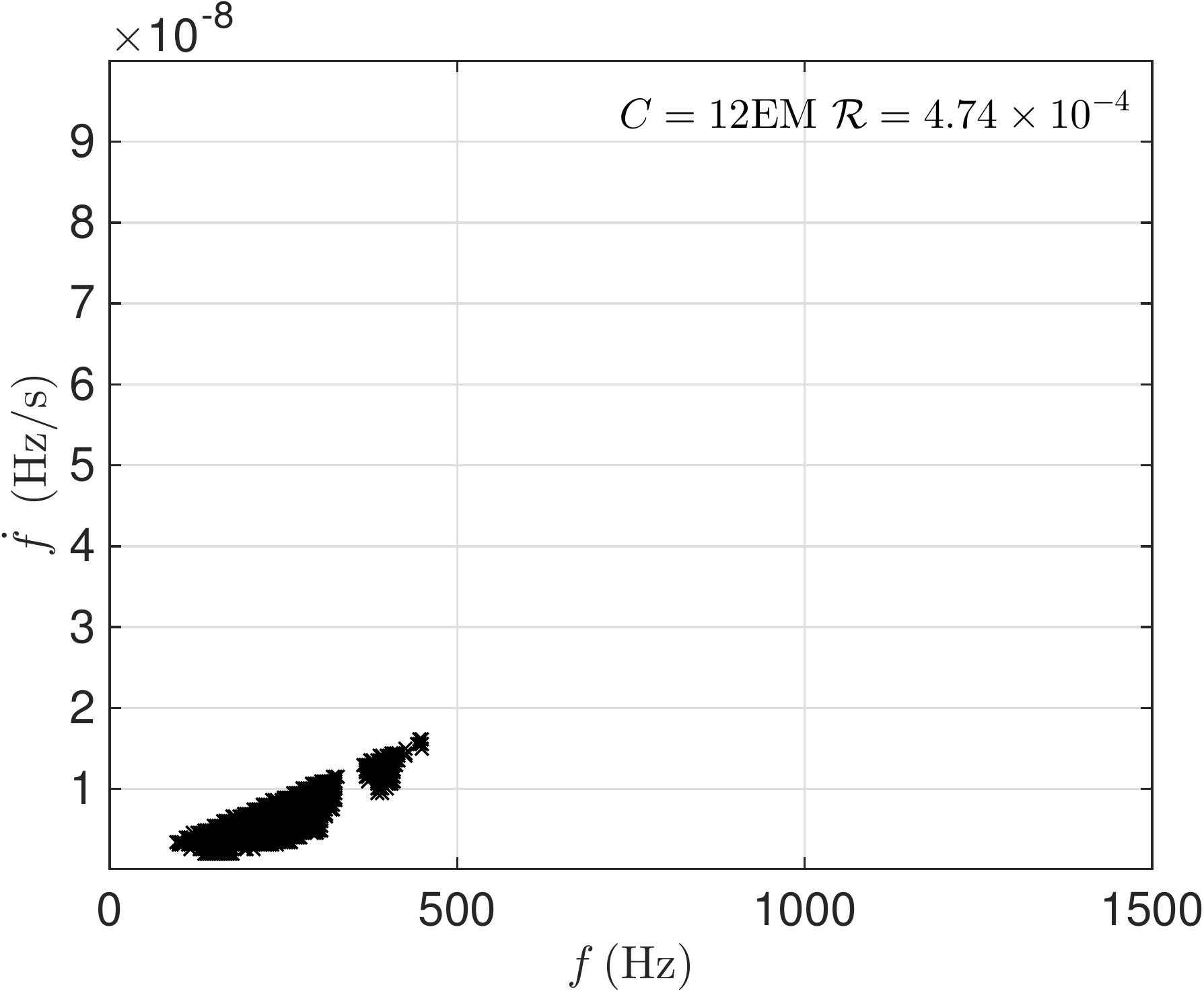}}}%
    \qquad
    \subfloat[Efficiency, 50 days]{{  \includegraphics[width=.20\linewidth]{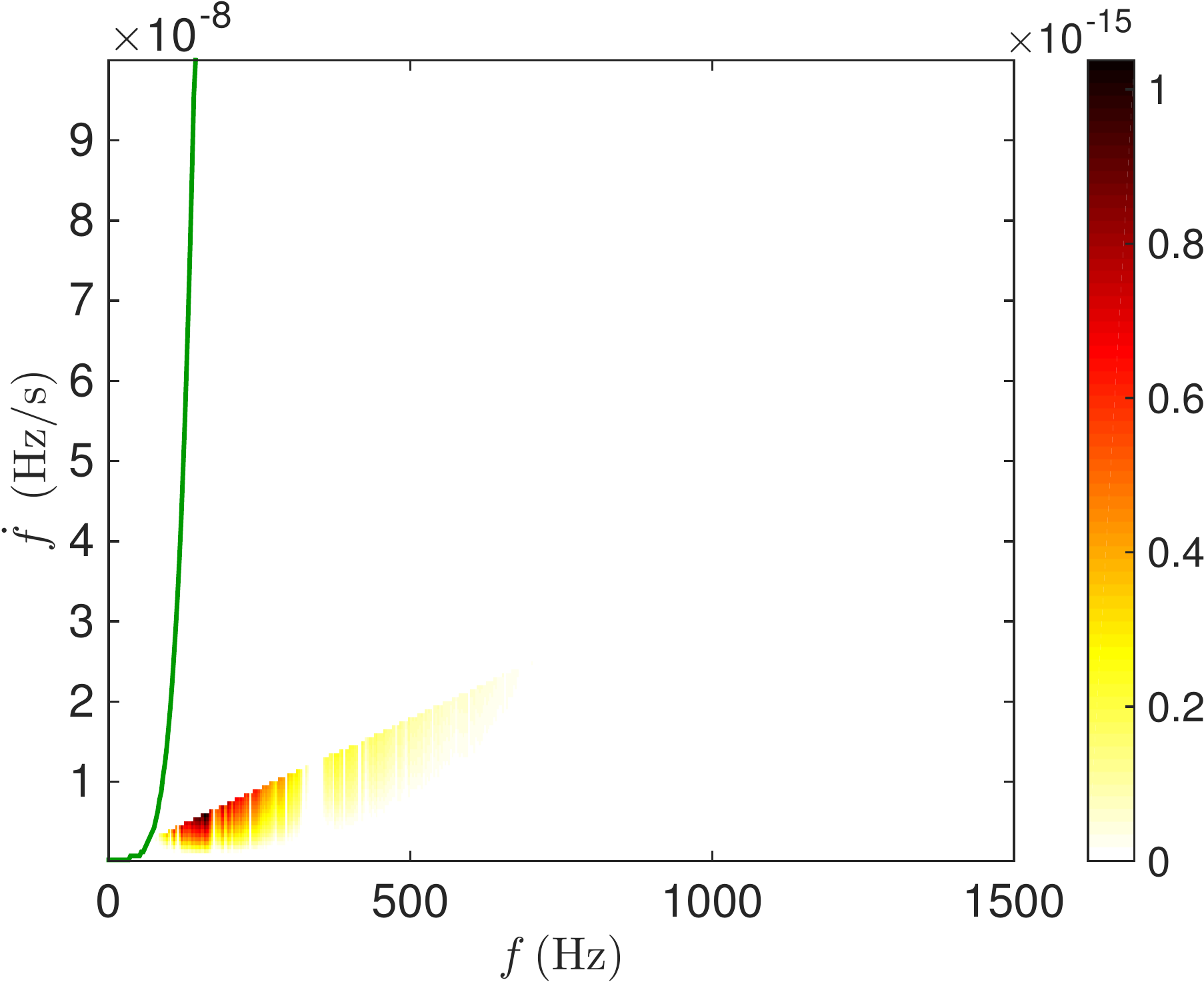}}}%
    \qquad
    \subfloat[Coverage, 50 days]{{  \includegraphics[width=.20\linewidth]{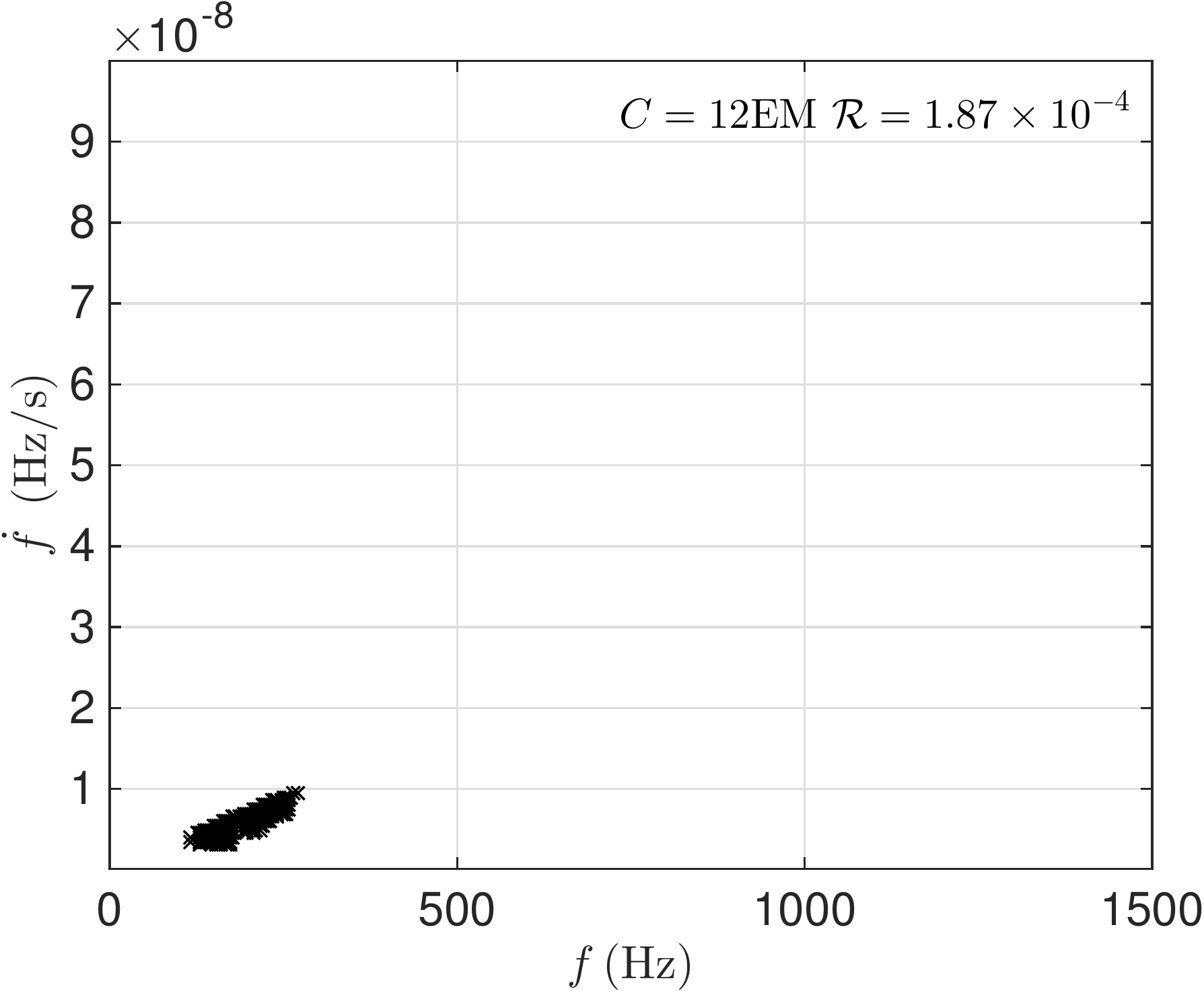}}}%
    \qquad
    \subfloat[Efficiency, 75 days]{{  \includegraphics[width=.20\linewidth]{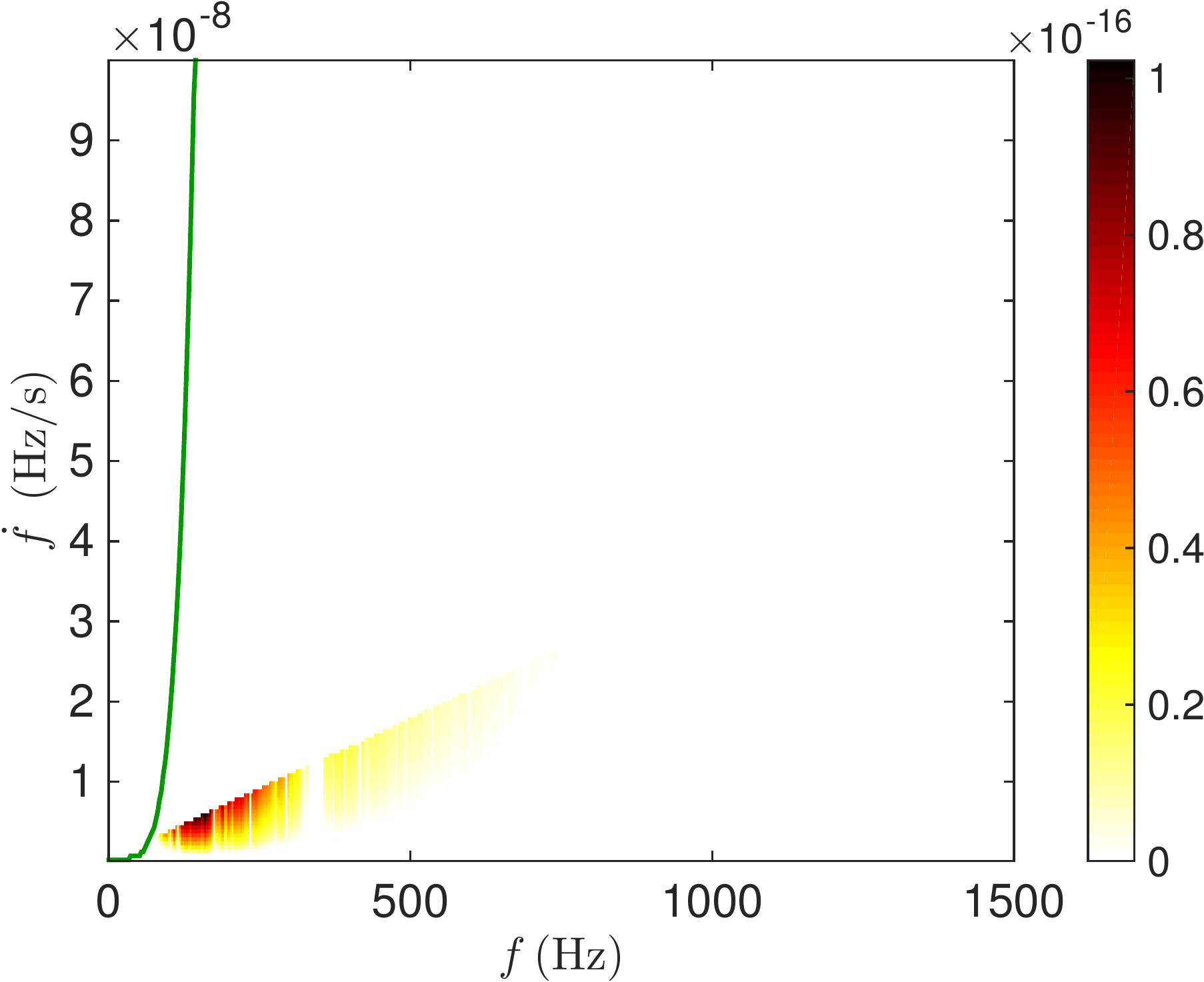}}}%
    \qquad
    \subfloat[Coverage, 75 days]{{  \includegraphics[width=.20\linewidth]{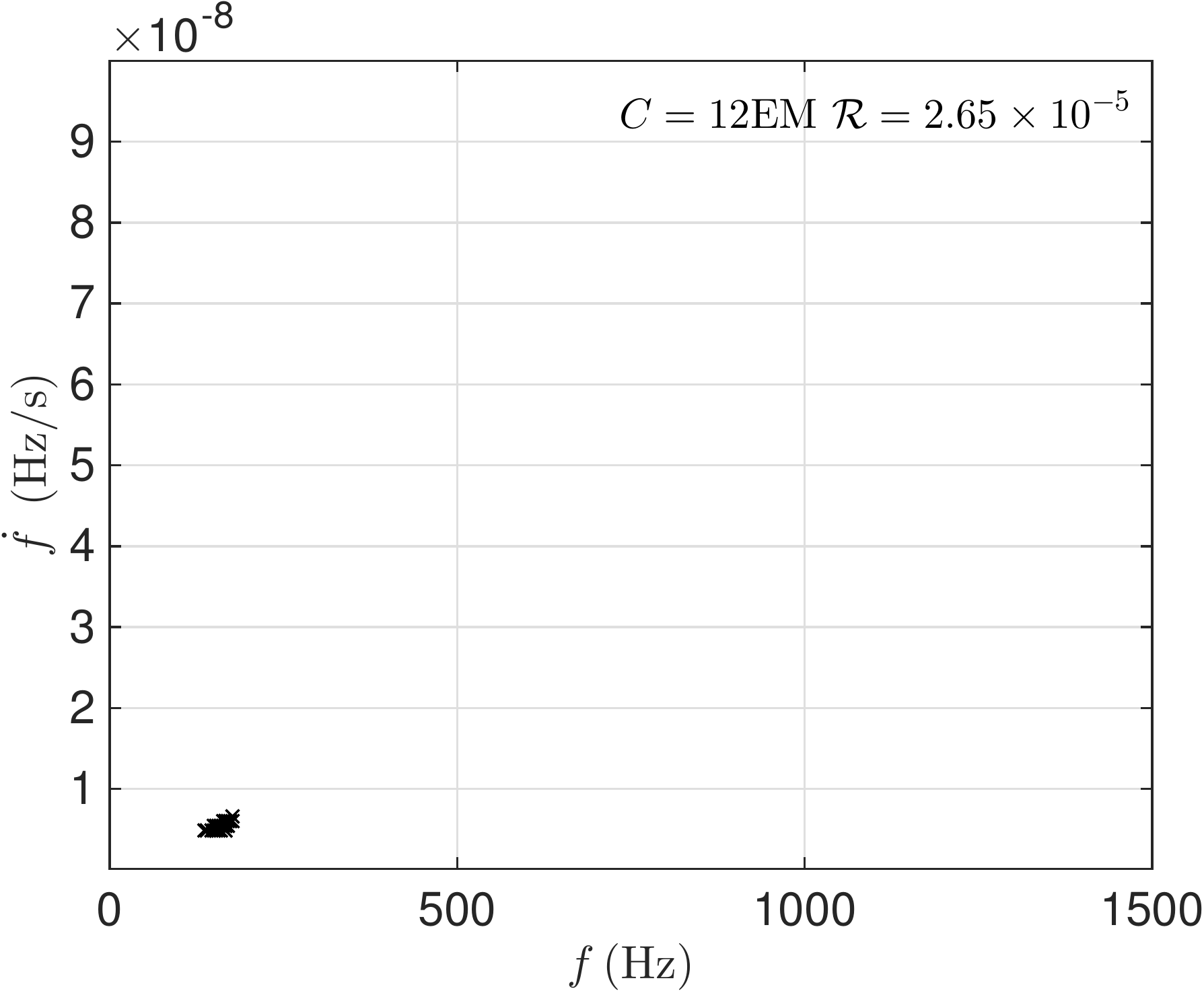}}}%
    
    \caption{Optimisation results for G350.1 at 4500 pc, 900 years old, assuming uniform and age-based priors, for various coherent search durations: 5, 10, 20, 30, 37.5, 50 and 75 days. The total computing budget is assumed to be 12 EM. }%
    \label{G3501_51020days}%
\end{figure*}

\begin{figure*}%
     \centering
     \subfloat[Coverage, cost: 12 EM]{{  \includegraphics[width=.45\linewidth]{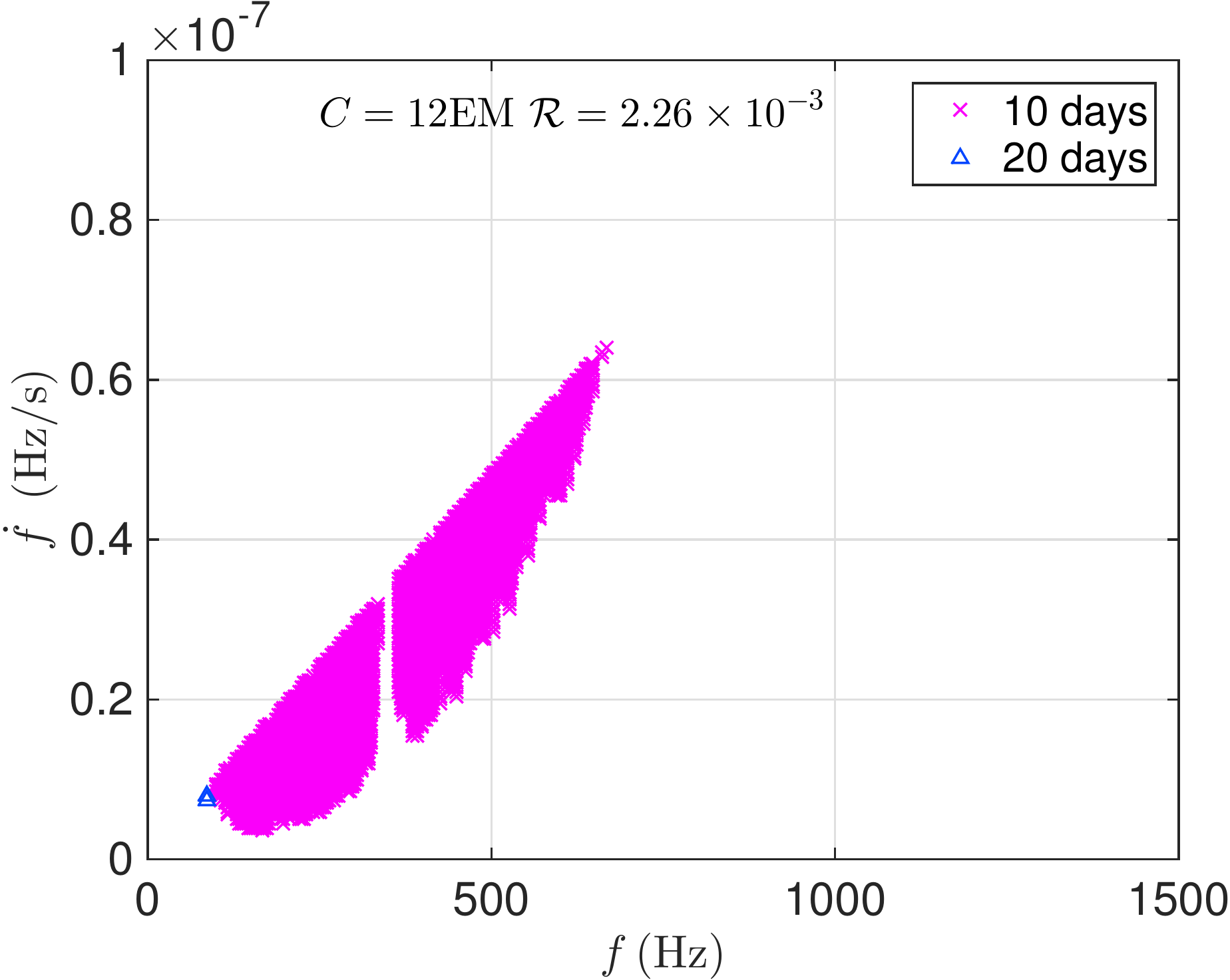}}}%
     \qquad
     \subfloat[Coverage,  cost: 24 EM]{{  \includegraphics[width=.45\linewidth]{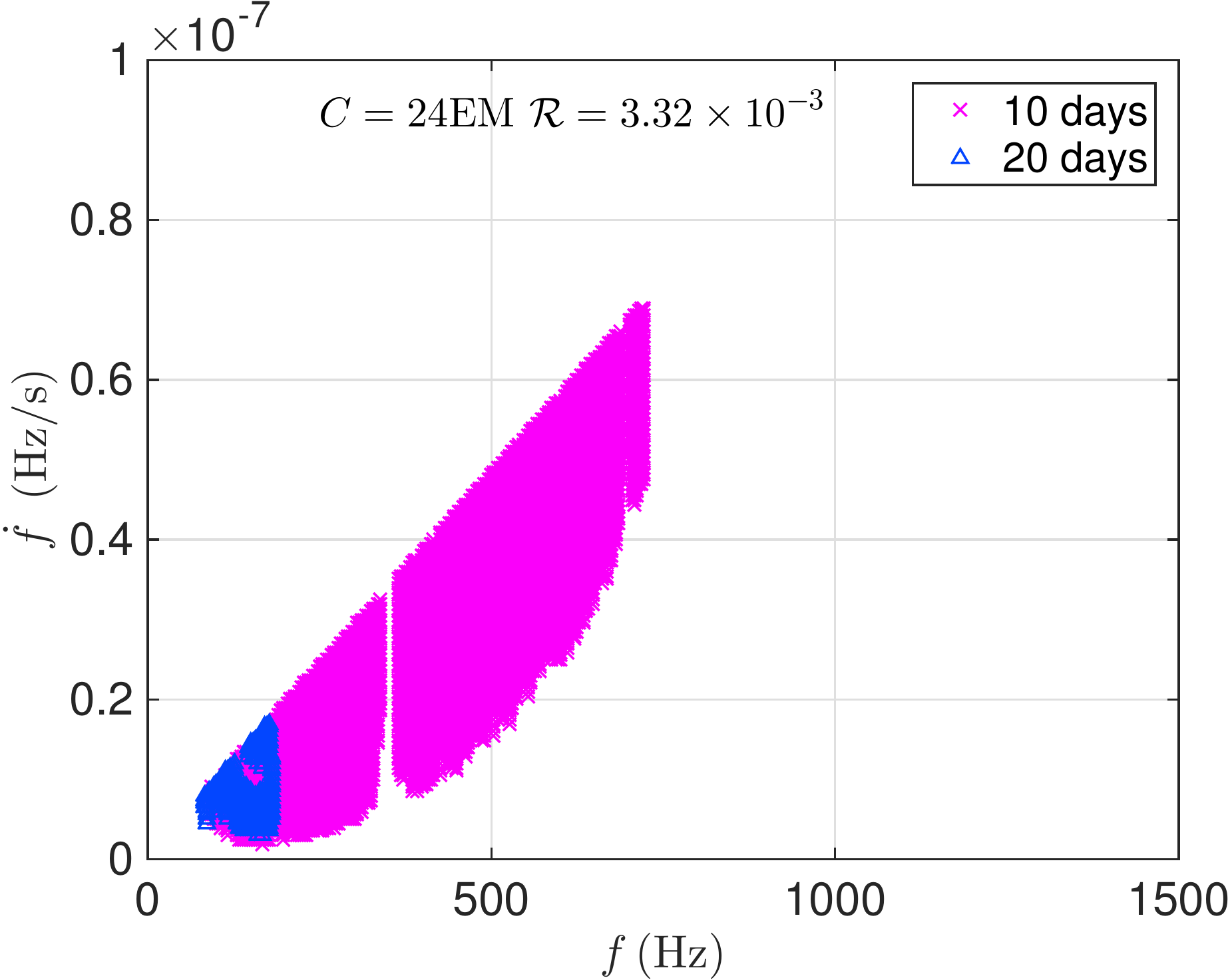}}}%
     \caption{Parameter space coverage for Cas A at 3500 pc, 330 years old, assuming uniform and age-based priors and optimizing over the 7  search set-ups also considered above at 12 EM (left plot) and 24 EM (right plot).}%
     \label{CasA_best_age}%
 \end{figure*}

  \begin{figure*}%
     \centering
     \subfloat[Coverage, cost: 12 EM]{{  \includegraphics[width=.45\linewidth]{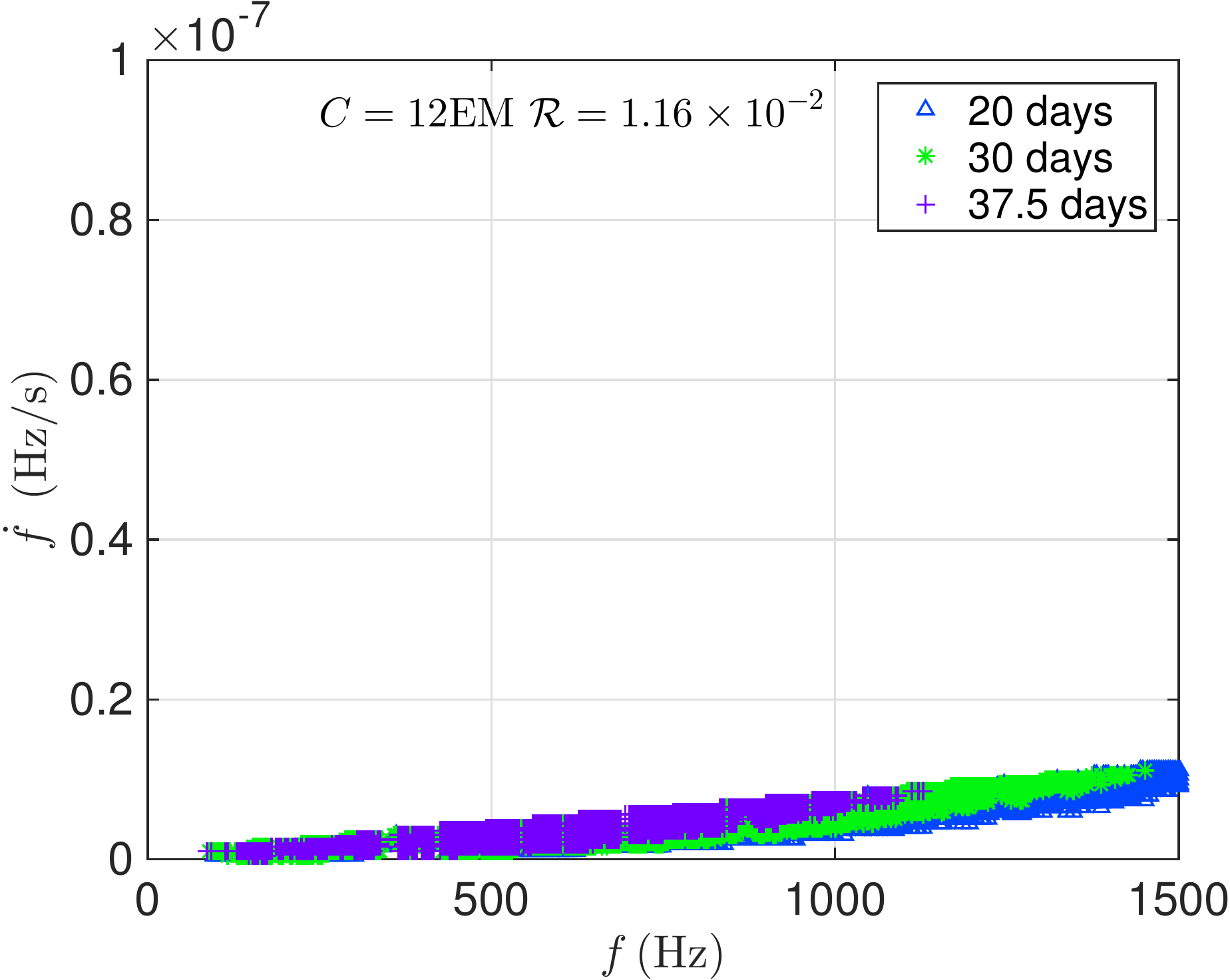}}}%
     \qquad
     \subfloat[Coverage,  cost: 24 EM]{{  \includegraphics[width=.45\linewidth]{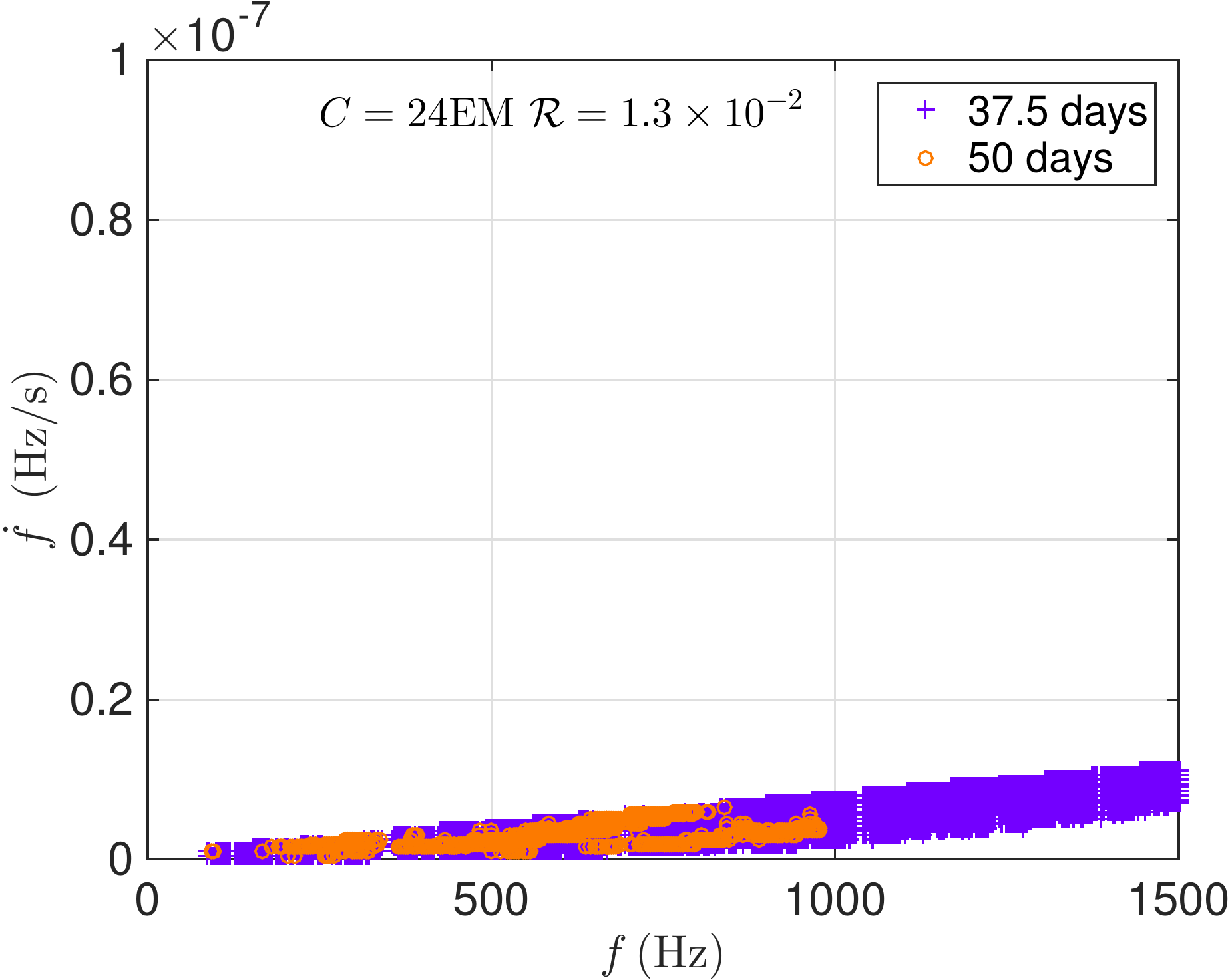}}}%
     \caption{Parameter space coverage for Vela Jr at 750 pc, 4300 years old, assuming uniform and age-based priors and optimizing over the 7  search set-ups also considered above at 12 EM (left plot) and 24 EM (right plot).}%
     \label{2662_best_FO}%
 \end{figure*}

 \begin{figure*}%
     \centering
     \subfloat[Coverage, cost: 12 EM]{{  \includegraphics[width=.45\linewidth]{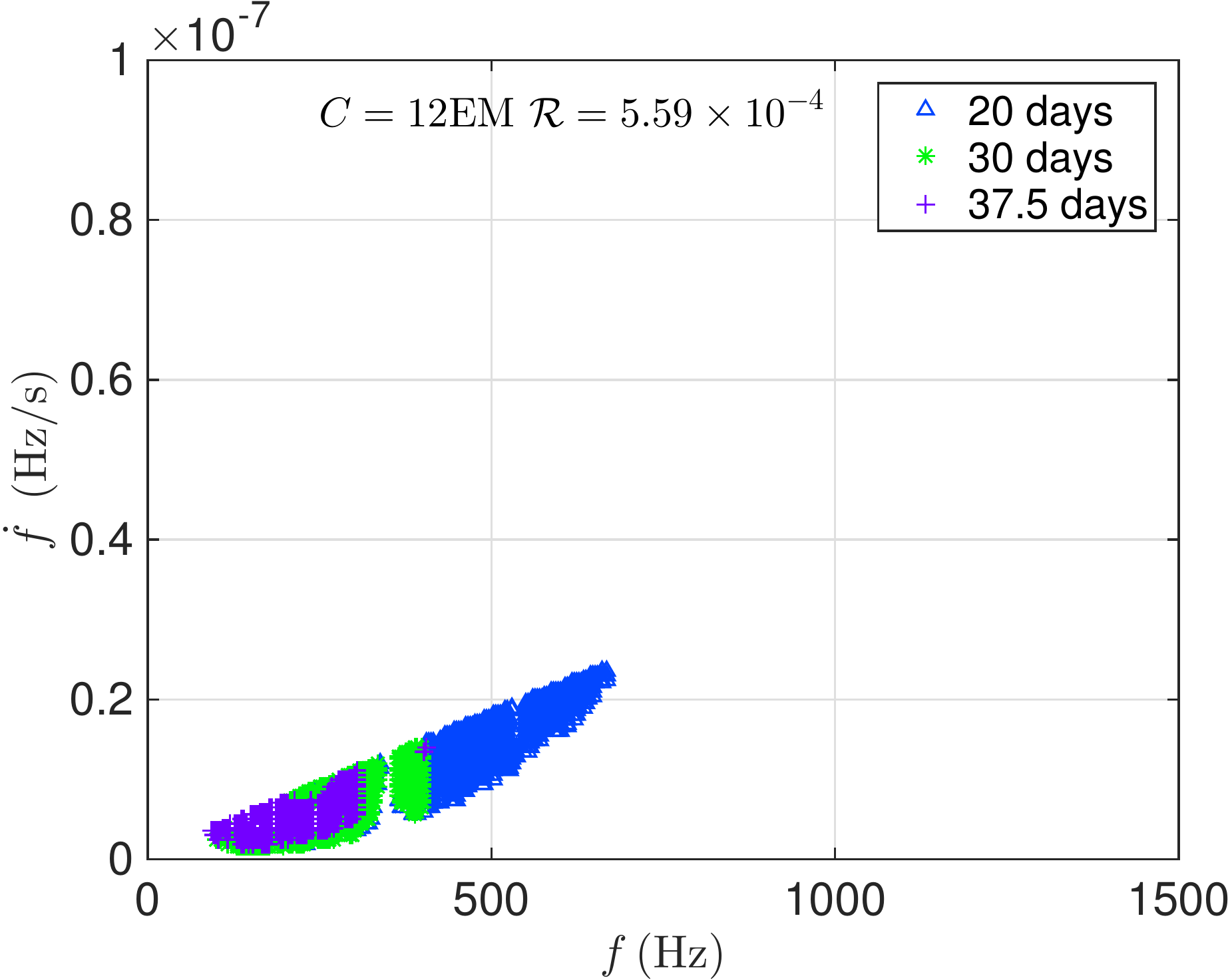}}}%
     \qquad
     \subfloat[Coverage,  cost: 24 EM]{{  \includegraphics[width=.45\linewidth]{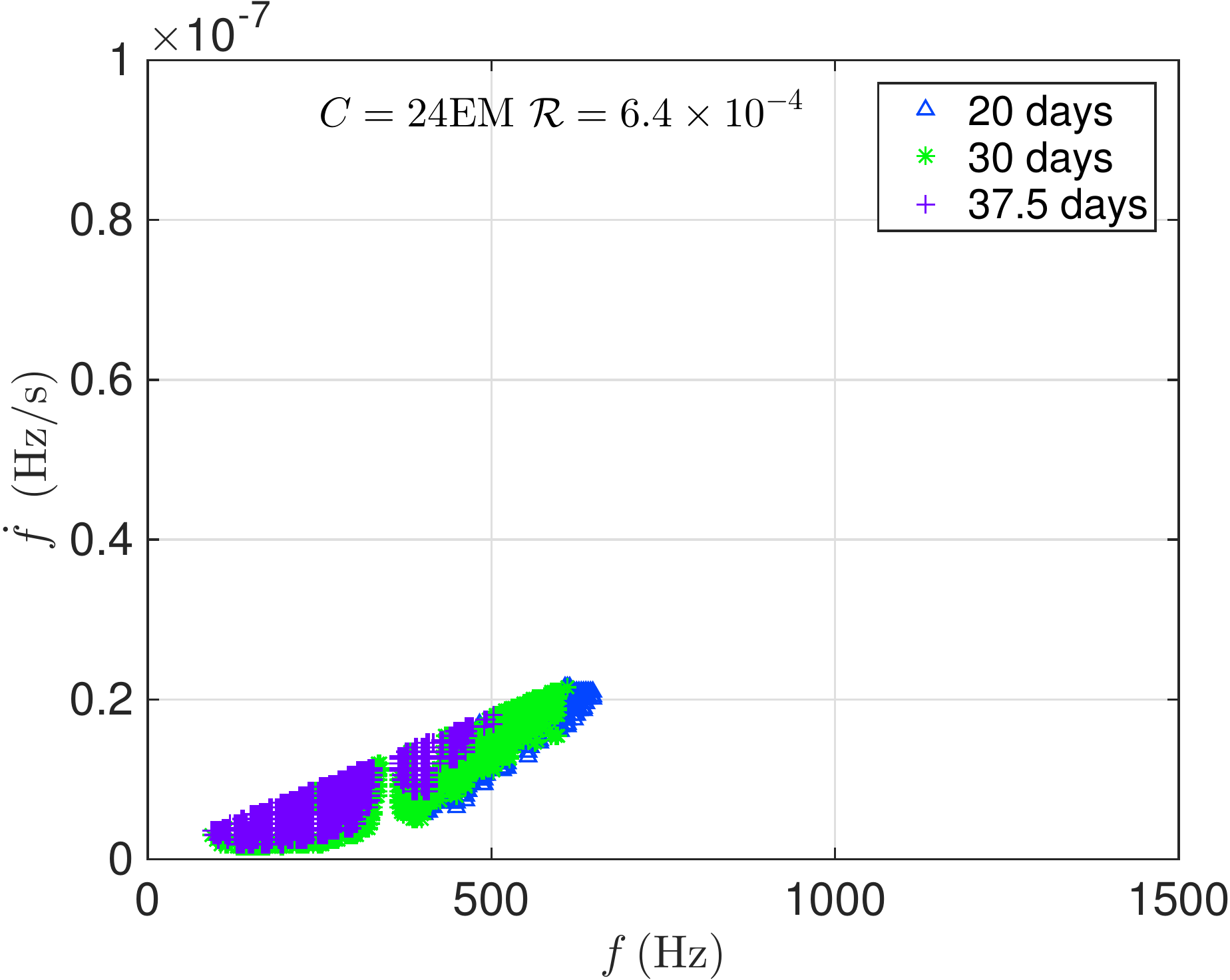}}}%
     \caption{Parameter space coverage for G350.1 at 4500 pc, 900 years old, assuming uniform and age-based priors and optimizing over the 7  search set-ups also considered above at 12 EM (left plot) and 24 EM (right plot).}%
     \label{3501_best_age}%
 \end{figure*}



\begin{figure*}%
    \centering
    \subfloat[Efficiency(lg), 5 days]{{  \includegraphics[width=.20\linewidth]{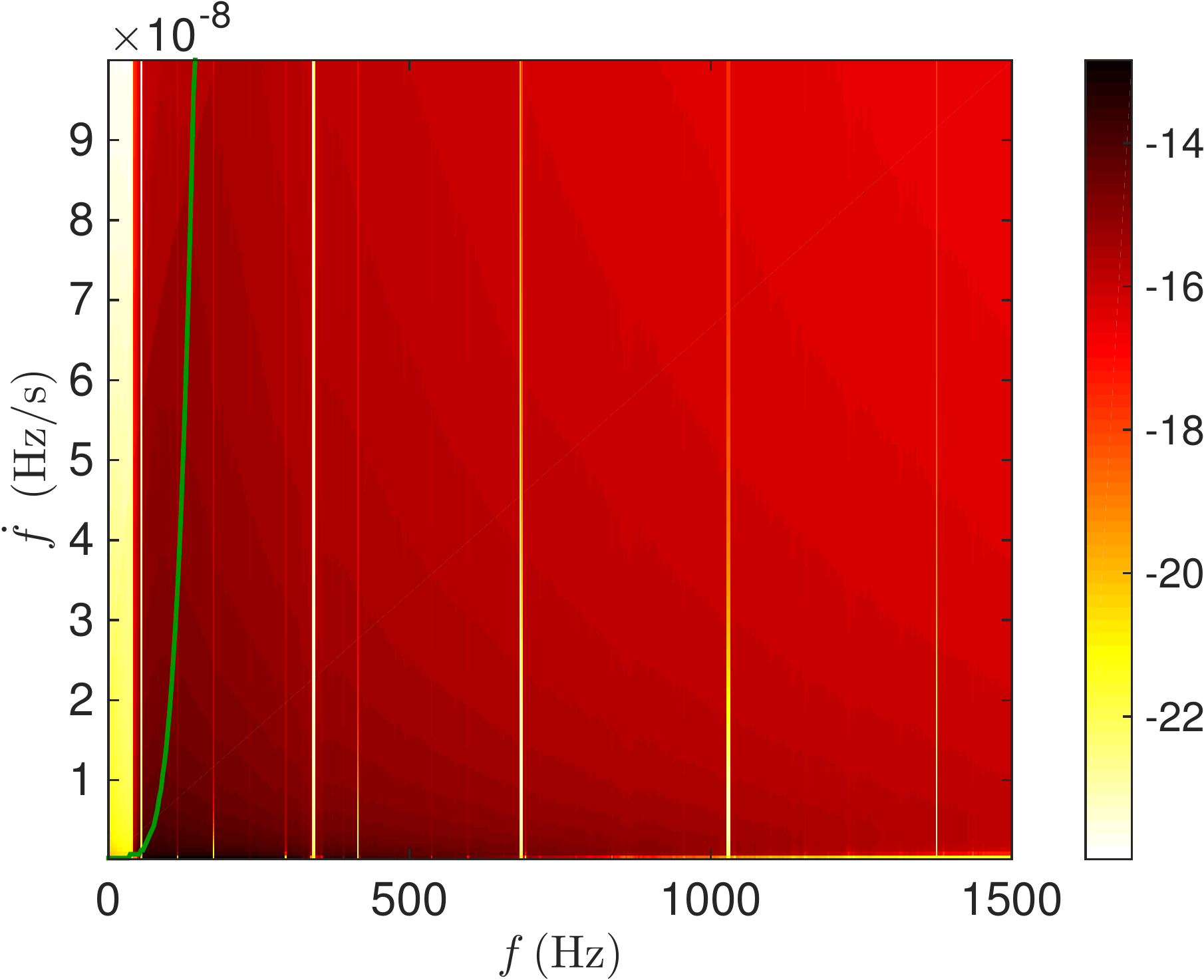}}}%
    \qquad
    \subfloat[Coverage, 5 days]{{  \includegraphics[width=.20\linewidth]{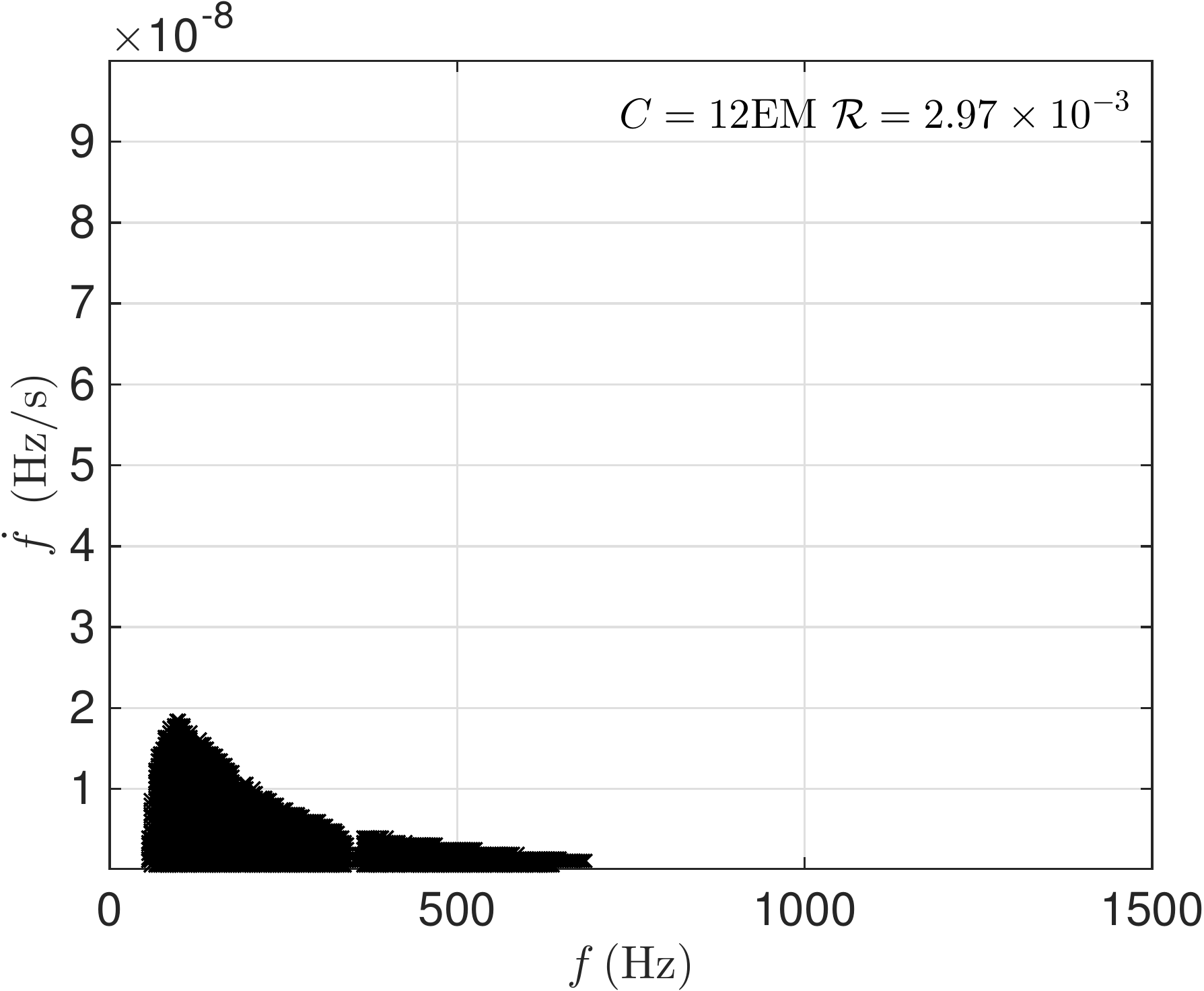}}}%
    \qquad
    \subfloat[Efficiency(lg), 10 days]{{  \includegraphics[width=.20\linewidth]{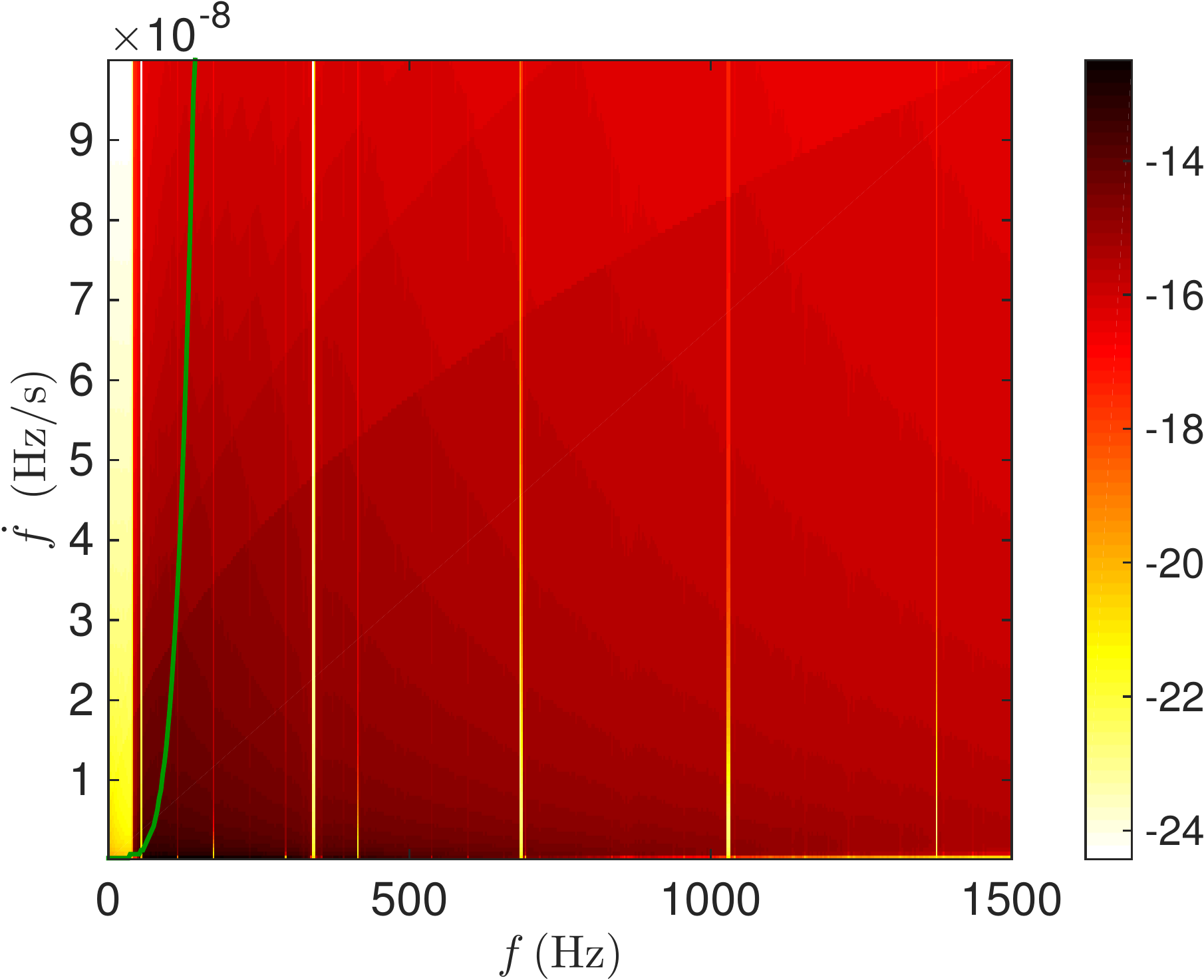}}}%
    \qquad
    \subfloat[Coverage, 10 days]{{  \includegraphics[width=.20\linewidth]{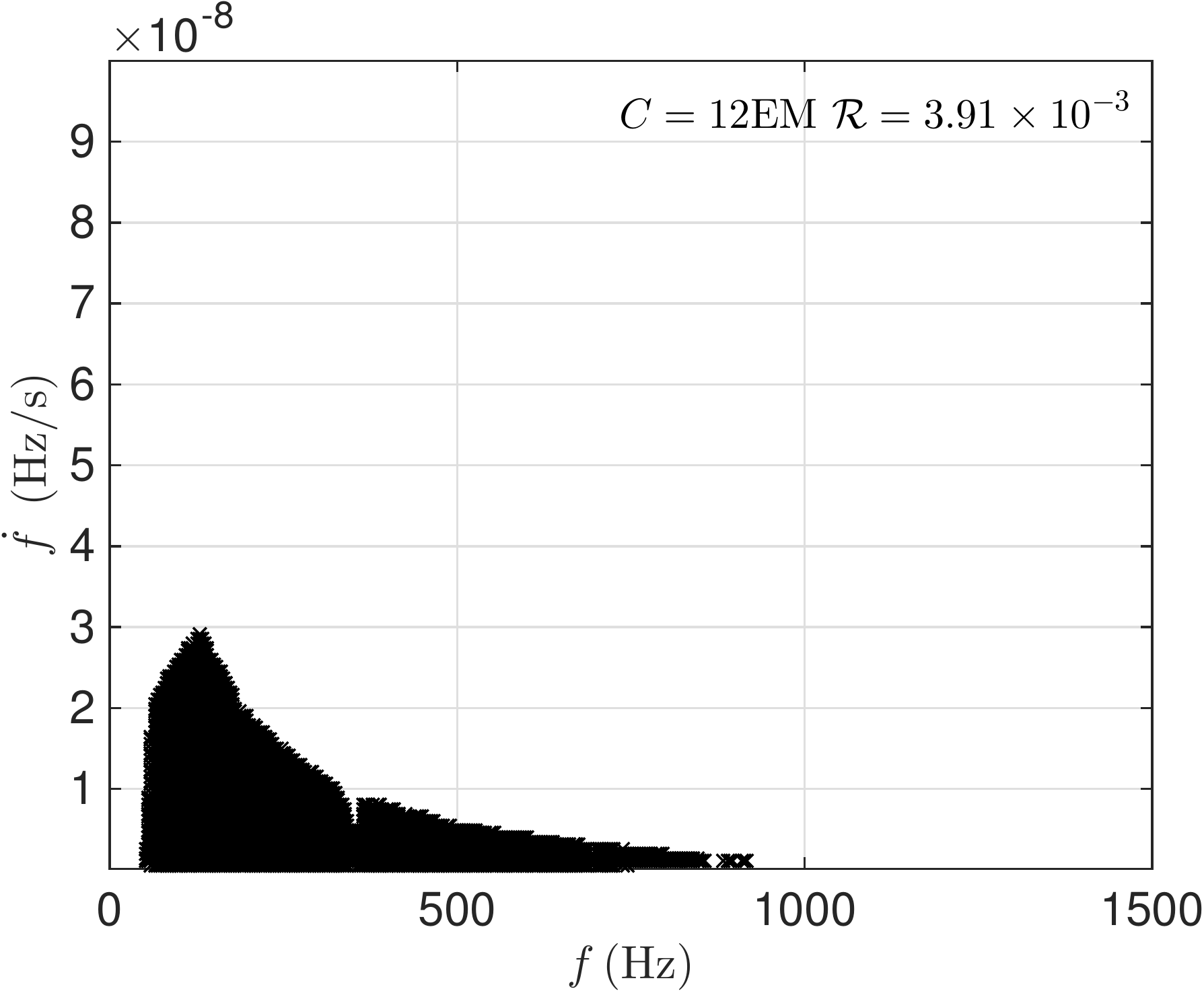}}}%
    \qquad
    \subfloat[Efficiency(lg), 20 days]{{  \includegraphics[width=.20\linewidth]{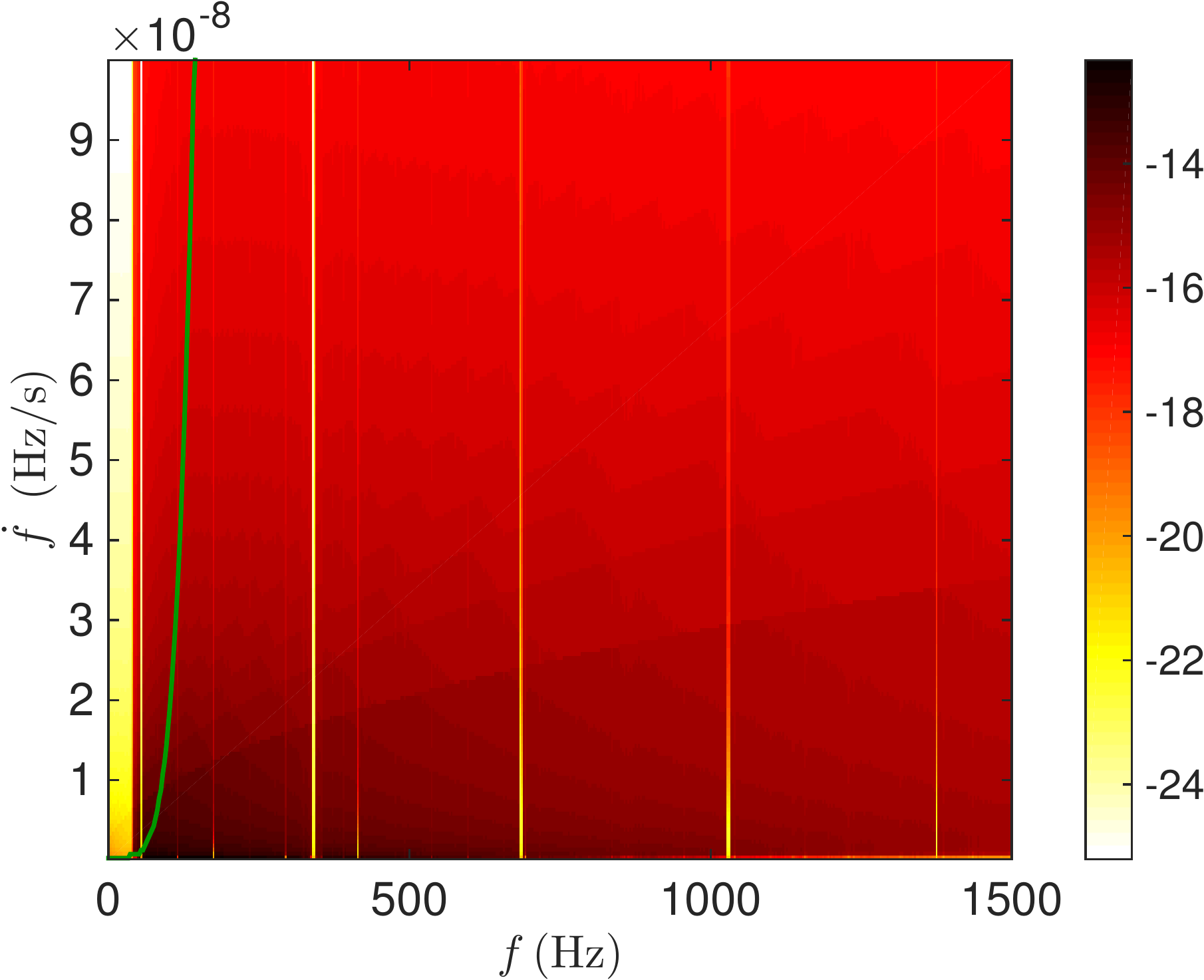}}}%
    \qquad
    \subfloat[Coverage, 20 days]{{  \includegraphics[width=.20\linewidth]{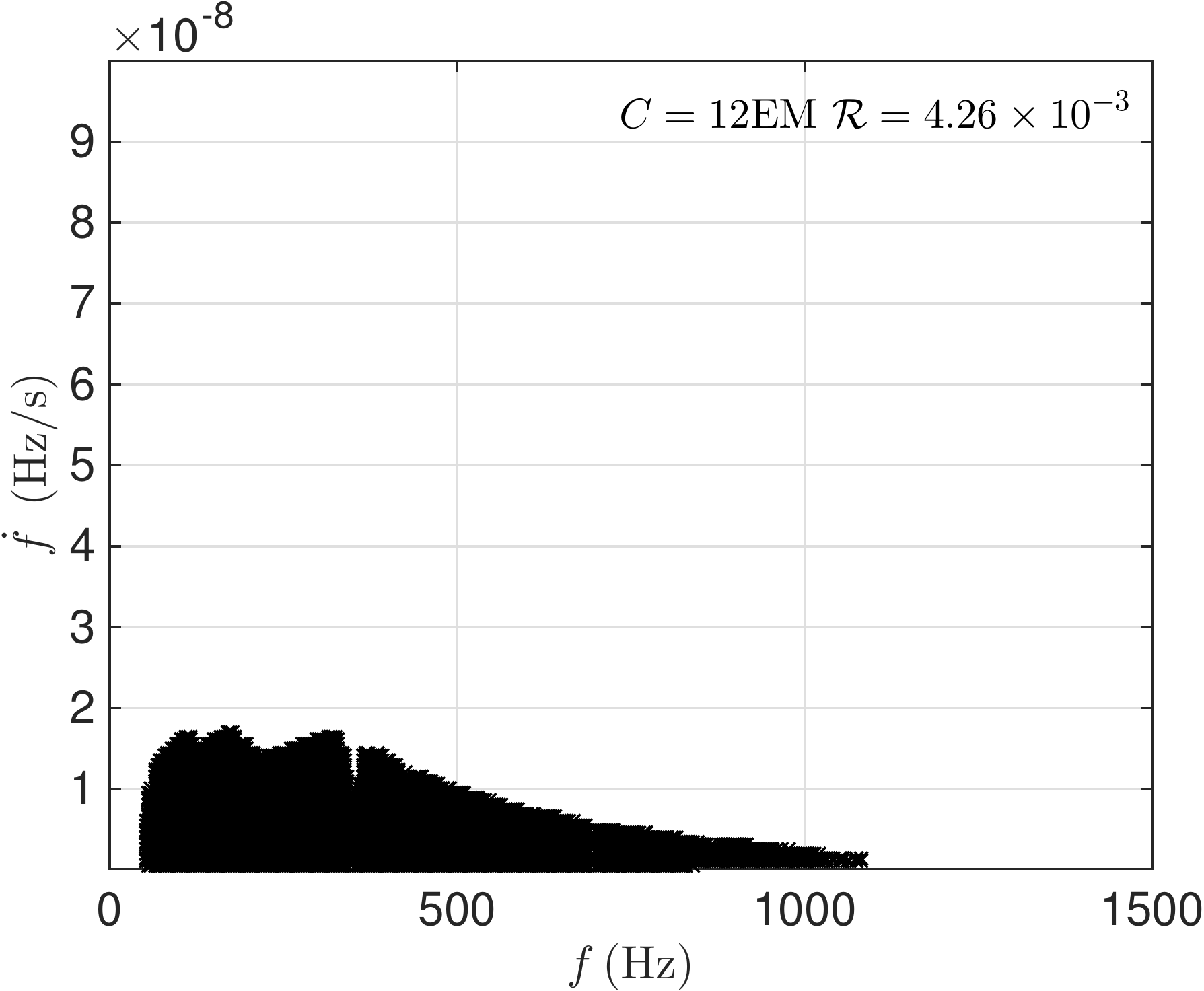}}}%
    \qquad
    \subfloat[Efficiency(lg), 30 days]{{  \includegraphics[width=.20\linewidth]{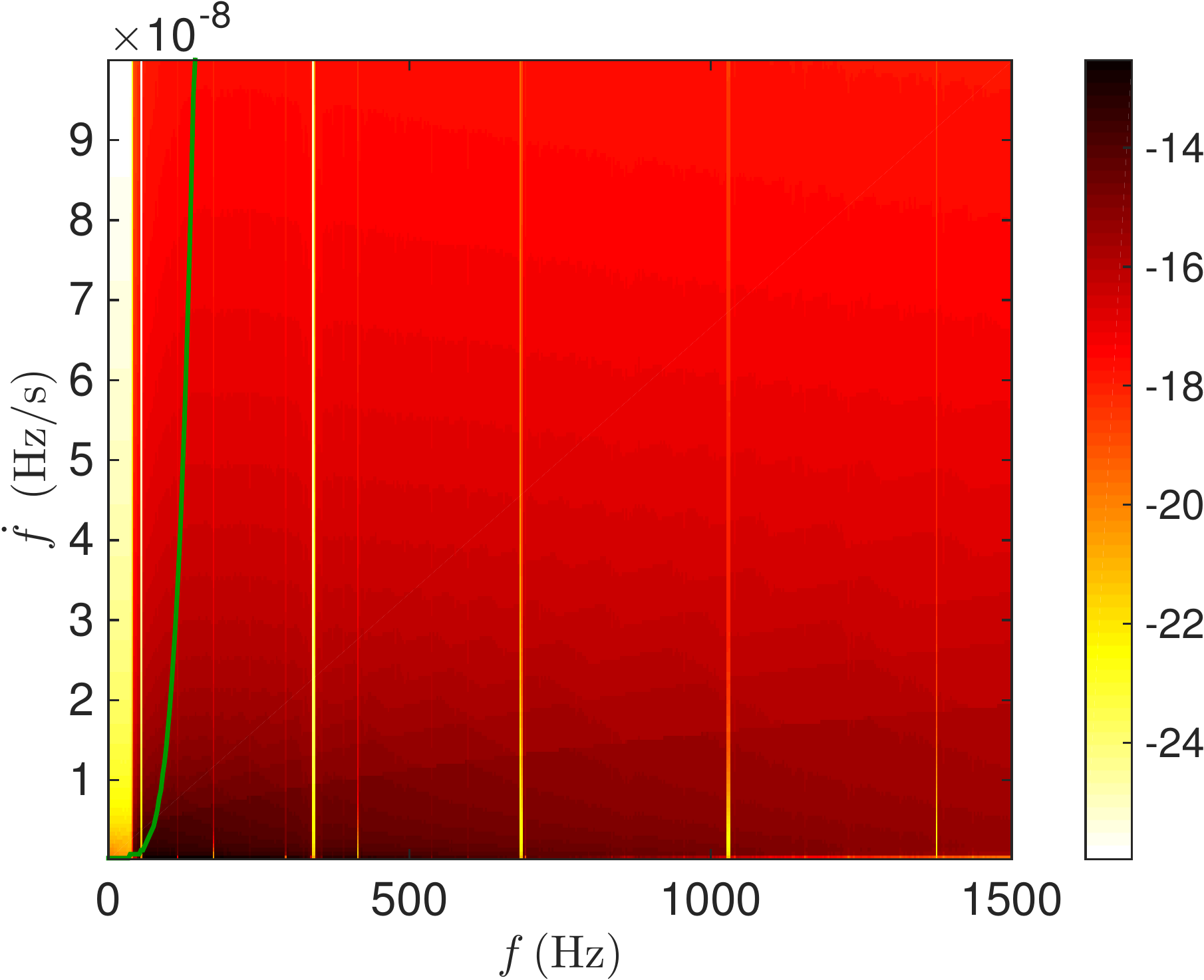}}}%
    \qquad
    \subfloat[Coverage, 30 days]{{  \includegraphics[width=.20\linewidth]{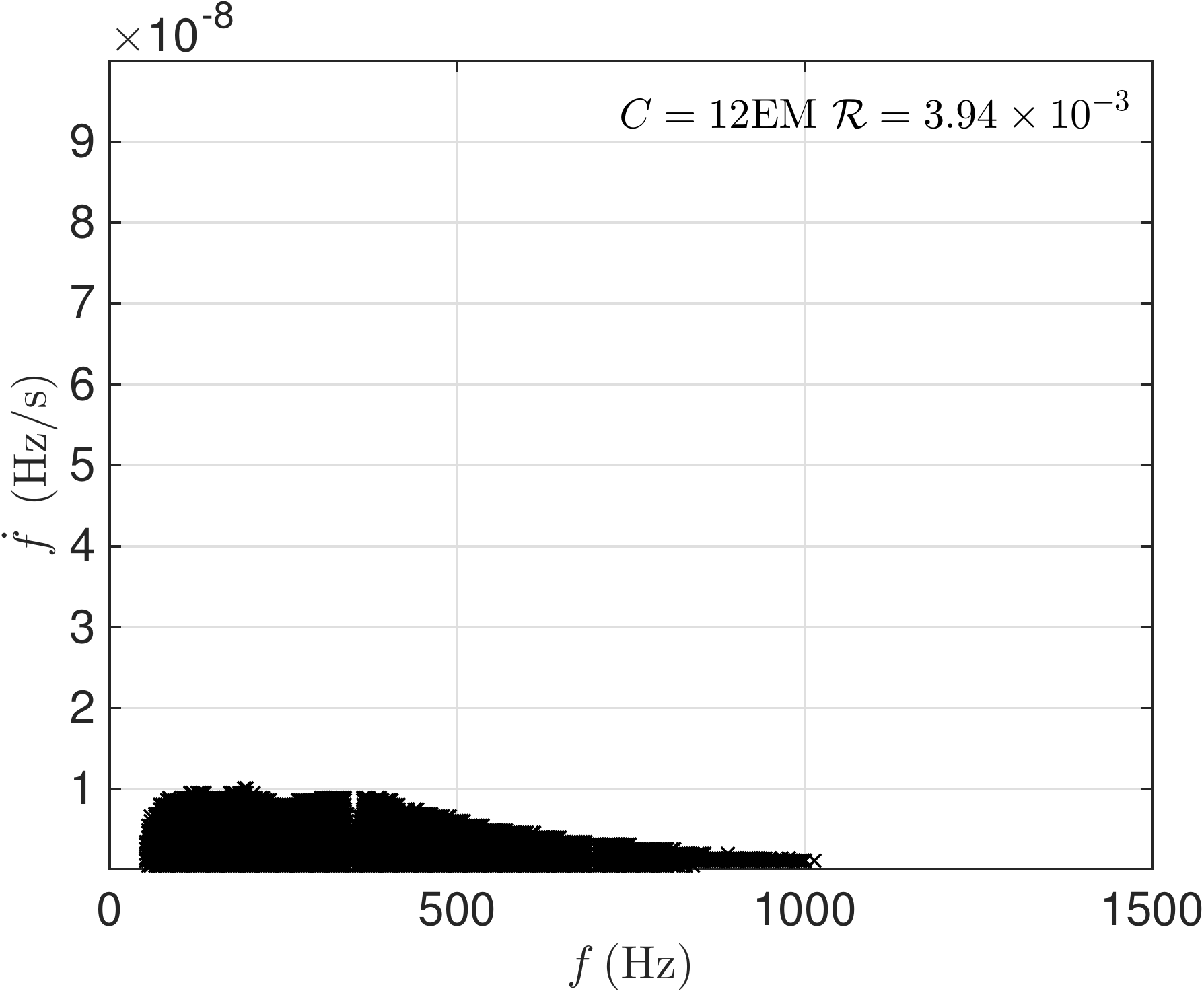}}}%
    \qquad
    \subfloat[Efficiency(lg), 37.5 days]{{  \includegraphics[width=.20\linewidth]{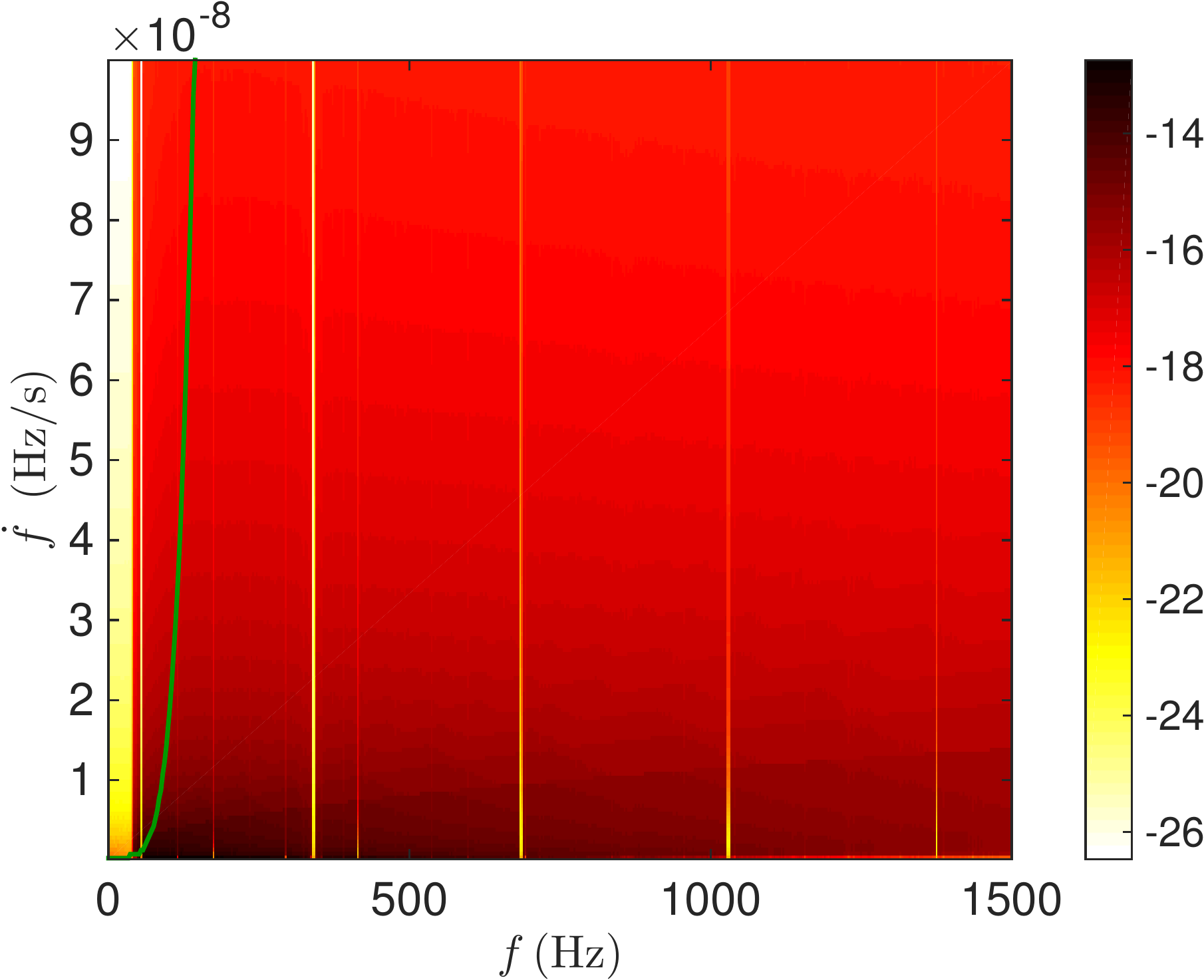}}}%
    \qquad
    \subfloat[Coverage, 37.5 days]{{  \includegraphics[width=.20\linewidth]{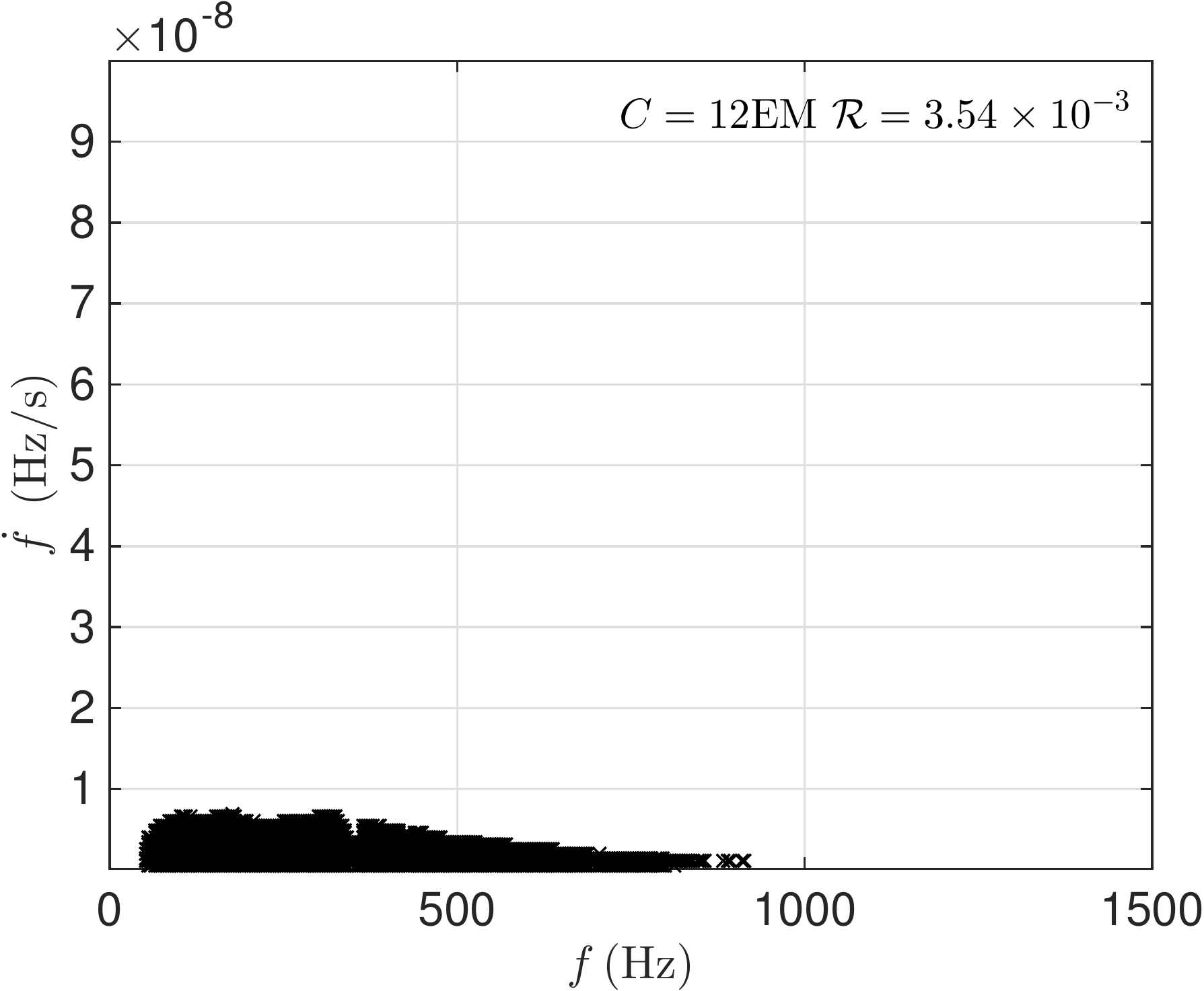}}}%
    \qquad
    \subfloat[Efficiency(lg), 50 days]{{  \includegraphics[width=.20\linewidth]{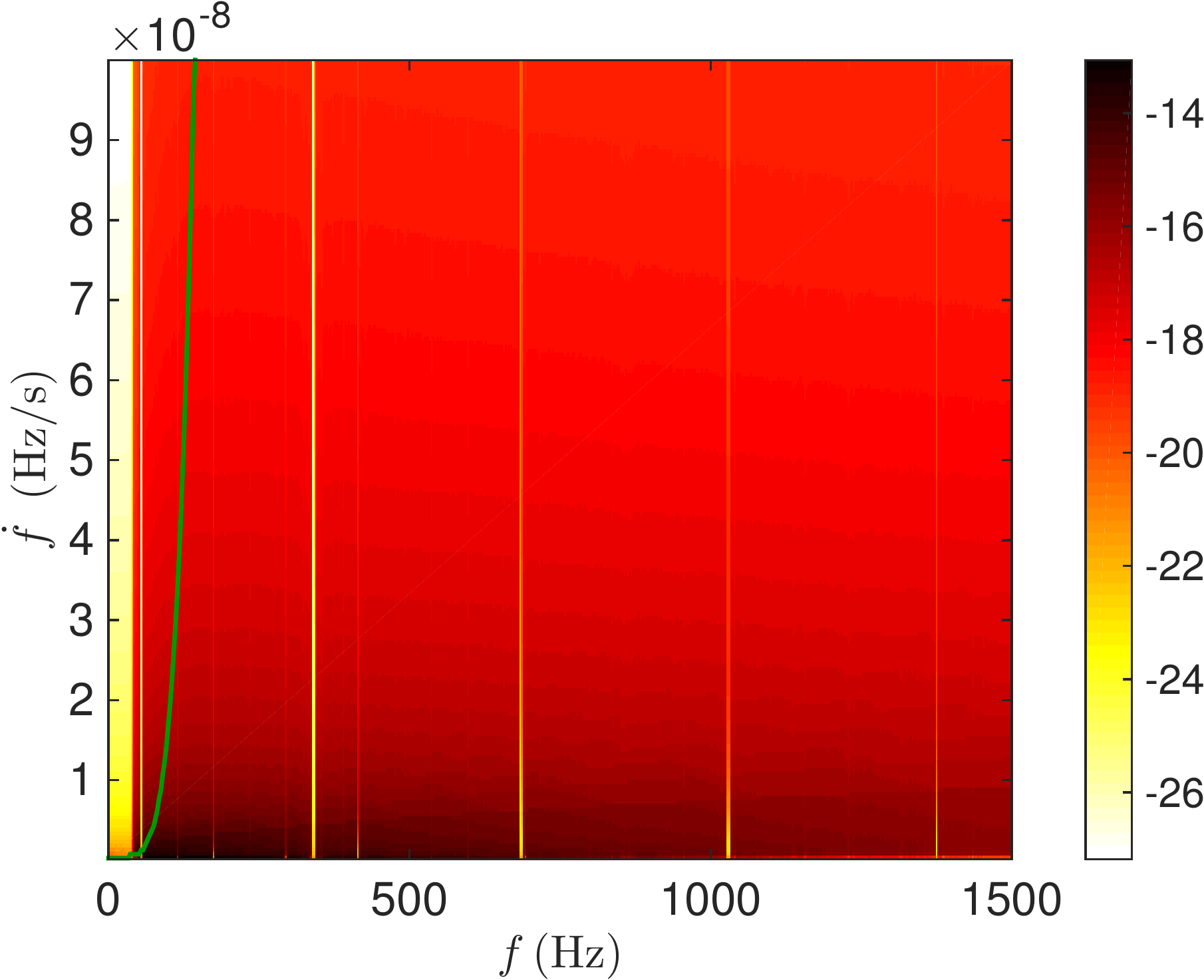}}}%
    \qquad
    \subfloat[Coverage, 50 days]{{  \includegraphics[width=.20\linewidth]{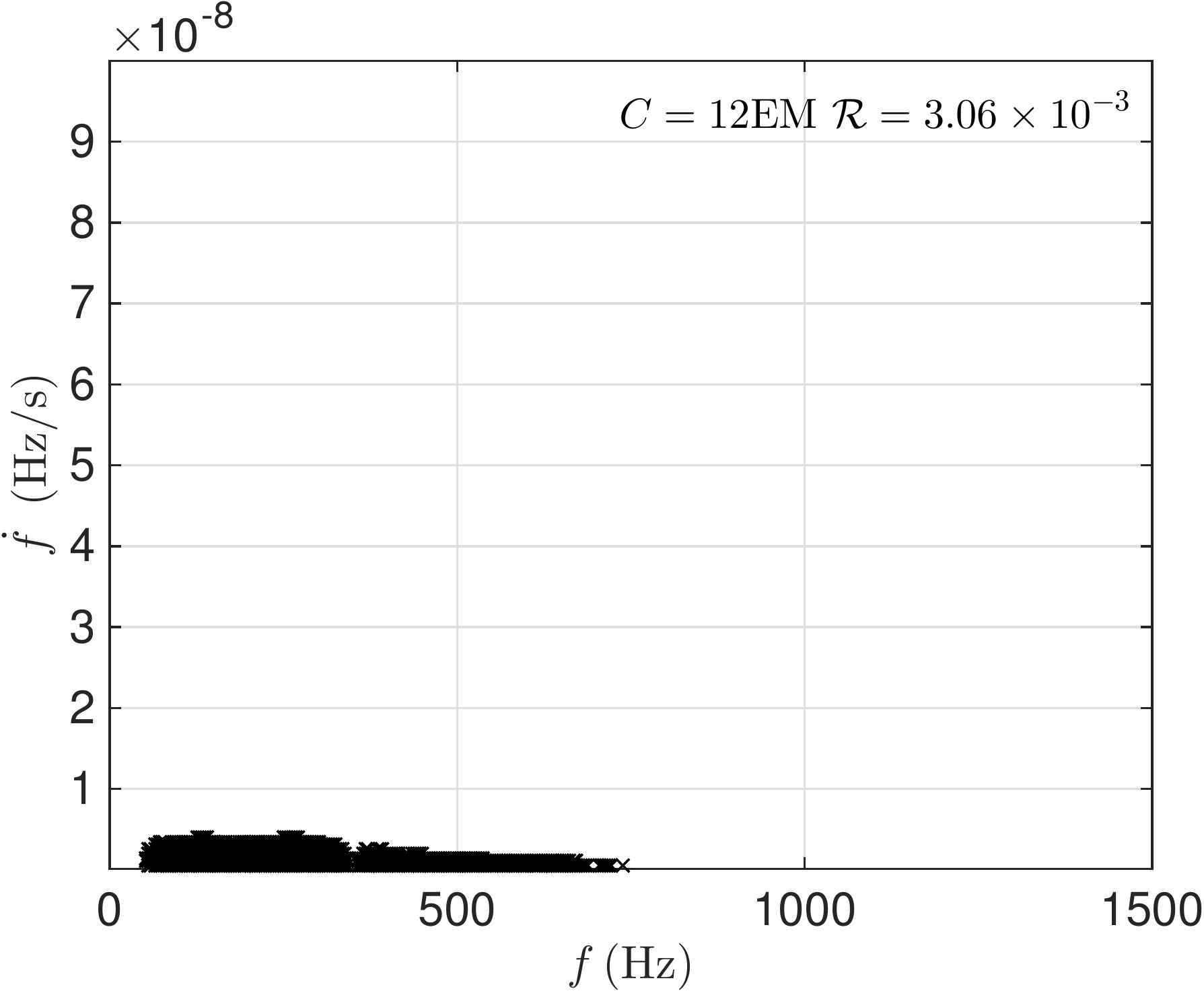}}}%
    \qquad
    \subfloat[Efficiency(lg), 75 days]{{  \includegraphics[width=.20\linewidth]{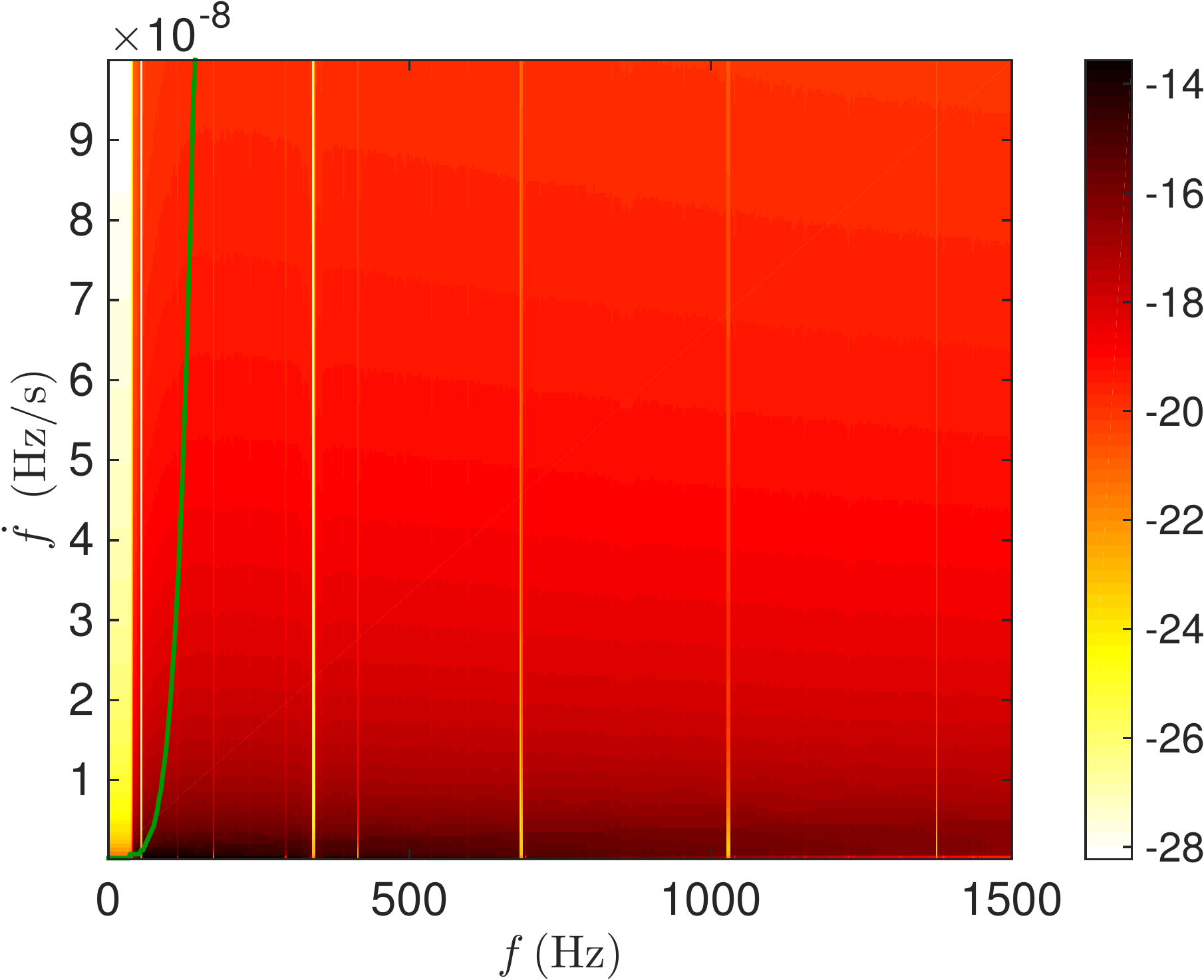}}}%
    \qquad
    \subfloat[Coverage, 75 days]{{  \includegraphics[width=.20\linewidth]{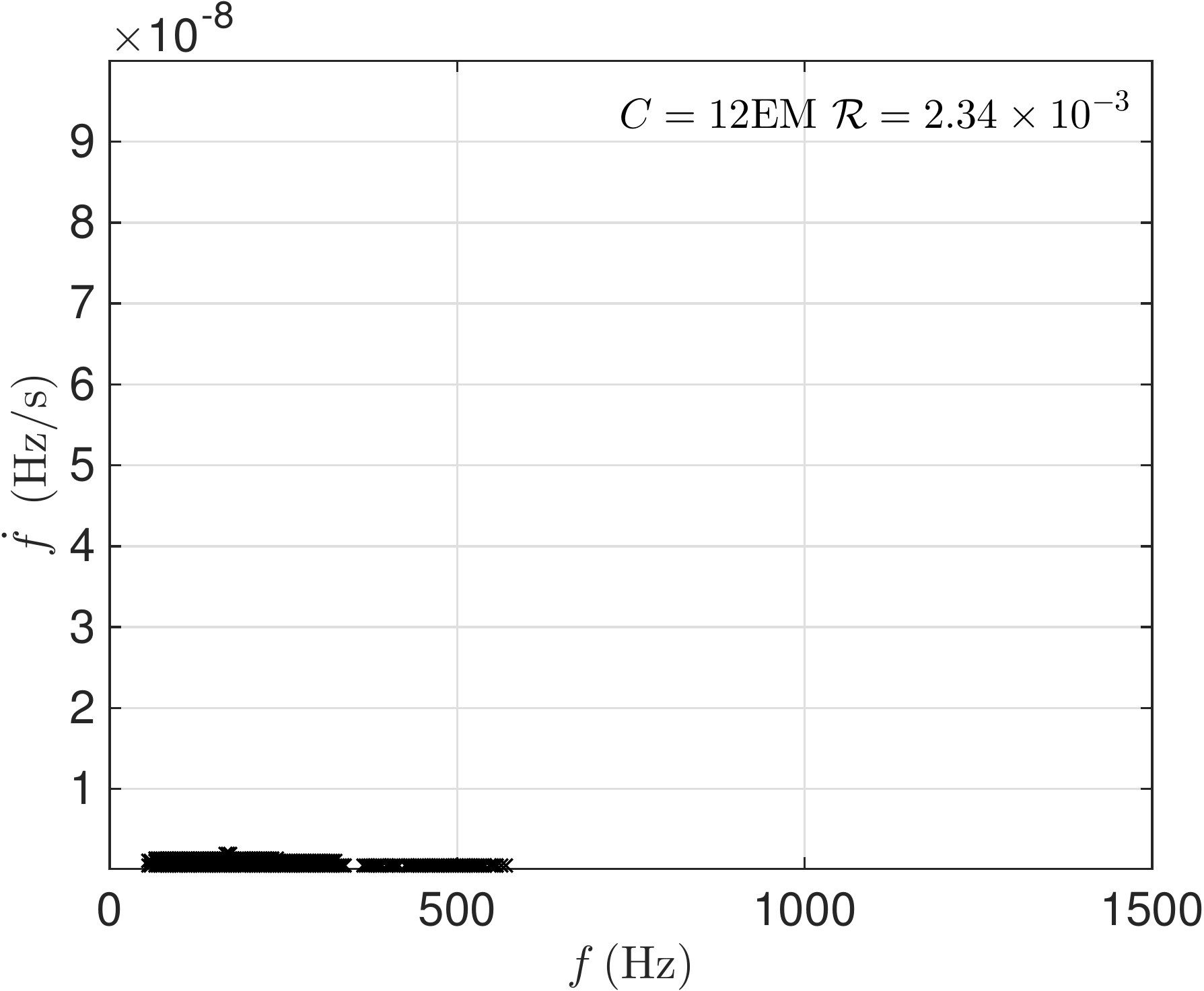}}}%
     \caption{Optimisation results for Vela Jr at 200 pc, assuming log-uniform and distance-based priors, for various coherent search durations: 5, 10, 20, 30, 37.5, 50 and 75 days. The total computing budget is assumed to be 12 EM.}%
    \label{G2662_51020days_noage_log}%
\end{figure*}

\begin{figure*}%
    \centering
    \subfloat[Efficiency(lg), 5 days]{{  \includegraphics[width=.20\linewidth]{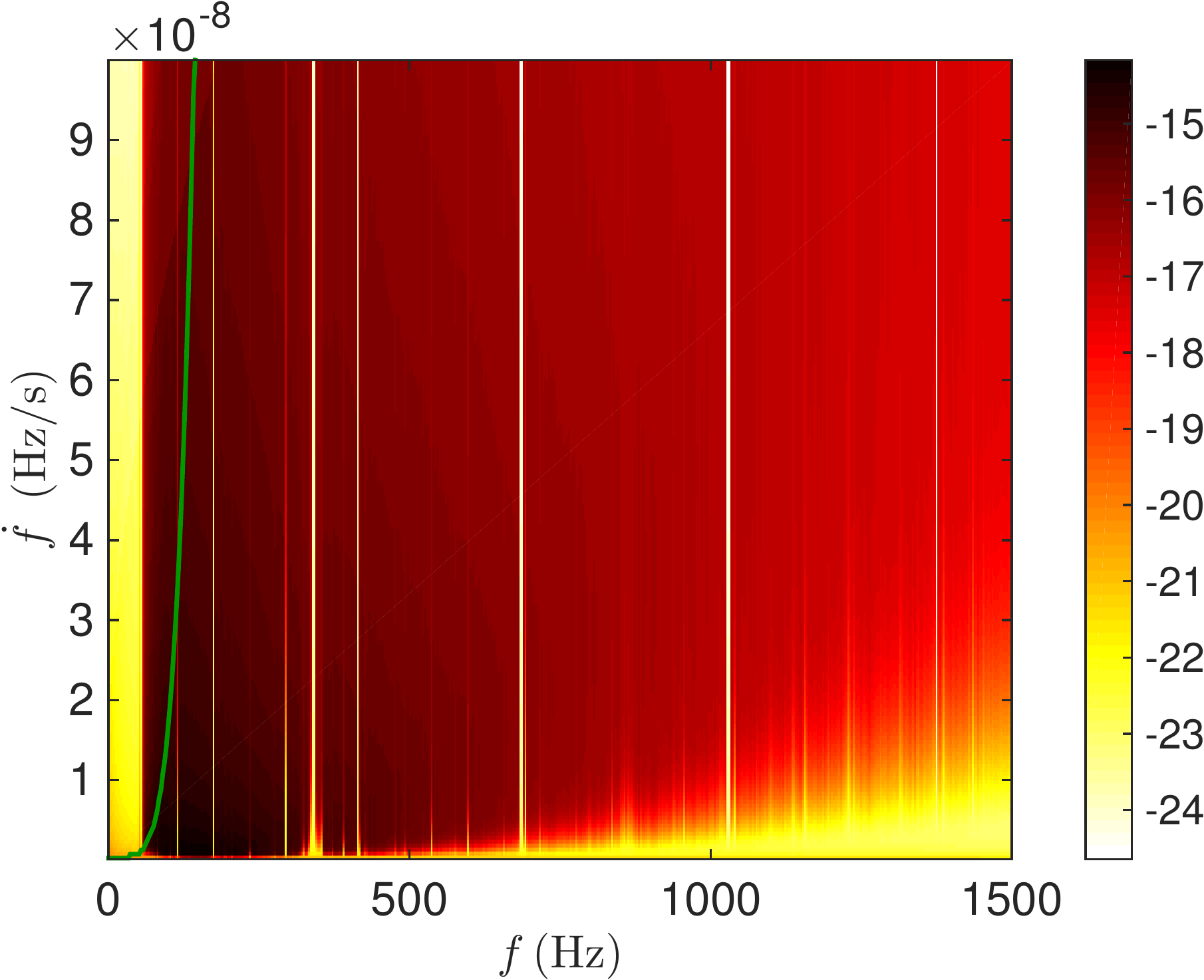}}}%
    \qquad
    \subfloat[Coverage, 5 days]{{  \includegraphics[width=.20\linewidth]{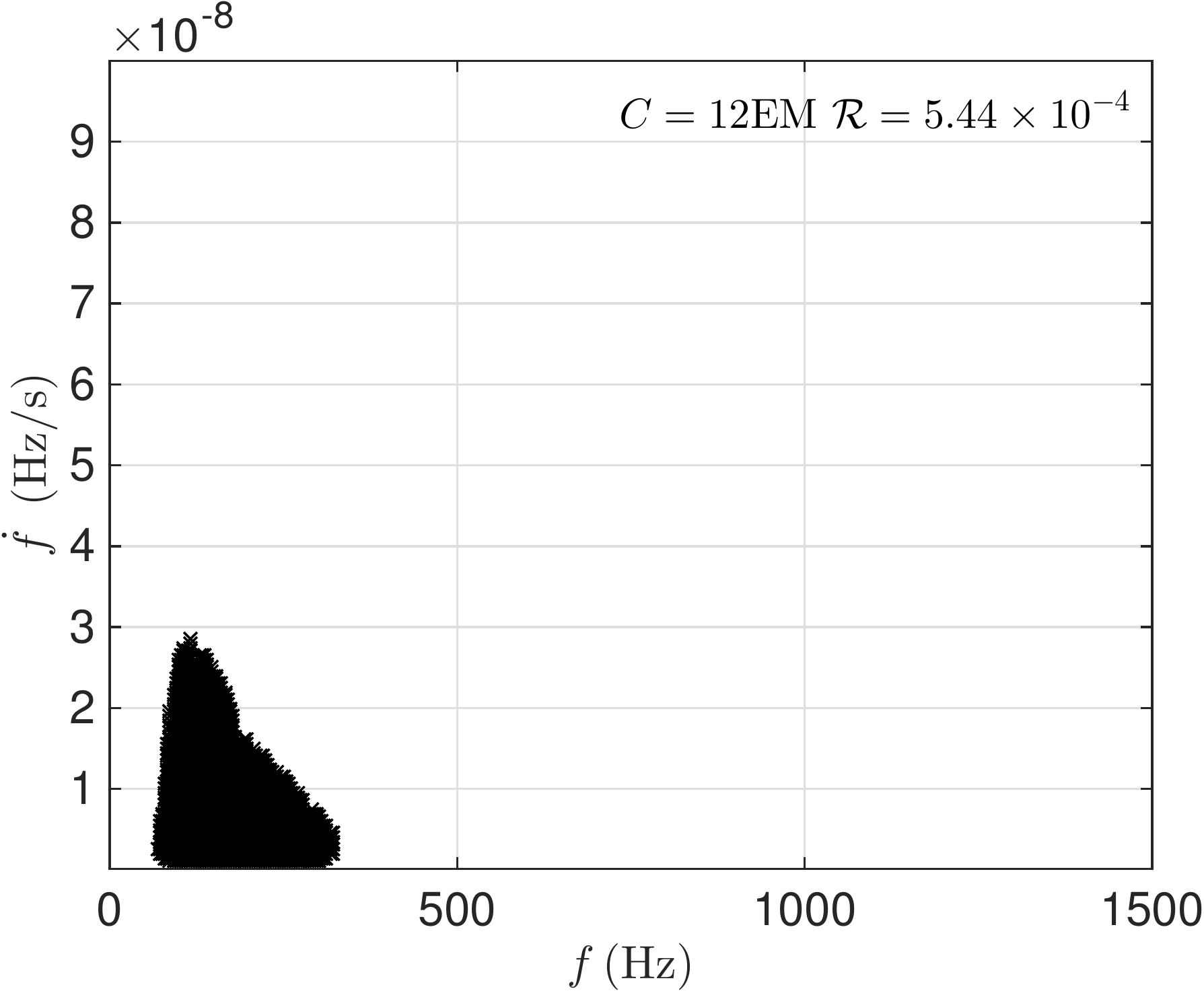}}}%
    \qquad
    \subfloat[Efficiency(lg), 10 days]{{  \includegraphics[width=.20\linewidth]{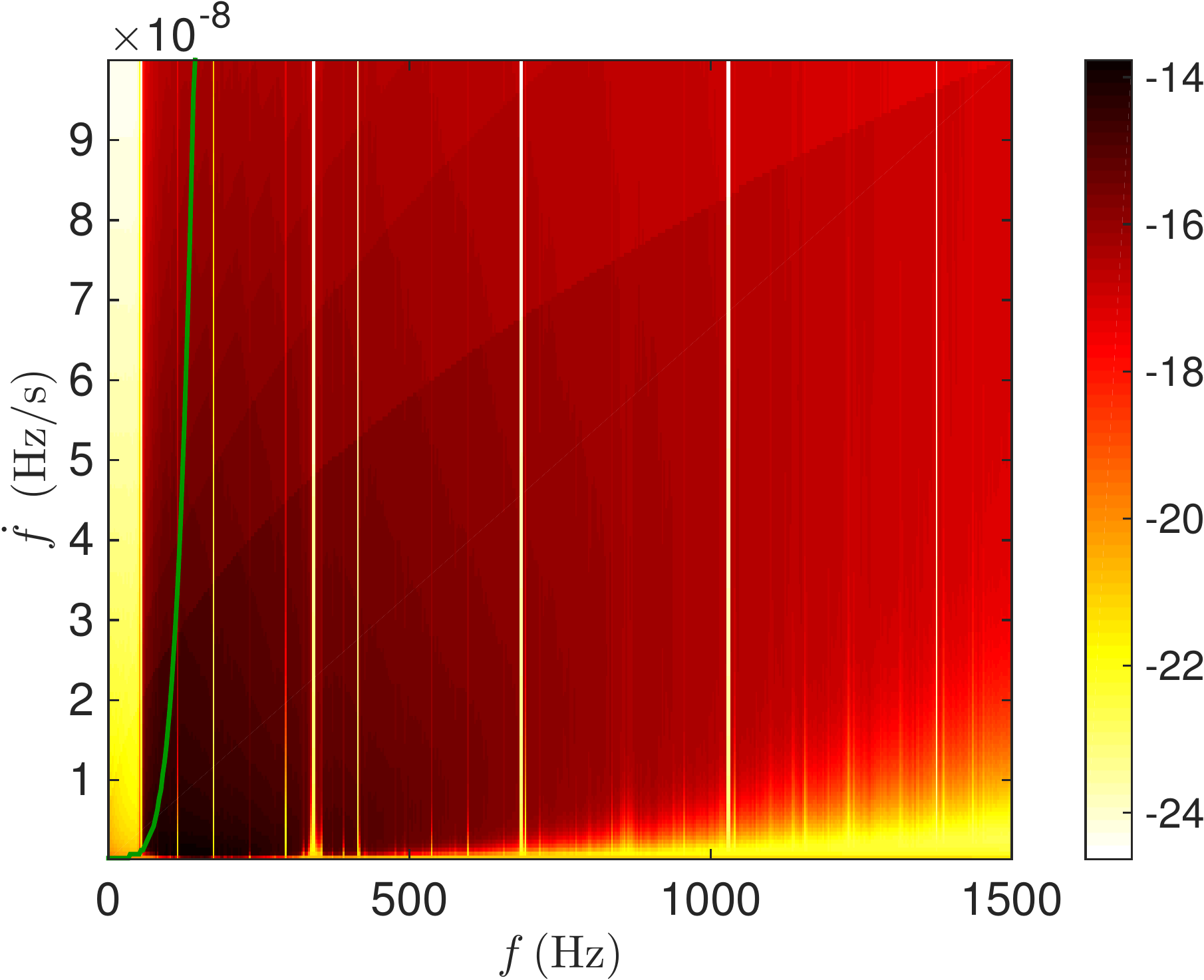}}}%
    \qquad
    \subfloat[Coverage, 10 days]{{  \includegraphics[width=.20\linewidth]{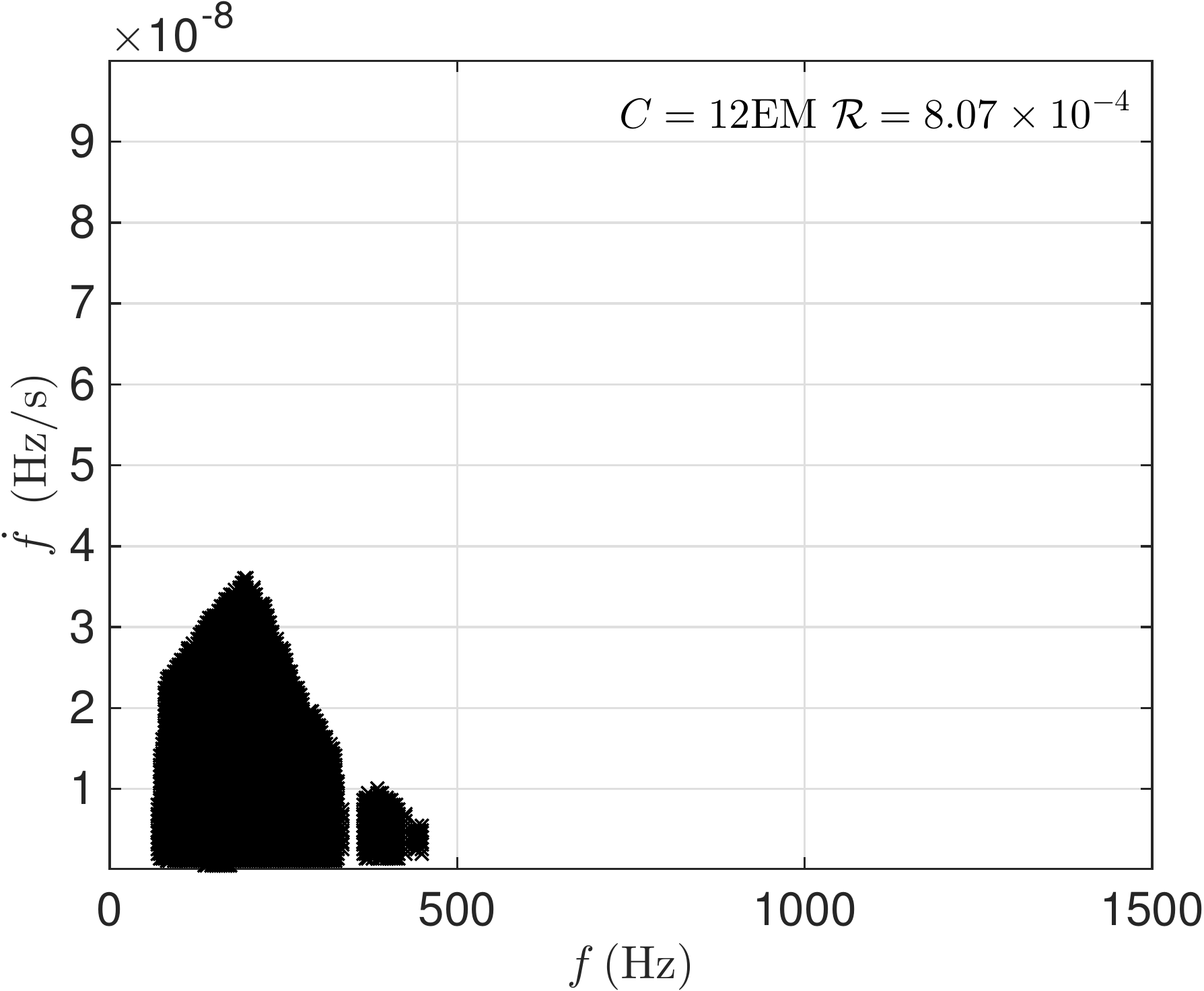}}}%
    \qquad
    \subfloat[Efficiency(lg), 20 days]{{  \includegraphics[width=.20\linewidth]{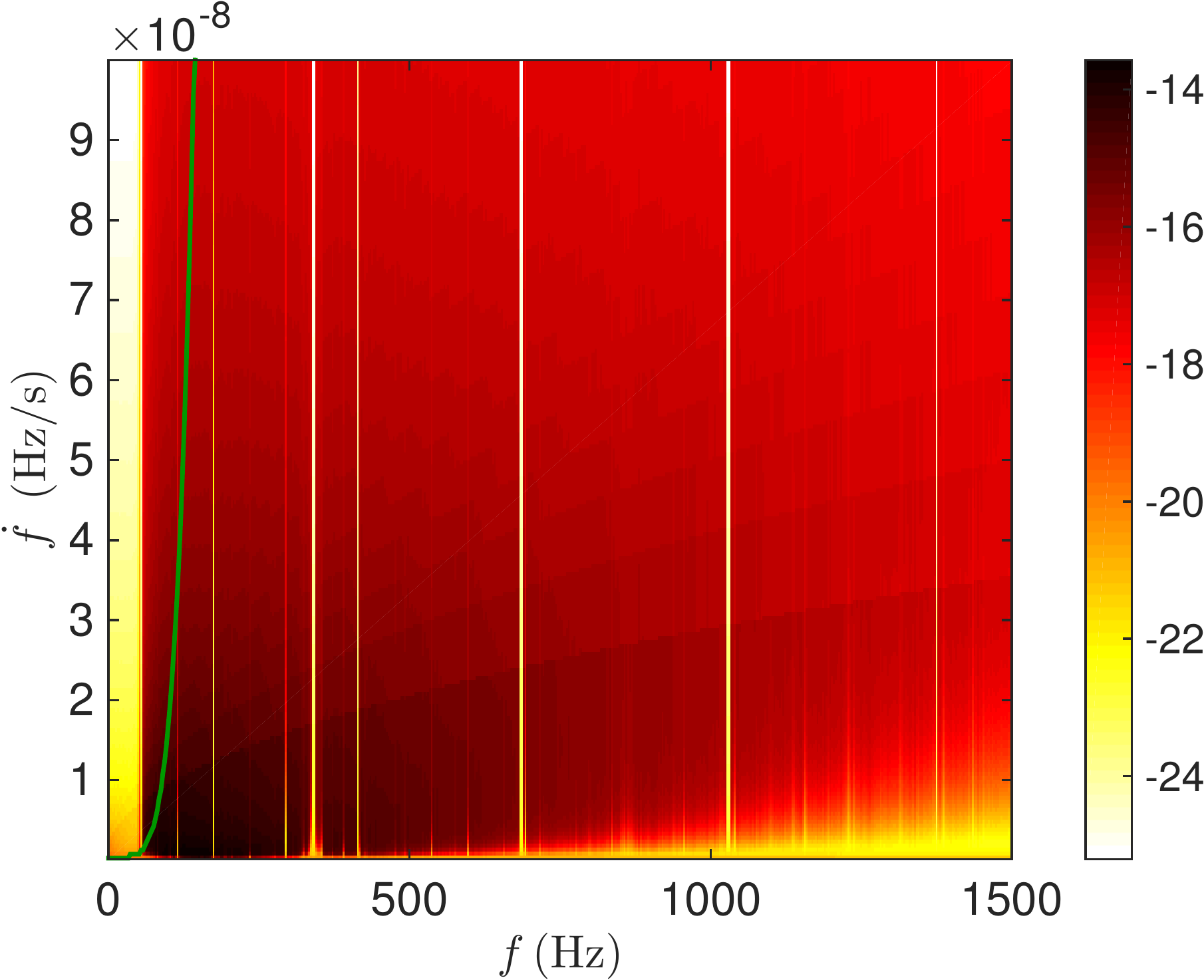}}}%
    \qquad
    \subfloat[Coverage, 20 days]{{  \includegraphics[width=.20\linewidth]{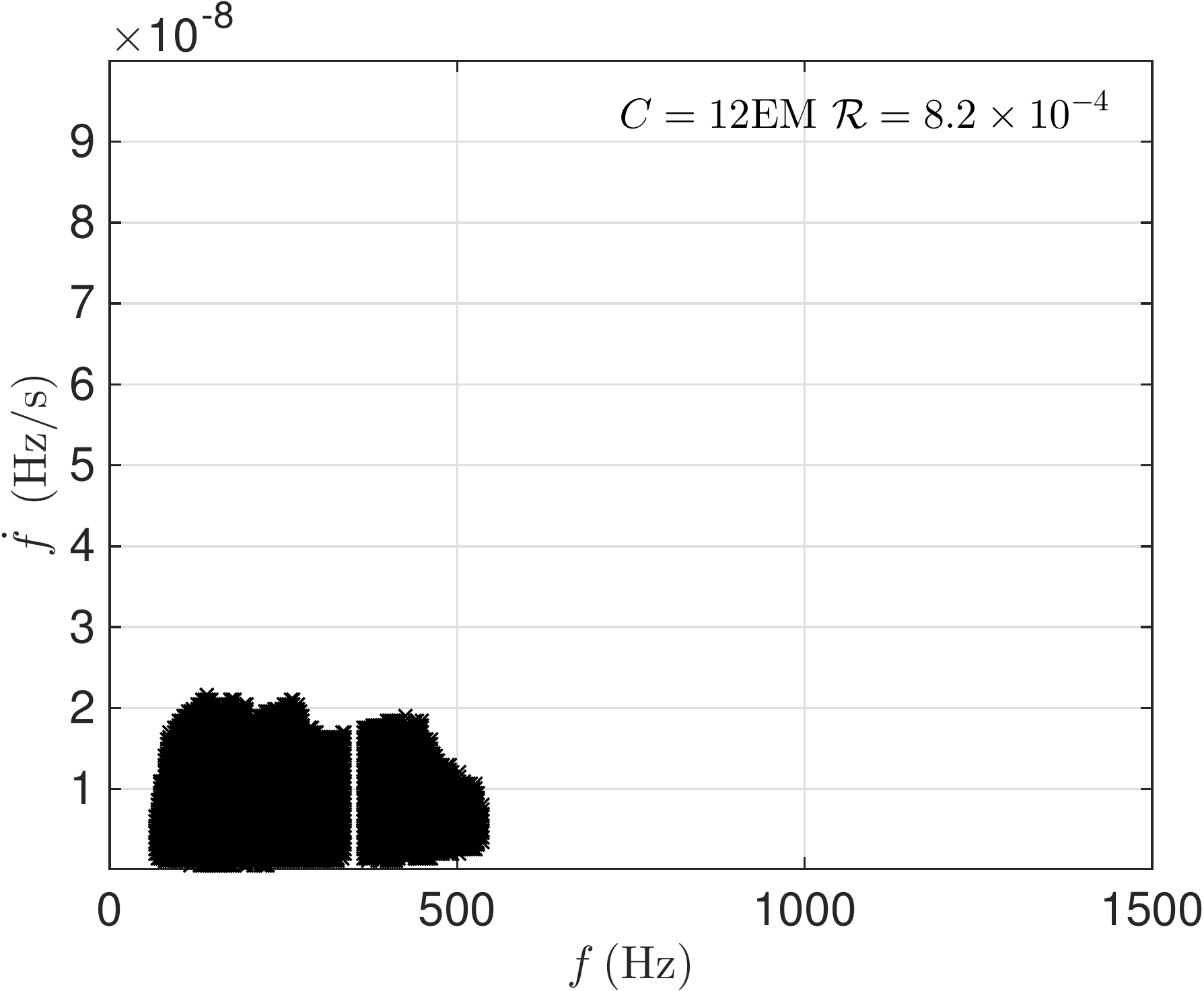}}}%
    \qquad
    \subfloat[Efficiency(lg), 30 days]{{  \includegraphics[width=.20\linewidth]{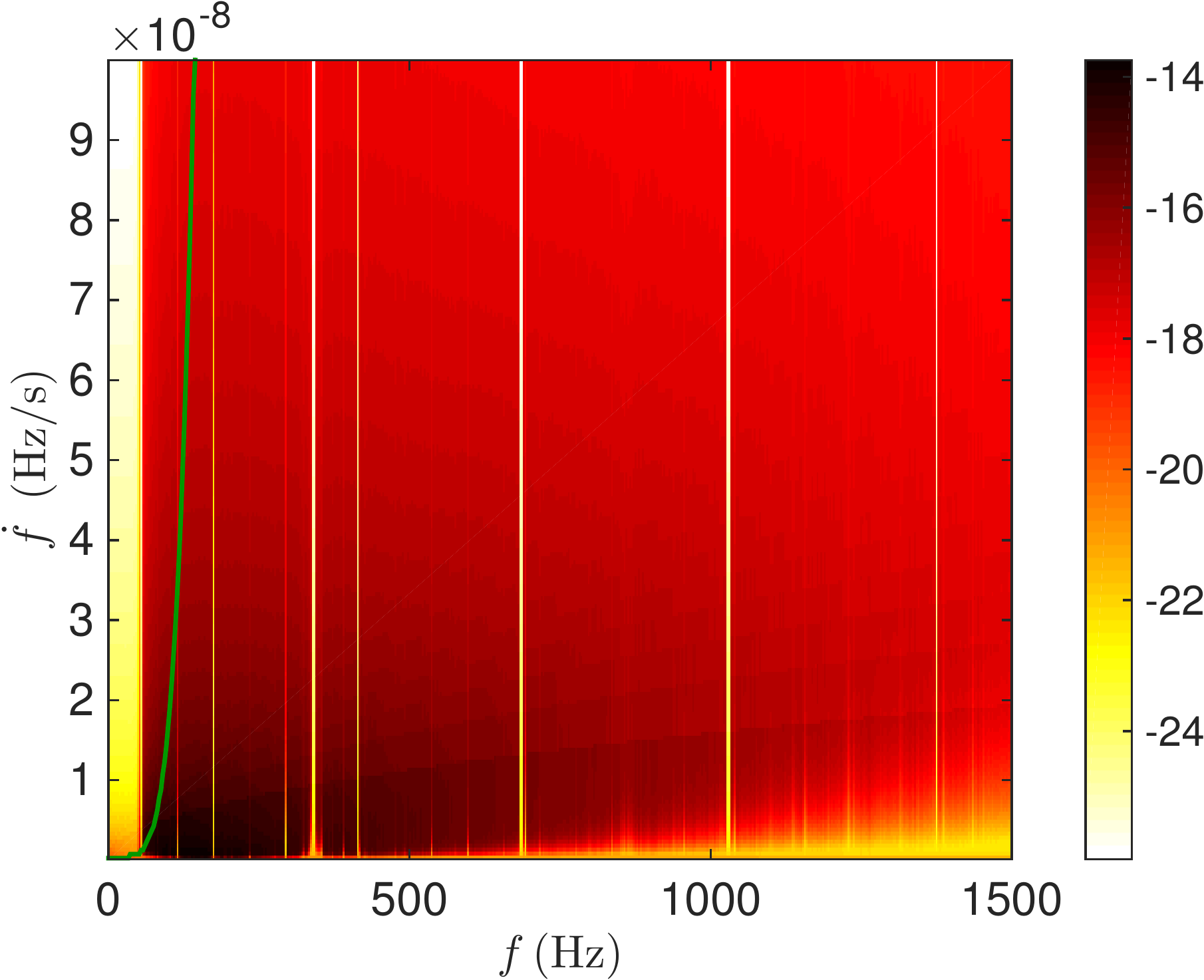}}}%
    \qquad
    \subfloat[Coverage, 30 days]{{  \includegraphics[width=.20\linewidth]{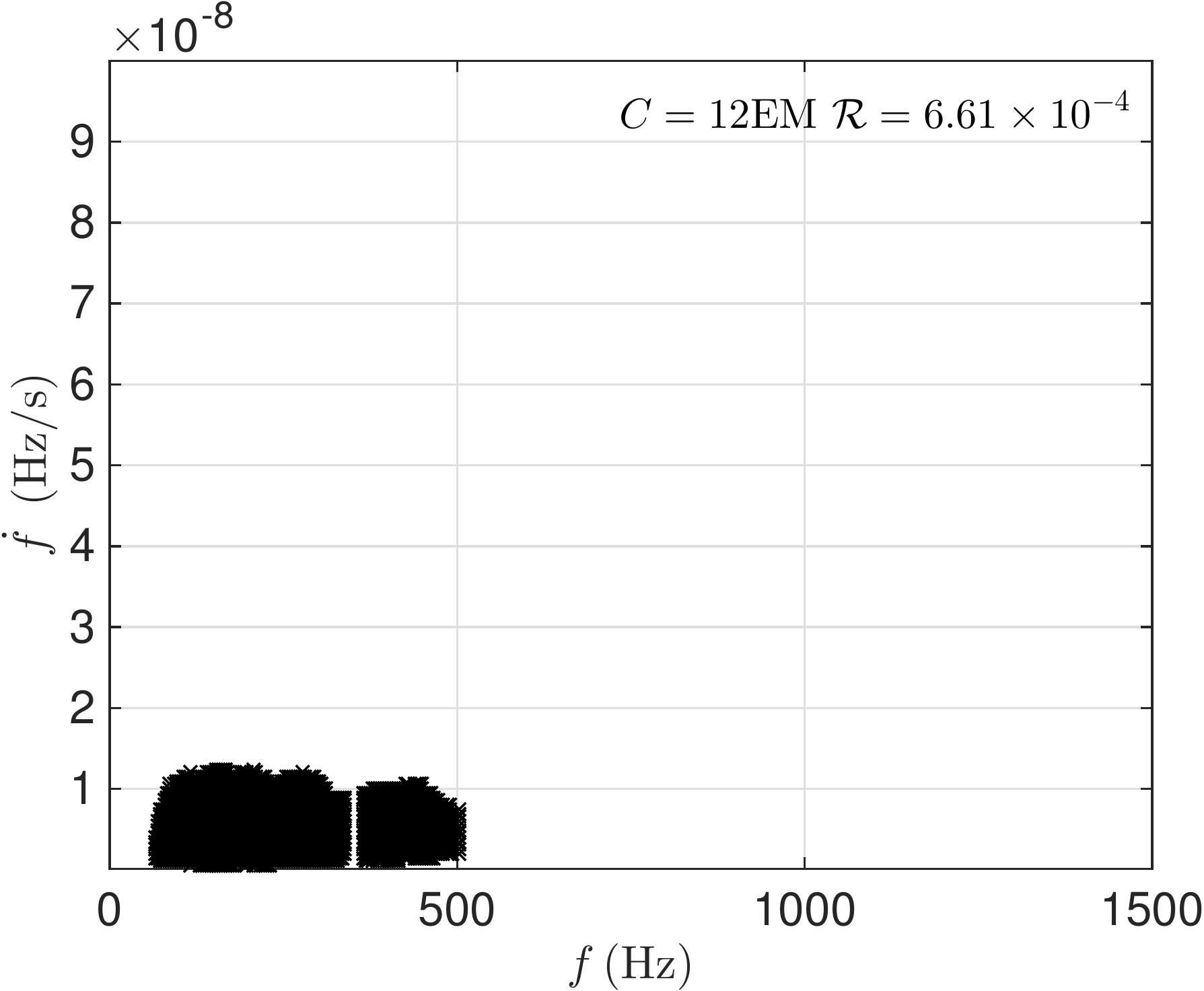}}}%
    \qquad
    \subfloat[Efficiency(lg), 37.5 days]{{  \includegraphics[width=.20\linewidth]{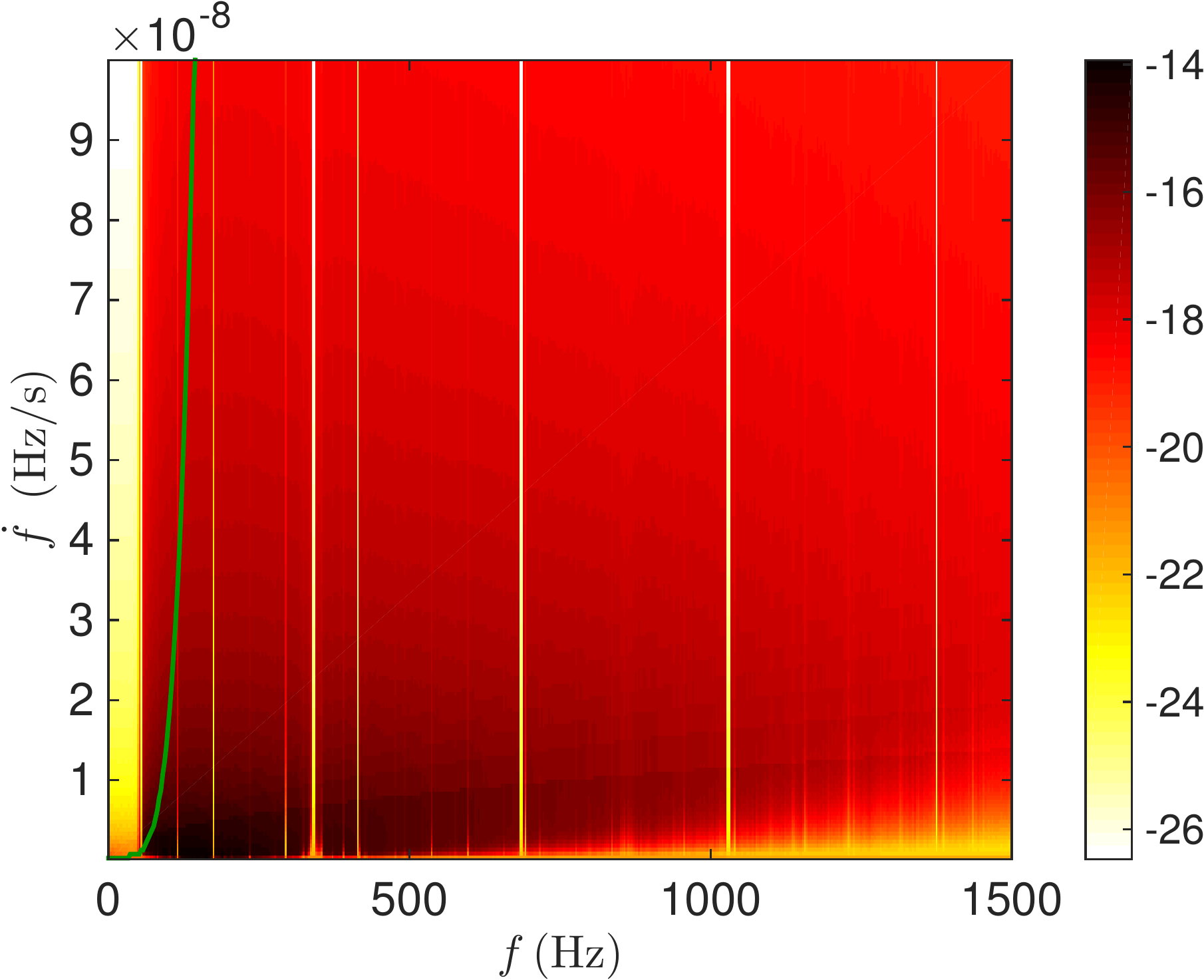}}}%
    \qquad
    \subfloat[Coverage, 37.5 days]{{  \includegraphics[width=.20\linewidth]{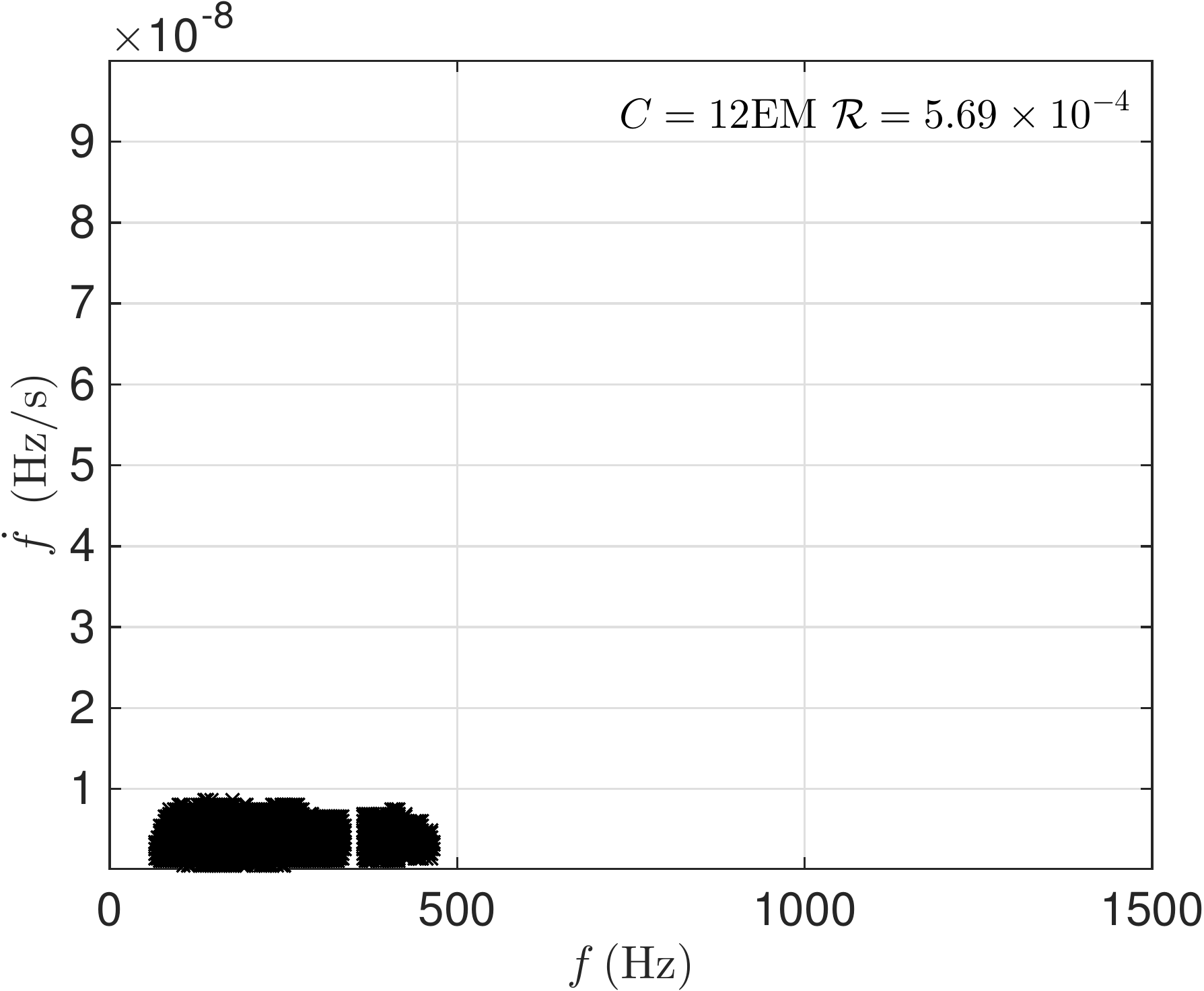}}}%
    \qquad
    \subfloat[Efficiency(lg), 50 days]{{  \includegraphics[width=.20\linewidth]{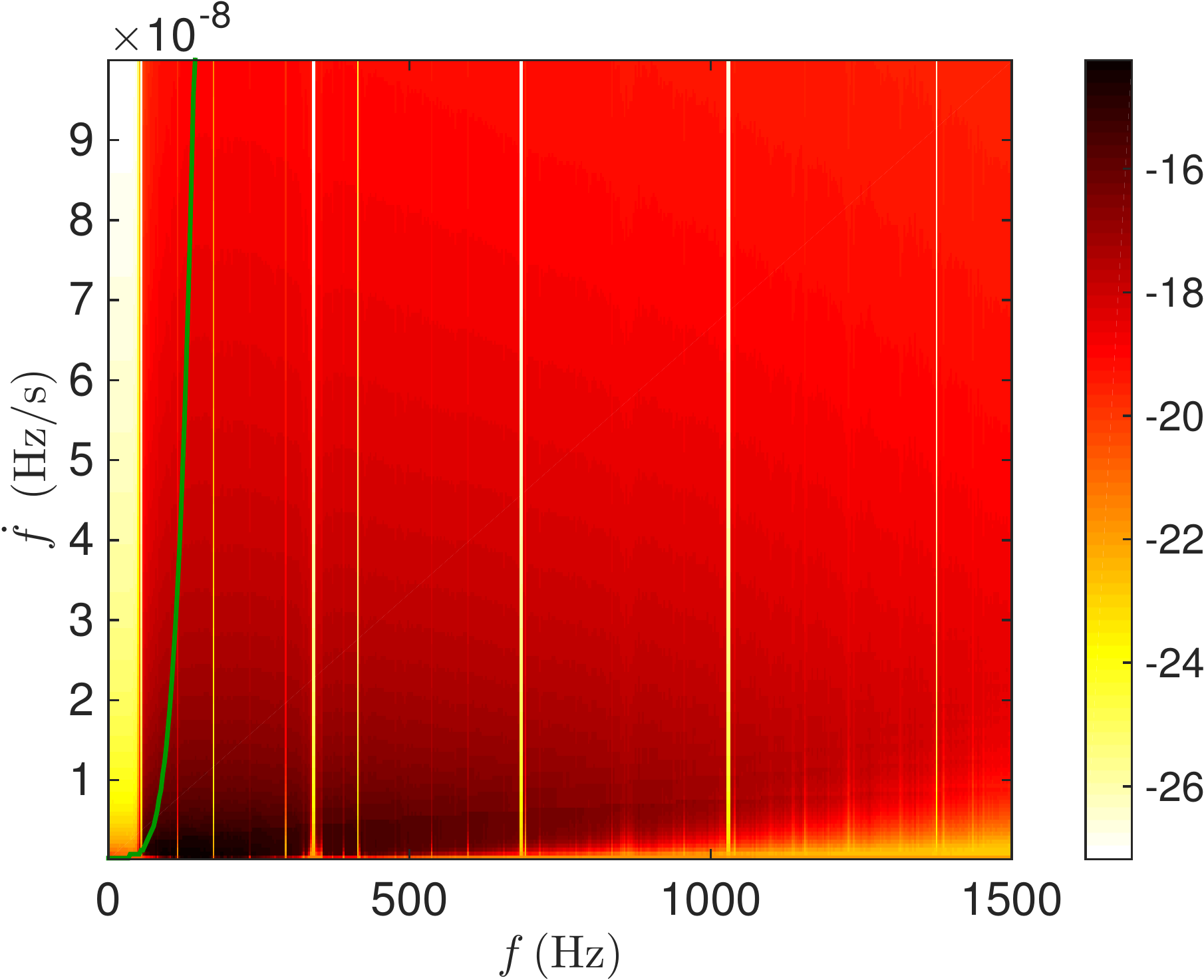}}}%
    \qquad
    \subfloat[Coverage, 50 days]{{  \includegraphics[width=.20\linewidth]{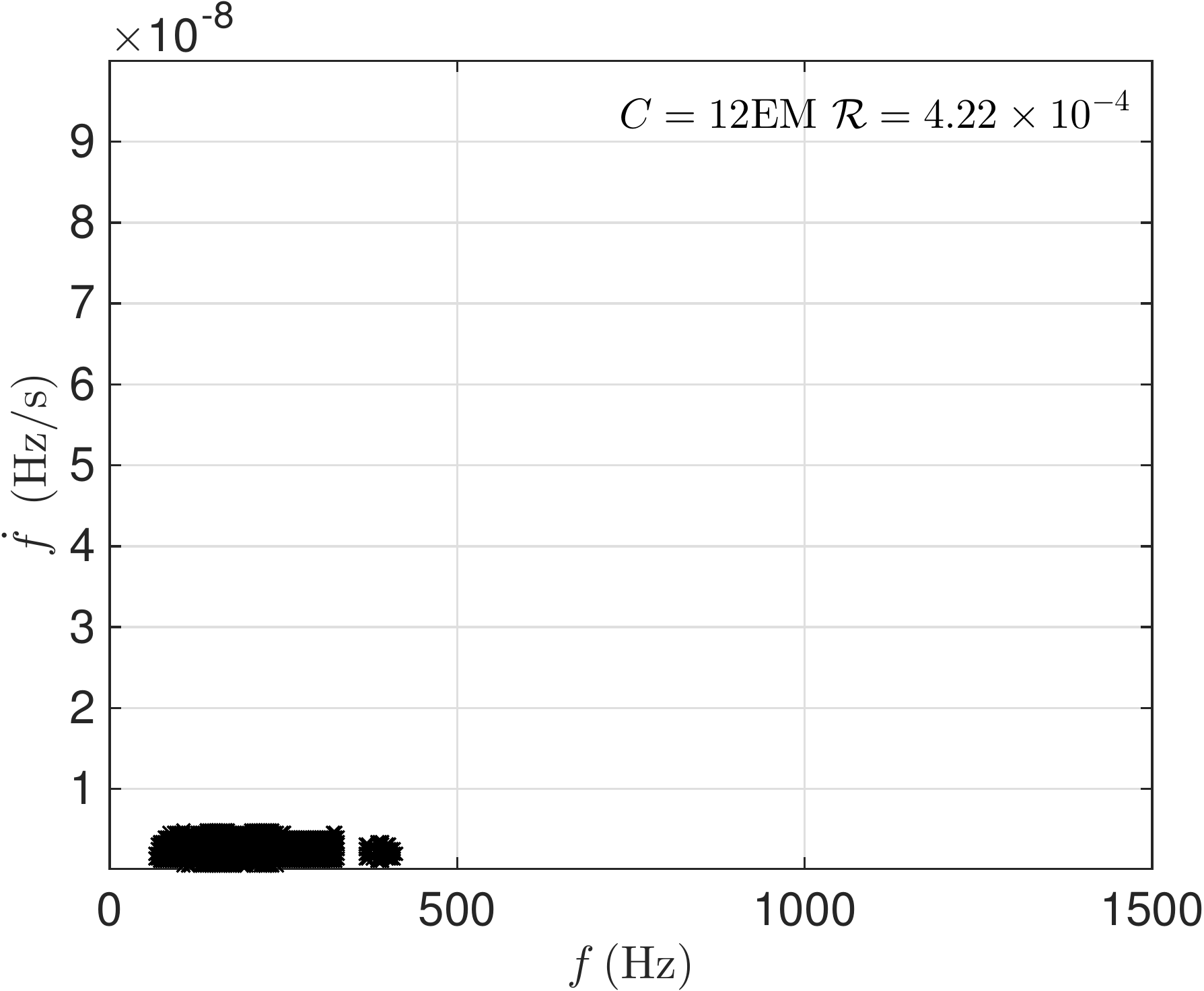}}}%
    \qquad
    \subfloat[Efficiency(lg), 75 days]{{  \includegraphics[width=.20\linewidth]{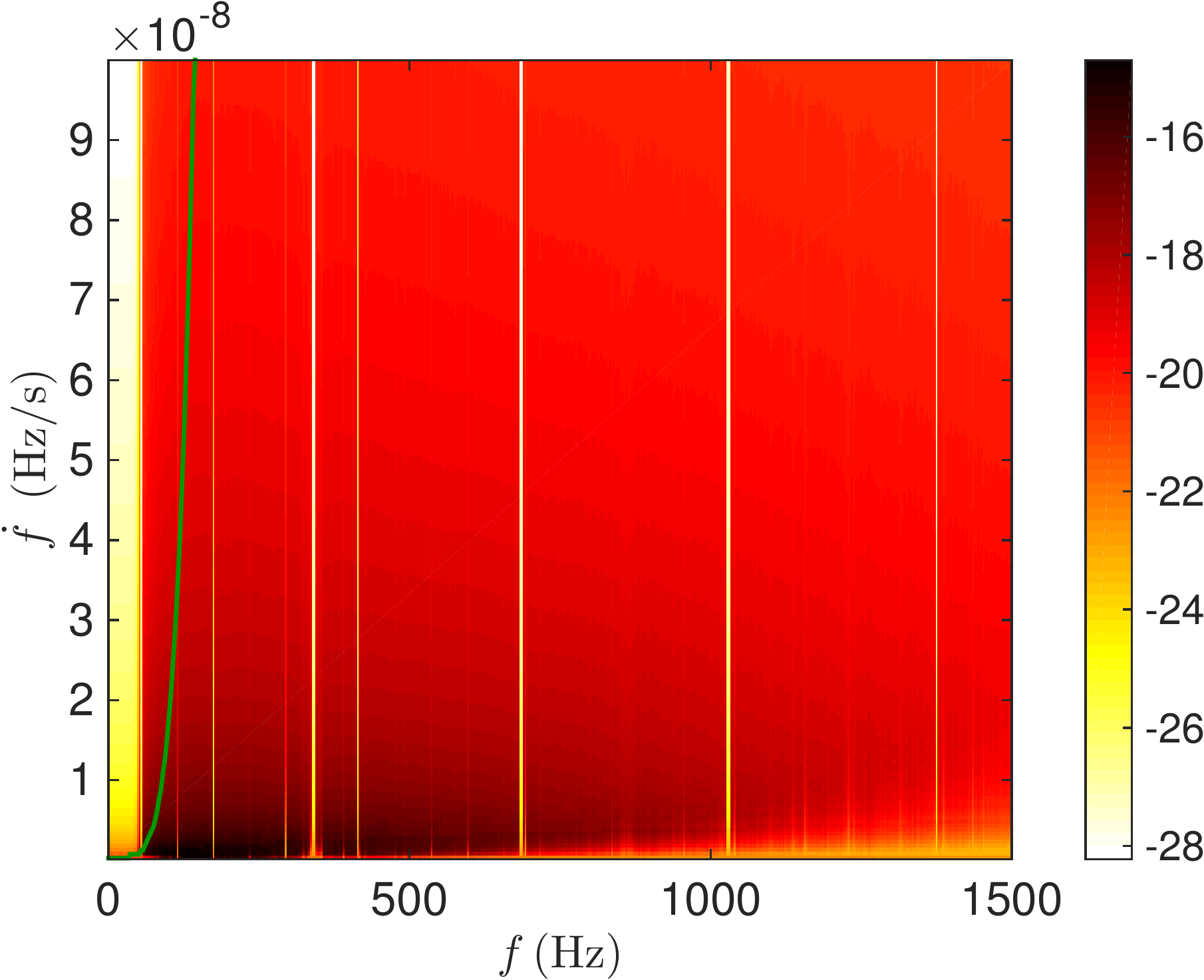}}}%
    \qquad
    \subfloat[Coverage, 75 days]{{  \includegraphics[width=.20\linewidth]{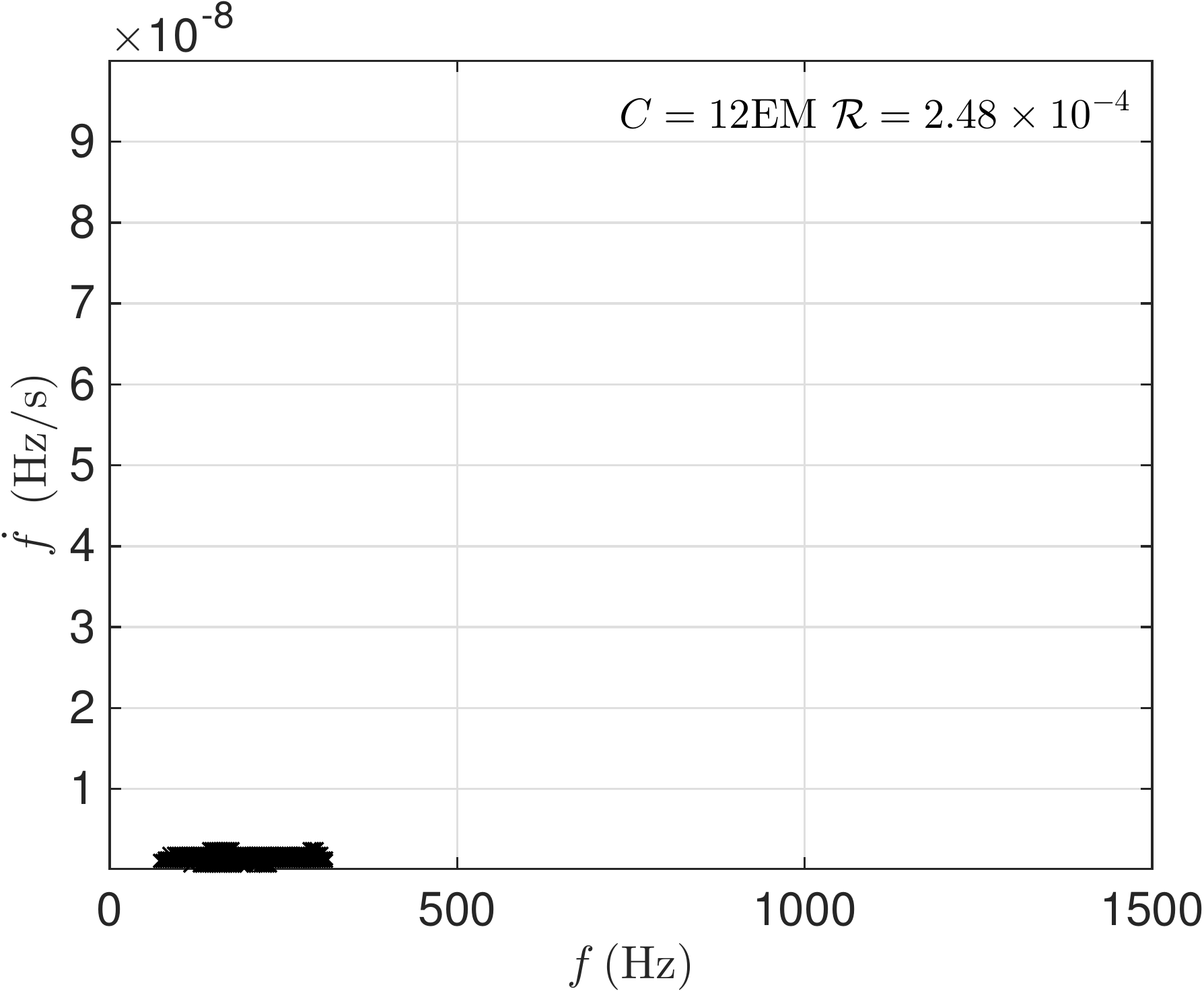}}}%
     \caption{Optimisation results for G347.3 at 1300 pc assuming log-uniform and distance-based priors, for various coherent search durations: 5, 10, 20, 30, 37.5, 50 and 75 days. The total computing budget is assumed to be 12 EM.}%
    \label{G3473_51020days_noage_log}%
\end{figure*}

\begin{figure*}%
    \centering
    \subfloat[Efficiency(lg), 5 days]{{  \includegraphics[width=.20\linewidth]{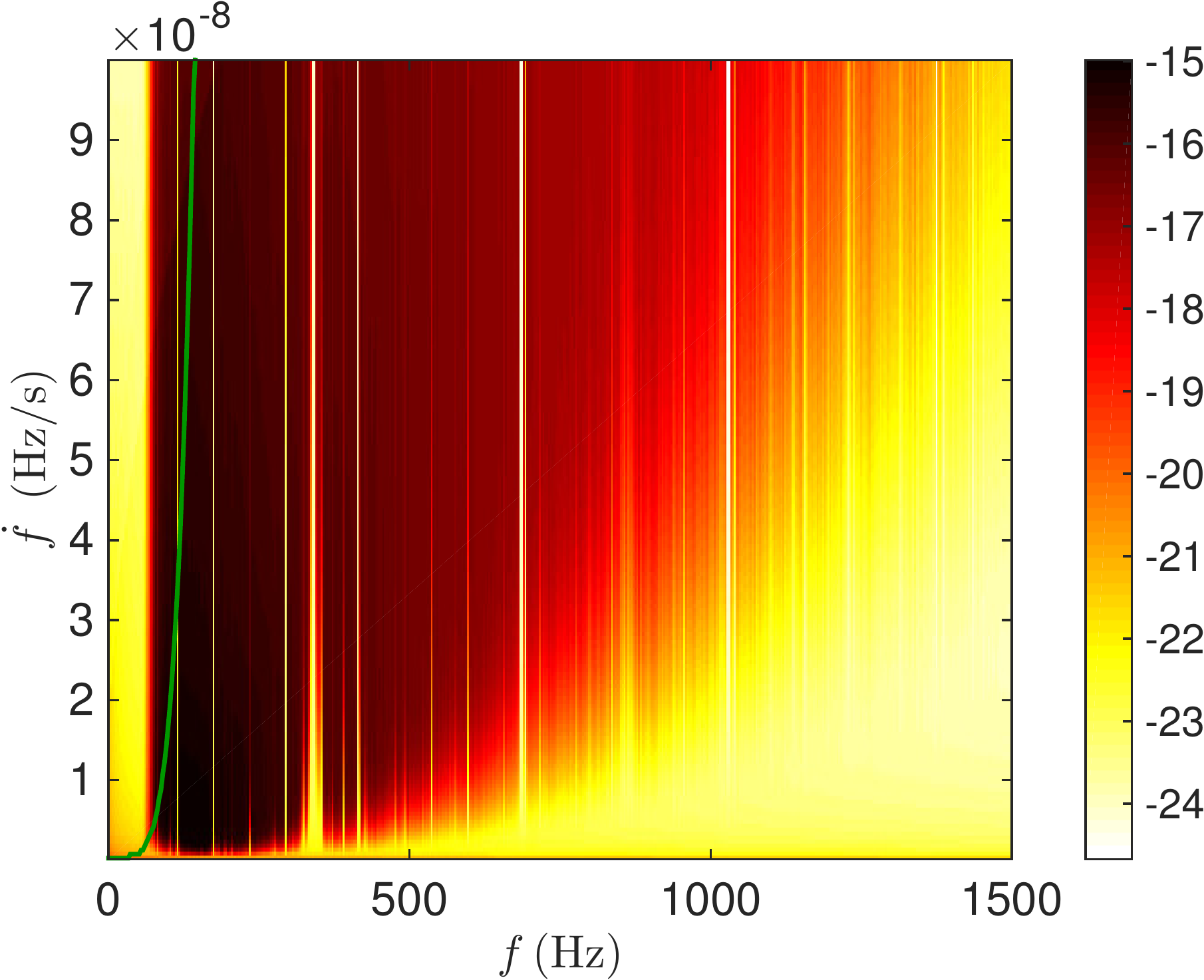}}}%
    \qquad
    \subfloat[Coverage, 5 days]{{  \includegraphics[width=.20\linewidth]{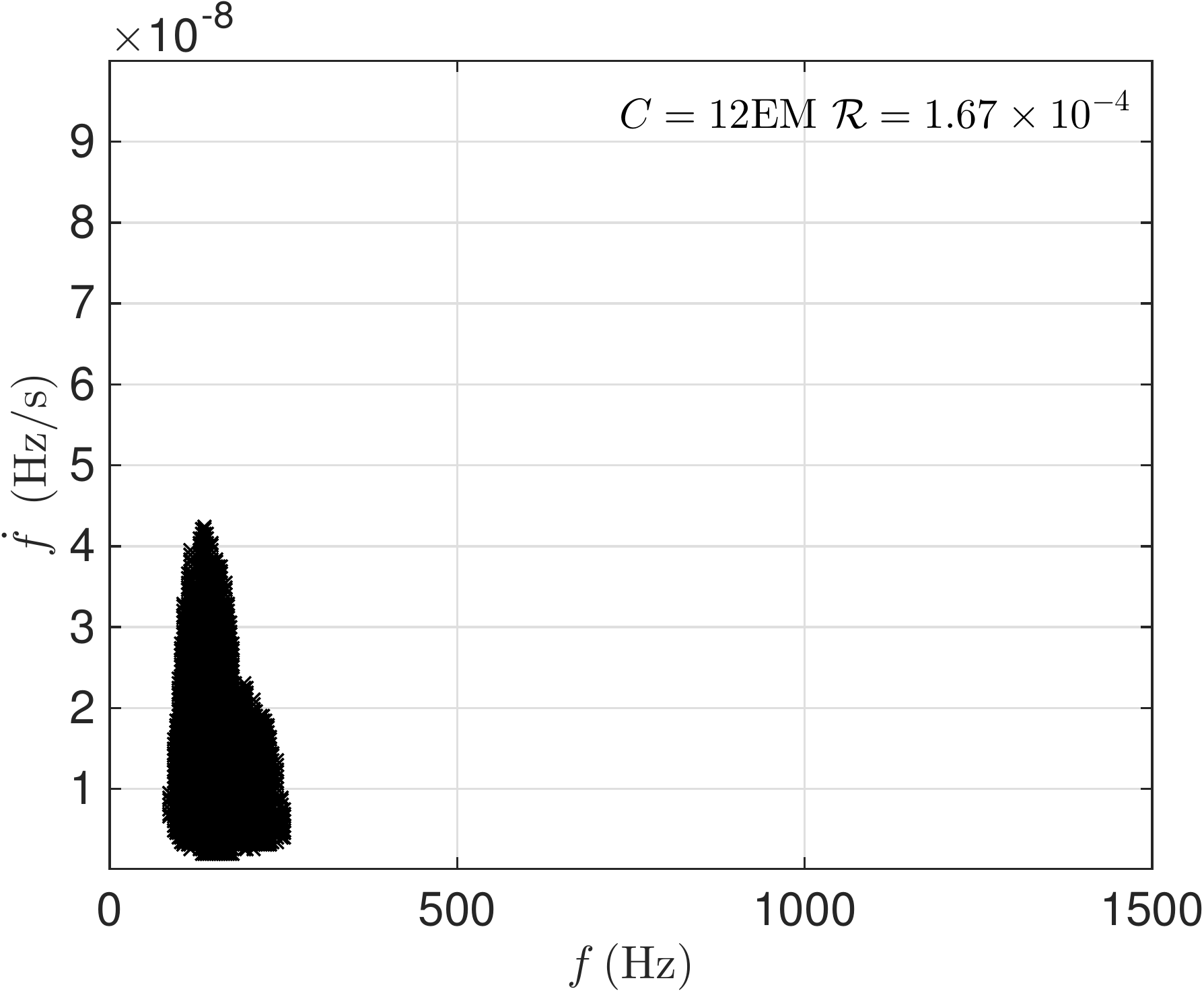}}}%
    \qquad
    \subfloat[Efficiency(lg), 10 days]{{  \includegraphics[width=.20\linewidth]{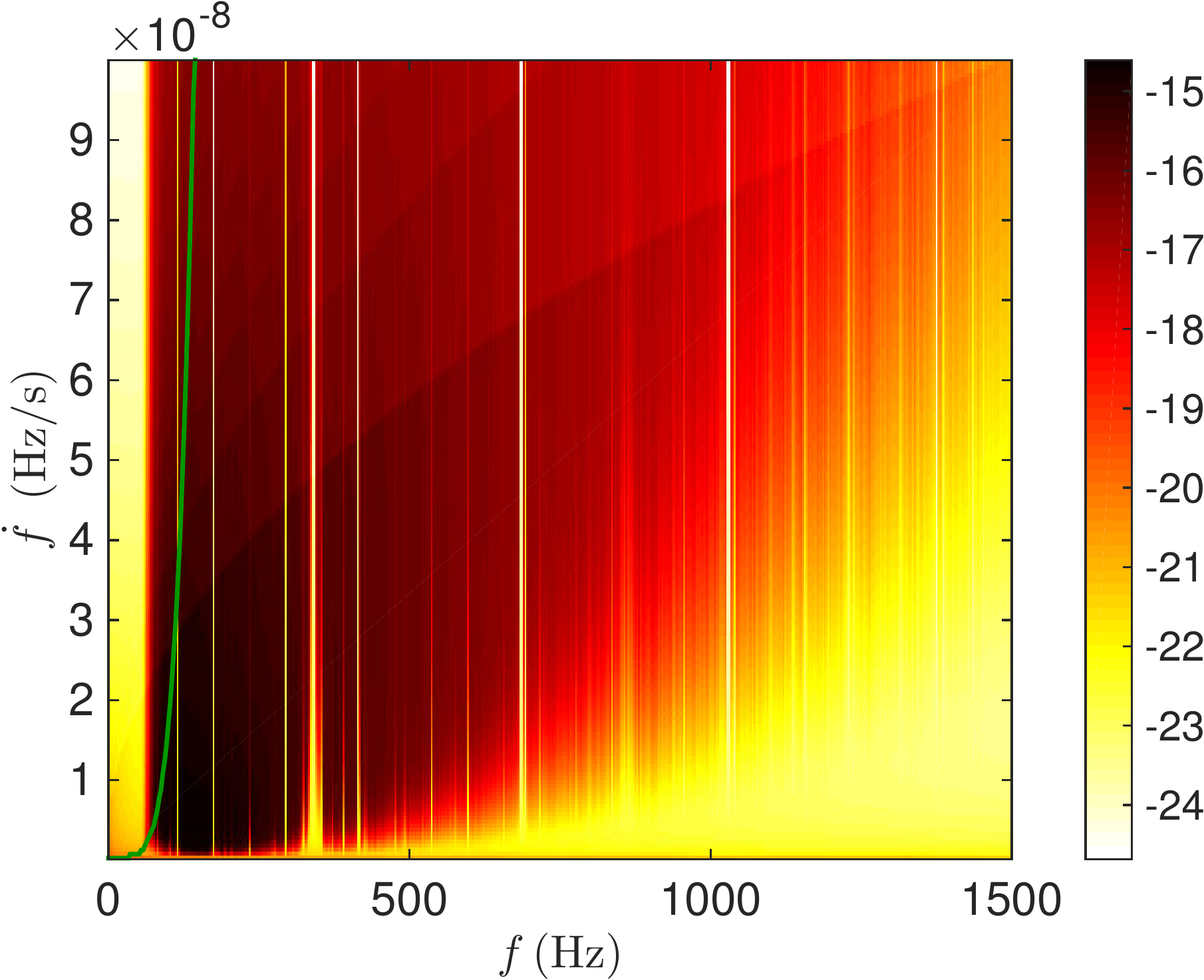}}}%
    \qquad
    \subfloat[Coverage, 10 days]{{  \includegraphics[width=.20\linewidth]{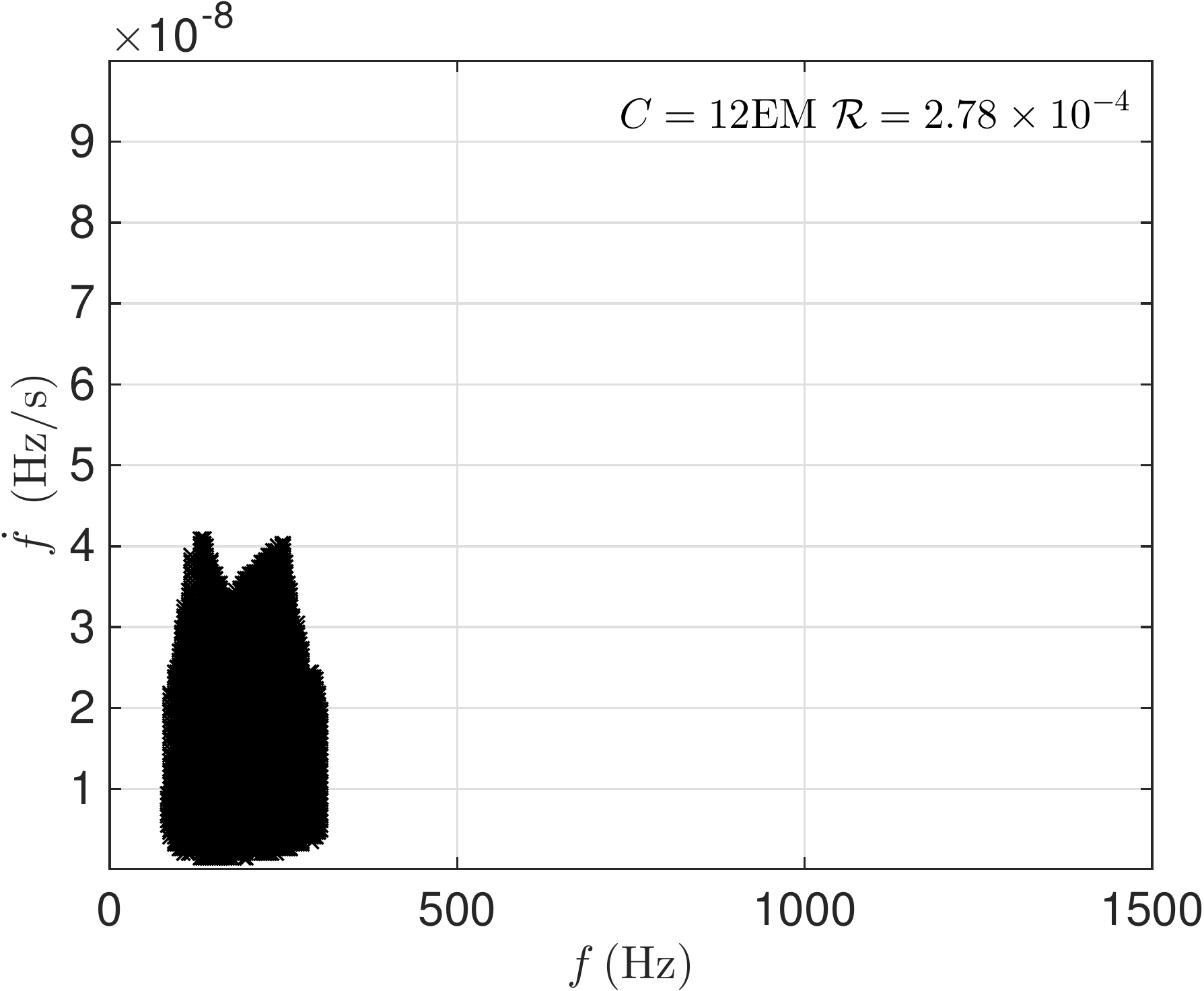}}}%
    \qquad
    \subfloat[Efficiency(lg), 20 days]{{  \includegraphics[width=.20\linewidth]{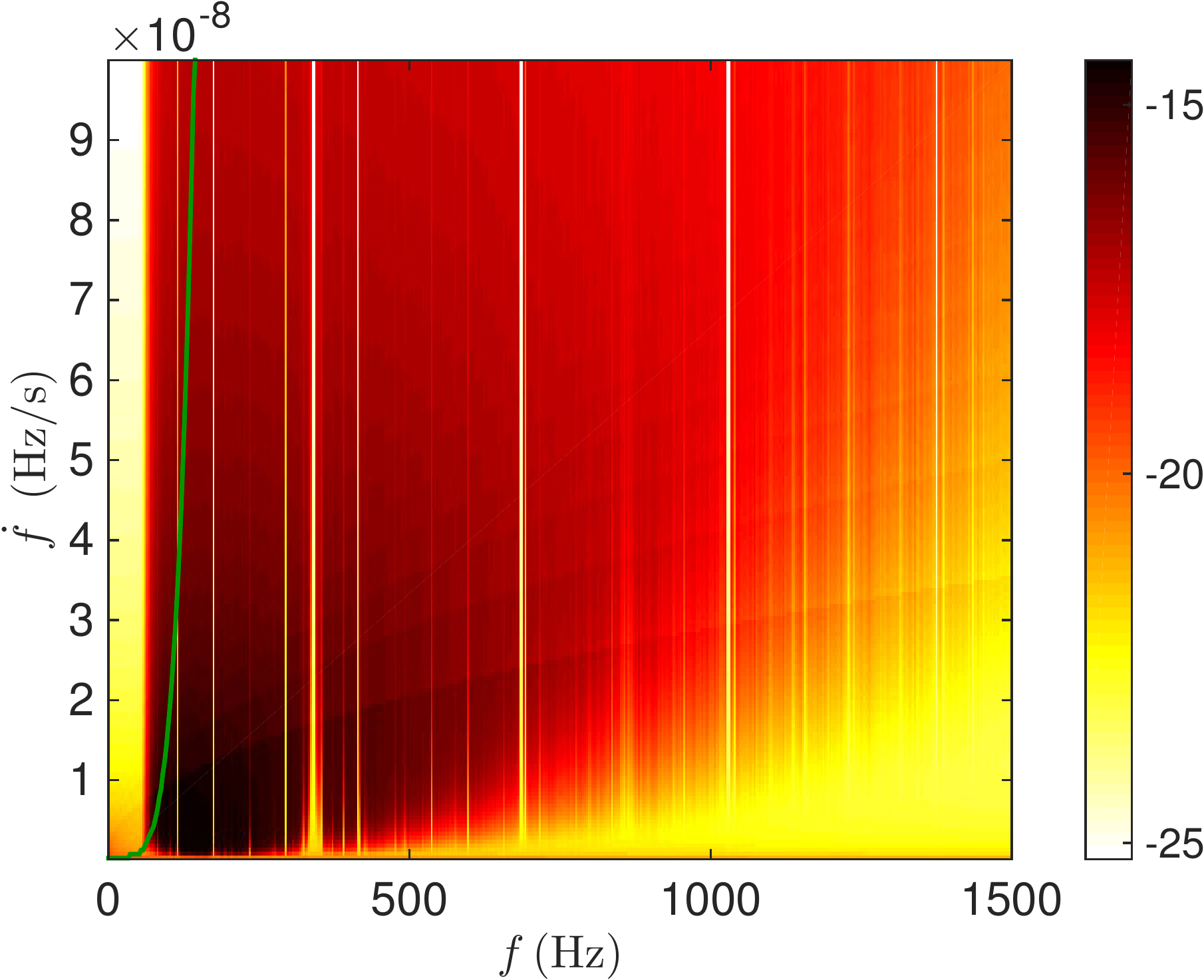}}}%
    \qquad
    \subfloat[Coverage, 20 days]{{  \includegraphics[width=.20\linewidth]{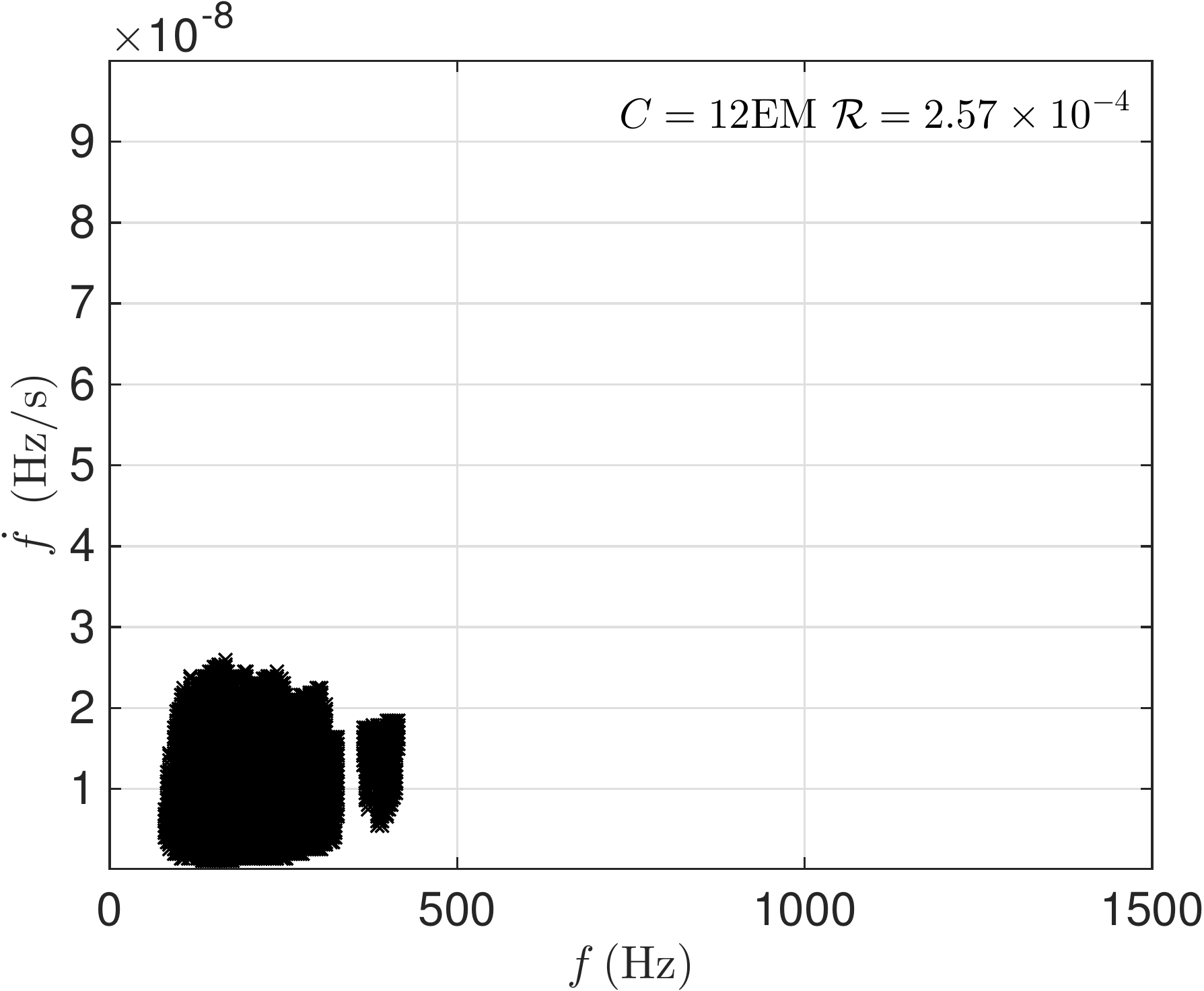}}}%
    \qquad
    \subfloat[Efficiency(lg), 30 days]{{  \includegraphics[width=.20\linewidth]{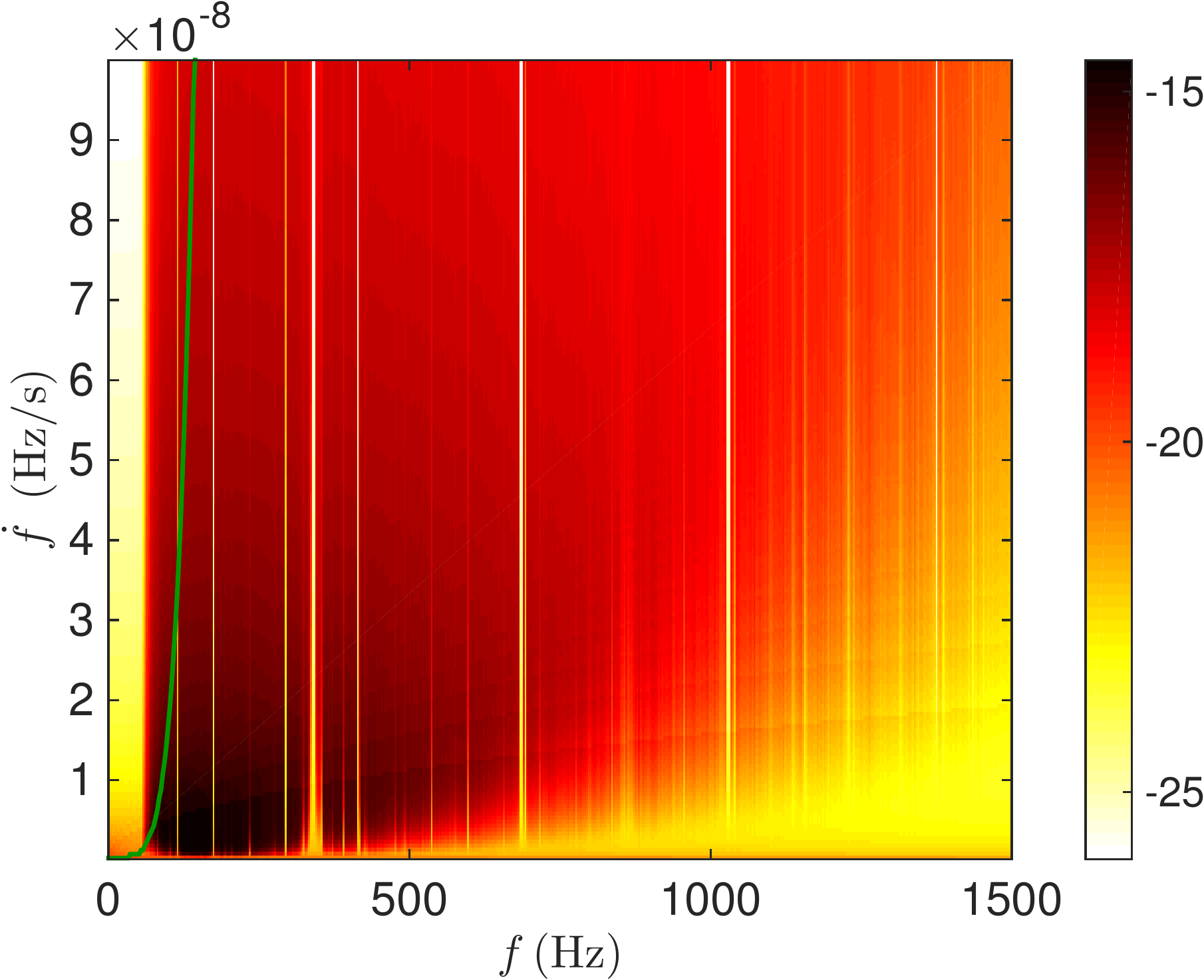}}}%
    \qquad
    \subfloat[Coverage, 30 days]{{  \includegraphics[width=.20\linewidth]{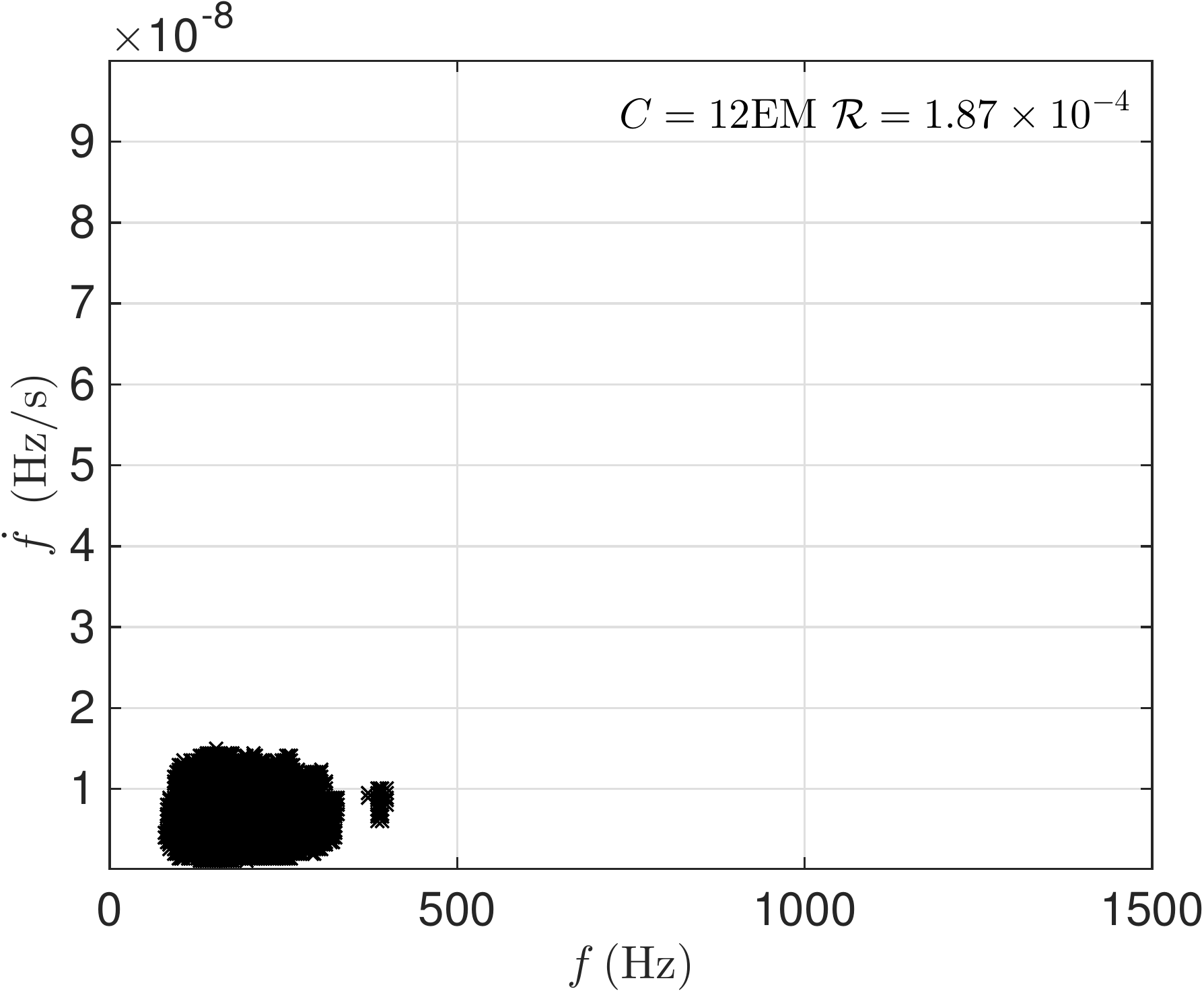}}}%
    \qquad
    \subfloat[Efficiency(lg), 37.5 days]{{  \includegraphics[width=.20\linewidth]{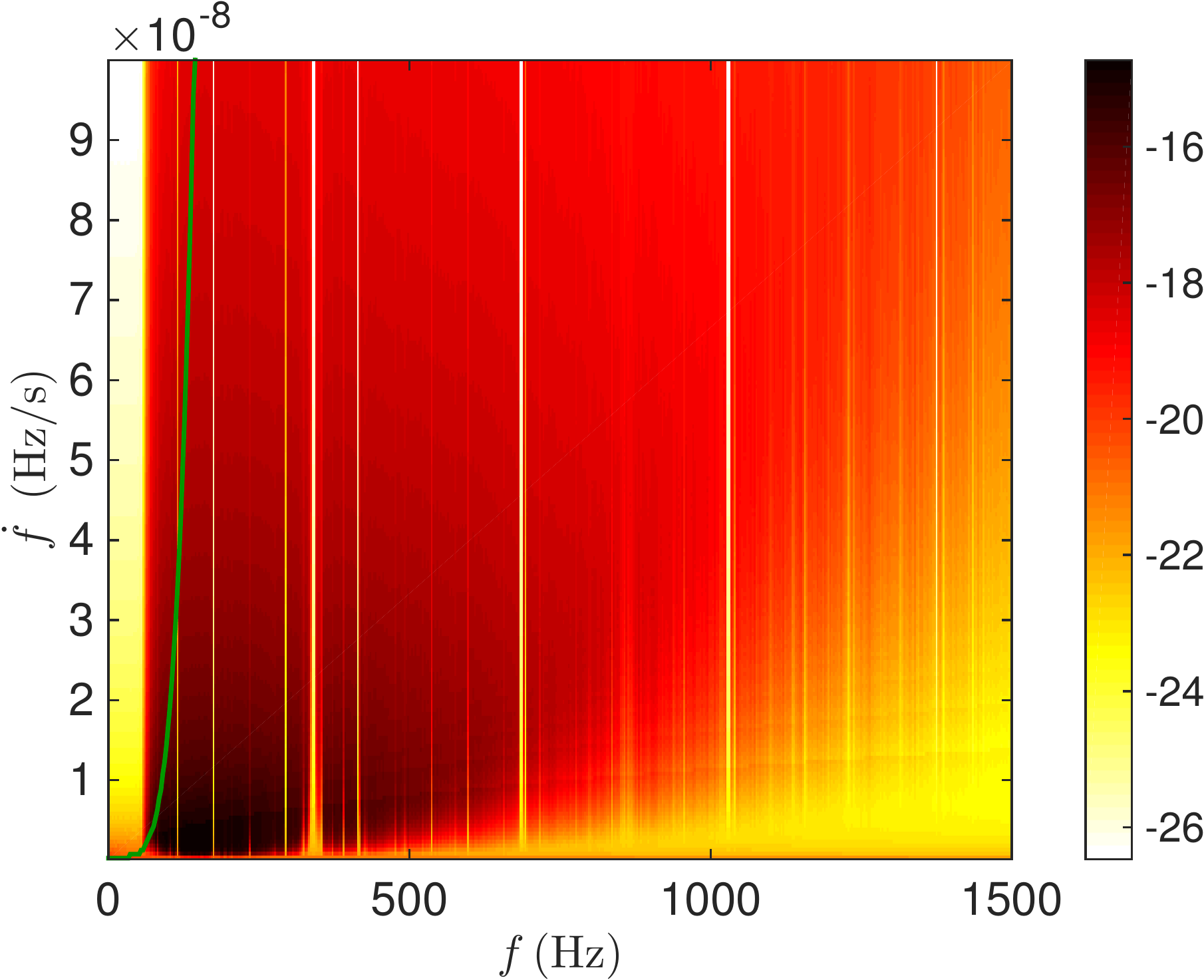}}}%
    \qquad
    \subfloat[Coverage, 37.5 days]{{  \includegraphics[width=.20\linewidth]{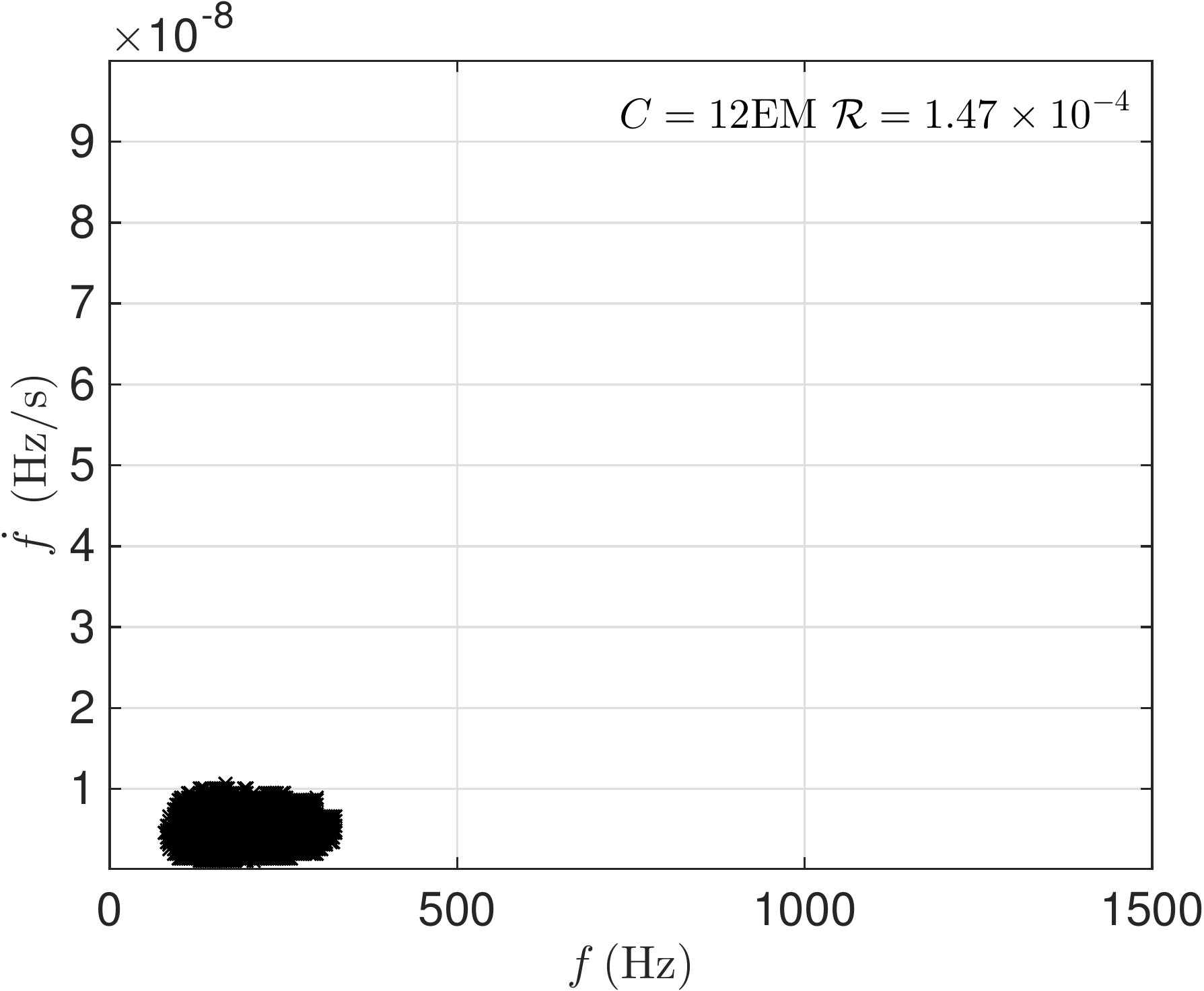}}}%
    \qquad
    \subfloat[Efficiency(lg), 50 days]{{  \includegraphics[width=.20\linewidth]{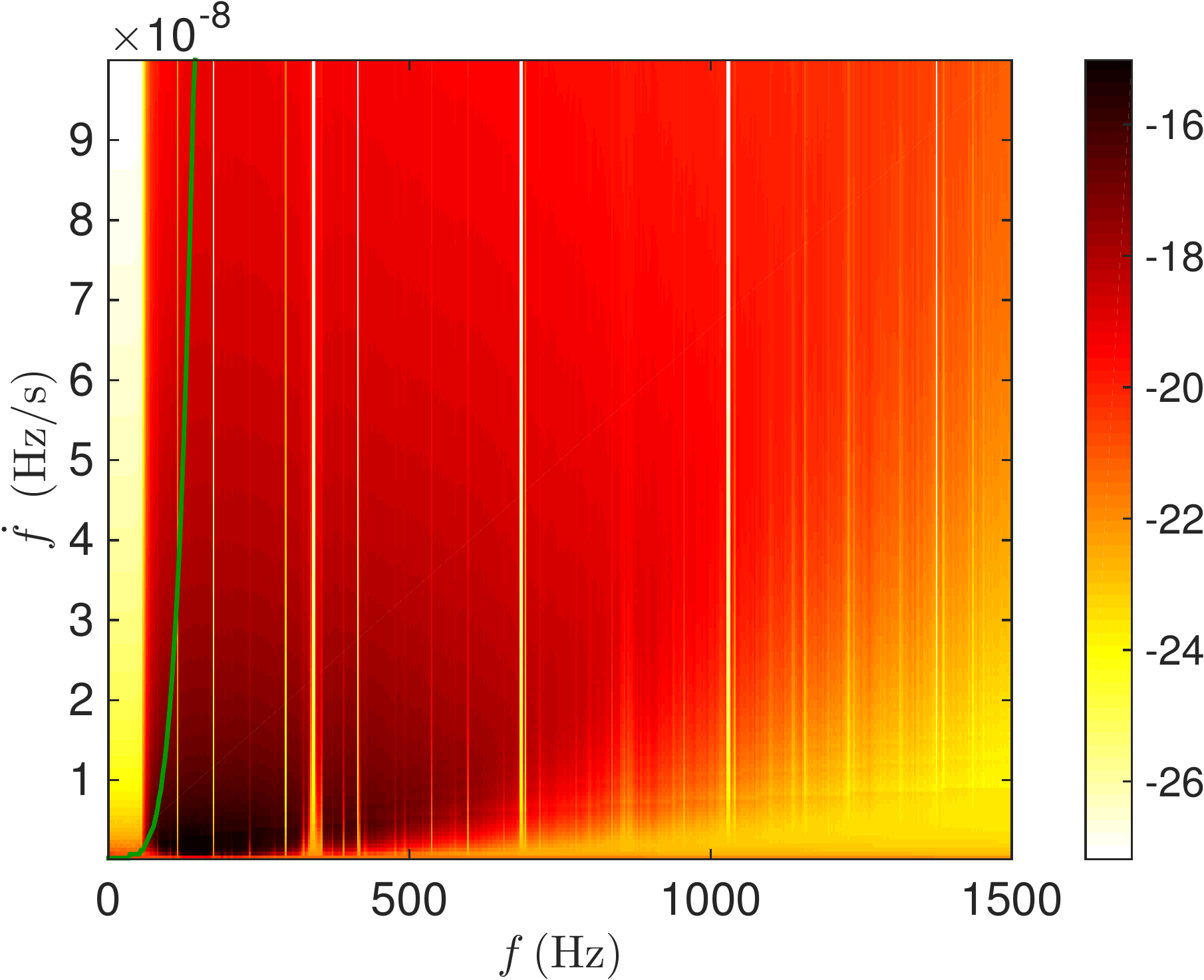}}}%
    \qquad
    \subfloat[Coverage, 50 days]{{  \includegraphics[width=.20\linewidth]{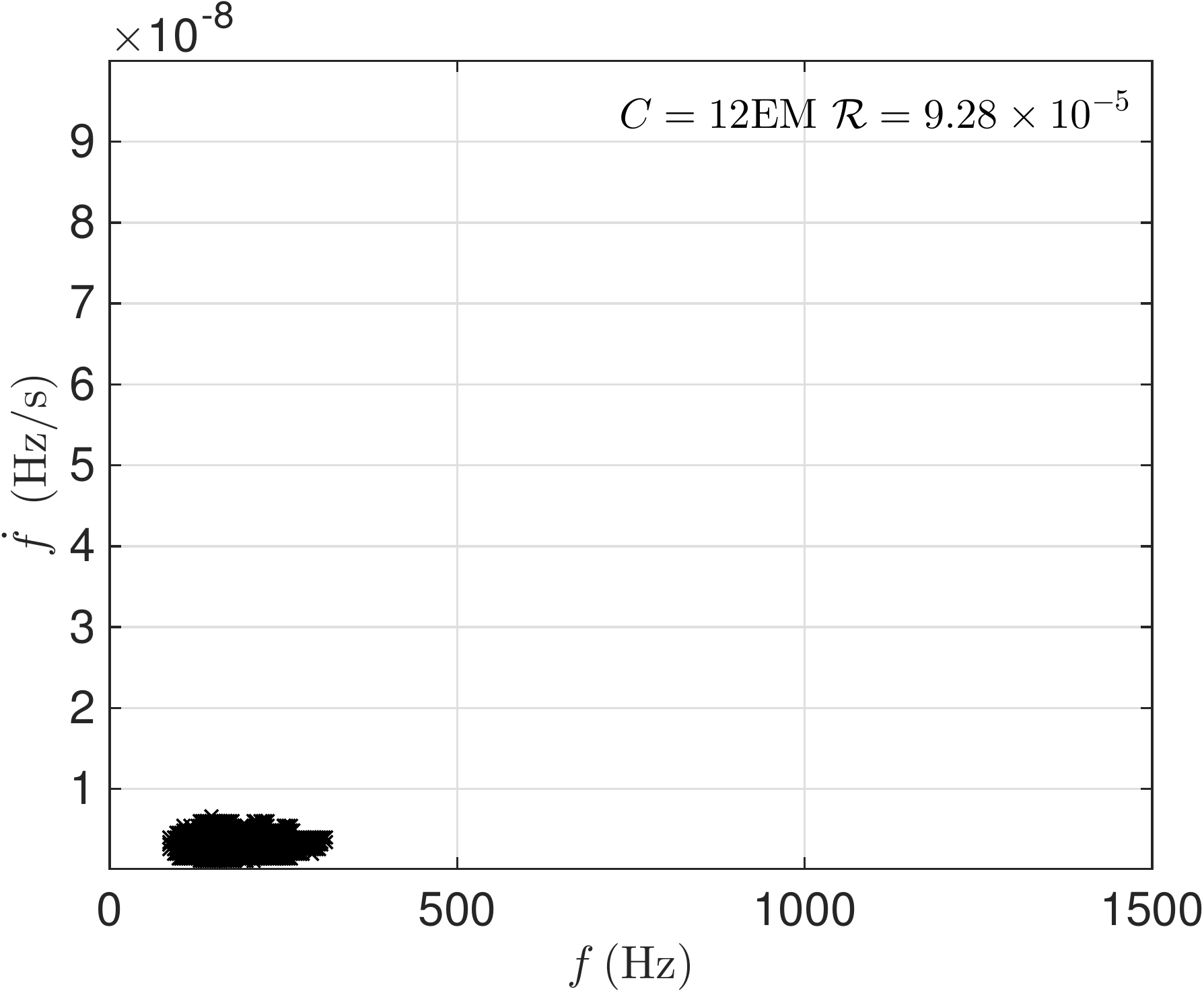}}}%
    \qquad
    \subfloat[Efficiency(lg), 75 days]{{  \includegraphics[width=.20\linewidth]{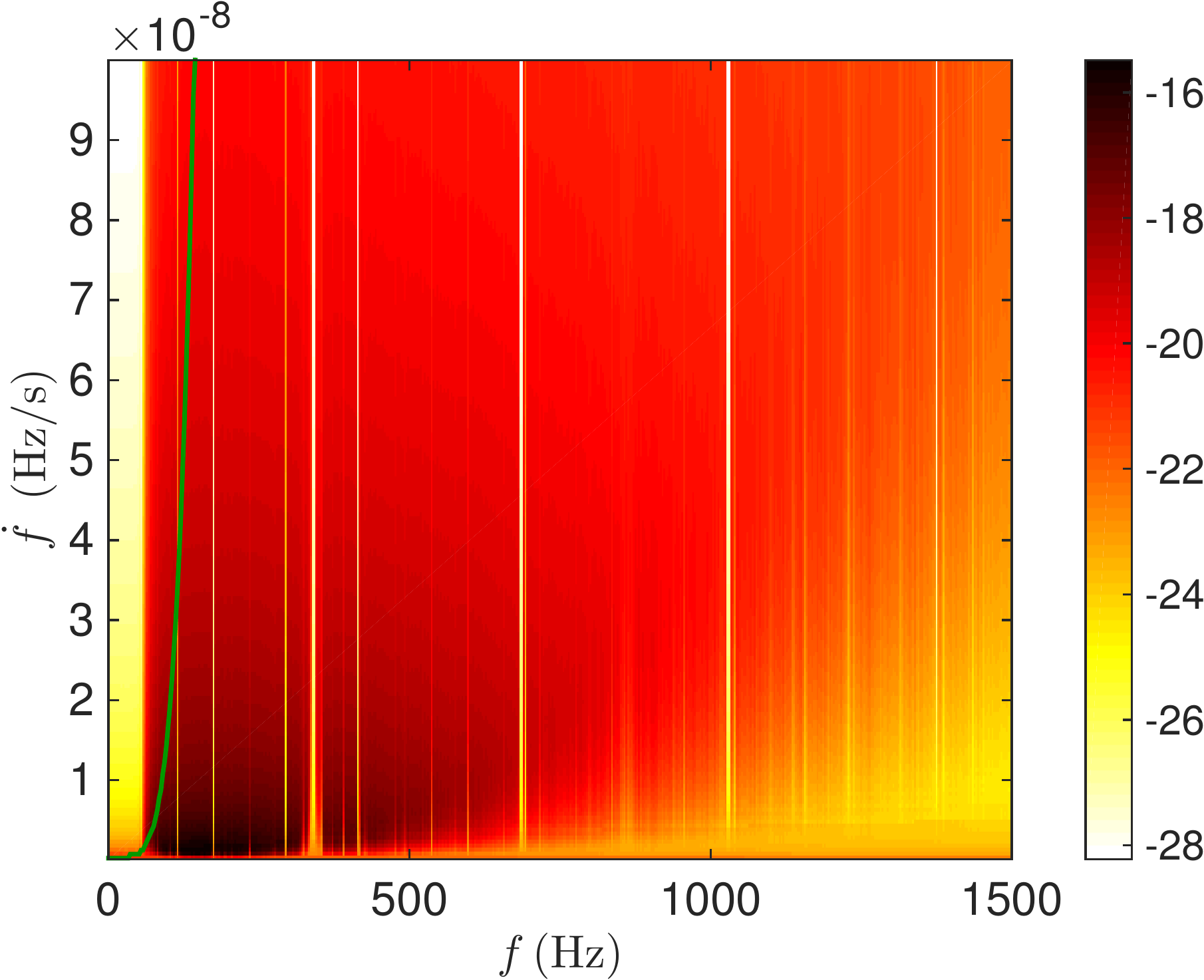}}}%
    \qquad
    \subfloat[Coverage, 75 days]{{  \includegraphics[width=.20\linewidth]{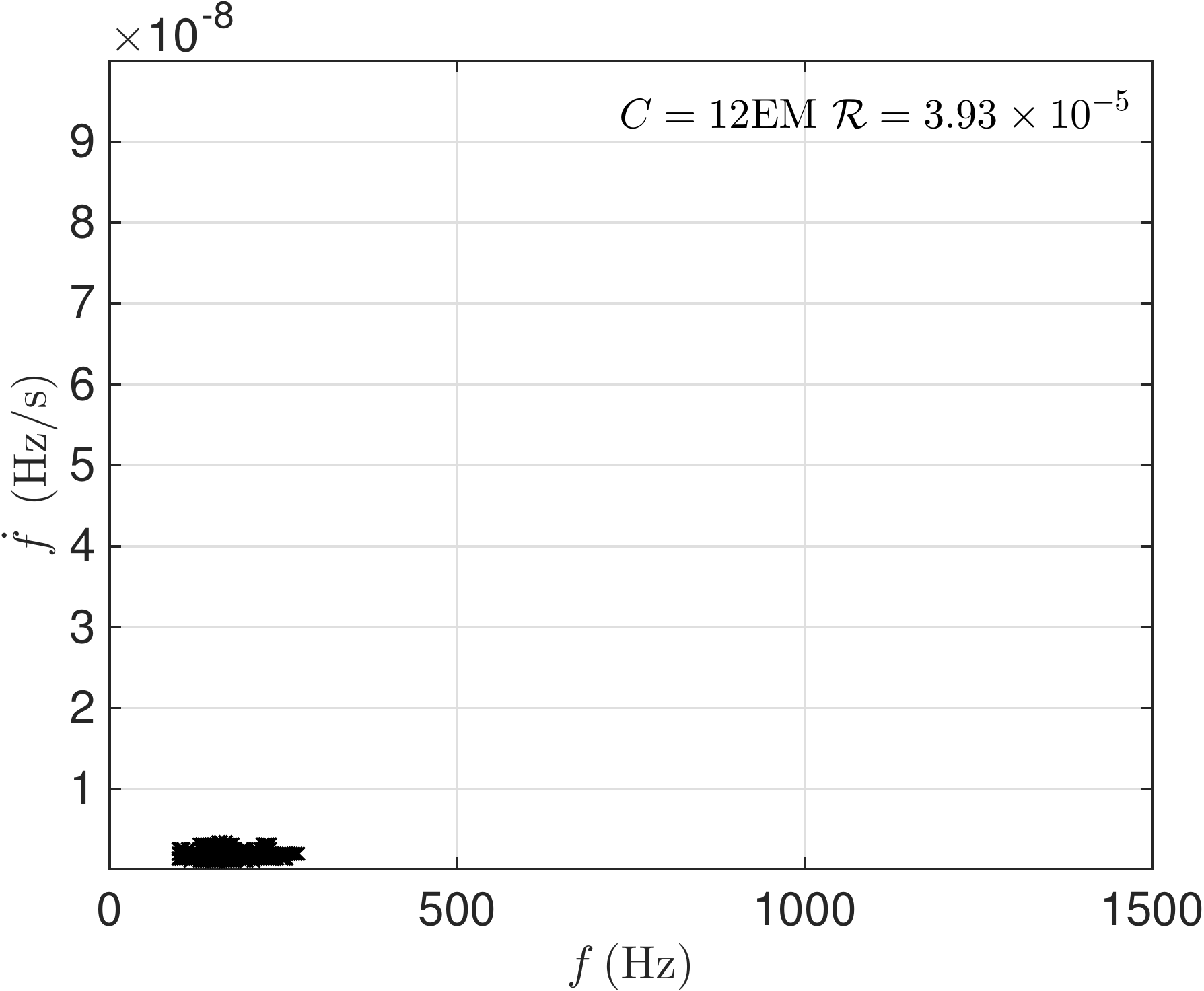}}}%
     \caption{Optimisation results for Cas A at 3500 pc assuming log-uniform and distance-based priors, for various coherent search durations: 5, 10, 20, 30, 37.5, 50 and 75 days. The total computing budget is assumed to be 12 EM.}%
    \label{CasA_51020days_noage_log}%
\end{figure*}

\begin{figure*}%
    \centering
    \subfloat[Coverage, cost: 12 EM]{{  \includegraphics[width=.45\linewidth]{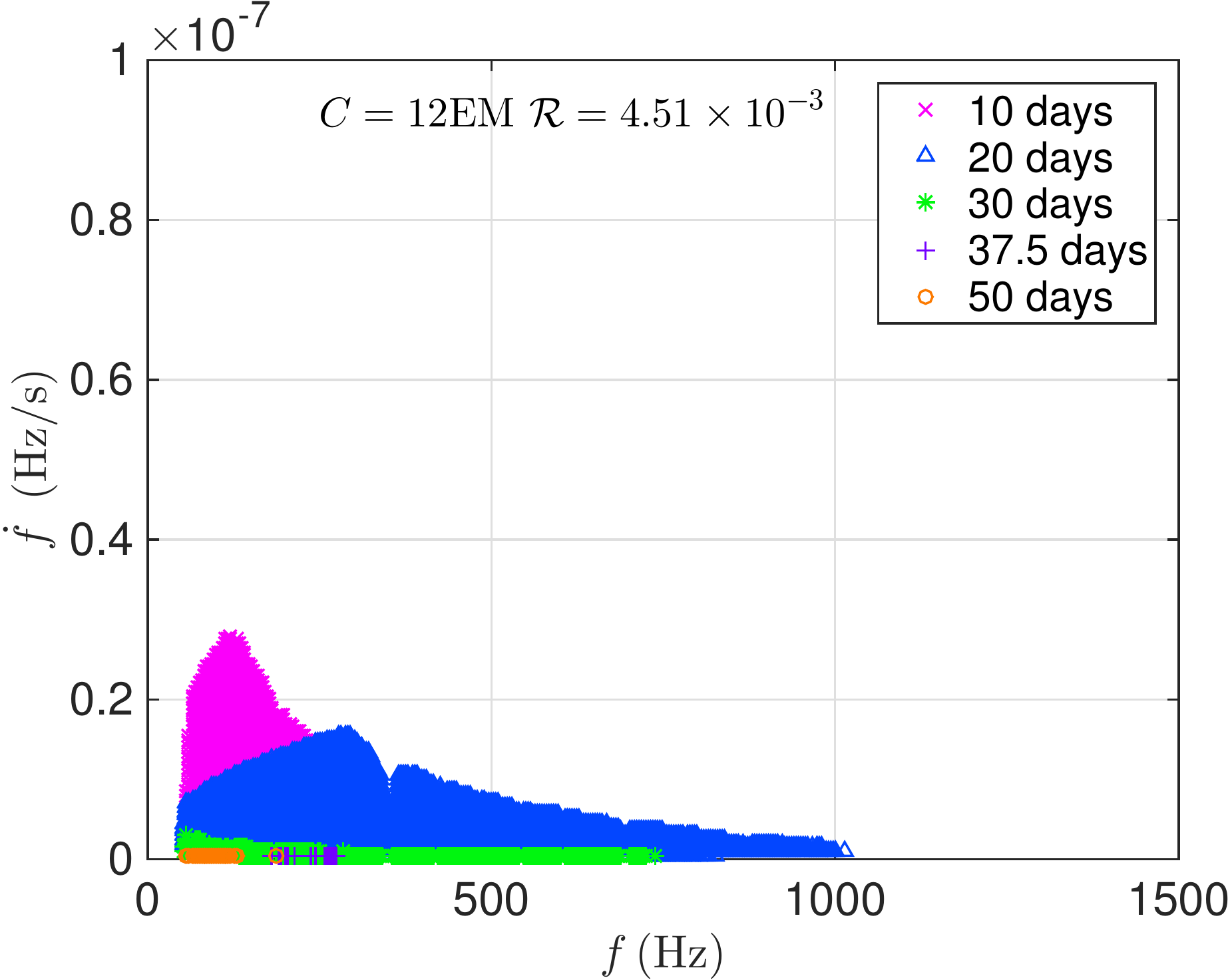}}}%
    \qquad
    \subfloat[Coverage,  cost: 24 EM]{{  \includegraphics[width=.45\linewidth]{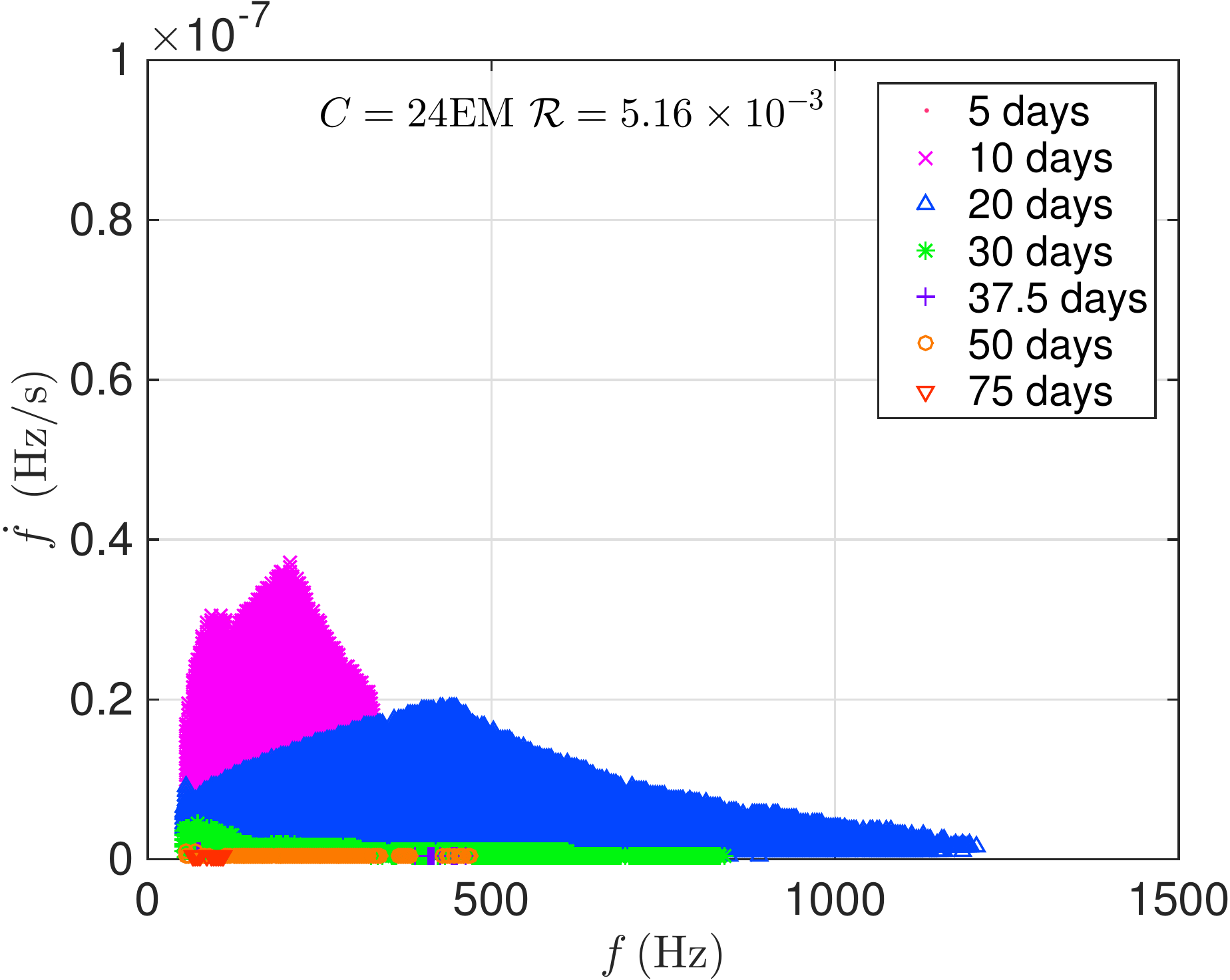}}}%
    \caption{Parameter space coverage for Vela Jr at 200 pc, assuming log-uniform and distance-based priors and optimizing over the 7  search set-ups also considered above at 12 EM (left plot) and 24 EM (right plot).}%
    \label{G2662_best_noage_shortdis_log}%
\end{figure*}

\begin{figure*}%
    \centering
    \subfloat[Coverage, cost: 12 EM]{{  \includegraphics[width=.45\linewidth]{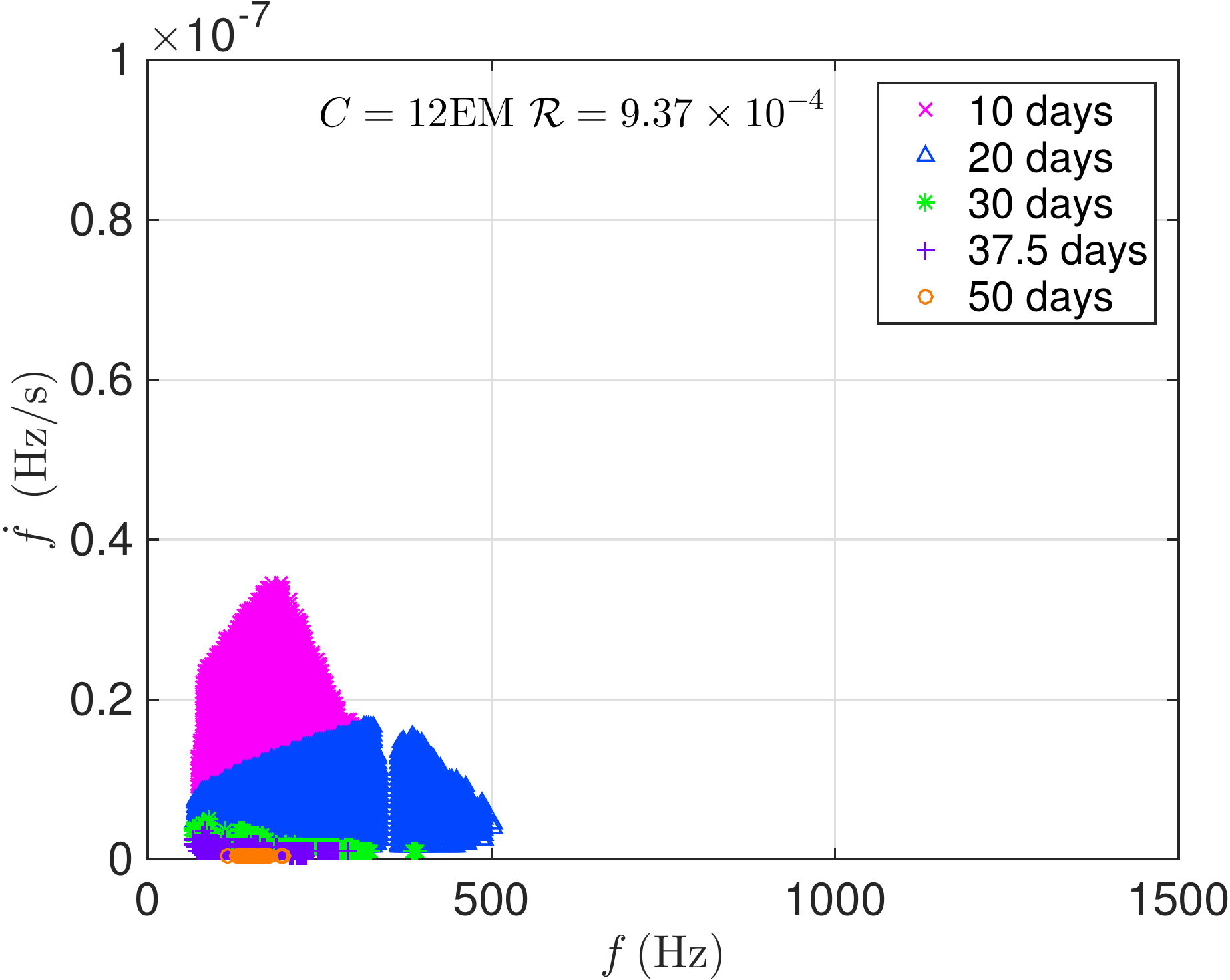}}}%
    \qquad
    \subfloat[Coverage,  cost: 24 EM]{{  \includegraphics[width=.45\linewidth]{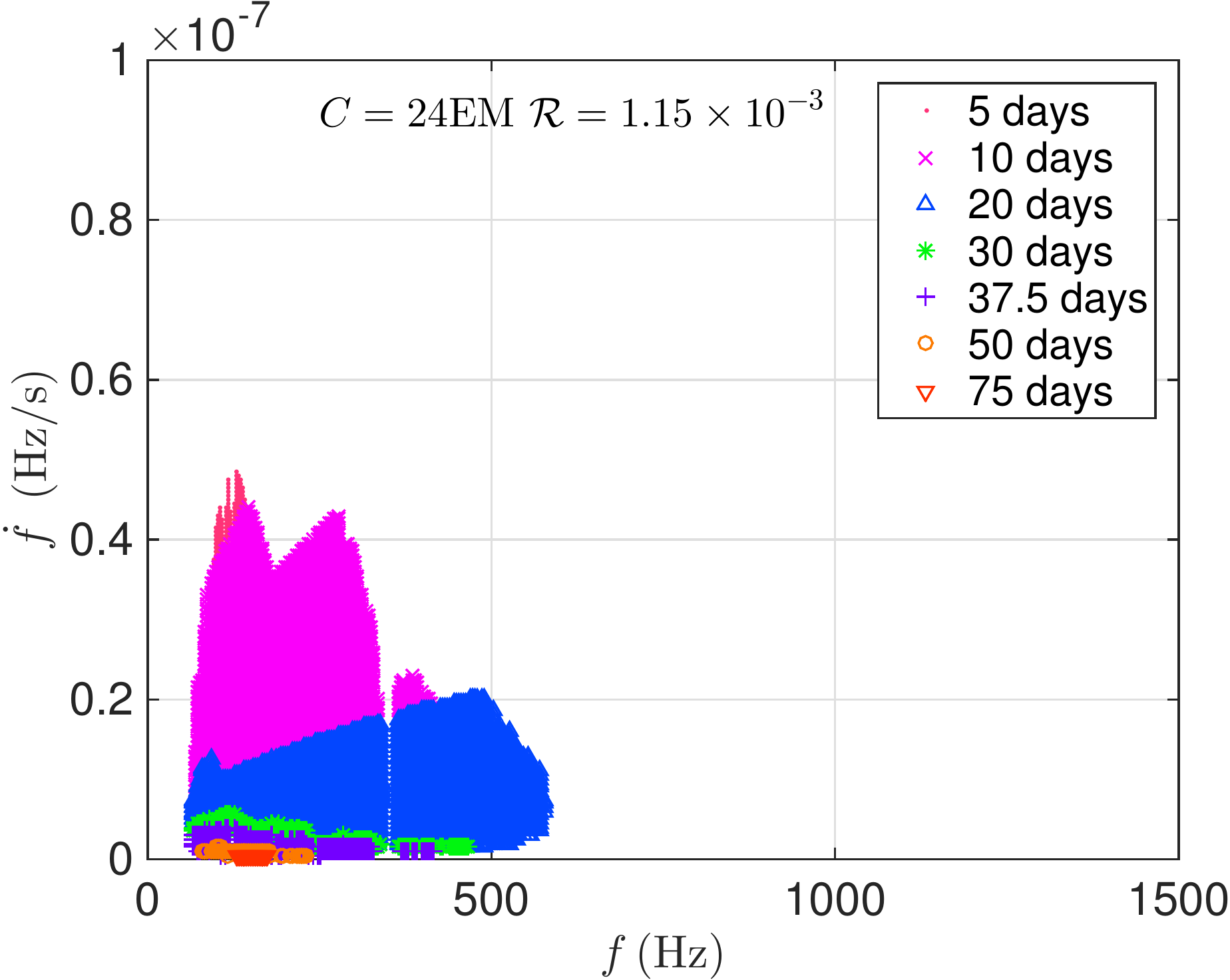}}}%
   \caption{Parameter space coverage for G347.3 at 1300 pc, assuming log-uniform and distance-based priors and optimizing over the 7  search set-ups also considered above at 12 EM (left plot) and 24 EM (right plot).}%
    \label{G3473_best_noage_log}%
\end{figure*}

\begin{figure*}%
    \centering
    \subfloat[Coverage, cost: 12 EM]{{  \includegraphics[width=.45\linewidth]{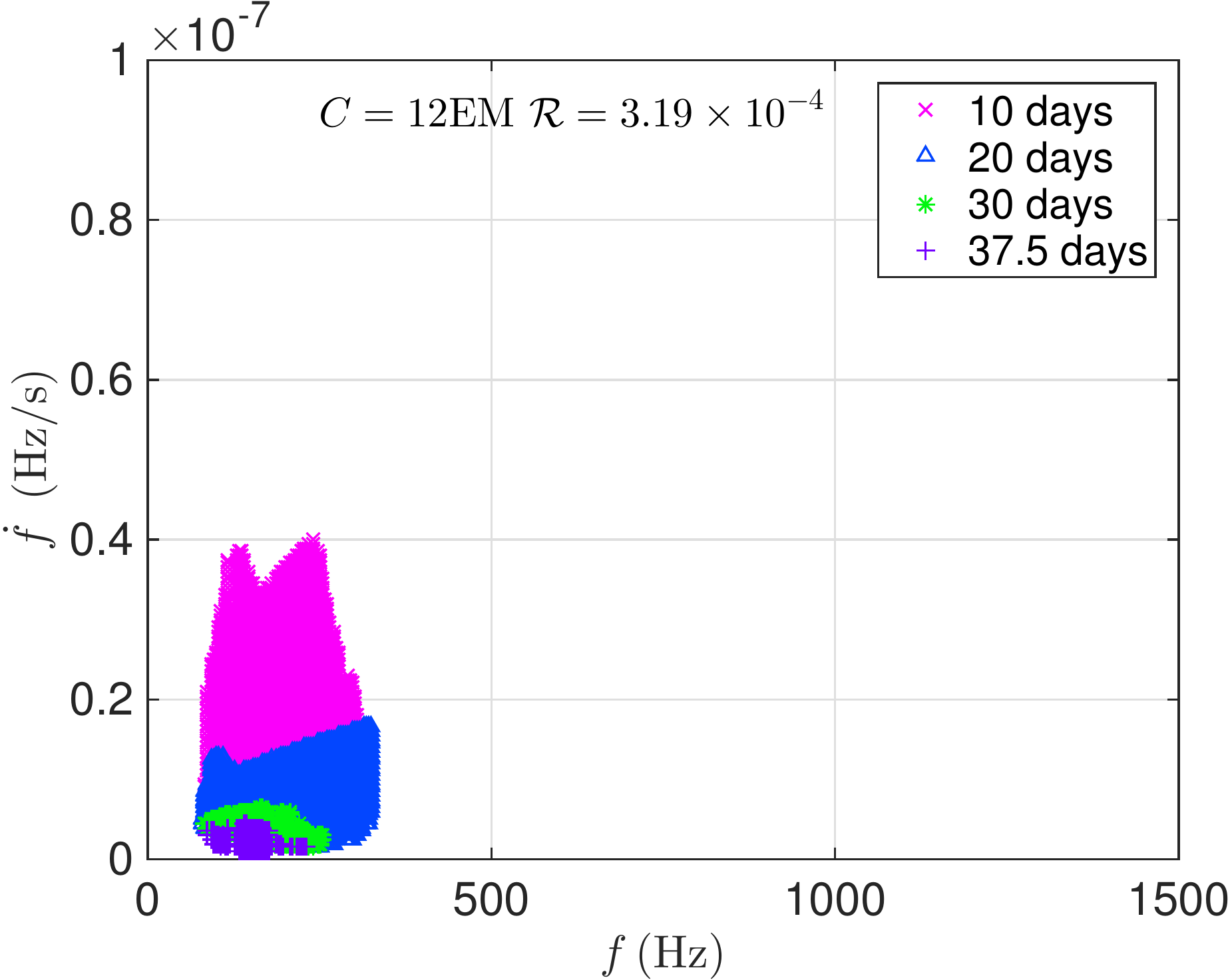}}}%
    \qquad
    \subfloat[Coverage,  cost: 24 EM]{{  \includegraphics[width=.45\linewidth]{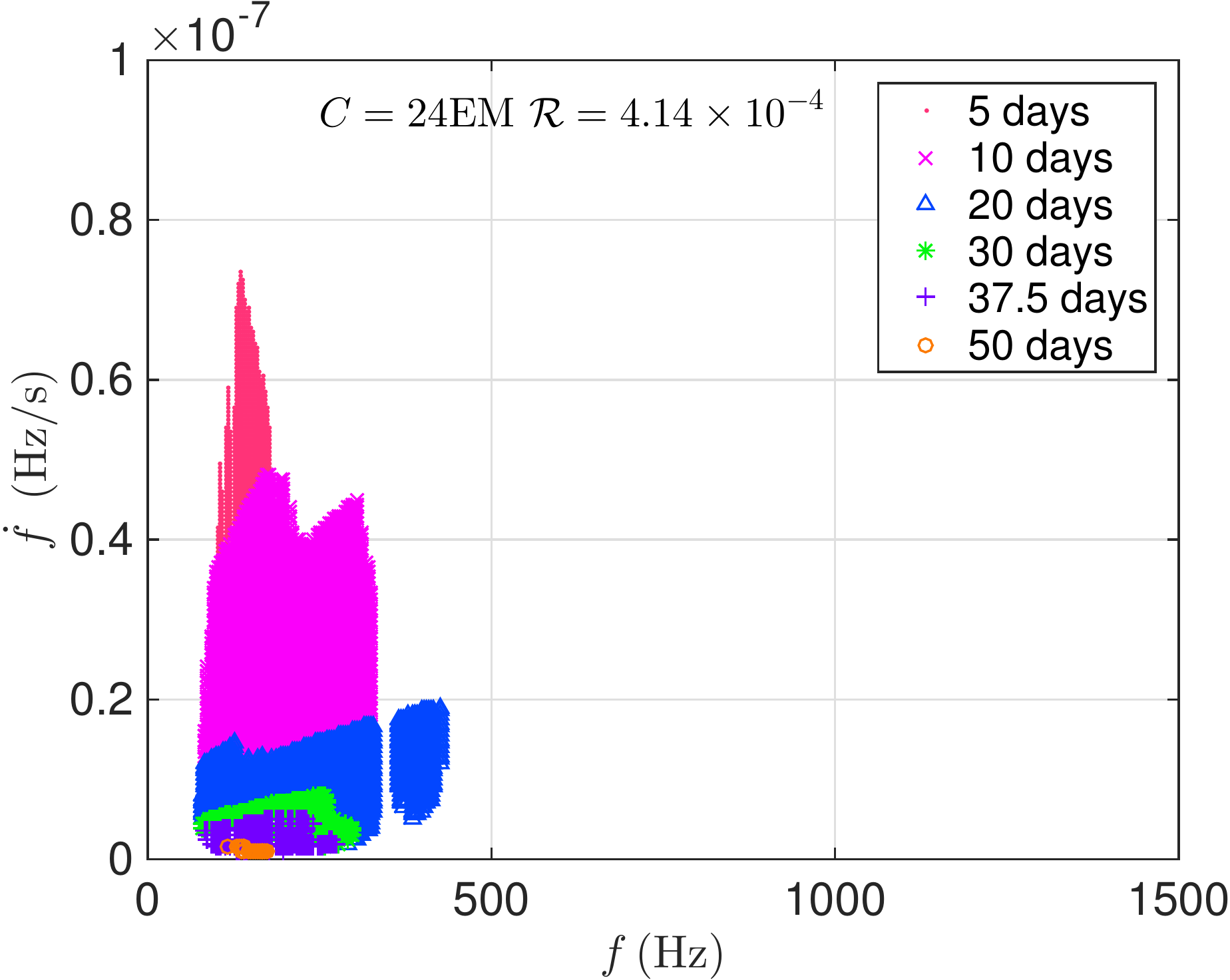}}}%
    \caption{Parameter space coverage for Cas A at 3500 pc, assuming log-uniform and distance-based priors and optimizing over the 7  search set-ups also considered above at 12 EM (left plot) and 24 EM (right plot).}%
    \label{CasA_best_noage_log}%
\end{figure*}

\begin{figure*}%
    \centering
    \subfloat[Coverage of 3 sources, Vela Jr at 200 pc, Cost 12 EM ]{{  \includegraphics[width=.45\linewidth]{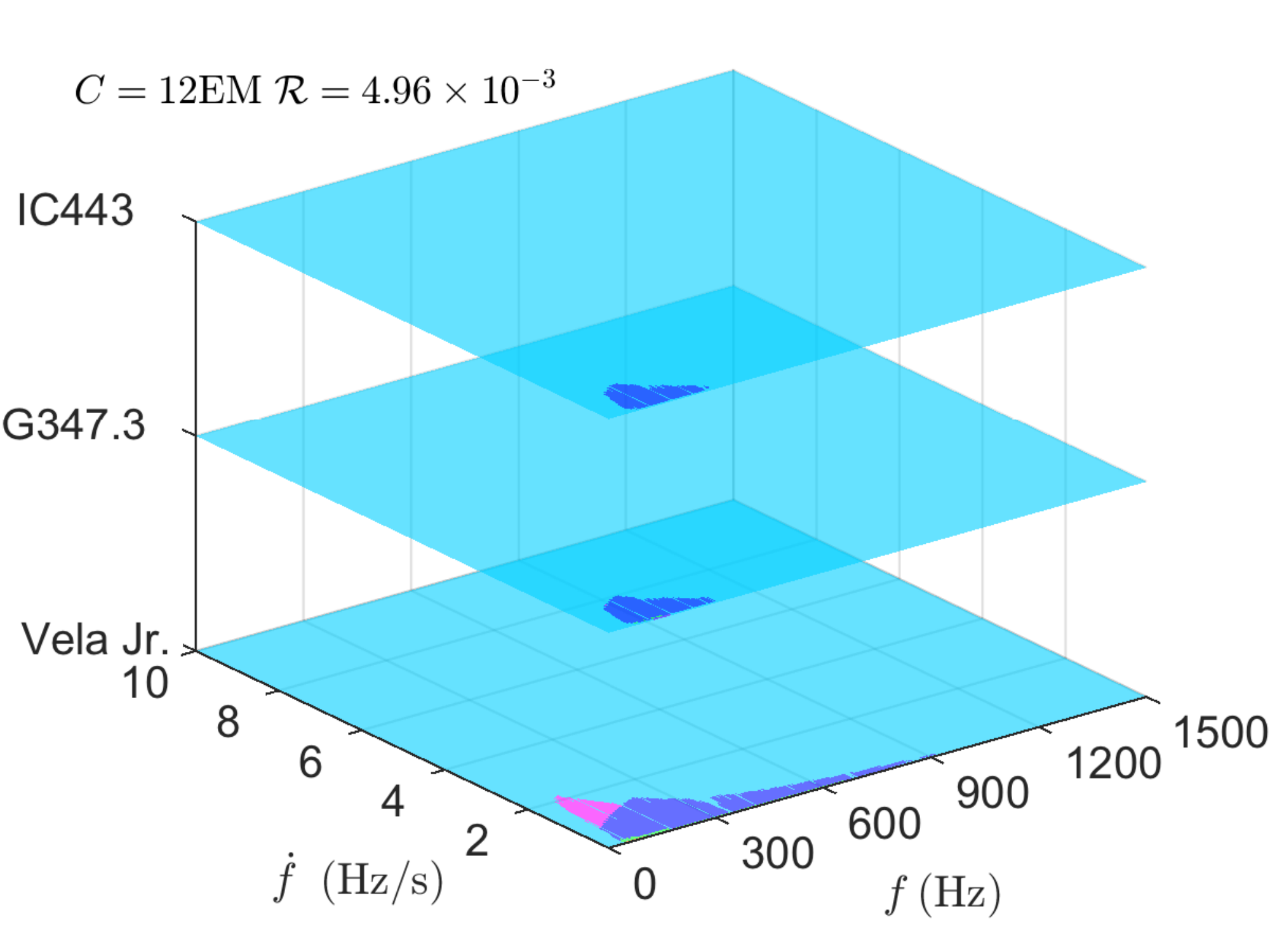}}}%
    \qquad
    \subfloat[Coverage of 3 sources, Vela Jr at 750 pc, Cost 12 EM ]{{  \includegraphics[width=.45\linewidth]{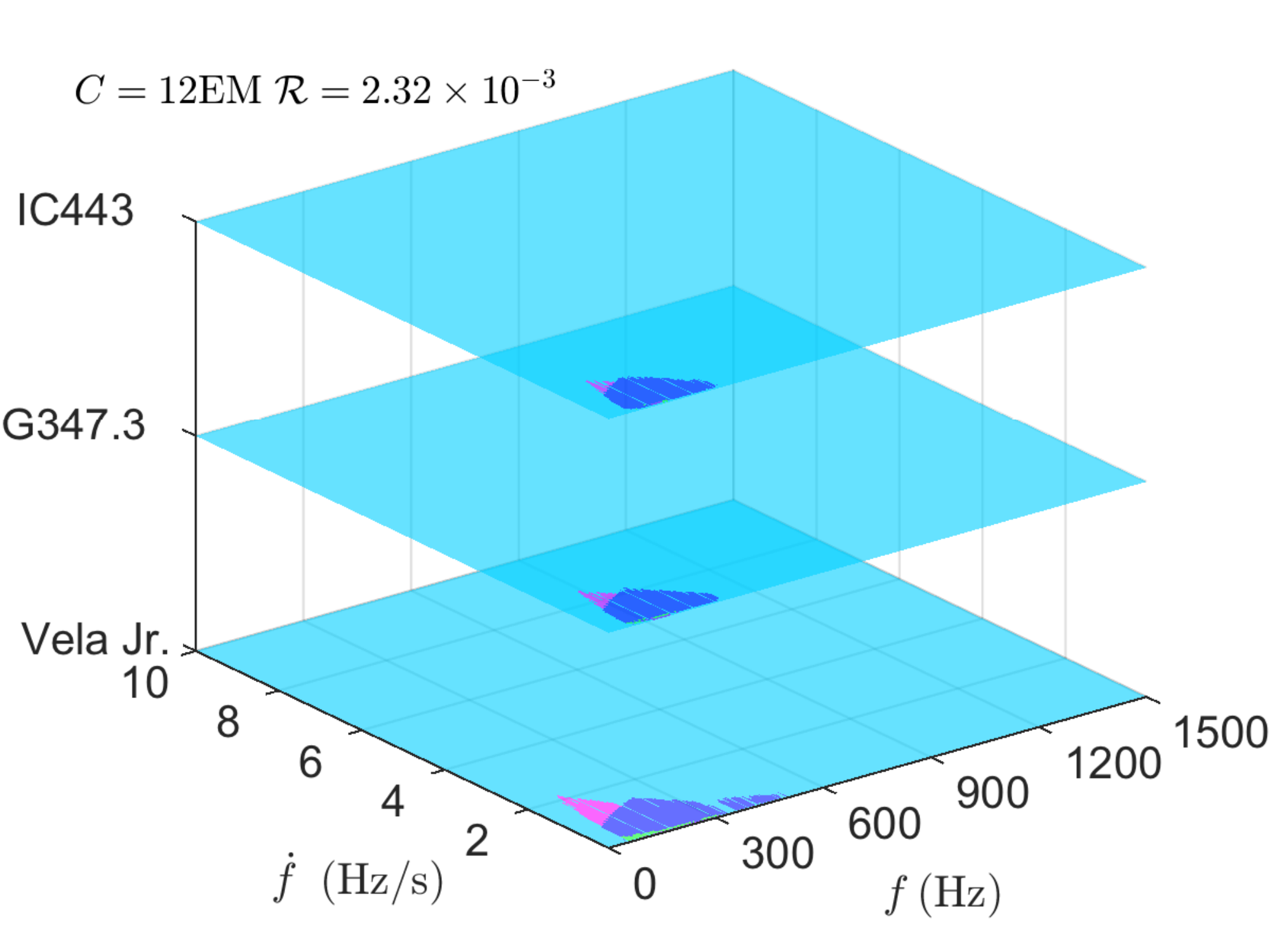}}}%
    \qquad
    \subfloat[Coverage of 3 sources, Vela Jr at 200 pc, Cost 24 EM ]{{  \includegraphics[width=.45\linewidth]{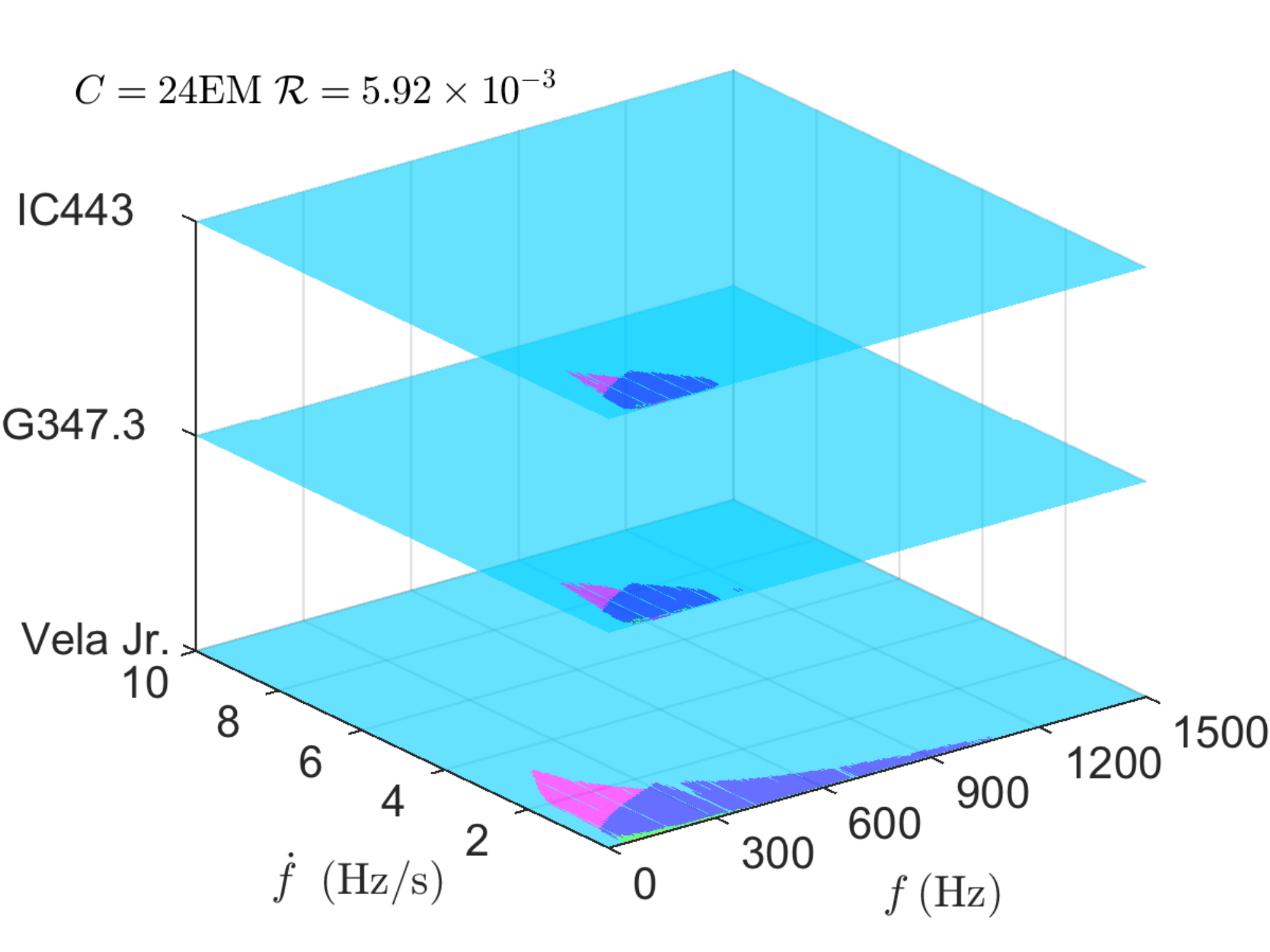}}}%
    \qquad
    \subfloat[Coverage of 3 sources, Vela Jr at 750 pc, Cost 24 EM ]{{  \includegraphics[width=.45\linewidth]{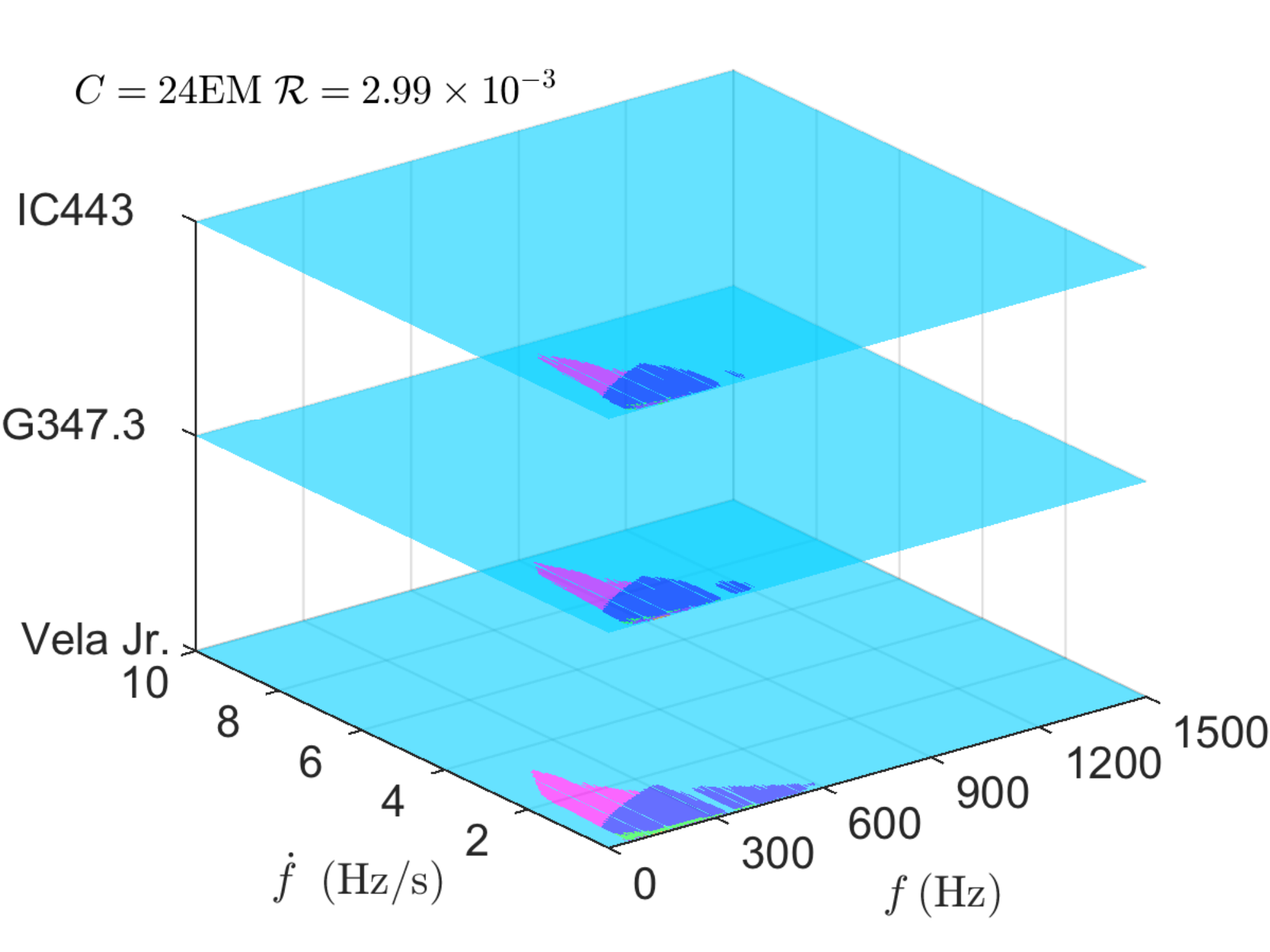}}}%
    \qquad
    \subfloat[Coverage of 3 sources, Vela Jr at 200 pc, Cost 48 EM ]{{  \includegraphics[width=.45\linewidth]{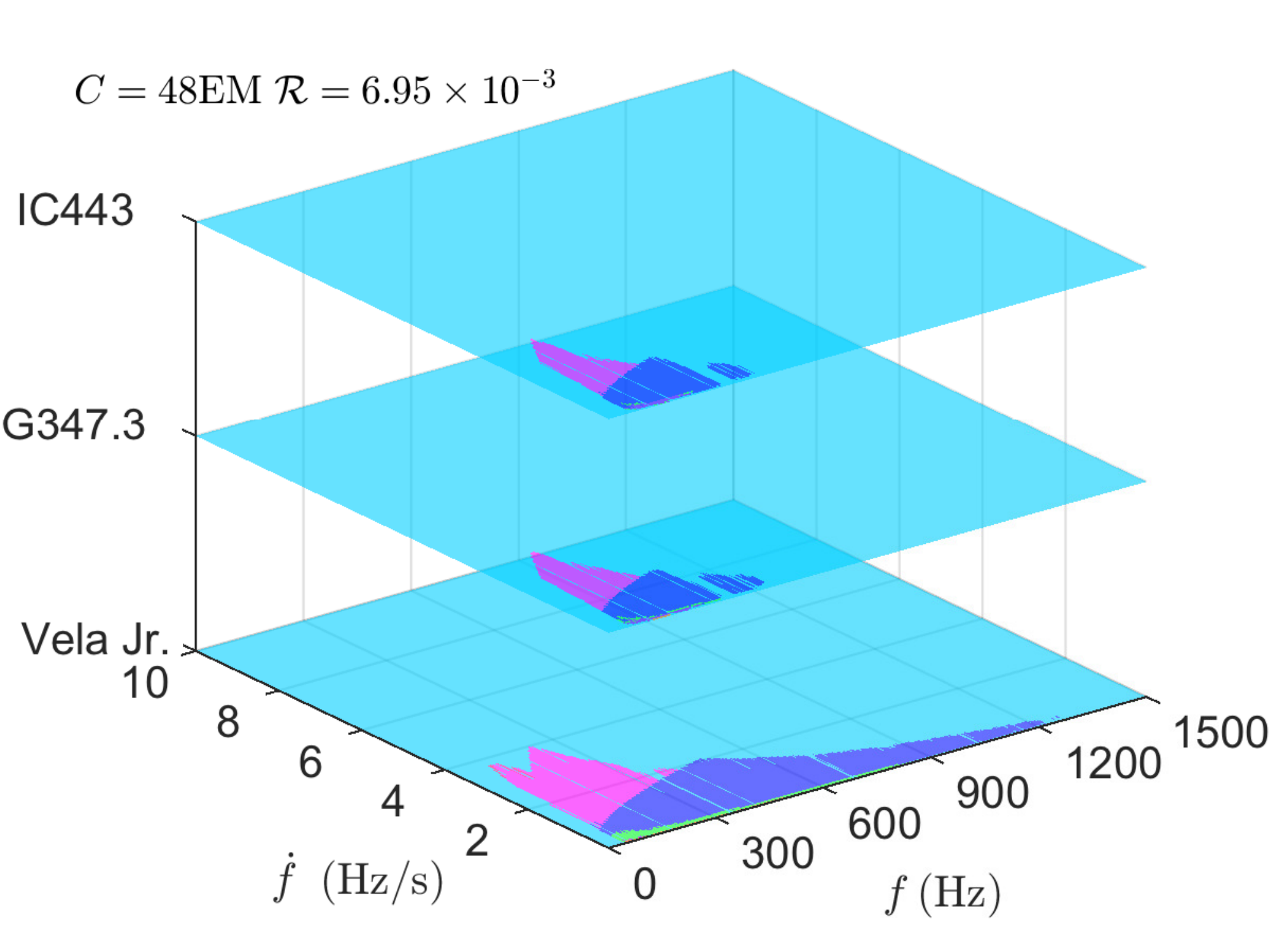}}}%
    \qquad
    \subfloat[Coverage of 3 sources, Vela Jr at 750 pc, Cost 48 EM ]{{  \includegraphics[width=.45\linewidth]{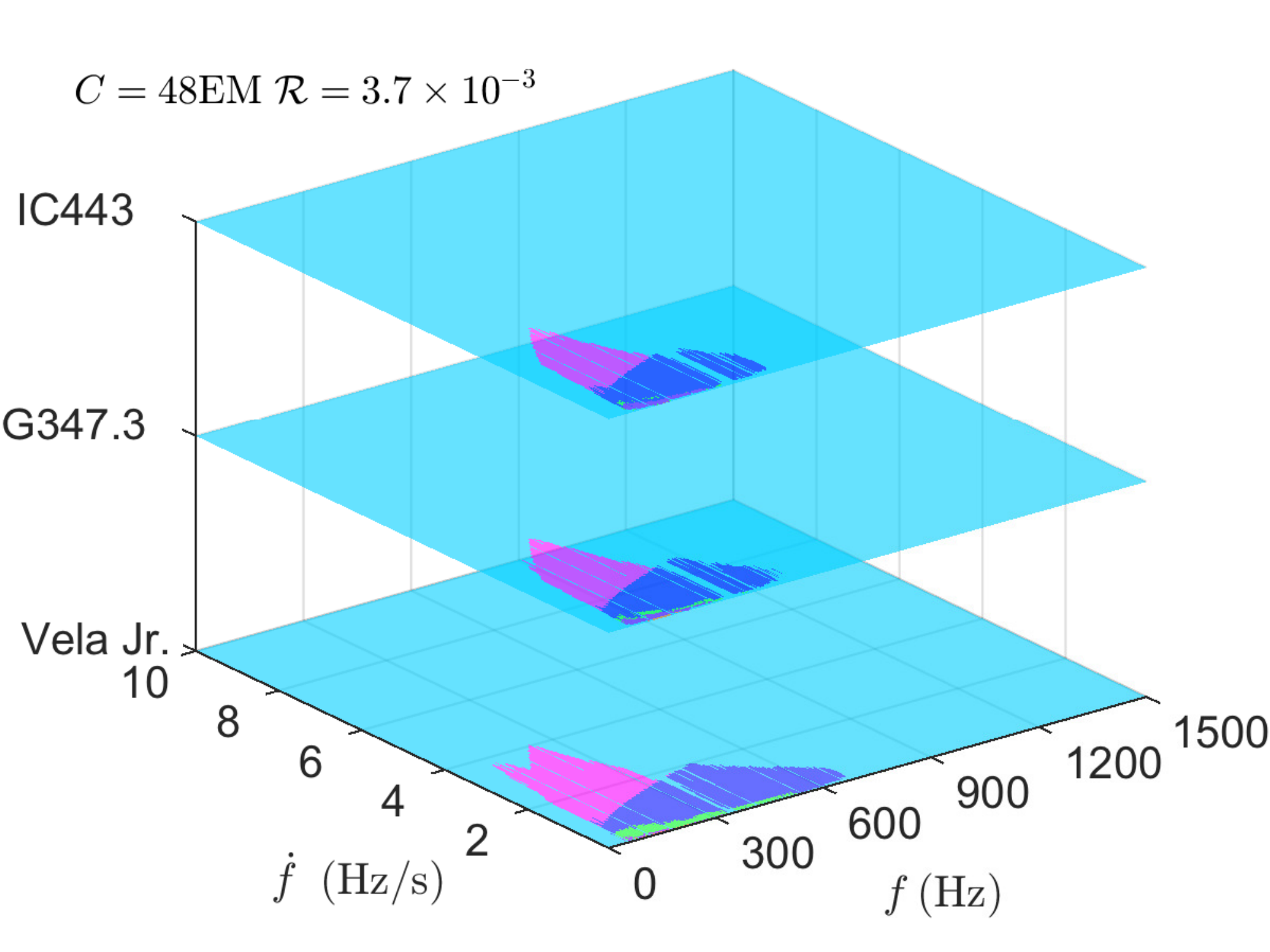}}}%
    
    \caption{Parameter space coverage assuming log-uniform and distance-based priors and optimizing over the 7 search set-ups also considered above and over the three closest targets (left plots: Vela Jr at 200 pc, G347.3, IC443 and right plots: Vela Jr at 750 pc, G347.3, IC443) at 12, 24 and 48 EMs.}%
    \label{all_noage_log}%
\end{figure*}



\begin{figure*}%
    \centering
    \subfloat[Efficiency(lg), 5 days]{{  \includegraphics[width=.20\linewidth]{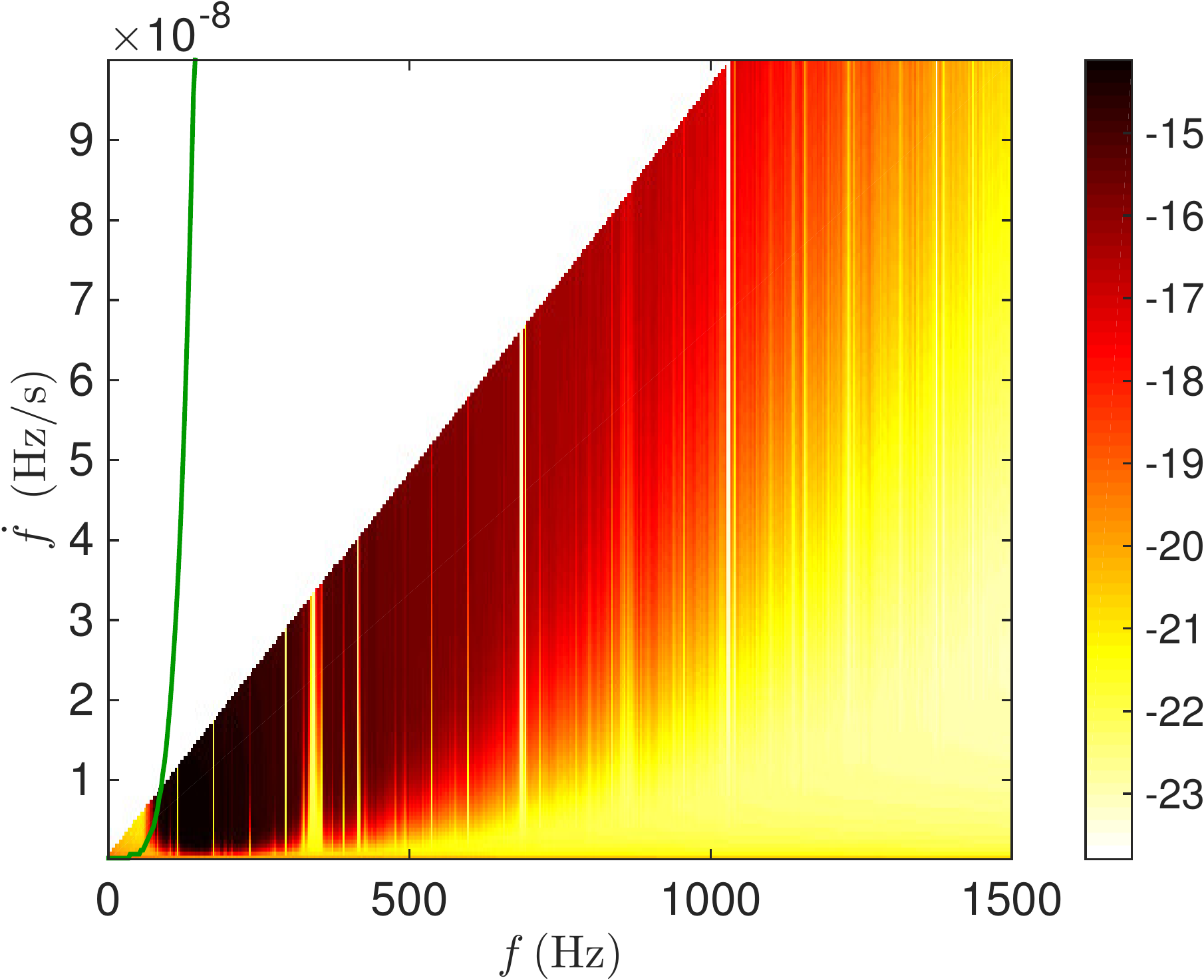}}}%
    \qquad
    \subfloat[Coverage, 5 days]{{  \includegraphics[width=.20\linewidth]{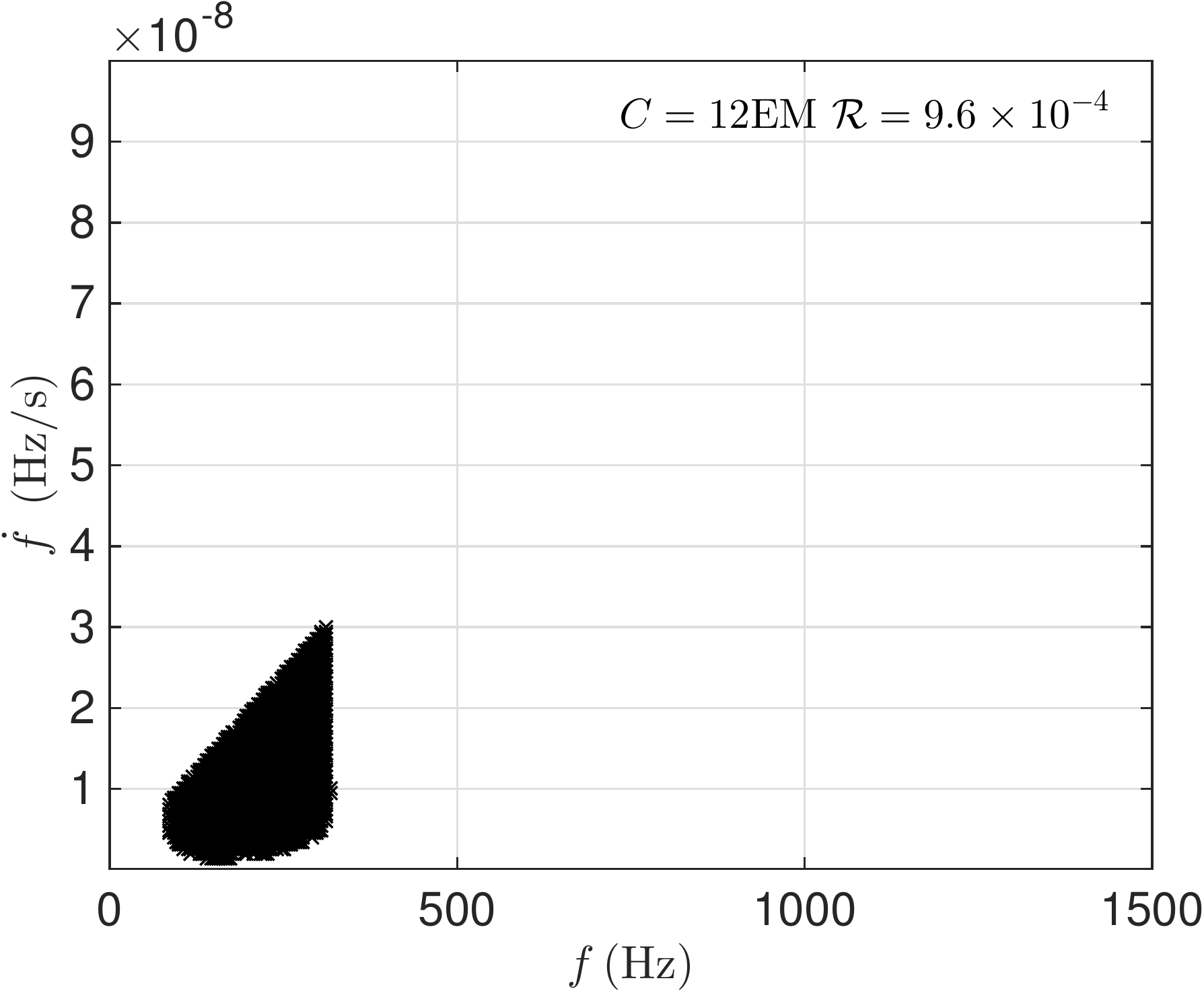}}}%
    \qquad
    \subfloat[Efficiency(lg), 10 days]{{  \includegraphics[width=.20\linewidth]{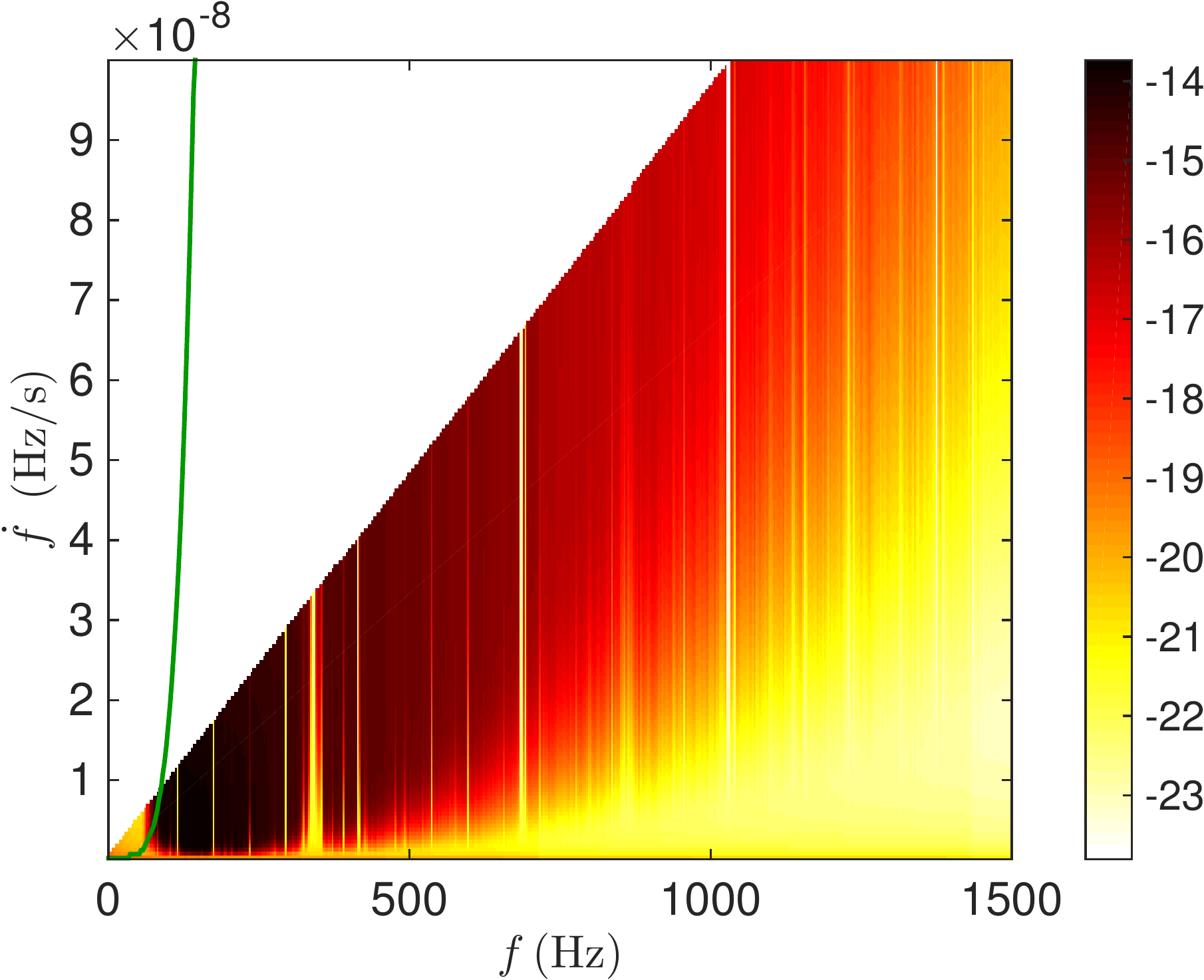}}}%
    \qquad
    \subfloat[Coverage, 10 days]{{  \includegraphics[width=.20\linewidth]{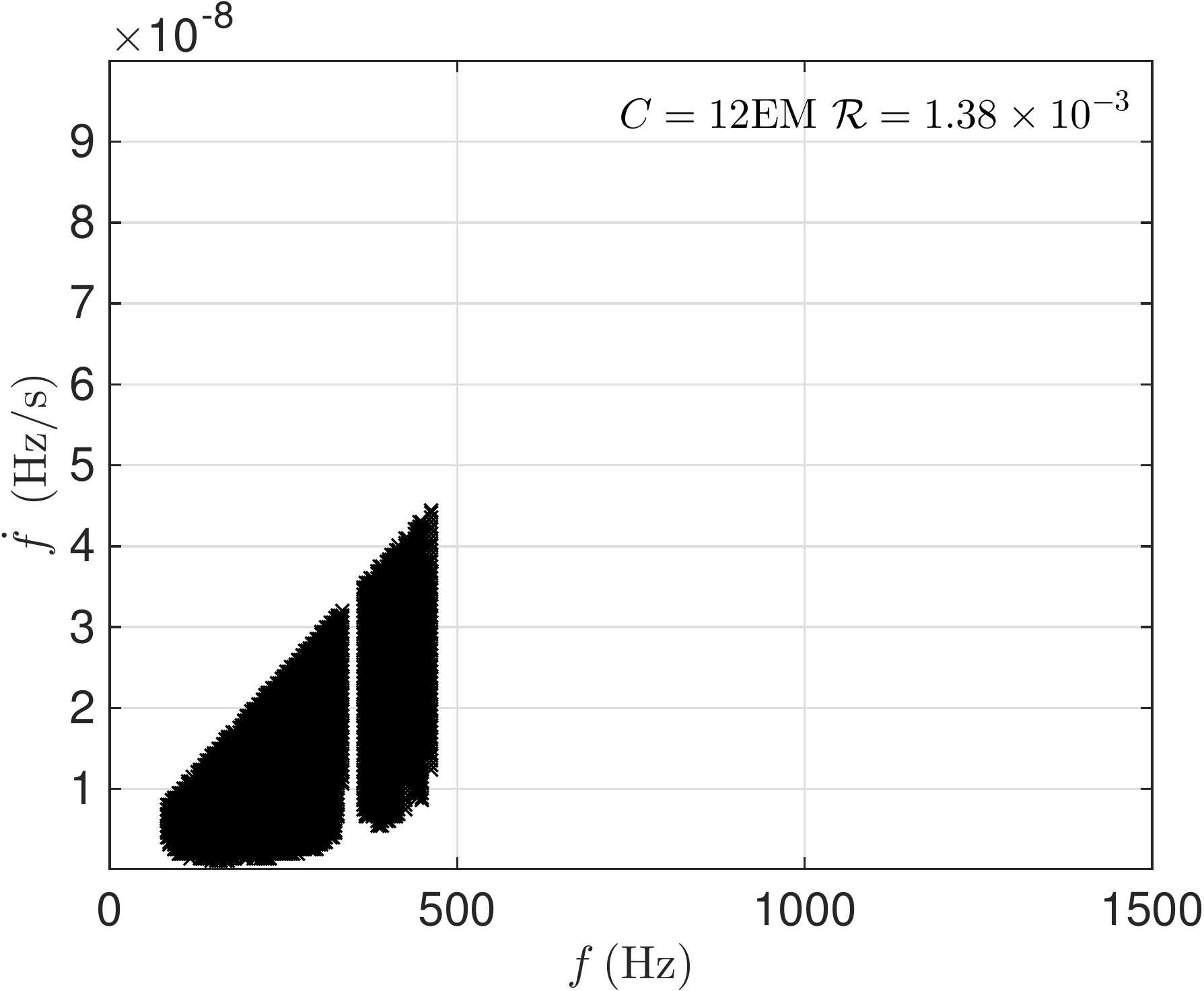}}}%
    \qquad
    \subfloat[Efficiency(lg), 20 days]{{  \includegraphics[width=.20\linewidth]{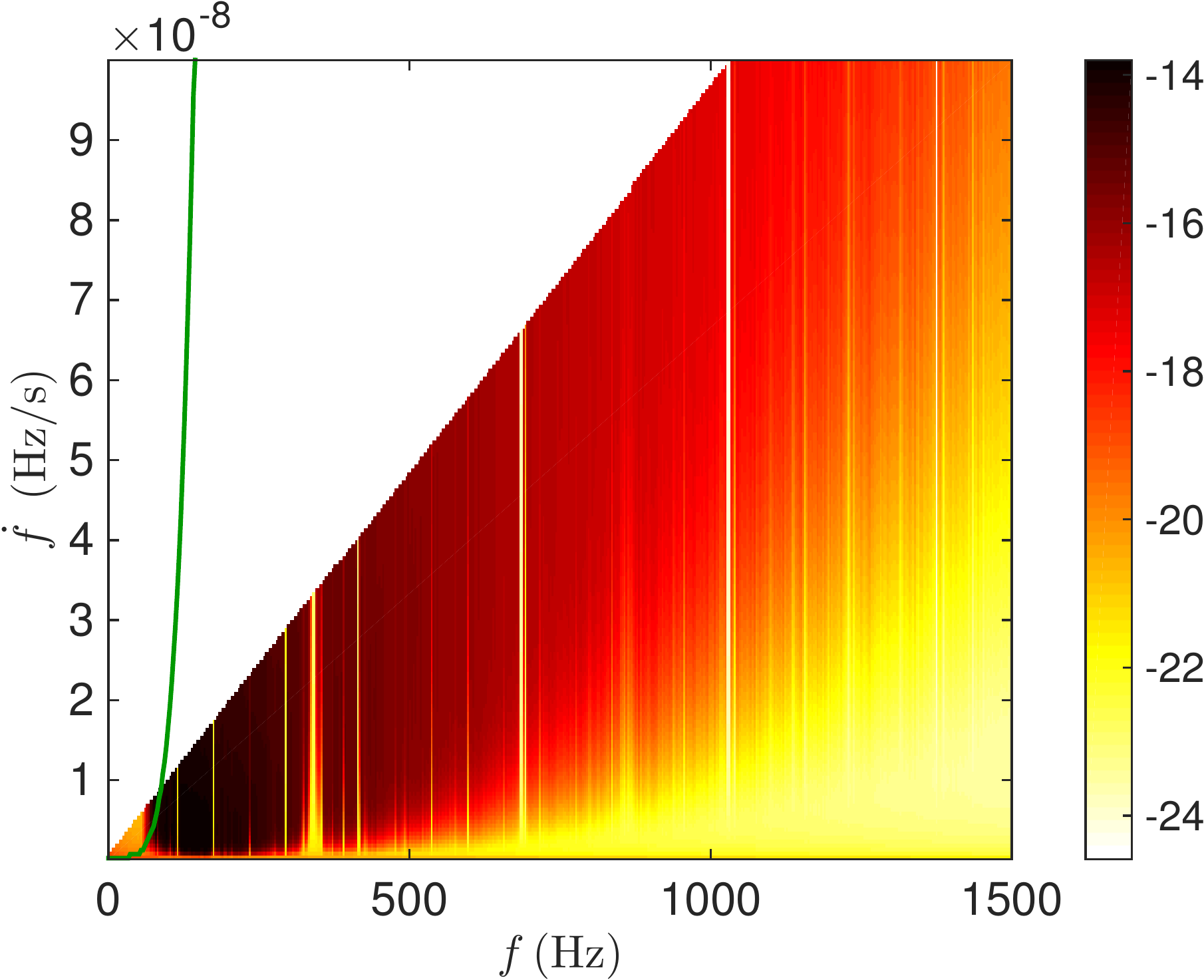}}}%
    \qquad
    \subfloat[Coverage, 20 days]{{  \includegraphics[width=.20\linewidth]{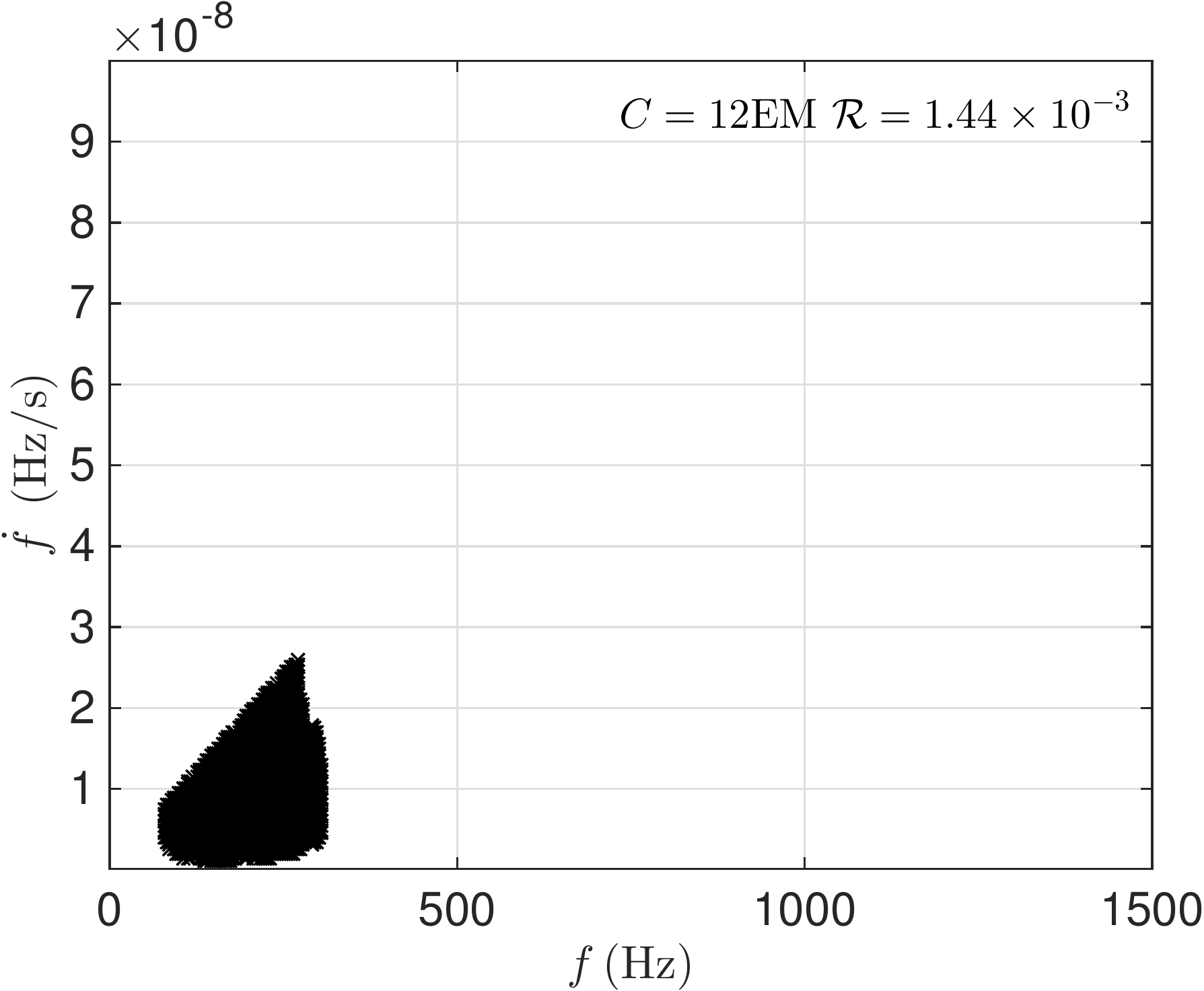}}}%
    \qquad
    \subfloat[Efficiency(lg), 30 days]{{  \includegraphics[width=.20\linewidth]{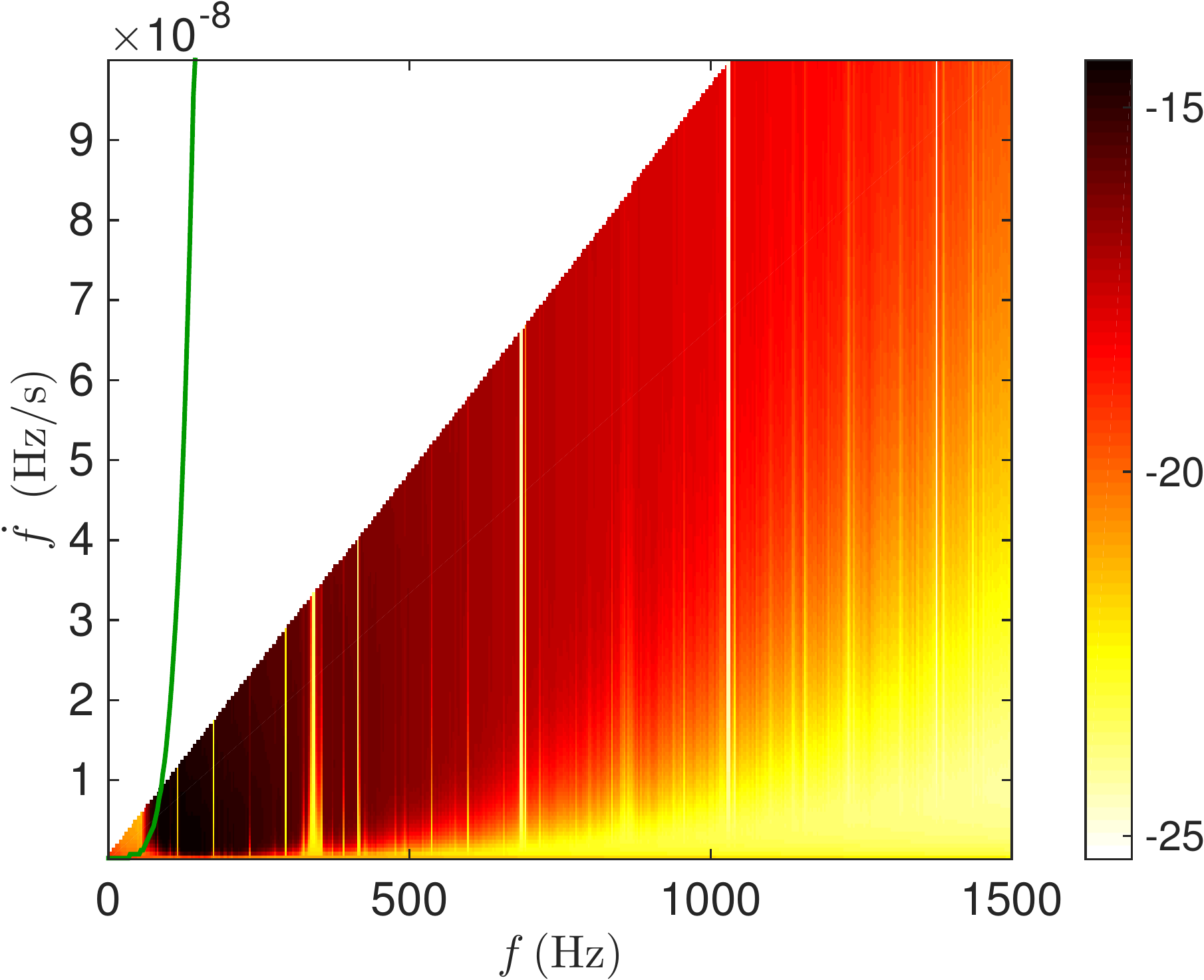}}}%
    \qquad
    \subfloat[Coverage, 30 days]{{  \includegraphics[width=.20\linewidth]{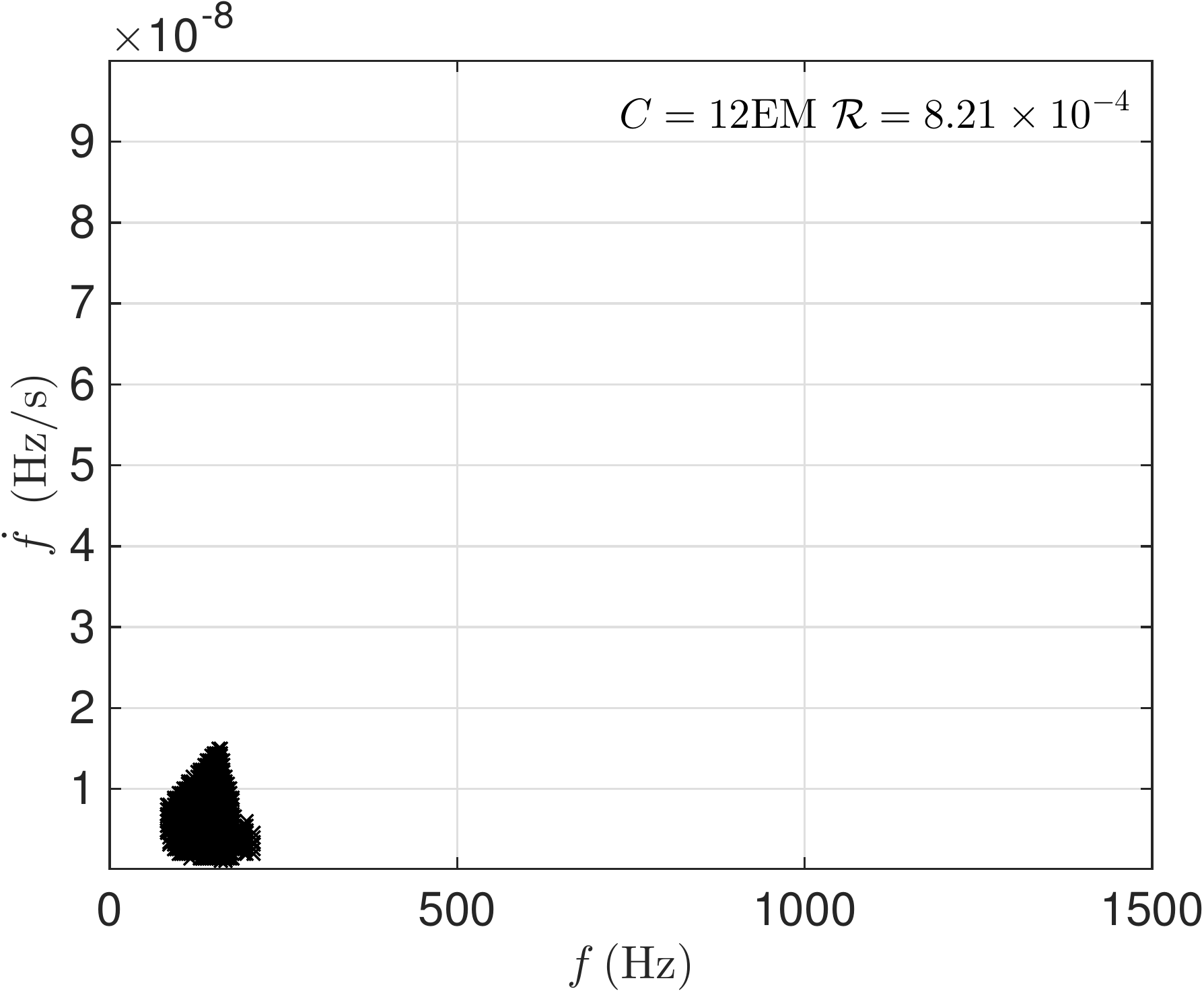}}}%
    \qquad
    \subfloat[Efficiency(lg), 37.5 days]{{  \includegraphics[width=.20\linewidth]{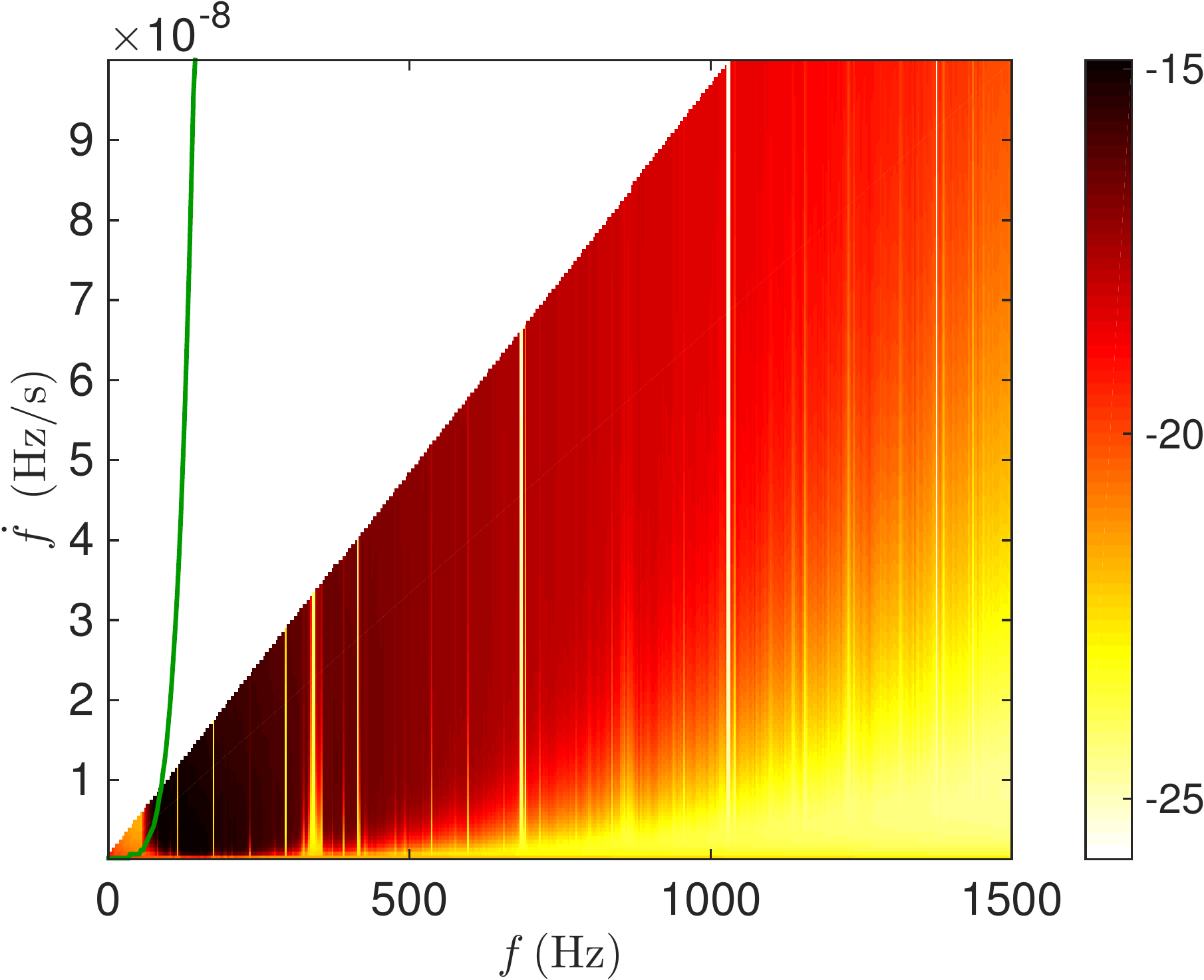}}}%
    \qquad
    \subfloat[Coverage, 37.5 days]{{  \includegraphics[width=.20\linewidth]{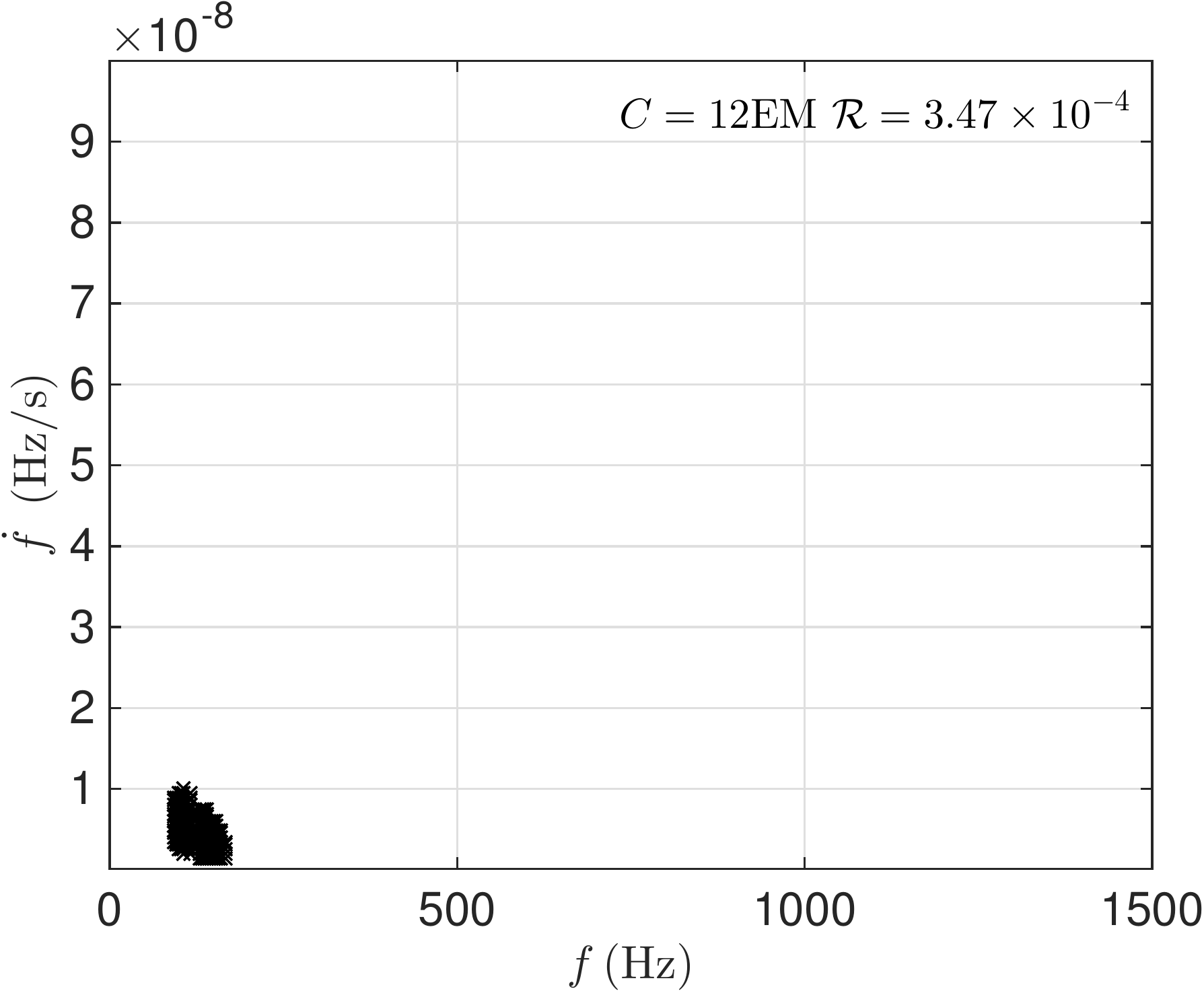}}}%
    \qquad
    \subfloat[Efficiency(lg), 50 days]{{  \includegraphics[width=.20\linewidth]{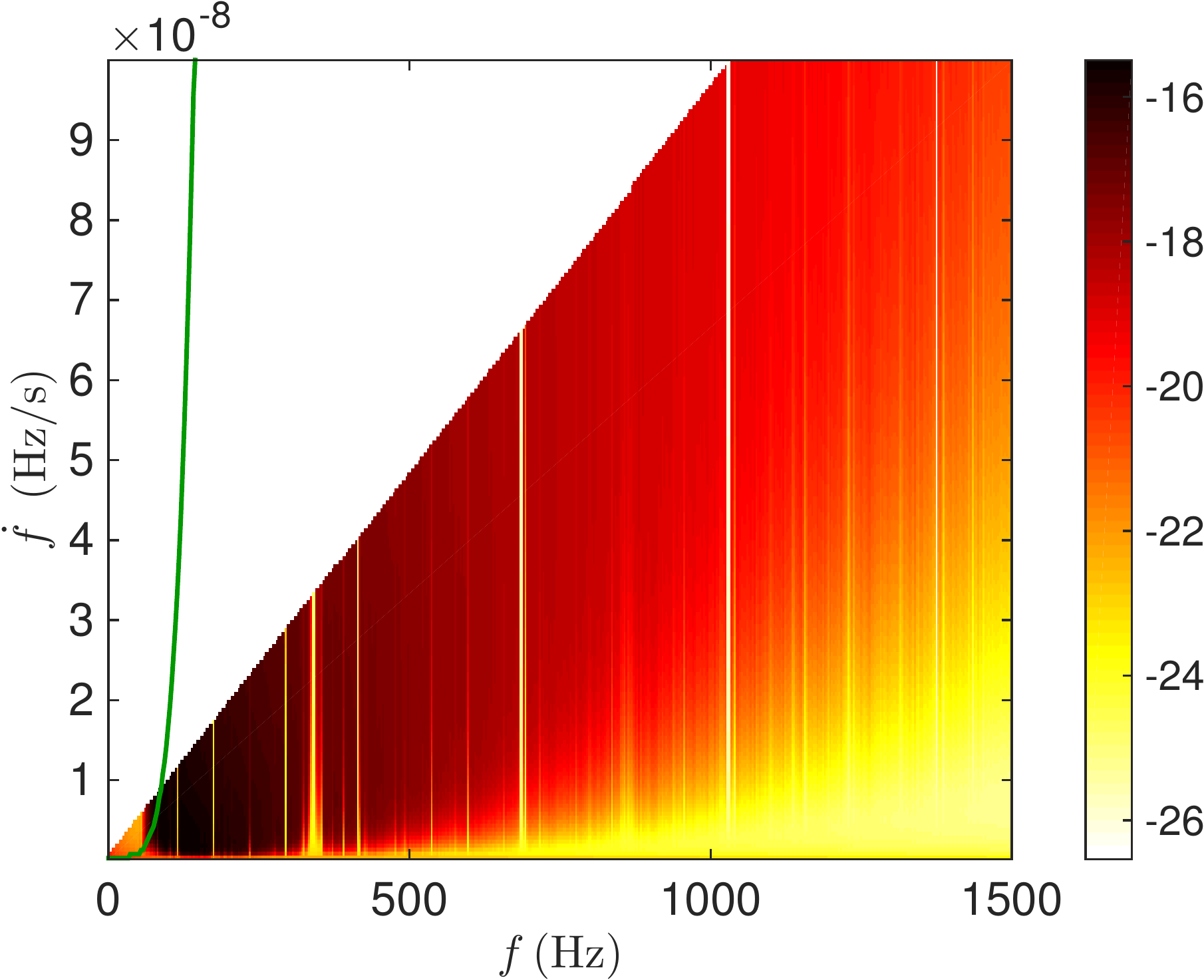}}}%
    \qquad
    \subfloat[Coverage, 50 days]{{  \includegraphics[width=.20\linewidth]{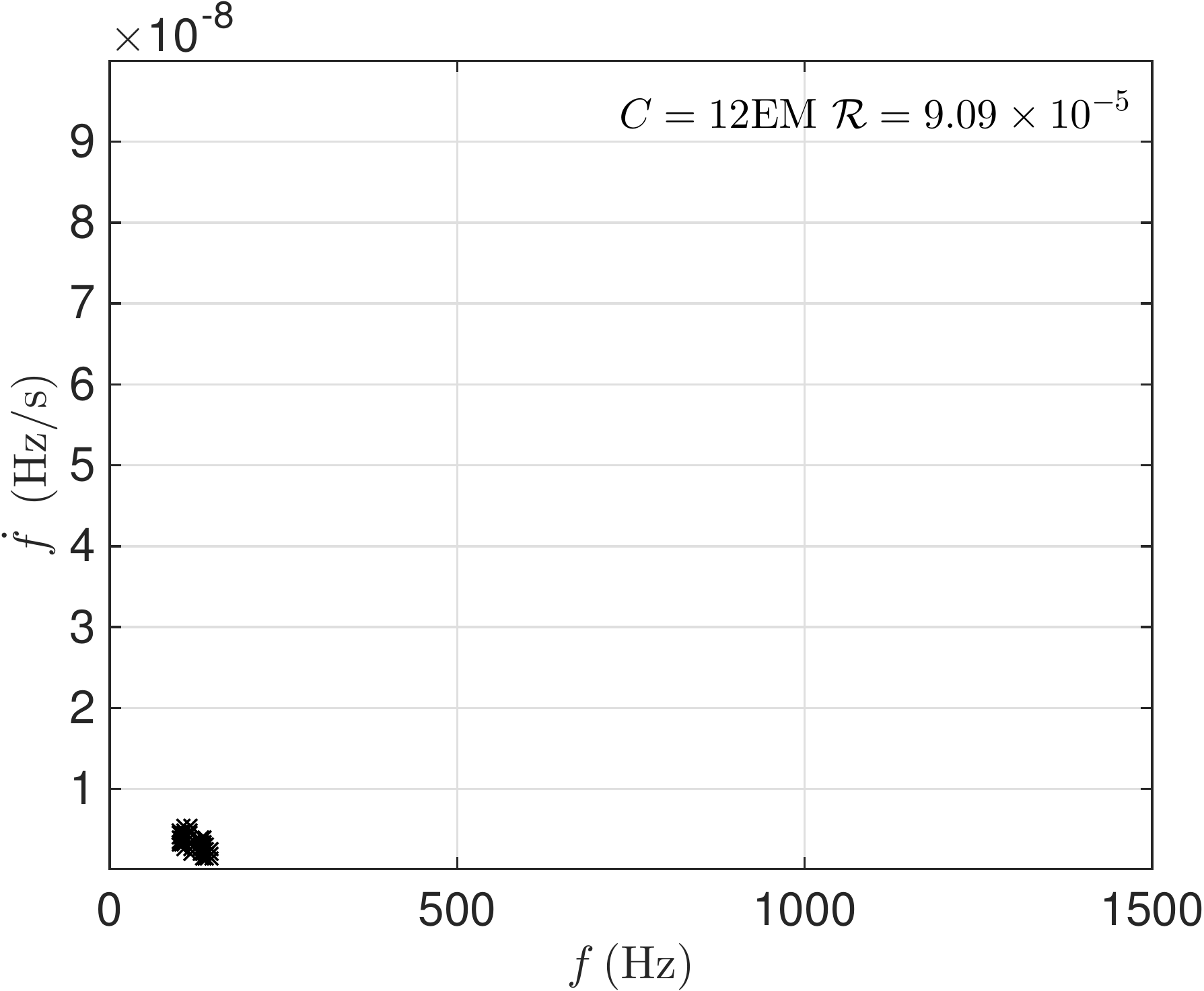}}}%
    \qquad
    \subfloat[Efficiency(lg), 75 days]{{  \includegraphics[width=.20\linewidth]{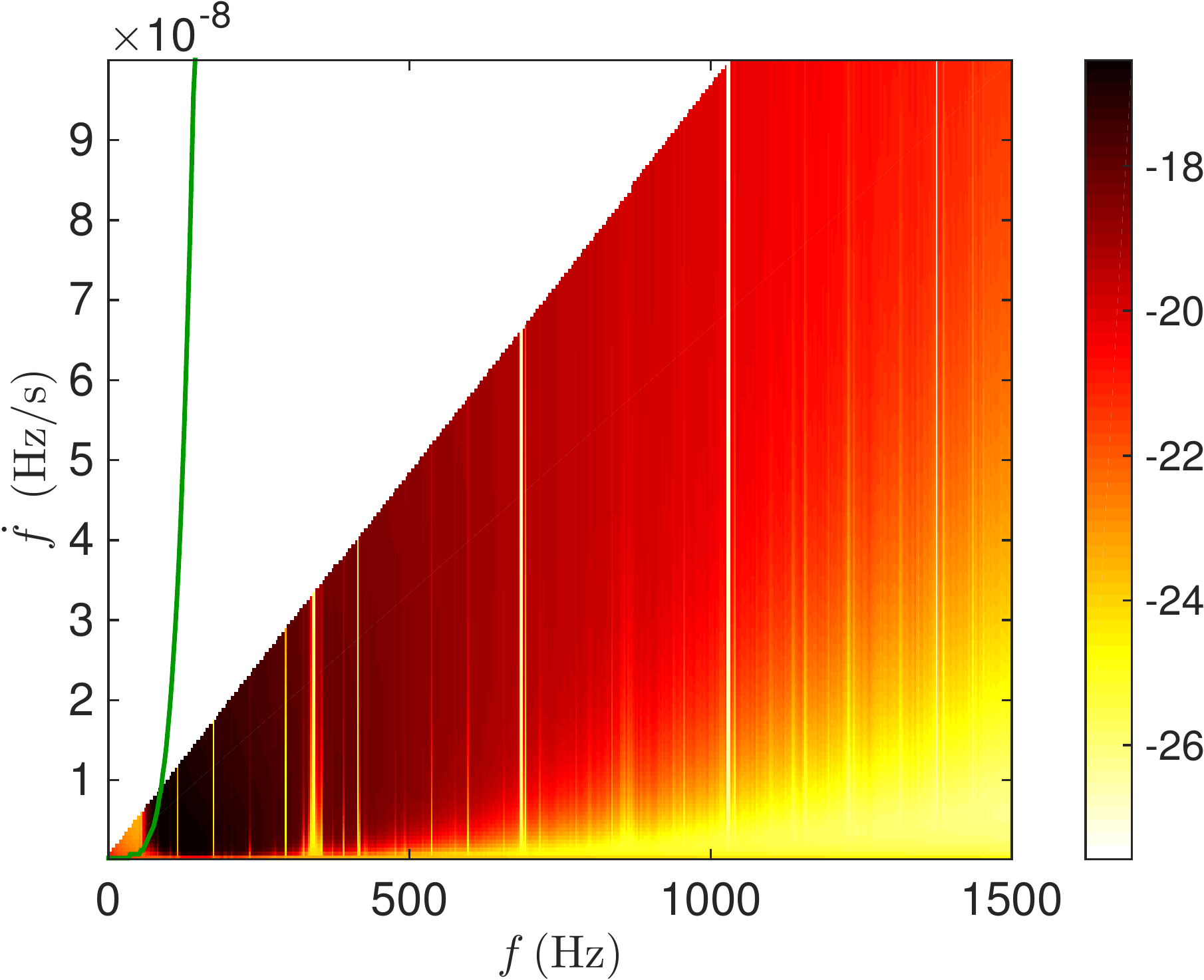}}}%
    \qquad
    \subfloat[Coverage, 75 days]{{  \includegraphics[width=.20\linewidth]{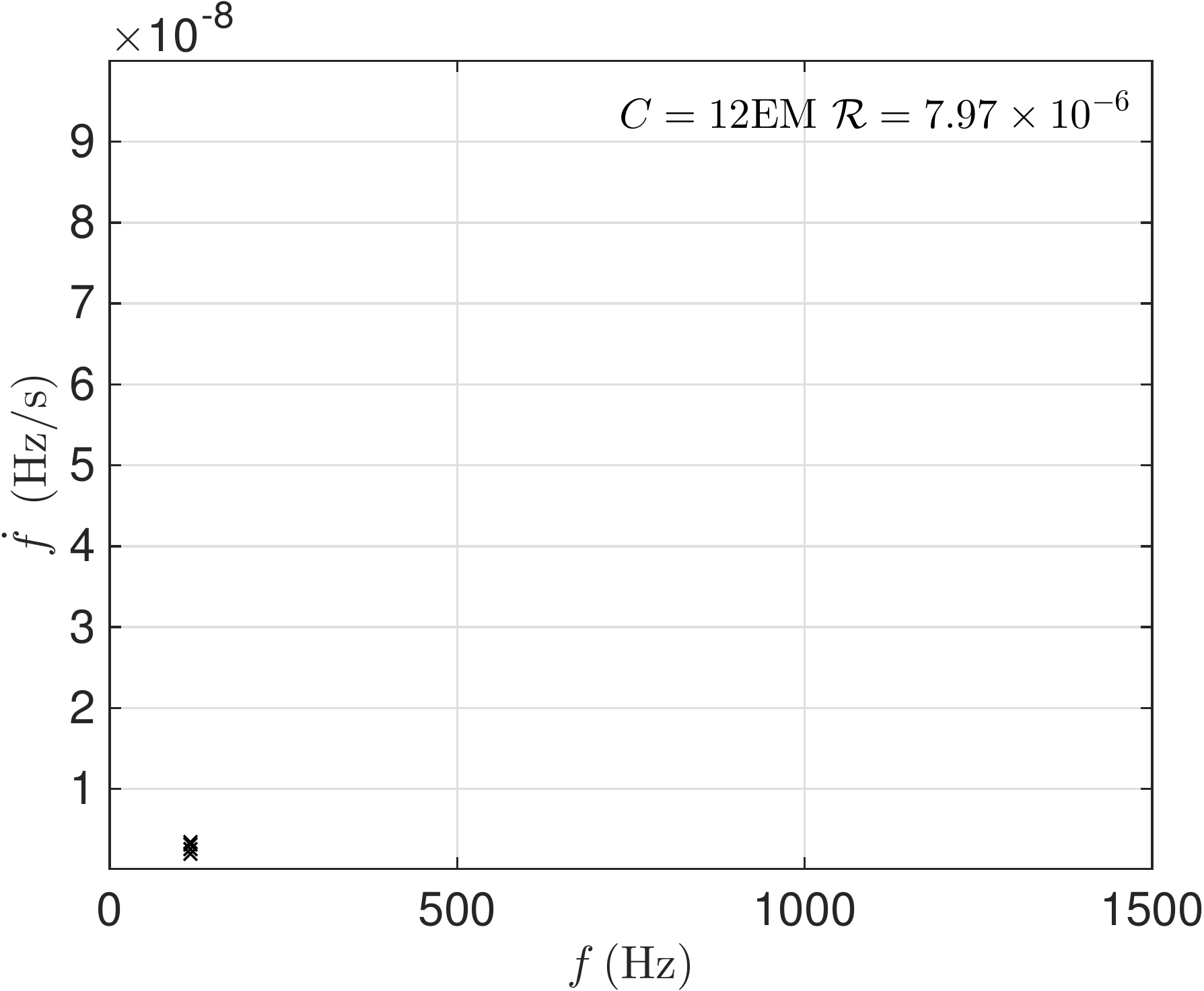}}}%
    
   \caption{Optimisation results for Cas A at 3500 pc, 330 years old, assuming log-uniform and age-based priors, for various coherent search durations: 5, 10, 20, 30, 37.5, 50 and 75 days. The total computing budget is assumed to be 12 EM.}%
    \label{CasA_51020days_age_log}%
\end{figure*}

\begin{figure*}%
    \centering
    \subfloat[Efficiency(lg), 5 days]{{  \includegraphics[width=.20\linewidth]{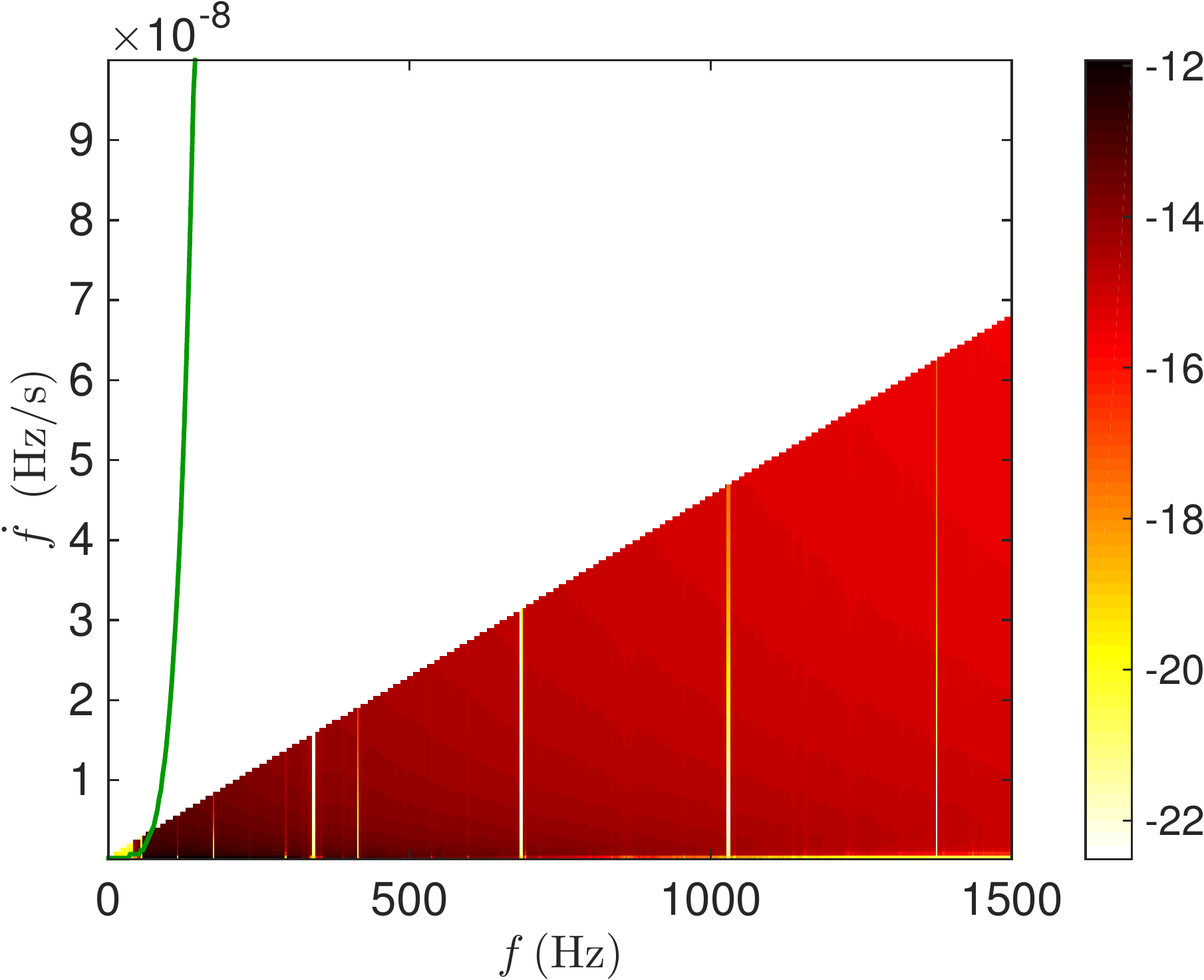}}}%
    \qquad
    \subfloat[Coverage, 5 days]{{  \includegraphics[width=.20\linewidth]{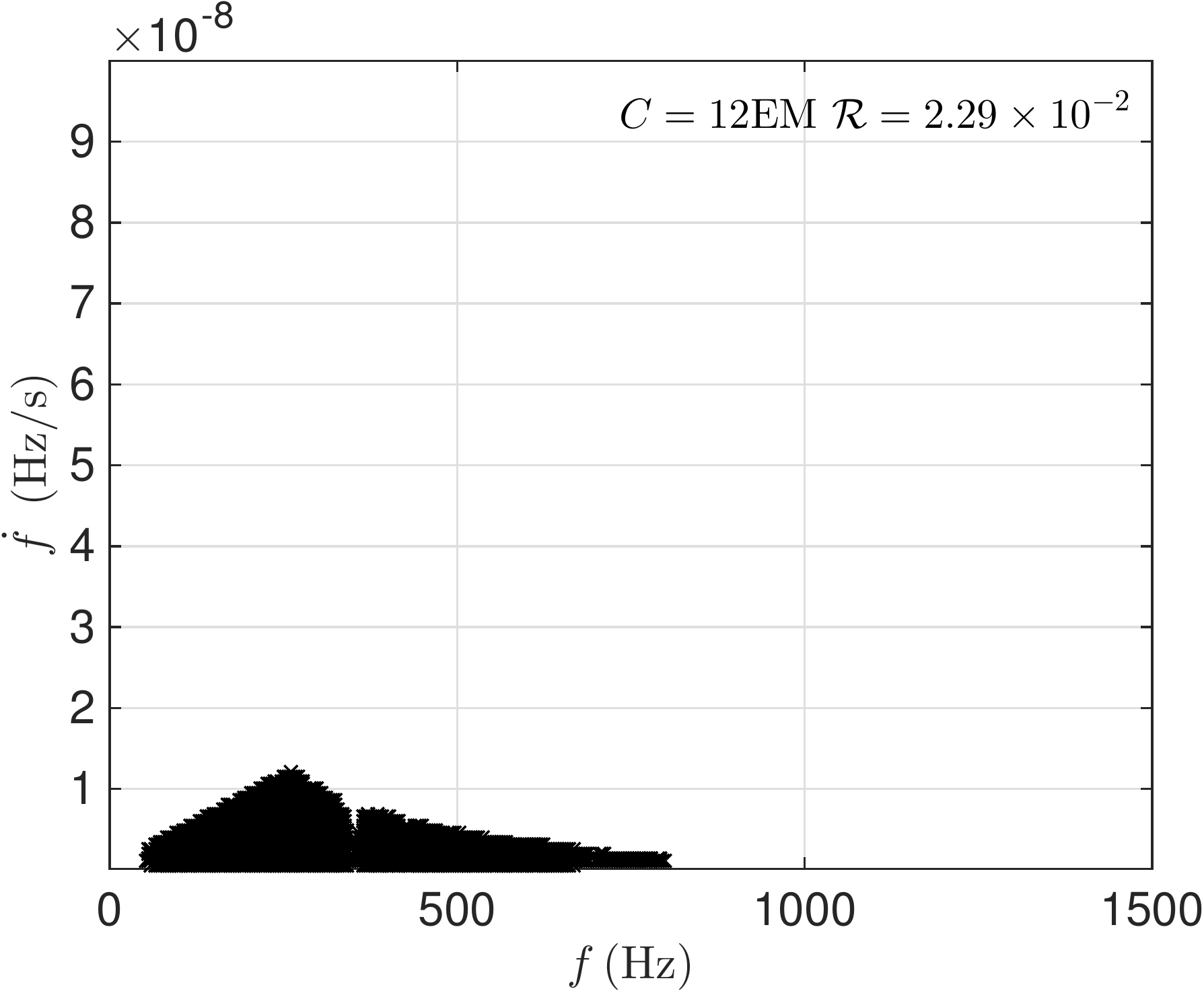}}}%
    \qquad
    \subfloat[Efficiency(lg), 10 days]{{  \includegraphics[width=.20\linewidth]{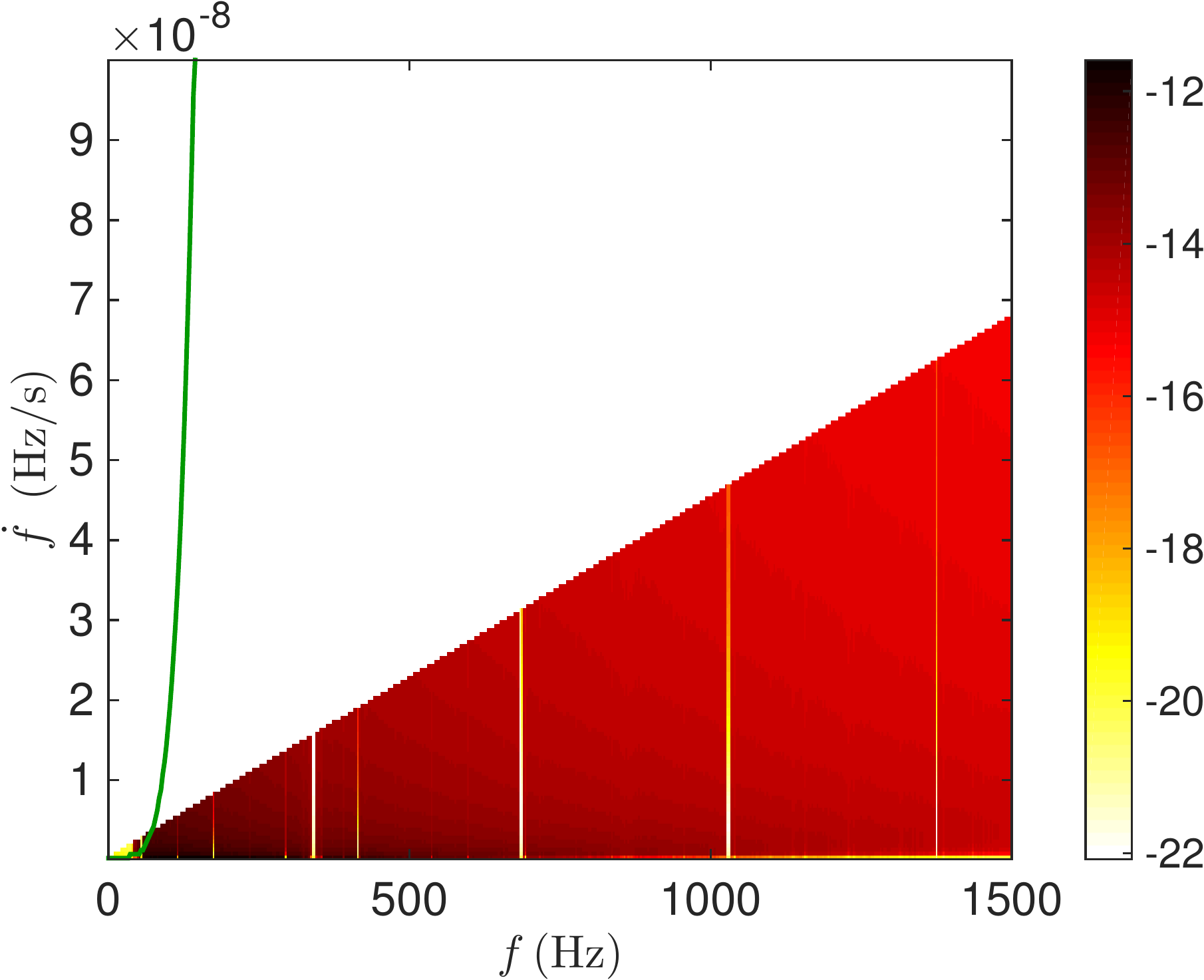}}}%
    \qquad
    \subfloat[Coverage, 10 days]{{  \includegraphics[width=.20\linewidth]{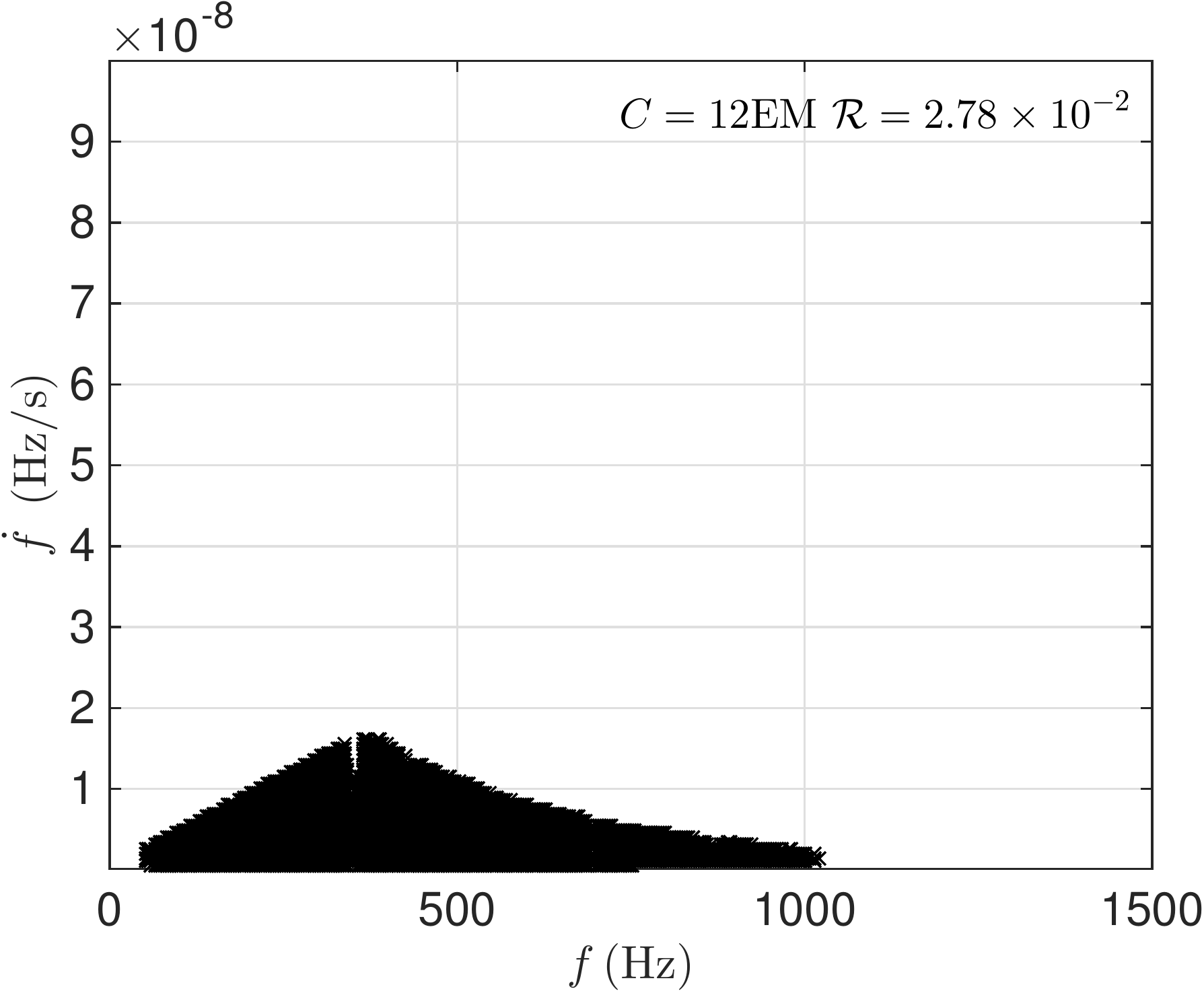}}}%
    \qquad
    \subfloat[Efficiency(lg), 20 days]{{  \includegraphics[width=.20\linewidth]{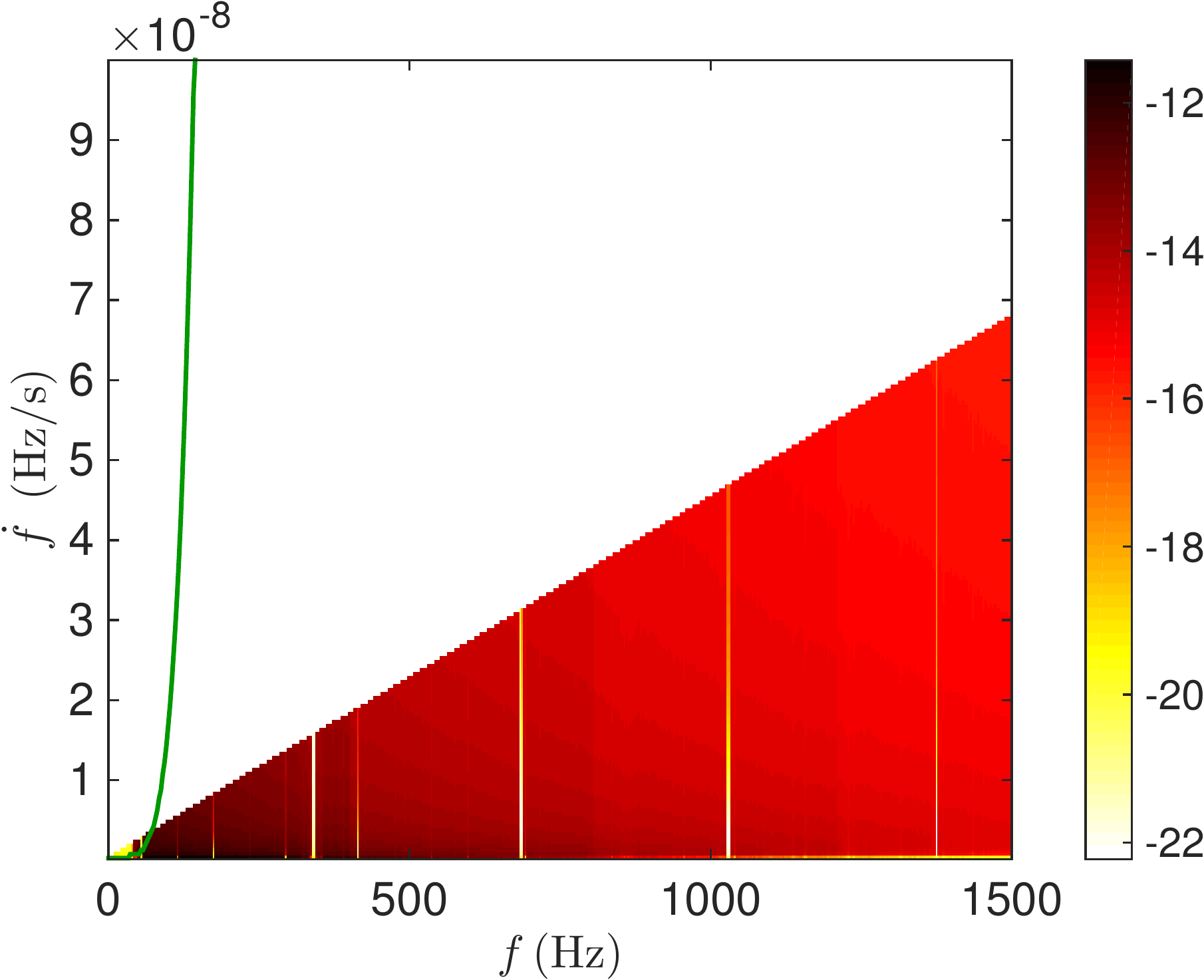}}}%
    \qquad
    \subfloat[Coverage, 20 days]{{  \includegraphics[width=.20\linewidth]{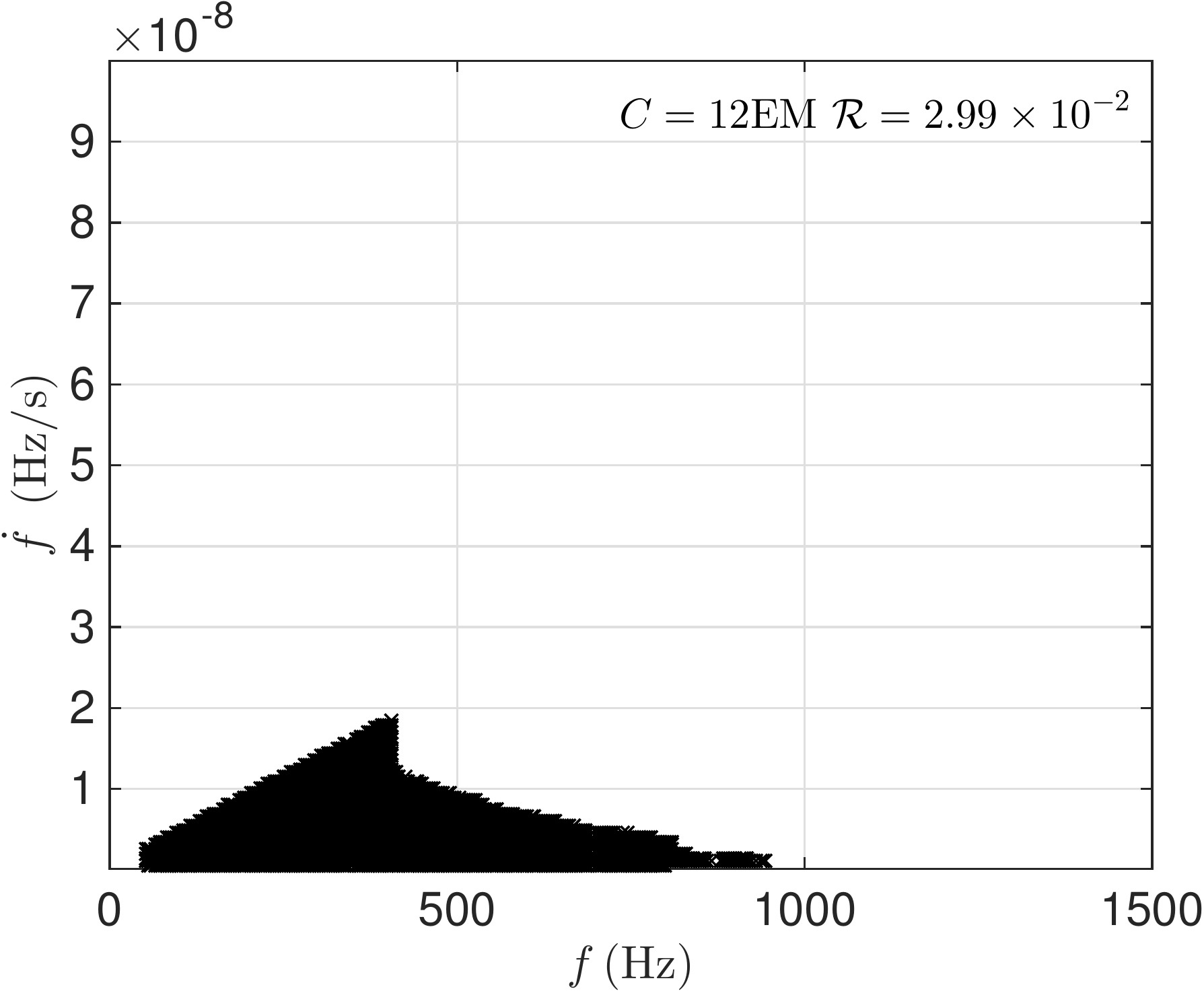}}}%
    \qquad
    \subfloat[Efficiency(lg), 30 days]{{  \includegraphics[width=.20\linewidth]{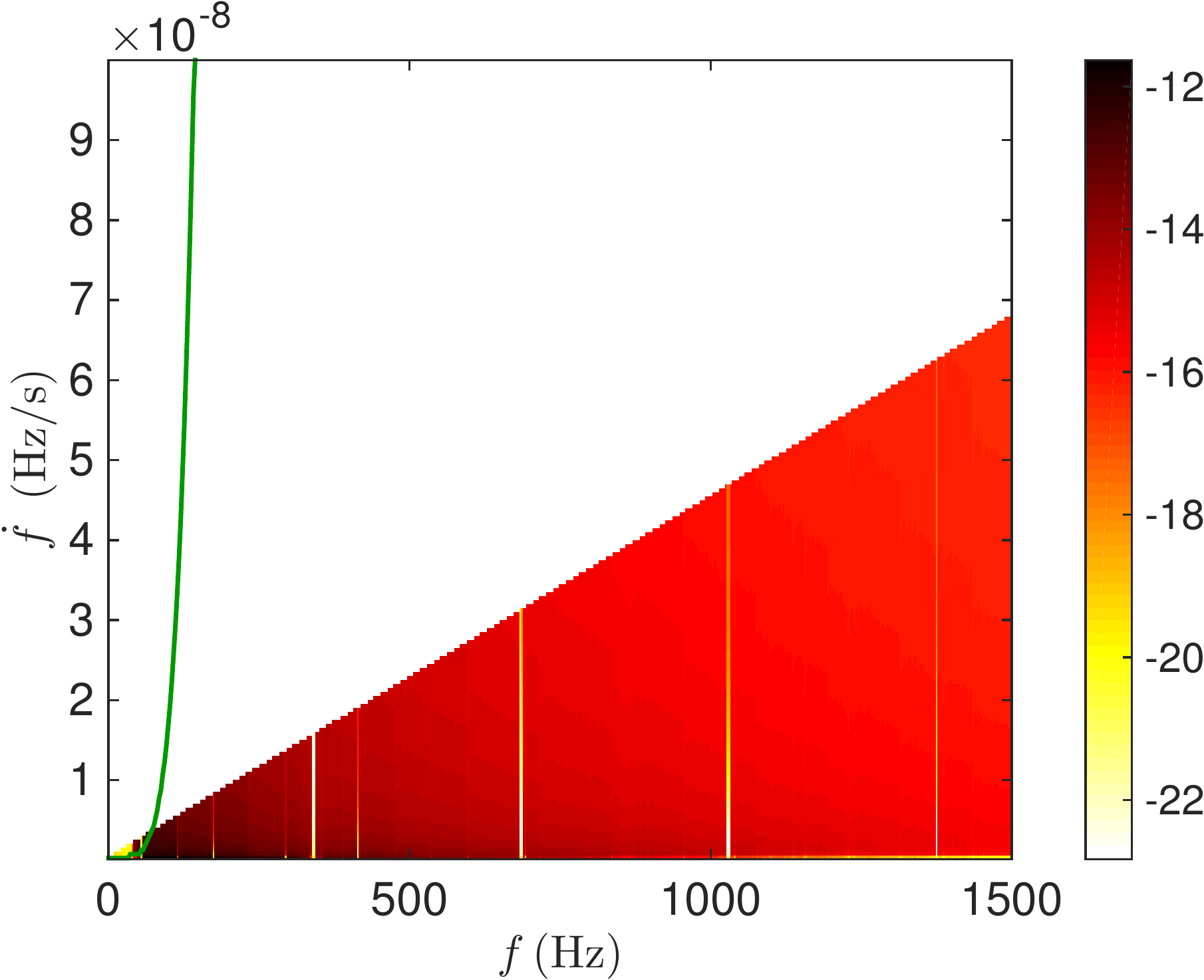}}}%
    \qquad
    \subfloat[Coverage, 30 days]{{  \includegraphics[width=.20\linewidth]{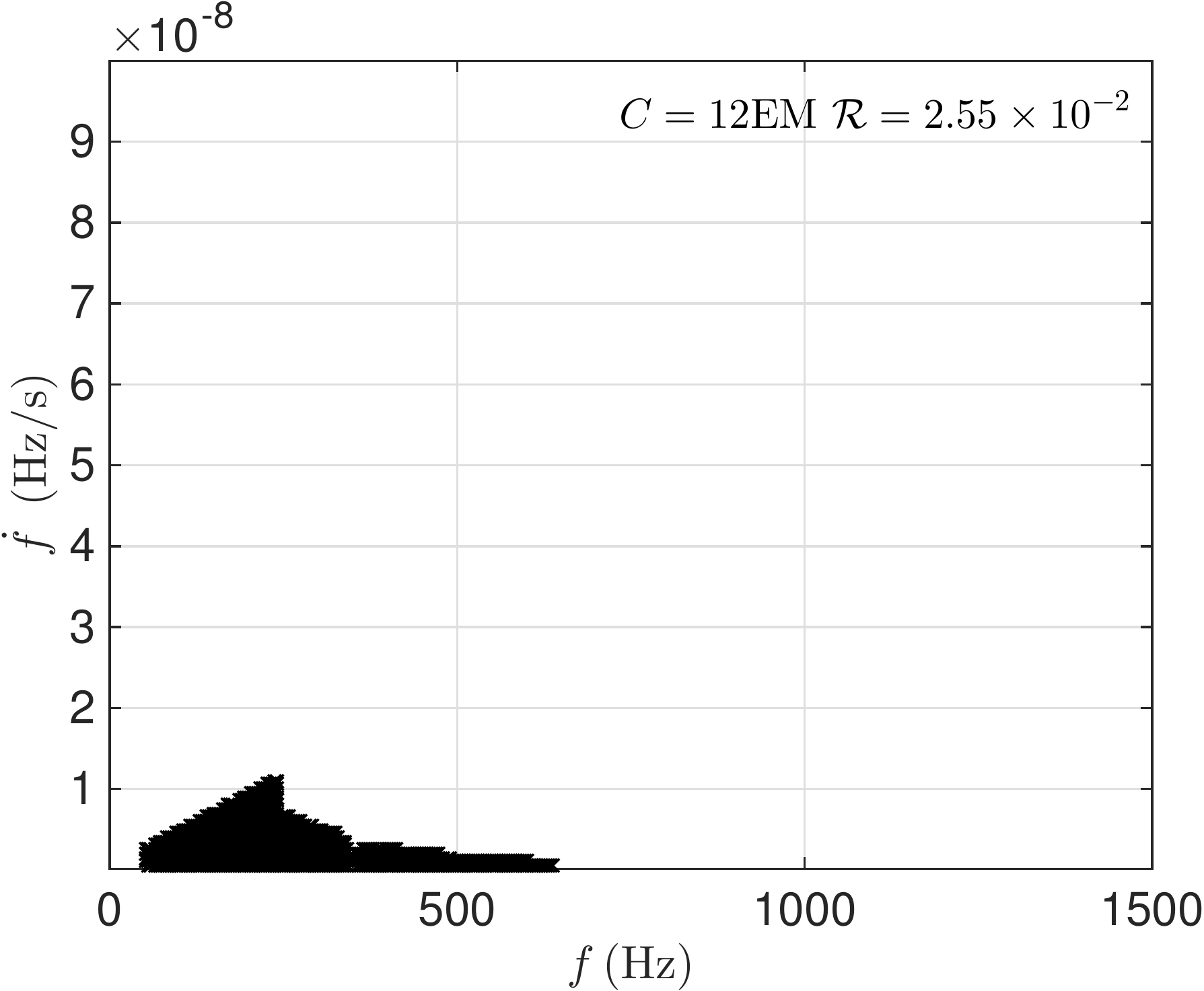}}}%
    \qquad
    \subfloat[Efficiency(lg), 37.5 days]{{  \includegraphics[width=.20\linewidth]{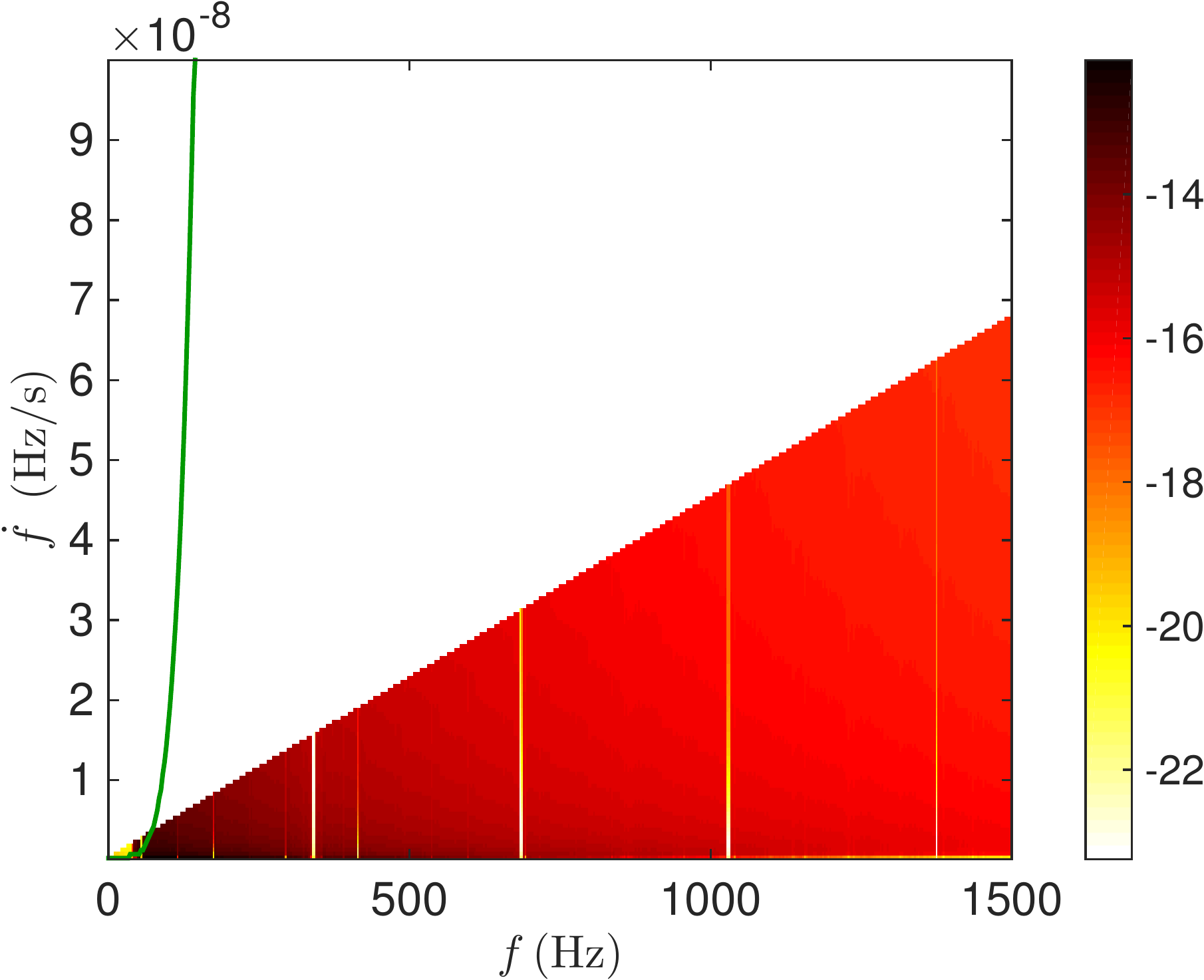}}}%
    \qquad
    \subfloat[Coverage, 37.5 days]{{  \includegraphics[width=.20\linewidth]{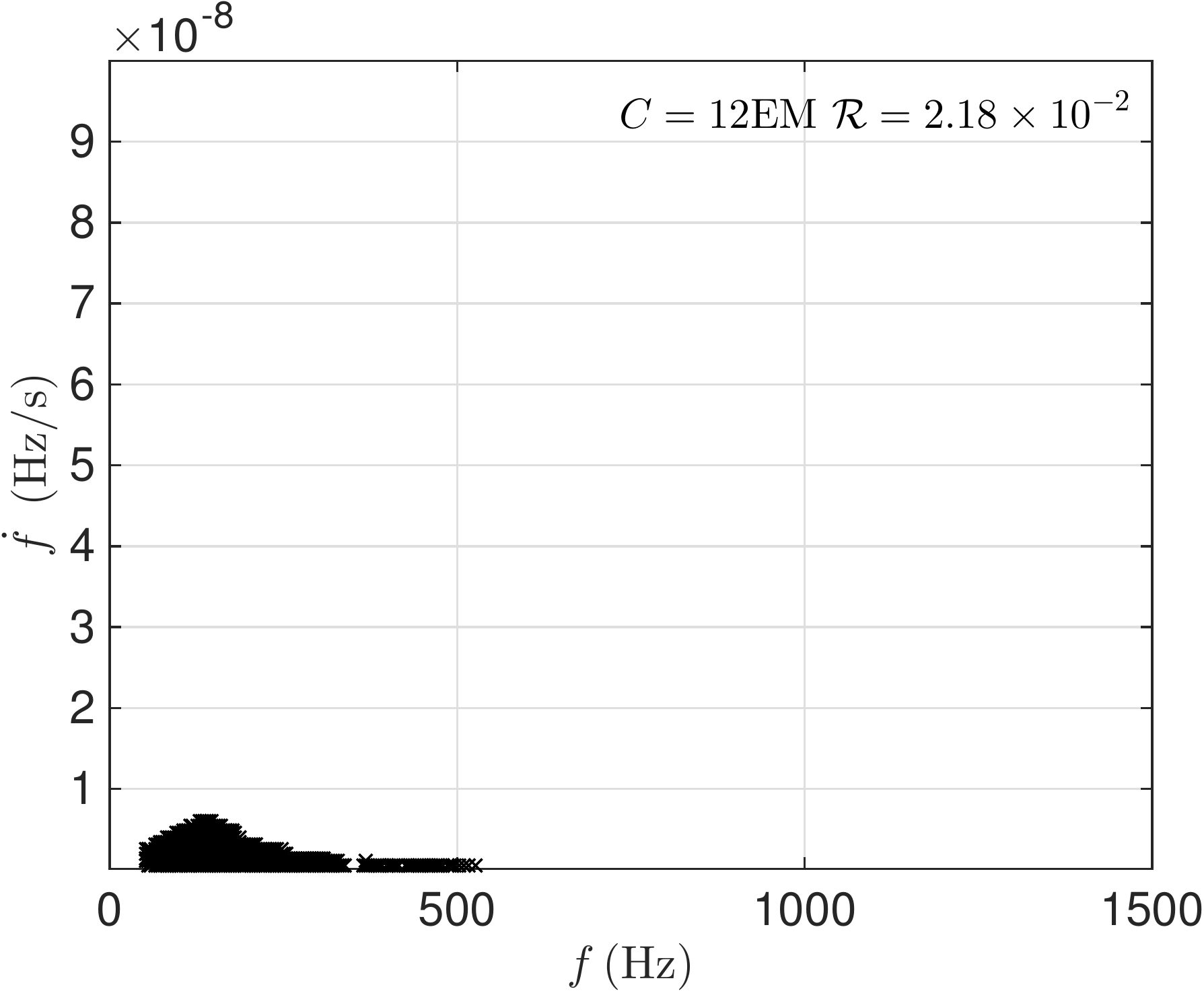}}}%
    \qquad
    \subfloat[Efficiency(lg), 50 days]{{  \includegraphics[width=.20\linewidth]{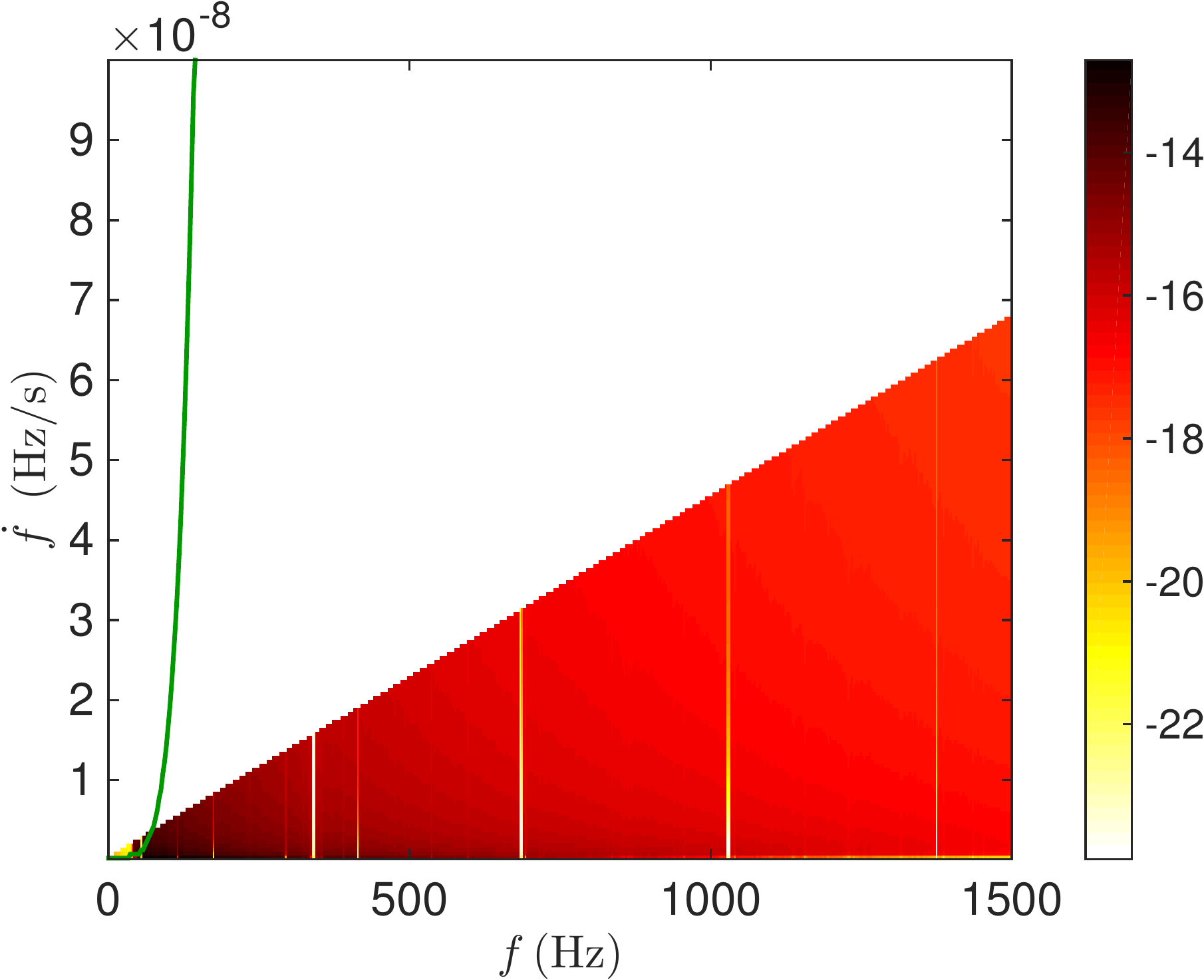}}}%
    \qquad
    \subfloat[Coverage, 50 days]{{  \includegraphics[width=.20\linewidth]{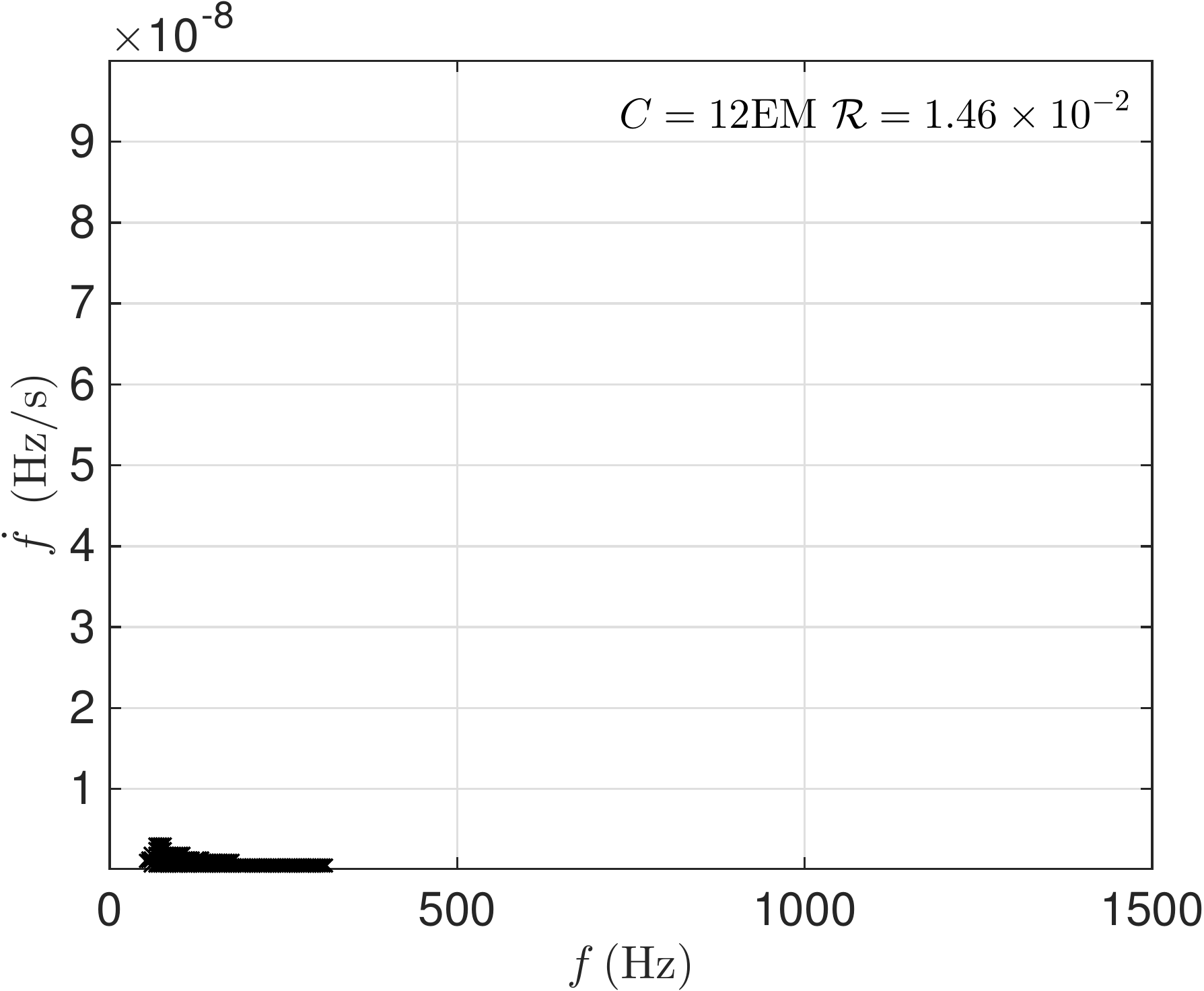}}}%
    \qquad
    \subfloat[Efficiency(lg), 75 days]{{  \includegraphics[width=.20\linewidth]{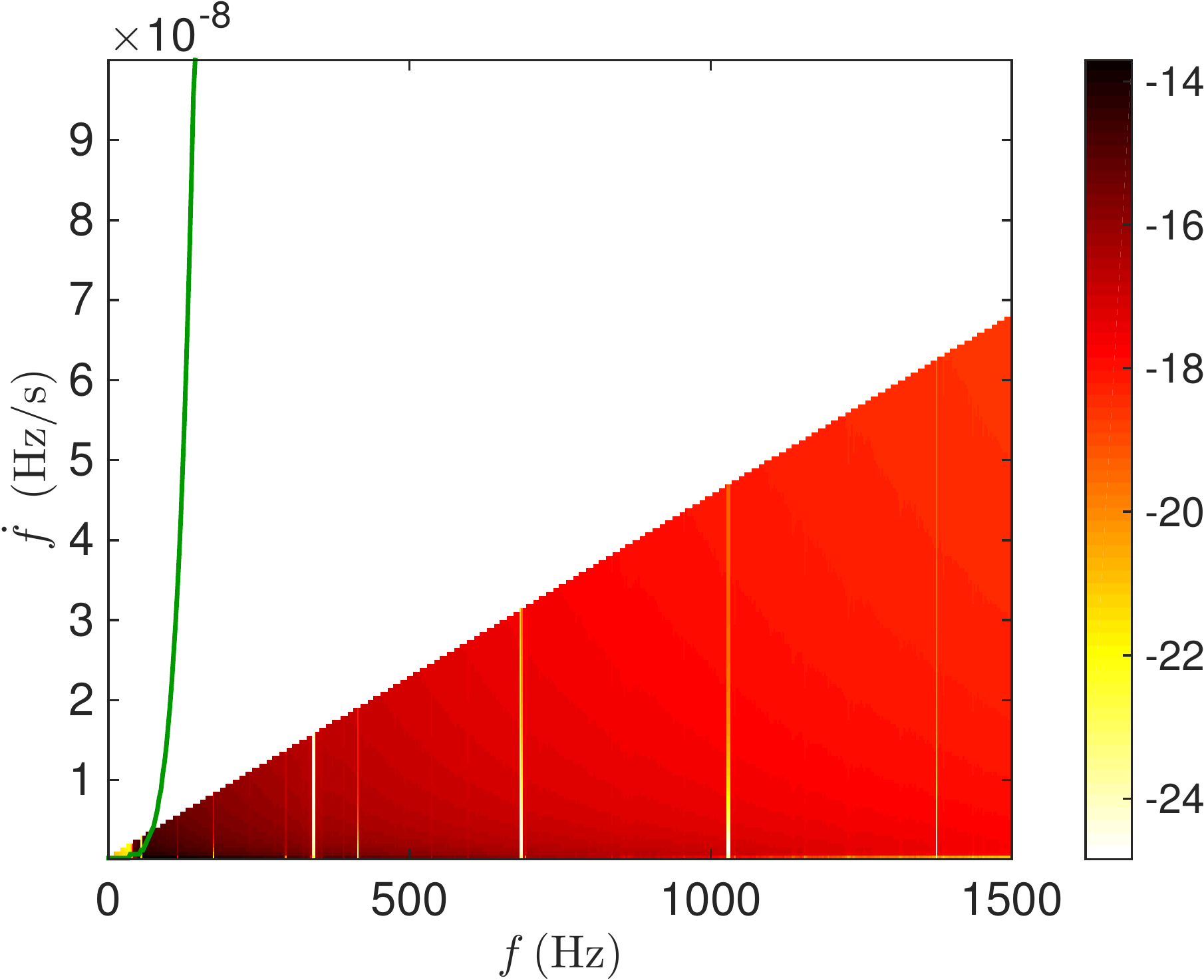}}}%
    \qquad
    \subfloat[Coverage, 75 days]{{  \includegraphics[width=.20\linewidth]{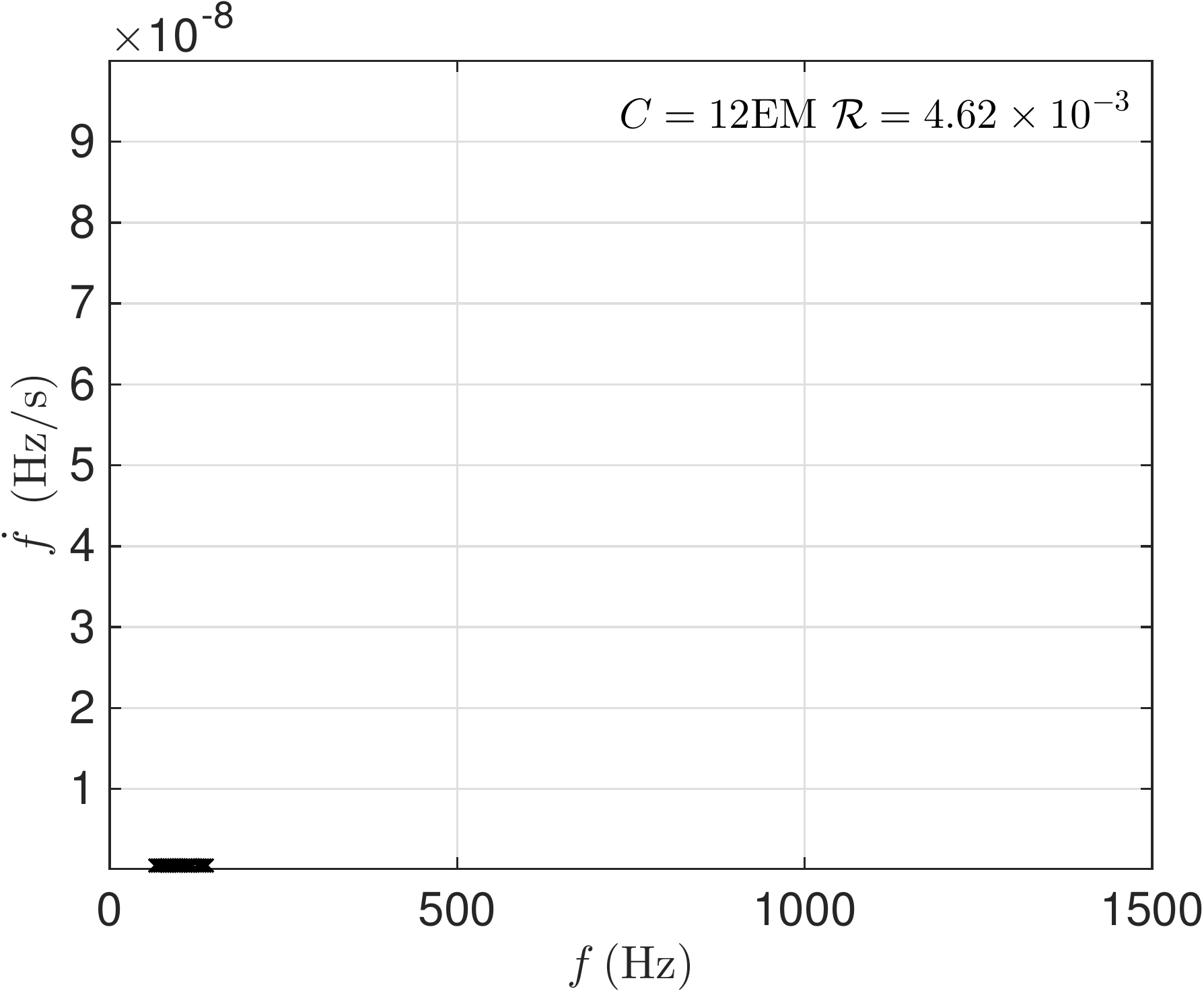}}}%
 
     \caption{Optimisation results for Vela Jr, at 200 pc and 700 years old (close and young, CY), assuming log-uniform and age-based priors, for various coherent search durations: 5, 10, 20, 30, 37.5, 50 and 75 days. The total computing budget is assumed to be 12 EM.}%
    \label{G2662_51020days_shortage_log}%
\end{figure*}

\begin{figure*}%
    \centering
    \subfloat[Efficiency(lg), 5 days]{{  \includegraphics[width=.20\linewidth]{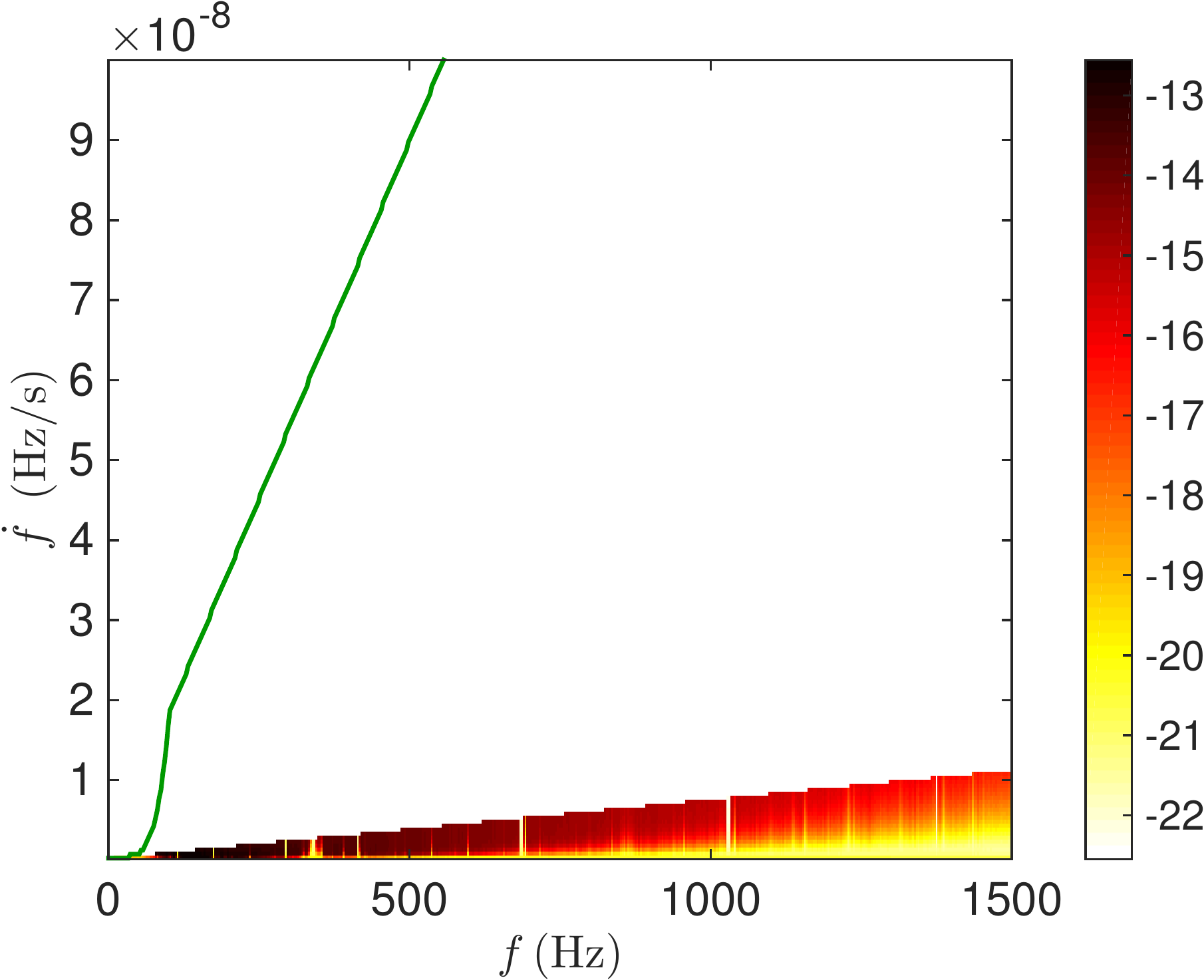}}}%
    \qquad
    \subfloat[Coverage, 5 days]{{  \includegraphics[width=.20\linewidth]{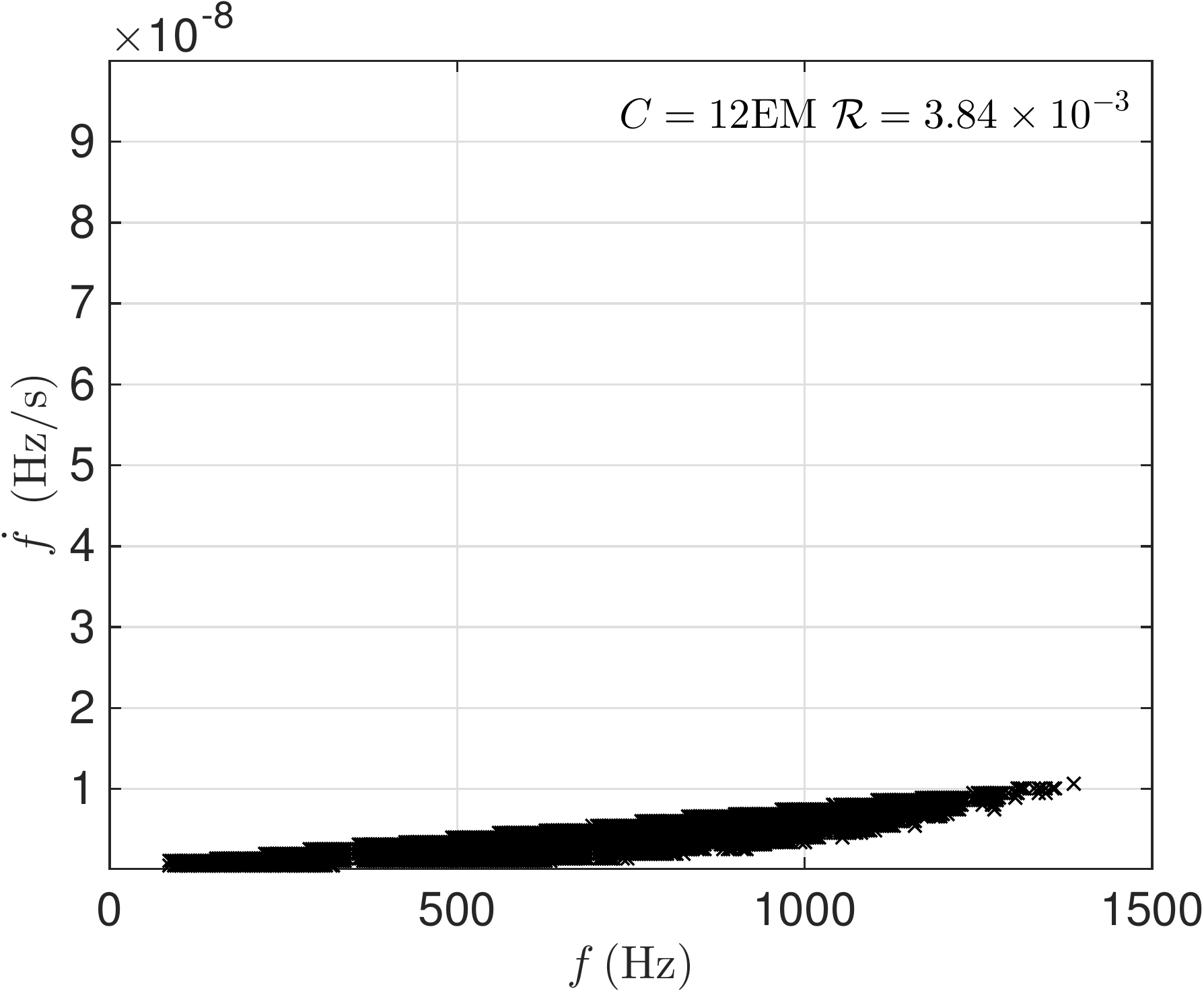}}}%
    \qquad
    \subfloat[Efficiency(lg), 10 days]{{  \includegraphics[width=.20\linewidth]{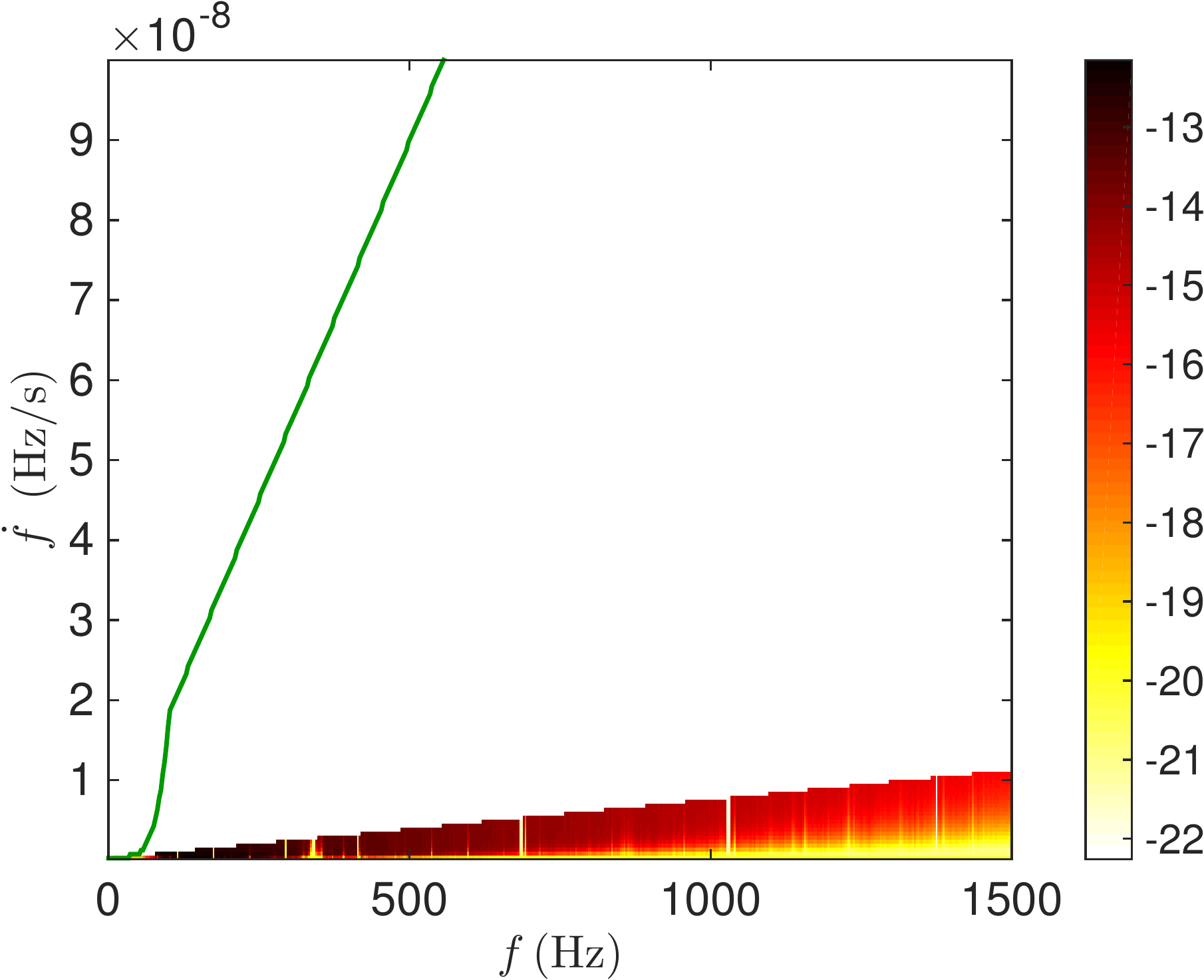}}}%
    \qquad
    \subfloat[Coverage, 10 days]{{  \includegraphics[width=.20\linewidth]{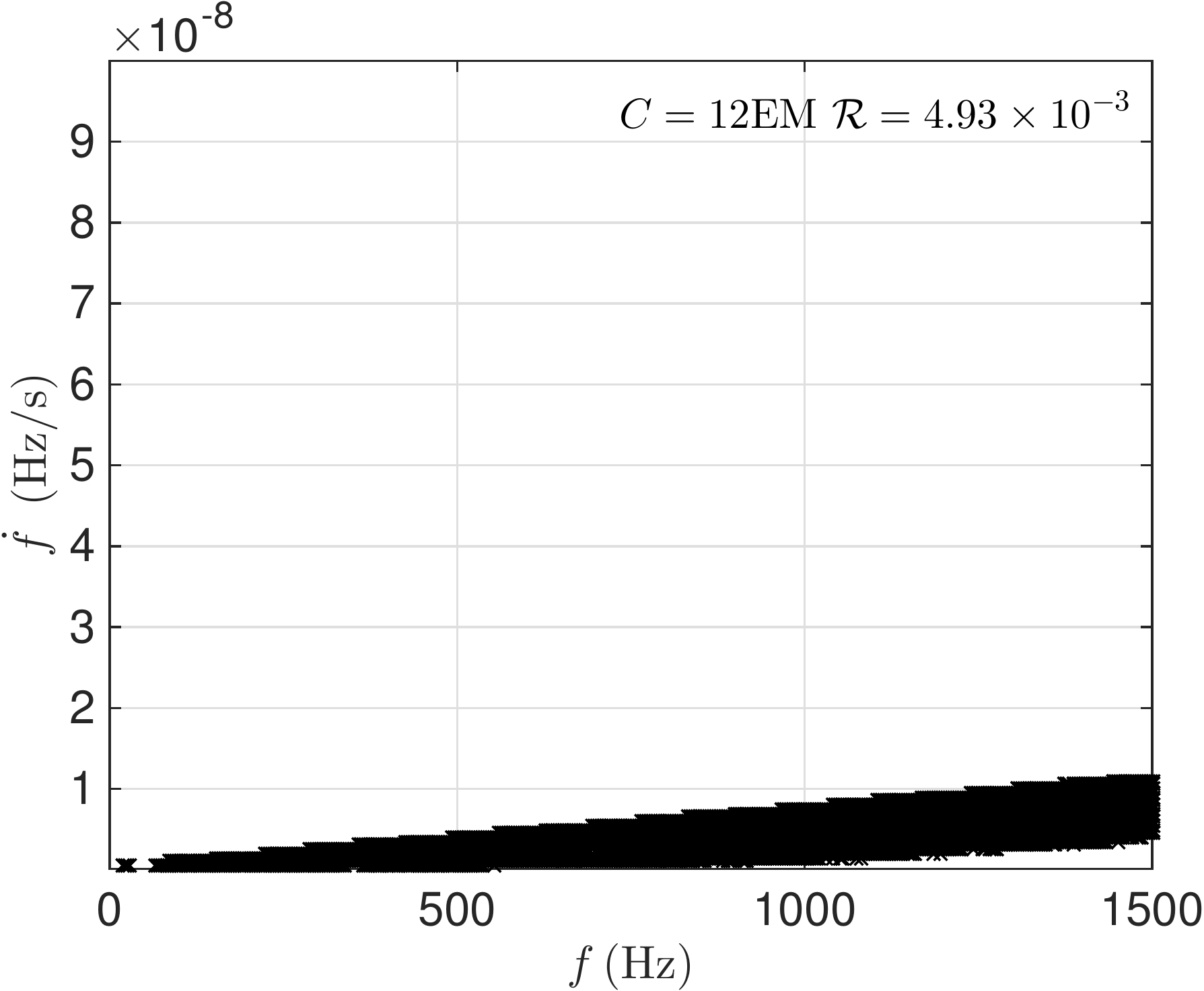}}}%
    \qquad
    \subfloat[Efficiency(lg), 20 days]{{  \includegraphics[width=.20\linewidth]{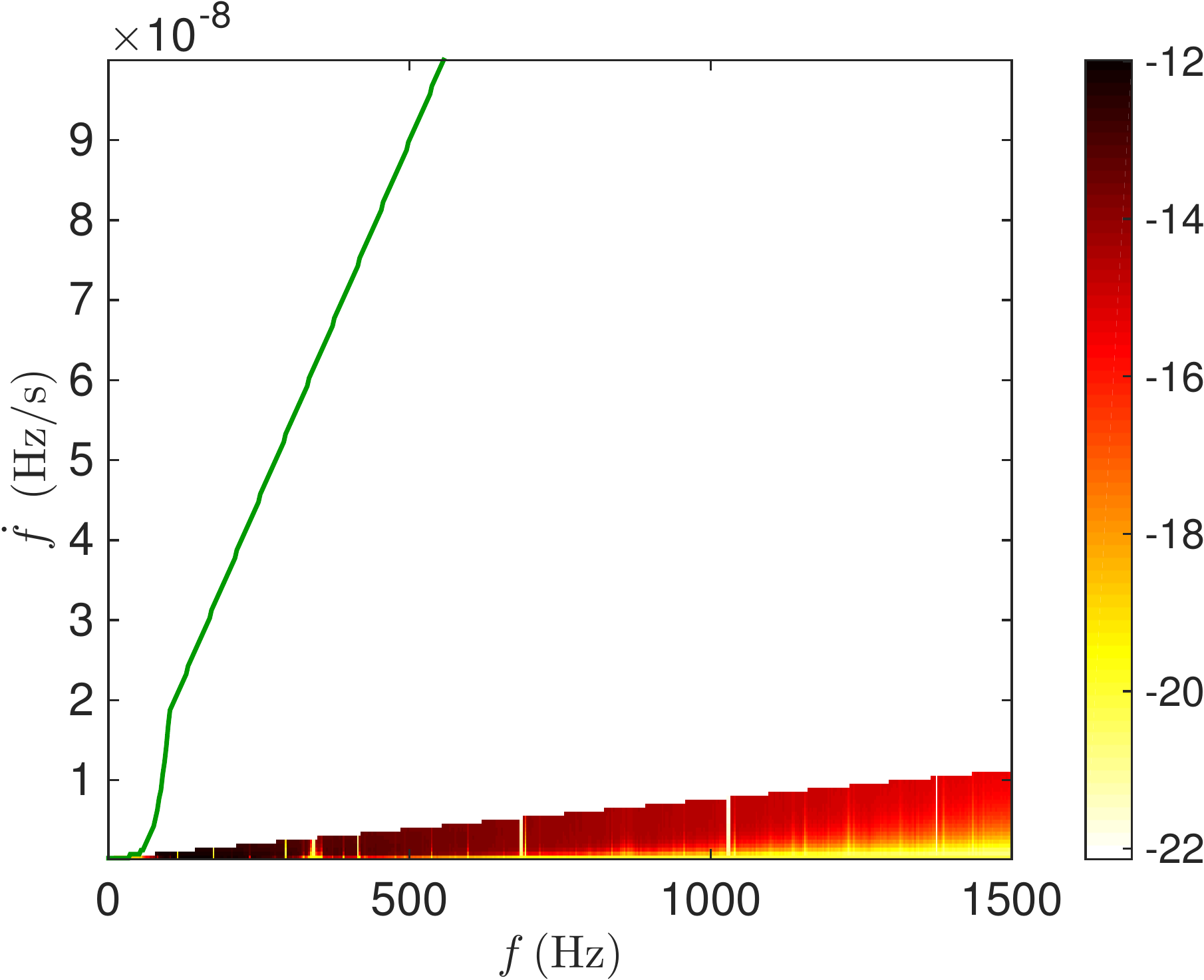}}}%
    \qquad
    \subfloat[Coverage, 20 days]{{  \includegraphics[width=.20\linewidth]{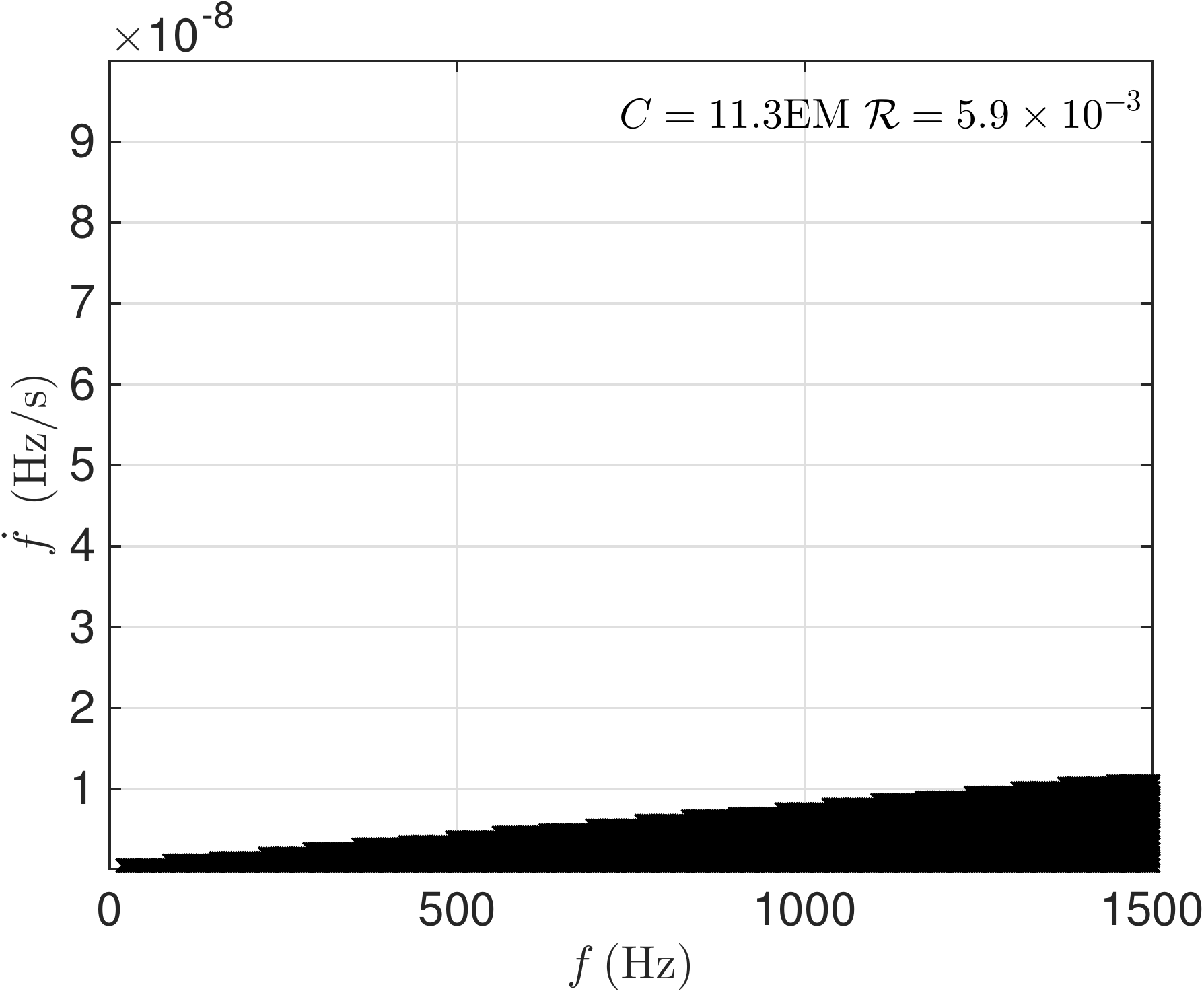}}}%
    \qquad
    \subfloat[Efficiency(lg), 30 days]{{  \includegraphics[width=.20\linewidth]{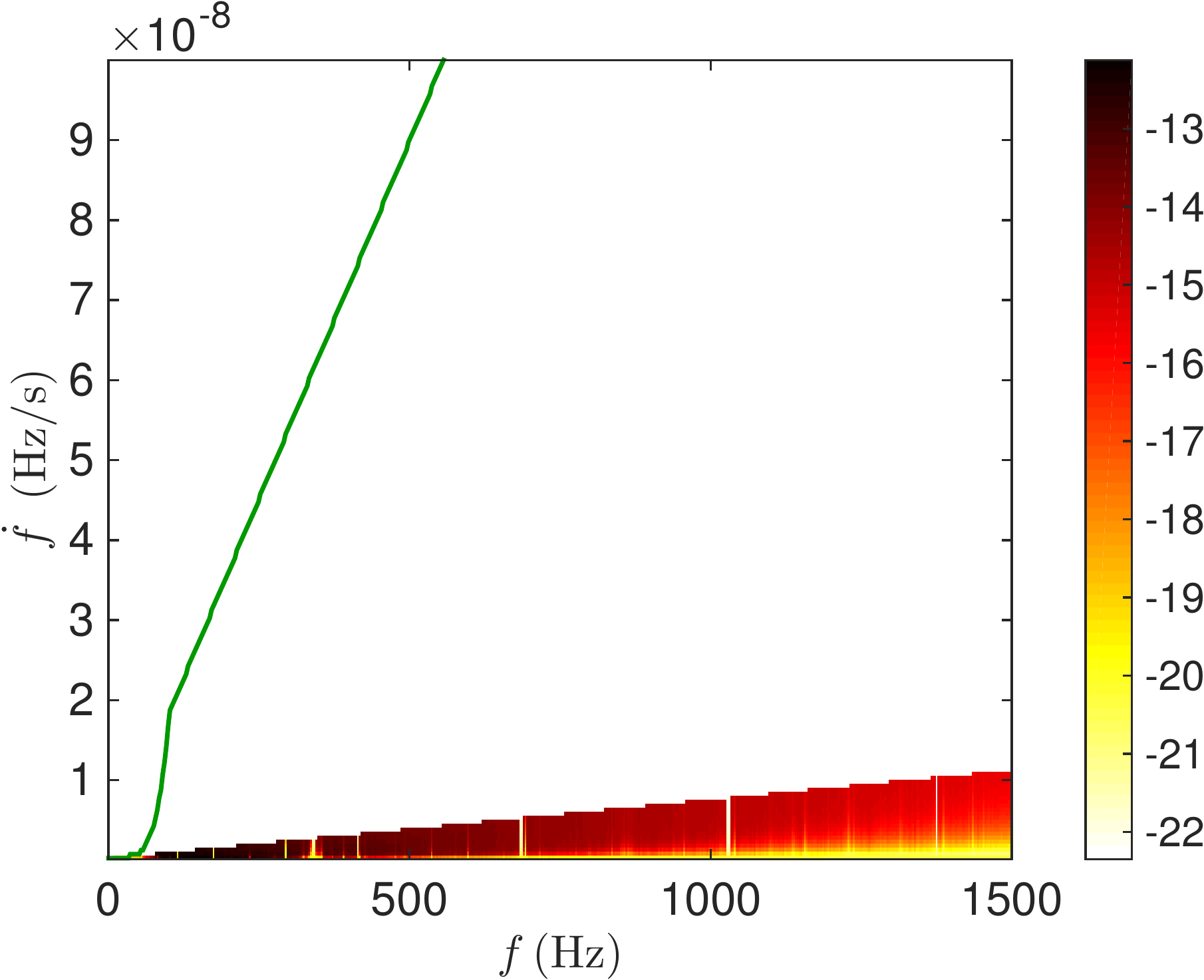}}}%
    \qquad
    \subfloat[Coverage, 30 days]{{  \includegraphics[width=.20\linewidth]{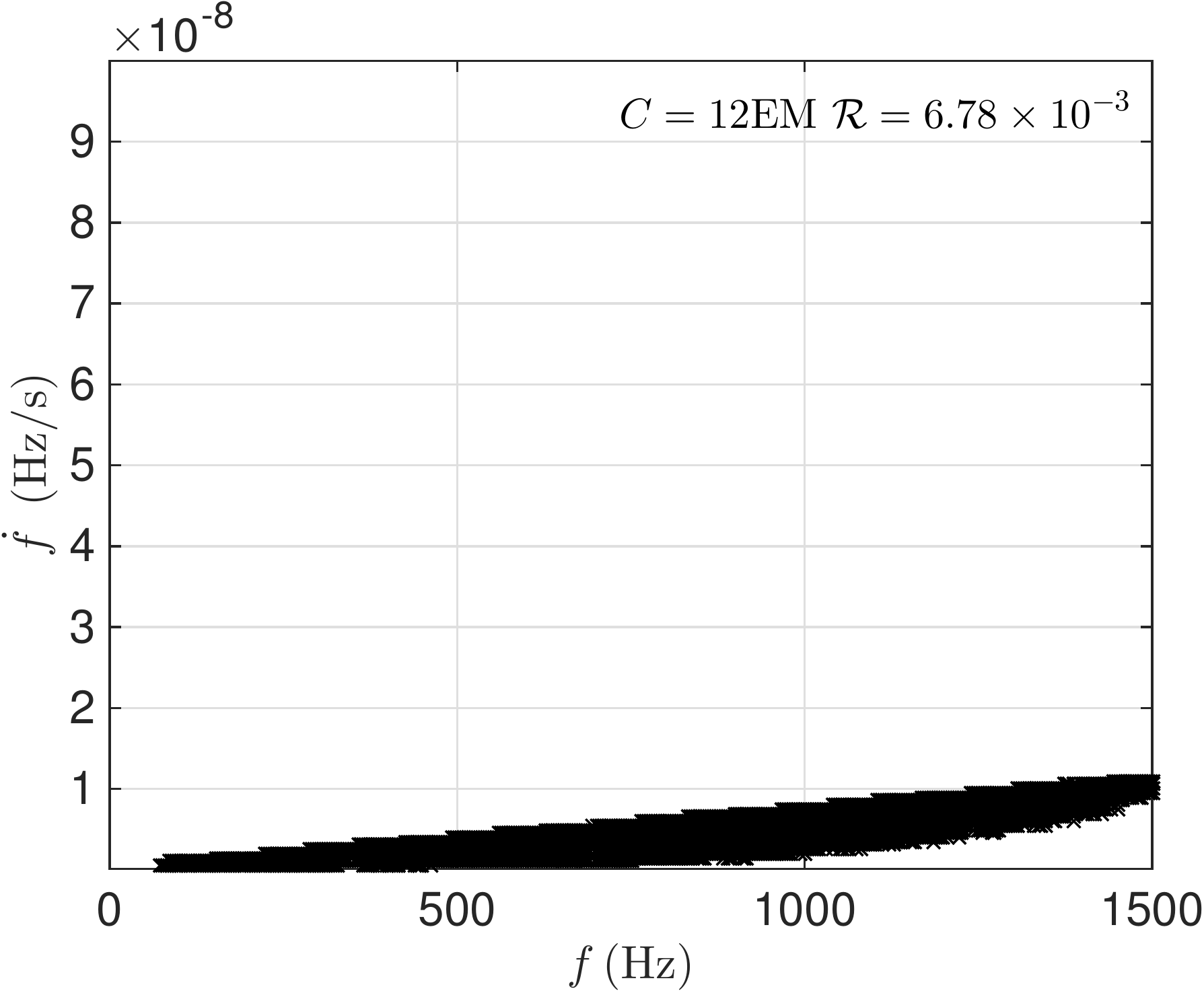}}}%
    \qquad
    \subfloat[Efficiency(lg), 37.5 days]{{  \includegraphics[width=.20\linewidth]{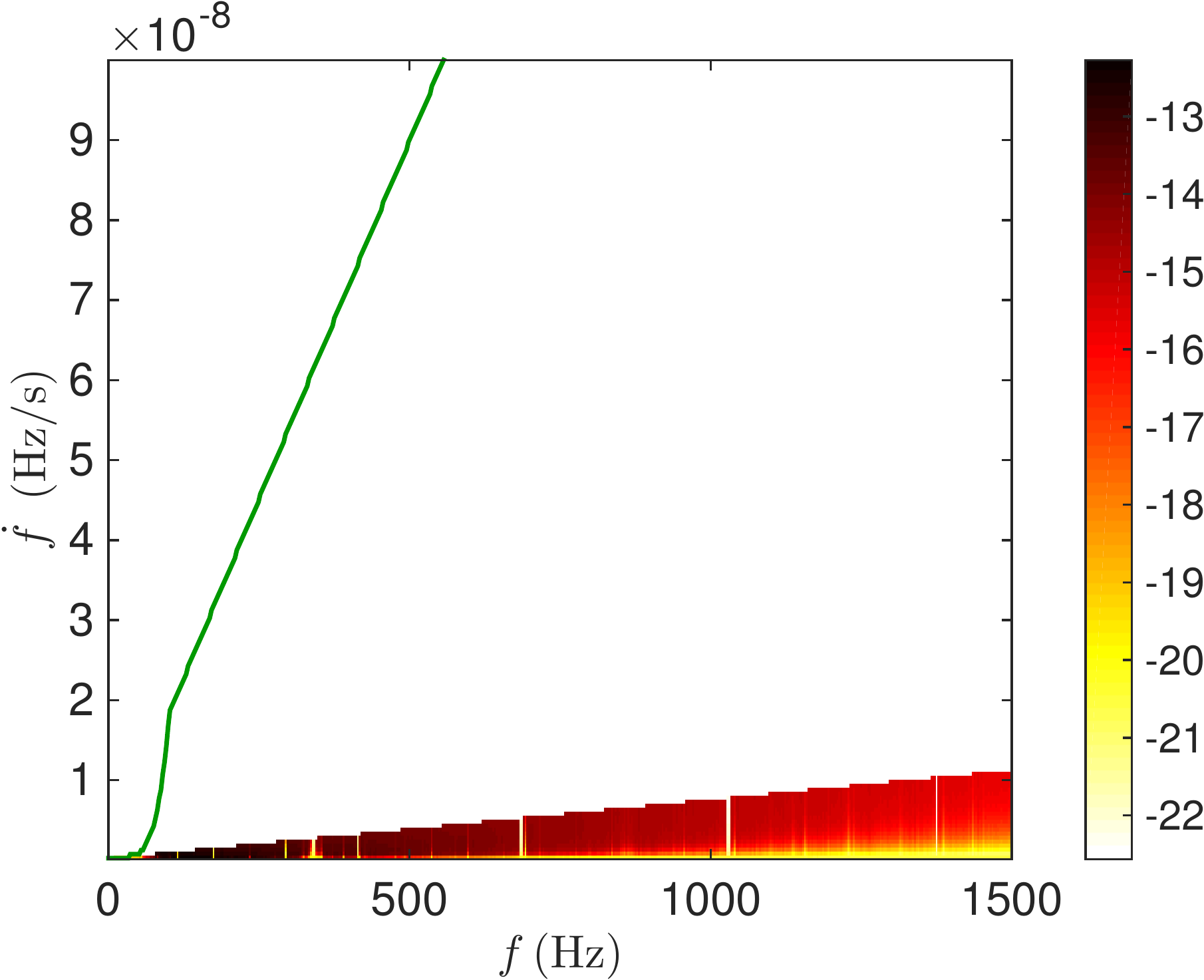}}}%
    \qquad
    \subfloat[Coverage, 37.5 days]{{  \includegraphics[width=.20\linewidth]{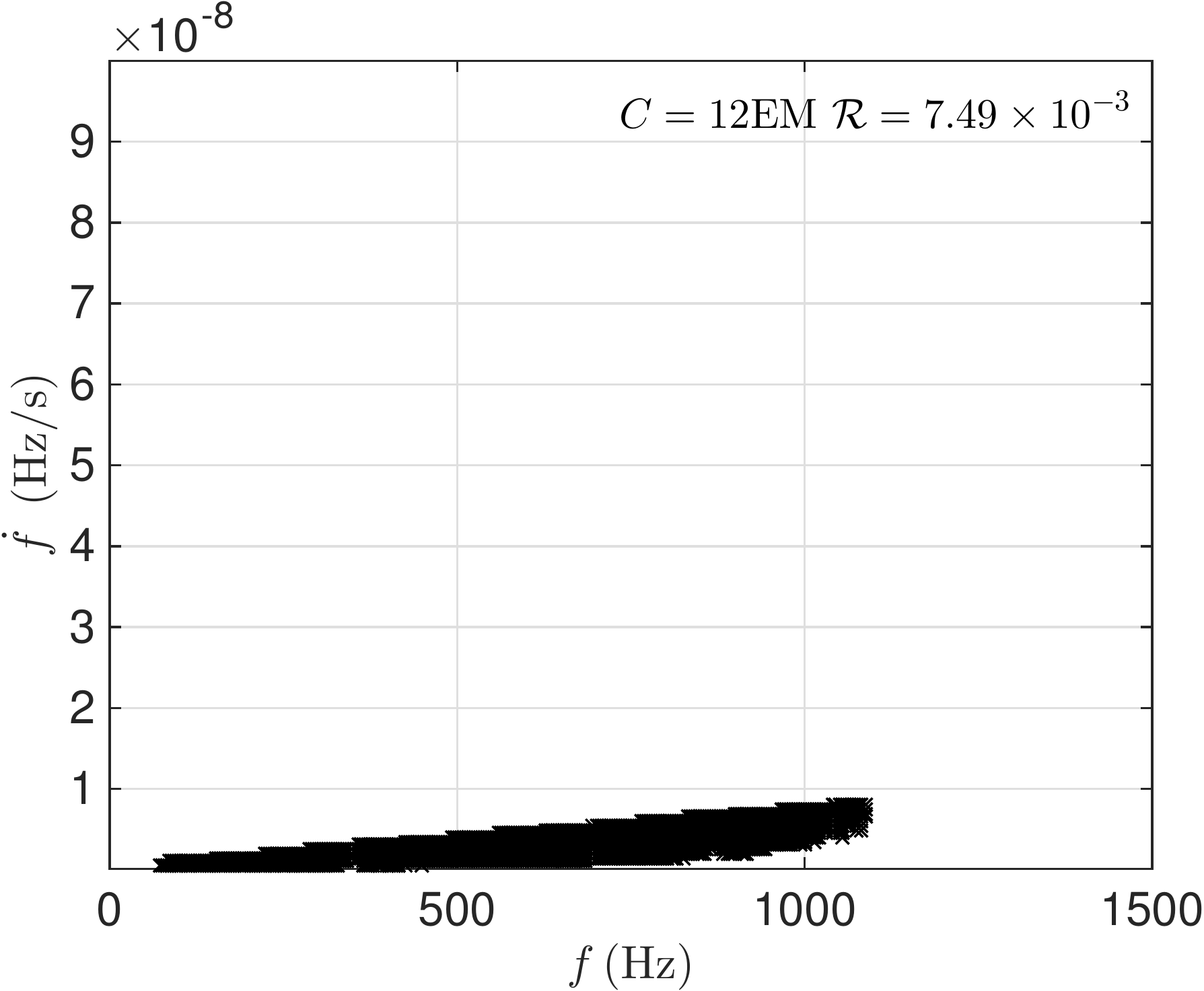}}}%
    \qquad
    \subfloat[Efficiency(lg), 50 days]{{  \includegraphics[width=.20\linewidth]{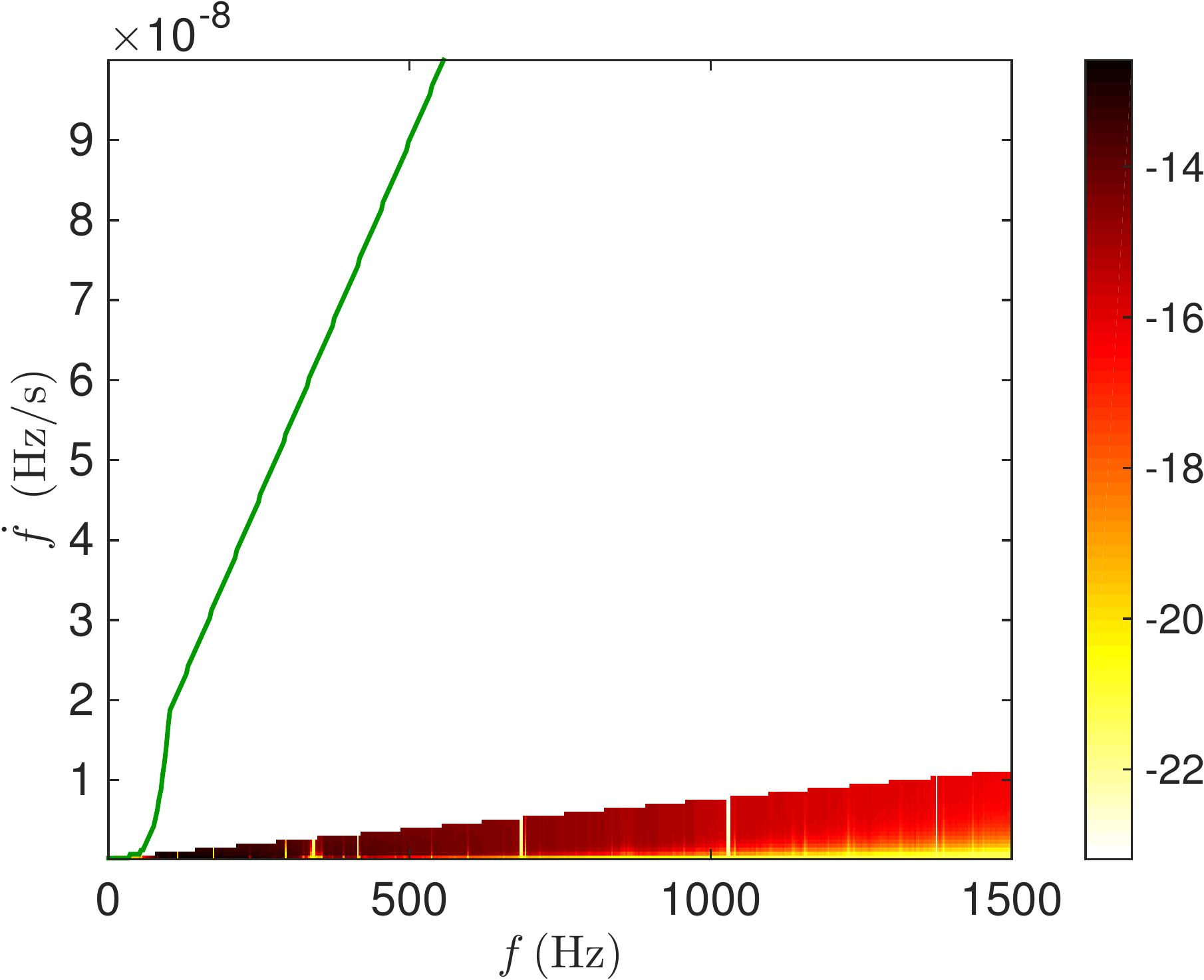}}}%
    \qquad
    \subfloat[Coverage, 50 days]{{  \includegraphics[width=.20\linewidth]{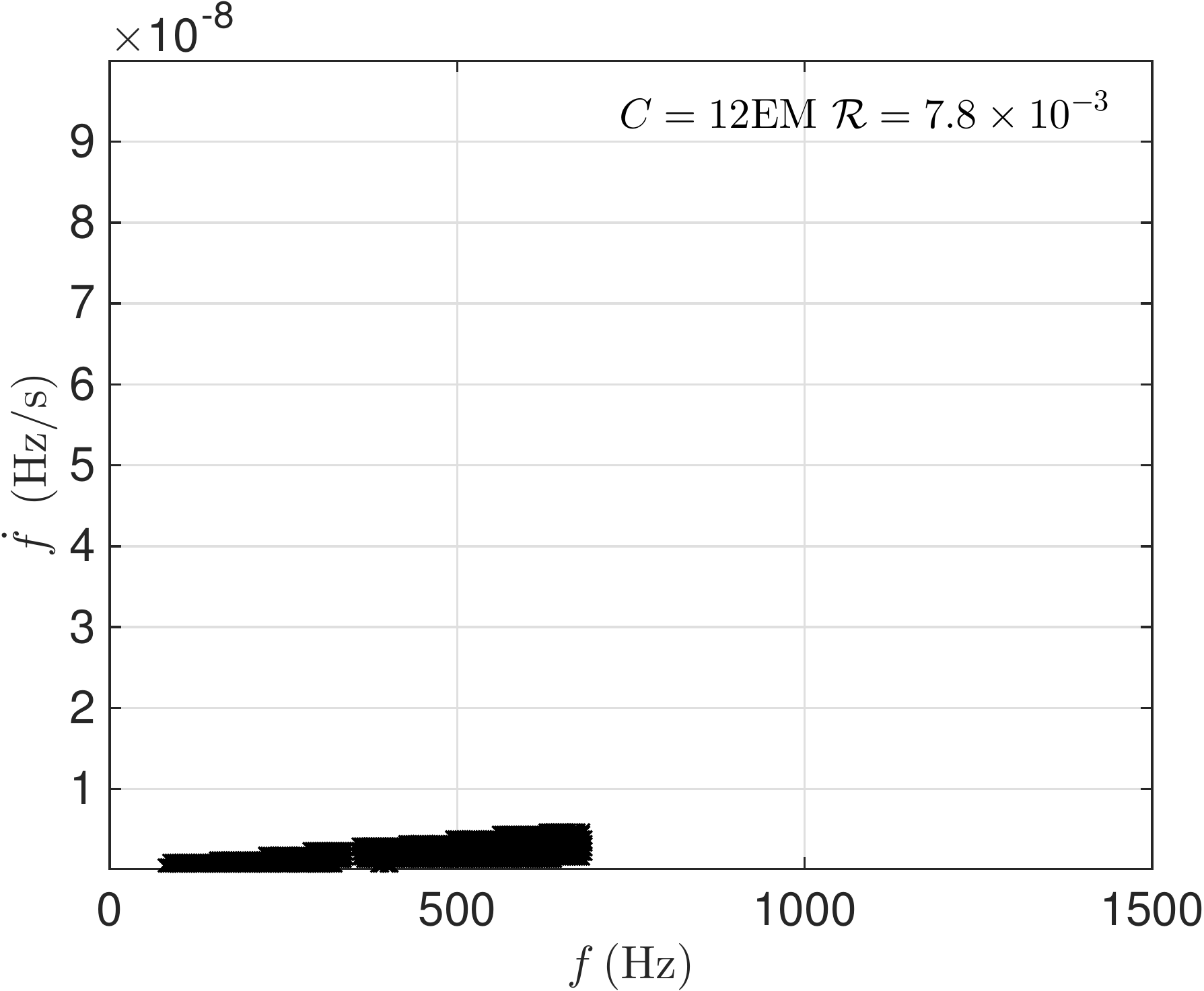}}}%
    \qquad
    \subfloat[Efficiency(lg), 75 days]{{  \includegraphics[width=.20\linewidth]{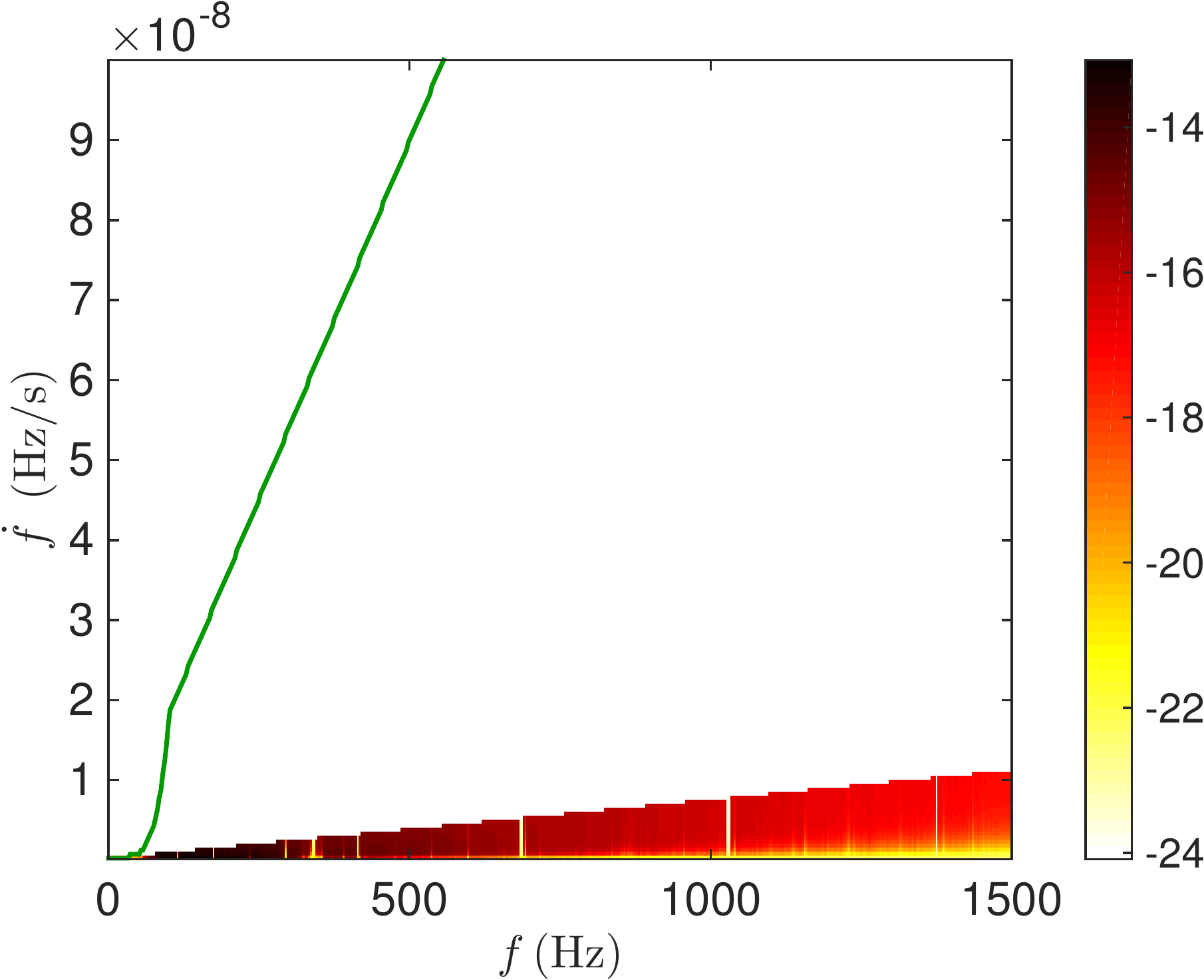}}}%
    \qquad
    \subfloat[Coverage, 75 days]{{  \includegraphics[width=.20\linewidth]{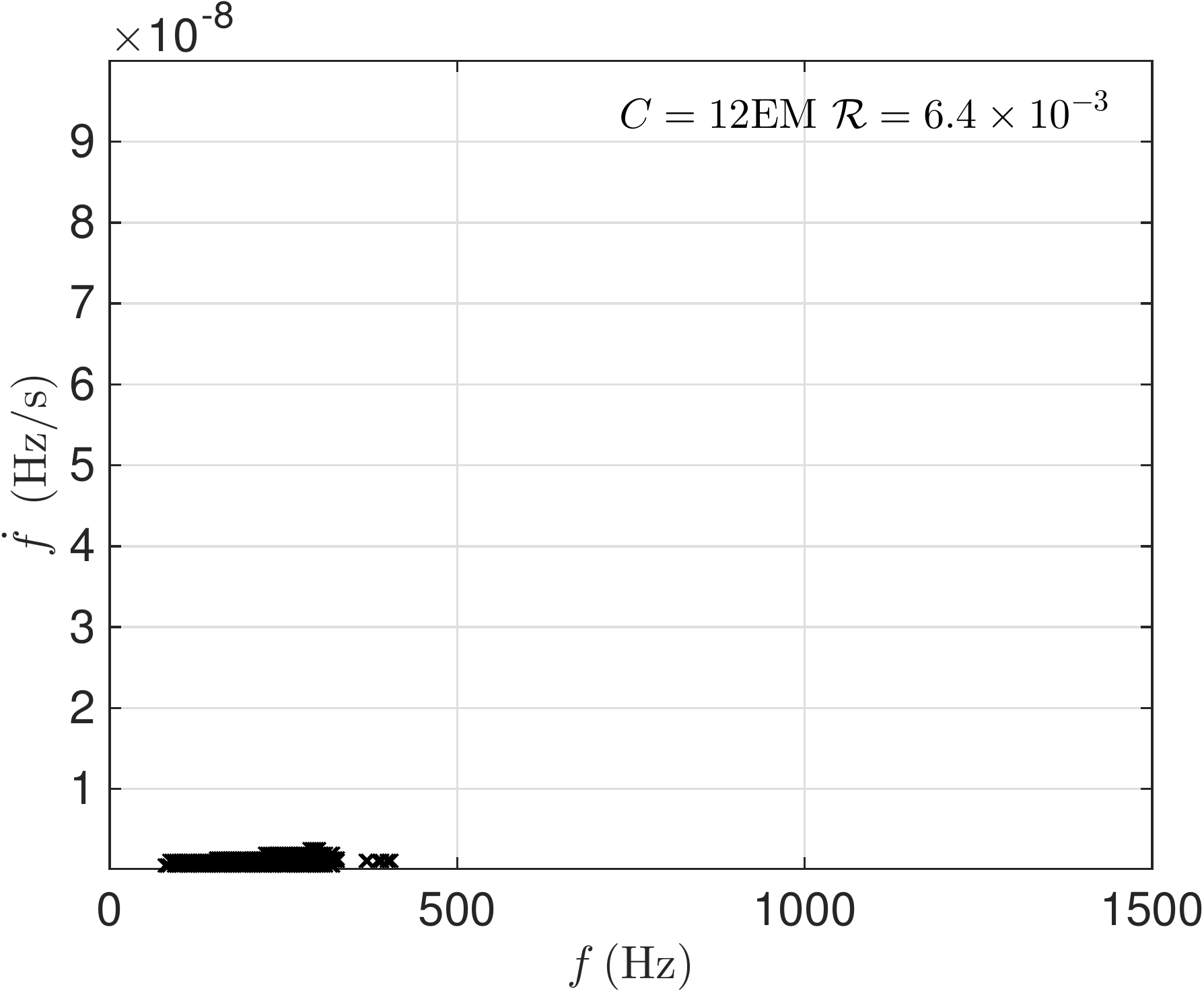}}}%
    \caption{Optimisation results for Vela Jr at 750 pc and 4300 years old (far and old, FO) assuming log-uniform and age-based priors, for various coherent search durations: 5, 10, 20, 30, 37.5, 50 and 75 days. The total computing budget is assumed to be 12 EM. }%
    \label{G2662_51020days_longage_longdist_log}%
\end{figure*}

\begin{figure*}%
    \centering
    \subfloat[Efficiency(lg), 5 days]{{  \includegraphics[width=.20\linewidth]{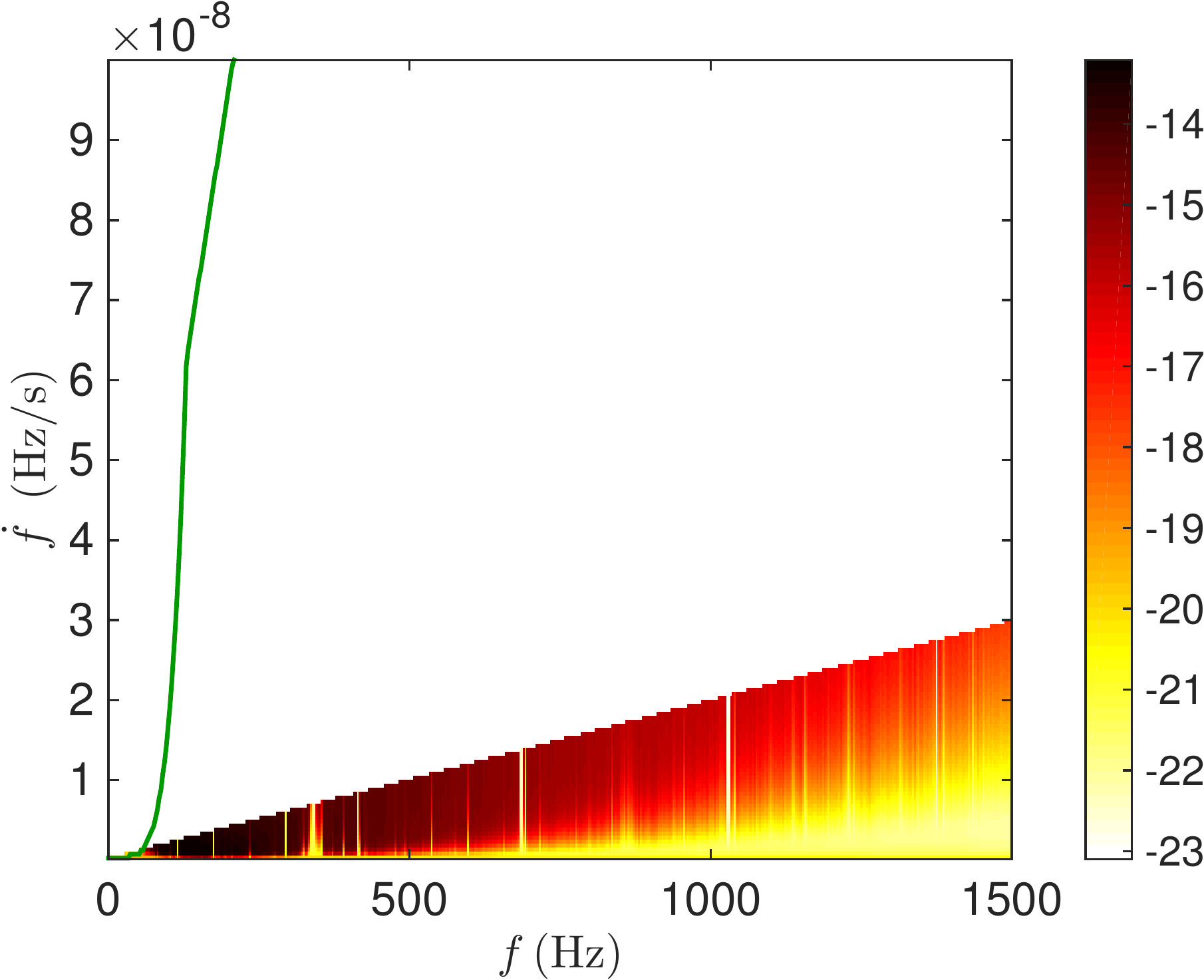}}}%
    \qquad
    \subfloat[Coverage, 5 days]{{  \includegraphics[width=.20\linewidth]{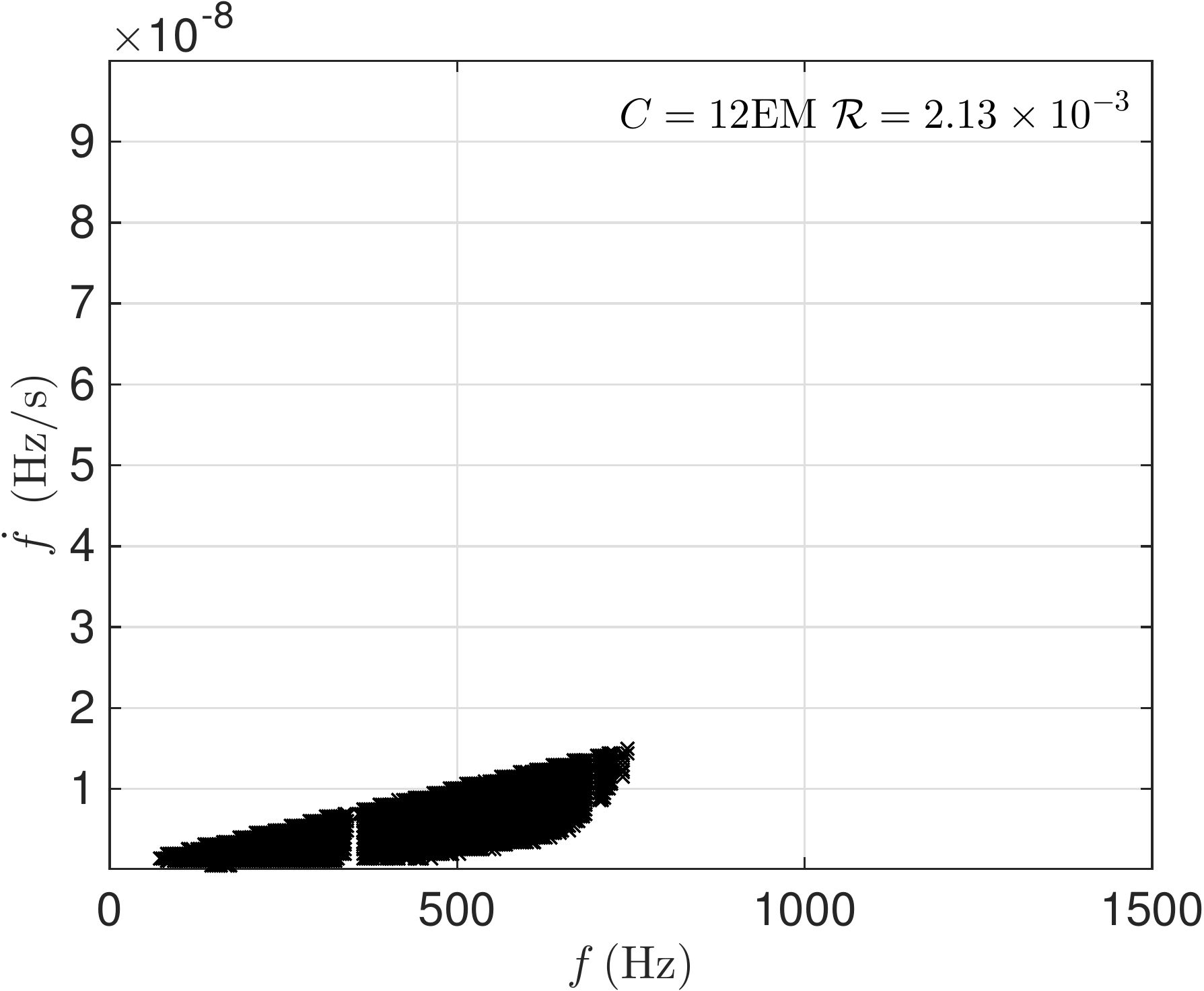}}}%
    \qquad
    \subfloat[Efficiency(lg), 10 days]{{  \includegraphics[width=.20\linewidth]{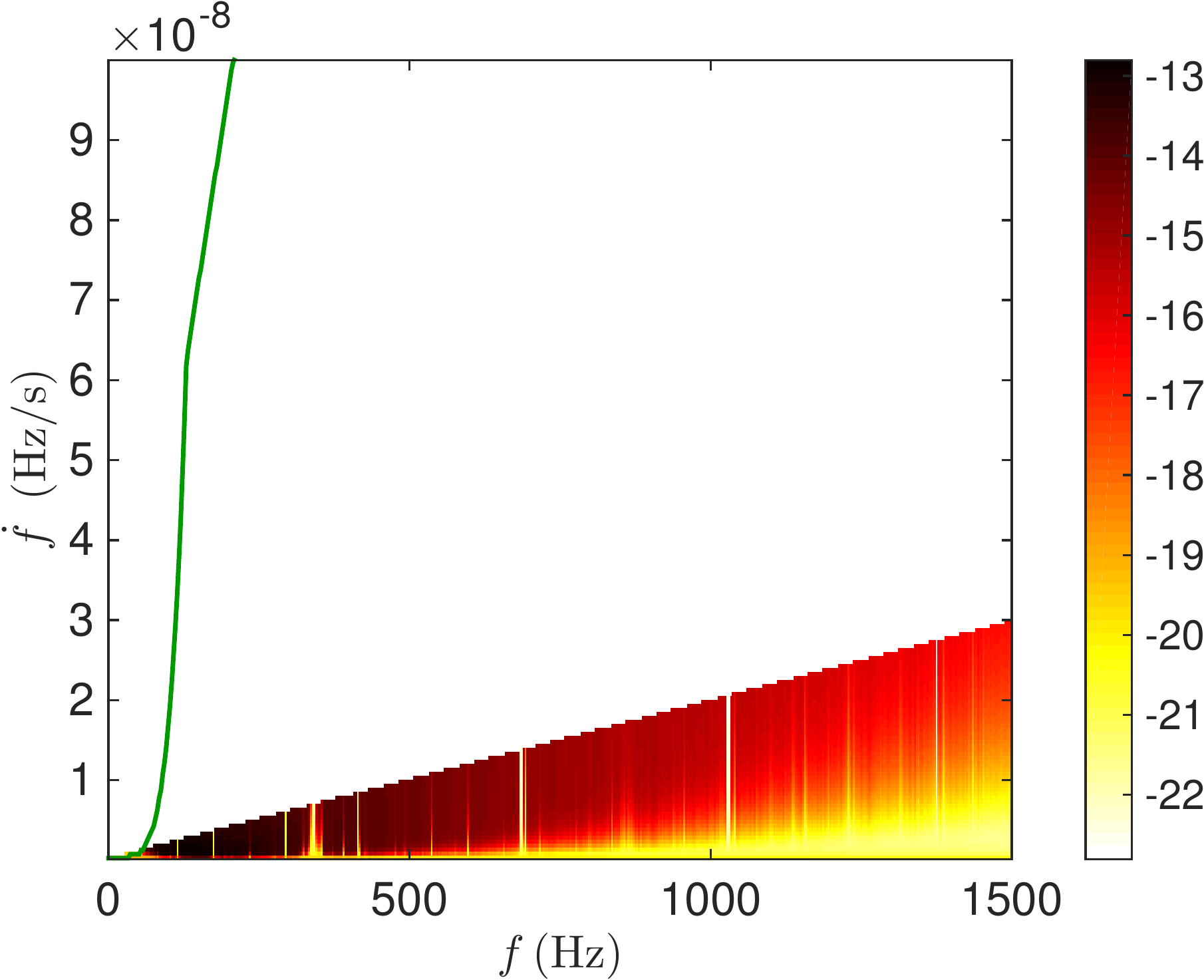}}}%
    \qquad
    \subfloat[Coverage, 10 days]{{  \includegraphics[width=.20\linewidth]{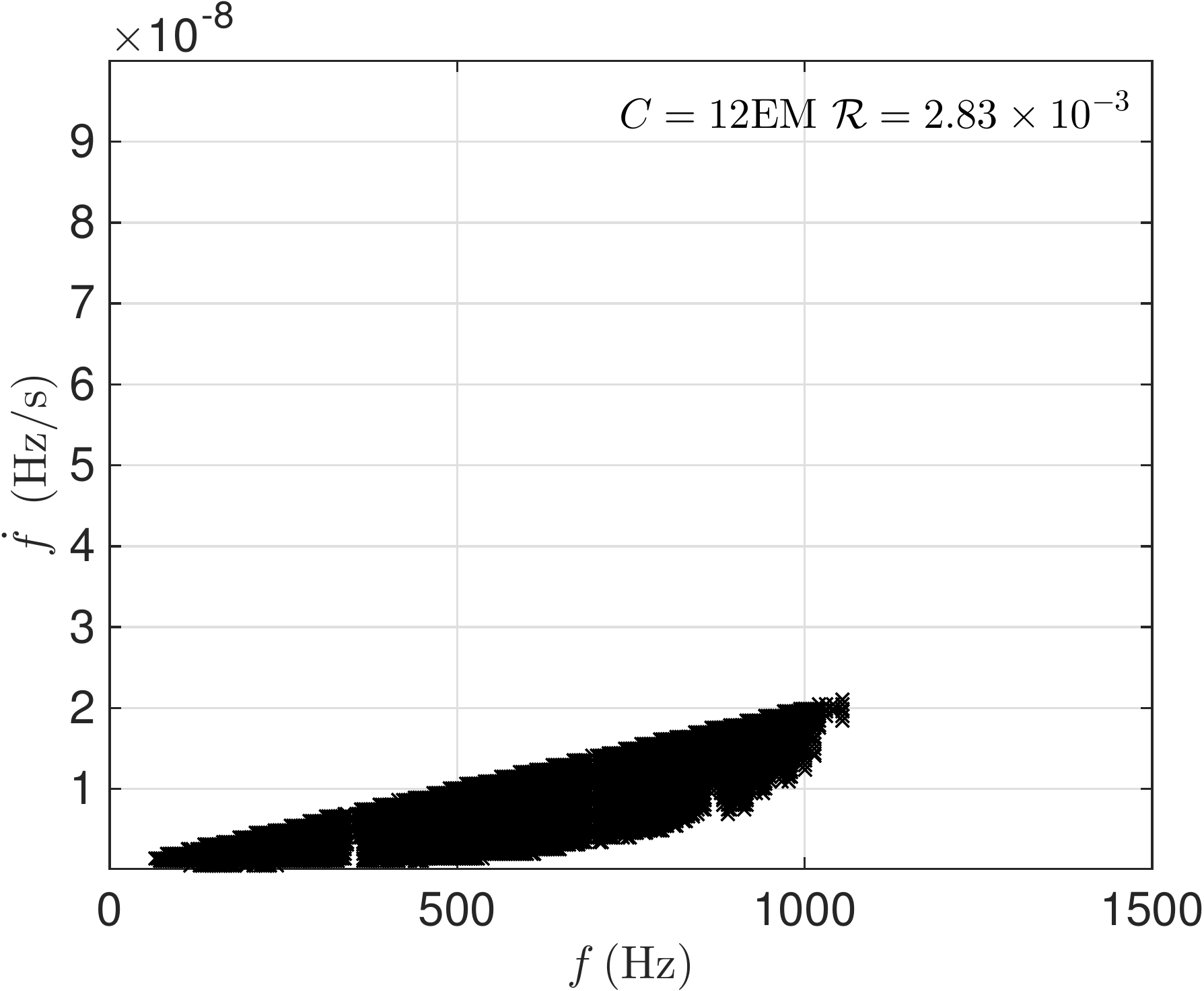}}}%
    \qquad
    \subfloat[Efficiency(lg), 20 days]{{  \includegraphics[width=.20\linewidth]{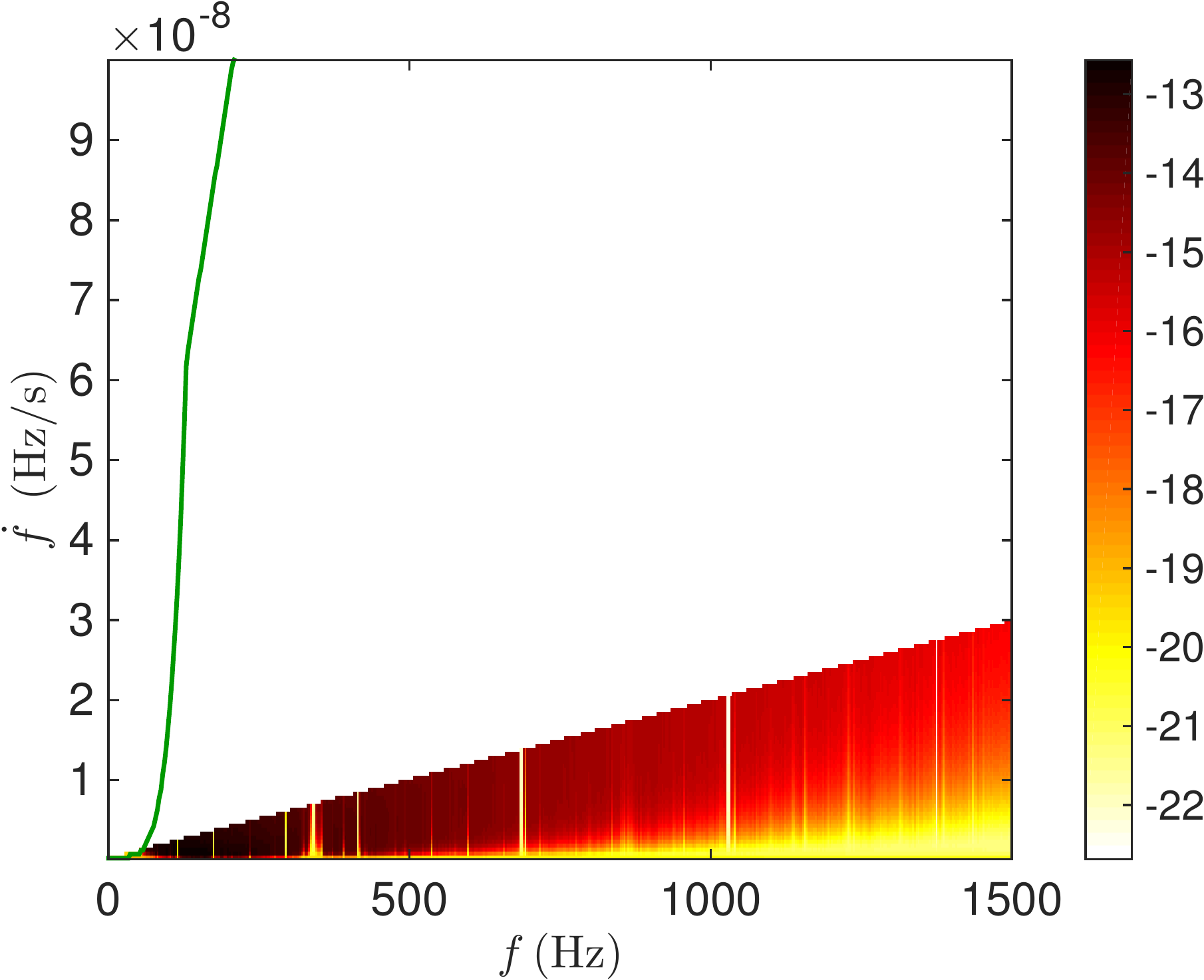}}}%
    \qquad
    \subfloat[Coverage, 20 days]{{  \includegraphics[width=.20\linewidth]{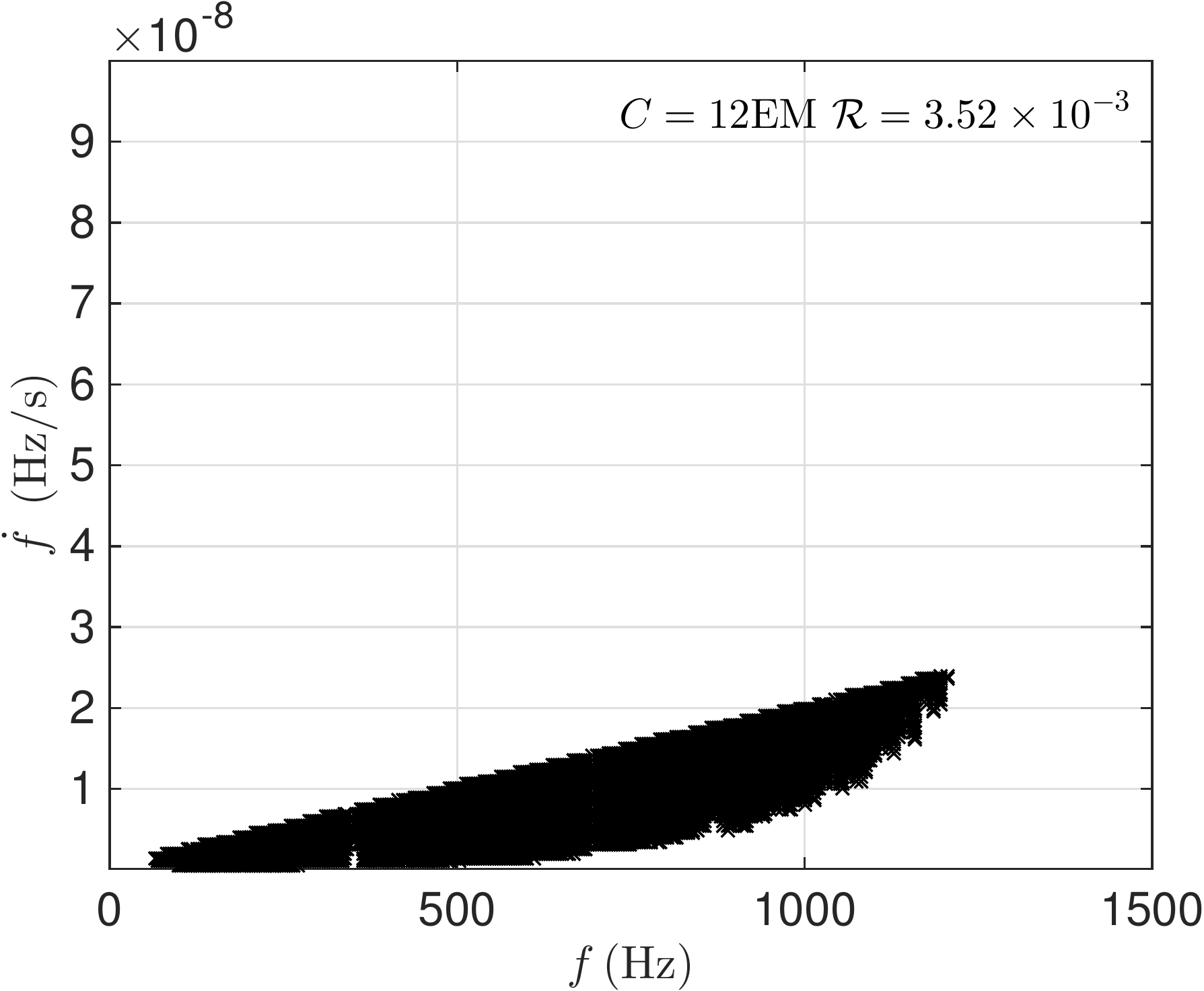}}}%
    \qquad
    \subfloat[Efficiency(lg), 30 days]{{  \includegraphics[width=.20\linewidth]{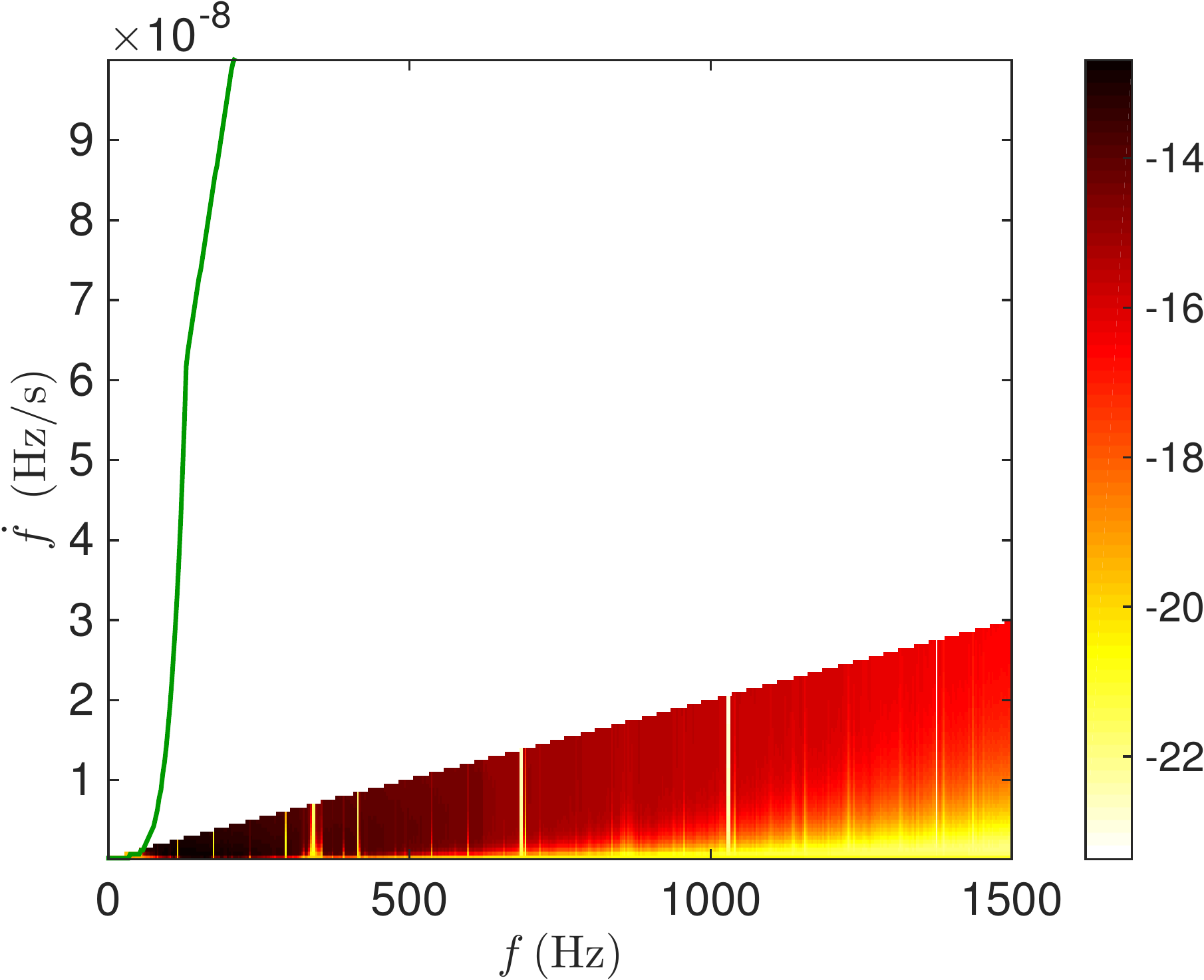}}}%
    \qquad
    \subfloat[Coverage, 30 days]{{  \includegraphics[width=.20\linewidth]{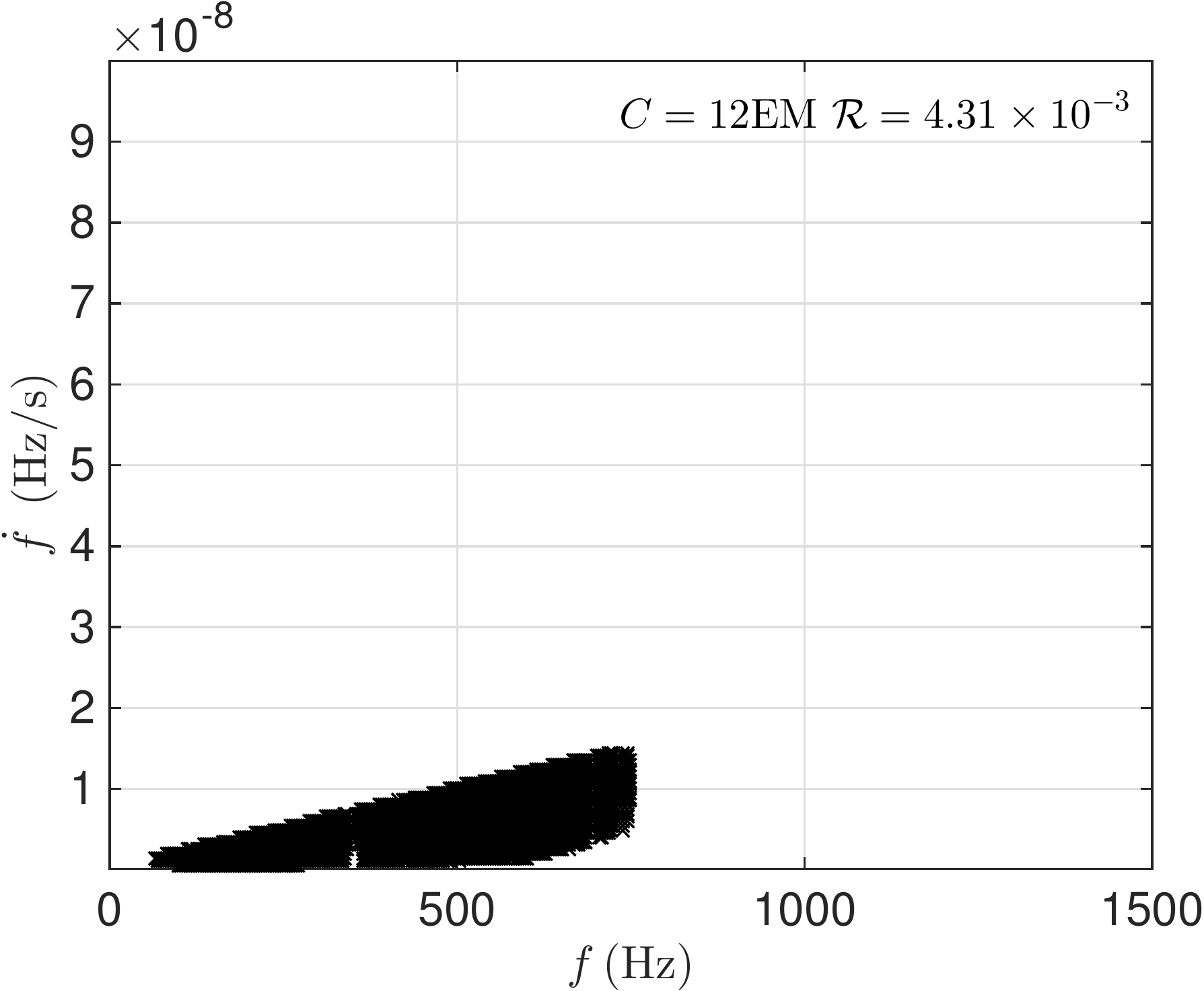}}}%
    \qquad
    \subfloat[Efficiency(lg), 37.5 days]{{  \includegraphics[width=.20\linewidth]{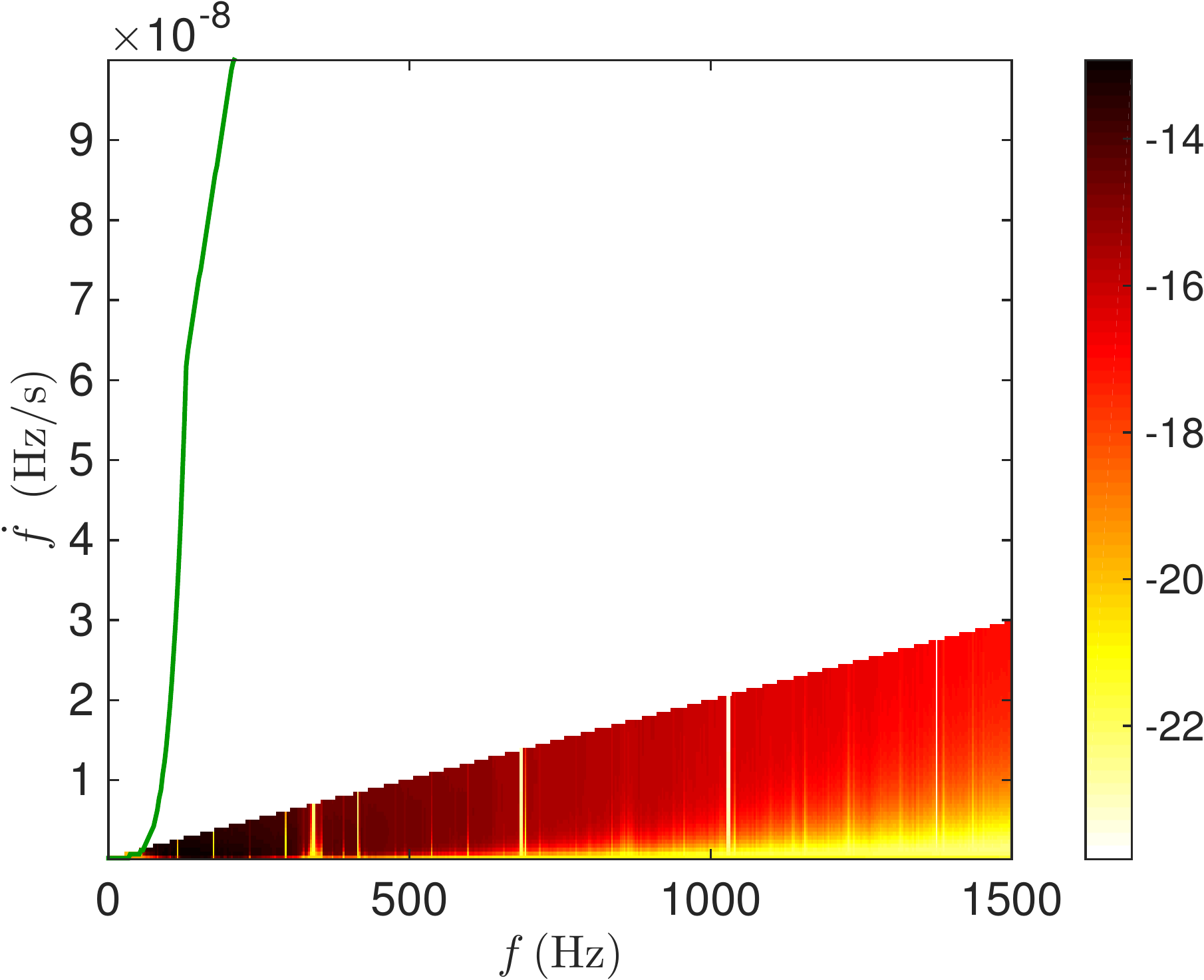}}}%
    \qquad
    \subfloat[Coverage, 37.5 days]{{  \includegraphics[width=.20\linewidth]{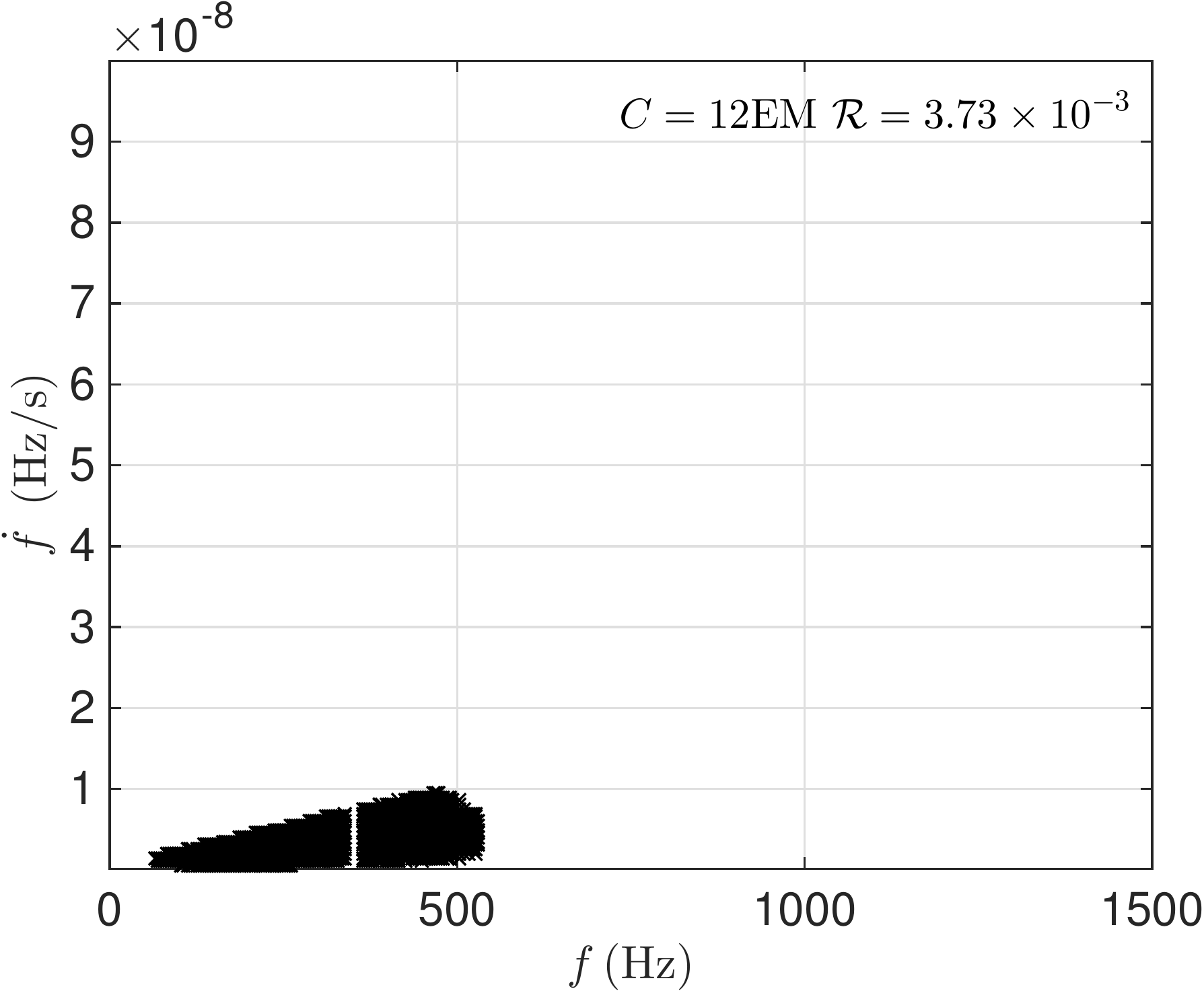}}}%
    \qquad
    \subfloat[Efficiency(lg), 50 days]{{  \includegraphics[width=.20\linewidth]{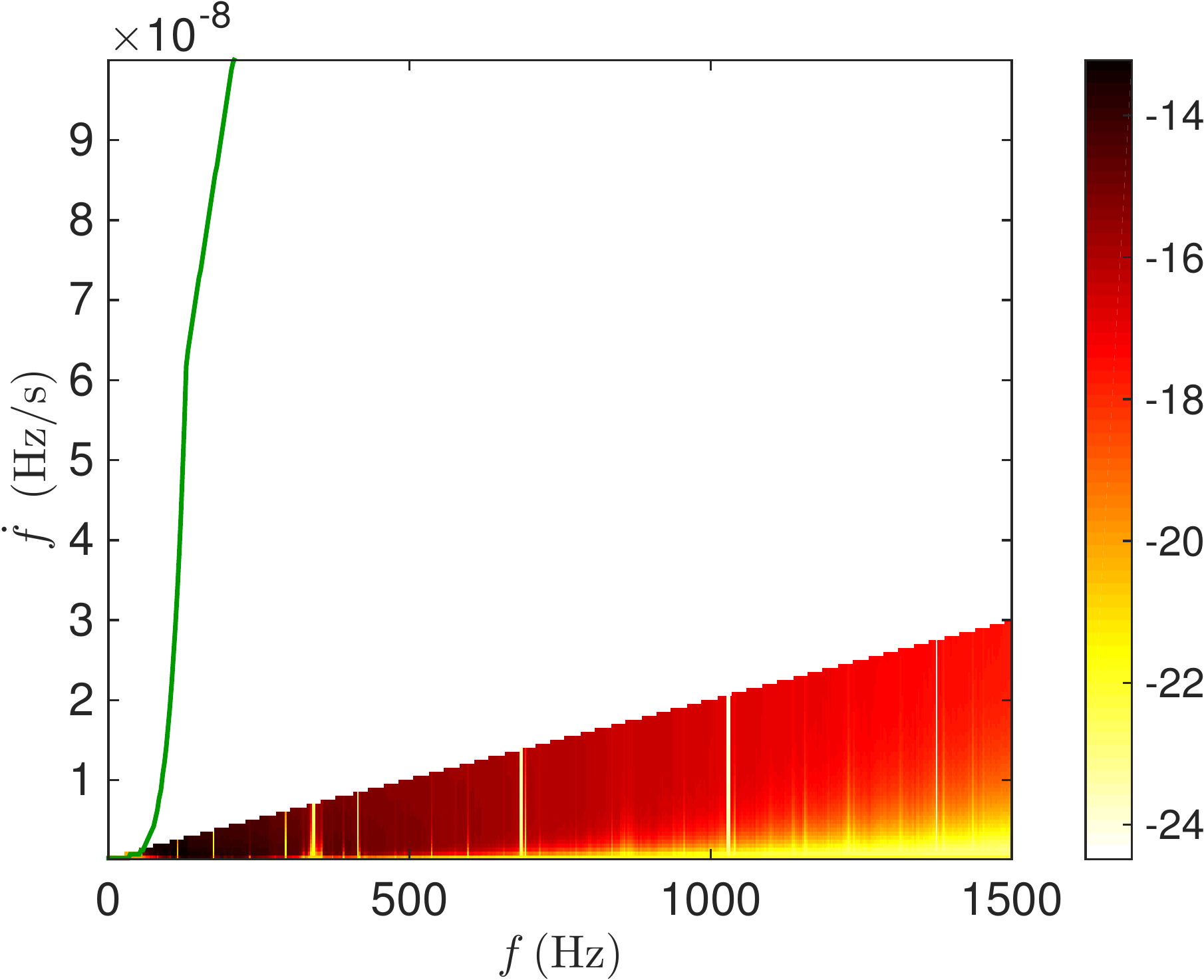}}}%
    \qquad
    \subfloat[Coverage, 50 days]{{  \includegraphics[width=.20\linewidth]{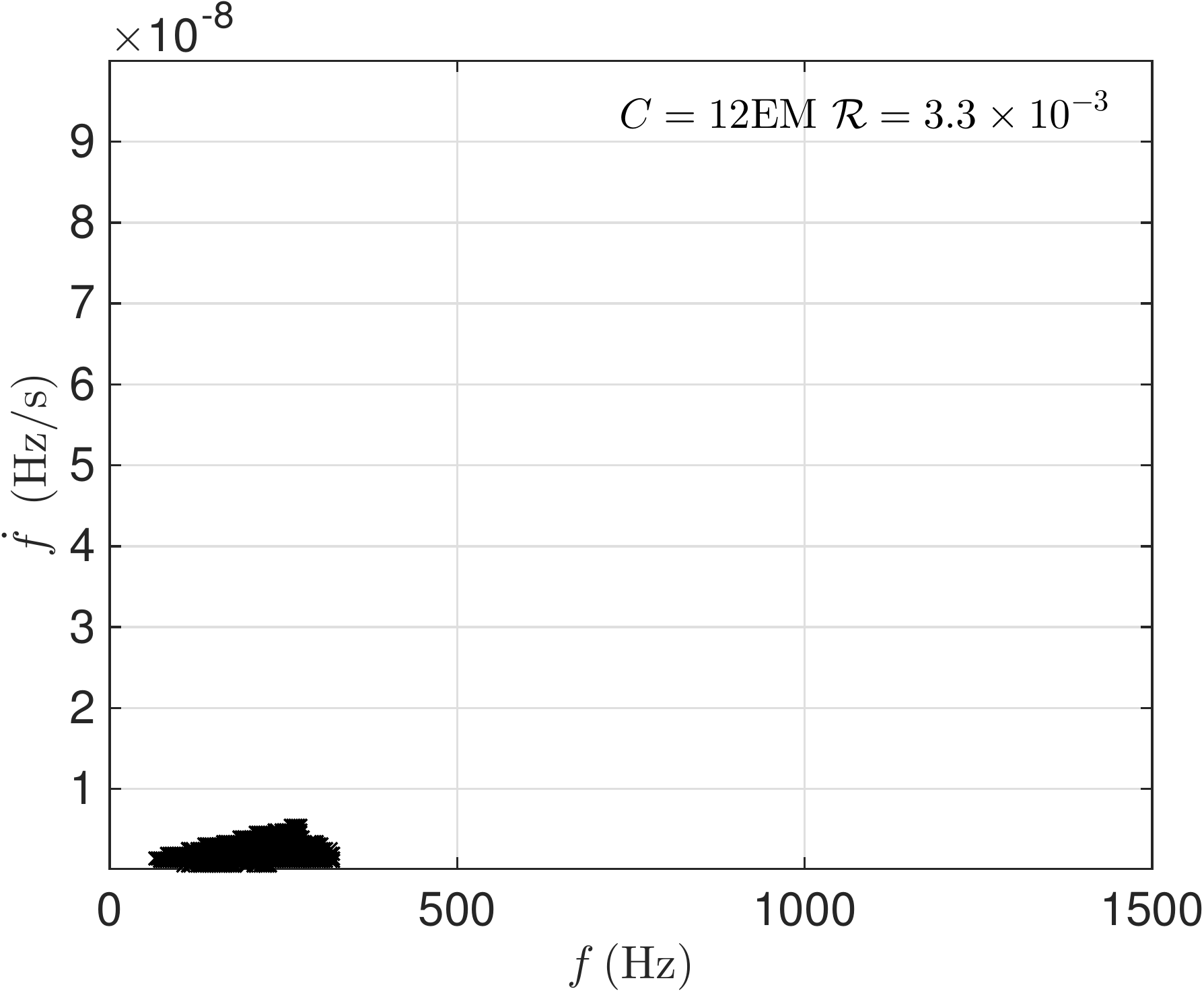}}}%
    \qquad
    \subfloat[Efficiency(lg), 75 days]{{  \includegraphics[width=.20\linewidth]{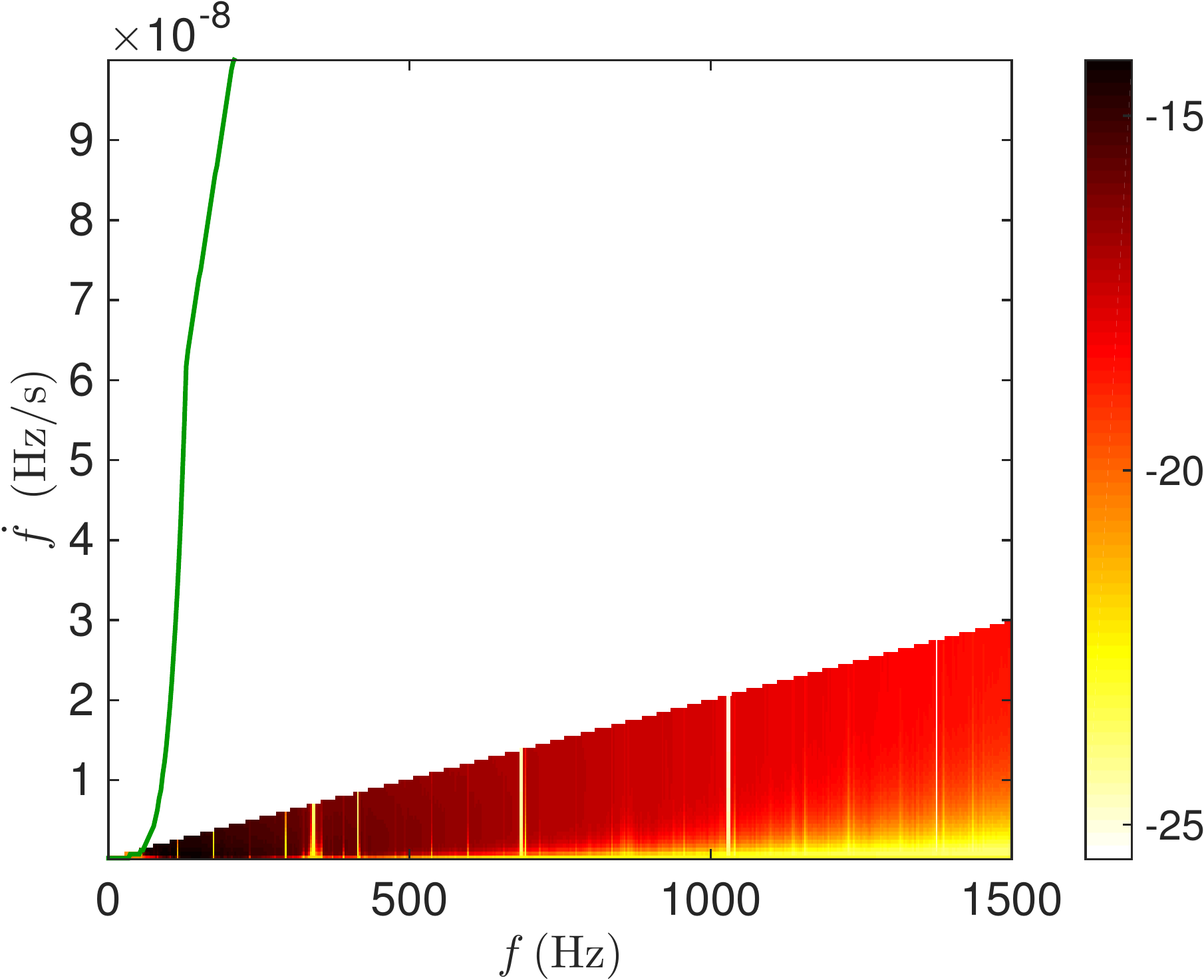}}}%
    \qquad
    \subfloat[Coverage, 75 days]{{  \includegraphics[width=.20\linewidth]{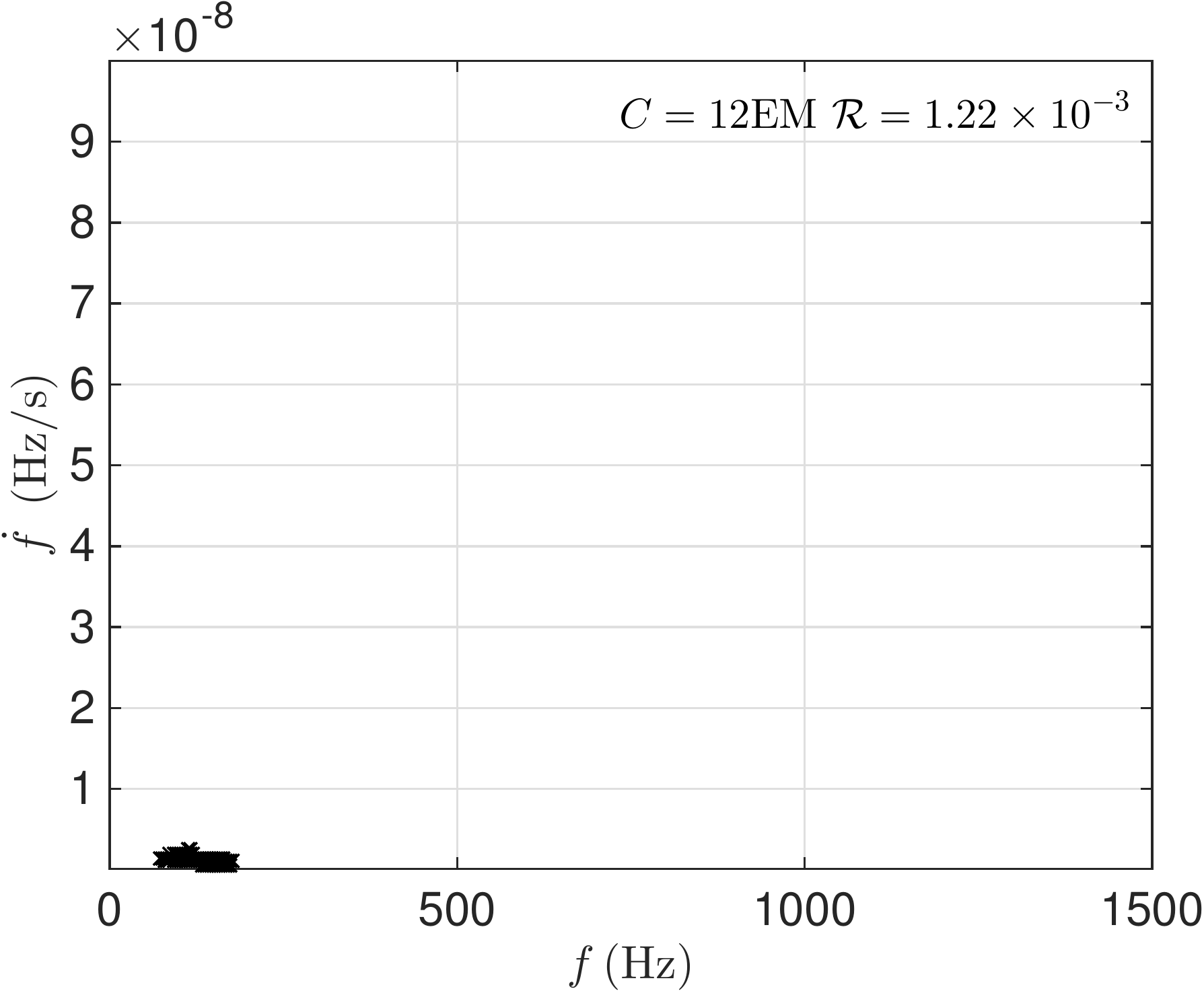}}}%

    \caption{Optimisation results for G347.3 Cas A at 1300 pc, 1600 years old, assuming log-uniform and age-based priors, for various coherent search durations: 5, 10, 20, 30, 37.5, 50 and 75 days. The total computing budget is assumed to be 12 EM.}%
    \label{G3473_51020days_shortage_log}%
\end{figure*}

\FloatBarrier

\begin{figure*}%
    \centering
    \subfloat[Efficiency(lg), 5 days]{{  \includegraphics[width=.20\linewidth]{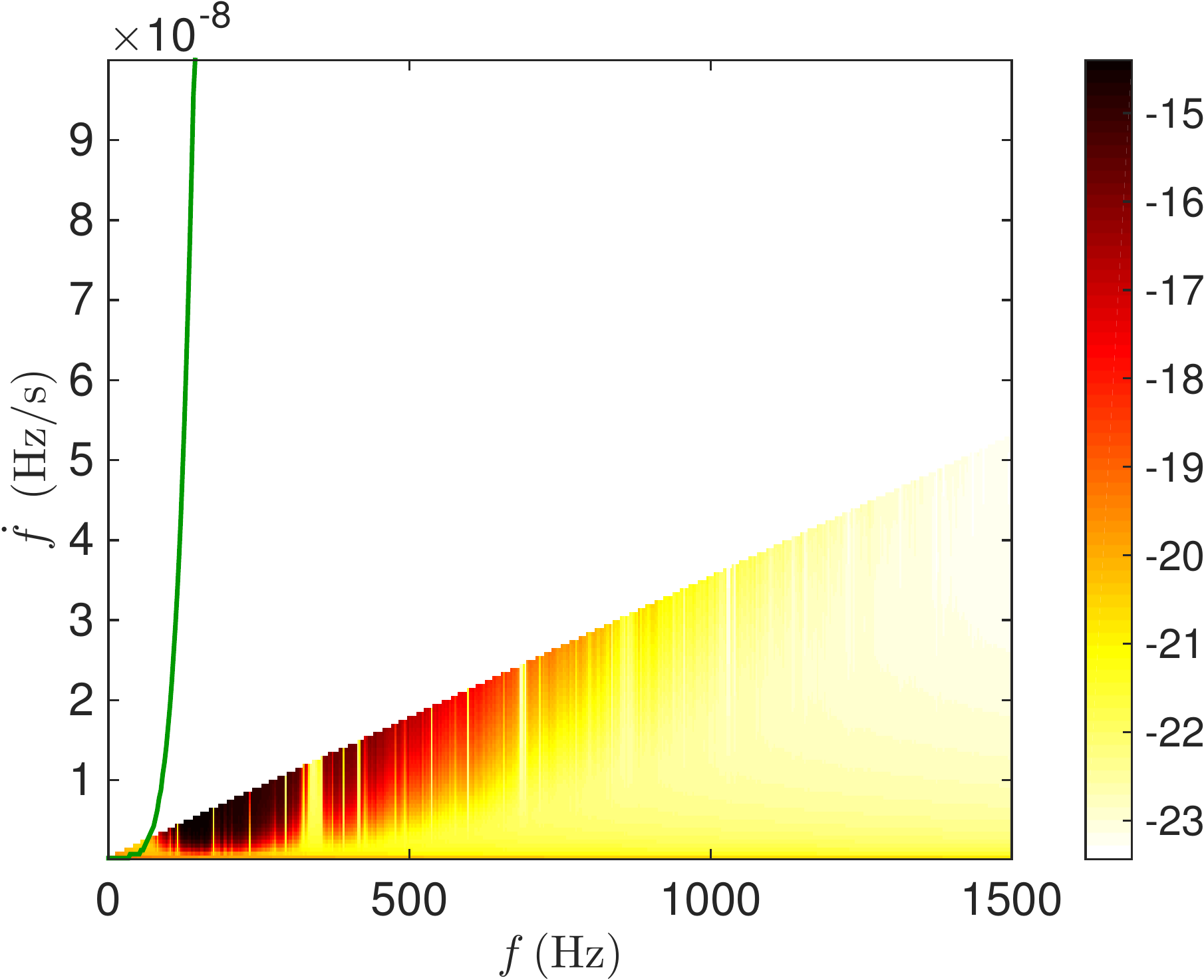}}}%
    \qquad
    \subfloat[Coverage, 5 days]{{  \includegraphics[width=.20\linewidth]{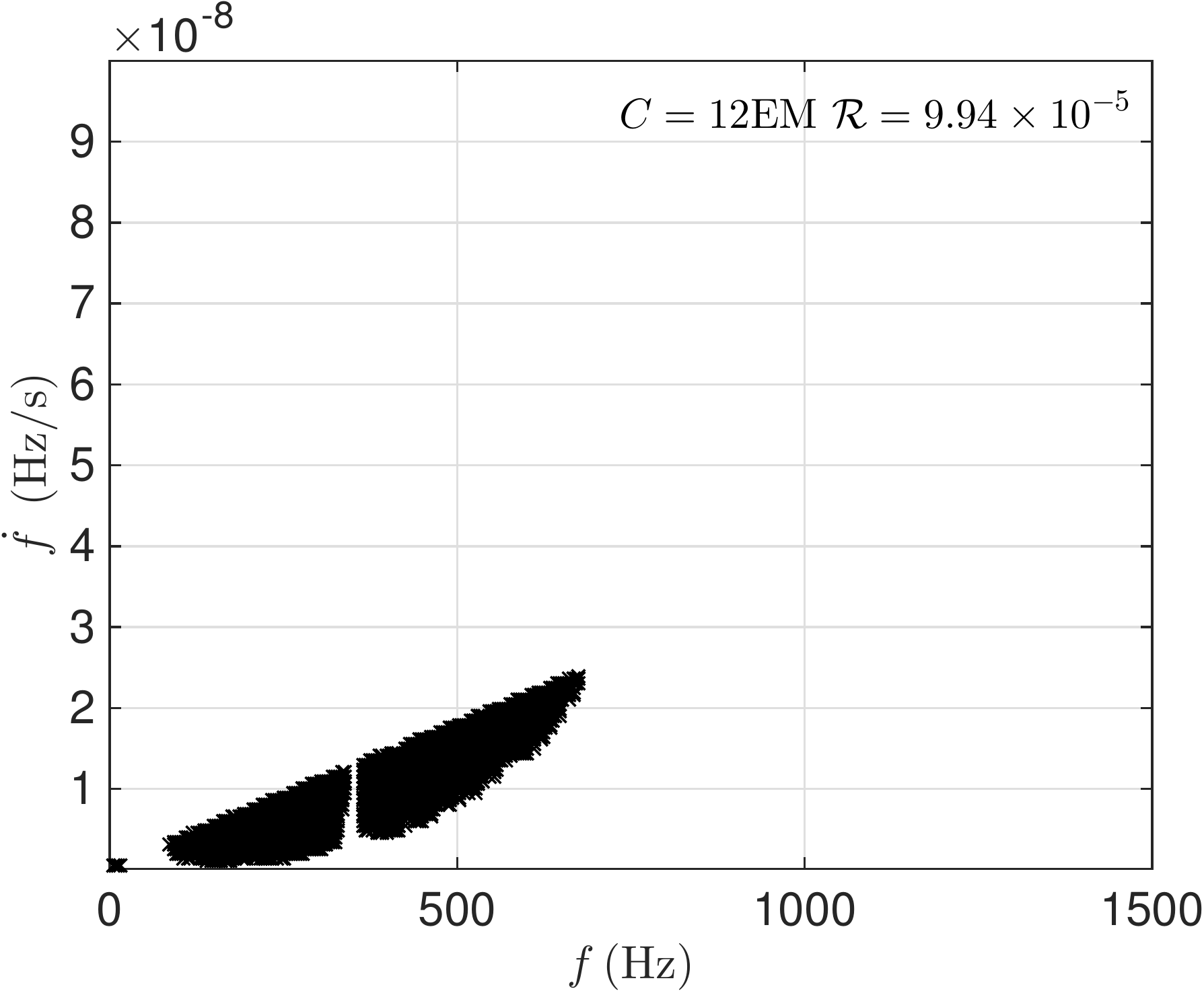}}}%
    \qquad
    \subfloat[Efficiency(lg), 10 days]{{  \includegraphics[width=.20\linewidth]{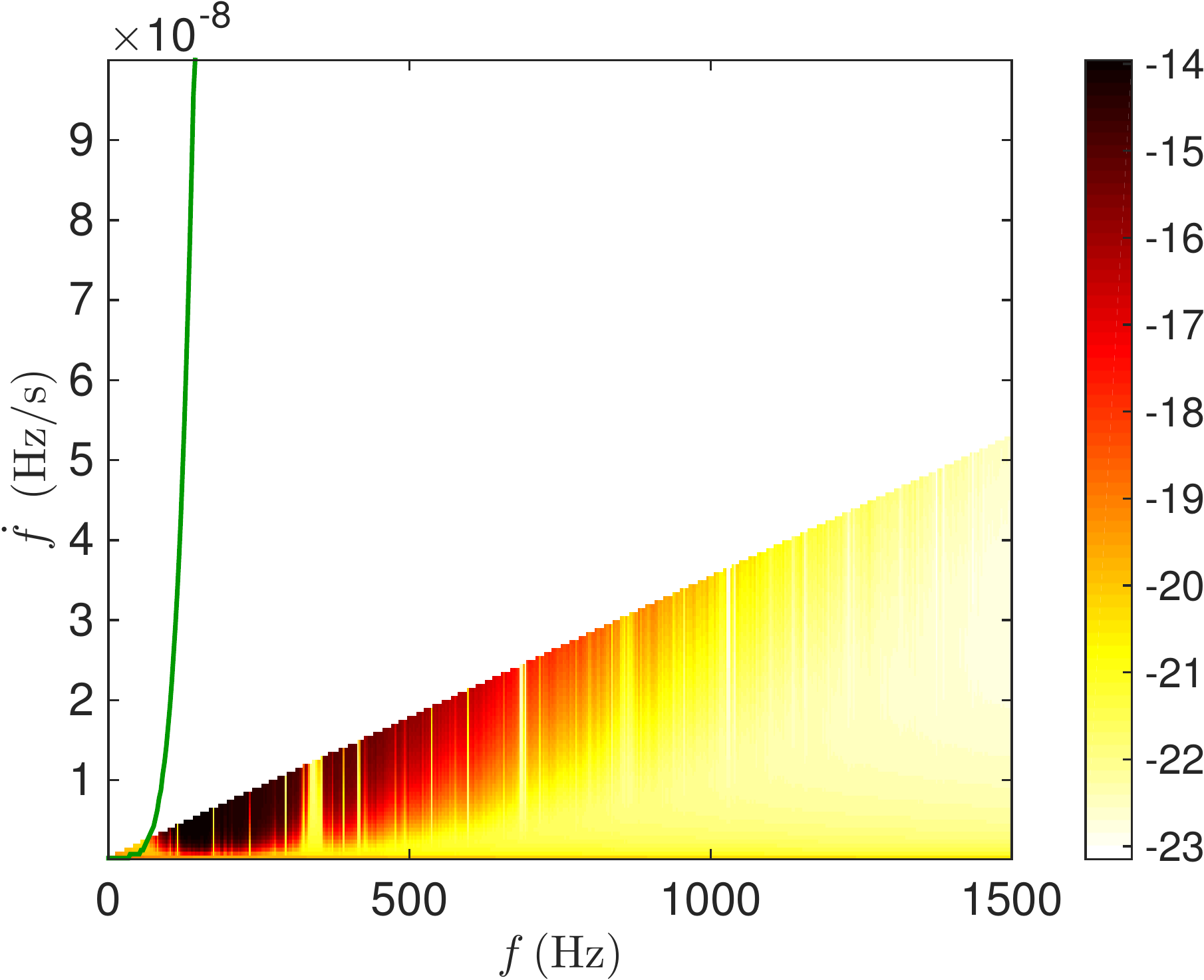}}}%
    \qquad
    \subfloat[Coverage, 10 days]{{  \includegraphics[width=.20\linewidth]{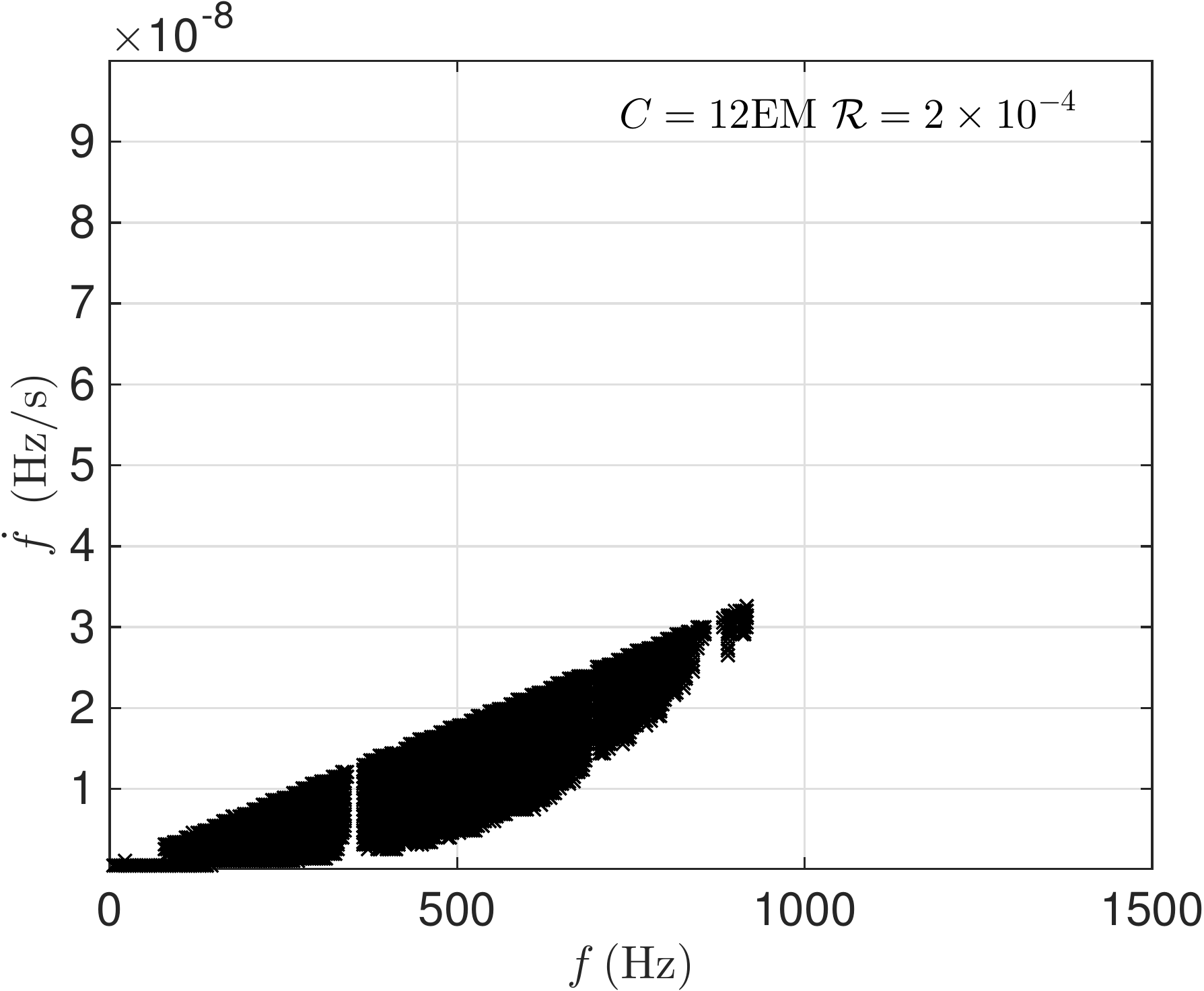}}}%
    \qquad
    \subfloat[Efficiency(lg), 20 days]{{  \includegraphics[width=.20\linewidth]{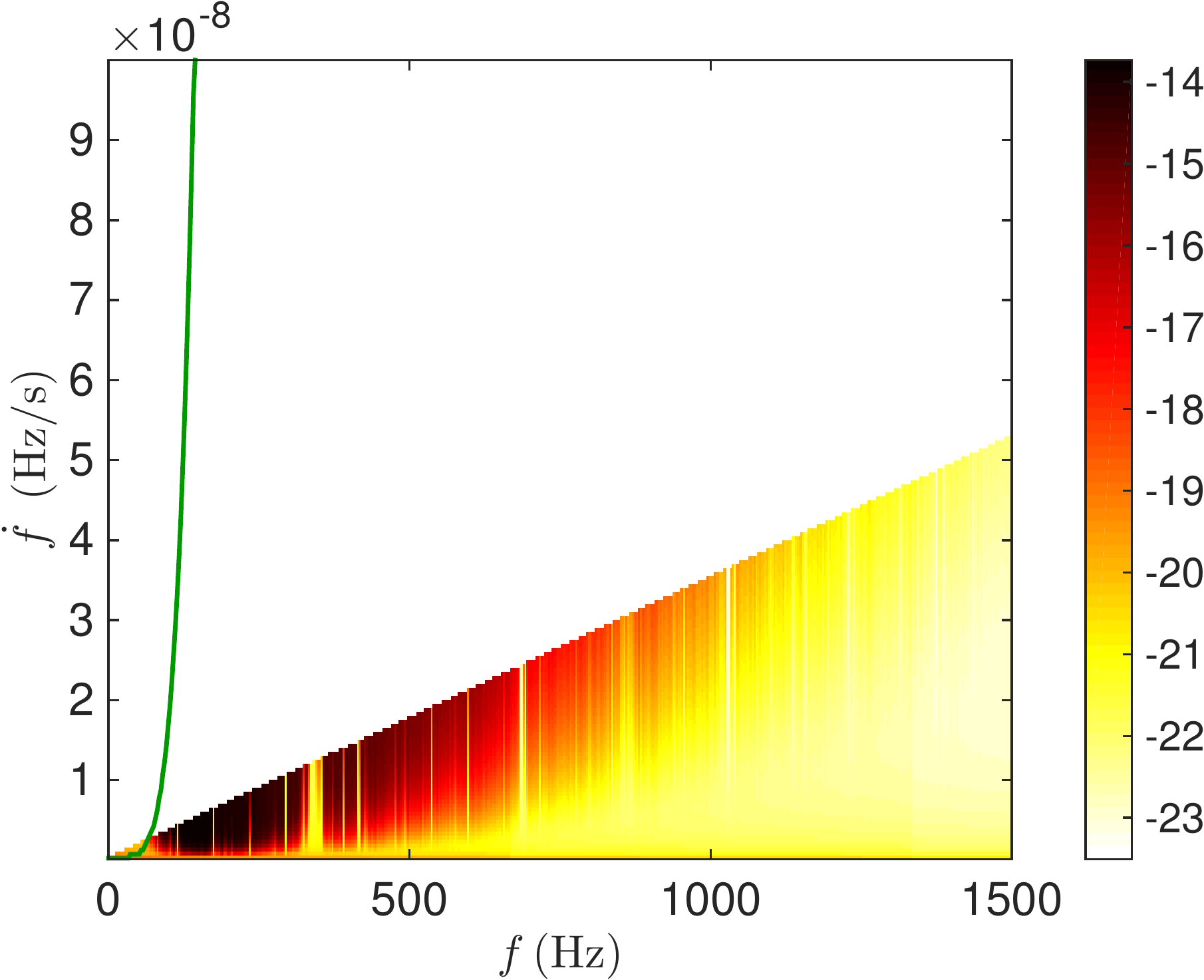}}}%
    \qquad
    \subfloat[Coverage, 20 days]{{  \includegraphics[width=.20\linewidth]{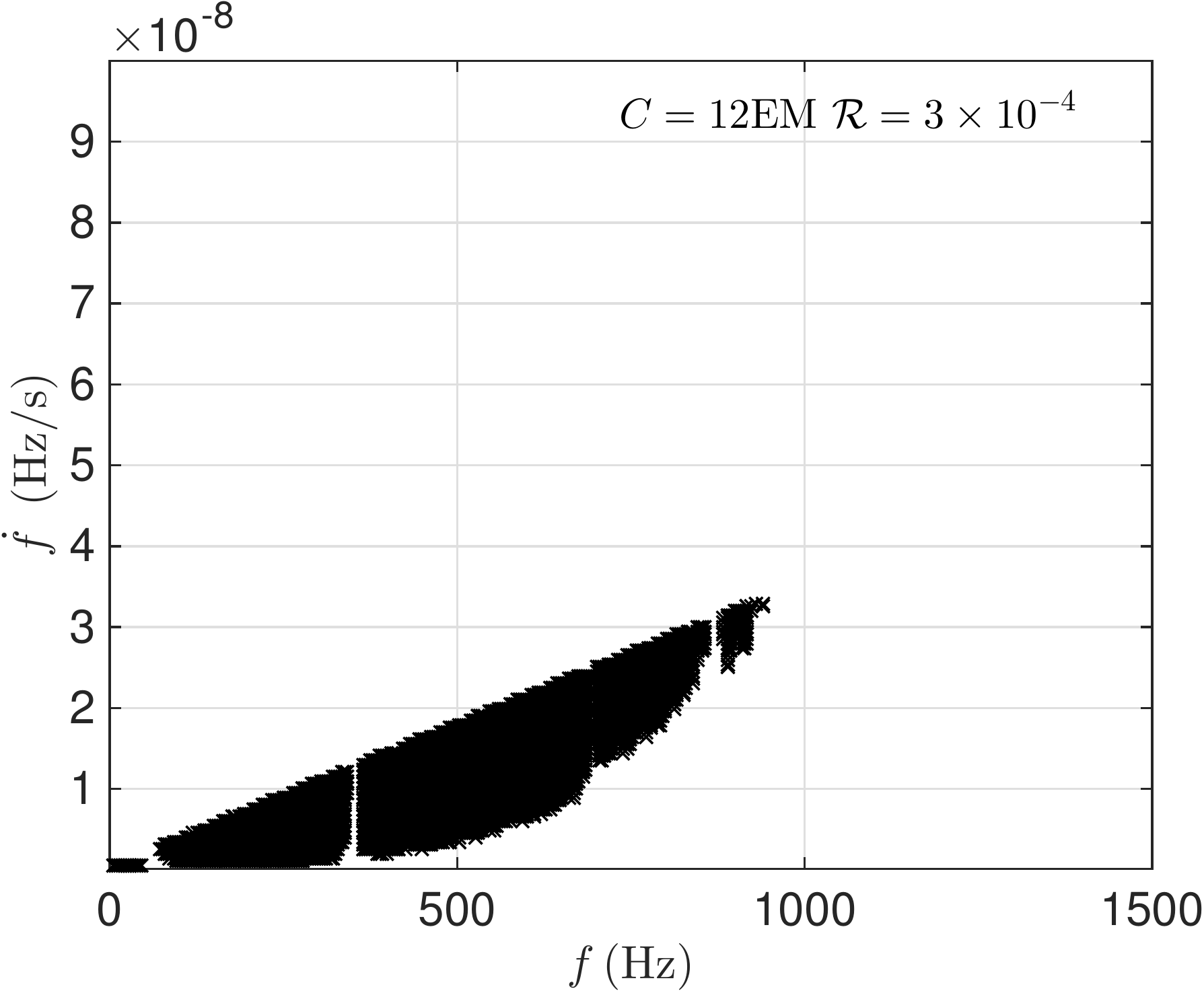}}}%
    \qquad
    \subfloat[Efficiency(lg), 30 days]{{  \includegraphics[width=.20\linewidth]{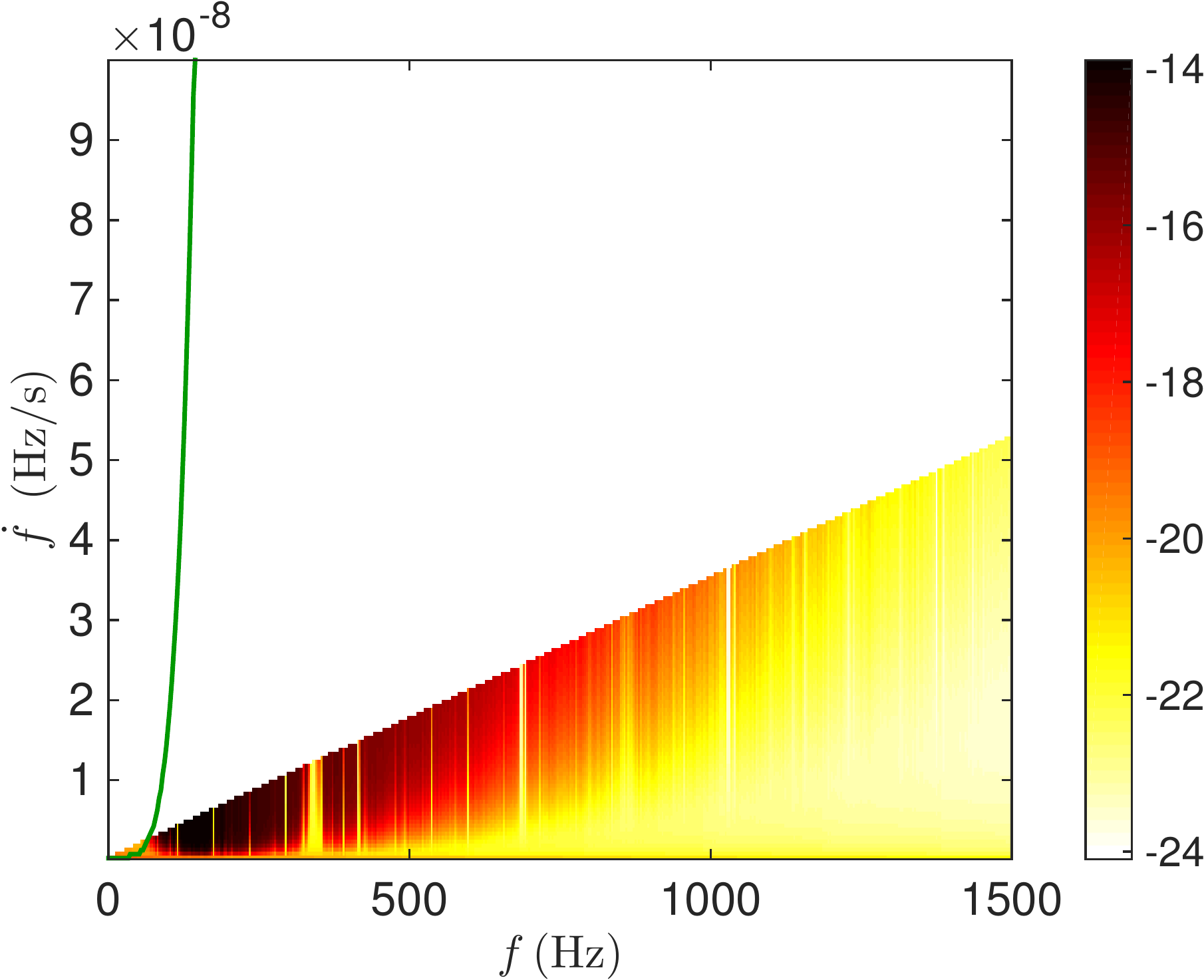}}}%
    \qquad
    \subfloat[Coverage, 30 days]{{  \includegraphics[width=.20\linewidth]{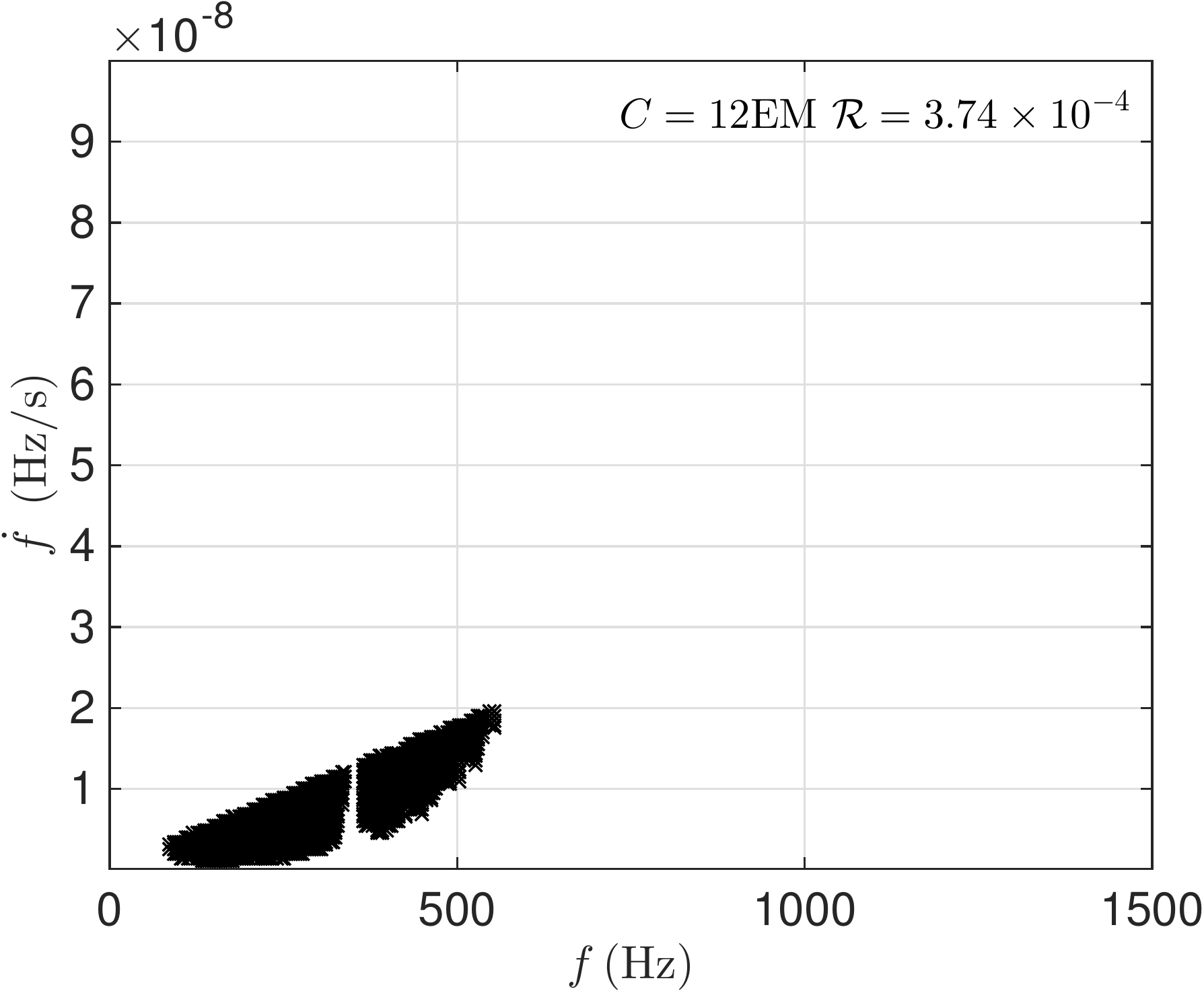}}}%
    \qquad
    \subfloat[Efficiency(lg), 37.5 days]{{  \includegraphics[width=.20\linewidth]{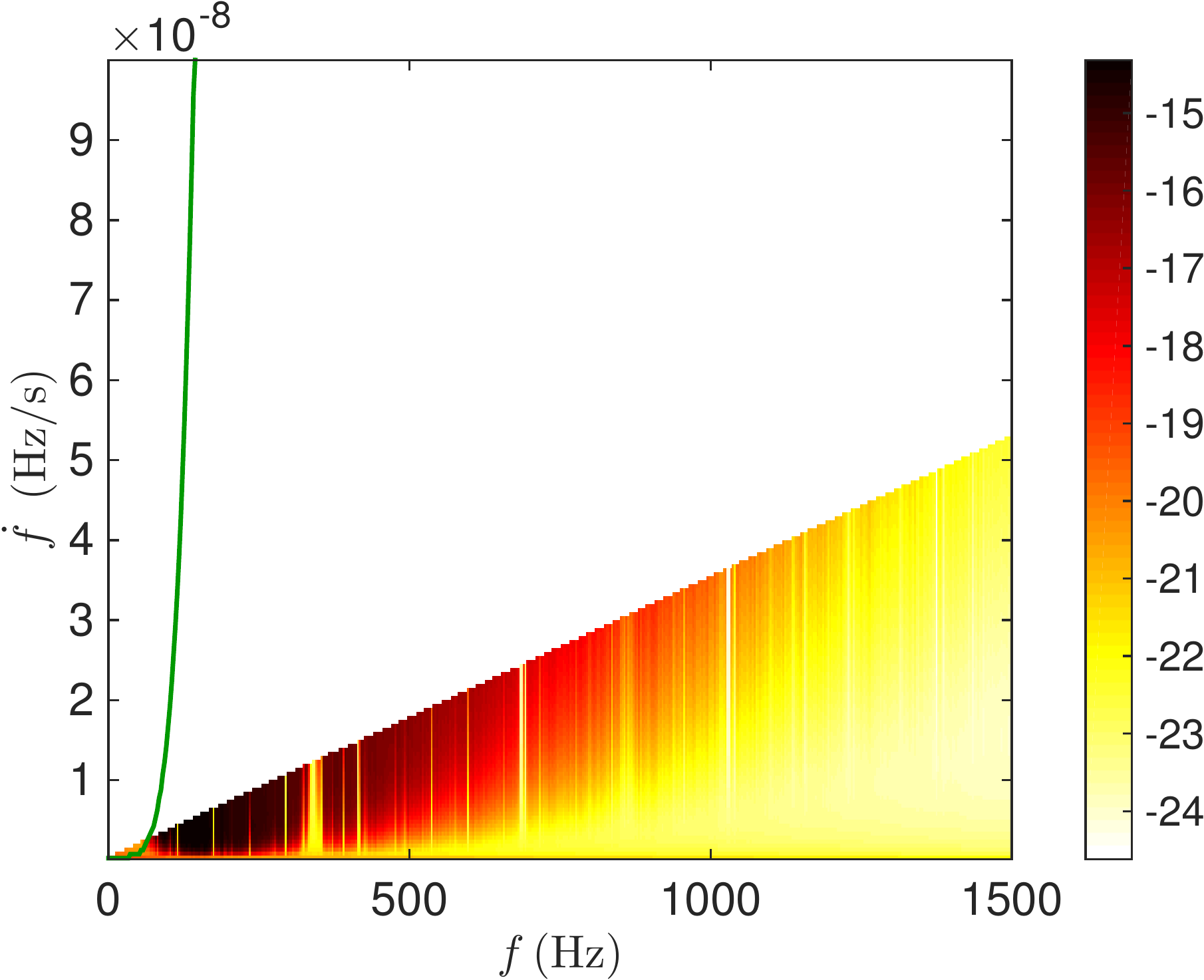}}}%
    \qquad
    \subfloat[Coverage, 37.5 days]{{  \includegraphics[width=.20\linewidth]{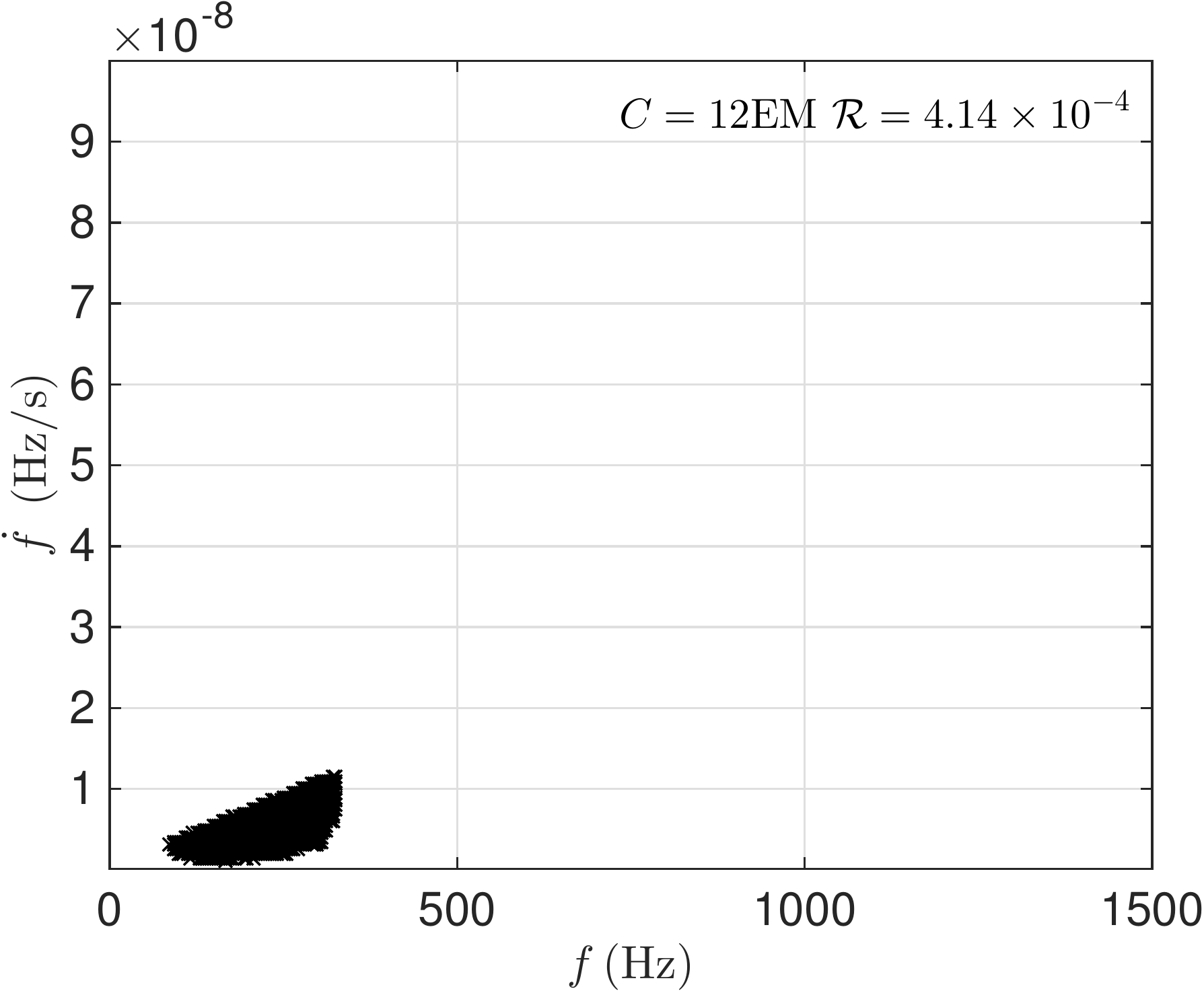}}}%
    \qquad
    \subfloat[Efficiency(lg), 50 days]{{  \includegraphics[width=.20\linewidth]{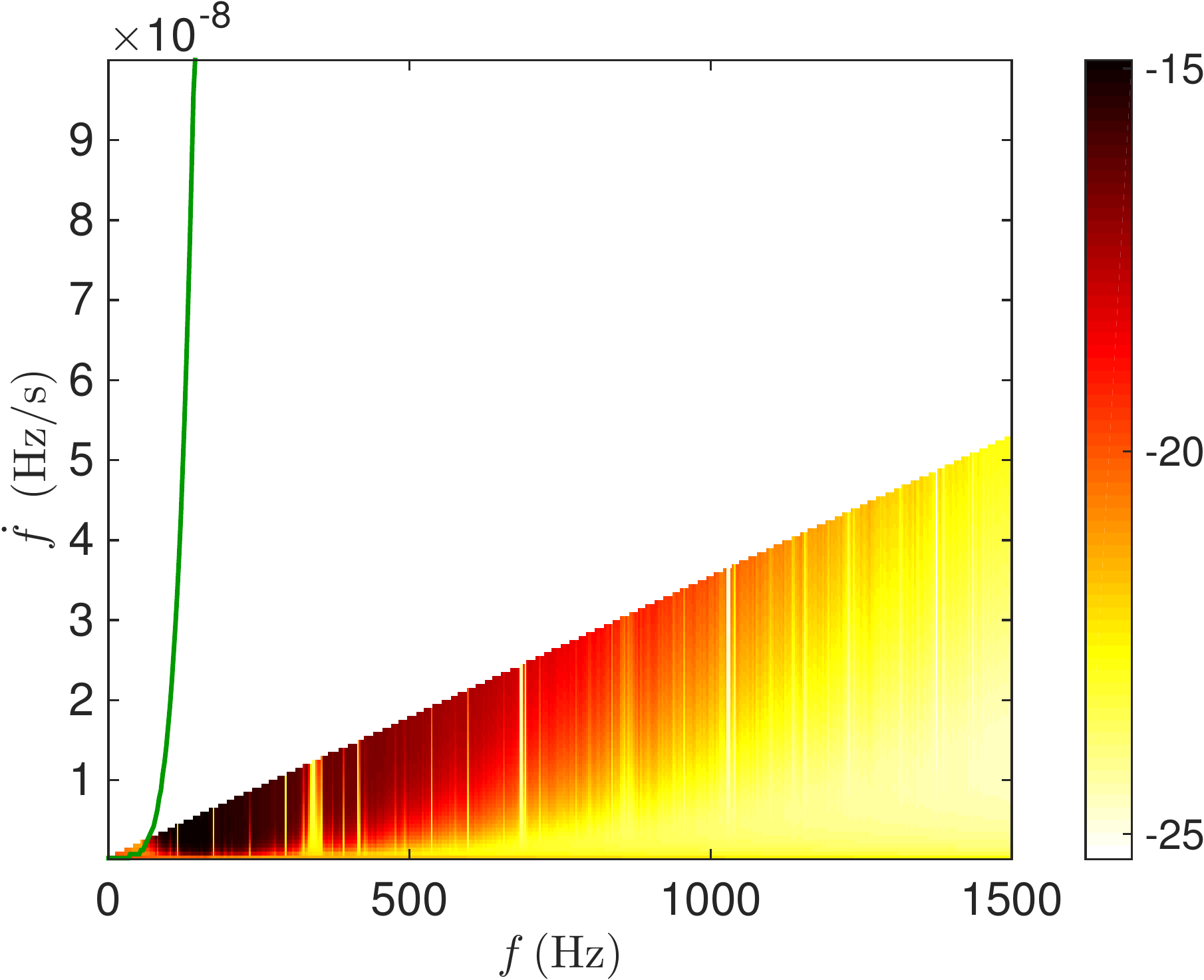}}}%
    \qquad
    \subfloat[Coverage, 50 days]{{  \includegraphics[width=.20\linewidth]{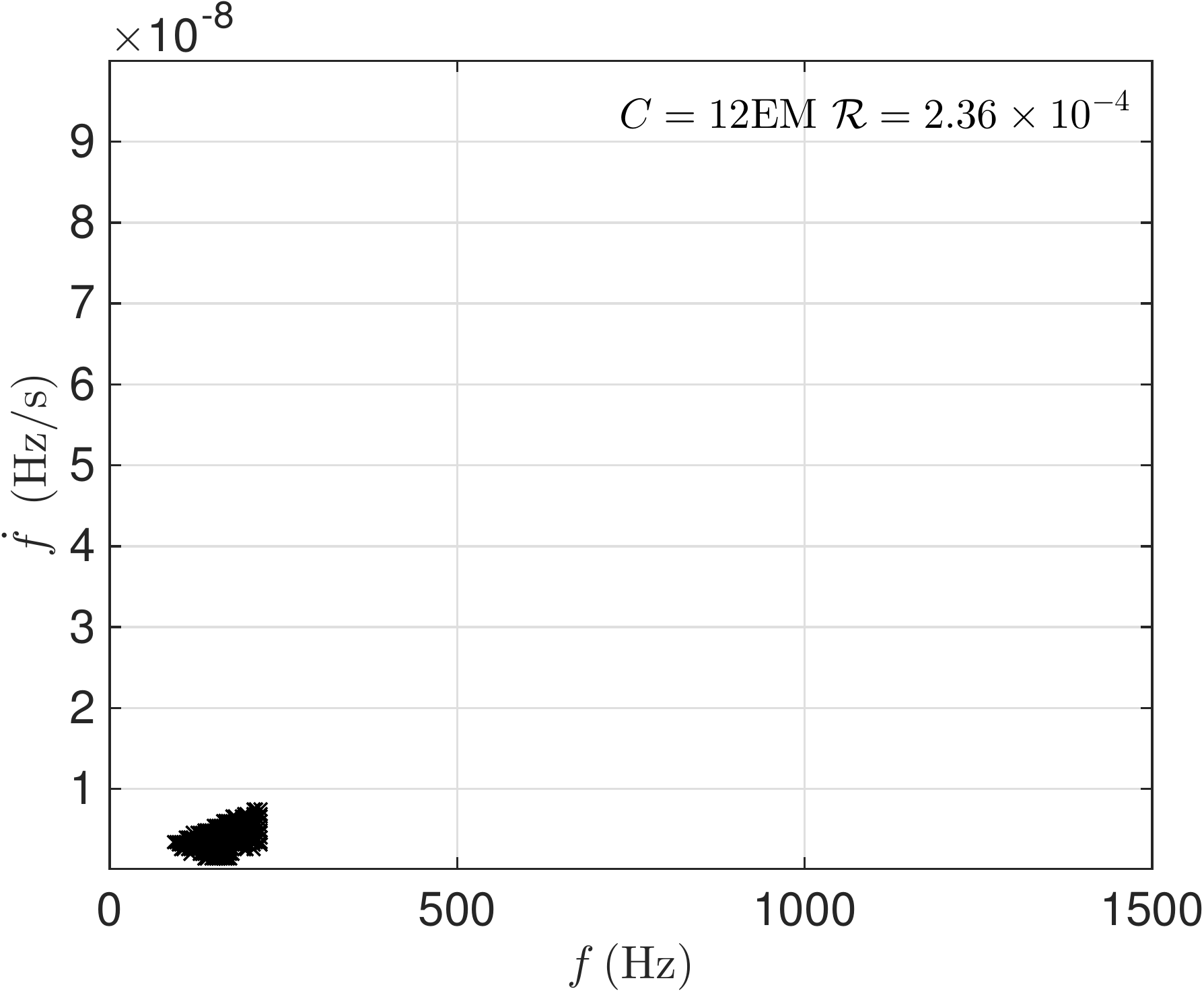}}}%
    \qquad
    \subfloat[Efficiency(lg), 75 days]{{  \includegraphics[width=.20\linewidth]{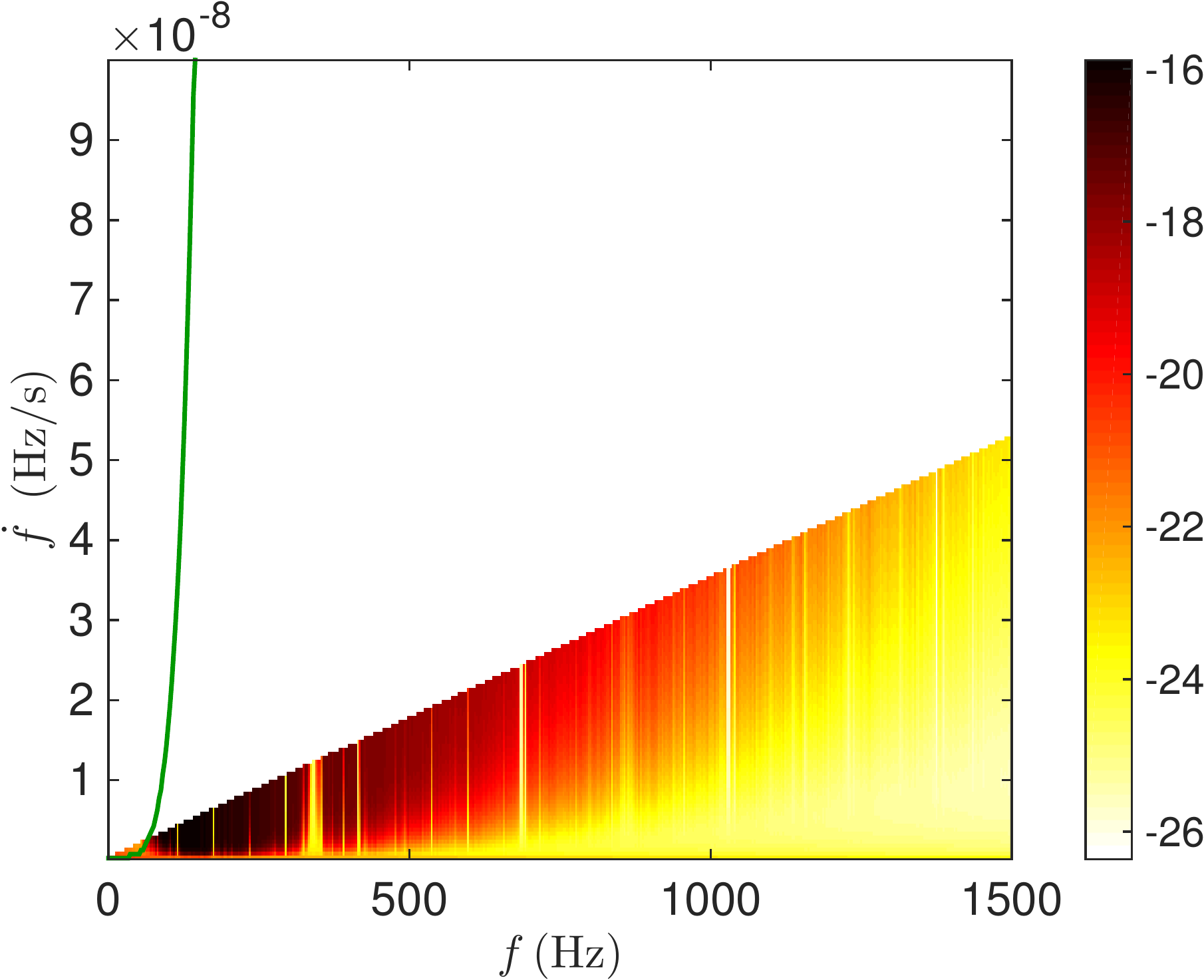}}}%
    \qquad
    \subfloat[Coverage, 75 days]{{  \includegraphics[width=.20\linewidth]{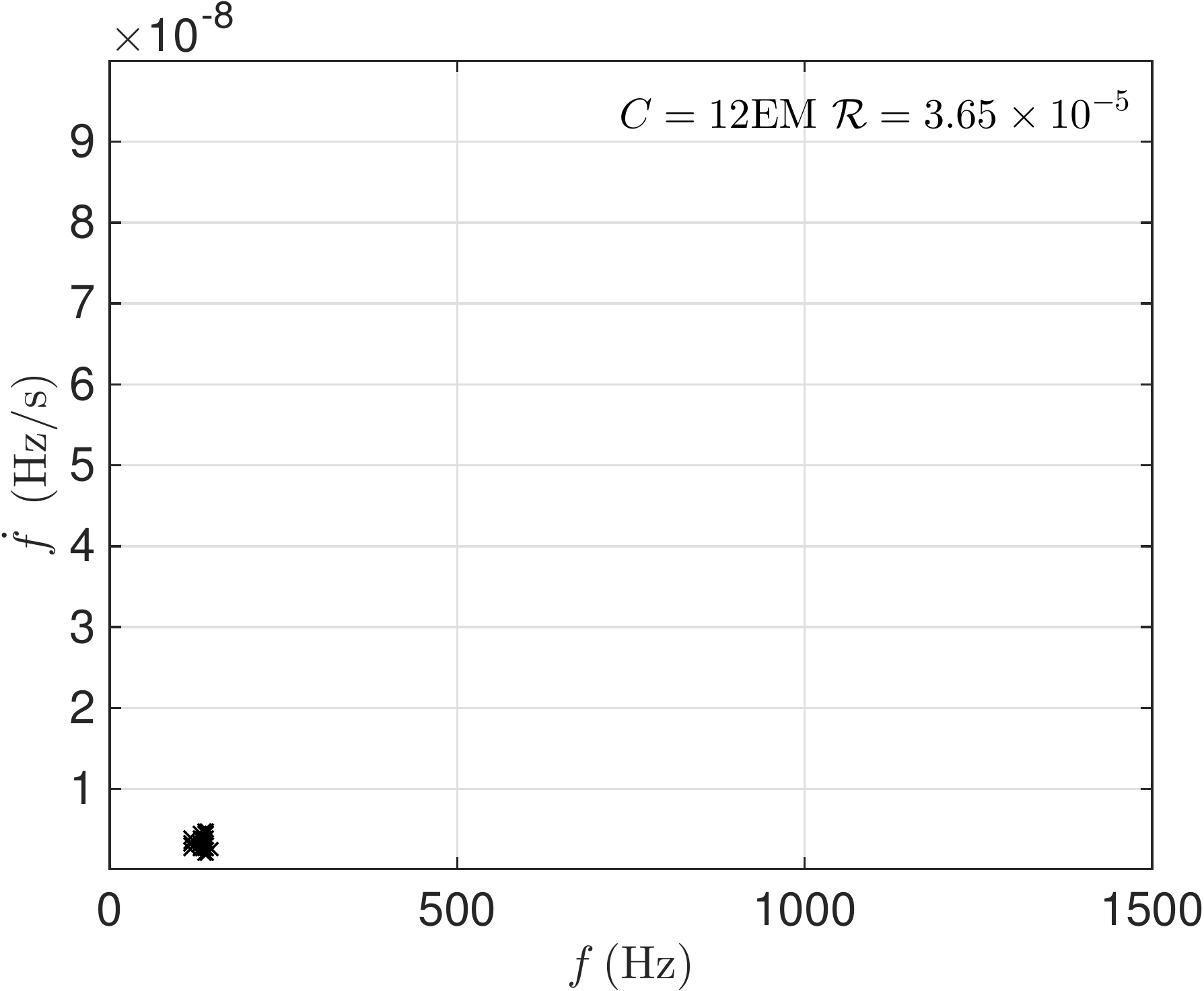}}}%
    
    \caption{Optimisation results for G350.1 at 4500 pc, 900 years old, assuming log-uniform and age-based priors, for various coherent search durations: 5, 10, 20, 30, 37.5, 50 and 75 days. The total computing budget is assumed to be 12 EM.}%
    \label{G3501_51020days_log}%
\end{figure*}

\begin{figure*}%
     \centering
     \subfloat[Coverage, cost: 12 EM]{{  \includegraphics[width=.45\linewidth]{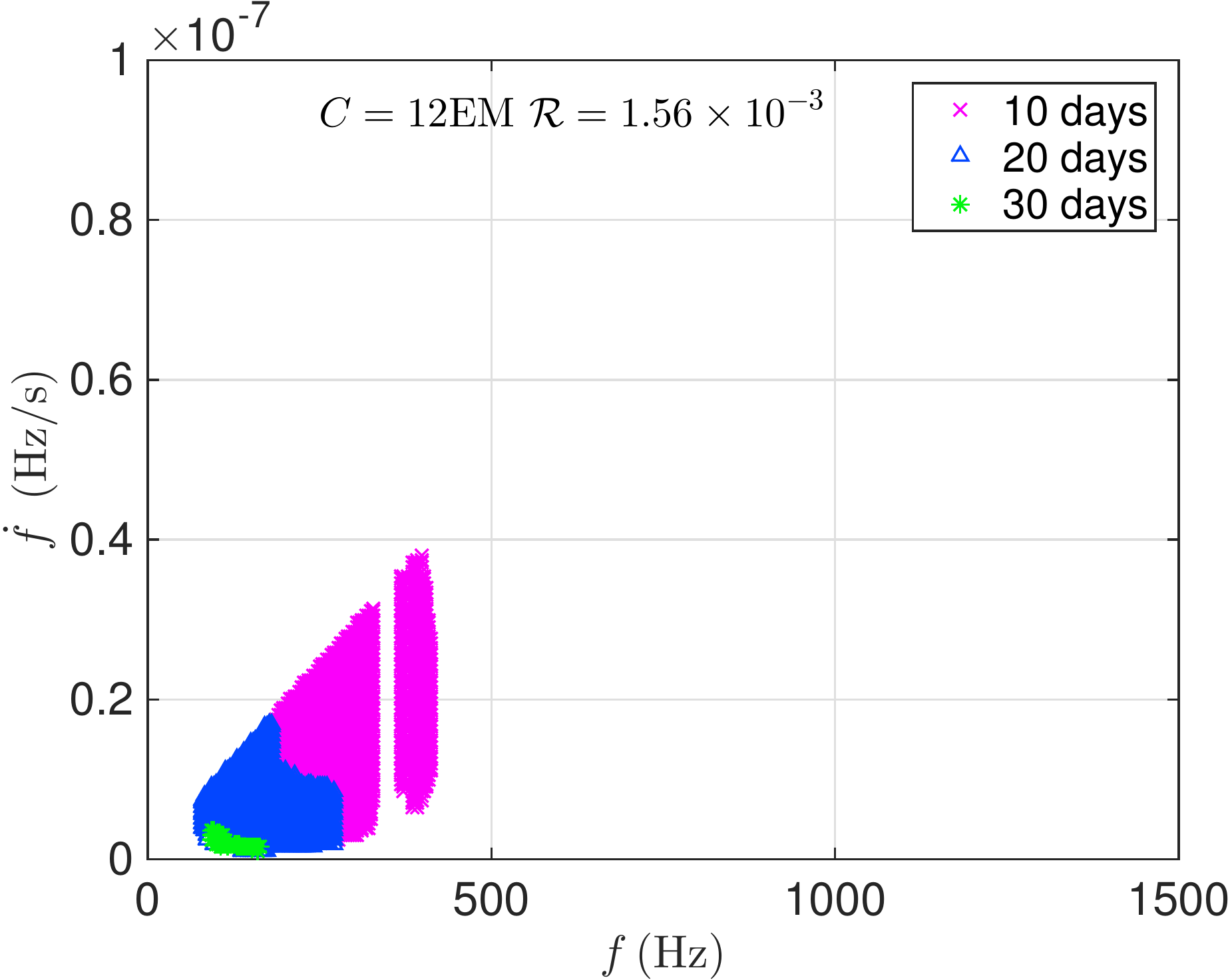}}}%
     \qquad
     \subfloat[Coverage,  cost: 24 EM]{{  \includegraphics[width=.45\linewidth]{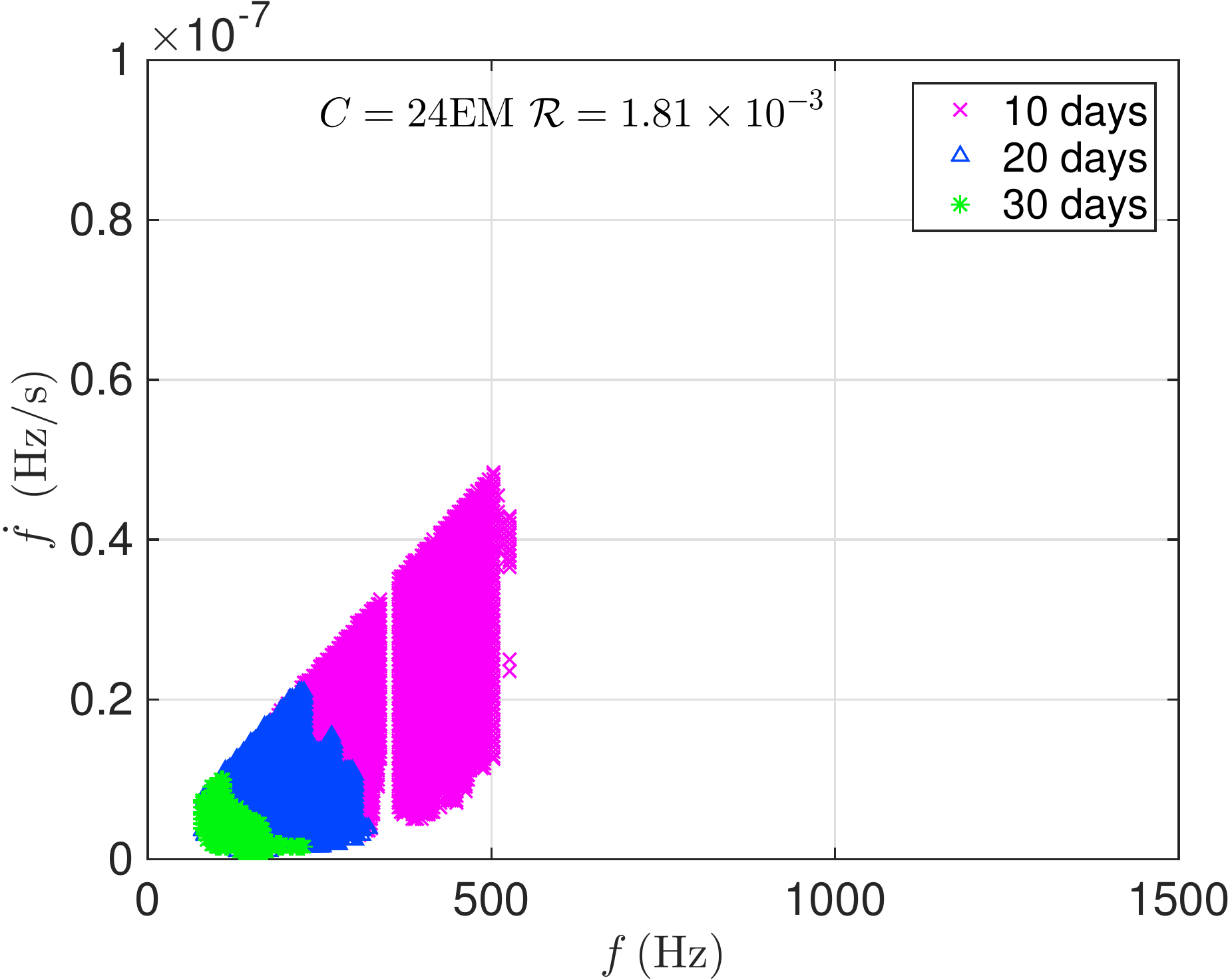}}}%
     \caption{Parameter space coverage for Cas A at 3500 pc, 330 years old, assuming log-uniform and age-based priors and optimizing over the 7  search set-ups also considered above at 12 EM (left plot) and 24 EM (right plot). }
     \label{CasA_best_age_log}%
 \end{figure*}

 \begin{figure*}%
     \centering
     \subfloat[Coverage, cost: 12 EM]{{  \includegraphics[width=.45\linewidth]{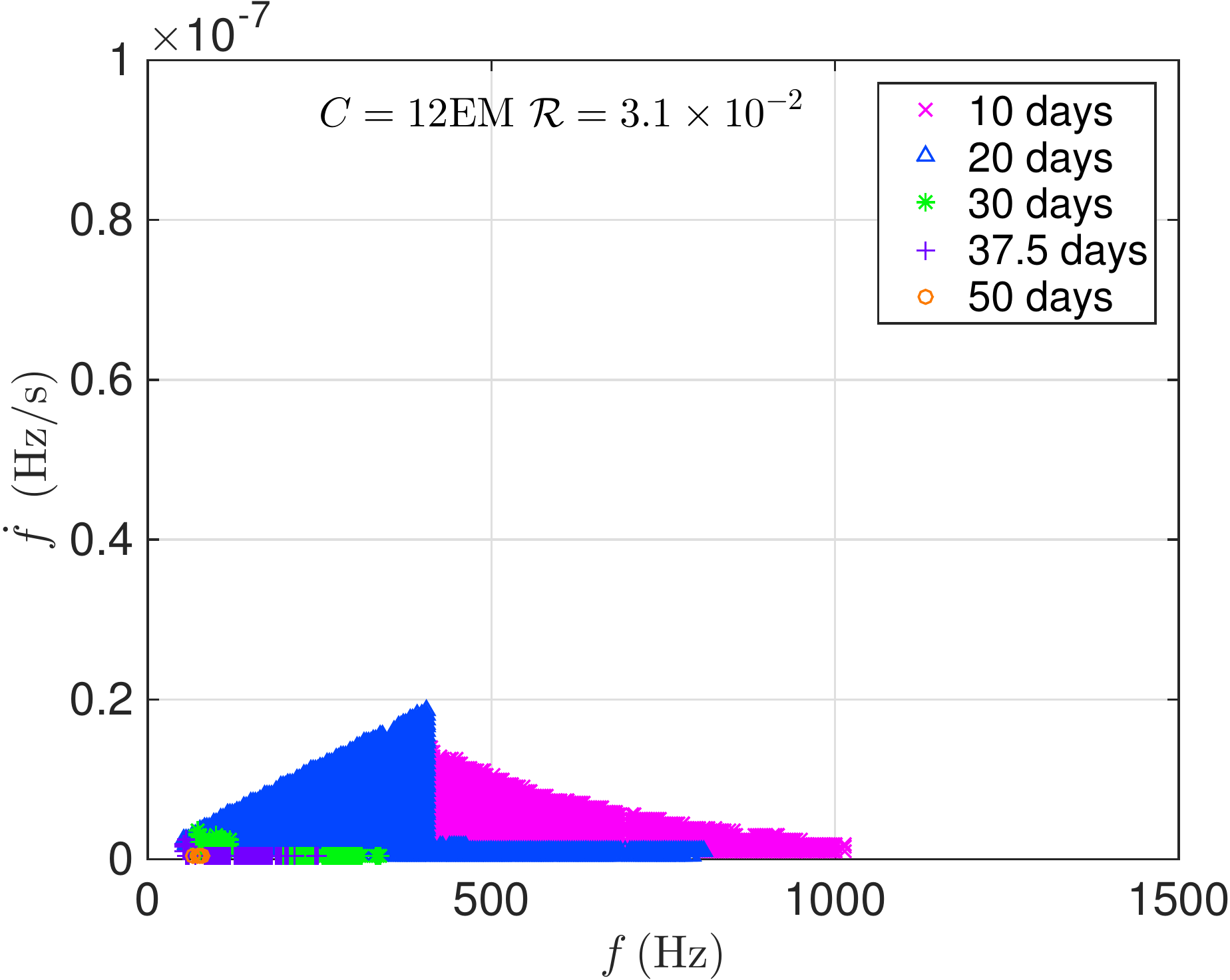}}}%
     \qquad
     \subfloat[Coverage,  cost: 24 EM]{{  \includegraphics[width=.45\linewidth]{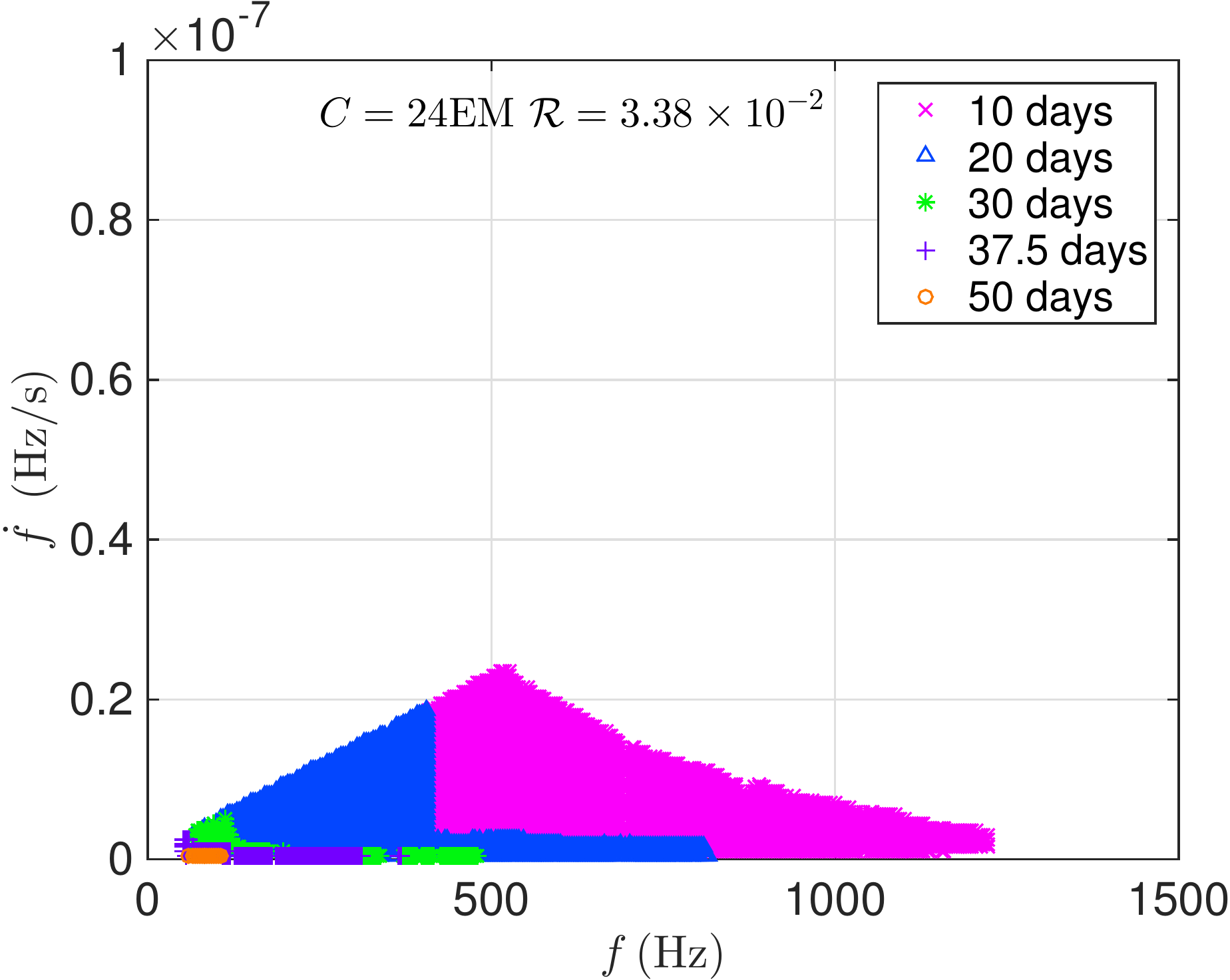}}}%
    \caption{Parameter space coverage for Vela Jr at 200 pc, 700 years old, assuming log-uniform and age-based priors and optimizing over the 7  search set-ups also considered above at 12 EM (left plot) and 24 EM (right plot).}%
     \label{2662_best_age_log}%
 \end{figure*}

 \begin{figure*}%
     \centering
     \subfloat[Coverage, cost: 12 EM]{{  \includegraphics[width=.45\linewidth]{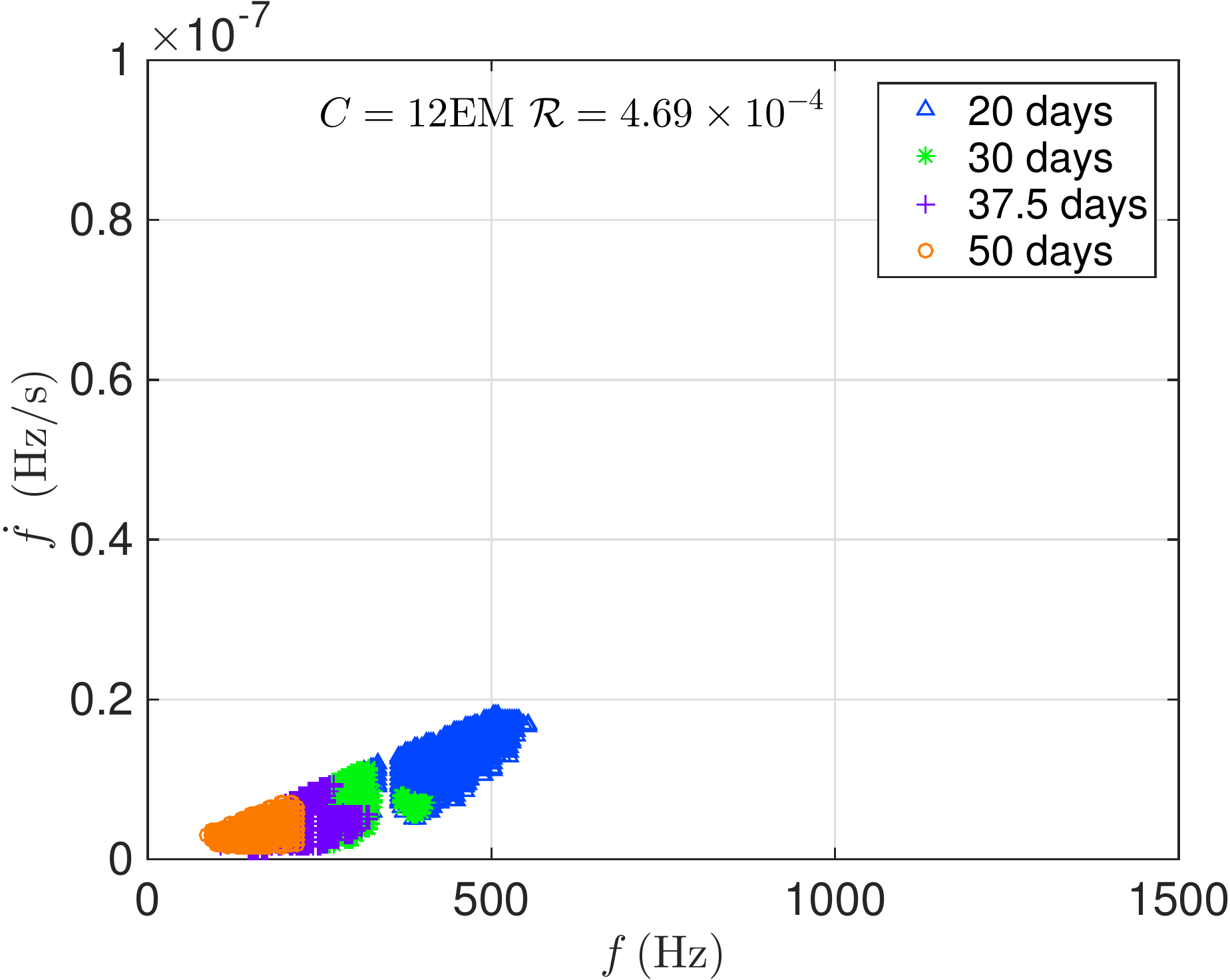}}}%
     \qquad
     \subfloat[Coverage,  cost: 24 EM]{{  \includegraphics[width=.45\linewidth]{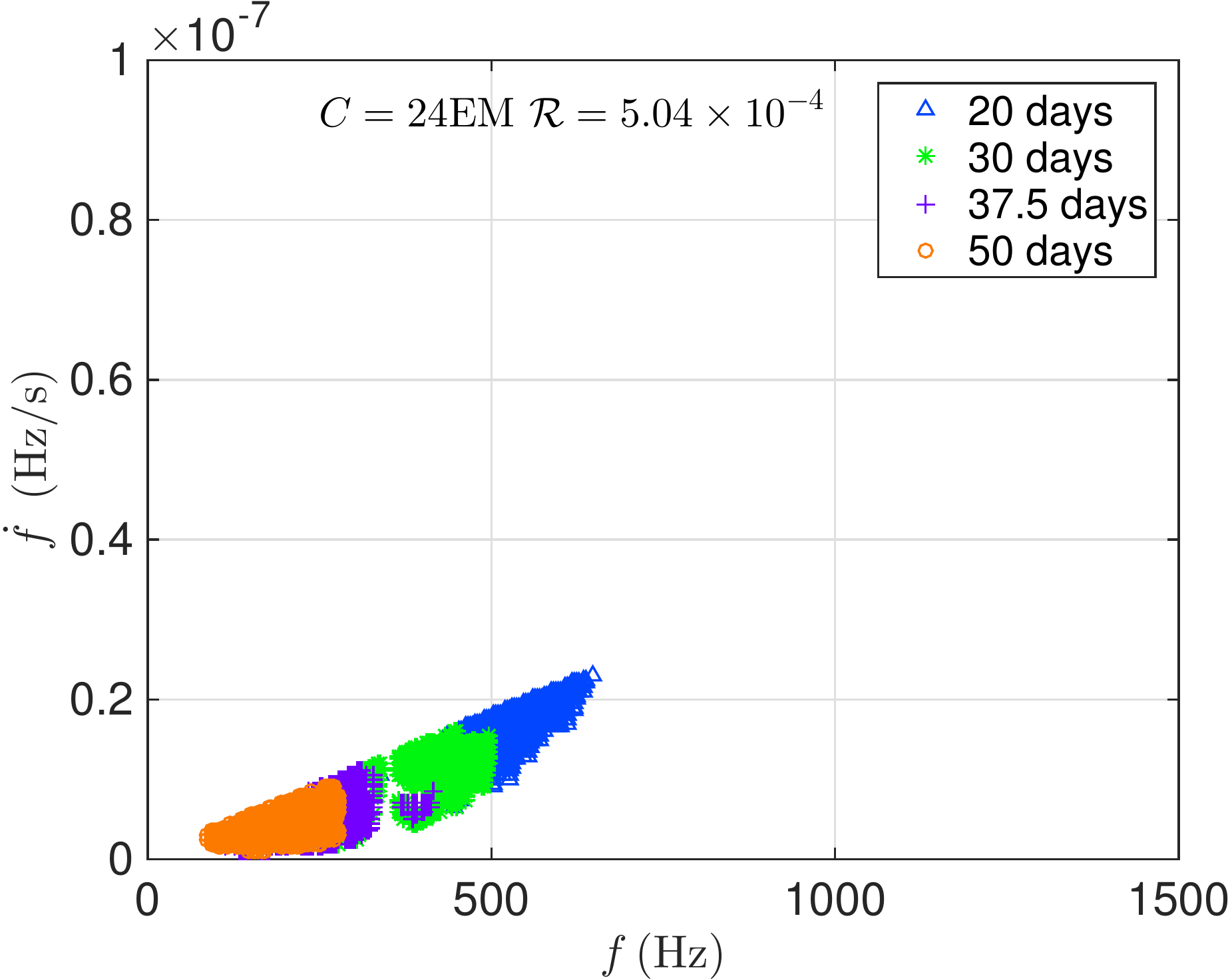}}}%
     \caption{Parameter space coverage for G350.1 at 4500 pc, 900 years old, log-assuming uniform and age-based priors and optimizing over the 7  search set-ups also considered above at 12 EM (left plot) and 24 EM (right plot).}%
     \label{3501_best_age_log}%
 \end{figure*}

\end{document}